\RequirePackage[dvipsnames,prologue]{xcolor}

\newif\ifarxiv
\newif\ifnotes

\arxivtrue
\notestrue

\ifarxiv
\documentclass[acmsmall,nonacm]{acmart}
\else
\documentclass[acmsmall,review,anonymous]{acmart}
\fi

\acmConference[PL'18]{ACM SIGPLAN Conference on Programming Languages}{January 01--03, 2018}{New York, NY, USA}
\acmYear{2018}
\acmISBN{} 
\acmDOI{} 
\startPage{1}
\setcopyright{none}
\usepackage{microtype}                

\usepackage{float}                    
\usepackage{placeins}                 

\usepackage{hyperref}                 
\hypersetup{
    colorlinks,
    linkcolor={red!50!black},
    citecolor={blue!50!black},
    urlcolor={blue!80!black}
}

\usepackage{amsmath}                  
\usepackage{amsfonts}                 
\usepackage{amssymb}                  
\usepackage{amsthm}                   
\usepackage{mathtools}                

\usepackage[dvipsnames]{xcolor}       

\usepackage{natbib}                   

\usepackage{stmaryrd}                 
\usepackage{pifont}                   
\usepackage{textcomp}                 

\usepackage{centernot}                

\usepackage{stackengine}              
\stackMath

\usepackage{nicefrac}                 

\usepackage{array}                    
\usepackage{tabularx}                 

\usepackage{framed}                   

\usepackage{lipsum}                   

\usepackage{enumitem}                 

\usepackage{mathpartir}               

\usepackage{geometry}                 

\newcommand{\colorTEXT}{black}
\newcommand{\colorMATH}{black!20!blue}
\newcommand{\colorSYNTAX}{black!55!green}

\newcommand{\slashedrel}[1]{\mathrel{\centernot{#1}}}

\newcommand{\note}[1]{{\color{red}#1}}

\usepackage{comment}
\usepackage{xspace}
\usepackage{pgfplots}
\usepackage{listings}
\usepackage[normalem]{ulem}
\usepackage{subcaption}
\usepackage{wrapfig}

\ifnotes
\else
\renewcommand{\note}[1]{}
\fi

\definecolor{cbsafeABright}{RGB}{8,72,145}
\definecolor{cbsafeADark}{RGB}{109,36,150}

\definecolor{cbsafeBBright}{RGB}{85,119,13}
\definecolor{cbsafeBDark}{RGB}{109,70,17}

\definecolor{cbsafeCBright}{RGB}{150,48,89}
\definecolor{cbsafeCDark}{RGB}{70,24,48}

\newcommand{\colorMATHM}{cbsafeABright}
\newcommand{\colorSYNTAXM}{cbsafeADark!80!black}
\newcommand\colorMATHS{cbsafeBBright}
\newcommand\colorSYNTAXS{cbsafeBDark}
\newcommand\colorMATHP{cbsafeCBright}
\newcommand\colorSYNTAXP{cbsafeCDark}

\newcommand{\dfuzz}{\emph{DFuzz}\xspace}
\newcommand{\fuzz}{\emph{Fuzz}\xspace}
\newcommand{\system}{\textsc{Duet}\xspace}
\newcommand{\hoaresq}{\emph{HOARe$^2$}\xspace}

\renewcommand{\colorMATH}{\colorMATHM}
\renewcommand{\colorSYNTAX}{\colorSYNTAXM}

\newcommand{\overbracketarg}[2]{\overbracket{#2}^{#1}}

\newcommand{\rmspc}{\vspace*{-2mm}}

\renewcommand\MathparLineskip{}

\newcommand\smallerplz\small

\author{Joseph P. Near}
\affiliation{University of Vermont}
\email{jnear@uvm.edu}

\author{David Darais}
\affiliation{University of Vermont}
\email{David.Darais@uvm.edu}

\author{Chike Abuah}
\affiliation{University of Vermont}
\email{cabuah@uvm.edu}

\author{Tim Stevens}
\affiliation{University of Vermont}
\email{Timothy.Stevens@uvm.edu}

\author{Pranav Gaddamadugu}
\affiliation{University of California, Berkeley}
\email{pranavsaig@berkeley.edu}

\author{Lun Wang}
\affiliation{University of California, Berkeley}
\email{wanglun@berkeley.edu}

\author{Neel Somani}
\affiliation{University of California, Berkeley}
\email{neel@berkeley.edu}

\author{Mu Zhang}
\affiliation{University of Utah}
\email{muzhang@cs.utah.edu}

\author{Nikhil Sharma}
\affiliation{University of California, Berkeley}
\email{ennsharma@berkeley.edu}

\author{Alex Shan}
\affiliation{University of California, Berkeley}
\email{alexshan@berkeley.edu}

\author{Dawn Song}
\affiliation{University of California, Berkeley}
\email{dawnsong@cs.berkeley.edu}

\makeatletter
\let\@authorsaddresses\@empty
\makeatother

\title[\system: An Expressive Higher-order Language and Linear Type System for
Statically Enforcing Differential Privacy]{\system: An Expressive Higher-order
Language and Linear Type System for Statically Enforcing Differential Privacy}

\begin{document}

\begin{abstract}
  During the past decade, differential privacy has become the gold standard for
  protecting the privacy of individuals. However, verifying that a particular
  program provides differential privacy often remains a manual task to be
  completed by an expert in the field. Language-based techniques have been
  proposed for fully automating proofs of differential privacy via type system
  design, however these results have lagged behind advances in
  differentially-private algorithms, leaving a noticeable gap in programs which
  can be automatically verified while also providing state-of-the-art bounds on
  privacy.
  

  We propose \system, an expressive higher-order language, linear type system
  and tool for automatically verifying differential privacy of general-purpose
  higher-order programs. In addition to general purpose programming, \system
  supports encoding machine learning algorithms such as stochastic gradient
  descent, as well as common auxiliary data analysis tasks such as
  clipping, normalization and hyperparameter tuning--each of which are
  particularly challenging to encode in a statically verified differential
  privacy framework.
  
  We present a core design of the \system language and linear type system, and
  complete key proofs about privacy for well-typed programs. We then show how
  to extend \system to support realistic machine learning applications and
  recent variants of differential privacy which result in improved accuracy for
  many practical differentially private algorithms. Finally, we implement
  several differentially private machine learning algorithms in \system which
  have never before been automatically verified by a language-based tool, and
  we present experimental results which demonstrate the benefits of \system's
  language design in terms of accuracy of trained machine learning models.
\end{abstract}

\maketitle

\definecolor{col1}{RGB}{91, 192, 235}
\definecolor{col2}{RGB}{249, 89, 52}
\definecolor{col3}{RGB}{155, 197, 61}
\definecolor{col4}{RGB}{253, 231, 76}
\definecolor{col5}{RGB}{250, 191, 73}
\definecolor{col6}{RGB}{100, 100, 100}

\section{Introduction}

Advances in big data and machine learning have achieved large-scale societal
impact over the past decade. This impact is accompanied by a growing demand for
data collection, aggregation and analysis at scale. This resulting explosion in
the amount of data collected by organizations, however, has raised important
new security and privacy concerns.

Differential privacy~\cite{dworkdifferential2006, dwork2006calibrating,
privacybook} is a promising technique for addressing these issues. Differential
privacy allows general statistical analysis of data while protecting \emph{data
about individuals} with a strong formal guarantee of privacy.
Because of its desirable formal guarantees, differential privacy has received
increased attention, with ongoing real-world deployments at organizations
including Google~\cite{erlingsson2014rappor}, Apple~\cite{apple}, and the US
Census~\cite{machanavajjhala2008privacy,haney2017utility}. A number of systems
for performing differentially private data analytics have been built and
demonstrated to be effective~\cite{mcsherry2009privacy,
proserpio2014calibrating, johnson2018towards, es_paper_arxiv, roy2010airavat, narayan2012djoin,
mohan2012gupt}.

Differential privacy plays an increasingly important role in machine
learning, as recent work has shown that a trained model can leak information
about data it was trained on~\cite{mod-inv1, mod-inv2,mem-inf1}.
Differential privacy provides a robust solution to this problem, and as a
result, a number of differentially private algorithms have been developed for
machine learning~\cite{cms11, SCS, bst14, ttz16, PSGD,
friedman2016differential, DPDL, papernot2016semi}.

Few practical approaches exist, however, for automatically proving that a
\emph{general-purpose program} satisfies differential privacy---an increasingly
desirable goal, since many machine learning pipelines are expressed as programs
that combine existing algorithms with custom code.
Enforcing differential privacy for a new program currently requires a new,
manually-written privacy proof.
This process is arduous, error-prone, and must be performed by an expert in differential privacy (and re-performed, each time the program is modified).




We present \system, a programming language, type system and tool for
expressing and statically verifying privacy-preserving
programs. \system supports (1) general purpose programming features
like compound datatypes and higher-order functions, (2) library
functions for matrix-based computations, 
and (3) multiple state-of-the-art
variants of differential privacy--{{\color{\colorMATH}\ensuremath{(\epsilon ,\delta )}}}-differential
privacy~\cite{privacybook}, R\'enyi differential
privacy~\cite{mironov2017renyi}, zero-concentrated differential
privacy (zCDP)~\cite{bun2016concentrated}, and truncated-concentrated
differential privacy (tCDP)~\cite{bun2018composable}--and can be easily
extended to new ones. \system strikes a strategic balance between
generality, practicality, extensibility, and precision of computed
privacy bounds.


The design of \system consists of \emph{two} separate, mutually embedded
languages, each with its own type system. The \emph{sensitivity language} uses
linear types {{\color{\colorTEXT}\textit{with}}} metric scaling (as in \fuzz~\cite{reed2010distance}) to
bound function sensitivity. The \emph{privacy language} uses linear types
{{\color{\colorTEXT}\textit{without}}} metric scaling (novel in \system) to compose differentially private
computations. Disallowing the use of scaling in the privacy language is essential
to encode more advanced variants of differential privacy (like {{\color{\colorMATH}\ensuremath{(\epsilon ,\delta )}}}) in a
linear type system.



Linear typing~\cite{di-linearlogic, Girard:1987:LL:35356.35357} is a
good fit for both privacy and sensitivity analysis, because resources
are tracked per-variable and combined additively.
In particular, our linear typing approach to {{\color{\colorTEXT}\textit{privacy}}} allows for
independent privacy costs for multiple function arguments, a feature
shared by Fuzz and DFuzz (which only support pure
{{\color{\colorMATH}\ensuremath{\epsilon }}}-differential privacy), but not supported by prior type systems for
{{\color{\colorMATH}\ensuremath{(\epsilon ,\delta )}}}-differential privacy. This limitation of prior work is due to
the treatment of privacy as a computational ``effect''--a property
of the output, embodied in an indexed monad--as opposed to our
treatment of privacy as a ``co-effect''--a property of the context,
embodied in linear typing.

Our main idea is to co-design {{\color{\colorTEXT}\textit{two}}} separate languages for privacy and
sensitivity, and our main insight is that a linear type system can (1) model
more powerful variants of differential privacy (like {{\color{\colorMATH}\ensuremath{(\epsilon ,\delta )}}}) when strengthened
to disallow scaling, and (2) interact seamlessly with a sensitivity-type
system which does allow scaling. Each language embeds inside the other, and the
privacy mechanisms of the underlying privacy definition ({e.g.} the Gaussian
mechanism~\cite{privacybook}) form the interface between the two languages.
Both languages use similar syntax and identical types. The two
languages aid type checking, the proof of type soundness, and our
implementation of type inference; programmers need not be intimately aware of
the multi-language design.

In addition to basic differential-privacy primitives like the Gaussian
mechanism, we provide a core language design for matrix-based data analysis
tasks, such as aggregation, clipping and gradients. Key challenges that we
overcome in our design are how each of these features compose in terms of
function sensitivity, and how to statically track bounds on vector norms (due
to clipping, for the purposes of privacy)---and each in a way that is general
enough to support a wide range of useful applications.





We demonstrate the usefulness of \system by implementing and verifying several
differentially private machine learning algorithms from the literature,
including private stochastic gradient descent~\cite{bst14} and private
Frank-Wolfe~\cite{ttz16}, among many others. We also implement a variant of
stochastic gradient descent suitable for deep learning. For each of these
algorithms, no prior work has demonstrated an automatic verification of
differential privacy, and \system is able to automatically infer privacy bounds
that equal \emph{and in some cases improve upon} previously published manual
privacy proofs.

We have implemented a typechecker and interpreter for \system, and we use these
to perform an empirical evaluation comparing the accuracy of models trained
using our implementations. Although the ``punchline'' of the empirical results
are unsurprising due to known advantages of the differential
privacy definitions used ({e.g.}, that using recent variants like zero-concentrated differential
privacy results in improved accuracy), our results show the
extent of the accuracy improvements for specific algorithms and further reinforce the idea that choosing the best definition consistently results in
substantially better accuracy of the trained model.

\paragraph{Contributions.}
In summary, we make the following contributions:

\begin{itemize}[topsep=0.5mm,leftmargin=4mm]
\itemsep1.0mm
\item We present \system, a novel language, linear type system and tool for
  expressing and automatically verifying differentially private programs.
  Unlike previous work, \system supports a combination of (1) general purpose,
  higher order programming, (2) advanced definitions of differential privacy,
  (3) independent tracking of privacy costs for multiple function arguments,
  and (4) auxiliary differentially-private data analysis tasks such as
  clipping, normalization, and hyperparameter tuning.

\item We formalize \system's type system and semantics, and complete key proofs
  about privacy of well-typed programs.

\item We demonstrate a battery of case studies consisting of medium-sized,
  real-world, differentially private machine learning algorithms which are
  successfully verified with optimal (or near-optimal) privacy bounds. In
  some cases, \system infers privacy bounds which improve on the best
  previously published manually-verified result.
  
\item We conduct an experimental evaluation to demonstrate \system's
  feasibility in practice by training two machine learning algorithms on
  several non-toy real-world datasets using \system's interpreter. These
  results demonstrate the effect of improved privacy bounds on the
  accuracy of the trained models.
\end{itemize}


\section{Preliminaries}

\subsection{Background: Differential Privacy}

This section briefly summarizes the basics of differential privacy. See
Dwork and Roth's reference~\cite{privacybook} for a detailed
description.
Differential privacy considers sensitive input data represented by a vector {{\color{\colorMATH}\ensuremath{x
\in  D^{n}}}}, in which {{\color{\colorMATH}\ensuremath{x_{i}}}} represents the data contributed by user {{\color{\colorMATH}\ensuremath{i}}}. The
\emph{distance} between two inputs {{\color{\colorMATH}\ensuremath{x, y \in  D^{n}}}} is {{\color{\colorMATH}\ensuremath{d(x,y) = |\{ i | x_{i} \neq  y_{i}\} |}}}. Two inputs {{\color{\colorMATH}\ensuremath{x,y}}} are
\emph{neighbors} if {{\color{\colorMATH}\ensuremath{d(x,y) = 1}}}, {i.e.}, if they differ in only one index. A randomized mechanism {{\color{\colorMATH}\ensuremath{{\mathcal{K}} : D^{n} \rightarrow  {\mathbb{R}}^{d}}}}
preserves {{\color{\colorMATH}\ensuremath{(\epsilon ,\delta )}}}-differential privacy if for any neighbors {{\color{\colorMATH}\ensuremath{x, y \in  D^{n}}}} and any
set {{\color{\colorMATH}\ensuremath{ S }}} of possible outputs:
\vspace*{-0.25em}\begingroup\color{\colorMATH}\begin{gather*}{{\color{\colorMATH}\ensuremath{\operatorname{Pr}}}}[{\mathcal{K}}(x)\in  S] \leq  e^{\epsilon } {{\color{\colorMATH}\ensuremath{\operatorname{Pr}}}}[{\mathcal{K}}(y)\in  S] + \delta  \vspace*{-1em}\end{gather*}\endgroup 
The main idea is that when {{\color{\colorMATH}\ensuremath{\epsilon }}} and {{\color{\colorMATH}\ensuremath{\delta }}} are very small, then the resulting
output distributions will be very close, and therefore nearly
indistinguishible.

The {{\color{\colorMATH}\ensuremath{\epsilon }}} parameter, also called the \emph{privacy budget}, controls the strength
of the privacy guarantee. The {{\color{\colorMATH}\ensuremath{\delta }}} parameter allows for a non-zero
probability that the guarantee fails, and is typically set to a negligible
value. The case when {{\color{\colorMATH}\ensuremath{\delta  = 0}}} is called \emph{pure} or
{{\color{\colorMATH}\ensuremath{\epsilon }}}-differential privacy; the case when {{\color{\colorMATH}\ensuremath{\delta  > 0}}} is called
\emph{approximate} or {{\color{\colorMATH}\ensuremath{(\epsilon , \delta )}}}-differential privacy.
Typical values for {{\color{\colorMATH}\ensuremath{\epsilon }}} and {{\color{\colorMATH}\ensuremath{\delta }}} are {{\color{\colorMATH}\ensuremath{\epsilon  \in  [0.1-10]}}} and {{\color{\colorMATH}\ensuremath{\delta  = \frac{1}{n^{2}}}}}
where {{\color{\colorMATH}\ensuremath{n}}} is the number of input entries~\cite{privacybook}.

\paragraph{Function sensitivity.}
A function's \emph{sensitivity} is the amount its output can change
when its input changes. For example, the function {{\color{\colorMATH}\ensuremath{f(x) = x + x}}} has a
sensitivity of 2, since increasing or decreasing its input by 1 has
the effect of increasing or decreasing its output by 2. A real-valued
function {{\color{\colorMATH}\ensuremath{f}}} is called \emph{{{\color{\colorMATH}\ensuremath{n}}}-sensitive} if {{\color{\colorMATH}\ensuremath{\max_{x,y:|x-y|\leq 1}
|f(x)-f(y)| = n}}}.

This idea can be generalized to vector-valued functions. The
\emph{global {{\color{\colorMATH}\ensuremath{L1}}} sensitivity} of a query {{\color{\colorMATH}\ensuremath{f:D^{n} \rightarrow  {\mathbb{R}}^{d}}}} is written
{{\color{\colorMATH}\ensuremath{GS_{f}}}} and defined {{\color{\colorMATH}\ensuremath{GS_{f} = \max_{x,y:d(x,y)=1} |f(x)-f(y)|_{1}}}} where
{{\color{\colorMATH}\ensuremath{|\underline{\hspace{0.66em}\vspace*{5ex}}-\underline{\hspace{0.66em}\vspace*{5ex}}|_{1}}}} is the {{\color{\colorMATH}\ensuremath{L1}}} norm ({i.e.}, sum of pointwise distances between
elements). The {{\color{\colorMATH}\ensuremath{L2}}} sensitivity is analogous, using the {{\color{\colorMATH}\ensuremath{L2}}} norm.



\paragraph{Differential privacy mechanisms.}
Two basic differential privacy mechanisms are the \emph{Laplace
mechanism}~\cite{dwork2006calibrating}, which preserves {{\color{\colorMATH}\ensuremath{(\epsilon ,0)}}}-differential
privacy, and the \emph{Gaussian mechanism}~\cite{privacybook}, which preserves
{{\color{\colorMATH}\ensuremath{(\epsilon , \delta )}}}-differential privacy. For a function {{\color{\colorMATH}\ensuremath{f: D^{n} \rightarrow  {\mathbb{R}}^{d}}}} with {{\color{\colorMATH}\ensuremath{L1}}} sensitivity of {{\color{\colorMATH}\ensuremath{\Delta _{1}}}}, the Laplace mechanism adds
noise drawn from {{\color{\colorMATH}\ensuremath{{{\color{\colorMATH}\ensuremath{\operatorname{Lap}}}}(\frac{\Delta _{1}}{\epsilon })}}} to each element of the
output. For {{\color{\colorMATH}\ensuremath{f}}} with {{\color{\colorMATH}\ensuremath{L2}}} sensitivity of {{\color{\colorMATH}\ensuremath{\Delta _{2}}}}, the Gaussian mechanism
adds noise drawn from {{\color{\colorMATH}\ensuremath{{\mathcal{N}}(0, \frac{2\Delta _{2}^{2}\ln (1.25/\delta )}{\epsilon ^{2}})}}} to each element.

%
%
%
%
%
%
%

The exponential mechanism~\cite{mcsherry2007mechanism} selects an element of a
set based on the scores assigned to each element by a \emph{scoring function}.
Let {{\color{\colorMATH}\ensuremath{u: D^{n} \times  {\mathcal{R}} \rightarrow  {\mathbb{R}}}}} be a scoring function with {{\color{\colorMATH}\ensuremath{L1}}} sensitivity {{\color{\colorMATH}\ensuremath{\Delta }}}. The
mechanism selects and outputs an element {{\color{\colorMATH}\ensuremath{r \in  {\mathcal{R}}}}} with probability
proportional to {{\color{\colorMATH}\ensuremath{\exp(\frac{\epsilon  u(x, r)}{2\Delta })}}}, and preserves
{{\color{\colorMATH}\ensuremath{(\epsilon , 0)}}}-differential privacy.

%
%

\paragraph{Composition.}
A key property of differential privacy is that differentially private
computations \emph{compose}. The sequential composition theorem says that if
{{\color{\colorMATH}\ensuremath{{\mathcal{M}}_{1}}}} and {{\color{\colorMATH}\ensuremath{{\mathcal{M}}_{2}}}} satisfy {{\color{\colorMATH}\ensuremath{(\epsilon , \delta )}}}-differential privacy, then their combination
satisfies {{\color{\colorMATH}\ensuremath{(2\epsilon , 2\delta )}}}-differential privacy. 

%

Tighter bounds on privacy cost can be achieved using the advanced composition theorem~\cite{privacybook}, 
at the expense of increasing {{\color{\colorMATH}\ensuremath{\delta }}}.
The advanced composition theorem says that for {{\color{\colorMATH}\ensuremath{0 < \epsilon ^{\prime} < 1}}} and {{\color{\colorMATH}\ensuremath{\delta ^{\prime}
> 0}}}, the class of {{\color{\colorMATH}\ensuremath{(\epsilon , \delta )}}}-differentially private mechanisms
satisfies {{\color{\colorMATH}\ensuremath{(\epsilon ^{\prime}, k\delta  + \delta ^{\prime})}}}-differential privacy under {{\color{\colorMATH}\ensuremath{k}}}-fold
adaptive composition (e.g. a loop with {{\color{\colorMATH}\ensuremath{k}}} iterations) for {{\color{\colorMATH}\ensuremath{\epsilon ^{\prime} = 2 \epsilon  \sqrt {2 k \ln (1/\delta ^{\prime})}}}}.

The moments accountant was introduced by Talwar et al.~\cite{DPDL} specifically for stochastic gradient descent in deep learning applications. It provides tight bounds on privacy loss in iterative applications of the Gaussian mechanism, as in SGD. The R\'enyi differential privacy and zero-concentrated differential privacy generalize the ideas behind the moments accountant.

\begin{figure*}
  \footnotesize
\centering

\vspace*{-1em}\end{gather*}\endgroup 
  \vspace{-1.5em}
  \caption{Variants of Differential Privacy}
  \vspace{-0.5em}
  \label{fig:typing_variants}
\end{figure*}

\paragraph{Variants of differential privacy.}
In addition to {{\color{\colorMATH}\ensuremath{\epsilon }}} and {{\color{\colorMATH}\ensuremath{(\epsilon , \delta )}}}-differential privacy, other variants of differential privacy with significant benefits have recently been developed. Three examples are R\'enyi differential privacy (RDP)~\cite{mironov2017renyi}, zero-concentrated differential privacy (zCDP)~\cite{bun2016concentrated}, and truncated concentrated differential privacy (tCDP)~\cite{bun2018composable}. Each one has different privacy parameters and a different form of sequential composition, summarized in Figure~\ref{fig:typing_variants}. The basic mechanism for RDP and zCDP is the Gaussian mechanism; tCDP uses a novel \emph{sinh-normal} mechanism~\cite{bun2018composable} which decays more quickly in its tails. All three can be converted to {{\color{\colorMATH}\ensuremath{(\epsilon ,\delta )}}}-differential privacy, allowing them to be compared and composed with each other.
These three variants provide asymptotically tight bounds on privacy cost under composition, while at the same time eliminating the ``catastrophic'' privacy failure that can occur with probability {{\color{\colorMATH}\ensuremath{\delta }}} under {{\color{\colorMATH}\ensuremath{(\epsilon ,\delta )}}}-differential privacy.


\paragraph{Group privacy.}
Differential privacy is normally used to protect the privacy of
individuals, but it turns out that protection for an individual also
translates to (weaker) protection for \emph{groups} of individuals. A
mechanism which provides pure {{\color{\colorMATH}\ensuremath{\epsilon }}}-differential privacy for individuals
also provides {{\color{\colorMATH}\ensuremath{k\epsilon }}}-differential privacy for groups of size
{{\color{\colorMATH}\ensuremath{k}}}~\cite{privacybook}. Group privacy also exists for {{\color{\colorMATH}\ensuremath{(\epsilon ,
\delta )}}}-differential privacy, RDP, zCDP, and tCDP, but the scaling of the
privacy parameters is nonlinear.

\subsection{Related Work}

Language-based approaches for differential privacy fall into two categories: approaches based on type systems, and those based on program logics. Barthe et al.~\cite{barthe2016programming} provide a survey. The type-system based approaches are most related to our work, but program-logic-based approaches have also received considerable attention in recent years~\cite{barthe2016proving, barthe2016advanced, barthe2013beyond, barthe2013probabilistic, sato2016approximate, sato2019approximate}.

\paragraph{Linear Type Systems}
Type-system-based solutions to proving that a program adheres to differential
privacy began with Reed and Pierce's \fuzz language~\cite{reed2010distance},
which is based on linear typing. \fuzz, as well as subsequent work based on
linear types aided by SMT solvers~\cite{dfuzz}, supports type inference of
privacy bounds with type-level dependency and higher-order composition of
programs. However, these systems only support the original and most basic
variant of differential privacy called \emph{{{\color{\colorMATH}\ensuremath{\epsilon }}}-differential privacy}. 
More recent variants, like {{\color{\colorMATH}\ensuremath{(\epsilon , \delta )}}}-differential
privacy~\cite{privacybook} and others~\cite{mironov2017renyi,
  bun2016concentrated, bun2018composable}, improve on {{\color{\colorMATH}\ensuremath{\epsilon }}}-differential
privacy by providing vastly more accurate answers for the same amount
of privacy ``cost'' (at the expense of introducing a negligible chance of failure).

As described by Azevedo de Amorim et
al.~\cite{DBLP:journals/corr/abs-1807-05091}, encoding
{{\color{\colorMATH}\ensuremath{(\epsilon ,\delta )}}}-differential privacy in linear type systems like \fuzz is
particularly challenging because these systems place restrictions on
the interpretation of the linear function space, and
{{\color{\colorMATH}\ensuremath{(\epsilon ,\delta )}}}-differential privacy does not satisfy these restrictions. In
particular, using \fuzz requires that the desired notion of privacy
can be recovered from an instantiation of function sensitivity for an
appropriately defined metric on probabilistic functions. No such
metric can be defined for {{\color{\colorMATH}\ensuremath{(\epsilon ,\delta )}}}-differential privacy, preventing a
straightforward interpretation of linear functions as
{{\color{\colorMATH}\ensuremath{(\epsilon ,\delta )}}}-differentially private functions.

In their work, Azevedo de Amorim et
al.~\cite{DBLP:journals/corr/abs-1807-05091} define a \emph{path construction} to
encode non-linear scaling via an indexed probability monad, which can be used to
extend \fuzz with support for arbitrary relational properties (including {{\color{\colorMATH}\ensuremath{(\epsilon ,
\delta )}}}-differential privacy). However, this approach (1) internalizes the use of
\emph{group privacy}~\cite{privacybook} which in many cases provides
sub-optimal bounds on privacy cost--and (2) is unable to provide privacy bounds
for more than one input to a function--a useful capability of the original \fuzz
language, and a necessary feature to obtain optimal privacy bounds for
multi-argument functions.

\paragraph{Higher-order Relational Type Systems} Following the initial work on
linear typing for differential privacy~\cite{reed2010distance}, a parallel line
of work~\cite{barthe2015higher,barthe2016differentially} leverages {{\color{\colorTEXT}\textit{relational
refinement types}}} aided by SMT solvers in order to support type-level
dependency of privacy parameters (\`a la DFuzz~\cite{dfuzz}) in addition to more
powerful variants of differential privacy such as {{\color{\colorMATH}\ensuremath{(\epsilon ,\delta )}}}-differential privacy.
These approaches support {{\color{\colorMATH}\ensuremath{(\epsilon ,\delta )}}}-differential privacy, but did not support usable type inference until a recently proposed heuristic
bi-directional type system~\cite{DBLP:journals/corr/abs-1812-05067}. Although a
direct case study of bidirectional type inference for relational refinement
types has not yet been applied to differential privacy, the possibility of such
a system appears promising. 

The overall technique for supporting {{\color{\colorMATH}\ensuremath{(\epsilon ,\delta )}}}-differential privacy in these
relational refinement type systems is similar to (and predates) Azevedo de
Amorim et al.--privacy cost is tracked through an ``effect'' type, embodied by an
indexed monad. It is this ``effect''-based treatment of privacy cost that
fundamentally limits these type system to not support multi-arity functions,
resulting in non-optimal privacy bounds for some programs.

\paragraph{First-order Relational Type Systems}
Yet another approach is LightDP which uses a light-weight relational type
system to verify {{\color{\colorMATH}\ensuremath{(\epsilon ,\delta )}}}-differential privacy bounds of first-order imperative
programs~\cite{zhang2017lightdp}, and is suitable for verifying low-level
implementations of differentially-private mechanisms. A notable achievement
of this work is a lightweight, automated verification of the Sparse
Vector Technique~\cite{privacybook} (SVT). However, LightDP is not suitable for
sensitivity analysis, an important component of differentially-private
algorithm design. Differential privacy mechanisms often require knowledge of (or
place restrictions on) function sensitivity of arguments to the mechanism. In
principle, a language like \fuzz could be combined with LightDP to fully verify
both an application which uses SVT, as well as the implementation of SVT itself.

\paragraph{Type Systems Enriched with Program Logics}
At a high level, Fuzzi~\cite{fuzzi} has a similar aim to \system:
supporting differential privacy for general-purpose programs and
supporting recent variants of differential privacy. \system is
designed primarily as a fully-automated type system with a rich set of
primitives for vector-based and higher-order programming; low-level
mechanisms in \system are opaque and trusted. On the other hand, Fuzzi
is designed for general-purpose programming, low-level mechanism
implementation, and their combination; however, to achieve this, Fuzzi
has less support for higher-order programming and automation in
typechecking.

\subsection{Our Approach}

\newcommand\TFuzz{Fuzz~\cite{reed2010distance}}
\newcommand\TDFuzz{DFuzz~\cite{dfuzz}}
\newcommand\TPatMet{PathC~\cite{DBLP:journals/corr/abs-1807-05091}}
\newcommand\THoaresq{\hoaresq~\cite{barthe2015higher}}
\newcommand\TBiRelTC{BiRelTC~\cite{DBLP:journals/corr/abs-1812-05067}}
\newcommand\TLightDP{LightDP~\cite{zhang2017lightdp}}
\newcommand\TFuzzi{Fuzzi~\cite{fuzzi}}
\begin{figure}
{\footnotesize
\begin{tabular}{l||c|c|c|c|c|c|c|c|c
}            & SA & HO & DT         & MA & Rel-ext & {{\color{\colorMATH}\ensuremath{\epsilon }}}-DP & {{\color{\colorMATH}\ensuremath{(\epsilon ,\delta )}}}-DP & R\'enyi/zCDP/tCDP  & SVT-imp
\cr\hline  \TFuzz     & \checkmark   & \checkmark   & \text{\ding{55}}          & \checkmark   & \text{\ding{55}}       & \checkmark       & \text{\ding{55}}          & \text{\ding{55}}                & \text{\ding{55}}
\cr\hline  \TDFuzz    & \checkmark   & \checkmark   & \checkmark           & \checkmark   & \text{\ding{55}}       & \checkmark       & \text{\ding{55}}          & \text{\ding{55}}                & \text{\ding{55}}
\cr\hline  \TPatMet   & \checkmark   & \checkmark   & \text{\ding{55}}$^{1}$       & \text{\ding{55}}  & \checkmark        & \checkmark       & \checkmark           & \checkmark $^{2}$             & \text{\ding{55}}
\cr\hline  \THoaresq  & \checkmark   & \checkmark   & \checkmark           & \text{\ding{55}}  & \checkmark        & \checkmark       & \checkmark           & \checkmark $^{2}$             & \text{\ding{55}}
\cr\hline  \TLightDP  & \text{\ding{55}}  & \text{\ding{55}}  & \checkmark           & \checkmark   & \text{\ding{55}}       & \checkmark       & \checkmark           & \checkmark $^{2}$             & \checkmark 
\cr\hline  \TFuzzi    & \checkmark   & \text{\ding{55}}  & \text{\ding{55}}$^{1}$       & \checkmark   & \checkmark        & \checkmark       & \checkmark           & \checkmark                 & \checkmark 
\cr\hline  \system    & \checkmark   & \checkmark   & \checkmark           & \checkmark   & \text{\ding{55}}       & \checkmark       & \checkmark           & \checkmark                 & \text{\ding{55}}
\end{tabular}
}
\vspace{-1.0em}
\caption{\footnotesize
  Legend: 
    SA = capable of sensitivity analysis;
    HO = support for higher order programming, program composition, and
      compound datatypes; 
    DT = support for dependently typed privacy bounds;
    MA = support for distinct privacy bounds of multiple input arguments;
    Rel-ext = supports extensions to support non-differential-privacy relations;
    {{\color{\colorMATH}\ensuremath{\epsilon }}}-DP = supports {{\color{\colorMATH}\ensuremath{\epsilon }}}-differential-privacy;
    {{\color{\colorMATH}\ensuremath{(\epsilon ,\delta )}}}-DP = supports {{\color{\colorMATH}\ensuremath{(\epsilon ,\delta )}}}-differential-privacy;
    R\'enyi/zCDP/tCDP: supports R\'enyi, zero-concentrated and
      truncated concentrated differential privacy;
    SVT-imp: supports verified {{\color{\colorTEXT}\textit{implementation}}} of the sparse vector technique.
    1: This limitation is not fundamental and could be supported by simple
       extension to underlying type theory. 
    2: Not described in prior work, but could be achieved through a trivial
       extension to existing support for {{\color{\colorMATH}\ensuremath{(\epsilon ,\delta )}}}-differential privacy.
  }
\vspace{-0.5em}
\label{fig:comparison}
\end{figure}

We show the strengths and limitations of \system in relation to approaches from
prior work in Figure~\ref{fig:comparison}. In particular, strengths of Duet
{w.r.t.} prior work are: (1) \system supports sensitivity analysis in
combination with higher order programming, program composition, and compound
datatypes, building on ideas from \fuzz (SA+HO); (2) \system supports
type-level dependency on values, which enables differentially private
algorithms to be verified {w.r.t.} symbolic privacy parameters, building on
ideas from DFuzz (DT); (3) \system supports calculation of independent privacy
costs for multiple program arguments via a novel approach (MA); and (4) \system
supports {{\color{\colorMATH}\ensuremath{(\epsilon ,\delta )}}}-differential privacy---in addition to other recent powerful
variants, such as R\'enyi, zero-concentrated and truncated concentrated
differential privacy---via a novel approach ({{\color{\colorMATH}\ensuremath{(\epsilon ,\delta )}}}-DP, R\'enyi/ZC/TC)).

In striking this balance, \system comes with known limitations: (1) \system is
not easy to extend with new relational properties (Rel-ext); and (2) \system is
not suitable for verifying implementations of low-level mechanisms, such as the
implementation of advanced composition, gradient operations, and the
sparse-vector technique (SVT-imp).

\section{\system: A Language for Privacy}

This section describes the syntax, type system and formal properties of
\system. Our design of \system is the result of two key insights. 
\begin{enumerate}[leftmargin=*]\item  {{\color{\colorTEXT}\textit{Linear typing, when restricted to disallow scaling, can be a powerful
   foundation for enforcing {{\color{\colorMATH}\ensuremath{(\epsilon ,\delta )}}}-differential privacy.}}} 
   Privacy bounds in {{\color{\colorMATH}\ensuremath{(\epsilon ,\delta )}}}-differential privacy do not scale linearly,
   and cannot be accurately modeled by linear type systems which permit
   unrestricted scaling.
\item  {{\color{\colorTEXT}\textit{Sensitivity and privacy cost are distinct properties, and warrant distinct
   type systems to enforce them.}}} Our design for \system is a co-design of two
   distinct, mutually embedded languages: one for sensitivity which leverages
   linear typing {{\color{\colorTEXT}\textit{with}}} scaling a la \fuzz, and one for privacy which leverages
   linear typing {{\color{\colorTEXT}\textit{without}}} scaling and is novel in this work. 
\end{enumerate}
Before describing the syntax, semantics and types for each of \system's two
languages, we first provide some context which motivates each design decision
made. We do this through several small examples and type signatures drawn from
state-of-the-art type systems such as \fuzz~\cite{reed2010distance},
\hoaresq~\cite{barthe2015higher} and Azevedo de Amorim et al's {{\color{\colorTEXT}\textit{path
construction}}}~\cite{DBLP:journals/corr/abs-1807-05091}.

\subsection{Design Challenges}

\paragraph{Higher-Order Programming}

An important design goal of Duet is to support sensitivity analysis of
higher-order, general purpose programs. Prior work (\fuzz and \hoaresq) has
demonstrated exactly this, and we build on their techniques. In \fuzz, the
types for the higher-order {{\color{\colorMATH}\ensuremath{\operatorname{map}}}} function and a list of reals named {{\color{\colorMATH}\ensuremath{xs}}} looks
like this:
\vspace*{-0.25em}\begingroup\color{\colorMATH}\begin{gather*}\begin{array}{rcl
  } {{\color{\colorMATH}\ensuremath{\operatorname{map}}}} &{}\mathrel{:}{}& (\tau _{1} \multimap _{s} \tau _{2}) \multimap _{\infty } {{\color{\colorMATH}\ensuremath{\operatorname{list}}}}\hspace*{0.33em}\tau _{1} \multimap _{s} {{\color{\colorMATH}\ensuremath{\operatorname{list}}}}\hspace*{0.33em}\tau _{2}
  \cr  {{\color{\colorMATH}\ensuremath{xs}}} &{}\mathrel{:}{}& {{\color{\colorMATH}\ensuremath{\operatorname{list}}}}\hspace*{0.33em}{\mathbb{R}}
  \end{array}
\vspace*{-1em}\end{gather*}\endgroup 
The type of {{\color{\colorMATH}\ensuremath{\operatorname{map}}}} reads: ``Take as a first argument an {{\color{\colorMATH}\ensuremath{s}}}-sensitive function
from {{\color{\colorMATH}\ensuremath{\tau _{1}}}} to {{\color{\colorMATH}\ensuremath{\tau _{2}}}} which {{\color{\colorMATH}\ensuremath{\operatorname{map}}}} is allowed to use as many times as it wants. Take
as second argument a list of {{\color{\colorMATH}\ensuremath{\tau _{1}}}}, and return a result list of {{\color{\colorMATH}\ensuremath{\tau _{2}}}} which is
{{\color{\colorMATH}\ensuremath{s}}}-sensitive in the list of {{\color{\colorMATH}\ensuremath{\tau _{1}}}}.'' Two programs that use {{\color{\colorMATH}\ensuremath{\operatorname{map}}}} might look like
this:
\vspace*{-0.25em}\begingroup\color{\colorMATH}\begin{gather*}\begin{array}{l@{\hspace*{1.00em}\hspace*{1.00em}\hspace*{1.00em}}r
  } {{\color{\colorMATH}\ensuremath{\operatorname{map}}}}\hspace*{0.33em}(\lambda \hspace*{0.33em}x \rightarrow  x + 1)\hspace*{0.33em}{{\color{\colorMATH}\ensuremath{\operatorname{xs}}}} & {{\color{\colorTEXT}\textit{(1)}}}
  \cr  {{\color{\colorMATH}\ensuremath{\operatorname{map}}}}\hspace*{0.33em}(\lambda \hspace*{0.33em}x \rightarrow  x + x)\hspace*{0.33em}{{\color{\colorMATH}\ensuremath{\operatorname{xs}}}} & {{\color{\colorTEXT}\textit{(2)}}}
  \end{array}
\vspace*{-1em}\end{gather*}\endgroup 
The \fuzz type system reports that {{\color{\colorTEXT}\textit{(1)}}} is {{\color{\colorMATH}\ensuremath{1}}}-sensitive in {{\color{\colorMATH}\ensuremath{xs}}}, and that
{{\color{\colorTEXT}\textit{(2)}}} is {{\color{\colorMATH}\ensuremath{2}}}-sensitive in {{\color{\colorMATH}\ensuremath{xs}}}. To arrive at this conclusion, the \fuzz type
checker is essentially counting how many times {{\color{\colorMATH}\ensuremath{x}}} is used in the body of the
lambda, and type soundness for \fuzz means that these counts correspond to the
semantic property of function sensitivity.

In \hoaresq the type for {{\color{\colorMATH}\ensuremath{\operatorname{map}}}} is instead:
\vspace*{-0.25em}\begingroup\color{\colorMATH}\begin{gather*}\begin{array}{rcl
  } {{\color{\colorMATH}\ensuremath{\operatorname{map}}}} &{}\mathrel{:}{}& (\forall  s^{\prime}.\hspace*{0.33em} \{ x \mathrel{:: } \tau _{1} \mathrel{|} {\mathfrak{D}} _{\tau _{1}}(x_{\vartriangleleft },x_{\vartriangleright }) \leq  s^{\prime}\}  \rightarrow  \{ y \mathrel{:: } \tau _{2} \mathrel{|} {\mathfrak{D}} _{\tau _{2}}(y_{\vartriangleleft },y_{\vartriangleright }) \leq  s\mathrel{\mathord{\cdotp }}s^{\prime}\} ) 
  \cr        &{}\rightarrow {}& \forall  s^{\prime}.\hspace*{0.33em} \{ xs \mathrel{:: } list\hspace*{0.33em}\tau _{1} \mathrel{|} {\mathfrak{D}} _{(list\hspace*{0.33em}\tau _{1})}(xs_{\vartriangleleft },xs_{\vartriangleright }) \leq  s^{\prime}\}  \rightarrow  \{ y \mathrel{:: } list\hspace*{0.33em}\tau _{2} \mathrel{|} {\mathfrak{D}} _{(list\hspace*{0.33em}\tau _{2})}(ys_{\vartriangleleft },ys_{\vartriangleright }) \leq  s\mathrel{\mathord{\cdotp }}s^{\prime}\} 
  \end{array}
\vspace*{-1em}\end{gather*}\endgroup 
This type for {{\color{\colorMATH}\ensuremath{\operatorname{map}}}} means the same thing as the \fuzz type shown above, and \hoaresq likewise
reports that {{\color{\colorTEXT}\textit{(1)}}} is {{\color{\colorMATH}\ensuremath{1}}}-sensitive and {{\color{\colorTEXT}\textit{(2)}}} is {{\color{\colorMATH}\ensuremath{2}}}-sensitive, each in {{\color{\colorMATH}\ensuremath{xs}}},
and where {{\color{\colorMATH}\ensuremath{{\mathfrak{D}} _{\tau }}}} is some family of distance metrics indexed by types {{\color{\colorMATH}\ensuremath{\tau }}}. To
arrive at this conclusion, \hoaresq generates relational verification conditions
(where, {e.g.}, {{\color{\colorMATH}\ensuremath{x_{\vartriangleleft }}}} is drawn from a hypothetical ``first/left run'' of the
program, and {{\color{\colorMATH}\ensuremath{x_{\vartriangleright }}}} is drawn from a hypothetical ``second/right run'' of the
program) which are discharged by an external solver ({e.g.}, SMT). In this
approach, sensitivity is not concluded via an interpretation of a purely
{{\color{\colorTEXT}\textit{syntactic}}} type system ({e.g.}, linear typing in \fuzz), rather the relational
{{\color{\colorTEXT}\textit{semantic}}} property of sensitivity (and its scaling) is embedded directly in
the relational refinements of higher-order function types.

In designing \system, we follow the design of \fuzz in that programs adhere to
a linear type discipline, {i.e.}, the mechanics of our type system is based on
counting variables and (in some cases) scaling, and we prove a soundness
theorem that says well-typed programs are guaranteed to be sensitive/private
programs. Our type for {{\color{\colorMATH}\ensuremath{\operatorname{map}}}} is identical to the one shown above for \fuzz.

\paragraph{Non-Linear Scaling}

\fuzz encodes an {{\color{\colorMATH}\ensuremath{\epsilon }}}-differentially private function as an {{\color{\colorMATH}\ensuremath{\epsilon }}}-sensitive
function which returns a monadic type {{\color{\colorMATH}\ensuremath{\bigcirc \hspace*{0.33em}\tau }}}. The Laplace differential privacy
mechanism is then encoded in \fuzz as an {{\color{\colorMATH}\ensuremath{\epsilon }}}-sensitive function from {{\color{\colorMATH}\ensuremath{{\mathbb{R}}}}} to
{{\color{\colorMATH}\ensuremath{\bigcirc \hspace*{0.33em}{\mathbb{R}}}}}:
\vspace*{-0.25em}\begingroup\color{\colorMATH}\begin{gather*}\begin{array}{l
  } {{\color{\colorMATH}\ensuremath{\operatorname{laplace}}}} \mathrel{:} {\mathbb{R}} \multimap _{{{\color{\colorMATH}\ensuremath{\epsilon }}}} \bigcirc \hspace*{0.33em}{\mathbb{R}}
  \end{array}
\vspace*{-1em}\end{gather*}\endgroup 
Because the metric on distributions for pure {{\color{\colorMATH}\ensuremath{\epsilon }}}-differential privacy scales
linearly, {{\color{\colorMATH}\ensuremath{\operatorname{laplace}}}} can be applied to a {{\color{\colorMATH}\ensuremath{2}}}-sensitive argument to achieve
{{\color{\colorMATH}\ensuremath{2\epsilon }}}-differential privacy, {e.g.}:
\vspace*{-0.25em}\begingroup\color{\colorMATH}\begin{gather*}\begin{array}{l
  } {{\color{\colorMATH}\ensuremath{\operatorname{laplace}}}}\hspace*{0.33em}(x + x)
  \end{array}
\vspace*{-1em}\end{gather*}\endgroup 
gives {{\color{\colorMATH}\ensuremath{2\epsilon }}}-differential privacy for {{\color{\colorMATH}\ensuremath{x}}}. Adding more advanced variants of
differential privacy like {{\color{\colorMATH}\ensuremath{(\epsilon ,\delta )}}} to \fuzz has proved challenging because these
variants do not scale linearly. Azevedo de Amorim et al's {{\color{\colorTEXT}\textit{path construction}}}
successfully adds {{\color{\colorMATH}\ensuremath{(\epsilon ,\delta )}}}-differential privacy to \fuzz by tracking privacy
``cost'' as an index on the monadic type operator {{\color{\colorMATH}\ensuremath{\bigcirc _{\epsilon ,\delta }}}}. However, in order to
interpret a function application like the one shown, the group privacy
property for {{\color{\colorMATH}\ensuremath{(\epsilon ,\delta )}}}-differential privacy must be used, which results in undesirable non-linear scaling
of the privacy cost. The derived bound for this program using group privacy
(for {{\color{\colorMATH}\ensuremath{k=2}}}) is not {{\color{\colorMATH}\ensuremath{(2\epsilon ,2\delta )}}} but {{\color{\colorMATH}\ensuremath{(2\epsilon ,2e^{\epsilon }\delta )}}}~\cite{privacybook}. As a result, achieving a desired
{{\color{\colorMATH}\ensuremath{\epsilon }}} and {{\color{\colorMATH}\ensuremath{\delta }}} by treating an {{\color{\colorMATH}\ensuremath{s}}}-sensitive function as {{\color{\colorMATH}\ensuremath{1}}}-sensitive and
leveraging group privacy requires adding much more noise than simply applying
the Gaussian mechanism with a sensitivity of {{\color{\colorMATH}\ensuremath{s}}}.


In \hoaresq, the use of scaling which might warrant the use of group privacy is
explicitly disallowed in the stated relational refinement type. This is in
contrast to sensitivity, which likewise must explicitly allow arbitrary
scaling. The type for {{\color{\colorMATH}\ensuremath{\operatorname{gauss}}}} in \hoaresq (the analogous mechanism to {{\color{\colorMATH}\ensuremath{\operatorname{laplace}}}}
in the {{\color{\colorMATH}\ensuremath{(\epsilon ,\delta )}}}-differential privacy setting) is written:
\vspace*{-0.25em}\begingroup\color{\colorMATH}\begin{gather*}\begin{array}{l
  } {{\color{\colorMATH}\ensuremath{\operatorname{gauss}}}} \mathrel{:} \{ x \mathrel{:: } {\mathbb{R}} \mathrel{|} {\mathfrak{D}} _{{\mathbb{R}}}(x_{\vartriangleleft },x_{\vartriangleright }) \leq  1 \}  \rightarrow  {\textbf{M}} _{\epsilon ,\delta }\hspace*{0.33em}{\mathbb{R}}
  \end{array}
\vspace*{-1em}\end{gather*}\endgroup 
Notice the assumed sensitivity of {{\color{\colorMATH}\ensuremath{x}}} to be bounded by {{\color{\colorMATH}\ensuremath{1}}}, not some arbitrary
{{\color{\colorMATH}\ensuremath{s^{\prime}}}} to be scaled in the output refinement (as was seen in the type for {{\color{\colorMATH}\ensuremath{\operatorname{map}}}}
in \hoaresq above). In this way, \hoaresq is able to restrict {{\color{\colorTEXT}\textit{uses}}} of {{\color{\colorMATH}\ensuremath{\operatorname{gauss}}}} to
strictly {{\color{\colorMATH}\ensuremath{1}}}-sensitive arguments, a restriction that is not possible in a pure
linear type system where arbitrary program composition is allowed and
interpreted via scaling.

In \system, we co-design two languages which are mutually embedded inside one
another. The {\begingroup\renewcommand\colorMATH{\colorMATHS}\renewcommand\colorSYNTAX{\colorSYNTAXS}{{\color{\colorMATH}\ensuremath{{{\color{\colorMATH}\ensuremath{\operatorname{sensitivity}}}}}}}\endgroup } language is nearly identical to \fuzz, supports
arbitrary scaling, and is never interpreted to mean privacy. The {\begingroup\renewcommand\colorMATH{\colorMATHP}\renewcommand\colorSYNTAX{\colorSYNTAXP}{{\color{\colorMATH}\ensuremath{{{\color{\colorMATH}\ensuremath{\operatorname{privacy}}}}}}}\endgroup }
language is also linearly typed, but restricts function call parameters to be
strictly {{\color{\colorMATH}\ensuremath{1}}}-sensitive---a property established in the {\begingroup\renewcommand\colorMATH{\colorMATHS}\renewcommand\colorSYNTAX{\colorSYNTAXS}{{\color{\colorMATH}\ensuremath{{{\color{\colorMATH}\ensuremath{\operatorname{sensitivity}}}}}}}\endgroup }
fragment. The {{\color{\colorMATH}\ensuremath{\operatorname{gauss}}}} mechanism in \system is (essentially) given the type:
\vspace*{-0.25em}\begingroup\color{\colorMATH}\begin{gather*}\begin{array}{l
  } {{\color{\colorMATH}\ensuremath{\operatorname{gauss}}}} \mathrel{:} {\mathbb{R}}@\langle \epsilon ,\delta \rangle  \multimap ^{*} {\mathbb{R}}
  \end{array}
\vspace*{-1em}\end{gather*}\endgroup 
where {{\color{\colorMATH}\ensuremath{\multimap ^{*}}}} is the function space in \system's privacy language, and the
annotation {{\color{\colorMATH}\ensuremath{@\langle \epsilon ,\delta \rangle }}} tracks the privacy cost of that argument following a
linear typing discipline.

\paragraph{Multiple Private Parameters}
Both \hoaresq and the {{\color{\colorTEXT}\textit{path construction}}} track {{\color{\colorMATH}\ensuremath{(\epsilon ,\delta )}}}-differential privacy
via an indexed monadic type, notated {{\color{\colorMATH}\ensuremath{{\textbf{M}}_{\epsilon ,\delta }}}} and {{\color{\colorMATH}\ensuremath{\bigcirc _{\epsilon ,\delta }}}} respectively. E.g., a
program that returns an {{\color{\colorMATH}\ensuremath{(\epsilon ,\delta )}}}-differentially private real number has the type
{{\color{\colorMATH}\ensuremath{{\textbf{M}}_{\epsilon ,\delta }({\mathbb{R}})}}} in \hoaresq. These monadic approaches to privacy inherently follow
an ``effect'' type discipline, and as a result the monad index must track the
{{\color{\colorTEXT}\textit{sum total of all privacy costs to any parameter}}}. For example, a small program
that takes two parameters, applies a mechanism to enforce differential privacy
for each parameter, and adds them together, will report a double-counting of
privacy cost. {E.g.}, in this \hoaresq program (translated to Haskell-ish
``do''-notation):
\vspace*{-0.25em}\begingroup\color{\colorMATH}\begin{gather*}\begin{array}{l
  } {{\color{\colorSYNTAX}\texttt{let}}}\hspace*{0.33em}f = {{\color{\colorSYNTAX}\texttt{\ensuremath{\lambda }}}}\hspace*{0.33em}x\hspace*{0.33em}y \rightarrow  {{\color{\colorSYNTAX}\texttt{do}}}\hspace*{0.33em}\{ \hspace*{0.33em}r_{1} \leftarrow  gauss_{\epsilon ,\delta }\hspace*{0.33em}x \mathrel{;} r_{2} \leftarrow  gauss_{\epsilon ,\delta }\hspace*{0.33em}y \mathrel{;} return\hspace*{0.33em}(r_{1} + r_{2})\hspace*{0.33em}\} 
  \end{array}
\vspace*{-1em}\end{gather*}\endgroup 
The type of {{\color{\colorMATH}\ensuremath{f}}} in \hoaresq reports that it costs {{\color{\colorMATH}\ensuremath{(2\epsilon ,2\delta )}}} privacy:
\vspace*{-0.25em}\begingroup\color{\colorMATH}\begin{gather*}\begin{array}{l
  } f \mathrel{:} \{ x \mathrel{:: } {\mathbb{R}} \mathrel{|} {\mathfrak{D}} _{{\mathbb{R}}}(x_{\vartriangleleft },x_{\vartriangleright }) \leq  1\}  \rightarrow  \{ y \mathrel{:: } {\mathbb{R}} \mathrel{|} {\mathfrak{D}} _{{\mathbb{R}}}(y_{\vartriangleleft },y_{\vartriangleright }) \leq  1\}  \rightarrow  {\textbf{M}}_{2\epsilon ,2\delta }\hspace*{0.33em}{\mathbb{R}}
  \end{array}
\vspace*{-1em}\end{gather*}\endgroup 
This bound is too conservative in many cases: it is the best bound in the case
that {{\color{\colorMATH}\ensuremath{f}}} is applied to the same variable for both arguments ({e.g.}, in
{{\color{\colorMATH}\ensuremath{f\hspace*{0.33em}a\hspace*{0.33em}a}}}), however, if {{\color{\colorMATH}\ensuremath{f}}} is applied to {{\color{\colorTEXT}\textit{different}}} variables ({e.g.}, in
{{\color{\colorMATH}\ensuremath{f\hspace*{0.33em}a\hspace*{0.33em}b}}}) then a privacy cost of {{\color{\colorMATH}\ensuremath{(2\epsilon ,2\delta )}}} is still claimed, interpreted as {{\color{\colorTEXT}\textit{for
either or both variables}}} {{\color{\colorMATH}\ensuremath{2\epsilon ,2\delta }}} privacy is consumed. A better accounting of
privacy in this second case should report {{\color{\colorMATH}\ensuremath{(\epsilon ,\delta )}}}-differential
privacy {{\color{\colorTEXT}\textit{independently}}} for both variables {{\color{\colorMATH}\ensuremath{a}}} and {{\color{\colorMATH}\ensuremath{b}}}, and such accounting is
not possible in either \hoaresq or the {{\color{\colorTEXT}\textit{path construction}}}.

In \system, we track privacy following a {{\color{\colorTEXT}\textit{co-effect}}} discipline (linear typing
without scaling), as opposed to an effect discipline, in order to distinguish
privacy costs independently for each variable. The type of the above program in
\system is:
\vspace*{-0.25em}\begingroup\color{\colorMATH}\begin{gather*}\begin{array}{l
  } f \mathrel{:} ({\mathbb{R}}@\langle \epsilon ,\delta \rangle ,{\mathbb{R}}@\langle \epsilon ,\delta \rangle ) \multimap ^{*} {\mathbb{R}}
  \end{array}
\vspace*{-1em}\end{gather*}\endgroup 
indicating that {{\color{\colorMATH}\ensuremath{f}}} ``costs'' {{\color{\colorMATH}\ensuremath{(\epsilon ,\delta )}}} for each parameter independently, and only
when {{\color{\colorMATH}\ensuremath{f}}} is called with two identical variables as arguments are they combined
as {{\color{\colorMATH}\ensuremath{(2\epsilon ,2\delta )}}}.

Due to limitations of linear logic in the absence of scaling, {\begingroup\renewcommand\colorMATH{\colorMATHP}\renewcommand\colorSYNTAX{\colorSYNTAXP}{{\color{\colorMATH}\ensuremath{{{\color{\colorMATH}\ensuremath{\operatorname{privacy}}}}}}}\endgroup }
lambdas must be multi-argument in the core design of \system---they cannot be
recovered by single-argument lambdas. As a consequence, our {\begingroup\renewcommand\colorMATH{\colorMATHP}\renewcommand\colorSYNTAX{\colorSYNTAXP}{{\color{\colorMATH}\ensuremath{{{\color{\colorMATH}\ensuremath{\operatorname{privacy}}}}}}}\endgroup }
language is not Cartesian closed.

\subsection{\system by Example}
\label{sec:by_example}

\paragraph{Sensitivity.}
\system consists of two languages: one for tracking sensitivities
(typeset in {\begingroup\renewcommand\colorMATH{\colorMATHS}\renewcommand\colorSYNTAX{\colorSYNTAXS}{{\color{\colorMATH}\ensuremath{{{\color{\colorMATH}\ensuremath{\operatorname{green}}}}}}}\endgroup }), and one for tracking privacy cost (typeset
in {\begingroup\renewcommand\colorMATH{\colorMATHP}\renewcommand\colorSYNTAX{\colorSYNTAXP}{{\color{\colorMATH}\ensuremath{{{\color{\colorMATH}\ensuremath{\operatorname{red}}}}}}}\endgroup }). The sensitivity language is similar to that of
DFuzz~\cite{dfuzz}; its typing rules track the sensitivity of each
variable by annotating the context. For example, the expression
{\begingroup\renewcommand\colorMATH{\colorMATHS}\renewcommand\colorSYNTAX{\colorSYNTAXS}{{\color{\colorMATH}\ensuremath{{\begingroup\renewcommand\colorMATH{\colorMATHM}\renewcommand\colorSYNTAX{\colorSYNTAXM}{{\color{\colorMATH}\ensuremath{x}}}\endgroup } + {\begingroup\renewcommand\colorMATH{\colorMATHM}\renewcommand\colorSYNTAX{\colorSYNTAXM}{{\color{\colorMATH}\ensuremath{x}}}\endgroup }}}}\endgroup } is 2-sensitive in {\begingroup\renewcommand\colorMATH{\colorMATHM}\renewcommand\colorSYNTAX{\colorSYNTAXM}{{\color{\colorMATH}\ensuremath{x}}}\endgroup }; the typing rules in
Figure~\ref{fig:typing} allow us to conclude:
\vspace*{-0.25em}\begingroup\color{\colorMATH}\begin{gather*} \{  x:_{2} {{\color{\colorSYNTAX}\texttt{\ensuremath{{\mathbb{R}}}}}} \}  \vdash  {\begingroup\renewcommand\colorMATH{\colorMATHS}\renewcommand\colorSYNTAX{\colorSYNTAXS}{{\color{\colorMATH}\ensuremath{{\begingroup\renewcommand\colorMATH{\colorMATHM}\renewcommand\colorSYNTAX{\colorSYNTAXM}{{\color{\colorMATH}\ensuremath{x}}}\endgroup } + {\begingroup\renewcommand\colorMATH{\colorMATHM}\renewcommand\colorSYNTAX{\colorSYNTAXM}{{\color{\colorMATH}\ensuremath{x}}}\endgroup }}}}\endgroup } \mathrel{:} {{\color{\colorSYNTAX}\texttt{\ensuremath{{\mathbb{R}}}}}} \vspace*{-1em}\end{gather*}\endgroup 
\noindent In this case, the context {\begingroup\renewcommand\colorMATH{\colorMATHM}\renewcommand\colorSYNTAX{\colorSYNTAXM}{{\color{\colorMATH}\ensuremath{\{  x:_{2} {{\color{\colorSYNTAX}\texttt{\ensuremath{{\mathbb{R}}}}}} \} }}}\endgroup } tells us that
the expression is 2-sensitive in {\begingroup\renewcommand\colorMATH{\colorMATHM}\renewcommand\colorSYNTAX{\colorSYNTAXM}{{\color{\colorMATH}\ensuremath{x}}}\endgroup }. The same idea works for
functions; for example:
\vspace*{-0.25em}\begingroup\color{\colorMATH}\begin{gather*} \emptyset \vdash  {\begingroup\renewcommand\colorMATH{\colorMATHS}\renewcommand\colorSYNTAX{\colorSYNTAXS}{{\color{\colorMATH}\ensuremath{ \lambda  {\begingroup\renewcommand\colorMATH{\colorMATHM}\renewcommand\colorSYNTAX{\colorSYNTAXM}{{\color{\colorMATH}\ensuremath{x \mathrel{:} {{\color{\colorSYNTAX}\texttt{\ensuremath{{\mathbb{R}}}}}}}}}\endgroup } \Rightarrow  {\begingroup\renewcommand\colorMATH{\colorMATHM}\renewcommand\colorSYNTAX{\colorSYNTAXM}{{\color{\colorMATH}\ensuremath{x}}}\endgroup } + {\begingroup\renewcommand\colorMATH{\colorMATHM}\renewcommand\colorSYNTAX{\colorSYNTAXM}{{\color{\colorMATH}\ensuremath{x}}}\endgroup } }}}\endgroup } \mathrel{:} {\begingroup\renewcommand\colorMATH{\colorMATHM}\renewcommand\colorSYNTAX{\colorSYNTAXM}{{\color{\colorMATH}\ensuremath{{{\color{\colorSYNTAX}\texttt{\ensuremath{{{\color{\colorMATH}\ensuremath{{{\color{\colorSYNTAX}\texttt{\ensuremath{{\mathbb{R}}}}}}}}} \multimap _{{\begingroup\renewcommand\colorMATH{\colorMATHS}\renewcommand\colorSYNTAX{\colorSYNTAXS}{{\color{\colorMATH}\ensuremath{2}}}\endgroup }} {{\color{\colorMATH}\ensuremath{{{\color{\colorSYNTAX}\texttt{\ensuremath{{\mathbb{R}}}}}}}}}}}}}}}}\endgroup }  \vspace*{-1em}\end{gather*}\endgroup 
\noindent Here, the context is empty; instead, the function's
sensitivity to its argument is encoded in an annotation on its type
(the {\begingroup\renewcommand\colorMATH{\colorMATHS}\renewcommand\colorSYNTAX{\colorSYNTAXS}{{\color{\colorMATH}\ensuremath{2}}}\endgroup } in {\begingroup\renewcommand\colorMATH{\colorMATHM}\renewcommand\colorSYNTAX{\colorSYNTAXM}{{\color{\colorMATH}\ensuremath{{{\color{\colorSYNTAX}\texttt{\ensuremath{{{\color{\colorMATH}\ensuremath{{{\color{\colorSYNTAX}\texttt{\ensuremath{{\mathbb{R}}}}}}}}} \multimap _{{\begingroup\renewcommand\colorMATH{\colorMATHS}\renewcommand\colorSYNTAX{\colorSYNTAXS}{{\color{\colorMATH}\ensuremath{2}}}\endgroup }} {{\color{\colorMATH}\ensuremath{{{\color{\colorSYNTAX}\texttt{\ensuremath{{\mathbb{R}}}}}}}}}}}}}}}}\endgroup }). Applying such a
function to an argument \emph{scales} the sensitivity of the argument
by the sensitivity of the function. This kind of scaling is
appropriate for sensitivities, and even has the correct effect for
higher-order functions. For example:
\vspace*{-0.25em}\begingroup\color{\colorMATH}\begin{gather*}
\begin{array}{l
} \{  y \mathrel{:}_{2} {{\color{\colorSYNTAX}\texttt{\ensuremath{{\mathbb{R}}}}}} \}  \vdash  {\begingroup\renewcommand\colorMATH{\colorMATHS}\renewcommand\colorSYNTAX{\colorSYNTAXS}{{\color{\colorMATH}\ensuremath{ (\lambda  {\begingroup\renewcommand\colorMATH{\colorMATHM}\renewcommand\colorSYNTAX{\colorSYNTAXM}{{\color{\colorMATH}\ensuremath{x \mathrel{:} {{\color{\colorSYNTAX}\texttt{\ensuremath{{\mathbb{R}}}}}}}}}\endgroup } \Rightarrow  {\begingroup\renewcommand\colorMATH{\colorMATHM}\renewcommand\colorSYNTAX{\colorSYNTAXM}{{\color{\colorMATH}\ensuremath{x}}}\endgroup } + {\begingroup\renewcommand\colorMATH{\colorMATHM}\renewcommand\colorSYNTAX{\colorSYNTAXM}{{\color{\colorMATH}\ensuremath{x}}}\endgroup })\hspace*{0.33em}{\begingroup\renewcommand\colorMATH{\colorMATHM}\renewcommand\colorSYNTAX{\colorSYNTAXM}{{\color{\colorMATH}\ensuremath{y}}}\endgroup } }}}\endgroup } \mathrel{:} {\begingroup\renewcommand\colorMATH{\colorMATHM}\renewcommand\colorSYNTAX{\colorSYNTAXM}{{\color{\colorMATH}\ensuremath{{{\color{\colorSYNTAX}\texttt{\ensuremath{{\mathbb{R}}}}}}}}}\endgroup }
\cr
\end{array}\vspace*{-1em}\end{gather*}\endgroup 
\paragraph{Privacy.}
Differentially private mechanisms like the Gaussian
mechanism~\cite{privacybook} specify how to add noise to a function
with a particular sensitivity in order to ensure differential
privacy. In \system, such mechanisms form the interface between the
sensitivity language and the privacy language. For example:
\vspace*{-0.25em}\begingroup\color{\colorMATH}\begin{gather*} \{  x \mathrel{:}_{{\begingroup\renewcommand\colorMATH{\colorMATHP}\renewcommand\colorSYNTAX{\colorSYNTAXP}{{\color{\colorMATH}\ensuremath{\epsilon , \delta }}}\endgroup }} {{\color{\colorSYNTAX}\texttt{\ensuremath{{\mathbb{R}}}}}} \}  
\vdash  {\begingroup\renewcommand\colorMATH{\colorMATHP}\renewcommand\colorSYNTAX{\colorSYNTAXP}{{\color{\colorMATH}\ensuremath{{{\color{\colorSYNTAX}\texttt{gauss}}}[{{\color{\colorSYNTAX}\texttt{\ensuremath{{\mathbb{R}}}}}}^{+}[{\begingroup\renewcommand\colorMATH{\colorMATHM}\renewcommand\colorSYNTAX{\colorSYNTAXM}{{\color{\colorMATH}\ensuremath{2.0}}}\endgroup }], {\begingroup\renewcommand\colorMATH{\colorMATHM}\renewcommand\colorSYNTAX{\colorSYNTAXM}{{\color{\colorMATH}\ensuremath{\epsilon }}}\endgroup }, {\begingroup\renewcommand\colorMATH{\colorMATHM}\renewcommand\colorSYNTAX{\colorSYNTAXM}{{\color{\colorMATH}\ensuremath{\delta }}}\endgroup }]\hspace*{0.33em}{<}{\begingroup\renewcommand\colorMATH{\colorMATHM}\renewcommand\colorSYNTAX{\colorSYNTAXM}{{\color{\colorMATH}\ensuremath{x}}}\endgroup }{>}\hspace*{0.33em}\{  {\begingroup\renewcommand\colorMATH{\colorMATHS}\renewcommand\colorSYNTAX{\colorSYNTAXS}{{\color{\colorMATH}\ensuremath{{\begingroup\renewcommand\colorMATH{\colorMATHM}\renewcommand\colorSYNTAX{\colorSYNTAXM}{{\color{\colorMATH}\ensuremath{x}}}\endgroup } + {\begingroup\renewcommand\colorMATH{\colorMATHM}\renewcommand\colorSYNTAX{\colorSYNTAXM}{{\color{\colorMATH}\ensuremath{x}}}\endgroup }}}}\endgroup } \}  }}}\endgroup } \mathrel{:} {{\color{\colorSYNTAX}\texttt{\ensuremath{{\mathbb{R}}}}}}
\vspace*{-1em}\end{gather*}\endgroup 
\noindent In a {{\color{\colorMATH}\ensuremath{{\begingroup\renewcommand\colorMATH{\colorMATHP}\renewcommand\colorSYNTAX{\colorSYNTAXP}{{\color{\colorMATH}\ensuremath{{{\color{\colorSYNTAX}\texttt{gauss}}}}}}\endgroup }}}} expression, the first three elements
(inside the square brackets) represent the maximum allowed sensitivity
of variables in the expression's body, and the desired privacy
parameters {{\color{\colorMATH}\ensuremath{\epsilon }}} and {{\color{\colorMATH}\ensuremath{\delta }}}. The fourth element (here, {{\color{\colorMATH}\ensuremath{{\begingroup\renewcommand\colorMATH{\colorMATHP}\renewcommand\colorSYNTAX{\colorSYNTAXP}{{\color{\colorMATH}\ensuremath{{<}{\begingroup\renewcommand\colorMATH{\colorMATHM}\renewcommand\colorSYNTAX{\colorSYNTAXM}{{\color{\colorMATH}\ensuremath{x}}}\endgroup }{>}}}}\endgroup }}}})
is a list of variables whose privacy we are interested in
tracking. Variables not in this list will be assigned infinite privacy
cost. 

The value of the {{\color{\colorMATH}\ensuremath{{\begingroup\renewcommand\colorMATH{\colorMATHP}\renewcommand\colorSYNTAX{\colorSYNTAXP}{{\color{\colorMATH}\ensuremath{{{\color{\colorSYNTAX}\texttt{gauss}}}}}}\endgroup }}}} expression is the value of its fifth
element (the ``body''), plus enough noise to ensure the desired level
of privacy. The body of a {{\color{\colorMATH}\ensuremath{{\begingroup\renewcommand\colorMATH{\colorMATHP}\renewcommand\colorSYNTAX{\colorSYNTAXP}{{\color{\colorMATH}\ensuremath{{{\color{\colorSYNTAX}\texttt{gauss}}}}}}\endgroup }}}} expression is a sensitivity
expression, and the {{\color{\colorMATH}\ensuremath{{\begingroup\renewcommand\colorMATH{\colorMATHP}\renewcommand\colorSYNTAX{\colorSYNTAXP}{{\color{\colorMATH}\ensuremath{{{\color{\colorSYNTAX}\texttt{gauss}}}}}}\endgroup }}}} expression is well-typed only if its
body typechecks in a context assigning a sensitivity to each variable
of interest which does not exceed the maximum allowed sensitivity. For
example, the expression {\begingroup\renewcommand\colorMATH{\colorMATHP}\renewcommand\colorSYNTAX{\colorSYNTAXP}{{\color{\colorMATH}\ensuremath{{{\color{\colorSYNTAX}\texttt{gauss}}}[{{\color{\colorSYNTAX}\texttt{\ensuremath{{\mathbb{R}}}}}}^{+}[{\begingroup\renewcommand\colorMATH{\colorMATHM}\renewcommand\colorSYNTAX{\colorSYNTAXM}{{\color{\colorMATH}\ensuremath{1.0}}}\endgroup }], {\begingroup\renewcommand\colorMATH{\colorMATHM}\renewcommand\colorSYNTAX{\colorSYNTAXM}{{\color{\colorMATH}\ensuremath{\epsilon }}}\endgroup },
{\begingroup\renewcommand\colorMATH{\colorMATHM}\renewcommand\colorSYNTAX{\colorSYNTAXM}{{\color{\colorMATH}\ensuremath{\delta }}}\endgroup }]\hspace*{0.33em}{<}{\begingroup\renewcommand\colorMATH{\colorMATHM}\renewcommand\colorSYNTAX{\colorSYNTAXM}{{\color{\colorMATH}\ensuremath{x}}}\endgroup }{>}\hspace*{0.33em}\{  {\begingroup\renewcommand\colorMATH{\colorMATHS}\renewcommand\colorSYNTAX{\colorSYNTAXS}{{\color{\colorMATH}\ensuremath{{\begingroup\renewcommand\colorMATH{\colorMATHM}\renewcommand\colorSYNTAX{\colorSYNTAXM}{{\color{\colorMATH}\ensuremath{x}}}\endgroup } + {\begingroup\renewcommand\colorMATH{\colorMATHM}\renewcommand\colorSYNTAX{\colorSYNTAXM}{{\color{\colorMATH}\ensuremath{x}}}\endgroup }}}}\endgroup } \}  }}}\endgroup } is not well-typed, because
{\begingroup\renewcommand\colorMATH{\colorMATHS}\renewcommand\colorSYNTAX{\colorSYNTAXS}{{\color{\colorMATH}\ensuremath{{\begingroup\renewcommand\colorMATH{\colorMATHM}\renewcommand\colorSYNTAX{\colorSYNTAXM}{{\color{\colorMATH}\ensuremath{x}}}\endgroup } + {\begingroup\renewcommand\colorMATH{\colorMATHM}\renewcommand\colorSYNTAX{\colorSYNTAXM}{{\color{\colorMATH}\ensuremath{x}}}\endgroup }}}}\endgroup } is 2-sensitive in {{\color{\colorMATH}\ensuremath{x}}}, but the maximum allowed
sensitivity is 1.

Privacy expressions like the example above are typed under a
\emph{privacy context} which records privacy cost for individual
variables. The context for this example ({{\color{\colorMATH}\ensuremath{\{  x \mathrel{:}_{{\begingroup\renewcommand\colorMATH{\colorMATHP}\renewcommand\colorSYNTAX{\colorSYNTAXP}{{\color{\colorMATH}\ensuremath{\epsilon , \delta }}}\endgroup }} {{\color{\colorSYNTAX}\texttt{\ensuremath{{\mathbb{R}}}}}} \} }}})
says that the expression provides {{\color{\colorMATH}\ensuremath{(\epsilon , \delta )}}}-differential privacy for
the variable {{\color{\colorMATH}\ensuremath{x}}}. Tracking privacy costs using a co-effect discipline
allows precise tracking of the privacy cost for programs with multiple
inputs:
\vspace*{-0.25em}\begingroup\color{\colorMATH}\begin{gather*} \{  x \mathrel{:}_{{\begingroup\renewcommand\colorMATH{\colorMATHP}\renewcommand\colorSYNTAX{\colorSYNTAXP}{{\color{\colorMATH}\ensuremath{\epsilon , \delta }}}\endgroup }} {{\color{\colorSYNTAX}\texttt{\ensuremath{{\mathbb{R}}}}}}, y \mathrel{:}_{{\begingroup\renewcommand\colorMATH{\colorMATHP}\renewcommand\colorSYNTAX{\colorSYNTAXP}{{\color{\colorMATH}\ensuremath{\epsilon , \delta }}}\endgroup }} {{\color{\colorSYNTAX}\texttt{\ensuremath{{\mathbb{R}}}}}} \}  
\vdash  {\begingroup\renewcommand\colorMATH{\colorMATHP}\renewcommand\colorSYNTAX{\colorSYNTAXP}{{\color{\colorMATH}\ensuremath{{{\color{\colorSYNTAX}\texttt{gauss}}}[{{\color{\colorSYNTAX}\texttt{\ensuremath{{\mathbb{R}}}}}}^{+}[{\begingroup\renewcommand\colorMATH{\colorMATHM}\renewcommand\colorSYNTAX{\colorSYNTAXM}{{\color{\colorMATH}\ensuremath{1.0}}}\endgroup }], {\begingroup\renewcommand\colorMATH{\colorMATHM}\renewcommand\colorSYNTAX{\colorSYNTAXM}{{\color{\colorMATH}\ensuremath{\epsilon }}}\endgroup }, {\begingroup\renewcommand\colorMATH{\colorMATHM}\renewcommand\colorSYNTAX{\colorSYNTAXM}{{\color{\colorMATH}\ensuremath{\delta }}}\endgroup }]\hspace*{0.33em}{<}{\begingroup\renewcommand\colorMATH{\colorMATHM}\renewcommand\colorSYNTAX{\colorSYNTAXM}{{\color{\colorMATH}\ensuremath{x,y}}}\endgroup }{>}\hspace*{0.33em}\{  {\begingroup\renewcommand\colorMATH{\colorMATHS}\renewcommand\colorSYNTAX{\colorSYNTAXS}{{\color{\colorMATH}\ensuremath{{\begingroup\renewcommand\colorMATH{\colorMATHM}\renewcommand\colorSYNTAX{\colorSYNTAXM}{{\color{\colorMATH}\ensuremath{x}}}\endgroup } + {\begingroup\renewcommand\colorMATH{\colorMATHM}\renewcommand\colorSYNTAX{\colorSYNTAXM}{{\color{\colorMATH}\ensuremath{y}}}\endgroup }}}}\endgroup } \}  }}}\endgroup } \mathrel{:} {{\color{\colorSYNTAX}\texttt{\ensuremath{{\mathbb{R}}}}}}
\vspace*{-1em}\end{gather*}\endgroup 
The {{\color{\colorTEXT}\textsc{\scriptsize Bind}}} rule encodes the sequential composition property of
differential privacy. For example:
\vspace*{-0.25em}\begingroup\color{\colorMATH}\begin{gather*}

\vspace*{-1em}\end{gather*}\endgroup 
\noindent In the example on the left, the Gaussian mechanism is
applied to {{\color{\colorMATH}\ensuremath{x}}} twice, so the total privacy cost for {{\color{\colorMATH}\ensuremath{x}}} is {{\color{\colorMATH}\ensuremath{(2\epsilon ,
2\delta )}}}. In the example on the right, {{\color{\colorMATH}\ensuremath{x}}} and {{\color{\colorMATH}\ensuremath{y}}} are each used once, and
their privacy costs are tracked separately. The {{\color{\colorTEXT}\textsc{\scriptsize Return}}} rule provides
a second interface between the sensitivity and privacy languages: a
{\begingroup\renewcommand\colorMATH{\colorMATHP}\renewcommand\colorSYNTAX{\colorSYNTAXP}{{\color{\colorMATH}\ensuremath{return}}}\endgroup } expression is part of the privacy language, but its
argument is a sensitivity expression. The value of a {\begingroup\renewcommand\colorMATH{\colorMATHP}\renewcommand\colorSYNTAX{\colorSYNTAXP}{{\color{\colorMATH}\ensuremath{return}}}\endgroup }
expression is exactly the value of its argument, so the variables used
in its argument are assigned \emph{infinite} privacy cost. {\begingroup\renewcommand\colorMATH{\colorMATHP}\renewcommand\colorSYNTAX{\colorSYNTAXP}{{\color{\colorMATH}\ensuremath{return}}}\endgroup }
expressions are therefore typically used to compute on values which
are \emph{already} differentially private (like {{\color{\colorMATH}\ensuremath{v_{1}}}} and {{\color{\colorMATH}\ensuremath{v_{2}}}} above),
since infinite privacy cost is not a problem in that case.

\paragraph{Gradient descent.}
Machine learning problems are typically defined in terms of a \emph{loss function} {{\color{\colorMATH}\ensuremath{{\mathcal{L}}(\theta ; X, y)}}} on a \emph{model} {{\color{\colorMATH}\ensuremath{\theta }}}, \emph{training samples} {{\color{\colorMATH}\ensuremath{X = (x_{1}, x_{2}, {.}\hspace{-1pt}{.}\hspace{-1pt}{.}, x_{n})}}} (in which each sample is typically represented as a \emph{feature vector}) and corresponding \emph{labels} {{\color{\colorMATH}\ensuremath{y = (y_{1}, y_{2}, {.}\hspace{-1pt}{.}\hspace{-1pt}{.}, y_{n}}}}) (i.e. the prediction target). The training task is to find a model {{\color{\colorMATH}\ensuremath{\hat \theta }}} which minimizes the loss on the training samples (i.e. {{\color{\colorMATH}\ensuremath{\hat \theta  = {{\color{\colorMATH}\ensuremath{\operatorname{argmin}}}}_{\theta } {\mathcal{L}}(\theta ; X, y)}}}.

One solution to the training task is \emph{gradient descent}, which starts with an initial guess for {{\color{\colorMATH}\ensuremath{\theta }}} and iteratively moves in the direction of an improved {{\color{\colorMATH}\ensuremath{\theta }}} until the current setting is close to {{\color{\colorMATH}\ensuremath{\hat \theta }}}. To determine which direction to move, the algorithm evaluates the \emph{gradient} of the loss, which yields a vector representing the direction of greatest \emph{increase} in {{\color{\colorMATH}\ensuremath{{\mathcal{L}}(\theta ; X, y)}}}. Then, the algorithm moves in the \emph{opposite} direction.

To ensure differential privacy for gradient-based algorithms, we need
to bound the sensitivity of the gradient computation. The gradients
for many kinds of convex loss functions are
\emph{$1$-Lipschitz}~\cite{PSGD}: if each sample in {{\color{\colorMATH}\ensuremath{X = (x_{1}, {.}\hspace{-1pt}{.}\hspace{-1pt}{.}, x_{n})}}}
has bounded {{\color{\colorMATH}\ensuremath{L2}}} norm (i.e. {{\color{\colorMATH}\ensuremath{\| x_{i}\| _{2} \leq  1}}}), then for all models {{\color{\colorMATH}\ensuremath{\theta }}} and
labelings {{\color{\colorMATH}\ensuremath{y}}}, the gradient {{\color{\colorMATH}\ensuremath{{\mathrel{\nabla }}(\theta ; X, y)}}} has {{\color{\colorMATH}\ensuremath{L2}}} sensitivity bounded
by 1. For now, we will assume the existence of a function called
{{\color{\colorMATH}\ensuremath{{\begingroup\renewcommand\colorMATH{\colorMATHS}\renewcommand\colorSYNTAX{\colorSYNTAXS}{{\color{\colorMATH}\ensuremath{{{\color{\colorMATH}\ensuremath{\operatorname{gradient}}}}}}}\endgroup }}}} with this property (more details in
Section~\ref{sec:machine_learning}).
\vspace*{-0.25em}\begingroup\color{\colorMATH}\begin{gather*}
  \begin{array}{rcl@{\hspace*{1.00em}\hspace*{1.00em}}l
  }  {\begingroup\renewcommand\colorMATH{\colorMATHS}\renewcommand\colorSYNTAX{\colorSYNTAXS}{{\color{\colorMATH}\ensuremath{{{\color{\colorMATH}\ensuremath{\operatorname{gradient}}}}}}}\endgroup }    &{}\mathrel{:}{}& \multicolumn{2}{l}{{\begingroup\renewcommand\colorMATH{\colorMATHM}\renewcommand\colorSYNTAX{\colorSYNTAXM}{{\color{\colorMATH}\ensuremath{{{\color{\colorSYNTAX}\texttt{\ensuremath{{\mathbb{M}}_{L2}^{{{\color{\colorSYNTAX}\texttt{U}}}}[{{\color{\colorMATH}\ensuremath{1}}},{{\color{\colorMATH}\ensuremath{n}}}]\hspace*{0.33em}{{\color{\colorSYNTAX}\texttt{\ensuremath{{\mathbb{R}}}}}} 
                                         \multimap _{{\begingroup\renewcommand\colorMATH{\colorMATHS}\renewcommand\colorSYNTAX{\colorSYNTAXS}{{\color{\colorMATH}\ensuremath{{{\color{\colorSYNTAX}\texttt{\ensuremath{\infty }}}}}}}\endgroup }} {\mathbb{M}}_{L\infty }^{{{\color{\colorSYNTAX}\texttt{U}}}}[{{\color{\colorMATH}\ensuremath{m}}},{{\color{\colorMATH}\ensuremath{n}}}]\hspace*{0.33em}{\mathbb{D}} 
                                         \multimap _{{{\color{\colorMATH}\ensuremath{\frac{1}{m}}}}} {\mathbb{M}}_{L\infty }^{{{\color{\colorSYNTAX}\texttt{U}}}}[{{\color{\colorMATH}\ensuremath{m}}},{{\color{\colorMATH}\ensuremath{1}}}]\hspace*{0.33em}{\mathbb{D}} 
                                         \multimap _{{{\color{\colorMATH}\ensuremath{\frac{1}{m}}}}} {\mathbb{M}}_{L2}^{{{\color{\colorSYNTAX}\texttt{U}}}}[{{\color{\colorMATH}\ensuremath{1}}},{{\color{\colorMATH}\ensuremath{n}}}]\hspace*{0.33em}{{\color{\colorSYNTAX}\texttt{\ensuremath{{\mathbb{R}}}}}}}}}}}}}\endgroup }}
\end{array}
\vspace*{-1em}\end{gather*}\endgroup 
The function's arguments are the current {{\color{\colorMATH}\ensuremath{\theta }}}, a {{\color{\colorMATH}\ensuremath{m \times  n}}} matrix {{\color{\colorMATH}\ensuremath{X}}}
containing {{\color{\colorMATH}\ensuremath{n}}} training samples, and a {{\color{\colorMATH}\ensuremath{1 \times  n}}} matrix {{\color{\colorMATH}\ensuremath{y}}} containing
the corresponding labels. In Duet, the type {{\color{\colorMATH}\ensuremath{{\mathbb{M}}_{L\infty }^{{{\color{\colorSYNTAX}\texttt{U}}}}[{{\color{\colorMATH}\ensuremath{m}}},{{\color{\colorMATH}\ensuremath{n}}}]\hspace*{0.33em}{\mathbb{D}}}}}
represents a {{\color{\colorMATH}\ensuremath{m \times  n}}} matrix of \emph{discrete} real numbers;
neighboring matrices of this type differ arbitrarily in a single
row. The function's output is a new {{\color{\colorMATH}\ensuremath{\theta }}} of type
{{\color{\colorMATH}\ensuremath{{\mathbb{M}}_{L2}^{{{\color{\colorSYNTAX}\texttt{U}}}}[{{\color{\colorMATH}\ensuremath{1}}},{{\color{\colorMATH}\ensuremath{n}}}]\hspace*{0.33em}{{\color{\colorSYNTAX}\texttt{\ensuremath{{\mathbb{R}}}}}}}}}, representing a matrix of real numbers with
bounded {{\color{\colorMATH}\ensuremath{L2}}} sensitivity (see Section~\ref{sec:machine_learning} for
details on matrix types). We can use the {{\color{\colorMATH}\ensuremath{{{\color{\colorMATH}\ensuremath{\operatorname{gradient}}}}}}} function to
implement a differentially private gradient descent algorithm:
\vspace*{-0.25em}\begingroup\color{\colorMATH}\begin{gather*}
\begingroup\renewcommand\colorMATH{\colorMATHP}\renewcommand\colorSYNTAX{\colorSYNTAXP}\begingroup\color{\colorMATH}
\begin{array}{l
} {{\color{\colorMATH}\ensuremath{\operatorname{noisy-gradient-descent}}}}({\begingroup\renewcommand\colorMATH{\colorMATHM}\renewcommand\colorSYNTAX{\colorSYNTAXM}{{\color{\colorMATH}\ensuremath{X}}}\endgroup },{\begingroup\renewcommand\colorMATH{\colorMATHM}\renewcommand\colorSYNTAX{\colorSYNTAXM}{{\color{\colorMATH}\ensuremath{y}}}\endgroup },{\begingroup\renewcommand\colorMATH{\colorMATHM}\renewcommand\colorSYNTAX{\colorSYNTAXM}{{\color{\colorMATH}\ensuremath{k}}}\endgroup },{\begingroup\renewcommand\colorMATH{\colorMATHM}\renewcommand\colorSYNTAX{\colorSYNTAXM}{{\color{\colorMATH}\ensuremath{\epsilon }}}\endgroup },{\begingroup\renewcommand\colorMATH{\colorMATHM}\renewcommand\colorSYNTAX{\colorSYNTAXM}{{\color{\colorMATH}\ensuremath{\delta }}}\endgroup }) \triangleq 
\cr  \hspace*{1.00em} {{\color{\colorSYNTAX}\texttt{\ensuremath{{{\color{\colorSYNTAX}\texttt{let}}}\hspace*{0.33em}{\begingroup\renewcommand\colorMATH{\colorMATHM}\renewcommand\colorSYNTAX{\colorSYNTAXM}{{\color{\colorMATH}\ensuremath{\theta _{0}}}}\endgroup } = {\begingroup\renewcommand\colorMATH{\colorMATHS}\renewcommand\colorSYNTAX{\colorSYNTAXS}{{\color{\colorMATH}\ensuremath{{{\color{\colorMATH}\ensuremath{\operatorname{zeros}}}}\hspace*{0.33em}({{\color{\colorMATH}\ensuremath{\operatorname{cols}}}}\hspace*{0.33em}{\begingroup\renewcommand\colorMATH{\colorMATHM}\renewcommand\colorSYNTAX{\colorSYNTAXM}{{\color{\colorMATH}\ensuremath{X_{1}}}}\endgroup })}}}\endgroup }\hspace*{0.33em}{{\color{\colorSYNTAX}\texttt{in}}}}}}}
\cr
\end{array}\endgroup \endgroup 
\vspace*{-1em}\end{gather*}\endgroup 
\noindent The arguments to our algorithm are the training data ({{\color{\colorMATH}\ensuremath{X}}}
and {{\color{\colorMATH}\ensuremath{y}}}), the desired number of iterations {{\color{\colorMATH}\ensuremath{k}}}, and the privacy
parameters {{\color{\colorMATH}\ensuremath{\epsilon }}} and {{\color{\colorMATH}\ensuremath{\delta }}}. The first line constructs an initial model
{{\color{\colorMATH}\ensuremath{\theta _{0}}}} consisting of zeros for all parameters. Lines 2-4 represent the
iterative part of the algorithm: {{\color{\colorMATH}\ensuremath{k}}} times, compute the gradient of
the loss on {{\color{\colorMATH}\ensuremath{X}}} and {{\color{\colorMATH}\ensuremath{y}}} with respect to the current model, add noise
to the gradient using the Gaussian mechanism, and subtract the
gradient from the current model (thus moving in the opposite direction
of the gradient) to improve the model.

The typing rules presented in Figure~\ref{fig:typing} allow us to
derive a privacy bound for this algorithm which is equivalent to
manual proof of Bassily et al.~\cite{BST}. Based on the type of the
{{\color{\colorMATH}\ensuremath{{\begingroup\renewcommand\colorMATH{\colorMATHS}\renewcommand\colorSYNTAX{\colorSYNTAXS}{{\color{\colorMATH}\ensuremath{{{\color{\colorMATH}\ensuremath{\operatorname{gradient}}}}}}}\endgroup }}}} function, the {{\color{\colorMATH}\ensuremath{{{\color{\colorSYNTAX}\texttt{\ensuremath{\multimap }}}}{{\color{\colorTEXT}\textsc{\scriptsize -E}}}}}} rule allows us to conclude
that the gradient operation is {{\color{\colorMATH}\ensuremath{\frac{1}{m}}}}-sensitive in the training
data, which is reflected by the sensitivity annotations in the
context:
\vspace*{-0.25em}\begingroup\color{\colorMATH}\begin{gather*}
\begin{array}{l
} \{  \theta  \mathrel{:}_{\infty } \tau _{1},
     X \mathrel{:}_{\frac{1}{m}} \tau _{2},
     y \mathrel{:}_{\frac{1}{m}} \tau _{3} \} 
     \vdash  {\begingroup\renewcommand\colorMATH{\colorMATHM}\renewcommand\colorSYNTAX{\colorSYNTAXM}{{\color{\colorMATH}\ensuremath{{\begingroup\renewcommand\colorMATH{\colorMATHS}\renewcommand\colorSYNTAX{\colorSYNTAXS}{{\color{\colorMATH}\ensuremath{{{\color{\colorMATH}\ensuremath{\operatorname{gradient}}}}}}}\endgroup }\hspace*{0.33em}\theta \hspace*{0.33em}X\hspace*{0.33em}y }}}\endgroup } \mathrel{:} {\mathbb{M}}_{L2}^{{{\color{\colorSYNTAX}\texttt{U}}}}[{{\color{\colorMATH}\ensuremath{1}}},{{\color{\colorMATH}\ensuremath{n}}}]\hspace*{0.33em}{{\color{\colorSYNTAX}\texttt{\ensuremath{{\mathbb{R}}}}}}
\smallskip
\cr  {{\color{\colorTEXT}\textit{where}}}\hspace*{0.33em}\tau _{1} = {\mathbb{M}}_{L2}^{{{\color{\colorSYNTAX}\texttt{U}}}}[{{\color{\colorMATH}\ensuremath{1}}},{{\color{\colorMATH}\ensuremath{n}}}]\hspace*{0.33em}{{\color{\colorSYNTAX}\texttt{\ensuremath{{\mathbb{R}}}}}}
\cr  \phantom{{{\color{\colorTEXT}\textit{where}}}\hspace*{0.33em}} \tau _{2} = {\mathbb{M}}_{L\infty }^{{{\color{\colorSYNTAX}\texttt{U}}}}[{{\color{\colorMATH}\ensuremath{m}}},{{\color{\colorMATH}\ensuremath{n}}}]\hspace*{0.33em}{\mathbb{D}}
\cr  \phantom{{{\color{\colorTEXT}\textit{where}}}\hspace*{0.33em}} \tau _{3} = {\mathbb{M}}_{L\infty }^{{{\color{\colorSYNTAX}\texttt{U}}}}[{{\color{\colorMATH}\ensuremath{m}}},{{\color{\colorMATH}\ensuremath{1}}}]\hspace*{0.33em}{\mathbb{D}}
   \end{array}\vspace*{-1em}\end{gather*}\endgroup 
   \noindent Next, the {{\color{\colorMATH}\ensuremath{{{\color{\colorTEXT}\textsc{\scriptsize MGauss}}}}}} rule represents the use of the
   Gaussian mechanism, and transitions from the sensitivity language
   (implementing the gradient) to the privacy language (in which we
   use the noisy gradient). The rule allows us to conclude that since
   the sensitivity of the gradient computation is {{\color{\colorMATH}\ensuremath{\frac{1}{m}}}}, our
   use of the Gaussian mechanism satisfies {{\color{\colorMATH}\ensuremath{(\epsilon , \delta )}}}-differential
   privacy. This context is a \emph{privacy} context, and its
   annotations represent privacy costs rather than sensitivities.
\vspace*{-0.25em}\begingroup\color{\colorMATH}\begin{gather*} \{  \theta  \mathrel{:}_{\infty } \tau _{1},
     X \mathrel{:}_{\langle \epsilon , \delta \rangle } \tau _{2},
     y \mathrel{:}_{\langle \epsilon , \delta \rangle } \tau _{3} \} 
   \vdash  {{\color{\colorSYNTAX}\texttt{mgauss}}}[ {\begingroup\renewcommand\colorMATH{\colorMATHM}\renewcommand\colorSYNTAX{\colorSYNTAXM}{{\color{\colorMATH}\ensuremath{\frac{1}{m}}}}\endgroup }, {\begingroup\renewcommand\colorMATH{\colorMATHM}\renewcommand\colorSYNTAX{\colorSYNTAXM}{{\color{\colorMATH}\ensuremath{\epsilon }}}\endgroup } , {\begingroup\renewcommand\colorMATH{\colorMATHM}\renewcommand\colorSYNTAX{\colorSYNTAXM}{{\color{\colorMATH}\ensuremath{\delta }}}\endgroup } ]\hspace*{0.33em}{<}{\begingroup\renewcommand\colorMATH{\colorMATHM}\renewcommand\colorSYNTAX{\colorSYNTAXM}{{\color{\colorMATH}\ensuremath{X}}}\endgroup },{\begingroup\renewcommand\colorMATH{\colorMATHM}\renewcommand\colorSYNTAX{\colorSYNTAXM}{{\color{\colorMATH}\ensuremath{y}}}\endgroup }{>}\hspace*{0.33em}\{  {\begingroup\renewcommand\colorMATH{\colorMATHM}\renewcommand\colorSYNTAX{\colorSYNTAXM}{{\color{\colorMATH}\ensuremath{{\begingroup\renewcommand\colorMATH{\colorMATHS}\renewcommand\colorSYNTAX{\colorSYNTAXS}{{\color{\colorMATH}\ensuremath{{{\color{\colorMATH}\ensuremath{\operatorname{gradient}}}}}}}\endgroup }\hspace*{0.33em}\theta \hspace*{0.33em}X\hspace*{0.33em}y }}}\endgroup } \}  \mathrel{:} {\mathbb{M}}_{L2}^{{{\color{\colorSYNTAX}\texttt{U}}}}[{{\color{\colorMATH}\ensuremath{1}}},{{\color{\colorMATH}\ensuremath{n}}}]\hspace*{0.33em}{{\color{\colorSYNTAX}\texttt{\ensuremath{{\mathbb{R}}}}}} \vspace*{-1em}\end{gather*}\endgroup 
   \noindent Finally, the {{\color{\colorMATH}\ensuremath{{{\color{\colorTEXT}\textsc{\scriptsize Loop}}}}}} rule for advanced composition
   allows us to derive a bound on the total privacy cost of the
   iterative algorithm, based on the number of times the loop runs:
\vspace*{-0.25em}\begingroup\color{\colorMATH}\begin{gather*}\begin{array}{l
} \{  \theta  \mathrel{:}_{\infty } \tau _{1},
   X \mathrel{:}_{\langle \epsilon ^{\prime}, k\delta +\delta ^{\prime}\rangle } \tau _{2},
   y \mathrel{:}_{\langle \epsilon ^{\prime}, k\delta +\delta ^{\prime}\rangle } \tau _{3} \} 
 \vdash  {{\color{\colorSYNTAX}\texttt{\ensuremath{{{\color{\colorSYNTAX}\texttt{loop}}}[{\begingroup\renewcommand\colorMATH{\colorMATHM}\renewcommand\colorSYNTAX{\colorSYNTAXM}{{\color{\colorMATH}\ensuremath{\delta ^{\prime}}}}\endgroup }]\hspace*{0.33em}{\begingroup\renewcommand\colorMATH{\colorMATHM}\renewcommand\colorSYNTAX{\colorSYNTAXM}{{\color{\colorMATH}\ensuremath{k}}}\endgroup }\hspace*{0.33em}{{\color{\colorSYNTAX}\texttt{on}}}\hspace*{0.33em}{\begingroup\renewcommand\colorMATH{\colorMATHM}\renewcommand\colorSYNTAX{\colorSYNTAXM}{{\color{\colorMATH}\ensuremath{\theta _{0}}}}\endgroup }\hspace*{0.33em}{<}{\begingroup\renewcommand\colorMATH{\colorMATHM}\renewcommand\colorSYNTAX{\colorSYNTAXM}{{\color{\colorMATH}\ensuremath{X_{1}}}}\endgroup },{\begingroup\renewcommand\colorMATH{\colorMATHM}\renewcommand\colorSYNTAX{\colorSYNTAXM}{{\color{\colorMATH}\ensuremath{y}}}\endgroup }{>}\hspace*{0.33em}\{  {\begingroup\renewcommand\colorMATH{\colorMATHM}\renewcommand\colorSYNTAX{\colorSYNTAXM}{{\color{\colorMATH}\ensuremath{t}}}\endgroup }, {\begingroup\renewcommand\colorMATH{\colorMATHM}\renewcommand\colorSYNTAX{\colorSYNTAXM}{{\color{\colorMATH}\ensuremath{\theta }}}\endgroup } \Rightarrow  {.}\hspace{-1pt}{.}\hspace{-1pt}{.} \}  }}}}\mathrel{:} {\mathbb{M}}_{L2}^{{{\color{\colorSYNTAX}\texttt{U}}}}[{{\color{\colorMATH}\ensuremath{1}}},{{\color{\colorMATH}\ensuremath{n}}}]\hspace*{0.33em}{{\color{\colorSYNTAX}\texttt{\ensuremath{{\mathbb{R}}}}}}
\cr  {{\color{\colorTEXT}\textit{where}}}\hspace*{0.33em}\epsilon ^{\prime} =  2\epsilon \sqrt {2k\log (1/\delta ^{\prime})} 
\end{array}\vspace*{-1em}\end{gather*}\endgroup 
\paragraph{Variants of Differential Privacy.}
The typing rules presented in Figure~\ref{fig:typing} are specific to
{{\color{\colorMATH}\ensuremath{(\epsilon , \delta )}}}-differential privacy, but the same framework can be easily
extended to support the other variants described in
Figure~\ref{fig:typing_variants}.
New variants can be supported by making three simple changes: (1)
Modify the \emph{privacy cost} syntax {{\color{\colorMATH}\ensuremath{{\begingroup\renewcommand\colorMATH{\colorMATHP}\renewcommand\colorSYNTAX{\colorSYNTAXP}{{\color{\colorMATH}\ensuremath{p}}}\endgroup }}}} to describe the privacy
parameters of the new variant; (2) Modify the sum operator {\begingroup\renewcommand\colorMATH{\colorMATHP}\renewcommand\colorSYNTAX{\colorSYNTAXP}{{\color{\colorMATH}\ensuremath{\underline{\hspace{0.66em}\vspace*{5ex}}{+}\underline{\hspace{0.66em}\vspace*{5ex}}}}}\endgroup }
to reflect sequential composition in the new variant; and (3) Modify
the typing for basic mechanisms (e.g. {{\color{\colorMATH}\ensuremath{{\begingroup\renewcommand\colorMATH{\colorMATHP}\renewcommand\colorSYNTAX{\colorSYNTAXP}{{\color{\colorMATH}\ensuremath{{{\color{\colorSYNTAX}\texttt{gauss}}}}}}\endgroup }}}}) to reflect
corresponding mechanisms in the new
variant. The extended version of this paper includes typing rules for the
variants in Figure~\ref{fig:typing_variants}.

As an example, considering the following variant of the noisy gradient
descent algorithm presented earlier, but with {{\color{\colorMATH}\ensuremath{\rho }}}-zCDP instead of {{\color{\colorMATH}\ensuremath{(\epsilon ,
\delta )}}}-differential privacy. There are only two differences: the {{\color{\colorMATH}\ensuremath{{{\color{\colorSYNTAX}\texttt{loop}}}}}}
construct under zCDP has no {{\color{\colorMATH}\ensuremath{\delta ^{\prime}}}} parameter, since standard composition
yields tight bounds, and the {{\color{\colorMATH}\ensuremath{{{\color{\colorSYNTAX}\texttt{mgauss}}}}}} construct has a single privacy
parameter ({{\color{\colorMATH}\ensuremath{\rho }}}) instead of {{\color{\colorMATH}\ensuremath{\epsilon }}} and {{\color{\colorMATH}\ensuremath{\delta }}}.
\vspace*{-0.25em}\begingroup\color{\colorMATH}\begin{gather*}
\begingroup\renewcommand\colorMATH{\colorMATHP}\renewcommand\colorSYNTAX{\colorSYNTAXP}\begingroup\color{\colorMATH}
\begin{array}{l
} {{\color{\colorMATH}\ensuremath{\operatorname{noisy-gradient-descent}}}}({\begingroup\renewcommand\colorMATH{\colorMATHM}\renewcommand\colorSYNTAX{\colorSYNTAXM}{{\color{\colorMATH}\ensuremath{X}}}\endgroup },{\begingroup\renewcommand\colorMATH{\colorMATHM}\renewcommand\colorSYNTAX{\colorSYNTAXM}{{\color{\colorMATH}\ensuremath{y}}}\endgroup },{\begingroup\renewcommand\colorMATH{\colorMATHM}\renewcommand\colorSYNTAX{\colorSYNTAXM}{{\color{\colorMATH}\ensuremath{k}}}\endgroup },{\begingroup\renewcommand\colorMATH{\colorMATHM}\renewcommand\colorSYNTAX{\colorSYNTAXM}{{\color{\colorMATH}\ensuremath{\rho }}}\endgroup }) \triangleq 
\cr  \hspace*{1.00em} {{\color{\colorSYNTAX}\texttt{\ensuremath{{{\color{\colorSYNTAX}\texttt{let}}}\hspace*{0.33em}{\begingroup\renewcommand\colorMATH{\colorMATHM}\renewcommand\colorSYNTAX{\colorSYNTAXM}{{\color{\colorMATH}\ensuremath{\theta _{0}}}}\endgroup } = {\begingroup\renewcommand\colorMATH{\colorMATHS}\renewcommand\colorSYNTAX{\colorSYNTAXS}{{\color{\colorMATH}\ensuremath{{{\color{\colorMATH}\ensuremath{\operatorname{zeros}}}}\hspace*{0.33em}({{\color{\colorMATH}\ensuremath{\operatorname{cols}}}}\hspace*{0.33em}{\begingroup\renewcommand\colorMATH{\colorMATHM}\renewcommand\colorSYNTAX{\colorSYNTAXM}{{\color{\colorMATH}\ensuremath{X_{1}}}}\endgroup })}}}\endgroup }\hspace*{0.33em}{{\color{\colorSYNTAX}\texttt{in}}}}}}}
\cr
\end{array}\endgroup \endgroup 
\vspace*{-1em}\end{gather*}\endgroup 
Typechecking for this version proceeds in the same way as before, with
the modified typing rules; the resulting privacy context gives both
{{\color{\colorMATH}\ensuremath{X}}} and {{\color{\colorMATH}\ensuremath{y}}} a privacy cost of {{\color{\colorMATH}\ensuremath{k\rho }}}.

\paragraph{Mixing Variants.}
\system allows mixing variants of differential privacy in a single
program. For example, the total privacy cost of an algorithm is often
given in {{\color{\colorMATH}\ensuremath{(\epsilon , \delta )}}} form, to enable comparing the costs of different
algorithms; we can use this feature of \system to automatically derive
the cost of our zCDP-based gradient descent in terms of {{\color{\colorMATH}\ensuremath{\epsilon }}} and {{\color{\colorMATH}\ensuremath{\delta }}}.
\vspace*{-0.25em}\begingroup\color{\colorMATH}\begin{gather*}
\begingroup\renewcommand\colorMATH{\colorMATHP}\renewcommand\colorSYNTAX{\colorSYNTAXP}\begingroup\color{\colorMATH}
\begin{array}{l
} {{\color{\colorMATH}\ensuremath{\operatorname{noisy-gradient-descent}}}}({\begingroup\renewcommand\colorMATH{\colorMATHM}\renewcommand\colorSYNTAX{\colorSYNTAXM}{{\color{\colorMATH}\ensuremath{X}}}\endgroup },{\begingroup\renewcommand\colorMATH{\colorMATHM}\renewcommand\colorSYNTAX{\colorSYNTAXM}{{\color{\colorMATH}\ensuremath{y}}}\endgroup },{\begingroup\renewcommand\colorMATH{\colorMATHM}\renewcommand\colorSYNTAX{\colorSYNTAXM}{{\color{\colorMATH}\ensuremath{k}}}\endgroup },{\begingroup\renewcommand\colorMATH{\colorMATHM}\renewcommand\colorSYNTAX{\colorSYNTAXM}{{\color{\colorMATH}\ensuremath{\rho }}}\endgroup },{\begingroup\renewcommand\colorMATH{\colorMATHM}\renewcommand\colorSYNTAX{\colorSYNTAXM}{{\color{\colorMATH}\ensuremath{\delta }}}\endgroup }) \triangleq 
\cr  \hspace*{1.00em} {{\color{\colorSYNTAX}\texttt{\ensuremath{{{\color{\colorSYNTAX}\texttt{let}}}\hspace*{0.33em}{\begingroup\renewcommand\colorMATH{\colorMATHM}\renewcommand\colorSYNTAX{\colorSYNTAXM}{{\color{\colorMATH}\ensuremath{\theta _{0}}}}\endgroup } = {\begingroup\renewcommand\colorMATH{\colorMATHS}\renewcommand\colorSYNTAX{\colorSYNTAXS}{{\color{\colorMATH}\ensuremath{{{\color{\colorMATH}\ensuremath{\operatorname{zeros}}}}\hspace*{0.33em}({{\color{\colorMATH}\ensuremath{\operatorname{cols}}}}\hspace*{0.33em}{\begingroup\renewcommand\colorMATH{\colorMATHM}\renewcommand\colorSYNTAX{\colorSYNTAXM}{{\color{\colorMATH}\ensuremath{X_{1}}}}\endgroup })}}}\endgroup }\hspace*{0.33em}{{\color{\colorSYNTAX}\texttt{in}}}}}}}
\cr
\end{array}\endgroup \endgroup 
\vspace*{-1em}\end{gather*}\endgroup 
%
The {{\color{\colorMATH}\ensuremath{{\begingroup\renewcommand\colorMATH{\colorMATHP}\renewcommand\colorSYNTAX{\colorSYNTAXP}{{\color{\colorMATH}\ensuremath{{{\color{\colorSYNTAX}\texttt{ZCDP}}}\hspace*{0.33em}\{ {.}\hspace{-1pt}{.}\hspace{-1pt}{.}\} }}}\endgroup }}}} construct represents embedding a mechanism which
satisfies {{\color{\colorMATH}\ensuremath{\rho }}}-zCDP in another mechanism which provides {{\color{\colorMATH}\ensuremath{(\epsilon ,
\delta )}}}-differential privacy. The rule for typechecking this construct
encodes the property that if a mechanism satisfies {{\color{\colorMATH}\ensuremath{\rho }}}-zCDP, it also
satisfies {{\color{\colorMATH}\ensuremath{(\rho  + 2\sqrt {\rho \log (1/\delta )}, \delta )}}}-differential
privacy~\cite{bun2016concentrated}. Using this rule, we can derive a
total privacy cost for the gradient descent algorithm in terms of {{\color{\colorMATH}\ensuremath{\epsilon }}}
and {{\color{\colorMATH}\ensuremath{\delta }}}, but using the tight bound on composition that zCDP provides.
\vspace*{-0.25em}\begingroup\color{\colorMATH}\begin{gather*}\begin{array}{l
} \{  X \mathrel{:}_{\langle \epsilon ^{\prime}, \delta \rangle } \tau _{2},
  y \mathrel{:}_{\langle \epsilon ^{\prime}, \delta \rangle } \tau _{3},
  k \mathrel{:}_{\infty } {{\color{\colorSYNTAX}\texttt{\ensuremath{{\mathbb{R}}}}}}^{+}[k],
  \rho  \mathrel{:}_{\infty } {{\color{\colorSYNTAX}\texttt{\ensuremath{{\mathbb{R}}}}}}^{+}[\rho ] \} 
 \vdash  {\begingroup\renewcommand\colorMATH{\colorMATHP}\renewcommand\colorSYNTAX{\colorSYNTAXP}{{\color{\colorMATH}\ensuremath{{{\color{\colorMATH}\ensuremath{\operatorname{noisy-gradient-descent}}}}({\begingroup\renewcommand\colorMATH{\colorMATHM}\renewcommand\colorSYNTAX{\colorSYNTAXM}{{\color{\colorMATH}\ensuremath{X}}}\endgroup },{\begingroup\renewcommand\colorMATH{\colorMATHM}\renewcommand\colorSYNTAX{\colorSYNTAXM}{{\color{\colorMATH}\ensuremath{y}}}\endgroup },{\begingroup\renewcommand\colorMATH{\colorMATHM}\renewcommand\colorSYNTAX{\colorSYNTAXM}{{\color{\colorMATH}\ensuremath{k}}}\endgroup },{\begingroup\renewcommand\colorMATH{\colorMATHM}\renewcommand\colorSYNTAX{\colorSYNTAXM}{{\color{\colorMATH}\ensuremath{\rho }}}\endgroup },{\begingroup\renewcommand\colorMATH{\colorMATHM}\renewcommand\colorSYNTAX{\colorSYNTAXM}{{\color{\colorMATH}\ensuremath{\delta }}}\endgroup })}}}\endgroup } \mathrel{:} {\mathbb{M}}_{L2}^{{{\color{\colorSYNTAX}\texttt{U}}}}[{{\color{\colorMATH}\ensuremath{1}}},{{\color{\colorMATH}\ensuremath{n}}}]\hspace*{0.33em}{{\color{\colorSYNTAX}\texttt{\ensuremath{{\mathbb{R}}}}}}
\cr  {{\color{\colorTEXT}\textit{where}}}\hspace*{0.33em}\epsilon ^{\prime} =  k\rho  + 2\sqrt {k\rho  \log (1/\delta )}
\end{array}\vspace*{-1em}\end{gather*}\endgroup 
We might also want to nest these conversions. For example, when the
dimensionality of the training data is very small, the Laplace
mechanism might yield more accurate results than the Gaussian
mechanism (due to the shape of the distribution). To use the Laplace
mechanism in an iterative algorithm which satisfies zCDP, we can use
the fact that any {{\color{\colorMATH}\ensuremath{\epsilon }}}-differentially private mechanism also satisfies
{{\color{\colorMATH}\ensuremath{\frac{1}{2} \epsilon ^{2}}}}-zCDP; by nesting conversions, we can determine the
total cost of the algorithm in terms of {{\color{\colorMATH}\ensuremath{\epsilon }}} and {{\color{\colorMATH}\ensuremath{\delta }}}.
\vspace*{-0.25em}\begingroup\color{\colorMATH}\begin{gather*}
\begingroup\renewcommand\colorMATH{\colorMATHP}\renewcommand\colorSYNTAX{\colorSYNTAXP}\begingroup\color{\colorMATH}
\begin{array}{l
} {{\color{\colorMATH}\ensuremath{\operatorname{noisy-gradient-descent}}}}({\begingroup\renewcommand\colorMATH{\colorMATHM}\renewcommand\colorSYNTAX{\colorSYNTAXM}{{\color{\colorMATH}\ensuremath{X}}}\endgroup },{\begingroup\renewcommand\colorMATH{\colorMATHM}\renewcommand\colorSYNTAX{\colorSYNTAXM}{{\color{\colorMATH}\ensuremath{y}}}\endgroup },{\begingroup\renewcommand\colorMATH{\colorMATHM}\renewcommand\colorSYNTAX{\colorSYNTAXM}{{\color{\colorMATH}\ensuremath{k}}}\endgroup },{\begingroup\renewcommand\colorMATH{\colorMATHM}\renewcommand\colorSYNTAX{\colorSYNTAXM}{{\color{\colorMATH}\ensuremath{\epsilon }}}\endgroup },{\begingroup\renewcommand\colorMATH{\colorMATHM}\renewcommand\colorSYNTAX{\colorSYNTAXM}{{\color{\colorMATH}\ensuremath{\delta }}}\endgroup }) \triangleq 
\cr  \hspace*{1.00em} {{\color{\colorSYNTAX}\texttt{\ensuremath{{{\color{\colorSYNTAX}\texttt{let}}}\hspace*{0.33em}{\begingroup\renewcommand\colorMATH{\colorMATHM}\renewcommand\colorSYNTAX{\colorSYNTAXM}{{\color{\colorMATH}\ensuremath{\theta _{0}}}}\endgroup } = {\begingroup\renewcommand\colorMATH{\colorMATHS}\renewcommand\colorSYNTAX{\colorSYNTAXS}{{\color{\colorMATH}\ensuremath{{{\color{\colorMATH}\ensuremath{\operatorname{zeros}}}}\hspace*{0.33em}({{\color{\colorMATH}\ensuremath{\operatorname{cols}}}}\hspace*{0.33em}{\begingroup\renewcommand\colorMATH{\colorMATHM}\renewcommand\colorSYNTAX{\colorSYNTAXM}{{\color{\colorMATH}\ensuremath{X_{1}}}}\endgroup })}}}\endgroup }\hspace*{0.33em}{{\color{\colorSYNTAX}\texttt{in}}}}}}}
\cr
\end{array}\endgroup \endgroup 
\cr \begin{array}{l
} \{  X \mathrel{:}_{\langle \epsilon ^{\prime}, \delta \rangle } \tau _{2},
  y \mathrel{:}_{\langle \epsilon ^{\prime}, \delta \rangle } \tau _{3},
  k \mathrel{:}_{\infty } {{\color{\colorSYNTAX}\texttt{\ensuremath{{\mathbb{R}}}}}}^{+},
  \rho  \mathrel{:}_{\infty } {{\color{\colorSYNTAX}\texttt{\ensuremath{{\mathbb{R}}}}}}^{+} \} 
 \vdash  {\begingroup\renewcommand\colorMATH{\colorMATHP}\renewcommand\colorSYNTAX{\colorSYNTAXP}{{\color{\colorMATH}\ensuremath{{{\color{\colorMATH}\ensuremath{\operatorname{noisy-gradient-descent}}}}({\begingroup\renewcommand\colorMATH{\colorMATHM}\renewcommand\colorSYNTAX{\colorSYNTAXM}{{\color{\colorMATH}\ensuremath{X}}}\endgroup },{\begingroup\renewcommand\colorMATH{\colorMATHM}\renewcommand\colorSYNTAX{\colorSYNTAXM}{{\color{\colorMATH}\ensuremath{y}}}\endgroup },{\begingroup\renewcommand\colorMATH{\colorMATHM}\renewcommand\colorSYNTAX{\colorSYNTAXM}{{\color{\colorMATH}\ensuremath{k}}}\endgroup },{\begingroup\renewcommand\colorMATH{\colorMATHM}\renewcommand\colorSYNTAX{\colorSYNTAXM}{{\color{\colorMATH}\ensuremath{\epsilon }}}\endgroup },{\begingroup\renewcommand\colorMATH{\colorMATHM}\renewcommand\colorSYNTAX{\colorSYNTAXM}{{\color{\colorMATH}\ensuremath{\delta }}}\endgroup })}}}\endgroup } \mathrel{:} {\mathbb{M}}_{L2}^{{{\color{\colorSYNTAX}\texttt{U}}}}[{{\color{\colorMATH}\ensuremath{1}}},{{\color{\colorMATH}\ensuremath{n}}}]\hspace*{0.33em}{{\color{\colorSYNTAX}\texttt{\ensuremath{{\mathbb{R}}}}}}
\cr  {{\color{\colorTEXT}\textit{where}}}\hspace*{0.33em}\epsilon ^{\prime} =  \frac{1}{2}k\epsilon ^{2} + 2\sqrt {\frac{1}{2}k\epsilon ^{2} \log (1/\delta )}
\end{array}\vspace*{-1em}\end{gather*}\endgroup 
Such nestings are sometimes useful in practice: in
Section~\ref{sec:examples}, we will define a variant of the Private
Frank-Wolfe algorithm which uses the exponential mechanism (which
satisfies {{\color{\colorMATH}\ensuremath{\epsilon }}}-differential privacy) in a loop for which composition is
performed with zCDP, and report the total privacy cost in terms of {{\color{\colorMATH}\ensuremath{\epsilon }}}
and {{\color{\colorMATH}\ensuremath{\delta }}}.


\paragraph{Contextual Modal Types}

A new problem arises in the design of \system governing the interaction of
{\begingroup\renewcommand\colorMATH{\colorMATHS}\renewcommand\colorSYNTAX{\colorSYNTAXS}{{\color{\colorMATH}\ensuremath{{{\color{\colorMATH}\ensuremath{\operatorname{sensitivity}}}}}}}\endgroup } and {\begingroup\renewcommand\colorMATH{\colorMATHP}\renewcommand\colorSYNTAX{\colorSYNTAXP}{{\color{\colorMATH}\ensuremath{{{\color{\colorMATH}\ensuremath{\operatorname{privacy}}}}}}}\endgroup } languages: in general---and for very good
reasons which are detailed in the next section---let-binding intermediate results
in the {\begingroup\renewcommand\colorMATH{\colorMATHP}\renewcommand\colorSYNTAX{\colorSYNTAXP}{{\color{\colorMATH}\ensuremath{{{\color{\colorMATH}\ensuremath{\operatorname{privacy}}}}}}}\endgroup } language doesn't always preserve typeability. Not only is
let-binding intermediate results desirable for code readability, it can often
be {{\color{\colorTEXT}\textit{essential}}} in order to achieve desirable performance. Consider a loop body which
performs an expensive operation that does not depend on
the inner-loop parameter:
\vspace*{-0.25em}\begingroup\color{\colorMATH}\begin{gather*}\begin{array}{l
  } \lambda \hspace*{0.33em}xs\hspace*{0.33em}\theta _{0} \rightarrow  \begin{array}[t]{l
               } {{\color{\colorSYNTAX}\texttt{loop}}}\hspace*{0.33em}k\hspace*{0.33em}{{\color{\colorSYNTAX}\texttt{times}}}\hspace*{0.33em}{{\color{\colorSYNTAX}\texttt{on}}}\hspace*{0.33em}\theta _{0}\hspace*{0.33em}\{ \hspace*{0.33em}\theta  \rightarrow  
               \cr  \hspace*{1.00em}\hspace*{1.00em} {{\color{\colorMATH}\ensuremath{\operatorname{gauss}}}}_{\epsilon ,\delta }\hspace*{0.33em}(f\hspace*{0.33em}({{\color{\colorMATH}\ensuremath{\operatorname{expensive}}}}\hspace*{0.33em}xs)\hspace*{0.33em}\theta ))\hspace*{0.33em}\} 
               \end{array}
  \end{array}
\vspace*{-1em}\end{gather*}\endgroup 
A simple refactoring achieves much better performance:
\vspace*{-0.25em}\begingroup\color{\colorMATH}\begin{gather*}\begin{array}{l
  } \lambda \hspace*{0.33em}xs\hspace*{0.33em}\theta _{0} \rightarrow  \begin{array}[t]{l
               } {{\color{\colorSYNTAX}\texttt{let}}}\hspace*{0.33em}{{\color{\colorMATH}\ensuremath{\operatorname{temp}}}} = {{\color{\colorMATH}\ensuremath{\operatorname{expensive}}}}\hspace*{0.33em}xs\hspace*{0.33em}{{\color{\colorSYNTAX}\texttt{in}}}
               \cr  {{\color{\colorSYNTAX}\texttt{loop}}}\hspace*{0.33em}k\hspace*{0.33em}{{\color{\colorSYNTAX}\texttt{times}}}\hspace*{0.33em}{{\color{\colorSYNTAX}\texttt{on}}}\hspace*{0.33em}\theta _{0}\hspace*{0.33em}\{ \hspace*{0.33em}\theta  \rightarrow  
               \cr  \hspace*{1.00em}\hspace*{1.00em} {{\color{\colorMATH}\ensuremath{\operatorname{gauss}}}}_{\epsilon ,\delta }\hspace*{0.33em}(f\hspace*{0.33em}{{\color{\colorMATH}\ensuremath{\operatorname{temp}}}}\hspace*{0.33em}\theta )\hspace*{0.33em}\} 
               \end{array}
  \end{array}
\vspace*{-1em}\end{gather*}\endgroup  
However instead of providing {{\color{\colorMATH}\ensuremath{(\epsilon ,\delta )}}}-differential privacy for {{\color{\colorMATH}\ensuremath{xs}}}, as was the
case before the refactor, the new program provides {{\color{\colorMATH}\ensuremath{(\epsilon ,\delta )}}}-differential privacy
for {{\color{\colorMATH}\ensuremath{{{\color{\colorMATH}\ensuremath{\operatorname{temp}}}}}}}---an intermediate variable we don't care about---and makes no guarantees
of privacy for {{\color{\colorMATH}\ensuremath{xs}}}.

To accommodate this pattern we borrow ideas from {{\color{\colorTEXT}\textit{contextual modal type
theory}}}~\cite{Nanevski:2008:CMT:1352582.1352591} to allow ``boxing'' a
sensitivity context, and ``unboxing'' that context at a later time. In terms of
differential privacy, the argument that the above loop is differentially
private relies on the fact that {{\color{\colorMATH}\ensuremath{{{\color{\colorMATH}\ensuremath{\operatorname{temp}}}} \equiv  {{\color{\colorMATH}\ensuremath{\operatorname{expensive}}}}(xs)}}} is {{\color{\colorMATH}\ensuremath{1}}}-sensitive in
{{\color{\colorMATH}\ensuremath{xs}}} (assuming {{\color{\colorMATH}\ensuremath{\operatorname{expensive}}}} is {{\color{\colorMATH}\ensuremath{1}}}-sensitive), a property which is lost by
the typing rule for {{\color{\colorSYNTAX}\texttt{let}}} in the privacy language. We therefore ``box'' this
sensitivity information outside the loop, and ``unbox'' it inside the loop, like
so:
\vspace*{-0.25em}\begingroup\color{\colorMATH}\begin{gather*}\begin{array}{l
  } \lambda \hspace*{0.33em}xs\hspace*{0.33em}\theta _{0} \rightarrow  \begin{array}[t]{l
               } {{\color{\colorSYNTAX}\texttt{let}}}\hspace*{0.33em}{{\color{\colorMATH}\ensuremath{\operatorname{temp}}}} = {{\color{\colorSYNTAX}\texttt{box}}}\hspace*{0.33em}({{\color{\colorMATH}\ensuremath{\operatorname{expensive}}}}\hspace*{0.33em}xs)\hspace*{0.33em}{{\color{\colorSYNTAX}\texttt{in}}}
               \cr  {{\color{\colorSYNTAX}\texttt{loop}}}\hspace*{0.33em}k\hspace*{0.33em}{{\color{\colorSYNTAX}\texttt{times}}}\hspace*{0.33em}{{\color{\colorSYNTAX}\texttt{on}}}\hspace*{0.33em}\theta _{0}\hspace*{0.33em}\{ \hspace*{0.33em}\theta  \rightarrow  
               \cr  \hspace*{1.00em}\hspace*{1.00em} {{\color{\colorMATH}\ensuremath{\operatorname{gauss}}}}_{\epsilon ,\delta }\hspace*{0.33em}(f\hspace*{0.33em}({{\color{\colorSYNTAX}\texttt{unbox}}}\hspace*{0.33em}{{\color{\colorMATH}\ensuremath{\operatorname{temp}}}})\hspace*{0.33em}\theta )\hspace*{0.33em}\} 
               \end{array}
  \end{array}
\vspace*{-1em}\end{gather*}\endgroup  
In this example, the type of {{\color{\colorMATH}\ensuremath{{{\color{\colorMATH}\ensuremath{\operatorname{temp}}}}}}} is a {{\color{\colorMATH}\ensuremath{\square [xs@1]\hspace*{0.33em}{{\color{\colorSYNTAX}\texttt{data}}}}}} (a ``box of data
{{\color{\colorMATH}\ensuremath{1}}}-sensitive in {{\color{\colorMATH}\ensuremath{xs}}}'') indicating that when unboxed, {{\color{\colorMATH}\ensuremath{{{\color{\colorMATH}\ensuremath{\operatorname{temp}}}}}}} will report
{{\color{\colorMATH}\ensuremath{1}}}-sensitivity {w.r.t} {{\color{\colorMATH}\ensuremath{xs}}}, not {{\color{\colorMATH}\ensuremath{{{\color{\colorMATH}\ensuremath{\operatorname{temp}}}}}}}. {{\color{\colorMATH}\ensuremath{f}}} is then able to make good on its
promise to {{\color{\colorMATH}\ensuremath{\operatorname{gauss}}}} that the result of {{\color{\colorMATH}\ensuremath{f}}} is {{\color{\colorMATH}\ensuremath{1}}}-sensitive in {{\color{\colorMATH}\ensuremath{xs}}} (assuming
{{\color{\colorMATH}\ensuremath{f}}} is {{\color{\colorMATH}\ensuremath{1}}}-sensitive in its first argument), and {{\color{\colorMATH}\ensuremath{\operatorname{gauss}}}} properly reports its
privacy ``cost'' in terms of {{\color{\colorMATH}\ensuremath{xs}}}, not {{\color{\colorMATH}\ensuremath{{{\color{\colorMATH}\ensuremath{\operatorname{temp}}}}}}}.

We use exactly this pattern in many of our case studies, where {{\color{\colorMATH}\ensuremath{expensive}}}
is a pre-processing operation on the input data ({e.g.}, clipping or
normalizing), and {{\color{\colorMATH}\ensuremath{f}}} is a machine-learning training operation, such as
computing an improved model based on the current model {{\color{\colorMATH}\ensuremath{\theta }}} and the
pre-processed input data {{\color{\colorMATH}\ensuremath{temp}}}.

\subsection{\system Syntax \& Typing Rules}

\newcommand\FigureCoreLang{
\begin{figure}
\smallerplz
\begingroup\renewcommand\colorMATH{\colorMATHM}\renewcommand\colorSYNTAX{\colorSYNTAXM}
\vspace*{-0.25em}\begingroup\color{\colorMATH}\begin{gather*}\begin{tabularx}{\linewidth}{>{\centering\arraybackslash\(}X<{\)}}
   \hfill\hspace{0pt}
   \hfill\hspace{0pt}m,n\in {\mathbb{N}}
   \hfill\hspace{0pt}r\in {\mathbb{R}}
   \hfill\hspace{0pt}\dot r,\epsilon ,\delta \in {\mathbb{R}}^{+}
   \hfill\hspace{0pt}x,y\in {{\color{\colorMATH}\ensuremath{\operatorname{var}}}}
   \hfill\hspace{0pt}
   \hfill\hspace{0pt}
  \cr 
   \hfill\hspace{0pt}{\begingroup\renewcommand\colorMATH{\colorMATHS}\renewcommand\colorSYNTAX{\colorSYNTAXS}{{\color{\colorMATH}\ensuremath{s \in  {{\color{\colorMATH}\ensuremath{\operatorname{sens}}}} \mathrel{\Coloneqq } {\begingroup\renewcommand\colorMATH{\colorMATHM}\renewcommand\colorSYNTAX{\colorSYNTAXM}{{\color{\colorMATH}\ensuremath{\dot r}}}\endgroup } \mathrel{|} {{\color{\colorSYNTAX}\texttt{\ensuremath{\infty }}}}}}}\endgroup }
   \hfill\hspace{0pt}{\begingroup\renewcommand\colorMATH{\colorMATHP}\renewcommand\colorSYNTAX{\colorSYNTAXP}{{\color{\colorMATH}\ensuremath{p \in  {{\color{\colorMATH}\ensuremath{\operatorname{priv}}}} \mathrel{\Coloneqq } {{\color{\colorSYNTAX}\texttt{\ensuremath{{\begingroup\renewcommand\colorMATH{\colorMATHM}\renewcommand\colorSYNTAX{\colorSYNTAXM}{{\color{\colorMATH}\ensuremath{\epsilon }}}\endgroup },{\begingroup\renewcommand\colorMATH{\colorMATHM}\renewcommand\colorSYNTAX{\colorSYNTAXM}{{\color{\colorMATH}\ensuremath{\delta }}}\endgroup }}}}} \mathrel{|} {{\color{\colorSYNTAX}\texttt{\ensuremath{\infty }}}}}}}\endgroup }
   \hfill\hspace{0pt}
  \end{tabularx}
\cr \begin{array}{rclcl@{\hspace*{1.00em}}l
  } \tau      &{}\in {}& {{\color{\colorMATH}\ensuremath{\operatorname{type}}}}         &{}\mathrel{\Coloneqq }{}& {{\color{\colorSYNTAX}\texttt{\ensuremath{{\mathbb{N}}}}}} \mathrel{|} {{\color{\colorSYNTAX}\texttt{\ensuremath{{\mathbb{R}}}}}} \mathrel{|} {{\color{\colorSYNTAX}\texttt{\ensuremath{{\mathbb{N}}[{{\color{\colorMATH}\ensuremath{n}}}]}}}} \mathrel{|} {{\color{\colorSYNTAX}\texttt{\ensuremath{{\mathbb{R}}^{+}[{{\color{\colorMATH}\ensuremath{\dot r}}}]}}}} \mathrel{|} {{\color{\colorSYNTAX}\texttt{box}}}[{\begingroup\renewcommand\colorMATH{\colorMATHS}\renewcommand\colorSYNTAX{\colorSYNTAXS}{{\color{\colorMATH}\ensuremath{\Gamma _{s}}}}\endgroup }]\hspace*{0.33em}\tau        & {{\color{\colorTEXT}\textit{numeric and box}}}
  \cr        &{} {}&                &{}\mathrel{|}{}& {{\color{\colorSYNTAX}\texttt{\ensuremath{{{\color{\colorMATH}\ensuremath{\tau }}} \multimap _{{\begingroup\renewcommand\colorMATH{\colorMATHS}\renewcommand\colorSYNTAX{\colorSYNTAXS}{{\color{\colorMATH}\ensuremath{s}}}\endgroup }} {{\color{\colorMATH}\ensuremath{\tau }}}}}}} \mathrel{|} {{\color{\colorSYNTAX}\texttt{\ensuremath{({{\color{\colorMATH}\ensuremath{\tau }}}@{\begingroup\renewcommand\colorMATH{\colorMATHP}\renewcommand\colorSYNTAX{\colorSYNTAXP}{{\color{\colorMATH}\ensuremath{p}}}\endgroup },{.}\hspace{-1pt}{.}\hspace{-1pt}{.},{{\color{\colorMATH}\ensuremath{\tau }}}@{\begingroup\renewcommand\colorMATH{\colorMATHP}\renewcommand\colorSYNTAX{\colorSYNTAXP}{{\color{\colorMATH}\ensuremath{p}}}\endgroup }) \multimap ^{*} {{\color{\colorMATH}\ensuremath{\tau }}}}}}}                  & {{\color{\colorTEXT}\textit{functions}}}
  \cr  {\begingroup\renewcommand\colorMATH{\colorMATHS}\renewcommand\colorSYNTAX{\colorSYNTAXS}{{\color{\colorMATH}\ensuremath{\Gamma _{s}}}}\endgroup } &{}{\begingroup\renewcommand\colorMATH{\colorMATHS}\renewcommand\colorSYNTAX{\colorSYNTAXS}{{\color{\colorMATH}\ensuremath{\in }}}\endgroup }{}& {\begingroup\renewcommand\colorMATH{\colorMATHS}\renewcommand\colorSYNTAX{\colorSYNTAXS}{{\color{\colorMATH}\ensuremath{{{\color{\colorMATH}\ensuremath{\operatorname{tcxt}}}}_{s}}}}\endgroup } &{}{\begingroup\renewcommand\colorMATH{\colorMATHS}\renewcommand\colorSYNTAX{\colorSYNTAXS}{{\color{\colorMATH}\ensuremath{\triangleq }}}\endgroup }{}& {\begingroup\renewcommand\colorMATH{\colorMATHS}\renewcommand\colorSYNTAX{\colorSYNTAXS}{{\color{\colorMATH}\ensuremath{{\begingroup\renewcommand\colorMATH{\colorMATHM}\renewcommand\colorSYNTAX{\colorSYNTAXM}{{\color{\colorMATH}\ensuremath{{{\color{\colorMATH}\ensuremath{\operatorname{var}}}}}}}\endgroup } \rightharpoonup  {{\color{\colorMATH}\ensuremath{\operatorname{sens}}}} \times  {\begingroup\renewcommand\colorMATH{\colorMATHM}\renewcommand\colorSYNTAX{\colorSYNTAXM}{{\color{\colorMATH}\ensuremath{{{\color{\colorMATH}\ensuremath{\operatorname{type}}}}}}}\endgroup } \mathrel{\Coloneqq } \{ {\begingroup\renewcommand\colorMATH{\colorMATHM}\renewcommand\colorSYNTAX{\colorSYNTAXM}{{\color{\colorMATH}\ensuremath{x}}}\endgroup }\mathrel{:}_{s}{\begingroup\renewcommand\colorMATH{\colorMATHM}\renewcommand\colorSYNTAX{\colorSYNTAXM}{{\color{\colorMATH}\ensuremath{\tau }}}\endgroup },{.}\hspace{-1pt}{.}\hspace{-1pt}{.},{\begingroup\renewcommand\colorMATH{\colorMATHM}\renewcommand\colorSYNTAX{\colorSYNTAXM}{{\color{\colorMATH}\ensuremath{x}}}\endgroup }\mathrel{:}_{s}{\begingroup\renewcommand\colorMATH{\colorMATHM}\renewcommand\colorSYNTAX{\colorSYNTAXM}{{\color{\colorMATH}\ensuremath{\tau }}}\endgroup }\} }}}\endgroup }  & {{\color{\colorTEXT}\textit{sens. contexts}}}
  \cr  {\begingroup\renewcommand\colorMATH{\colorMATHP}\renewcommand\colorSYNTAX{\colorSYNTAXP}{{\color{\colorMATH}\ensuremath{\Gamma _{p}}}}\endgroup } &{}{\begingroup\renewcommand\colorMATH{\colorMATHP}\renewcommand\colorSYNTAX{\colorSYNTAXP}{{\color{\colorMATH}\ensuremath{\in }}}\endgroup }{}& {\begingroup\renewcommand\colorMATH{\colorMATHP}\renewcommand\colorSYNTAX{\colorSYNTAXP}{{\color{\colorMATH}\ensuremath{{{\color{\colorMATH}\ensuremath{\operatorname{tcxt}}}}_{p}}}}\endgroup } &{}{\begingroup\renewcommand\colorMATH{\colorMATHP}\renewcommand\colorSYNTAX{\colorSYNTAXP}{{\color{\colorMATH}\ensuremath{\triangleq }}}\endgroup }{}& {\begingroup\renewcommand\colorMATH{\colorMATHP}\renewcommand\colorSYNTAX{\colorSYNTAXP}{{\color{\colorMATH}\ensuremath{{\begingroup\renewcommand\colorMATH{\colorMATHM}\renewcommand\colorSYNTAX{\colorSYNTAXM}{{\color{\colorMATH}\ensuremath{{{\color{\colorMATH}\ensuremath{\operatorname{var}}}}}}}\endgroup } \rightharpoonup  {{\color{\colorMATH}\ensuremath{\operatorname{priv}}}} \times  {\begingroup\renewcommand\colorMATH{\colorMATHM}\renewcommand\colorSYNTAX{\colorSYNTAXM}{{\color{\colorMATH}\ensuremath{{{\color{\colorMATH}\ensuremath{\operatorname{type}}}}}}}\endgroup } \mathrel{\Coloneqq } \{ {\begingroup\renewcommand\colorMATH{\colorMATHM}\renewcommand\colorSYNTAX{\colorSYNTAXM}{{\color{\colorMATH}\ensuremath{x}}}\endgroup }\mathrel{:}_{p}{\begingroup\renewcommand\colorMATH{\colorMATHM}\renewcommand\colorSYNTAX{\colorSYNTAXM}{{\color{\colorMATH}\ensuremath{\tau }}}\endgroup },{.}\hspace{-1pt}{.}\hspace{-1pt}{.},{\begingroup\renewcommand\colorMATH{\colorMATHM}\renewcommand\colorSYNTAX{\colorSYNTAXM}{{\color{\colorMATH}\ensuremath{x}}}\endgroup }\mathrel{:}_{p}{\begingroup\renewcommand\colorMATH{\colorMATHM}\renewcommand\colorSYNTAX{\colorSYNTAXM}{{\color{\colorMATH}\ensuremath{\tau }}}\endgroup }\} }}}\endgroup }  & {{\color{\colorTEXT}\textit{priv. contexts}}}
  \end{array}
\cr \begingroup\renewcommand\colorMATH{\colorMATHS}\renewcommand\colorSYNTAX{\colorSYNTAXS}\begingroup\color{\colorMATH}
\begin{array}{rclcl@{\hspace*{1.00em}}l
  } e_{s} &{}\in {}& {{\color{\colorMATH}\ensuremath{\operatorname{exp}}}}_{s}    &{}\mathrel{\Coloneqq }{}& {{\color{\colorSYNTAX}\texttt{\ensuremath{{\mathbb{N}}[{\begingroup\renewcommand\colorMATH{\colorMATHM}\renewcommand\colorSYNTAX{\colorSYNTAXM}{{\color{\colorMATH}\ensuremath{n}}}\endgroup }]}}}} \mathrel{|} {{\color{\colorSYNTAX}\texttt{\ensuremath{{\mathbb{N}}[{\begingroup\renewcommand\colorMATH{\colorMATHM}\renewcommand\colorSYNTAX{\colorSYNTAXM}{{\color{\colorMATH}\ensuremath{\dot r}}}\endgroup }]}}}} \mathrel{|} {\begingroup\renewcommand\colorMATH{\colorMATHM}\renewcommand\colorSYNTAX{\colorSYNTAXM}{{\color{\colorMATH}\ensuremath{n}}}\endgroup } \mathrel{|} {\begingroup\renewcommand\colorMATH{\colorMATHM}\renewcommand\colorSYNTAX{\colorSYNTAXM}{{\color{\colorMATH}\ensuremath{r}}}\endgroup } \mathrel{|} {{\color{\colorSYNTAX}\texttt{real}}}\hspace*{0.33em}e                 & {{\color{\colorTEXT}\textit{numeric literals}}}
  \cr    &{} {}&            &{}\mathrel{|}{}& {{\color{\colorSYNTAX}\texttt{\ensuremath{{{\color{\colorMATH}\ensuremath{e}}}+{{\color{\colorMATH}\ensuremath{e}}}}}}} \mathrel{|} {{\color{\colorSYNTAX}\texttt{\ensuremath{{{\color{\colorMATH}\ensuremath{e}}}-{{\color{\colorMATH}\ensuremath{e}}}}}}} \mathrel{|} {{\color{\colorSYNTAX}\texttt{\ensuremath{{{\color{\colorMATH}\ensuremath{e}}}\mathrel{\mathord{\cdotp }}{{\color{\colorMATH}\ensuremath{e}}}}}}} \mathrel{|} {{\color{\colorSYNTAX}\texttt{\ensuremath{1/{{\color{\colorMATH}\ensuremath{e}}}}}}} \mathrel{|} e \mathrel{{{\color{\colorSYNTAX}\texttt{mod}}}} e        & {{\color{\colorTEXT}\textit{arithmetic}}}
  \cr
\end{array}
  \endgroup \endgroup 
\cr \begingroup\renewcommand\colorMATH{\colorMATHP}\renewcommand\colorSYNTAX{\colorSYNTAXP}\begingroup\color{\colorMATH}
  \begin{array}{rclcl@{\hspace*{1.00em}}l
  } e_{p}  &{}\in {}& {{\color{\colorMATH}\ensuremath{\operatorname{exp}}}}_{p}   &{}\mathrel{\Coloneqq }{}& {{\color{\colorSYNTAX}\texttt{return}}}\hspace*{0.33em}{\begingroup\renewcommand\colorMATH{\colorMATHS}\renewcommand\colorSYNTAX{\colorSYNTAXS}{{\color{\colorMATH}\ensuremath{e}}}\endgroup } \mathrel{|} {{\color{\colorSYNTAX}\texttt{\ensuremath{{\begingroup\renewcommand\colorMATH{\colorMATHM}\renewcommand\colorSYNTAX{\colorSYNTAXM}{{\color{\colorMATH}\ensuremath{x}}}\endgroup }\leftarrow {{\color{\colorMATH}\ensuremath{e}}} \mathrel{;} {{\color{\colorMATH}\ensuremath{e}}}}}}} \mathrel{|} {{\color{\colorSYNTAX}\texttt{\ensuremath{{\begingroup\renewcommand\colorMATH{\colorMATHS}\renewcommand\colorSYNTAX{\colorSYNTAXS}{{\color{\colorMATH}\ensuremath{e}}}\endgroup }({\begingroup\renewcommand\colorMATH{\colorMATHS}\renewcommand\colorSYNTAX{\colorSYNTAXS}{{\color{\colorMATH}\ensuremath{e}}}\endgroup },{.}\hspace{-1pt}{.}\hspace{-1pt}{.},{\begingroup\renewcommand\colorMATH{\colorMATHS}\renewcommand\colorSYNTAX{\colorSYNTAXS}{{\color{\colorMATH}\ensuremath{e}}}\endgroup })}}}}          & {{\color{\colorTEXT}\textit{ret/bind/priv. app.}}}
  \cr
\end{array}
  \endgroup \endgroup 
\vspace*{-1em}\end{gather*}\endgroup 
\endgroup 
\vspace{-1.5em}
\caption{Core Types and Terms}
\vspace{-0.5em}
\label{fig:syntax}
\end{figure}
}

\newcommand\FigureCoreTyping{
\begin{figure*}
\smallerplz
\begingroup\renewcommand\colorMATH{\colorMATHS}\renewcommand\colorSYNTAX{\colorSYNTAXS}
\vspace*{-0.25em}\begingroup\color{\colorMATH}\begin{gather*}\begin{tabularx}{\linewidth}{>{\centering\arraybackslash\(}X<{\)}}\hfill\hspace{0pt}\begingroup\color{\colorTEXT}\boxed{\begingroup\color{\colorMATH} \Gamma  \vdash  e \mathrel{:} {\begingroup\renewcommand\colorMATH{\colorMATHM}\renewcommand\colorSYNTAX{\colorSYNTAXM}{{\color{\colorMATH}\ensuremath{\tau }}}\endgroup } \endgroup}\endgroup \end{tabularx}\vspace*{-1em}\end{gather*}\endgroup 
\vspace{-1em}
\begingroup\color{\colorMATH}\begin{mathpar}\inferrule*[lab={{\color{\colorTEXT}\textsc{\scriptsize Nat}}}
  ]{ }{
     \vdash  {\begingroup\renewcommand\colorMATH{\colorMATHM}\renewcommand\colorSYNTAX{\colorSYNTAXM}{{\color{\colorMATH}\ensuremath{n}}}\endgroup } \mathrel{:} {\begingroup\renewcommand\colorMATH{\colorMATHM}\renewcommand\colorSYNTAX{\colorSYNTAXM}{{\color{\colorMATH}\ensuremath{{{\color{\colorSYNTAX}\texttt{\ensuremath{{\mathbb{N}}}}}}}}}\endgroup }
  }
\and\inferrule*[lab={{\color{\colorTEXT}\textsc{\scriptsize Real}}}
  ]{ }{
     \vdash  {\begingroup\renewcommand\colorMATH{\colorMATHM}\renewcommand\colorSYNTAX{\colorSYNTAXM}{{\color{\colorMATH}\ensuremath{r}}}\endgroup } \mathrel{:} {\begingroup\renewcommand\colorMATH{\colorMATHM}\renewcommand\colorSYNTAX{\colorSYNTAXM}{{\color{\colorMATH}\ensuremath{{{\color{\colorSYNTAX}\texttt{\ensuremath{{\mathbb{R}}}}}}}}}\endgroup }
  }
\and\inferrule*[lab={{\color{\colorTEXT}\textsc{\scriptsize Singleton Nat}}}
  ]{ }{
     \vdash  {{\color{\colorSYNTAX}\texttt{\ensuremath{{\mathbb{N}}[{\begingroup\renewcommand\colorMATH{\colorMATHM}\renewcommand\colorSYNTAX{\colorSYNTAXM}{{\color{\colorMATH}\ensuremath{n}}}\endgroup }]}}}} \mathrel{:} {\begingroup\renewcommand\colorMATH{\colorMATHM}\renewcommand\colorSYNTAX{\colorSYNTAXM}{{\color{\colorMATH}\ensuremath{{{\color{\colorSYNTAX}\texttt{\ensuremath{{\mathbb{N}}[{{\color{\colorMATH}\ensuremath{n}}}]}}}}}}}\endgroup }
  }
\and\inferrule*[lab={{\color{\colorTEXT}\textsc{\scriptsize Singleton Real}}}
  ]{ }{
     \vdash  {{\color{\colorSYNTAX}\texttt{\ensuremath{{\mathbb{R}}^{+}[{\begingroup\renewcommand\colorMATH{\colorMATHM}\renewcommand\colorSYNTAX{\colorSYNTAXM}{{\color{\colorMATH}\ensuremath{\dot r}}}\endgroup }]}}}} \mathrel{:} {\begingroup\renewcommand\colorMATH{\colorMATHM}\renewcommand\colorSYNTAX{\colorSYNTAXM}{{\color{\colorMATH}\ensuremath{{{\color{\colorSYNTAX}\texttt{\ensuremath{{\mathbb{R}}^{+}[{{\color{\colorMATH}\ensuremath{\dot r}}}]}}}}}}}\endgroup }
  }
\and\inferrule*[lab={{\color{\colorTEXT}\textsc{\scriptsize Real-S}}}
  ]{ \Gamma  \vdash  e \mathrel{:} {\begingroup\renewcommand\colorMATH{\colorMATHM}\renewcommand\colorSYNTAX{\colorSYNTAXM}{{\color{\colorMATH}\ensuremath{{{\color{\colorSYNTAX}\texttt{\ensuremath{{\mathbb{N}}[{{\color{\colorMATH}\ensuremath{n}}}]}}}}}}}\endgroup }
    }{
     \vdash  {{\color{\colorSYNTAX}\texttt{real}}}\hspace*{0.33em}e \mathrel{:} {\begingroup\renewcommand\colorMATH{\colorMATHM}\renewcommand\colorSYNTAX{\colorSYNTAXM}{{\color{\colorMATH}\ensuremath{{{\color{\colorSYNTAX}\texttt{\ensuremath{{\mathbb{R}}^{+}[{{\color{\colorMATH}\ensuremath{n}}}]}}}}}}}\endgroup }
  }
\and\inferrule*[lab={{\color{\colorTEXT}\textsc{\scriptsize Real-D}}}
  ]{ \Gamma  \vdash  e \mathrel{:} {\begingroup\renewcommand\colorMATH{\colorMATHM}\renewcommand\colorSYNTAX{\colorSYNTAXM}{{\color{\colorMATH}\ensuremath{{{\color{\colorSYNTAX}\texttt{\ensuremath{{\mathbb{N}}}}}}}}}\endgroup }
    }{
     \Gamma  \vdash  {{\color{\colorSYNTAX}\texttt{real}}}\hspace*{0.33em}e \mathrel{:} {\begingroup\renewcommand\colorMATH{\colorMATHM}\renewcommand\colorSYNTAX{\colorSYNTAXM}{{\color{\colorMATH}\ensuremath{{{\color{\colorSYNTAX}\texttt{\ensuremath{{\mathbb{R}}}}}}}}}\endgroup }
  }
\and\inferrule*[lab={{\color{\colorTEXT}\textsc{\scriptsize Times-DS}}}
  ]{ \Gamma _{1} \vdash  e_{1} \mathrel{:} {\begingroup\renewcommand\colorMATH{\colorMATHM}\renewcommand\colorSYNTAX{\colorSYNTAXM}{{\color{\colorMATH}\ensuremath{{{\color{\colorSYNTAX}\texttt{\ensuremath{{\mathbb{R}}}}}}}}}\endgroup }
  \\ \Gamma _{2} \vdash  e_{2} \mathrel{:} {\begingroup\renewcommand\colorMATH{\colorMATHM}\renewcommand\colorSYNTAX{\colorSYNTAXM}{{\color{\colorMATH}\ensuremath{{{\color{\colorSYNTAX}\texttt{\ensuremath{{\mathbb{R}}^{+}[{{\color{\colorMATH}\ensuremath{\dot r}}}]}}}}}}}\endgroup }
     }{
     {\begingroup\renewcommand\colorMATH{\colorMATHM}\renewcommand\colorSYNTAX{\colorSYNTAXM}{{\color{\colorMATH}\ensuremath{\dot r}}}\endgroup }\Gamma _{1} \vdash  {{\color{\colorSYNTAX}\texttt{\ensuremath{{{\color{\colorMATH}\ensuremath{e_{1}}}}\mathrel{\mathord{\cdotp }}{{\color{\colorMATH}\ensuremath{e_{2}}}}}}}} \mathrel{:} {\begingroup\renewcommand\colorMATH{\colorMATHM}\renewcommand\colorSYNTAX{\colorSYNTAXM}{{\color{\colorMATH}\ensuremath{\tau }}}\endgroup }
  }
\and\inferrule*[lab={{\color{\colorTEXT}\textsc{\scriptsize Mod-DS}}}
  ]{ \Gamma _{1} \vdash  e_{1} \mathrel{:} {\begingroup\renewcommand\colorMATH{\colorMATHM}\renewcommand\colorSYNTAX{\colorSYNTAXM}{{\color{\colorMATH}\ensuremath{{{\color{\colorSYNTAX}\texttt{\ensuremath{{\mathbb{R}}}}}}}}}\endgroup }
  \\ \Gamma _{2} \vdash  e_{2} \mathrel{:} {\begingroup\renewcommand\colorMATH{\colorMATHM}\renewcommand\colorSYNTAX{\colorSYNTAXM}{{\color{\colorMATH}\ensuremath{{{\color{\colorSYNTAX}\texttt{\ensuremath{{\mathbb{R}}^{+}[{{\color{\colorMATH}\ensuremath{\dot r}}}]}}}}}}}\endgroup }
     }{
     {}\rceil \Gamma _{1}\lceil {}^{{\begingroup\renewcommand\colorMATH{\colorMATHM}\renewcommand\colorSYNTAX{\colorSYNTAXM}{{\color{\colorMATH}\ensuremath{\dot r}}}\endgroup }} \vdash  {{\color{\colorSYNTAX}\texttt{\ensuremath{{{\color{\colorMATH}\ensuremath{e_{1}}}}\mathrel{{{\color{\colorSYNTAX}\texttt{mod}}}}{{\color{\colorMATH}\ensuremath{e_{2}}}}}}}} \mathrel{:} {\begingroup\renewcommand\colorMATH{\colorMATHM}\renewcommand\colorSYNTAX{\colorSYNTAXM}{{\color{\colorMATH}\ensuremath{\tau }}}\endgroup }
  }
\and\inferrule*[lab={{\color{\colorTEXT}\textsc{\scriptsize Var}}}
  ]{ }{
     \{ {\begingroup\renewcommand\colorMATH{\colorMATHM}\renewcommand\colorSYNTAX{\colorSYNTAXM}{{\color{\colorMATH}\ensuremath{x}}}\endgroup }\mathrel{:}_{{\begingroup\renewcommand\colorMATH{\colorMATHM}\renewcommand\colorSYNTAX{\colorSYNTAXM}{{\color{\colorMATH}\ensuremath{1}}}\endgroup }}{\begingroup\renewcommand\colorMATH{\colorMATHM}\renewcommand\colorSYNTAX{\colorSYNTAXM}{{\color{\colorMATH}\ensuremath{\tau }}}\endgroup }\}  \vdash  {\begingroup\renewcommand\colorMATH{\colorMATHM}\renewcommand\colorSYNTAX{\colorSYNTAXM}{{\color{\colorMATH}\ensuremath{x}}}\endgroup } \mathrel{:} {\begingroup\renewcommand\colorMATH{\colorMATHM}\renewcommand\colorSYNTAX{\colorSYNTAXM}{{\color{\colorMATH}\ensuremath{\tau }}}\endgroup }
  }
\and\inferrule*[lab={{\color{\colorTEXT}\textsc{\scriptsize Let}}}
  ]{ \Gamma _{1} \vdash  e_{1} \mathrel{:} {\begingroup\renewcommand\colorMATH{\colorMATHM}\renewcommand\colorSYNTAX{\colorSYNTAXM}{{\color{\colorMATH}\ensuremath{\tau _{1}}}}\endgroup }
  \\ \Gamma _{2}\uplus \{ {\begingroup\renewcommand\colorMATH{\colorMATHM}\renewcommand\colorSYNTAX{\colorSYNTAXM}{{\color{\colorMATH}\ensuremath{x}}}\endgroup }\mathrel{:}_{s}{\begingroup\renewcommand\colorMATH{\colorMATHM}\renewcommand\colorSYNTAX{\colorSYNTAXM}{{\color{\colorMATH}\ensuremath{\tau _{1}}}}\endgroup }\}  \vdash  e_{2} \mathrel{:} {\begingroup\renewcommand\colorMATH{\colorMATHM}\renewcommand\colorSYNTAX{\colorSYNTAXM}{{\color{\colorMATH}\ensuremath{\tau _{2}}}}\endgroup }
     }{
     s\Gamma _{1} + \Gamma _{2} \vdash  {{\color{\colorSYNTAX}\texttt{\ensuremath{{{\color{\colorSYNTAX}\texttt{let}}}\hspace*{0.33em}{\begingroup\renewcommand\colorMATH{\colorMATHM}\renewcommand\colorSYNTAX{\colorSYNTAXM}{{\color{\colorMATH}\ensuremath{x}}}\endgroup }={{\color{\colorMATH}\ensuremath{e_{1}}}}\hspace*{0.33em}{{\color{\colorSYNTAX}\texttt{in}}}\hspace*{0.33em}{{\color{\colorMATH}\ensuremath{e_{2}}}}}}}} \mathrel{:} {\begingroup\renewcommand\colorMATH{\colorMATHM}\renewcommand\colorSYNTAX{\colorSYNTAXM}{{\color{\colorMATH}\ensuremath{\tau _{2}}}}\endgroup }
  }
\and\inferrule*[flushleft,lab={{\color{\colorSYNTAX}\texttt{\ensuremath{\multimap }}}}{{\color{\colorTEXT}\textsc{\scriptsize -I}}}
  ]{ \Gamma \uplus \{ {\begingroup\renewcommand\colorMATH{\colorMATHM}\renewcommand\colorSYNTAX{\colorSYNTAXM}{{\color{\colorMATH}\ensuremath{x}}}\endgroup }\mathrel{:}_{s}{\begingroup\renewcommand\colorMATH{\colorMATHM}\renewcommand\colorSYNTAX{\colorSYNTAXM}{{\color{\colorMATH}\ensuremath{\tau _{1}}}}\endgroup }\}  \vdash  e \mathrel{:} {\begingroup\renewcommand\colorMATH{\colorMATHM}\renewcommand\colorSYNTAX{\colorSYNTAXM}{{\color{\colorMATH}\ensuremath{\tau _{2}}}}\endgroup }
     }{
     \Gamma  \vdash  ({{\color{\colorSYNTAX}\texttt{\ensuremath{\lambda \hspace*{0.33em}{\begingroup\renewcommand\colorMATH{\colorMATHM}\renewcommand\colorSYNTAX{\colorSYNTAXM}{{\color{\colorMATH}\ensuremath{x}}}\endgroup }\mathrel{:}{\begingroup\renewcommand\colorMATH{\colorMATHM}\renewcommand\colorSYNTAX{\colorSYNTAXM}{{\color{\colorMATH}\ensuremath{\tau _{1}}}}\endgroup }\Rightarrow {{\color{\colorMATH}\ensuremath{e}}}}}}}) \mathrel{:} {\begingroup\renewcommand\colorMATH{\colorMATHM}\renewcommand\colorSYNTAX{\colorSYNTAXM}{{\color{\colorMATH}\ensuremath{{{\color{\colorSYNTAX}\texttt{\ensuremath{{{\color{\colorMATH}\ensuremath{\tau _{1}}}} \multimap _{{\begingroup\renewcommand\colorMATH{\colorMATHS}\renewcommand\colorSYNTAX{\colorSYNTAXS}{{\color{\colorMATH}\ensuremath{s}}}\endgroup }} {{\color{\colorMATH}\ensuremath{\tau _{2}}}}}}}}}}}\endgroup }
  }
\and\inferrule*[flushleft,lab={{\color{\colorSYNTAX}\texttt{\ensuremath{\multimap }}}}{{\color{\colorTEXT}\textsc{\scriptsize -E}}}
  ]{ \Gamma _{1} \vdash  e_{1} \mathrel{:} {\begingroup\renewcommand\colorMATH{\colorMATHM}\renewcommand\colorSYNTAX{\colorSYNTAXM}{{\color{\colorMATH}\ensuremath{{{\color{\colorSYNTAX}\texttt{\ensuremath{{{\color{\colorMATH}\ensuremath{\tau _{1}}}} \multimap _{{\begingroup\renewcommand\colorMATH{\colorMATHS}\renewcommand\colorSYNTAX{\colorSYNTAXS}{{\color{\colorMATH}\ensuremath{s}}}\endgroup }} {{\color{\colorMATH}\ensuremath{\tau _{2}}}}}}}}}}}\endgroup }
  \\ \Gamma _{2} \vdash  e_{2} \mathrel{:} {\begingroup\renewcommand\colorMATH{\colorMATHM}\renewcommand\colorSYNTAX{\colorSYNTAXM}{{\color{\colorMATH}\ensuremath{\tau _{1}}}}\endgroup }
     }{
     \Gamma _{1} + s\Gamma _{2} \vdash  e_{1}\hspace*{0.33em}e_{2} \mathrel{:} {\begingroup\renewcommand\colorMATH{\colorMATHM}\renewcommand\colorSYNTAX{\colorSYNTAXM}{{\color{\colorMATH}\ensuremath{\tau _{2}}}}\endgroup }
  }
\and\inferrule*[flushleft,lab={{\color{\colorSYNTAX}\texttt{\ensuremath{\multimap ^{*}}}}}{{\color{\colorTEXT}\textsc{\scriptsize -I}}}
  ]{ {\begingroup\renewcommand\colorMATH{\colorMATHP}\renewcommand\colorSYNTAX{\colorSYNTAXP}{{\color{\colorMATH}\ensuremath{\Gamma \uplus \{ {\begingroup\renewcommand\colorMATH{\colorMATHM}\renewcommand\colorSYNTAX{\colorSYNTAXM}{{\color{\colorMATH}\ensuremath{x_{1}}}}\endgroup }\mathrel{:}_{p_{1}}{\begingroup\renewcommand\colorMATH{\colorMATHM}\renewcommand\colorSYNTAX{\colorSYNTAXM}{{\color{\colorMATH}\ensuremath{\tau _{1}}}}\endgroup },{.}\hspace{-1pt}{.}\hspace{-1pt}{.},{\begingroup\renewcommand\colorMATH{\colorMATHM}\renewcommand\colorSYNTAX{\colorSYNTAXM}{{\color{\colorMATH}\ensuremath{x_{n}}}}\endgroup }\mathrel{:}_{p_{n}}{\begingroup\renewcommand\colorMATH{\colorMATHM}\renewcommand\colorSYNTAX{\colorSYNTAXM}{{\color{\colorMATH}\ensuremath{\tau _{n}}}}\endgroup }\}  \vdash  e \mathrel{:} {\begingroup\renewcommand\colorMATH{\colorMATHM}\renewcommand\colorSYNTAX{\colorSYNTAXM}{{\color{\colorMATH}\ensuremath{\tau }}}\endgroup }}}}\endgroup }
     }{
     {}\rceil {\begingroup\renewcommand\colorMATH{\colorMATHP}\renewcommand\colorSYNTAX{\colorSYNTAXP}{{\color{\colorMATH}\ensuremath{\Gamma }}}\endgroup }\lceil {}^{{{\color{\colorSYNTAX}\texttt{\ensuremath{\infty }}}}} \vdash  
     ({{\color{\colorSYNTAX}\texttt{\ensuremath{p\lambda \hspace*{0.33em}({\begingroup\renewcommand\colorMATH{\colorMATHM}\renewcommand\colorSYNTAX{\colorSYNTAXM}{{\color{\colorMATH}\ensuremath{x_{1}}}}\endgroup }\mathrel{:}{\begingroup\renewcommand\colorMATH{\colorMATHM}\renewcommand\colorSYNTAX{\colorSYNTAXM}{{\color{\colorMATH}\ensuremath{\tau _{1}}}}\endgroup },{.}\hspace{-1pt}{.}\hspace{-1pt}{.},{\begingroup\renewcommand\colorMATH{\colorMATHM}\renewcommand\colorSYNTAX{\colorSYNTAXM}{{\color{\colorMATH}\ensuremath{x_{n}}}}\endgroup }\mathrel{:}{\begingroup\renewcommand\colorMATH{\colorMATHM}\renewcommand\colorSYNTAX{\colorSYNTAXM}{{\color{\colorMATH}\ensuremath{\tau _{n}}}}\endgroup }) \Rightarrow  {\begingroup\renewcommand\colorMATH{\colorMATHP}\renewcommand\colorSYNTAX{\colorSYNTAXP}{{\color{\colorMATH}\ensuremath{e}}}\endgroup }}}}})
     \mathrel{:} {\begingroup\renewcommand\colorMATH{\colorMATHM}\renewcommand\colorSYNTAX{\colorSYNTAXM}{{\color{\colorMATH}\ensuremath{{{\color{\colorSYNTAX}\texttt{\ensuremath{({{\color{\colorMATH}\ensuremath{\tau _{1}}}}@{\begingroup\renewcommand\colorMATH{\colorMATHP}\renewcommand\colorSYNTAX{\colorSYNTAXP}{{\color{\colorMATH}\ensuremath{p_{1}}}}\endgroup },{.}\hspace{-1pt}{.}\hspace{-1pt}{.},{{\color{\colorMATH}\ensuremath{\tau _{n}}}}@{\begingroup\renewcommand\colorMATH{\colorMATHP}\renewcommand\colorSYNTAX{\colorSYNTAXP}{{\color{\colorMATH}\ensuremath{p_{n}}}}\endgroup }) \multimap ^{*} {{\color{\colorMATH}\ensuremath{\tau }}}}}}}}}}\endgroup }
  }
\and\inferrule*[flushleft,lab={{\color{\colorTEXT}\textsc{\scriptsize Box-I}}}
  ]{ \Gamma  \vdash  e \mathrel{:} {\begingroup\renewcommand\colorMATH{\colorMATHM}\renewcommand\colorSYNTAX{\colorSYNTAXM}{{\color{\colorMATH}\ensuremath{\tau }}}\endgroup }
     }{
     \vdash  {{\color{\colorSYNTAX}\texttt{box}}}\hspace*{0.33em}e \mathrel{:} {\begingroup\renewcommand\colorMATH{\colorMATHM}\renewcommand\colorSYNTAX{\colorSYNTAXM}{{\color{\colorMATH}\ensuremath{{{\color{\colorSYNTAX}\texttt{box}}}[{\begingroup\renewcommand\colorMATH{\colorMATHS}\renewcommand\colorSYNTAX{\colorSYNTAXS}{{\color{\colorMATH}\ensuremath{\Gamma }}}\endgroup }]\hspace*{0.33em}\tau }}}\endgroup }
  }
\and\inferrule*[flushleft,lab={{\color{\colorTEXT}\textsc{\scriptsize Box-E}}}
  ]{ \Gamma  \vdash  e \mathrel{:} {\begingroup\renewcommand\colorMATH{\colorMATHM}\renewcommand\colorSYNTAX{\colorSYNTAXM}{{\color{\colorMATH}\ensuremath{{{\color{\colorSYNTAX}\texttt{box}}}[{\begingroup\renewcommand\colorMATH{\colorMATHS}\renewcommand\colorSYNTAX{\colorSYNTAXS}{{\color{\colorMATH}\ensuremath{\Gamma ^{\prime}}}}\endgroup }]\hspace*{0.33em}\tau }}}\endgroup }
     }{
     \Gamma  + \Gamma ^{\prime} \vdash  {{\color{\colorSYNTAX}\texttt{unbox}}}\hspace*{0.33em}e \mathrel{:} {\begingroup\renewcommand\colorMATH{\colorMATHM}\renewcommand\colorSYNTAX{\colorSYNTAXM}{{\color{\colorMATH}\ensuremath{\tau }}}\endgroup }
  }
\and\inferrule*[flushleft,lab={{\color{\colorTEXT}\textsc{\scriptsize Sub}}}
  ]{ \Gamma _{1} \vdash  e \mathrel{:} \tau 
  \\ \Gamma _{1} \leq  \Gamma _{2}
     }{
     \Gamma _{2} \vdash  e \mathrel{:} \tau 
  }
\end{mathpar}\endgroup 
\endgroup 
\begingroup\renewcommand\colorMATH{\colorMATHP}\renewcommand\colorSYNTAX{\colorSYNTAXP}
\vspace{-1em}
\vspace*{-0.25em}\begingroup\color{\colorMATH}\begin{gather*}\begin{tabularx}{\linewidth}{>{\centering\arraybackslash\(}X<{\)}}\hfill\hspace{0pt}\begingroup\color{\colorTEXT}\boxed{\begingroup\color{\colorMATH} \Gamma  \vdash  e \mathrel{:} {\begingroup\renewcommand\colorMATH{\colorMATHM}\renewcommand\colorSYNTAX{\colorSYNTAXM}{{\color{\colorMATH}\ensuremath{\tau }}}\endgroup } \endgroup}\endgroup \end{tabularx}\vspace*{-1em}\end{gather*}\endgroup 
\vspace{-2em}
\begingroup\color{\colorMATH}\begin{mathpar}\inferrule*[lab={{\color{\colorTEXT}\textsc{\scriptsize Return}}}
  ]{ {\begingroup\renewcommand\colorMATH{\colorMATHS}\renewcommand\colorSYNTAX{\colorSYNTAXS}{{\color{\colorMATH}\ensuremath{ \Gamma  \vdash  e \mathrel{:} {\begingroup\renewcommand\colorMATH{\colorMATHM}\renewcommand\colorSYNTAX{\colorSYNTAXM}{{\color{\colorMATH}\ensuremath{\tau }}}\endgroup } }}}\endgroup }
     }{
     {}\rceil {\begingroup\renewcommand\colorMATH{\colorMATHS}\renewcommand\colorSYNTAX{\colorSYNTAXS}{{\color{\colorMATH}\ensuremath{\Gamma }}}\endgroup }\lceil {}^{{{\color{\colorSYNTAX}\texttt{\ensuremath{\infty }}}}} \vdash  {{\color{\colorSYNTAX}\texttt{return}}}\hspace*{0.33em}{\begingroup\renewcommand\colorMATH{\colorMATHS}\renewcommand\colorSYNTAX{\colorSYNTAXS}{{\color{\colorMATH}\ensuremath{e}}}\endgroup } \mathrel{:} {\begingroup\renewcommand\colorMATH{\colorMATHM}\renewcommand\colorSYNTAX{\colorSYNTAXM}{{\color{\colorMATH}\ensuremath{\tau }}}\endgroup }
  }
\and\inferrule*[lab={{\color{\colorTEXT}\textsc{\scriptsize Bind}}}
  ]{ \Gamma _{1} \vdash  e_{1} \mathrel{:} {\begingroup\renewcommand\colorMATH{\colorMATHM}\renewcommand\colorSYNTAX{\colorSYNTAXM}{{\color{\colorMATH}\ensuremath{\tau _{1}}}}\endgroup }
  \\ \Gamma _{2}\uplus \{ {\begingroup\renewcommand\colorMATH{\colorMATHM}\renewcommand\colorSYNTAX{\colorSYNTAXM}{{\color{\colorMATH}\ensuremath{x}}}\endgroup }\mathrel{:}_{\infty }{\begingroup\renewcommand\colorMATH{\colorMATHM}\renewcommand\colorSYNTAX{\colorSYNTAXM}{{\color{\colorMATH}\ensuremath{\tau _{1}}}}\endgroup }\}  \vdash  e_{2} \mathrel{:} {\begingroup\renewcommand\colorMATH{\colorMATHM}\renewcommand\colorSYNTAX{\colorSYNTAXM}{{\color{\colorMATH}\ensuremath{\tau _{2}}}}\endgroup }
     }{
     \Gamma _{1} + \Gamma _{2} \vdash  {{\color{\colorSYNTAX}\texttt{\ensuremath{{\begingroup\renewcommand\colorMATH{\colorMATHM}\renewcommand\colorSYNTAX{\colorSYNTAXM}{{\color{\colorMATH}\ensuremath{x}}}\endgroup }\leftarrow {{\color{\colorMATH}\ensuremath{e_{1}}}}\mathrel{;}{{\color{\colorMATH}\ensuremath{e_{2}}}}}}}} \mathrel{:} {\begingroup\renewcommand\colorMATH{\colorMATHM}\renewcommand\colorSYNTAX{\colorSYNTAXM}{{\color{\colorMATH}\ensuremath{\tau _{2}}}}\endgroup }
  }
\and\inferrule*[flushleft,lab={{\color{\colorSYNTAX}\texttt{\ensuremath{\multimap ^{*}}}}}{{\color{\colorTEXT}\textsc{\scriptsize -E}}}
  ]{ {\begingroup\renewcommand\colorMATH{\colorMATHS}\renewcommand\colorSYNTAX{\colorSYNTAXS}{{\color{\colorMATH}\ensuremath{\Gamma  \vdash  e \mathrel{:} {\begingroup\renewcommand\colorMATH{\colorMATHM}\renewcommand\colorSYNTAX{\colorSYNTAXM}{{\color{\colorMATH}\ensuremath{{{\color{\colorSYNTAX}\texttt{\ensuremath{({{\color{\colorMATH}\ensuremath{\tau _{1}}}}@{\begingroup\renewcommand\colorMATH{\colorMATHP}\renewcommand\colorSYNTAX{\colorSYNTAXP}{{\color{\colorMATH}\ensuremath{p_{1}}}}\endgroup },{.}\hspace{-1pt}{.}\hspace{-1pt}{.},{{\color{\colorMATH}\ensuremath{\tau _{n}}}}@{\begingroup\renewcommand\colorMATH{\colorMATHP}\renewcommand\colorSYNTAX{\colorSYNTAXP}{{\color{\colorMATH}\ensuremath{p_{n}}}}\endgroup }) \multimap ^{*} {{\color{\colorMATH}\ensuremath{\tau }}}}}}}}}}\endgroup }}}}\endgroup }
  \\ {\begingroup\renewcommand\colorMATH{\colorMATHS}\renewcommand\colorSYNTAX{\colorSYNTAXS}{{\color{\colorMATH}\ensuremath{{}\rceil \Gamma _{1}\lceil {}^{{\begingroup\renewcommand\colorMATH{\colorMATHM}\renewcommand\colorSYNTAX{\colorSYNTAXM}{{\color{\colorMATH}\ensuremath{1}}}\endgroup }} \vdash  e_{1} \mathrel{:} {\begingroup\renewcommand\colorMATH{\colorMATHM}\renewcommand\colorSYNTAX{\colorSYNTAXM}{{\color{\colorMATH}\ensuremath{\tau _{1}}}}\endgroup }}}}\endgroup } \hspace*{1.00em}\mathrel{{\begingroup\renewcommand\colorMATH{\colorMATHM}\renewcommand\colorSYNTAX{\colorSYNTAXM}{{\color{\colorMATH}\ensuremath{{\mathord{\cdotp }}\hspace{-1pt}{\mathord{\cdotp }}\hspace{-1pt}{\mathord{\cdotp }}}}}\endgroup }}\hspace*{1.00em} {\begingroup\renewcommand\colorMATH{\colorMATHS}\renewcommand\colorSYNTAX{\colorSYNTAXS}{{\color{\colorMATH}\ensuremath{{}\rceil \Gamma _{n}\lceil {}^{{\begingroup\renewcommand\colorMATH{\colorMATHM}\renewcommand\colorSYNTAX{\colorSYNTAXM}{{\color{\colorMATH}\ensuremath{1}}}\endgroup }} \vdash  e_{n} \mathrel{:} {\begingroup\renewcommand\colorMATH{\colorMATHM}\renewcommand\colorSYNTAX{\colorSYNTAXM}{{\color{\colorMATH}\ensuremath{\tau _{n}}}}\endgroup }}}}\endgroup }
     }{
     {}\rceil {\begingroup\renewcommand\colorMATH{\colorMATHS}\renewcommand\colorSYNTAX{\colorSYNTAXS}{{\color{\colorMATH}\ensuremath{\Gamma }}}\endgroup }\lceil {}^{{{\color{\colorSYNTAX}\texttt{\ensuremath{\infty }}}}} + {}\rceil {\begingroup\renewcommand\colorMATH{\colorMATHS}\renewcommand\colorSYNTAX{\colorSYNTAXS}{{\color{\colorMATH}\ensuremath{\Gamma _{1}}}}\endgroup }\lceil {}^{p_{1}} + {\mathord{\cdotp }}\hspace{-1pt}{\mathord{\cdotp }}\hspace{-1pt}{\mathord{\cdotp }} + {}\rceil {\begingroup\renewcommand\colorMATH{\colorMATHS}\renewcommand\colorSYNTAX{\colorSYNTAXS}{{\color{\colorMATH}\ensuremath{\Gamma _{n}}}}\endgroup }\lceil {}^{p_{n}} \vdash  {{\color{\colorSYNTAX}\texttt{\ensuremath{{\begingroup\renewcommand\colorMATH{\colorMATHS}\renewcommand\colorSYNTAX{\colorSYNTAXS}{{\color{\colorMATH}\ensuremath{e}}}\endgroup }({\begingroup\renewcommand\colorMATH{\colorMATHS}\renewcommand\colorSYNTAX{\colorSYNTAXS}{{\color{\colorMATH}\ensuremath{e_{1}}}}\endgroup },{.}\hspace{-1pt}{.}\hspace{-1pt}{.},{\begingroup\renewcommand\colorMATH{\colorMATHS}\renewcommand\colorSYNTAX{\colorSYNTAXS}{{\color{\colorMATH}\ensuremath{e_{n}}}}\endgroup })}}}} \mathrel{:} {\begingroup\renewcommand\colorMATH{\colorMATHM}\renewcommand\colorSYNTAX{\colorSYNTAXM}{{\color{\colorMATH}\ensuremath{\tau }}}\endgroup }
  }
\and\inferrule*[flushleft,lab={{\color{\colorTEXT}\textsc{\scriptsize Loop}}} {{\color{\colorTEXT}\textit{\tiny (Advanced Composition)}}}
  ]{ {\begingroup\renewcommand\colorMATH{\colorMATHS}\renewcommand\colorSYNTAX{\colorSYNTAXS}{{\color{\colorMATH}\ensuremath{\Gamma _{1} \vdash  e_{1} \mathrel{:} {\begingroup\renewcommand\colorMATH{\colorMATHM}\renewcommand\colorSYNTAX{\colorSYNTAXM}{{\color{\colorMATH}\ensuremath{{{\color{\colorSYNTAX}\texttt{\ensuremath{{\mathbb{R}}^{+}[{{\color{\colorMATH}\ensuremath{\delta ^{\prime}}}}]}}}}}}}\endgroup }}}}\endgroup }
  \\ {\begingroup\renewcommand\colorMATH{\colorMATHS}\renewcommand\colorSYNTAX{\colorSYNTAXS}{{\color{\colorMATH}\ensuremath{\Gamma _{2} \vdash  e_{2} \mathrel{:} {\begingroup\renewcommand\colorMATH{\colorMATHM}\renewcommand\colorSYNTAX{\colorSYNTAXM}{{\color{\colorMATH}\ensuremath{{{\color{\colorSYNTAX}\texttt{\ensuremath{{\mathbb{N}}[{{\color{\colorMATH}\ensuremath{n}}}]}}}}}}}\endgroup }}}}\endgroup }
  \\ {\begingroup\renewcommand\colorMATH{\colorMATHS}\renewcommand\colorSYNTAX{\colorSYNTAXS}{{\color{\colorMATH}\ensuremath{\Gamma _{3} \vdash  e_{3} \mathrel{:} {\begingroup\renewcommand\colorMATH{\colorMATHM}\renewcommand\colorSYNTAX{\colorSYNTAXM}{{\color{\colorMATH}\ensuremath{\tau }}}\endgroup }}}}\endgroup }
  \\ \Gamma _{4} + \mathrlap{\hspace{-0.5pt}{}\rceil }\lfloor \Gamma _{4}^{\prime}\mathrlap{\hspace{-0.5pt}\rfloor }\lceil {}_{\{ {\begingroup\renewcommand\colorMATH{\colorMATHM}\renewcommand\colorSYNTAX{\colorSYNTAXM}{{\color{\colorMATH}\ensuremath{x_{1}^{\prime}}}}\endgroup },{.}\hspace{-1pt}{.}\hspace{-1pt}{.},{\begingroup\renewcommand\colorMATH{\colorMATHM}\renewcommand\colorSYNTAX{\colorSYNTAXM}{{\color{\colorMATH}\ensuremath{x_{n}^{\prime}}}}\endgroup }\} }^{{\begingroup\renewcommand\colorMATH{\colorMATHM}\renewcommand\colorSYNTAX{\colorSYNTAXM}{{\color{\colorMATH}\ensuremath{\epsilon }}}\endgroup },{\begingroup\renewcommand\colorMATH{\colorMATHM}\renewcommand\colorSYNTAX{\colorSYNTAXM}{{\color{\colorMATH}\ensuremath{\delta }}}\endgroup }} \uplus  \{ {\begingroup\renewcommand\colorMATH{\colorMATHM}\renewcommand\colorSYNTAX{\colorSYNTAXM}{{\color{\colorMATH}\ensuremath{x_{1}}}}\endgroup } \mathrel{:}_{\infty } {\begingroup\renewcommand\colorMATH{\colorMATHM}\renewcommand\colorSYNTAX{\colorSYNTAXM}{{\color{\colorMATH}\ensuremath{{{\color{\colorSYNTAX}\texttt{\ensuremath{{\mathbb{N}}}}}}}}}\endgroup },{\begingroup\renewcommand\colorMATH{\colorMATHM}\renewcommand\colorSYNTAX{\colorSYNTAXM}{{\color{\colorMATH}\ensuremath{x_{2}}}}\endgroup } \mathrel{:}_{\infty } {\begingroup\renewcommand\colorMATH{\colorMATHM}\renewcommand\colorSYNTAX{\colorSYNTAXM}{{\color{\colorMATH}\ensuremath{\tau }}}\endgroup }\}  \vdash  e_{4} \mathrel{:} {\begingroup\renewcommand\colorMATH{\colorMATHM}\renewcommand\colorSYNTAX{\colorSYNTAXM}{{\color{\colorMATH}\ensuremath{\tau }}}\endgroup }
     }{
     {}\rceil {\begingroup\renewcommand\colorMATH{\colorMATHS}\renewcommand\colorSYNTAX{\colorSYNTAXS}{{\color{\colorMATH}\ensuremath{\Gamma _{3}}}}\endgroup }\lceil {}^{{{\color{\colorSYNTAX}\texttt{\ensuremath{\infty }}}}} + {}\rceil \Gamma _{4}\lceil {}^{{{\color{\colorSYNTAX}\texttt{\ensuremath{\infty }}}}} 
     + \mathrlap{\hspace{-0.5pt}{}\rceil }\lfloor \Gamma _{4}^{\prime}\mathrlap{\hspace{-0.5pt}\rfloor }\lceil {}_{\{ {\begingroup\renewcommand\colorMATH{\colorMATHM}\renewcommand\colorSYNTAX{\colorSYNTAXM}{{\color{\colorMATH}\ensuremath{x_{1}^{\prime}}}}\endgroup },{.}\hspace{-1pt}{.}\hspace{-1pt}{.},{\begingroup\renewcommand\colorMATH{\colorMATHM}\renewcommand\colorSYNTAX{\colorSYNTAXM}{{\color{\colorMATH}\ensuremath{x_{n}^{\prime}}}}\endgroup }\} }^{{\begingroup\renewcommand\colorMATH{\colorMATHM}\renewcommand\colorSYNTAX{\colorSYNTAXM}{{\color{\colorMATH}\ensuremath{2\epsilon \sqrt {2n\ln (1/\delta ^{\prime})}}}}\endgroup },{\begingroup\renewcommand\colorMATH{\colorMATHM}\renewcommand\colorSYNTAX{\colorSYNTAXM}{{\color{\colorMATH}\ensuremath{\delta ^{\prime}+n\delta }}}\endgroup }} 
     \vdash  {{\color{\colorSYNTAX}\texttt{\ensuremath{{{\color{\colorSYNTAX}\texttt{loop}}}[{\begingroup\renewcommand\colorMATH{\colorMATHS}\renewcommand\colorSYNTAX{\colorSYNTAXS}{{\color{\colorMATH}\ensuremath{e_{1}}}}\endgroup }]\hspace*{0.33em}{\begingroup\renewcommand\colorMATH{\colorMATHS}\renewcommand\colorSYNTAX{\colorSYNTAXS}{{\color{\colorMATH}\ensuremath{e_{2}}}}\endgroup }\hspace*{0.33em}{{\color{\colorSYNTAX}\texttt{on}}}\hspace*{0.33em}{\begingroup\renewcommand\colorMATH{\colorMATHS}\renewcommand\colorSYNTAX{\colorSYNTAXS}{{\color{\colorMATH}\ensuremath{e_{3}}}}\endgroup }\hspace*{0.33em}{<}{\begingroup\renewcommand\colorMATH{\colorMATHM}\renewcommand\colorSYNTAX{\colorSYNTAXM}{{\color{\colorMATH}\ensuremath{x_{1}^{\prime}}}}\endgroup },{.}\hspace{-1pt}{.}\hspace{-1pt}{.},{\begingroup\renewcommand\colorMATH{\colorMATHM}\renewcommand\colorSYNTAX{\colorSYNTAXM}{{\color{\colorMATH}\ensuremath{x_{n}^{\prime}}}}\endgroup }{>}\hspace*{0.33em}\{ {\begingroup\renewcommand\colorMATH{\colorMATHM}\renewcommand\colorSYNTAX{\colorSYNTAXM}{{\color{\colorMATH}\ensuremath{x_{1}}}}\endgroup },{\begingroup\renewcommand\colorMATH{\colorMATHM}\renewcommand\colorSYNTAX{\colorSYNTAXM}{{\color{\colorMATH}\ensuremath{x_{2}}}}\endgroup } \Rightarrow  {{\color{\colorMATH}\ensuremath{e_{4}}}}\} }}}} \mathrel{:} {\begingroup\renewcommand\colorMATH{\colorMATHM}\renewcommand\colorSYNTAX{\colorSYNTAXM}{{\color{\colorMATH}\ensuremath{\tau }}}\endgroup }
  }
\and\inferrule*[lab={{\color{\colorTEXT}\textsc{\scriptsize Gauss}}}
  ]{ {\begingroup\renewcommand\colorMATH{\colorMATHS}\renewcommand\colorSYNTAX{\colorSYNTAXS}{{\color{\colorMATH}\ensuremath{\Gamma _{1} \vdash  e_{1} \mathrel{:} {\begingroup\renewcommand\colorMATH{\colorMATHM}\renewcommand\colorSYNTAX{\colorSYNTAXM}{{\color{\colorMATH}\ensuremath{{{\color{\colorSYNTAX}\texttt{\ensuremath{{\mathbb{R}}^{+}[{{\color{\colorMATH}\ensuremath{\dot r_{s}}}}]}}}}}}}\endgroup }}}}\endgroup }
  \\ {\begingroup\renewcommand\colorMATH{\colorMATHS}\renewcommand\colorSYNTAX{\colorSYNTAXS}{{\color{\colorMATH}\ensuremath{\Gamma _{2} \vdash  e_{2} \mathrel{:} {\begingroup\renewcommand\colorMATH{\colorMATHM}\renewcommand\colorSYNTAX{\colorSYNTAXM}{{\color{\colorMATH}\ensuremath{{{\color{\colorSYNTAX}\texttt{\ensuremath{{\mathbb{R}}^{+}[{{\color{\colorMATH}\ensuremath{\epsilon }}}]}}}}}}}\endgroup }}}}\endgroup }
  \\ {\begingroup\renewcommand\colorMATH{\colorMATHS}\renewcommand\colorSYNTAX{\colorSYNTAXS}{{\color{\colorMATH}\ensuremath{\Gamma _{3} \vdash  e_{3} \mathrel{:} {\begingroup\renewcommand\colorMATH{\colorMATHM}\renewcommand\colorSYNTAX{\colorSYNTAXM}{{\color{\colorMATH}\ensuremath{{{\color{\colorSYNTAX}\texttt{\ensuremath{{\mathbb{R}}^{+}[{{\color{\colorMATH}\ensuremath{\delta }}}]}}}}}}}\endgroup }}}}\endgroup }
  \\ {\begingroup\renewcommand\colorMATH{\colorMATHS}\renewcommand\colorSYNTAX{\colorSYNTAXS}{{\color{\colorMATH}\ensuremath{\Gamma _{4} + \mathrlap{\hspace{-0.5pt}{}\rceil }\lfloor \Gamma _{4}^{\prime}\mathrlap{\hspace{-0.5pt}\rfloor }\lceil {}_{\{ {\begingroup\renewcommand\colorMATH{\colorMATHM}\renewcommand\colorSYNTAX{\colorSYNTAXM}{{\color{\colorMATH}\ensuremath{x_{1}}}}\endgroup },{.}\hspace{-1pt}{.}\hspace{-1pt}{.},{\begingroup\renewcommand\colorMATH{\colorMATHM}\renewcommand\colorSYNTAX{\colorSYNTAXM}{{\color{\colorMATH}\ensuremath{x_{n}}}}\endgroup }\} }^{{\begingroup\renewcommand\colorMATH{\colorMATHM}\renewcommand\colorSYNTAX{\colorSYNTAXM}{{\color{\colorMATH}\ensuremath{\dot r_{s}}}}\endgroup }} \vdash  e_{4} \mathrel{:} {\begingroup\renewcommand\colorMATH{\colorMATHM}\renewcommand\colorSYNTAX{\colorSYNTAXM}{{\color{\colorMATH}\ensuremath{{{\color{\colorSYNTAX}\texttt{\ensuremath{{\mathbb{R}}}}}}}}}\endgroup }}}}\endgroup }
     }{
     {}\rceil {\begingroup\renewcommand\colorMATH{\colorMATHS}\renewcommand\colorSYNTAX{\colorSYNTAXS}{{\color{\colorMATH}\ensuremath{\Gamma _{4}}}}\endgroup }\lceil {}^{{{\color{\colorSYNTAX}\texttt{\ensuremath{\infty }}}}} + \mathrlap{\hspace{-0.5pt}{}\rceil }\lfloor {\begingroup\renewcommand\colorMATH{\colorMATHS}\renewcommand\colorSYNTAX{\colorSYNTAXS}{{\color{\colorMATH}\ensuremath{\Gamma _{4}^{\prime}}}}\endgroup }\mathrlap{\hspace{-0.5pt}\rfloor }\lceil {}_{\{ {\begingroup\renewcommand\colorMATH{\colorMATHM}\renewcommand\colorSYNTAX{\colorSYNTAXM}{{\color{\colorMATH}\ensuremath{x_{1}^{\prime}}}}\endgroup },{.}\hspace{-1pt}{.}\hspace{-1pt}{.},{\begingroup\renewcommand\colorMATH{\colorMATHM}\renewcommand\colorSYNTAX{\colorSYNTAXM}{{\color{\colorMATH}\ensuremath{x_{n}^{\prime}}}}\endgroup }\} }^{{\begingroup\renewcommand\colorMATH{\colorMATHM}\renewcommand\colorSYNTAX{\colorSYNTAXM}{{\color{\colorMATH}\ensuremath{\epsilon }}}\endgroup },{\begingroup\renewcommand\colorMATH{\colorMATHM}\renewcommand\colorSYNTAX{\colorSYNTAXM}{{\color{\colorMATH}\ensuremath{\delta }}}\endgroup }} 
     \vdash  {{\color{\colorSYNTAX}\texttt{\ensuremath{{{\color{\colorSYNTAX}\texttt{gauss}}}[{\begingroup\renewcommand\colorMATH{\colorMATHS}\renewcommand\colorSYNTAX{\colorSYNTAXS}{{\color{\colorMATH}\ensuremath{e_{1}}}}\endgroup },{\begingroup\renewcommand\colorMATH{\colorMATHS}\renewcommand\colorSYNTAX{\colorSYNTAXS}{{\color{\colorMATH}\ensuremath{e_{2}}}}\endgroup },{\begingroup\renewcommand\colorMATH{\colorMATHS}\renewcommand\colorSYNTAX{\colorSYNTAXS}{{\color{\colorMATH}\ensuremath{e_{3}}}}\endgroup }]\hspace*{0.33em}{<}{\begingroup\renewcommand\colorMATH{\colorMATHM}\renewcommand\colorSYNTAX{\colorSYNTAXM}{{\color{\colorMATH}\ensuremath{x_{1}^{\prime}}}}\endgroup },{.}\hspace{-1pt}{.}\hspace{-1pt}{.},{\begingroup\renewcommand\colorMATH{\colorMATHM}\renewcommand\colorSYNTAX{\colorSYNTAXM}{{\color{\colorMATH}\ensuremath{x_{n}^{\prime}}}}\endgroup }{>}\hspace*{0.33em}\{ {\begingroup\renewcommand\colorMATH{\colorMATHS}\renewcommand\colorSYNTAX{\colorSYNTAXS}{{\color{\colorMATH}\ensuremath{e_{4}}}}\endgroup }\} }}}} \mathrel{:} {\begingroup\renewcommand\colorMATH{\colorMATHM}\renewcommand\colorSYNTAX{\colorSYNTAXM}{{\color{\colorMATH}\ensuremath{{{\color{\colorSYNTAX}\texttt{\ensuremath{{\mathbb{R}}}}}}}}}\endgroup }
  }
\end{mathpar}\endgroup 
\endgroup 
\vspace{-1.5em}
\caption{Core Typing Rules}
\vspace{-0.5em}
\label{fig:typing}
\end{figure*}
}

\FigureCoreLang
\FigureCoreTyping

Figure~\ref{fig:syntax} shows a core subset of syntax for both languages. We
only present the privacy fragment for {{\color{\colorMATH}\ensuremath{(\epsilon ,\delta )}}}-differential privacy in the core
formalism, although support for other variants (and combined variants) is
straightforward as sketched in the previous section. See the extended version
of this paper for the complete presentation of the full language including all
advanced variants of differential privacy. We use color coding to distinguish
between the sensitivity language, privacy language, and shared syntax between
languages. The sensitivity and privacy languages share syntax for variables and
types, which are typeset in {{\color{\colorMATH}\ensuremath{{{\color{\colorMATH}\ensuremath{\operatorname{blue}}}}}}}. Expressions in the sensitivity language
are typeset in {\begingroup\renewcommand\colorMATH{\colorMATHS}\renewcommand\colorSYNTAX{\colorSYNTAXS}{{\color{\colorMATH}\ensuremath{{{\color{\colorMATH}\ensuremath{\operatorname{green}}}}}}}\endgroup }, while expressions in the privacy language are
typeset in {\begingroup\renewcommand\colorMATH{\colorMATHP}\renewcommand\colorSYNTAX{\colorSYNTAXP}{{\color{\colorMATH}\ensuremath{{{\color{\colorMATH}\ensuremath{\operatorname{red}}}}}}}\endgroup }.\footnote{Colors were chosen to minimize ambiguity for
colorblind persons following a colorblind-friendly palette:
\url{http://mkweb.bcgsc.ca/colorblind/img/colorblindness.palettes.png}}

Types {{\color{\colorMATH}\ensuremath{\tau }}} include base numeric types
{{\color{\colorSYNTAX}\texttt{\ensuremath{{\mathbb{N}}}}}} and {{\color{\colorSYNTAX}\texttt{\ensuremath{{\mathbb{R}}}}}} and their treatment is standard. We include singleton numeric
types {{\color{\colorSYNTAX}\texttt{\ensuremath{{\mathbb{N}}[{{\color{\colorMATH}\ensuremath{n}}}]}}}} and {{\color{\colorSYNTAX}\texttt{\ensuremath{{\mathbb{R}}^{+}[{{\color{\colorMATH}\ensuremath{\dot r}}}]}}}}; these types
classify runtime numeric values which are identical to the static index {{\color{\colorMATH}\ensuremath{n}}} or
{{\color{\colorMATH}\ensuremath{\dot r}}}, {e.g.}, {{\color{\colorSYNTAX}\texttt{\ensuremath{{\mathbb{N}}[{{\color{\colorMATH}\ensuremath{n}}}]}}}} is a type which exactly describes its
runtime value as the number {{\color{\colorMATH}\ensuremath{n}}}. Static reals only range over non-negative
values, and we write {{\color{\colorMATH}\ensuremath{\dot r}}} for elements of the non-negative reals {{\color{\colorSYNTAX}\texttt{\ensuremath{{\mathbb{R}}^{+}}}}}.
Singleton natural numbers are used primarily to construct matrices with some
statically known dimension, and to execute loops for some statically known
number of iterations. Singleton real numbers and are used primarily for
tracking sensitivity and privacy quantities.  Novel in \system is a ``boxed''
type {{\color{\colorSYNTAX}\texttt{\ensuremath{{{\color{\colorMATH}\ensuremath{\operatorname{box}}}}[{\begingroup\renewcommand\colorMATH{\colorMATHS}\renewcommand\colorSYNTAX{\colorSYNTAXS}{{\color{\colorMATH}\ensuremath{\Gamma _{s}}}}\endgroup }]\hspace*{0.33em}{{\color{\colorMATH}\ensuremath{\tau }}}}}}} which delays the ``payment'' of a value's sensitivity,
to be unboxed and ``paid for'' in a separate context. Boxing is discussed in more
detail later in this section. The sensitivity function space (a la Fuzz) is
written {{\color{\colorSYNTAX}\texttt{\ensuremath{{{\color{\colorMATH}\ensuremath{\tau _{1}}}} \multimap _{{\begingroup\renewcommand\colorMATH{\colorMATHS}\renewcommand\colorSYNTAX{\colorSYNTAXS}{{\color{\colorMATH}\ensuremath{s}}}\endgroup }} {{\color{\colorMATH}\ensuremath{\tau _{2}}}}}}}} and encodes an {\begingroup\renewcommand\colorMATH{\colorMATHS}\renewcommand\colorSYNTAX{\colorSYNTAXS}{{\color{\colorMATH}\ensuremath{s}}}\endgroup }-sensitive function from
{{\color{\colorMATH}\ensuremath{\tau _{1}}}} to {{\color{\colorMATH}\ensuremath{\tau _{2}}}}. The privacy function space (novel in \system) is written
{{\color{\colorSYNTAX}\texttt{\ensuremath{({{\color{\colorMATH}\ensuremath{\tau _{1}}}}@{\begingroup\renewcommand\colorMATH{\colorMATHP}\renewcommand\colorSYNTAX{\colorSYNTAXP}{{\color{\colorMATH}\ensuremath{p_{1}}}}\endgroup },{.}\hspace{-1pt}{.}\hspace{-1pt}{.},{{\color{\colorMATH}\ensuremath{\tau _{n}}}}@{\begingroup\renewcommand\colorMATH{\colorMATHP}\renewcommand\colorSYNTAX{\colorSYNTAXP}{{\color{\colorMATH}\ensuremath{p_{n}}}}\endgroup }) \multimap ^{*} {{\color{\colorMATH}\ensuremath{\tau }}}}}}} and encodes a multi-arity function that
preserves {\begingroup\renewcommand\colorMATH{\colorMATHP}\renewcommand\colorSYNTAX{\colorSYNTAXP}{{\color{\colorMATH}\ensuremath{p_{i}}}}\endgroup }-privacy for its {\begingroup\renewcommand\colorMATH{\colorMATHP}\renewcommand\colorSYNTAX{\colorSYNTAXP}{{\color{\colorMATH}\ensuremath{i}}}\endgroup }th argument. Privacy functions are
multi-arity because functions of multiple arguments cannot be recovered from
iterating functions over single arguments in the privacy language, as can be
done in the sensitivity language.

In our implementation and extended presentation of \system in the extended
version of this paper, we generalize the static representations of natural
numbers and reals to symbolic expression {{\color{\colorMATH}\ensuremath{\eta }}}, which may be arbitrary symbolic
polynomial formulas including variables. E.g., suppose {{\color{\colorMATH}\ensuremath{\epsilon }}} is a type-level
variable ranging over real numbers and {{\color{\colorMATH}\ensuremath{x {\mathrel{:}} {{\color{\colorSYNTAX}\texttt{\ensuremath{{\mathbb{N}}[{{\color{\colorMATH}\ensuremath{\epsilon }}}]}}}}}}}, then {{\color{\colorMATH}\ensuremath{2x {\mathrel{:}}
{{\color{\colorSYNTAX}\texttt{\ensuremath{{\mathbb{N}}[{{\color{\colorMATH}\ensuremath{2\epsilon }}}]}}}}}}}. Our type checker knows this is the same type as {{\color{\colorSYNTAX}\texttt{\ensuremath{{\mathbb{N}}[{{\color{\colorMATH}\ensuremath{\epsilon {+}\epsilon }}}]}}}}
using a custom solver we implemented but do not describe in this paper. Because
the typelevel representation of a natural number can be a variable, its value
is therefore not statically {{\color{\colorTEXT}\textit{determined}}}, rather it is statically {{\color{\colorTEXT}\textit{tracked}}} via
typelevel symbolic formulas.

Type contexts in the sensitivity language {\begingroup\renewcommand\colorMATH{\colorMATHS}\renewcommand\colorSYNTAX{\colorSYNTAXS}{{\color{\colorMATH}\ensuremath{\Gamma _{s}}}}\endgroup } track the \emph{sensitivity}
{\begingroup\renewcommand\colorMATH{\colorMATHS}\renewcommand\colorSYNTAX{\colorSYNTAXS}{{\color{\colorMATH}\ensuremath{s}}}\endgroup } of each free variable whereas in the privacy language {\begingroup\renewcommand\colorMATH{\colorMATHP}\renewcommand\colorSYNTAX{\colorSYNTAXP}{{\color{\colorMATH}\ensuremath{\Gamma _{p}}}}\endgroup } they track
\emph{privacy cost} {\begingroup\renewcommand\colorMATH{\colorMATHP}\renewcommand\colorSYNTAX{\colorSYNTAXP}{{\color{\colorMATH}\ensuremath{p}}}\endgroup }. Sensitivities are non-negative reals {{\color{\colorMATH}\ensuremath{\dot r}}} extended
with a distinguished infinity element {\begingroup\renewcommand\colorMATH{\colorMATHS}\renewcommand\colorSYNTAX{\colorSYNTAXS}{{\color{\colorMATH}\ensuremath{{{\color{\colorSYNTAX}\texttt{\ensuremath{\infty }}}}}}}\endgroup }, and privacy costs are specific
to the current privacy {{\color{\colorTEXT}\textit{mode}}}.
In the case of {{\color{\colorMATH}\ensuremath{(\epsilon ,\delta )}}}-differential privacy, {\begingroup\renewcommand\colorMATH{\colorMATHP}\renewcommand\colorSYNTAX{\colorSYNTAXP}{{\color{\colorMATH}\ensuremath{p}}}\endgroup } has the form
{\begingroup\renewcommand\colorMATH{\colorMATHP}\renewcommand\colorSYNTAX{\colorSYNTAXP}{{\color{\colorMATH}\ensuremath{{{\color{\colorSYNTAX}\texttt{\ensuremath{{{\color{\colorMATH}\ensuremath{\epsilon }}},{{\color{\colorMATH}\ensuremath{\delta }}}}}}}}}}\endgroup } or {\begingroup\renewcommand\colorMATH{\colorMATHP}\renewcommand\colorSYNTAX{\colorSYNTAXP}{{\color{\colorMATH}\ensuremath{{{\color{\colorSYNTAX}\texttt{\ensuremath{\infty }}}}}}}\endgroup } where {{\color{\colorMATH}\ensuremath{\epsilon }}} and {{\color{\colorMATH}\ensuremath{\delta }}} range over {{\color{\colorSYNTAX}\texttt{\ensuremath{{\mathbb{R}}^{+}}}}}.

We reuse notation conventions from Fuzz for manipulating contexts, {e.g.}, {\begingroup\renewcommand\colorMATH{\colorMATHS}\renewcommand\colorSYNTAX{\colorSYNTAXS}{{\color{\colorMATH}\ensuremath{\Gamma _{1}
{+} \Gamma _{2}}}}\endgroup } is partial and defined only when both contexts agree on the type of
each variable; adding contexts adds sensitivities pointwise, i.e., {\begingroup\renewcommand\colorMATH{\colorMATHS}\renewcommand\colorSYNTAX{\colorSYNTAXS}{{\color{\colorMATH}\ensuremath{\{ {\begingroup\renewcommand\colorMATH{\colorMATHM}\renewcommand\colorSYNTAX{\colorSYNTAXM}{{\color{\colorMATH}\ensuremath{x}}}\endgroup }
{\mathrel{:}_{s_{1}{+}s_{2}}} {\begingroup\renewcommand\colorMATH{\colorMATHM}\renewcommand\colorSYNTAX{\colorSYNTAXM}{{\color{\colorMATH}\ensuremath{\tau }}}\endgroup }\}  \in  \Gamma _{1} {+} \Gamma _{2}}}}\endgroup } when {\begingroup\renewcommand\colorMATH{\colorMATHS}\renewcommand\colorSYNTAX{\colorSYNTAXS}{{\color{\colorMATH}\ensuremath{\{ {\begingroup\renewcommand\colorMATH{\colorMATHM}\renewcommand\colorSYNTAX{\colorSYNTAXM}{{\color{\colorMATH}\ensuremath{x}}}\endgroup } {\mathrel{:}_{s_{1}}} {\begingroup\renewcommand\colorMATH{\colorMATHM}\renewcommand\colorSYNTAX{\colorSYNTAXM}{{\color{\colorMATH}\ensuremath{\tau }}}\endgroup }\}  \in  \Gamma _{1}}}}\endgroup } and {\begingroup\renewcommand\colorMATH{\colorMATHS}\renewcommand\colorSYNTAX{\colorSYNTAXS}{{\color{\colorMATH}\ensuremath{\{ {\begingroup\renewcommand\colorMATH{\colorMATHM}\renewcommand\colorSYNTAX{\colorSYNTAXM}{{\color{\colorMATH}\ensuremath{x}}}\endgroup }
{\mathrel{:}_{s_{2}}} {\begingroup\renewcommand\colorMATH{\colorMATHM}\renewcommand\colorSYNTAX{\colorSYNTAXM}{{\color{\colorMATH}\ensuremath{\tau }}}\endgroup }\}  \in  \Gamma _{2}}}}\endgroup }; and scaling contexts scales sensitivities pointwise,
i.e., {\begingroup\renewcommand\colorMATH{\colorMATHS}\renewcommand\colorSYNTAX{\colorSYNTAXS}{{\color{\colorMATH}\ensuremath{\{ {\begingroup\renewcommand\colorMATH{\colorMATHM}\renewcommand\colorSYNTAX{\colorSYNTAXM}{{\color{\colorMATH}\ensuremath{x}}}\endgroup } {\mathrel{:}_{ss^{\prime}}} {\begingroup\renewcommand\colorMATH{\colorMATHM}\renewcommand\colorSYNTAX{\colorSYNTAXM}{{\color{\colorMATH}\ensuremath{\tau }}}\endgroup }\}  \in  s\Gamma }}}\endgroup } when {\begingroup\renewcommand\colorMATH{\colorMATHS}\renewcommand\colorSYNTAX{\colorSYNTAXS}{{\color{\colorMATH}\ensuremath{\{ {\begingroup\renewcommand\colorMATH{\colorMATHM}\renewcommand\colorSYNTAX{\colorSYNTAXM}{{\color{\colorMATH}\ensuremath{x}}}\endgroup } {\mathrel{:}_{s^{\prime}}} {\begingroup\renewcommand\colorMATH{\colorMATHM}\renewcommand\colorSYNTAX{\colorSYNTAXM}{{\color{\colorMATH}\ensuremath{\tau }}}\endgroup }\}  \in  \Gamma }}}\endgroup }. 

We introduce a new operation not shown in prior work called {{\color{\colorTEXT}\textit{truncation}}} and
written {\begingroup\renewcommand\colorMATH{\colorMATHS}\renewcommand\colorSYNTAX{\colorSYNTAXS}{{\color{\colorMATH}\ensuremath{{}\rceil s_{1}\lceil {}^{s_{2}}}}}\endgroup } for truncating a sensitivity and {\begingroup\renewcommand\colorMATH{\colorMATHS}\renewcommand\colorSYNTAX{\colorSYNTAXS}{{\color{\colorMATH}\ensuremath{{}\rceil \Gamma \lceil {}^{s}}}}\endgroup } for truncating
a sensitivity context, which is pointwise truncation of sensitivities.
Sensitivity truncation {\begingroup\renewcommand\colorMATH{\colorMATHS}\renewcommand\colorSYNTAX{\colorSYNTAXS}{{\color{\colorMATH}\ensuremath{{}\rceil \underline{\hspace{0.66em}\vspace*{5ex}}\lceil {}^{s}}}}\endgroup } maps {{\color{\colorMATH}\ensuremath{0}}} to {{\color{\colorMATH}\ensuremath{0}}} and any other value to {\begingroup\renewcommand\colorMATH{\colorMATHS}\renewcommand\colorSYNTAX{\colorSYNTAXS}{{\color{\colorMATH}\ensuremath{s}}}\endgroup }:
\vspace*{-0.25em}\begingroup\color{\colorMATH}\begin{gather*}
\begin{tabularx}{\linewidth}{>{\centering\arraybackslash\(}X<{\)}}\hfill\hspace{0pt} {\begingroup\renewcommand\colorMATH{\colorMATHS}\renewcommand\colorSYNTAX{\colorSYNTAXS}{{\color{\colorMATH}\ensuremath{{}\rceil \underline{\hspace{0.66em}\vspace*{5ex}}\lceil {}^{\underline{\hspace{0.66em}\vspace*{5ex}}}}}}\endgroup } \in  {\begingroup\renewcommand\colorMATH{\colorMATHS}\renewcommand\colorSYNTAX{\colorSYNTAXS}{{\color{\colorMATH}\ensuremath{{{\color{\colorMATH}\ensuremath{\operatorname{sens}}}}}}}\endgroup } \times  {\begingroup\renewcommand\colorMATH{\colorMATHS}\renewcommand\colorSYNTAX{\colorSYNTAXS}{{\color{\colorMATH}\ensuremath{{{\color{\colorMATH}\ensuremath{\operatorname{sens}}}}}}}\endgroup } \rightarrow  {\begingroup\renewcommand\colorMATH{\colorMATHS}\renewcommand\colorSYNTAX{\colorSYNTAXS}{{\color{\colorMATH}\ensuremath{{{\color{\colorMATH}\ensuremath{\operatorname{sens}}}}}}}\endgroup }
  \hfill\hspace{0pt} {\begingroup\renewcommand\colorMATH{\colorMATHS}\renewcommand\colorSYNTAX{\colorSYNTAXS}{{\color{\colorMATH}\ensuremath{{}\rceil s_{1}\lceil {}^{s_{2}}}}}\endgroup } \triangleq  \left\{ \begin{array}{l@{\hspace*{1.00em}}l@{\hspace*{1.00em}}l
                     } 0      & {{\color{\colorTEXT}\textit{if}}} & {\begingroup\renewcommand\colorMATH{\colorMATHS}\renewcommand\colorSYNTAX{\colorSYNTAXS}{{\color{\colorMATH}\ensuremath{s_{1}}}}\endgroup } = 0
                     \cr  {\begingroup\renewcommand\colorMATH{\colorMATHS}\renewcommand\colorSYNTAX{\colorSYNTAXS}{{\color{\colorMATH}\ensuremath{s_{2}}}}\endgroup } & {{\color{\colorTEXT}\textit{if}}} & {\begingroup\renewcommand\colorMATH{\colorMATHS}\renewcommand\colorSYNTAX{\colorSYNTAXS}{{\color{\colorMATH}\ensuremath{s_{1}}}}\endgroup } \neq  0
                     \end{array}\right.
  \hfill\hspace{0pt}
\end{tabularx}
\vspace*{-1em}\end{gather*}\endgroup 
Truncation is defined analogously for privacies {\begingroup\renewcommand\colorMATH{\colorMATHP}\renewcommand\colorSYNTAX{\colorSYNTAXP}{{\color{\colorMATH}\ensuremath{{}\rceil p_{1}\lceil {}^{p_{2}}}}}\endgroup }, for converting
between sensitivities and privacies {\begingroup\renewcommand\colorMATH{\colorMATHP}\renewcommand\colorSYNTAX{\colorSYNTAXP}{{\color{\colorMATH}\ensuremath{{}\rceil {\begingroup\renewcommand\colorMATH{\colorMATHS}\renewcommand\colorSYNTAX{\colorSYNTAXS}{{\color{\colorMATH}\ensuremath{s}}}\endgroup }\lceil {}^{p}}}}\endgroup } and {\begingroup\renewcommand\colorMATH{\colorMATHS}\renewcommand\colorSYNTAX{\colorSYNTAXS}{{\color{\colorMATH}\ensuremath{{}\rceil {\begingroup\renewcommand\colorMATH{\colorMATHP}\renewcommand\colorSYNTAX{\colorSYNTAXP}{{\color{\colorMATH}\ensuremath{p}}}\endgroup }\lceil {}^{s}}}}\endgroup }, and also
for liftings of these operations pointwise over contexts {\begingroup\renewcommand\colorMATH{\colorMATHP}\renewcommand\colorSYNTAX{\colorSYNTAXP}{{\color{\colorMATH}\ensuremath{{}\rceil \Gamma \lceil {}^{p}}}}\endgroup },
{\begingroup\renewcommand\colorMATH{\colorMATHP}\renewcommand\colorSYNTAX{\colorSYNTAXP}{{\color{\colorMATH}\ensuremath{{}\rceil {\begingroup\renewcommand\colorMATH{\colorMATHS}\renewcommand\colorSYNTAX{\colorSYNTAXS}{{\color{\colorMATH}\ensuremath{\Gamma }}}\endgroup }\lceil {}^{p}}}}\endgroup } and {\begingroup\renewcommand\colorMATH{\colorMATHS}\renewcommand\colorSYNTAX{\colorSYNTAXS}{{\color{\colorMATH}\ensuremath{{}\rceil {\begingroup\renewcommand\colorMATH{\colorMATHP}\renewcommand\colorSYNTAX{\colorSYNTAXP}{{\color{\colorMATH}\ensuremath{\Gamma }}}\endgroup }\lceil {}^{s}}}}\endgroup }. Sensitivity truncation is used for typing
the modulus operator, and truncating between sensitivities and privacies is
always to {\begingroup\renewcommand\colorMATH{\colorMATHS}\renewcommand\colorSYNTAX{\colorSYNTAXS}{{\color{\colorMATH}\ensuremath{{{\color{\colorSYNTAX}\texttt{\ensuremath{\infty }}}}}}}\endgroup }/{\begingroup\renewcommand\colorMATH{\colorMATHP}\renewcommand\colorSYNTAX{\colorSYNTAXP}{{\color{\colorMATH}\ensuremath{{{\color{\colorSYNTAX}\texttt{\ensuremath{\infty }}}}}}}\endgroup } and appears frequently in typing rules that embed
sensitivity terms in privacy terms and vice versa.

The syntax and language features for both sensitivity and privacy languages are
discussed next alongside their typing rules. Figure~\ref{fig:typing} shows a
core subset of typing rules for both languages. In the typing rules, the
languages embed within each other---sensitivity typing contexts are transformed
into privacy contexts and vice versa. Type rules are written in logical style
with an explicit subsumption rule, although a purely algorithmic presentation
is possible (not shown) following ideas from Azevedo de Amorim et
al~\cite{dfuzz-impl} which serves as the basis for our implementation.

\subsection{Sensitivity Language}
\label{sec:sensitivitylang}

\system's sensitivity language is similar to that of DFuzz~\cite{dfuzz},
except that we extend it with significant new tools for machine learning in
Section~\ref{sec:machine_learning}. We do not present standard linear logic
connectives such as sums, additive products and multiplicative products (a la
Fuzz), or symbolic type-level expressions (a la DFuzz), although each are
implemented in our tool and described formally in the extended version of this
paper. We do not formalize or implement general recursive types in order to
ensure that all \system programs terminate. Including general recursive types
would be straightforward in \system (following the design of Fuzz), however
such a decision comes with known limitations. As described in Fuzz~\cite{reed2010distance},
requiring that all functions terminate is necessary in order to give both sound
and useful types to primitives like set-filter. The design space for the
combination of sensitivity types and nontermination is subtle, and discussed
extensively in prior work~\cite{reed2010distance,azevedo2017semantic}.

Typing for literal values is immediate ({{\color{\colorTEXT}\textsc{\scriptsize Nat}}}, {{\color{\colorTEXT}\textsc{\scriptsize Real}}}). Singleton values are
constructed using the same syntax as their types, and where the type level
representation is identical to the literal ({{\color{\colorTEXT}\textsc{\scriptsize Singleton Nat}}}, {{\color{\colorTEXT}\textsc{\scriptsize Singleton Real}}}).
Naturals can be converted to real numbers through the explicit conversion
operation {\begingroup\renewcommand\colorMATH{\colorMATHP}\renewcommand\colorSYNTAX{\colorSYNTAXP}{{\color{\colorMATH}\ensuremath{{{\color{\colorSYNTAX}\texttt{real}}}}}}\endgroup } ({{\color{\colorTEXT}\textsc{\scriptsize Real-S}}}, {{\color{\colorTEXT}\textsc{\scriptsize Real-D}}}). For the purposes of sensitivity
analysis, statically known numbers are considered constant, and as a
consequence any term that uses one is considered {{\color{\colorMATH}\ensuremath{0}}}-sensitive in the
statically known term. The result of this is that the sensitivity environment
{\begingroup\renewcommand\colorMATH{\colorMATHS}\renewcommand\colorSYNTAX{\colorSYNTAXS}{{\color{\colorMATH}\ensuremath{\Gamma }}}\endgroup } associated with the subterm at singleton type is dropped from the output
environment, e.g., in {{\color{\colorTEXT}\textsc{\scriptsize Real-S}}}. This dropping is justified by our metric space
interpretation {{\color{\colorMATH}\ensuremath{\llbracket {{\color{\colorSYNTAX}\texttt{\ensuremath{{\mathbb{N}}[{{\color{\colorMATH}\ensuremath{n}}}]}}}}\rrbracket }}} for statically known numbers as singleton sets
{{\color{\colorMATH}\ensuremath{\{ n\} }}}, and because for all {{\color{\colorMATH}\ensuremath{x,y \in  \llbracket {{\color{\colorSYNTAX}\texttt{\ensuremath{{\mathbb{N}}[{{\color{\colorMATH}\ensuremath{n}}}]}}}}\rrbracket }}}, {{\color{\colorMATH}\ensuremath{x = y}}} and therefore {{\color{\colorMATH}\ensuremath{|x - y|
= 0}}}.

Type rules for arithmetic operations are given in multiple variations,
depending on whether or not each argument is tracked statically or dynamically.
We show only the rule for multiplication when the left argument is dynamic and
the right argument is static ({{\color{\colorTEXT}\textsc{\scriptsize Times-DS}}}). The resulting sensitivity
environment reports the sensitivities of {{\color{\colorMATH}\ensuremath{e_{1}}}} scaled by {{\color{\colorMATH}\ensuremath{\dot r}}}---the statically
known value of {{\color{\colorMATH}\ensuremath{e_{2}}}}---and the sensitivities for {{\color{\colorMATH}\ensuremath{e_{2}}}} are not reported because its
value is fixed and cannot vary, as discussed above. When both arguments are
dynamic, the resulting sensitivity environment is {\begingroup\renewcommand\colorMATH{\colorMATHS}\renewcommand\colorSYNTAX{\colorSYNTAXS}{{\color{\colorMATH}\ensuremath{{{\color{\colorSYNTAX}\texttt{\ensuremath{\infty }}}}(\Gamma _{1} + \Gamma _{2})}}}\endgroup }, i.e.,
all potentially sensitive variables for each expression are bumped to infinity.
The modulus operation is similar to multiplication in that we have cases for
each variation of static or dynamic arguments, however the context is truncated
rather than scaled in the case of one singleton-typed parameter; we show only
this static-dynamic variant in the figure ({{\color{\colorTEXT}\textsc{\scriptsize Mod-DS}}}).

Typing for variables ({{\color{\colorTEXT}\textsc{\scriptsize Var}}}) and functions ({{\color{\colorTEXT}\textsc{\scriptsize {\begingroup\renewcommand\colorMATH{\colorMATHS}\renewcommand\colorSYNTAX{\colorSYNTAXS}{{\color{\colorMATH}\ensuremath{{{\color{\colorSYNTAX}\texttt{\ensuremath{\multimap }}}}}}}\endgroup }-I}}}, {{\color{\colorTEXT}\textsc{\scriptsize {\begingroup\renewcommand\colorMATH{\colorMATHS}\renewcommand\colorSYNTAX{\colorSYNTAXS}{{\color{\colorMATH}\ensuremath{{{\color{\colorSYNTAX}\texttt{\ensuremath{\multimap }}}}}}}\endgroup }-E}}}) is
the same as in Fuzz: variables are reported in the sensitivity environment with
sensitivity {{\color{\colorMATH}\ensuremath{1}}}; and closures are created by annotating the arrow with the
sensitivity {\begingroup\renewcommand\colorMATH{\colorMATHS}\renewcommand\colorSYNTAX{\colorSYNTAXS}{{\color{\colorMATH}\ensuremath{s}}}\endgroup } of the argument in the body, and by reporting the rest of the
sensitivities {\begingroup\renewcommand\colorMATH{\colorMATHS}\renewcommand\colorSYNTAX{\colorSYNTAXS}{{\color{\colorMATH}\ensuremath{\Gamma }}}\endgroup } from the function body as the sensitivity of whole closure
as a whole; and function application scales the argument by the function's
sensitivity {\begingroup\renewcommand\colorMATH{\colorMATHS}\renewcommand\colorSYNTAX{\colorSYNTAXS}{{\color{\colorMATH}\ensuremath{s}}}\endgroup }.

The first new ({w.r.t.} DFuzz) term in our sensitivity language is the privacy
lambda. Privacy lambdas are
multi-arity (as opposed to single-arity sensitivity lambdas) because the
privacy language does not support currying to recover multi-argument functions.
Privacy lambdas are created in the {{\color{\colorTEXT}\textit{sensitivity}}} language with
{\begingroup\renewcommand\colorMATH{\colorMATHS}\renewcommand\colorSYNTAX{\colorSYNTAXS}{{\color{\colorMATH}\ensuremath{{{\color{\colorSYNTAX}\texttt{\ensuremath{p\lambda \hspace*{0.33em}({\begingroup\renewcommand\colorMATH{\colorMATHM}\renewcommand\colorSYNTAX{\colorSYNTAXM}{{\color{\colorMATH}\ensuremath{x}}}\endgroup }\mathrel{:}{\begingroup\renewcommand\colorMATH{\colorMATHM}\renewcommand\colorSYNTAX{\colorSYNTAXM}{{\color{\colorMATH}\ensuremath{\tau }}}\endgroup },{.}\hspace{-1pt}{.}\hspace{-1pt}{.},{\begingroup\renewcommand\colorMATH{\colorMATHM}\renewcommand\colorSYNTAX{\colorSYNTAXM}{{\color{\colorMATH}\ensuremath{x}}}\endgroup }\mathrel{:}{\begingroup\renewcommand\colorMATH{\colorMATHM}\renewcommand\colorSYNTAX{\colorSYNTAXM}{{\color{\colorMATH}\ensuremath{\tau }}}\endgroup })\Rightarrow {\begingroup\renewcommand\colorMATH{\colorMATHP}\renewcommand\colorSYNTAX{\colorSYNTAXP}{{\color{\colorMATH}\ensuremath{e}}}\endgroup }}}}}}}}\endgroup } and applied in the
{{\color{\colorTEXT}\textit{privacy}}} language with {\begingroup\renewcommand\colorMATH{\colorMATHP}\renewcommand\colorSYNTAX{\colorSYNTAXP}{{\color{\colorMATH}\ensuremath{{{\color{\colorSYNTAX}\texttt{\ensuremath{{\begingroup\renewcommand\colorMATH{\colorMATHS}\renewcommand\colorSYNTAX{\colorSYNTAXS}{{\color{\colorMATH}\ensuremath{e}}}\endgroup }({\begingroup\renewcommand\colorMATH{\colorMATHS}\renewcommand\colorSYNTAX{\colorSYNTAXS}{{\color{\colorMATH}\ensuremath{e}}}\endgroup },{.}\hspace{-1pt}{.}\hspace{-1pt}{.},{\begingroup\renewcommand\colorMATH{\colorMATHS}\renewcommand\colorSYNTAX{\colorSYNTAXS}{{\color{\colorMATH}\ensuremath{e}}}\endgroup })}}}}}}}\endgroup }. 
The typing rule for privacy lambdas ({{\color{\colorSYNTAX}\texttt{\ensuremath{\multimap ^{*}}}}}{{\color{\colorTEXT}\textsc{\scriptsize -I}}}) types the body of the
lambda in a privacy type context extended with its formal parameters, and the
privacy cost of each parameter is annotated on its function argument type.
Unlike sensitivity lambdas, the privacy cost of variables in the closure
environment are not preserved in the resulting typing judgment. The reason for
this is twofold: (1) the final ``cost'' for variables in the closure environment
depends on how many times the closure is called, and in the absence of this
knowledge, we must conservatively assume that it could be called an infinite
number of times, and (2) the interpretation of an {\begingroup\renewcommand\colorMATH{\colorMATHS}\renewcommand\colorSYNTAX{\colorSYNTAXS}{{\color{\colorMATH}\ensuremath{{{\color{\colorSYNTAX}\texttt{\ensuremath{\infty }}}}}}}\endgroup }-sensitive function
coincides with that of an {\begingroup\renewcommand\colorMATH{\colorMATHP}\renewcommand\colorSYNTAX{\colorSYNTAXP}{{\color{\colorMATH}\ensuremath{{{\color{\colorSYNTAX}\texttt{\ensuremath{\infty }}}}}}}\endgroup }-private function, so we can soundly convert
between {\begingroup\renewcommand\colorMATH{\colorMATHP}\renewcommand\colorSYNTAX{\colorSYNTAXP}{{\color{\colorMATH}\ensuremath{{{\color{\colorSYNTAX}\texttt{\ensuremath{\infty }}}}}}}\endgroup }-privacy-cost and {\begingroup\renewcommand\colorMATH{\colorMATHS}\renewcommand\colorSYNTAX{\colorSYNTAXS}{{\color{\colorMATH}\ensuremath{{{\color{\colorSYNTAX}\texttt{\ensuremath{\infty }}}}}}}\endgroup }-sensitivity contexts freely using
truncation.

The final two new terms in our sensitivity language are introduction and
elimination forms for ``boxes'' ({{\color{\colorTEXT}\textsc{\scriptsize Box-I}}} and {{\color{\colorTEXT}\textsc{\scriptsize Box-E}}}). Boxes have no operational
behavior and are purely a type-level mechanism for tracking sensitivity. 
The rules for box introduction capture the sensitivity context of the
expression, and the rule for box elimination pays for that cost at a later
time. Boxes are reminiscent of {{\color{\colorTEXT}\textit{contextual modal type
theory}}}~\cite{Nanevski:2008:CMT:1352582.1352591}---they allow temporary capture
of a linear context via boxing---thereby deferring its payment---and
re-introduction of the context at later time via unboxing. In a linear type
system that supports scaling, this boxing would not be necessary, but it
becomes necessary in our system to achieve the desired operational behavior
when interacting with the privacy language, which does not support scaling.
{E.g.}, in many of our examples we perform some pre-processing on the database
parameter (such as clipping) and then use this parameter in the body of a loop.
Without boxing, the only way to achieve the desired semantics is to re-clip the
input (a deterministic operation) every time around the loop---boxing allows you
to clip on the outside of the loop and remember that privacy costs should be
``billed'' to the initial input.

\subsection{Privacy Language}
\label{sec:privacylang}

\system's privacy language is designed specifically to enable the composition
of individual differentially private computations. It has a linear type system,
but unlike the sensitivity language, annotations instead track
privacy cost, and the privacy language \emph{does not allow scaling} of these
annotations, that is, the notation {\begingroup\renewcommand\colorMATH{\colorMATHP}\renewcommand\colorSYNTAX{\colorSYNTAXP}{{\color{\colorMATH}\ensuremath{p\Gamma }}}\endgroup } is not used and cannot be defined.
Syntax {\begingroup\renewcommand\colorMATH{\colorMATHP}\renewcommand\colorSYNTAX{\colorSYNTAXP}{{\color{\colorMATH}\ensuremath{{{\color{\colorSYNTAX}\texttt{return}}}\hspace*{0.33em}e}}}\endgroup } and {\begingroup\renewcommand\colorMATH{\colorMATHP}\renewcommand\colorSYNTAX{\colorSYNTAXP}{{\color{\colorMATH}\ensuremath{{{\color{\colorSYNTAX}\texttt{\ensuremath{{\begingroup\renewcommand\colorMATH{\colorMATHM}\renewcommand\colorSYNTAX{\colorSYNTAXM}{{\color{\colorMATH}\ensuremath{x}}}\endgroup }{\leftarrow }{{\color{\colorMATH}\ensuremath{e}}}{\mathrel{;}}{{\color{\colorMATH}\ensuremath{e}}}}}}}}}}\endgroup } (pronounced ``bind'') are
standard from Fuzz, as are their typing rules ({{\color{\colorTEXT}\textsc{\scriptsize Return}}}, {{\color{\colorTEXT}\textsc{\scriptsize Bind}}}), except for
our explicit conversion from a sensitivity context {\begingroup\renewcommand\colorMATH{\colorMATHS}\renewcommand\colorSYNTAX{\colorSYNTAXS}{{\color{\colorMATH}\ensuremath{\Gamma }}}\endgroup } to a privacy context
{\begingroup\renewcommand\colorMATH{\colorMATHP}\renewcommand\colorSYNTAX{\colorSYNTAXP}{{\color{\colorMATH}\ensuremath{\Gamma }}}\endgroup } by truncation to infinity in the conclusion of {{\color{\colorTEXT}\textsc{\scriptsize Return}}}. {{\color{\colorTEXT}\textsc{\scriptsize Bind}}} encodes
exactly the post-processing property of differential privacy---it allows {{\color{\colorMATH}\ensuremath{{\begingroup\renewcommand\colorMATH{\colorMATHP}\renewcommand\colorSYNTAX{\colorSYNTAXP}{{\color{\colorMATH}\ensuremath{e_{2}}}}\endgroup }}}}
to use the value computed by {{\color{\colorMATH}\ensuremath{{\begingroup\renewcommand\colorMATH{\colorMATHP}\renewcommand\colorSYNTAX{\colorSYNTAXP}{{\color{\colorMATH}\ensuremath{e_{1}}}}\endgroup }}}} any number of times after paying for it
once.

Privacy application {\begingroup\renewcommand\colorMATH{\colorMATHP}\renewcommand\colorSYNTAX{\colorSYNTAXP}{{\color{\colorMATH}\ensuremath{{{\color{\colorSYNTAX}\texttt{\ensuremath{{\begingroup\renewcommand\colorMATH{\colorMATHS}\renewcommand\colorSYNTAX{\colorSYNTAXS}{{\color{\colorMATH}\ensuremath{e}}}\endgroup }({\begingroup\renewcommand\colorMATH{\colorMATHS}\renewcommand\colorSYNTAX{\colorSYNTAXS}{{\color{\colorMATH}\ensuremath{e}}}\endgroup },{.}\hspace{-1pt}{.}\hspace{-1pt}{.},{\begingroup\renewcommand\colorMATH{\colorMATHS}\renewcommand\colorSYNTAX{\colorSYNTAXS}{{\color{\colorMATH}\ensuremath{e}}}\endgroup })}}}}}}}\endgroup } applies a privacy function
({\begingroup\renewcommand\colorMATH{\colorMATHP}\renewcommand\colorSYNTAX{\colorSYNTAXP}{{\color{\colorMATH}\ensuremath{{{\color{\colorSYNTAX}\texttt{\ensuremath{p\lambda }}}}}}}\endgroup }, created in the sensitivity language) to a sequence of
{{\color{\colorTEXT}\textit{1-sensitivity}}} arguments---the sensitivity is enforced by the typing rule. The
type rule ({\begingroup\renewcommand\colorMATH{\colorMATHM}\renewcommand\colorSYNTAX{\colorSYNTAXM}{{\color{\colorMATH}\ensuremath{{{\color{\colorSYNTAX}\texttt{\ensuremath{\multimap ^{*}}}}}}}}\endgroup }{{\color{\colorTEXT}\textsc{\scriptsize -E}}}) checks that the first term produces a privacy
function and applies its privacy costs to function arguments which are
restricted by the type system to be 1-sensitive. We use truncation in
well-typed hypothesis for {\begingroup\renewcommand\colorMATH{\colorMATHS}\renewcommand\colorSYNTAX{\colorSYNTAXS}{{\color{\colorMATH}\ensuremath{e_{1}}}}\endgroup } {.}\hspace{-1pt}{.}\hspace{-1pt}{.} {\begingroup\renewcommand\colorMATH{\colorMATHS}\renewcommand\colorSYNTAX{\colorSYNTAXS}{{\color{\colorMATH}\ensuremath{e_{n}}}}\endgroup } to encode the restriction that the
argument must be 1-sensitive. This restriction is crucial for type
soundness---arbitrary terms cannot be given tight privacy bounds statically due
to the lack of a tight scaling operation in the model for {{\color{\colorMATH}\ensuremath{(\epsilon ,\delta )}}}-differential
privacy. The same is true for other advanced variants of differential privacy.

The {\begingroup\renewcommand\colorMATH{\colorMATHP}\renewcommand\colorSYNTAX{\colorSYNTAXP}{{\color{\colorMATH}\ensuremath{{{\color{\colorSYNTAX}\texttt{loop}}}}}}\endgroup } expression is for loop iteration fixed to a statically known
number of iterations. The syntax includes a list of variables
({\begingroup\renewcommand\colorMATH{\colorMATHP}\renewcommand\colorSYNTAX{\colorSYNTAXP}{{\color{\colorMATH}\ensuremath{{{\color{\colorSYNTAX}\texttt{\ensuremath{{<}{\begingroup\renewcommand\colorMATH{\colorMATHM}\renewcommand\colorSYNTAX{\colorSYNTAXM}{{\color{\colorMATH}\ensuremath{x}}}\endgroup },{.}\hspace{-1pt}{.}\hspace{-1pt}{.},{\begingroup\renewcommand\colorMATH{\colorMATHM}\renewcommand\colorSYNTAX{\colorSYNTAXM}{{\color{\colorMATH}\ensuremath{x}}}\endgroup }{>}}}}}}}}\endgroup }) to indicate which variables should be considered
when calculating final privacy costs, as explained shortly. The typing rule
({{\color{\colorTEXT}\textsc{\scriptsize Loop}}}) encodes advanced composition for {{\color{\colorMATH}\ensuremath{(\epsilon ,\delta )}}}-differential privacy. {\begingroup\renewcommand\colorMATH{\colorMATHS}\renewcommand\colorSYNTAX{\colorSYNTAXS}{{\color{\colorMATH}\ensuremath{e_{1}}}}\endgroup }
is the {{\color{\colorMATH}\ensuremath{\delta ^{\prime}}}} parameter to the advanced composition bound and {\begingroup\renewcommand\colorMATH{\colorMATHS}\renewcommand\colorSYNTAX{\colorSYNTAXS}{{\color{\colorMATH}\ensuremath{e_{2}}}}\endgroup } is the
number of loop iterations---each of these values must be statically known,
which we encode with singleton types (a la DFuzz). Statically known values are
fixed and their sensitivities do not appear in the resulting context. {\begingroup\renewcommand\colorMATH{\colorMATHS}\renewcommand\colorSYNTAX{\colorSYNTAXS}{{\color{\colorMATH}\ensuremath{e_{3}}}}\endgroup } is
the initial value passed to the loop, and for which no claim is made of
privacy, indicated by truncation to infinity. {\begingroup\renewcommand\colorMATH{\colorMATHP}\renewcommand\colorSYNTAX{\colorSYNTAXP}{{\color{\colorMATH}\ensuremath{e_{4}}}}\endgroup } is a loop body with free
variables {\begingroup\renewcommand\colorMATH{\colorMATHM}\renewcommand\colorSYNTAX{\colorSYNTAXM}{{\color{\colorMATH}\ensuremath{x_{1}}}}\endgroup } and {\begingroup\renewcommand\colorMATH{\colorMATHM}\renewcommand\colorSYNTAX{\colorSYNTAXM}{{\color{\colorMATH}\ensuremath{x_{2}}}}\endgroup } which will be iterated {\begingroup\renewcommand\colorMATH{\colorMATHS}\renewcommand\colorSYNTAX{\colorSYNTAXS}{{\color{\colorMATH}\ensuremath{e_{2}}}}\endgroup } times with the first
variable bound to the iteration index, and the second variable bound to the
loop state, where {\begingroup\renewcommand\colorMATH{\colorMATHS}\renewcommand\colorSYNTAX{\colorSYNTAXS}{{\color{\colorMATH}\ensuremath{e_{3}}}}\endgroup } is used as the starting value. The loop body {\begingroup\renewcommand\colorMATH{\colorMATHP}\renewcommand\colorSYNTAX{\colorSYNTAXP}{{\color{\colorMATH}\ensuremath{e_{4}}}}\endgroup } is
checked in a privacy context {\begingroup\renewcommand\colorMATH{\colorMATHP}\renewcommand\colorSYNTAX{\colorSYNTAXP}{{\color{\colorMATH}\ensuremath{\Gamma _{4} +
\mathrlap{\hspace{-0.5pt}{}\rceil }\lfloor \Gamma _{4}^{\prime}\mathrlap{\hspace{-0.5pt}\rfloor }\lceil {}^{{{\color{\colorSYNTAX}\texttt{\ensuremath{{\begingroup\renewcommand\colorMATH{\colorMATHM}\renewcommand\colorSYNTAX{\colorSYNTAXM}{{\color{\colorMATH}\ensuremath{\epsilon }}}\endgroup },{\begingroup\renewcommand\colorMATH{\colorMATHM}\renewcommand\colorSYNTAX{\colorSYNTAXM}{{\color{\colorMATH}\ensuremath{\delta }}}\endgroup }}}}}}_{\{ {\begingroup\renewcommand\colorMATH{\colorMATHM}\renewcommand\colorSYNTAX{\colorSYNTAXM}{{\color{\colorMATH}\ensuremath{x_{1}^{\prime}}}}\endgroup },{.}\hspace{-1pt}{.}\hspace{-1pt}{.},{\begingroup\renewcommand\colorMATH{\colorMATHM}\renewcommand\colorSYNTAX{\colorSYNTAXM}{{\color{\colorMATH}\ensuremath{x_{n}^{\prime}}}}\endgroup }\} }}}}\endgroup }, shorthand for
{\begingroup\renewcommand\colorMATH{\colorMATHP}\renewcommand\colorSYNTAX{\colorSYNTAXP}{{\color{\colorMATH}\ensuremath{{}\rceil {}\lfloor \Gamma _{4}^{\prime}\rfloor _{\{ {\begingroup\renewcommand\colorMATH{\colorMATHM}\renewcommand\colorSYNTAX{\colorSYNTAXM}{{\color{\colorMATH}\ensuremath{x_{1}^{\prime}}}}\endgroup },{.}\hspace{-1pt}{.}\hspace{-1pt}{.},{\begingroup\renewcommand\colorMATH{\colorMATHM}\renewcommand\colorSYNTAX{\colorSYNTAXM}{{\color{\colorMATH}\ensuremath{x_{n}^{\prime}}}}\endgroup }\} }\lceil {}^{{{\color{\colorSYNTAX}\texttt{\ensuremath{{\begingroup\renewcommand\colorMATH{\colorMATHM}\renewcommand\colorSYNTAX{\colorSYNTAXM}{{\color{\colorMATH}\ensuremath{\epsilon }}}\endgroup },{\begingroup\renewcommand\colorMATH{\colorMATHM}\renewcommand\colorSYNTAX{\colorSYNTAXM}{{\color{\colorMATH}\ensuremath{\delta }}}\endgroup }}}}}}}}}\endgroup } where
{\begingroup\renewcommand\colorMATH{\colorMATHP}\renewcommand\colorSYNTAX{\colorSYNTAXP}{{\color{\colorMATH}\ensuremath{\lfloor \Gamma _{4}^{\prime}\rfloor _{\{ {\begingroup\renewcommand\colorMATH{\colorMATHM}\renewcommand\colorSYNTAX{\colorSYNTAXM}{{\color{\colorMATH}\ensuremath{x_{1}^{\prime}}}}\endgroup },{.}\hspace{-1pt}{.}\hspace{-1pt}{.},{\begingroup\renewcommand\colorMATH{\colorMATHM}\renewcommand\colorSYNTAX{\colorSYNTAXM}{{\color{\colorMATH}\ensuremath{x_{n}^{\prime}}}}\endgroup }\} }}}}\endgroup } is a context restricted to only the variables
{\begingroup\renewcommand\colorMATH{\colorMATHM}\renewcommand\colorSYNTAX{\colorSYNTAXM}{{\color{\colorMATH}\ensuremath{x_{1}^{\prime},{.}\hspace{-1pt}{.}\hspace{-1pt}{.},x_{n}^{\prime}}}}\endgroup }. The {\begingroup\renewcommand\colorMATH{\colorMATHP}\renewcommand\colorSYNTAX{\colorSYNTAXP}{{\color{\colorMATH}\ensuremath{{{\color{\colorSYNTAX}\texttt{\ensuremath{{\begingroup\renewcommand\colorMATH{\colorMATHM}\renewcommand\colorSYNTAX{\colorSYNTAXM}{{\color{\colorMATH}\ensuremath{\epsilon }}}\endgroup },{\begingroup\renewcommand\colorMATH{\colorMATHM}\renewcommand\colorSYNTAX{\colorSYNTAXM}{{\color{\colorMATH}\ensuremath{\delta }}}\endgroup }}}}}}}}\endgroup } is an upper bound on the privacy cost of the
variables {\begingroup\renewcommand\colorMATH{\colorMATHM}\renewcommand\colorSYNTAX{\colorSYNTAXM}{{\color{\colorMATH}\ensuremath{x_{i}^{\prime}}}}\endgroup } in the loop body, and the resulting privacy bound is
restricted to only those variables. This allows variables for which the
programmer is not interested in tracking privacy to appear in {\begingroup\renewcommand\colorMATH{\colorMATHP}\renewcommand\colorSYNTAX{\colorSYNTAXP}{{\color{\colorMATH}\ensuremath{\Gamma _{4}}}}\endgroup } in the
premise, and the rule's conclusion makes no claims about privacy for these
variables. We make use of this feature in all of our examples programs. 

The {\begingroup\renewcommand\colorMATH{\colorMATHP}\renewcommand\colorSYNTAX{\colorSYNTAXP}{{\color{\colorMATH}\ensuremath{{{\color{\colorSYNTAX}\texttt{gauss}}}}}}\endgroup } expression is a {{\color{\colorTEXT}\textit{mechanism}}} of {{\color{\colorMATH}\ensuremath{(\epsilon ,\delta )}}}-differential privacy;
other mechanisms are used for other privacy variants. Like the
loop expression, mechanism expressions take a list of variables to indicate
which variables should be considered in the final privacy cost. The typing rule
({{\color{\colorTEXT}\textsc{\scriptsize Gauss}}}) is similar in spirit to {{\color{\colorTEXT}\textsc{\scriptsize Loop}}}: it takes parameters to the mechanism
which must be statically known (encoded as singleton types), a list of
variables to consider for the purposes of the resulting privacy bound, and a
term {\begingroup\renewcommand\colorMATH{\colorMATHP}\renewcommand\colorSYNTAX{\colorSYNTAXP}{{\color{\colorMATH}\ensuremath{{{\color{\colorSYNTAX}\texttt{\ensuremath{\{ {{\color{\colorMATH}\ensuremath{{\begingroup\renewcommand\colorMATH{\colorMATHS}\renewcommand\colorSYNTAX{\colorSYNTAXS}{{\color{\colorMATH}\ensuremath{e}}}\endgroup }}}}\} }}}}}}}\endgroup } for which there is a bound {{\color{\colorMATH}\ensuremath{\dot r}}} on the sensitivity of
free variables {\begingroup\renewcommand\colorMATH{\colorMATHM}\renewcommand\colorSYNTAX{\colorSYNTAXM}{{\color{\colorMATH}\ensuremath{x_{1},{.}\hspace{-1pt}{.}\hspace{-1pt}{.},x_{n}}}}\endgroup }. The resulting privacy guarantee is that the term in
brackets {\begingroup\renewcommand\colorMATH{\colorMATHP}\renewcommand\colorSYNTAX{\colorSYNTAXP}{{\color{\colorMATH}\ensuremath{{{\color{\colorSYNTAX}\texttt{\ensuremath{\{ {{\color{\colorMATH}\ensuremath{{\begingroup\renewcommand\colorMATH{\colorMATHS}\renewcommand\colorSYNTAX{\colorSYNTAXS}{{\color{\colorMATH}\ensuremath{e}}}\endgroup }}}}\} }}}}}}}\endgroup } is {\begingroup\renewcommand\colorMATH{\colorMATHP}\renewcommand\colorSYNTAX{\colorSYNTAXP}{{\color{\colorMATH}\ensuremath{{{\color{\colorSYNTAX}\texttt{\ensuremath{{\begingroup\renewcommand\colorMATH{\colorMATHM}\renewcommand\colorSYNTAX{\colorSYNTAXM}{{\color{\colorMATH}\ensuremath{\epsilon }}}\endgroup },{\begingroup\renewcommand\colorMATH{\colorMATHM}\renewcommand\colorSYNTAX{\colorSYNTAXM}{{\color{\colorMATH}\ensuremath{\delta }}}\endgroup }}}}}}}}\endgroup } differentially private.
Whereas {\begingroup\renewcommand\colorMATH{\colorMATHP}\renewcommand\colorSYNTAX{\colorSYNTAXP}{{\color{\colorMATH}\ensuremath{{{\color{\colorSYNTAX}\texttt{loop}}}}}}\endgroup } and advanced composition consider a \emph{privacy term}
loop body with an upper bound on \emph{privacy leakage}, {\begingroup\renewcommand\colorMATH{\colorMATHP}\renewcommand\colorSYNTAX{\colorSYNTAXP}{{\color{\colorMATH}\ensuremath{{{\color{\colorSYNTAX}\texttt{gauss}}}}}}\endgroup } considers
a \emph{sensitivity term} body with an upper bound on its \emph{sensitivity}.

\subsection{Metatheory}

We denote sensitivity language terms {\begingroup\renewcommand\colorMATH{\colorMATHS}\renewcommand\colorSYNTAX{\colorSYNTAXS}{{\color{\colorMATH}\ensuremath{e \in  {{\color{\colorMATH}\ensuremath{\operatorname{exp}}}}}}}\endgroup } into total, functional,
linear maps between metric spaces---the same model as the terminating fragment
of Fuzz. Every term in our language terminates by design, which dramatically
simplifies our models and proofs.  This restriction poses no issues in
implementing most differentially private machine learning algorithms, because
such algorithms typically terminate in a statically determined number of loop iterations in
order to achieve a particular privacy cost. 

Types in \system denote metric spaces, as in Fuzz. We notate metric spaces {{\color{\colorMATH}\ensuremath{D}}},
their underlying carrier set {{\color{\colorMATH}\ensuremath{\| D\| }}}, and their distance metric {{\color{\colorMATH}\ensuremath{|x-y|_{D}}}}, or
{{\color{\colorMATH}\ensuremath{|x-y|}}} where {{\color{\colorMATH}\ensuremath{D}}} can be inferred from context.
Sensitivity typing judgments {\begingroup\renewcommand\colorMATH{\colorMATHS}\renewcommand\colorSYNTAX{\colorSYNTAXS}{{\color{\colorMATH}\ensuremath{\Gamma  \vdash  e \mathrel{:} {\begingroup\renewcommand\colorMATH{\colorMATHM}\renewcommand\colorSYNTAX{\colorSYNTAXM}{{\color{\colorMATH}\ensuremath{\tau }}}\endgroup }}}}\endgroup } denote linear maps from a scaled
cartesian product interpretation of {\begingroup\renewcommand\colorMATH{\colorMATHS}\renewcommand\colorSYNTAX{\colorSYNTAXS}{{\color{\colorMATH}\ensuremath{\Gamma }}}\endgroup }:
\begingroup\renewcommand\colorMATH{\colorMATHS}\renewcommand\colorSYNTAX{\colorSYNTAXS}
\vspace*{-0.25em}\begingroup\color{\colorMATH}\begin{gather*} \llbracket \overbracketarg \Gamma {\{ {\begingroup\renewcommand\colorMATH{\colorMATHM}\renewcommand\colorSYNTAX{\colorSYNTAXM}{{\color{\colorMATH}\ensuremath{x_{1}}}}\endgroup }{\mathrel{:}}_{s_{1}}{\begingroup\renewcommand\colorMATH{\colorMATHM}\renewcommand\colorSYNTAX{\colorSYNTAXM}{{\color{\colorMATH}\ensuremath{\tau _{1}}}}\endgroup }{,}{.}\hspace{-1pt}{.}\hspace{-1pt}{.}{,}{\begingroup\renewcommand\colorMATH{\colorMATHM}\renewcommand\colorSYNTAX{\colorSYNTAXM}{{\color{\colorMATH}\ensuremath{x_{n}}}}\endgroup }{\mathrel{:}}_{s_{n}}{\begingroup\renewcommand\colorMATH{\colorMATHM}\renewcommand\colorSYNTAX{\colorSYNTAXM}{{\color{\colorMATH}\ensuremath{\tau _{n}}}}\endgroup }\} } \vdash  {\begingroup\renewcommand\colorMATH{\colorMATHM}\renewcommand\colorSYNTAX{\colorSYNTAXM}{{\color{\colorMATH}\ensuremath{\tau }}}\endgroup } \rrbracket  \triangleq  {\begingroup\renewcommand\colorMATH{\colorMATHM}\renewcommand\colorSYNTAX{\colorSYNTAXM}{{\color{\colorMATH}\ensuremath{{!}_{{\begingroup\renewcommand\colorMATH{\colorMATHS}\renewcommand\colorSYNTAX{\colorSYNTAXS}{{\color{\colorMATH}\ensuremath{s_{1}}}}\endgroup }}\llbracket \tau _{1}\rrbracket  \otimes  {\mathord{\cdotp }}\hspace{-1pt}{\mathord{\cdotp }}\hspace{-1pt}{\mathord{\cdotp }} \otimes  {!}_{{\begingroup\renewcommand\colorMATH{\colorMATHS}\renewcommand\colorSYNTAX{\colorSYNTAXS}{{\color{\colorMATH}\ensuremath{s_{n}}}}\endgroup }}\llbracket \tau _{n}\rrbracket  \multimap  \llbracket \tau \rrbracket }}}\endgroup }
\vspace*{-1em}\end{gather*}\endgroup 
\endgroup 
Although we do not make metric space scaling explicit in our syntax (for the
purposes of effective type inference, a la DFuzz~\cite{dfuzz-impl}), scaling
becomes apparent explicitly in our model. Privacy judgments {\begingroup\renewcommand\colorMATH{\colorMATHP}\renewcommand\colorSYNTAX{\colorSYNTAXP}{{\color{\colorMATH}\ensuremath{\Gamma  \vdash  e \mathrel{:} {\begingroup\renewcommand\colorMATH{\colorMATHM}\renewcommand\colorSYNTAX{\colorSYNTAXM}{{\color{\colorMATH}\ensuremath{\tau }}}\endgroup }}}}\endgroup }
denote {{\color{\colorTEXT}\textit{probabilistic, privacy preserving maps}}} from an {{\color{\colorTEXT}\textit{unscaled}}} product
interpretation of {\begingroup\renewcommand\colorMATH{\colorMATHP}\renewcommand\colorSYNTAX{\colorSYNTAXP}{{\color{\colorMATH}\ensuremath{\Gamma }}}\endgroup }:
\begingroup\renewcommand\colorMATH{\colorMATHP}\renewcommand\colorSYNTAX{\colorSYNTAXP}
\vspace*{-0.25em}\begingroup\color{\colorMATH}\begin{gather*} \llbracket \overbracketarg \Gamma {\{ {\begingroup\renewcommand\colorMATH{\colorMATHM}\renewcommand\colorSYNTAX{\colorSYNTAXM}{{\color{\colorMATH}\ensuremath{x_{1}}}}\endgroup }{\mathrel{:}}_{p_{1}}{\begingroup\renewcommand\colorMATH{\colorMATHM}\renewcommand\colorSYNTAX{\colorSYNTAXM}{{\color{\colorMATH}\ensuremath{\tau _{1}}}}\endgroup }{,}{.}\hspace{-1pt}{.}\hspace{-1pt}{.}{,}{\begingroup\renewcommand\colorMATH{\colorMATHM}\renewcommand\colorSYNTAX{\colorSYNTAXM}{{\color{\colorMATH}\ensuremath{x_{n}}}}\endgroup }{\mathrel{:}}_{p_{n}}{\begingroup\renewcommand\colorMATH{\colorMATHM}\renewcommand\colorSYNTAX{\colorSYNTAXM}{{\color{\colorMATH}\ensuremath{\tau _{n}}}}\endgroup }\} } \vdash  {\begingroup\renewcommand\colorMATH{\colorMATHM}\renewcommand\colorSYNTAX{\colorSYNTAXM}{{\color{\colorMATH}\ensuremath{\tau }}}\endgroup } \rrbracket  \triangleq  {\begingroup\renewcommand\colorMATH{\colorMATHM}\renewcommand\colorSYNTAX{\colorSYNTAXM}{{\color{\colorMATH}\ensuremath{(\llbracket \tau _{1}\rrbracket @{\begingroup\renewcommand\colorMATH{\colorMATHP}\renewcommand\colorSYNTAX{\colorSYNTAXP}{{\color{\colorMATH}\ensuremath{p_{1}}}}\endgroup }{,}{.}\hspace{-1pt}{.}\hspace{-1pt}{.}{,}\llbracket \tau _{n}\rrbracket @{\begingroup\renewcommand\colorMATH{\colorMATHP}\renewcommand\colorSYNTAX{\colorSYNTAXP}{{\color{\colorMATH}\ensuremath{p_{n}}}}\endgroup }) \multimap ^{*} \| \llbracket \tau \rrbracket \| }}}\endgroup }
\vspace*{-1em}\end{gather*}\endgroup 
\endgroup 
The multi-arity {{\color{\colorMATH}\ensuremath{(\epsilon ,\delta )}}}-differential-privacy-preserving map is defined:
\vspace*{-0.25em}\begingroup\color{\colorMATH}\begin{gather*} \begin{array}{l
   } (D_{1}@(\epsilon _{1},\delta _{1}){,}{.}\hspace{-1pt}{.}\hspace{-1pt}{.}{,}D_{n}@(\epsilon _{n},\delta _{n})) \multimap ^{*} X \triangleq  
   \cr  \hspace*{1.00em}\hspace*{1.00em} \begin{array}[t]{rcl
         } &{}\{ {}&  f \in  \| D_{1}\|  \times  {\mathord{\cdotp }}\hspace{-1pt}{\mathord{\cdotp }}\hspace{-1pt}{\mathord{\cdotp }} \times  \| D_{n}\|  \rightarrow  {\mathcal{D}}(X)
         \cr  &{}\mathrel{|}{}& |x_{i} - y|_{D_{i}} \leq  1 \Rightarrow  {{\color{\colorMATH}\ensuremath{\operatorname{Pr}}}}[f(x_{1}{,}{.}\hspace{-1pt}{.}\hspace{-1pt}{.}{,}x_{i}{,}{.}\hspace{-1pt}{.}\hspace{-1pt}{.}{,}x_{n}) = d] \leq  e^{\epsilon _{i}}{{\color{\colorMATH}\ensuremath{\operatorname{Pr}}}}[f(x_{1}{,}{.}\hspace{-1pt}{.}\hspace{-1pt}{.}{,}y{,}{.}\hspace{-1pt}{.}\hspace{-1pt}{.}{,}x_{n}) = d] + \delta _{i} \hspace*{0.33em}\} 
         \end{array}
   \end{array}
\vspace*{-1em}\end{gather*}\endgroup 
where {{\color{\colorMATH}\ensuremath{{\mathcal{D}}(X)}}} is a distribution over elements in {{\color{\colorMATH}\ensuremath{X}}}.

We give a full semantic account of typing in the extended version of this
paper, as well as prove key type soundness lemmas, many of which appeal to
well-known differential privacy proofs from the literature.

The final soundness theorem, proven by induction over typing derivations, is
that the denotations for well-typed open terms {\begingroup\renewcommand\colorMATH{\colorMATHS}\renewcommand\colorSYNTAX{\colorSYNTAXS}{{\color{\colorMATH}\ensuremath{e_{s}}}}\endgroup } and {\begingroup\renewcommand\colorMATH{\colorMATHP}\renewcommand\colorSYNTAX{\colorSYNTAXP}{{\color{\colorMATH}\ensuremath{e_{p}}}}\endgroup } in well-typed
environments {\begingroup\renewcommand\colorMATH{\colorMATHS}\renewcommand\colorSYNTAX{\colorSYNTAXS}{{\color{\colorMATH}\ensuremath{\gamma _{s}}}}\endgroup } and {\begingroup\renewcommand\colorMATH{\colorMATHP}\renewcommand\colorSYNTAX{\colorSYNTAXP}{{\color{\colorMATH}\ensuremath{\gamma _{p}}}}\endgroup } are
contained in the denotation of their typing contexts {\begingroup\renewcommand\colorMATH{\colorMATHS}\renewcommand\colorSYNTAX{\colorSYNTAXS}{{\color{\colorMATH}\ensuremath{\Gamma _{s} \vdash  {\begingroup\renewcommand\colorMATH{\colorMATHM}\renewcommand\colorSYNTAX{\colorSYNTAXM}{{\color{\colorMATH}\ensuremath{\tau }}}\endgroup }}}}\endgroup } and {\begingroup\renewcommand\colorMATH{\colorMATHP}\renewcommand\colorSYNTAX{\colorSYNTAXP}{{\color{\colorMATH}\ensuremath{\Gamma _{p} \vdash 
{\begingroup\renewcommand\colorMATH{\colorMATHM}\renewcommand\colorSYNTAX{\colorSYNTAXM}{{\color{\colorMATH}\ensuremath{\tau }}}\endgroup }}}}\endgroup }.
\begin{theorem} \ 
  \noindent
  \begin{enumerate}
    \item If {\begingroup\renewcommand\colorMATH{\colorMATHP}\renewcommand\colorSYNTAX{\colorSYNTAXP}{{\color{\colorMATH}\ensuremath{\Gamma _{p} \vdash  e_{p} \mathrel{:} {\begingroup\renewcommand\colorMATH{\colorMATHM}\renewcommand\colorSYNTAX{\colorSYNTAXM}{{\color{\colorMATH}\ensuremath{\tau }}}\endgroup }}}}\endgroup } and {\begingroup\renewcommand\colorMATH{\colorMATHP}\renewcommand\colorSYNTAX{\colorSYNTAXP}{{\color{\colorMATH}\ensuremath{\Gamma _{p} \vdash  \gamma _{p}}}}\endgroup } then {\begingroup\renewcommand\colorMATH{\colorMATHM}\renewcommand\colorSYNTAX{\colorSYNTAXM}{{\color{\colorMATH}\ensuremath{{\begingroup\renewcommand\colorMATH{\colorMATHP}\renewcommand\colorSYNTAX{\colorSYNTAXP}{{\color{\colorMATH}\ensuremath{\llbracket e_{p}\rrbracket ^{\gamma _{p}}}}}\endgroup } \in  {\begingroup\renewcommand\colorMATH{\colorMATHP}\renewcommand\colorSYNTAX{\colorSYNTAXP}{{\color{\colorMATH}\ensuremath{\llbracket \Gamma _{p} \vdash  {\begingroup\renewcommand\colorMATH{\colorMATHM}\renewcommand\colorSYNTAX{\colorSYNTAXM}{{\color{\colorMATH}\ensuremath{\tau }}}\endgroup }\rrbracket }}}\endgroup }}}}\endgroup }
    \item If {\begingroup\renewcommand\colorMATH{\colorMATHS}\renewcommand\colorSYNTAX{\colorSYNTAXS}{{\color{\colorMATH}\ensuremath{\Gamma _{s} \vdash  e_{s} \mathrel{:} {\begingroup\renewcommand\colorMATH{\colorMATHM}\renewcommand\colorSYNTAX{\colorSYNTAXM}{{\color{\colorMATH}\ensuremath{\tau }}}\endgroup }}}}\endgroup } and {\begingroup\renewcommand\colorMATH{\colorMATHS}\renewcommand\colorSYNTAX{\colorSYNTAXS}{{\color{\colorMATH}\ensuremath{\Gamma _{s} \vdash  \gamma _{s}}}}\endgroup } then {\begingroup\renewcommand\colorMATH{\colorMATHM}\renewcommand\colorSYNTAX{\colorSYNTAXM}{{\color{\colorMATH}\ensuremath{{\begingroup\renewcommand\colorMATH{\colorMATHS}\renewcommand\colorSYNTAX{\colorSYNTAXS}{{\color{\colorMATH}\ensuremath{\llbracket e_{s}\rrbracket ^{\gamma _{s}}}}}\endgroup } \in  {\begingroup\renewcommand\colorMATH{\colorMATHS}\renewcommand\colorSYNTAX{\colorSYNTAXS}{{\color{\colorMATH}\ensuremath{\llbracket \Gamma _{s} \vdash  {\begingroup\renewcommand\colorMATH{\colorMATHM}\renewcommand\colorSYNTAX{\colorSYNTAXM}{{\color{\colorMATH}\ensuremath{\tau }}}\endgroup }\rrbracket }}}\endgroup }}}}\endgroup }
  \end{enumerate}
\end{theorem}
\noindent
A corollary is that any well-typed privacy lambda function satisfies
{{\color{\colorMATH}\ensuremath{(\epsilon ,\delta )}}}-differential privacy for each of its arguments {w.r.t.} that argument's
privacy annotation used in typing.

\section{Language Tools for Machine Learning}
\newcommand\FigureMatrixLang{
\begin{figure}
\smallerplz
\begingroup\renewcommand\colorMATH{\colorMATHM}\renewcommand\colorSYNTAX{\colorSYNTAXM}
\vspace*{-0.25em}\begingroup\color{\colorMATH}\begin{gather*}\begin{tabularx}{\linewidth}{>{\centering\arraybackslash\(}X<{\)}}
   \hfill\hspace{0pt}\ell      \in  {{\color{\colorMATH}\ensuremath{\operatorname{norm}}}}         \mathrel{\Coloneqq } {{\color{\colorSYNTAX}\texttt{\ensuremath{L1}}}} \mathrel{|} {{\color{\colorSYNTAX}\texttt{\ensuremath{L2}}}} \mathrel{|} {{\color{\colorSYNTAX}\texttt{\ensuremath{L\infty }}}}
   \hfill\hspace{0pt}c     \in  {{\color{\colorMATH}\ensuremath{\operatorname{clip}}}}         \mathrel{\Coloneqq } \ell  \mathrel{|} {{\color{\colorSYNTAX}\texttt{\ensuremath{U}}}}
   \hfill\hspace{0pt}
  \end{tabularx}
\cr \tau  \in  {{\color{\colorMATH}\ensuremath{\operatorname{type}}}} \mathrel{\coloneqq } {.}\hspace{-1pt}{.}\hspace{-1pt}{.} \mathrel{|} {{\color{\colorSYNTAX}\texttt{\ensuremath{{\mathbb{D}}}}}} \mathrel{|} {{\color{\colorSYNTAX}\texttt{\ensuremath{{{\color{\colorSYNTAX}\texttt{idx}}}[{{\color{\colorMATH}\ensuremath{n}}}]}}}} \mathrel{|} {{\color{\colorSYNTAX}\texttt{\ensuremath{{\mathbb{M}}_{{{\color{\colorMATH}\ensuremath{\ell }}}}^{{{\color{\colorMATH}\ensuremath{c}}}}[{{\color{\colorMATH}\ensuremath{n}}},{{\color{\colorMATH}\ensuremath{n}}}]\hspace*{0.33em}{{\color{\colorMATH}\ensuremath{\tau }}}}}}} \hspace*{1.00em}\hspace*{1.00em}\hspace*{1.00em}\hspace*{1.00em} {{\color{\colorTEXT}\textit{matrix types}}}
\cr \begingroup\renewcommand\colorMATH{\colorMATHS}\renewcommand\colorSYNTAX{\colorSYNTAXS}\begingroup\color{\colorMATH}
  \begin{array}{rclcl@{\hspace*{1.00em}}l
  } g  &{}\in {}& {{\color{\colorMATH}\ensuremath{\operatorname{grad}}}}      &{}\mathrel{\Coloneqq }{}& {{\color{\colorSYNTAX}\texttt{\ensuremath{ LS }}}} \mathrel{|} {{\color{\colorSYNTAX}\texttt{\ensuremath{ LR }}}} \mathrel{|} {{\color{\colorSYNTAX}\texttt{\ensuremath{ GR }}}} \mathrel{|} {{\color{\colorSYNTAX}\texttt{\ensuremath{ HSVM }}}}                          & {{\color{\colorTEXT}\textit{gradients}}} 
  \cr  e  &{}\in {}& {{\color{\colorMATH}\ensuremath{\operatorname{exp}}}}       &{}\mathrel{\Coloneqq }{}& {.}\hspace{-1pt}{.}\hspace{-1pt}{.} \mathrel{|} mo                                                               & {{\color{\colorTEXT}\textit{matrix operations}}}
  \cr  mo &{}\in {}& {{\color{\colorMATH}\ensuremath{\operatorname{matrix-op}}}} &{}\mathrel{\Coloneqq }{}& {{\color{\colorMATH}\ensuremath{\operatorname{mcreate}}}}_{{\begingroup\renewcommand\colorMATH{\colorMATHM}\renewcommand\colorSYNTAX{\colorSYNTAXM}{{\color{\colorMATH}\ensuremath{\ell }}}\endgroup }} \mathrel{|} \underline{\hspace{0.66em}\vspace*{5ex}}{\#}[\underline{\hspace{0.66em}\vspace*{5ex}}{,}\underline{\hspace{0.66em}\vspace*{5ex}}] \mathrel{|} \underline{\hspace{0.66em}\vspace*{5ex}}{\#}[\underline{\hspace{0.66em}\vspace*{5ex}}{,}\underline{\hspace{0.66em}\vspace*{5ex}}\mapsto \underline{\hspace{0.66em}\vspace*{5ex}}]                           & 
  \cr     &{} {}&             &{}\mathrel{|}{}& {{\color{\colorMATH}\ensuremath{\operatorname{rows}}}} \mathrel{|} {{\color{\colorMATH}\ensuremath{\operatorname{cols}}}} \mathrel{|} {{\color{\colorMATH}\ensuremath{\operatorname{convert}}}} \mathrel{|} {{\color{\colorMATH}\ensuremath{\operatorname{clip}}}}^{{\begingroup\renewcommand\colorMATH{\colorMATHM}\renewcommand\colorSYNTAX{\colorSYNTAXM}{{\color{\colorMATH}\ensuremath{\ell }}}\endgroup }}                          & 
  \cr     &{} {}&             &{}\mathrel{|}{}& {{\color{\colorMATH}\ensuremath{\operatorname{map}}}} \mathrel{|} {{\color{\colorMATH}\ensuremath{\operatorname{map-row}}}} \mathrel{|} {{\color{\colorMATH}\ensuremath{\operatorname{fld-row}}}}                                        & 
  \cr     &{} {}&             &{}\mathrel{|}{}& {{\color{\colorMATH}\ensuremath{\operatorname{L{\mathrel{\nabla }}}}}}_{{\begingroup\renewcommand\colorMATH{\colorMATHM}\renewcommand\colorSYNTAX{\colorSYNTAXM}{{\color{\colorMATH}\ensuremath{\ell }}}\endgroup }}^{g}[\underline{\hspace{0.66em}\vspace*{5ex}}\mathrel{;}\underline{\hspace{0.66em}\vspace*{5ex}},\underline{\hspace{0.66em}\vspace*{5ex}}] \mathrel{|} {{\color{\colorMATH}\ensuremath{\operatorname{U{\mathrel{\nabla }}}}}}[\underline{\hspace{0.66em}\vspace*{5ex}}\mathrel{;}\underline{\hspace{0.66em}\vspace*{5ex}},\underline{\hspace{0.66em}\vspace*{5ex}}]                                    & 
  \end{array}
  \endgroup \endgroup 
\cr \begingroup\renewcommand\colorMATH{\colorMATHP}\renewcommand\colorSYNTAX{\colorSYNTAXP}\begingroup\color{\colorMATH}
  \begin{array}{rclcl@{\hspace*{1.00em}}l
  } e  &{}\in {}& {{\color{\colorMATH}\ensuremath{\operatorname{exp}}}}   &{}\mathrel{\Coloneqq }{}& {.}\hspace{-1pt}{.}\hspace{-1pt}{.} \mathrel{|} {{\color{\colorSYNTAX}\texttt{\ensuremath{{{\color{\colorSYNTAX}\texttt{mgauss}}}[{\begingroup\renewcommand\colorMATH{\colorMATHS}\renewcommand\colorSYNTAX{\colorSYNTAXS}{{\color{\colorMATH}\ensuremath{e}}}\endgroup },{\begingroup\renewcommand\colorMATH{\colorMATHS}\renewcommand\colorSYNTAX{\colorSYNTAXS}{{\color{\colorMATH}\ensuremath{e}}}\endgroup },{\begingroup\renewcommand\colorMATH{\colorMATHS}\renewcommand\colorSYNTAX{\colorSYNTAXS}{{\color{\colorMATH}\ensuremath{e}}}\endgroup }]\hspace*{0.33em}{<}{\begingroup\renewcommand\colorMATH{\colorMATHM}\renewcommand\colorSYNTAX{\colorSYNTAXM}{{\color{\colorMATH}\ensuremath{x}}}\endgroup },{.}\hspace{-1pt}{.}\hspace{-1pt}{.},{\begingroup\renewcommand\colorMATH{\colorMATHM}\renewcommand\colorSYNTAX{\colorSYNTAXM}{{\color{\colorMATH}\ensuremath{x}}}\endgroup }{>}\hspace*{0.33em}\{ {\begingroup\renewcommand\colorMATH{\colorMATHS}\renewcommand\colorSYNTAX{\colorSYNTAXS}{{\color{\colorMATH}\ensuremath{e}}}\endgroup }\} }}}}          & {{\color{\colorTEXT}\textit{matrix gaussian}}}
  \cr     &{} {}&         &{}\mathrel{|}{}& {{\color{\colorSYNTAX}\texttt{\ensuremath{{{\color{\colorSYNTAX}\texttt{exponential}}}[{\begingroup\renewcommand\colorMATH{\colorMATHS}\renewcommand\colorSYNTAX{\colorSYNTAXS}{{\color{\colorMATH}\ensuremath{e}}}\endgroup }{,}{\begingroup\renewcommand\colorMATH{\colorMATHS}\renewcommand\colorSYNTAX{\colorSYNTAXS}{{\color{\colorMATH}\ensuremath{e}}}\endgroup }]\hspace*{0.33em}{\begingroup\renewcommand\colorMATH{\colorMATHS}\renewcommand\colorSYNTAX{\colorSYNTAXS}{{\color{\colorMATH}\ensuremath{e}}}\endgroup }\hspace*{0.33em}{<}{\begingroup\renewcommand\colorMATH{\colorMATHM}\renewcommand\colorSYNTAX{\colorSYNTAXM}{{\color{\colorMATH}\ensuremath{x}}}\endgroup },{.}\hspace{-1pt}{.}\hspace{-1pt}{.},{\begingroup\renewcommand\colorMATH{\colorMATHM}\renewcommand\colorSYNTAX{\colorSYNTAXM}{{\color{\colorMATH}\ensuremath{x}}}\endgroup }{>}\hspace*{0.33em}\{ {\begingroup\renewcommand\colorMATH{\colorMATHM}\renewcommand\colorSYNTAX{\colorSYNTAXM}{{\color{\colorMATH}\ensuremath{x}}}\endgroup }\Rightarrow {\begingroup\renewcommand\colorMATH{\colorMATHS}\renewcommand\colorSYNTAX{\colorSYNTAXS}{{\color{\colorMATH}\ensuremath{e}}}\endgroup }\} }}}}   & {{\color{\colorTEXT}\textit{exp. mechanism}}}
  \end{array}
  \endgroup \endgroup 
\vspace*{-1em}\end{gather*}\endgroup 
\vspace{-1.5em}
\endgroup 
\caption{Matrix Types and Terms}
\label{fig:matrix_lang}
\vspace{-0.5em}
\end{figure}
}

\newcommand\FigureMatrixTyping{
\begin{figure*}
\smallerplz
\vspace*{-0.25em}\begingroup\color{\colorMATH}\begin{gather*}
\begin{tabularx}{\linewidth}{>{\centering\arraybackslash\(}X<{\)}}\hfill\hspace{0pt}\ell      \in  {{\color{\colorMATH}\ensuremath{\operatorname{norm}}}}         \mathrel{\Coloneqq } {{\color{\colorSYNTAX}\texttt{\ensuremath{L1}}}} \mathrel{|} {{\color{\colorSYNTAX}\texttt{\ensuremath{L2}}}} \mathrel{|} {{\color{\colorSYNTAX}\texttt{\ensuremath{L\infty }}}}
  \hfill\hspace{0pt}c     \in  {{\color{\colorMATH}\ensuremath{\operatorname{clip}}}}         \mathrel{\Coloneqq } \ell  \mathrel{|} {{\color{\colorSYNTAX}\texttt{\ensuremath{U}}}}
  \hfill\hspace{0pt}\tau      \in  {{\color{\colorMATH}\ensuremath{\operatorname{type}}}}          \mathrel{\coloneqq } {.}\hspace{-1pt}{.}\hspace{-1pt}{.} \mathrel{|} {{\color{\colorSYNTAX}\texttt{\ensuremath{{\mathbb{D}}}}}} \mathrel{|} {{\color{\colorSYNTAX}\texttt{\ensuremath{{{\color{\colorSYNTAX}\texttt{idx}}}[{{\color{\colorMATH}\ensuremath{n}}}]}}}} \mathrel{|} {{\color{\colorSYNTAX}\texttt{\ensuremath{{\mathbb{M}}_{{{\color{\colorMATH}\ensuremath{\ell }}}}^{{{\color{\colorMATH}\ensuremath{c}}}}[{{\color{\colorMATH}\ensuremath{n}}},{{\color{\colorMATH}\ensuremath{n}}}]\hspace*{0.33em}{{\color{\colorMATH}\ensuremath{\tau }}}}}}}
  \hfill\hspace{0pt}
\cr 
\cr
\end{tabularx}
\vspace*{-1em}\end{gather*}\endgroup 
\vspace{-2.0em}
\begingroup\renewcommand\colorMATH{\colorMATHP}\renewcommand\colorSYNTAX{\colorSYNTAXP}
\vspace*{-0.25em}\begingroup\color{\colorMATH}\begin{gather*}\begin{tabularx}{\linewidth}{>{\centering\arraybackslash\(}X<{\)}}\hfill\hspace{0pt}\begingroup\color{\colorTEXT}\boxed{\begingroup\color{\colorMATH} \Gamma  \vdash  e \mathrel{:} \tau  \endgroup}\endgroup \end{tabularx}\vspace*{-1em}\end{gather*}\endgroup 
\vspace{-2.0em}
\begingroup\color{\colorMATH}\begin{mathpar}\inferrule*[lab={{\color{\colorTEXT}\textsc{\scriptsize MGauss}}}
  ]{ {\begingroup\renewcommand\colorMATH{\colorMATHS}\renewcommand\colorSYNTAX{\colorSYNTAXS}{{\color{\colorMATH}\ensuremath{\Gamma _{1} \vdash  e_{1} \mathrel{:} {\begingroup\renewcommand\colorMATH{\colorMATHM}\renewcommand\colorSYNTAX{\colorSYNTAXM}{{\color{\colorMATH}\ensuremath{{{\color{\colorSYNTAX}\texttt{\ensuremath{{\mathbb{R}}^{+}[{{\color{\colorMATH}\ensuremath{\dot r}}}]}}}}}}}\endgroup }}}}\endgroup }
  \\ {\begingroup\renewcommand\colorMATH{\colorMATHS}\renewcommand\colorSYNTAX{\colorSYNTAXS}{{\color{\colorMATH}\ensuremath{\Gamma _{2} \vdash  e_{2} \mathrel{:} {\begingroup\renewcommand\colorMATH{\colorMATHM}\renewcommand\colorSYNTAX{\colorSYNTAXM}{{\color{\colorMATH}\ensuremath{{{\color{\colorSYNTAX}\texttt{\ensuremath{{\mathbb{R}}^{+}[{{\color{\colorMATH}\ensuremath{\epsilon }}}]}}}}}}}\endgroup }}}}\endgroup }
  \\ {\begingroup\renewcommand\colorMATH{\colorMATHS}\renewcommand\colorSYNTAX{\colorSYNTAXS}{{\color{\colorMATH}\ensuremath{\Gamma _{3} \vdash  e_{3} \mathrel{:} {\begingroup\renewcommand\colorMATH{\colorMATHM}\renewcommand\colorSYNTAX{\colorSYNTAXM}{{\color{\colorMATH}\ensuremath{{{\color{\colorSYNTAX}\texttt{\ensuremath{{\mathbb{R}}^{+}[{{\color{\colorMATH}\ensuremath{\delta }}}]}}}}}}}\endgroup }}}}\endgroup }
  \\ {\begingroup\renewcommand\colorMATH{\colorMATHS}\renewcommand\colorSYNTAX{\colorSYNTAXS}{{\color{\colorMATH}\ensuremath{\Gamma _{4} + \mathrlap{\hspace{-0.5pt}{}\rceil }\lfloor \Gamma _{5}\mathrlap{\hspace{-0.5pt}\rfloor }\lceil {}_{\{ {\begingroup\renewcommand\colorMATH{\colorMATHM}\renewcommand\colorSYNTAX{\colorSYNTAXM}{{\color{\colorMATH}\ensuremath{x_{1}}}}\endgroup },{.}\hspace{-1pt}{.}\hspace{-1pt}{.},{\begingroup\renewcommand\colorMATH{\colorMATHM}\renewcommand\colorSYNTAX{\colorSYNTAXM}{{\color{\colorMATH}\ensuremath{x_{n}}}}\endgroup }\} }^{{\begingroup\renewcommand\colorMATH{\colorMATHM}\renewcommand\colorSYNTAX{\colorSYNTAXM}{{\color{\colorMATH}\ensuremath{\dot r}}}\endgroup }} \vdash  e_{4} \mathrel{:} {\begingroup\renewcommand\colorMATH{\colorMATHM}\renewcommand\colorSYNTAX{\colorSYNTAXM}{{\color{\colorMATH}\ensuremath{{{\color{\colorSYNTAX}\texttt{\ensuremath{{\mathbb{M}}_{L2}^{{{\color{\colorMATH}\ensuremath{c}}}}[{{\color{\colorMATH}\ensuremath{m}}},{{\color{\colorMATH}\ensuremath{n}}}]\hspace*{0.33em}{\mathbb{R}}}}}}}}}\endgroup }}}}\endgroup }
     }{
     {}\rceil {\begingroup\renewcommand\colorMATH{\colorMATHS}\renewcommand\colorSYNTAX{\colorSYNTAXS}{{\color{\colorMATH}\ensuremath{\Gamma _{1} + \Gamma _{2} + \Gamma _{3}}}}\endgroup }\lceil {}^{{\begingroup\renewcommand\colorMATH{\colorMATHM}\renewcommand\colorSYNTAX{\colorSYNTAXM}{{\color{\colorMATH}\ensuremath{0}}}\endgroup },{\begingroup\renewcommand\colorMATH{\colorMATHM}\renewcommand\colorSYNTAX{\colorSYNTAXM}{{\color{\colorMATH}\ensuremath{0}}}\endgroup }} + {}\rceil {\begingroup\renewcommand\colorMATH{\colorMATHS}\renewcommand\colorSYNTAX{\colorSYNTAXS}{{\color{\colorMATH}\ensuremath{\Gamma _{4}}}}\endgroup }\lceil {}^{{{\color{\colorSYNTAX}\texttt{\ensuremath{\infty }}}}} + {}\rceil {\begingroup\renewcommand\colorMATH{\colorMATHS}\renewcommand\colorSYNTAX{\colorSYNTAXS}{{\color{\colorMATH}\ensuremath{\Gamma _{5}}}}\endgroup }\lceil {}^{{\begingroup\renewcommand\colorMATH{\colorMATHM}\renewcommand\colorSYNTAX{\colorSYNTAXM}{{\color{\colorMATH}\ensuremath{\epsilon }}}\endgroup },{\begingroup\renewcommand\colorMATH{\colorMATHM}\renewcommand\colorSYNTAX{\colorSYNTAXM}{{\color{\colorMATH}\ensuremath{\delta }}}\endgroup }} 
     \vdash  {{\color{\colorSYNTAX}\texttt{\ensuremath{{{\color{\colorSYNTAX}\texttt{mgauss}}}[{\begingroup\renewcommand\colorMATH{\colorMATHS}\renewcommand\colorSYNTAX{\colorSYNTAXS}{{\color{\colorMATH}\ensuremath{e_{1}}}}\endgroup },{\begingroup\renewcommand\colorMATH{\colorMATHS}\renewcommand\colorSYNTAX{\colorSYNTAXS}{{\color{\colorMATH}\ensuremath{e_{2}}}}\endgroup },{\begingroup\renewcommand\colorMATH{\colorMATHS}\renewcommand\colorSYNTAX{\colorSYNTAXS}{{\color{\colorMATH}\ensuremath{e_{3}}}}\endgroup }]\hspace*{0.33em}{<}{\begingroup\renewcommand\colorMATH{\colorMATHM}\renewcommand\colorSYNTAX{\colorSYNTAXM}{{\color{\colorMATH}\ensuremath{x_{1}}}}\endgroup },{.}\hspace{-1pt}{.}\hspace{-1pt}{.},{\begingroup\renewcommand\colorMATH{\colorMATHM}\renewcommand\colorSYNTAX{\colorSYNTAXM}{{\color{\colorMATH}\ensuremath{x_{n}}}}\endgroup }{>}\hspace*{0.33em}\{ {\begingroup\renewcommand\colorMATH{\colorMATHS}\renewcommand\colorSYNTAX{\colorSYNTAXS}{{\color{\colorMATH}\ensuremath{e_{4}}}}\endgroup }\} }}}} \mathrel{:} {\begingroup\renewcommand\colorMATH{\colorMATHM}\renewcommand\colorSYNTAX{\colorSYNTAXM}{{\color{\colorMATH}\ensuremath{{{\color{\colorSYNTAX}\texttt{\ensuremath{{\mathbb{M}}_{L\infty }^{U}[{{\color{\colorMATH}\ensuremath{m}}},{{\color{\colorMATH}\ensuremath{n}}}]\hspace*{0.33em}{\mathbb{R}}}}}}}}}\endgroup }
  }
\and\inferrule*[lab={{\color{\colorTEXT}\textsc{\scriptsize Exponential}}}
  ]{ {\begingroup\renewcommand\colorMATH{\colorMATHS}\renewcommand\colorSYNTAX{\colorSYNTAXS}{{\color{\colorMATH}\ensuremath{\Gamma _{1} \vdash  e_{1} \mathrel{:} {\begingroup\renewcommand\colorMATH{\colorMATHM}\renewcommand\colorSYNTAX{\colorSYNTAXM}{{\color{\colorMATH}\ensuremath{{{\color{\colorSYNTAX}\texttt{\ensuremath{{\mathbb{R}}^{+}[{{\color{\colorMATH}\ensuremath{\dot r}}}]}}}}}}}\endgroup }}}}\endgroup }
  \\ {\begingroup\renewcommand\colorMATH{\colorMATHS}\renewcommand\colorSYNTAX{\colorSYNTAXS}{{\color{\colorMATH}\ensuremath{\Gamma _{2} \vdash  e_{2} \mathrel{:} {\begingroup\renewcommand\colorMATH{\colorMATHM}\renewcommand\colorSYNTAX{\colorSYNTAXM}{{\color{\colorMATH}\ensuremath{{{\color{\colorSYNTAX}\texttt{\ensuremath{{\mathbb{R}}^{+}[{{\color{\colorMATH}\ensuremath{\epsilon }}}]}}}}}}}\endgroup }}}}\endgroup }
  \\ {\begingroup\renewcommand\colorMATH{\colorMATHS}\renewcommand\colorSYNTAX{\colorSYNTAXS}{{\color{\colorMATH}\ensuremath{\Gamma _{3} \vdash  e_{3} \mathrel{:} {\begingroup\renewcommand\colorMATH{\colorMATHM}\renewcommand\colorSYNTAX{\colorSYNTAXM}{{\color{\colorMATH}\ensuremath{{{\color{\colorSYNTAX}\texttt{\ensuremath{{\mathbb{M}}_{{{\color{\colorMATH}\ensuremath{\ell }}}}^{{{\color{\colorMATH}\ensuremath{c}}}}[{{\color{\colorMATH}\ensuremath{1}}},{{\color{\colorMATH}\ensuremath{m}}}]({{\color{\colorMATH}\ensuremath{\tau }}})}}}}}}}\endgroup }}}}\endgroup }
  \\ {\begingroup\renewcommand\colorMATH{\colorMATHS}\renewcommand\colorSYNTAX{\colorSYNTAXS}{{\color{\colorMATH}\ensuremath{\Gamma _{4} + \mathrlap{\hspace{-0.5pt}{}\rceil }\lfloor \Gamma _{5}\mathrlap{\hspace{-0.5pt}\rfloor }\lceil {}_{\{ {\begingroup\renewcommand\colorMATH{\colorMATHM}\renewcommand\colorSYNTAX{\colorSYNTAXM}{{\color{\colorMATH}\ensuremath{x_{1}}}}\endgroup },{.}\hspace{-1pt}{.}\hspace{-1pt}{.},{\begingroup\renewcommand\colorMATH{\colorMATHM}\renewcommand\colorSYNTAX{\colorSYNTAXM}{{\color{\colorMATH}\ensuremath{x_{n}}}}\endgroup }\} }^{{\begingroup\renewcommand\colorMATH{\colorMATHM}\renewcommand\colorSYNTAX{\colorSYNTAXM}{{\color{\colorMATH}\ensuremath{\dot r}}}\endgroup }}\uplus \{ {\begingroup\renewcommand\colorMATH{\colorMATHM}\renewcommand\colorSYNTAX{\colorSYNTAXM}{{\color{\colorMATH}\ensuremath{x}}}\endgroup }\mathrel{:}_{{{\color{\colorSYNTAX}\texttt{\ensuremath{\infty }}}}}{\begingroup\renewcommand\colorMATH{\colorMATHM}\renewcommand\colorSYNTAX{\colorSYNTAXM}{{\color{\colorMATH}\ensuremath{\tau }}}\endgroup }\}  \vdash  e_{4} \mathrel{:} {\begingroup\renewcommand\colorMATH{\colorMATHM}\renewcommand\colorSYNTAX{\colorSYNTAXM}{{\color{\colorMATH}\ensuremath{{{\color{\colorSYNTAX}\texttt{\ensuremath{{\mathbb{R}}}}}}}}}\endgroup } }}}\endgroup }
     }{
     {}\rceil {\begingroup\renewcommand\colorMATH{\colorMATHS}\renewcommand\colorSYNTAX{\colorSYNTAXS}{{\color{\colorMATH}\ensuremath{\Gamma _{1} + \Gamma _{2}}}}\endgroup }\lceil {}^{{\begingroup\renewcommand\colorMATH{\colorMATHM}\renewcommand\colorSYNTAX{\colorSYNTAXM}{{\color{\colorMATH}\ensuremath{0}}}\endgroup },{\begingroup\renewcommand\colorMATH{\colorMATHM}\renewcommand\colorSYNTAX{\colorSYNTAXM}{{\color{\colorMATH}\ensuremath{0}}}\endgroup }} + {}\rceil {\begingroup\renewcommand\colorMATH{\colorMATHS}\renewcommand\colorSYNTAX{\colorSYNTAXS}{{\color{\colorMATH}\ensuremath{\Gamma _{3} + \Gamma _{4}}}}\endgroup }\lceil {}^{{{\color{\colorSYNTAX}\texttt{\ensuremath{\infty }}}}} + {}\rceil {\begingroup\renewcommand\colorMATH{\colorMATHS}\renewcommand\colorSYNTAX{\colorSYNTAXS}{{\color{\colorMATH}\ensuremath{\Gamma _{5}}}}\endgroup }\lceil {}^{{\begingroup\renewcommand\colorMATH{\colorMATHM}\renewcommand\colorSYNTAX{\colorSYNTAXM}{{\color{\colorMATH}\ensuremath{\epsilon }}}\endgroup },{\begingroup\renewcommand\colorMATH{\colorMATHM}\renewcommand\colorSYNTAX{\colorSYNTAXM}{{\color{\colorMATH}\ensuremath{0}}}\endgroup }} 
     \vdash  {{\color{\colorSYNTAX}\texttt{\ensuremath{{{\color{\colorSYNTAX}\texttt{exponential}}}[{\begingroup\renewcommand\colorMATH{\colorMATHS}\renewcommand\colorSYNTAX{\colorSYNTAXS}{{\color{\colorMATH}\ensuremath{e_{1}}}}\endgroup },{\begingroup\renewcommand\colorMATH{\colorMATHS}\renewcommand\colorSYNTAX{\colorSYNTAXS}{{\color{\colorMATH}\ensuremath{e_{2}}}}\endgroup }]\hspace*{0.33em}{<}{\begingroup\renewcommand\colorMATH{\colorMATHM}\renewcommand\colorSYNTAX{\colorSYNTAXM}{{\color{\colorMATH}\ensuremath{x_{1}}}}\endgroup },{.}\hspace{-1pt}{.}\hspace{-1pt}{.},{\begingroup\renewcommand\colorMATH{\colorMATHM}\renewcommand\colorSYNTAX{\colorSYNTAXM}{{\color{\colorMATH}\ensuremath{x_{n}}}}\endgroup }{>}\hspace*{0.33em}{\begingroup\renewcommand\colorMATH{\colorMATHS}\renewcommand\colorSYNTAX{\colorSYNTAXS}{{\color{\colorMATH}\ensuremath{e_{3}}}}\endgroup }\hspace*{0.33em}\{ {\begingroup\renewcommand\colorMATH{\colorMATHM}\renewcommand\colorSYNTAX{\colorSYNTAXM}{{\color{\colorMATH}\ensuremath{x}}}\endgroup } \Rightarrow  {\begingroup\renewcommand\colorMATH{\colorMATHS}\renewcommand\colorSYNTAX{\colorSYNTAXS}{{\color{\colorMATH}\ensuremath{e_{4}}}}\endgroup }\} }}}} \mathrel{:} {\begingroup\renewcommand\colorMATH{\colorMATHM}\renewcommand\colorSYNTAX{\colorSYNTAXM}{{\color{\colorMATH}\ensuremath{\tau }}}\endgroup }
  }
\end{mathpar}\endgroup 
\vspace{-1.5em}
\endgroup 
\caption{Matrix Typing Rules}
\vspace{-0.5em}
\label{fig:matrix_typing}
\end{figure*}
}

\newcommand\FigureMatrixDistance{
\begin{figure}
  \smallerplz
\vspace*{-0.25em}\begingroup\color{\colorMATH}\begin{gather*}
\begin{tabularx}{\linewidth}{>{\centering\arraybackslash\(}X<{\)}}\begin{array}{c@{\hspace*{0.33em}}|@{\hspace*{0.33em}}c@{\hspace*{0.33em}}|@{\hspace*{0.33em}}rcl
  } {{\color{\colorTEXT}\textit{Domain}}}                     & {{\color{\colorTEXT}\textit{Carrier:}}}\hspace*{0.33em}X \in  {{\color{\colorMATH}\ensuremath{\operatorname{set}}}}  & \multicolumn{3}{c}{{{\color{\colorTEXT}\textit{Metric:}}}\hspace*{0.33em} |\underline{\hspace{0.66em}\vspace*{5ex}}-\underline{\hspace{0.66em}\vspace*{5ex}}|     \in  X \rightarrow  X \rightarrow  {\mathbb{R}}\uplus \{ \infty \} }
  \cr\hline  {{\color{\colorMATH}\ensuremath{\operatorname{real}}}}                       & {\mathbb{R}}                     & |r_{1} - r_{2}| &{}\triangleq {}& | r_{1} - r_{2} |_{{\mathbb{R}}}
  \cr  {{\color{\colorMATH}\ensuremath{\operatorname{data}}}}                       & {\mathbb{R}}                     & |r_{1} - r_{2}| &{}\triangleq {}& \left\{ \begin{array}{l
                                                                            } 0\hspace*{0.33em}{{\color{\colorTEXT}\textit{when}}}\hspace*{0.33em}r_{1}=r_{2}
                                                                            \cr  1\hspace*{0.33em}{{\color{\colorTEXT}\textit{when}}}\hspace*{0.33em}r_{1}\neq r_{2}
                                                                            \end{array}\right.
  \cr  {{\color{\colorMATH}\ensuremath{\operatorname{matrix}}}}[n_{1},n_{2}]_{L\infty }(D) & {\mathbb{M}}[n_{1},n_{2}](\| D\| )         & |m_{1} - m_{2}| &{}\triangleq {}& \sum \limits_{i} \max_{j} |m_{1}[i,j] - m_{2}[i,j]|_{D}
  \cr  {{\color{\colorMATH}\ensuremath{\operatorname{matrix}}}}[n_{1},n_{2}]_{L1}(D) & {\mathbb{M}}[n_{1},n_{2}](\| D\| )         & |m_{1} - m_{2}| &{}\triangleq {}& \sum \limits_{i,j}|m_{1}[i,j] - m_{2}[i,j]|_{D}
  \cr  {{\color{\colorMATH}\ensuremath{\operatorname{matrix}}}}[n_{1},n_{2}]_{L2}(D) & {\mathbb{M}}[n_{1},n_{2}](\| D\| )         & |m_{1} - m_{2}| &{}\triangleq {}& \sum \limits_{i} \sqrt {\sum \limits_{j} |m_{1}[i,j] - m_{2}[i,j]|_{D}^{2}}
  \end{array}               
\end{tabularx}
\vspace*{-1em}\end{gather*}\endgroup 
\vspace{-1.5em}
\caption{Distance Metrics for Matrices}
 \vspace{-0.5em}
\label{fig:matrix_distance}
\end{figure}
}

\label{sec:machine_learning}



Machine learning algorithms typically operate over a training set of
\emph{samples}, and implementations of these algorithms often
represent datasets using matrices. To express these algorithms,
\system includes a core matrix API which encodes sensitivity and
privacy properties of matrix operations. 

We add a matrix type {{\color{\colorSYNTAX}\texttt{\ensuremath{{\mathbb{M}}_{{{\color{\colorMATH}\ensuremath{\ell }}}}^{{{\color{\colorMATH}\ensuremath{c}}}}[{{\color{\colorMATH}\ensuremath{m}}},{{\color{\colorMATH}\ensuremath{n}}}]\hspace*{0.33em}{{\color{\colorMATH}\ensuremath{\tau }}}}}}}, encode vectors as
single-row matrices, and add typing rules for gradient computations
that encode desirable properties. We also introduce a type for matrix
indices {{\color{\colorMATH}\ensuremath{{{\color{\colorSYNTAX}\texttt{idx}}}[n]}}} for type-safe indexing.  These new types are shown
in Figure~\ref{fig:matrix_typing}, along with {\begingroup\renewcommand\colorMATH{\colorMATHS}\renewcommand\colorSYNTAX{\colorSYNTAXS}{{\color{\colorMATH}\ensuremath{{{\color{\colorMATH}\ensuremath{\operatorname{sensitivity}}}}}}}\endgroup }
operations on matrices---encoded as library functions because their
types can be encoded using existing connectives---and new matrix-level
differential {\begingroup\renewcommand\colorMATH{\colorMATHP}\renewcommand\colorSYNTAX{\colorSYNTAXP}{{\color{\colorMATH}\ensuremath{{{\color{\colorMATH}\ensuremath{\operatorname{privacy}}}}}}}\endgroup } mechanisms---encoded as primitive syntactic forms
because their types {{\color{\colorTEXT}\textit{cannot}}} be expressed using existing type-level
connectives.

In the matrix type {{\color{\colorSYNTAX}\texttt{\ensuremath{{\mathbb{M}}_{{{\color{\colorMATH}\ensuremath{\ell }}}}^{{{\color{\colorMATH}\ensuremath{c}}}}[{{\color{\colorMATH}\ensuremath{m}}},{{\color{\colorMATH}\ensuremath{n}}}]\hspace*{0.33em}{{\color{\colorMATH}\ensuremath{\tau }}}}}}}, the {{\color{\colorMATH}\ensuremath{m}}} and {{\color{\colorMATH}\ensuremath{n}}} parameters
refer to the number of rows and columns in the matrix, respectively. The {{\color{\colorMATH}\ensuremath{\ell }}}
parameter determines the distance metric used for the matrix metric for the
purposes of sensitivity analysis; the {{\color{\colorMATH}\ensuremath{c}}} parameter is used to specify a norm
bound on each row of the matrix, which will be useful when applying gradient
functions. 

\subsection{Distance Metrics for Matrices}

Differentially private machine learning algorithms typically move from one
distance metric on matrices and vectors to another as the algorithm progresses.
For example, two input training datasets are neighbors if they differ on
exactly one sample (i.e. one row of the matrix), but they may differ
arbitrarily in that row. After computing a gradient, the algorithm may consider
the {{\color{\colorMATH}\ensuremath{L2}}} sensitivity of the resulting vector---i.e. two gradients {{\color{\colorMATH}\ensuremath{g_{1}}}} and {{\color{\colorMATH}\ensuremath{g_{2}}}}
are neighbors if {{\color{\colorMATH}\ensuremath{\| g_{1}-g_{2}\| _{2} \leq  1}}}. These are very different notions of
distance---but the first is required by the definition of differential privacy,
and the second is required as a condition on the input to the Gaussian
mechanism.

The {{\color{\colorMATH}\ensuremath{\ell }}} annotation on matrix types in \system enables specifying the
desired notion of distance between \emph{rows}. The annotation is one
of {{\color{\colorMATH}\ensuremath{L\infty }}}, {{\color{\colorMATH}\ensuremath{L1}}}, or {{\color{\colorMATH}\ensuremath{L2}}}; an annotation of {{\color{\colorMATH}\ensuremath{L\infty }}}, for example, means that
the distance between two rows is equal to the {{\color{\colorMATH}\ensuremath{L\infty }}} norm of the
difference between the rows. The distance between two matrices is
always equal to the sum of the distances between rows. The distance
metric for the element datatype {{\color{\colorMATH}\ensuremath{\tau }}} determines the distance between
two corresponding elements, and the row metric {{\color{\colorMATH}\ensuremath{\ell }}} specifies how to
combine elementwise distances to determine the distance between two
rows.

Figure~\ref{fig:matrix_distance} presents the complete set of distance metrics
for matrices, as well as real numbers and the new domain {{\color{\colorMATH}\ensuremath{\operatorname{data}}}} for elements of
the {{\color{\colorSYNTAX}\texttt{\ensuremath{{\mathbb{D}}}}}} type, which is operationally a copy of {{\color{\colorSYNTAX}\texttt{\ensuremath{{\mathbb{R}}}}}} but with a discrete
distance metric. Many combinations are possible, including the following common
ones:
\noindent  \textbf{Ex. 1:} {{\color{\colorMATH}\ensuremath{|X-X^{\prime}|_{{{\color{\colorSYNTAX}\texttt{\ensuremath{{\mathbb{M}}_{L\infty }^{U}[{{\color{\colorMATH}\ensuremath{m}}},{{\color{\colorMATH}\ensuremath{n}}}]\hspace*{0.33em}{\mathbb{D}}}}}}} = \sum \limits_{i} \max_{j} |X_{i,j} - X_{i,j}^{\prime}|_{{\mathbb{D}}}}}}
\\\noindent  Distance is the \emph{number of rows on which {{\color{\colorMATH}\ensuremath{X}}} and {{\color{\colorMATH}\ensuremath{X^{\prime}}}} differ}; commonly
   used to describe neighboring input datasets.
\\\noindent  \textbf{Ex. 2:} {{\color{\colorMATH}\ensuremath{|X-X^{\prime}|_{{{\color{\colorSYNTAX}\texttt{\ensuremath{{\mathbb{M}}_{L1}^{U}[{{\color{\colorMATH}\ensuremath{m}}},{{\color{\colorMATH}\ensuremath{n}}}]\hspace*{0.33em}{\mathbb{R}}}}}}}  = \sum \limits_{i} \sum \limits_{j} |X_{i,j} - X_{i,j}^{\prime}|_{{{\color{\colorSYNTAX}\texttt{\ensuremath{{\mathbb{R}}}}}}}}}}
\\\noindent  Distance is the \emph{sum of elementwise differences}.
\\\noindent  \textbf{Ex. 3:} {{\color{\colorMATH}\ensuremath{|X-X^{\prime}|_{{{\color{\colorSYNTAX}\texttt{\ensuremath{{\mathbb{M}}_{L2}^{U}[{{\color{\colorMATH}\ensuremath{m}}},{{\color{\colorMATH}\ensuremath{n}}}]\hspace*{0.33em}{\mathbb{R}}}}}}} = \sum \limits_{i} \sqrt {\sum \limits_{j} |X_{i,j} - X_{i,j}^{\prime}|_{{{\color{\colorSYNTAX}\texttt{\ensuremath{{\mathbb{R}}}}}}}^{2}}}}}
\\\noindent  Distance is \emph{sum of the {{\color{\colorSYNTAX}\texttt{\ensuremath{L2}}}} norm of the differences between
   corresponding rows}.
\\\noindent  \textbf{Ex. 4:} {{\color{\colorMATH}\ensuremath{|X-X^{\prime}|_{{{\color{\colorSYNTAX}\texttt{\ensuremath{{\mathbb{M}}_{L2}^{U}[{{\color{\colorMATH}\ensuremath{1}}},{{\color{\colorMATH}\ensuremath{n}}}]\hspace*{0.33em}{\mathbb{R}}}}}}} = \sqrt {\sum \limits_{j} |X_{1,j} - X_{1,j}^{\prime}|_{{{\color{\colorSYNTAX}\texttt{\ensuremath{{\mathbb{R}}}}}}}^{2}}}}}
\\\noindent  Represents a vector; distance is {{\color{\colorSYNTAX}\texttt{\ensuremath{L2}}}} sensitivity for vectors, as required
   by the Gaussian mechanism.

\FigureMatrixDistance
\FigureMatrixTyping

These distance metrics are used in the types of library functions which operate
over matrices.

\subsection{Matrix Operations}

Figure~\ref{fig:matrix_typing} summarizes the matrix operations
available in \system's API. We focus on the non-standard operations
which are designed specifically for sensitivity or privacy
applications. For example, {{\color{\colorMATH}\ensuremath{{\begingroup\renewcommand\colorMATH{\colorMATHS}\renewcommand\colorSYNTAX{\colorSYNTAXS}{{\color{\colorMATH}\ensuremath{{{\color{\colorMATH}\ensuremath{\operatorname{fr-sens}}}}}}}\endgroup }}}} allows converting between
notions of distance between rows; when converting from {{\color{\colorMATH}\ensuremath{L2}}} to {{\color{\colorMATH}\ensuremath{L1}}},
the distance between two rows may increase by {{\color{\colorMATH}\ensuremath{\sqrt n}}} (by
Cauchy-Schwarz), so the corresponding version of {{\color{\colorMATH}\ensuremath{{\begingroup\renewcommand\colorMATH{\colorMATHS}\renewcommand\colorSYNTAX{\colorSYNTAXS}{{\color{\colorMATH}\ensuremath{{{\color{\colorMATH}\ensuremath{\operatorname{fr-sens}}}}}}}\endgroup }}}} has a
sensitivity annotation of {{\color{\colorMATH}\ensuremath{\sqrt n}}}.

{{\color{\colorMATH}\ensuremath{{\begingroup\renewcommand\colorMATH{\colorMATHS}\renewcommand\colorSYNTAX{\colorSYNTAXS}{{\color{\colorMATH}\ensuremath{{{\color{\colorMATH}\ensuremath{\operatorname{undisc}}}}}}}\endgroup }}}} allows converting from discrete to standard reals, and
is infinitely sensitive. {{\color{\colorMATH}\ensuremath{{\begingroup\renewcommand\colorMATH{\colorMATHS}\renewcommand\colorSYNTAX{\colorSYNTAXS}{{\color{\colorMATH}\ensuremath{{{\color{\colorMATH}\ensuremath{\operatorname{discf}}}}}}}\endgroup }}}} allows converting an infinitely
sensitive function which returns a real to a 1-sensitive function
returning a discrete real; we can recover a 1-sensitive function from
reals to discrete reals ({{\color{\colorMATH}\ensuremath{{{\color{\colorMATH}\ensuremath{\operatorname{disc}}}}\mathrel{:} {{\color{\colorSYNTAX}\texttt{\ensuremath{ {\mathbb{R}} \multimap _{{{\color{\colorMATH}\ensuremath{1}}}} {\mathbb{D}}}}}}}}}) by applying
{{\color{\colorMATH}\ensuremath{{\begingroup\renewcommand\colorMATH{\colorMATHS}\renewcommand\colorSYNTAX{\colorSYNTAXS}{{\color{\colorMATH}\ensuremath{{{\color{\colorSYNTAX}\texttt{discf}}}}}}\endgroup }}}} to the identity function.

{{\color{\colorMATH}\ensuremath{{\begingroup\renewcommand\colorMATH{\colorMATHP}\renewcommand\colorSYNTAX{\colorSYNTAXP}{{\color{\colorMATH}\ensuremath{{{\color{\colorMATH}\ensuremath{\operatorname{above-threshold}}}}}}}\endgroup }}}} encodes the \emph{Sparse Vector
  Technique}~\cite{privacybook}, discussed in
Section~\ref{sec:adaptive_clipping}. {{\color{\colorMATH}\ensuremath{{\begingroup\renewcommand\colorMATH{\colorMATHP}\renewcommand\colorSYNTAX{\colorSYNTAXP}{{\color{\colorMATH}\ensuremath{{{\color{\colorMATH}\ensuremath{\operatorname{pfld-rows}}}}}}}\endgroup }}}} encodes
parallel composition of privacy mechanisms, and is discussed in
Section~\ref{sec:parallel_minibatching}. {{\color{\colorMATH}\ensuremath{{\begingroup\renewcommand\colorMATH{\colorMATHP}\renewcommand\colorSYNTAX{\colorSYNTAXP}{{\color{\colorMATH}\ensuremath{{{\color{\colorMATH}\ensuremath{\operatorname{sample}}}}}}}\endgroup }}}} performs
random subsampling with privacy amplification, and is discussed in
Section~\ref{sec:minibatching}.

Gradients are computed using {\begingroup\renewcommand\colorMATH{\colorMATHS}\renewcommand\colorSYNTAX{\colorSYNTAXS}{{\color{\colorMATH}\ensuremath{{{\color{\colorMATH}\ensuremath{\operatorname{L{\mathrel{\nabla }}}}}}_{{\begingroup\renewcommand\colorMATH{\colorMATHM}\renewcommand\colorSYNTAX{\colorSYNTAXM}{{\color{\colorMATH}\ensuremath{\ell }}}\endgroup }}^{g}[\underline{\hspace{0.66em}\vspace*{5ex}};\underline{\hspace{0.66em}\vspace*{5ex}},\underline{\hspace{0.66em}\vspace*{5ex}}]}}}\endgroup } and
{\begingroup\renewcommand\colorMATH{\colorMATHS}\renewcommand\colorSYNTAX{\colorSYNTAXS}{{\color{\colorMATH}\ensuremath{{{\color{\colorMATH}\ensuremath{\operatorname{U{\mathrel{\nabla }}}}}}[\underline{\hspace{0.66em}\vspace*{5ex}};\underline{\hspace{0.66em}\vspace*{5ex}},\underline{\hspace{0.66em}\vspace*{5ex}}]}}}\endgroup }. The first represents an {{\color{\colorMATH}\ensuremath{\ell }}}-Lipschitz gradient
(typical in convex optimization problems like logistic regression)
like the {\begingroup\renewcommand\colorMATH{\colorMATHS}\renewcommand\colorSYNTAX{\colorSYNTAXS}{{\color{\colorMATH}\ensuremath{{{\color{\colorMATH}\ensuremath{\operatorname{gradient}}}}}}}\endgroup } function introduced in
Section~\ref{sec:by_example}; it is a 1-sensitive function which
produces a matrix of real numbers. The second represents a gradient
\emph{without} a known Lipschitz constant (typical in non-convex
optimization problems, including training neural networks); it
produces a matrix of discrete reals. We demonstrate applications of
both in Section~\ref{sec:examples}.

In order to produce a matrix with sensitivity bound {{\color{\colorSYNTAX}\texttt{L2}}}, {\begingroup\renewcommand\colorMATH{\colorMATHS}\renewcommand\colorSYNTAX{\colorSYNTAXS}{{\color{\colorMATH}\ensuremath{L{\mathrel{\nabla }}_{{\begingroup\renewcommand\colorMATH{\colorMATHM}\renewcommand\colorSYNTAX{\colorSYNTAXM}{{\color{\colorMATH}\ensuremath{{{\color{\colorSYNTAX}\texttt{L2}}}}}}\endgroup }}^{g}}}}\endgroup }
requires input of type {\begingroup\renewcommand\colorMATH{\colorMATHM}\renewcommand\colorSYNTAX{\colorSYNTAXM}{{\color{\colorMATH}\ensuremath{{{\color{\colorSYNTAX}\texttt{\ensuremath{{\mathbb{M}}_{{{\color{\colorMATH}\ensuremath{\ell }}}}^{L2}[{{\color{\colorMATH}\ensuremath{m}}},{{\color{\colorMATH}\ensuremath{n}}}]\hspace*{0.33em}{\mathbb{D}}}}}}}}}\endgroup } for any {{\color{\colorMATH}\ensuremath{\ell }}}. We obtain
such a matrix by \emph{clipping}, a common operation in differentially private
machine learning. Clipping scales each row of a matrix to ensure its {{\color{\colorMATH}\ensuremath{c}}} norm
(for {{\color{\colorMATH}\ensuremath{c \in  \{ L\infty , L1, L2\} }}}) is less than 1: 
\vspace*{-0.25em}\begingroup\color{\colorMATH}\begin{gather*}{{\color{\colorMATH}\ensuremath{\operatorname{clip}}}}^{c}\hspace*{0.33em}x_{i} \triangleq  \left\{ \begin{array}{l@{\hspace*{1.00em}}l@{\hspace*{1.00em}}l
                   } \frac{x_{i}}{\| x_{i}\| _{c}} &{}{{\color{\colorTEXT}\textit{if}}}{}& \| x_{i}\| _{c} > 1
                   \cr  x_{i}                 &{}{{\color{\colorTEXT}\textit{if}}}{}& \| x_{i}\| _{c} \leq  1
                   \end{array}\right.
\vspace*{-1em}\end{gather*}\endgroup 
The clipping process is encoded in \system as {{\color{\colorMATH}\ensuremath{\operatorname{clip}}}}
(Figure~\ref{fig:matrix_typing}), which introduces a new bound on the {{\color{\colorMATH}\ensuremath{c}}} norm
of its output.

\subsection{Vector-Valued Privacy Mechanisms}

Both the Laplace and Gaussian mechanisms are capable of operating directly over
vectors; the Laplace mechanism adds noise calibrated to the {{\color{\colorSYNTAX}\texttt{\ensuremath{L1}}}} sensitivity
of the vector, while the Gaussian mechanism uses its {{\color{\colorSYNTAX}\texttt{\ensuremath{L2}}}} sensitivity. With
the addition of matrices to \system, we can introduce typing rules for these
vector-valued mechanisms, using single-row matrices to represent vectors. We
present the typing rule for \textsc{MGauss} in Figure~\ref{fig:matrix_typing};
the rule for \textsc{MLaplace} is similar.
We also introduce a typing rule for the exponential mechanism, which picks one
element out of an input vector based on a sensitive scoring function
(Figure~\ref{fig:matrix_typing}, rule \textsc{Exponential}).

\section{Case Studies}
\label{sec:examples}

In this section, we demonstrate the use of \system to express and verify a number of different algorithms for differentially private machine learning. Our case studies fall into three broad categories: differentially private optimization algorithms (i.e. training algorithms), useful additions or modifications to those algorithms, and algorithms for data preprocessing or easing deployment. Our case studies are summarized in the following table.

\begin{center}
  \smallerplz
\begin{tabular}{l l l l}
  \textbf{Technique} & \textbf{Ref.} & \textbf{\S} & \textbf{Privacy Concept}\\
  \hline
  \textbf{Optimization Algorithms} & & & \\
  Noisy Gradient Descent & \cite{SCS, BST} & \ref{sec:noisy_gd} & Composition\\
  Gradient Descent w/ Output Perturbation & \cite{PSGD} & \ref{sec:gd_output} & Parallel Composition (sensitivity)\\
  Noisy Frank-Wolfe & \cite{ttz16} & \ref{sec:noisy_fw} & Exponential mechanism \\[1mm]

  \textbf{Variations on Gradient Descent} & & & \\
  Minibatching & \cite{BST} & \ref{sec:minibatching} & Amplification by subsampling \\
  Parallel-composition minibatching & --- & \ref{sec:parallel_minibatching} & Parallel composition\\
  Gradient clipping & \cite{DPDL} & \ref{sec:gradient_clipping} & Sensitivity bounds \\[1mm]

  \textbf{Preprocessing \& Deployment} & & & \\
  Hyperparameter tuning & \cite{chaudhuri2013stability} & \ref{sec:hyperparameters} & Exponential mechanism \\
  Adaptive clipping & --- & \ref{sec:adaptive_clipping} & Sparse Vector Technique \\
  Z-Score normalization & \cite{sklearn_normalization} & \ref{sec:normalization} & Composition \\[1mm]

  \textbf{Combining All of the Above} &  & \ref{sec:example_sys} & Composition \\
\end{tabular}
\end{center}

There are four basic approaches to differentially private convex optimization: input perturbation~\cite{cms11}, objective perturbation~\cite{cms11}, gradient perturbation~\cite{SCS, BST}, and output perturbation~\cite{cms11, PSGD}. Of these, the latter three are known to provide competitive accuracy, and the latter two (gradient perturbation and output perturbation) are the most widely used; our first two case studies verify these two techniques. Our third case study verifies the noisy Frank-Wolfe algorithm~\cite{ttz16}, a variant of gradient perturbation especially suited to high-dimensional datasets.

Our next three case studies demonstrate the use of \system to verify commonly-used variations on the above algorithms, including various kinds of minibatching and a gradient clipping approach used in deep learning.

Finally, we explore techniques for preprocessing input datasets so that the preconditions of the above algorithms are satisfied. Appropriate preprocessing is a vital step in practical deployments, both for providing privacy and ensuring accuracy of the final model, but practical preprocessing techniques have not received much attention in prior work. Due to space constraints, two of these case studies (hyperparameter tuning and Z-score normalization) appear in the extended version of this paper.

In Section~\ref{sec:example_sys}, we discuss the use of \system to combine all of these components---many of which leverage \emph{different} variants of differential privacy---to build a complete machine learning system. \system provides direct support for embedding one privacy variant in a program written using a different variant, and automatically converts the embedded variant to match the surrounding program. This ability is vital for designing practical systems, since recent variants provide tight composition, but some useful algorithms are only defined for pure-{{\color{\colorMATH}\ensuremath{\epsilon }}} or {{\color{\colorMATH}\ensuremath{(\epsilon ,\delta )}}}-differential privacy.

\subsection{Noisy Gradient Descent}
\label{sec:noisy_gd}

We begin with a fully-worked version of the differentially-private gradient descent algorithm from Section~\ref{sec:by_example}. This algorithm was first proposed by Song et al.~\cite{SCS} and later refined by Bassily et al.~\cite{BST}. Gradient descent is a simple but effective training algorithm in machine learning, and has been applied in a wide range of contexts, from simple linear models to deep neural networks.

The program below implements noisy gradient descent in \system
(without minibatching, though we will extend it with minibatching in
Section~\ref{sec:minibatching}). It performs {{\color{\colorMATH}\ensuremath{k}}} iterations of
gradient descent, starting from an initial guess {{\color{\colorMATH}\ensuremath{\theta _{0}}}} consisting of
all zeros. At each iteration, the algorithm computes a noisy gradient
using {\begingroup\renewcommand\colorMATH{\colorMATHP}\renewcommand\colorSYNTAX{\colorSYNTAXP}{{\color{\colorMATH}\ensuremath{{{\color{\colorMATH}\ensuremath{\operatorname{noisy-grad}}}}}}}\endgroup }, scales the gradient by the \emph{learning
  rate} {{\color{\colorMATH}\ensuremath{\eta }}}, and subtracts the result from the current model {{\color{\colorMATH}\ensuremath{\theta }}} to
arrive at the updated model.


\vspace{-1.0em}



To ensure differential privacy, this definition uses \emph{gradient
  perturbation}---adding noise directly to the gradient in each
iteration. The {{\color{\colorMATH}\ensuremath{{\begingroup\renewcommand\colorMATH{\colorMATHP}\renewcommand\colorSYNTAX{\colorSYNTAXP}{{\color{\colorMATH}\ensuremath{{{\color{\colorSYNTAX}\texttt{mgauss}}}}}}\endgroup }}}} construct is used in the definition of
{{\color{\colorMATH}\ensuremath{{\begingroup\renewcommand\colorMATH{\colorMATHP}\renewcommand\colorSYNTAX{\colorSYNTAXP}{{\color{\colorMATH}\ensuremath{{{\color{\colorMATH}\ensuremath{\operatorname{noisy-grad}}}}}}}\endgroup }}}} to add the right amount of noise.
Under {{\color{\colorMATH}\ensuremath{(\epsilon , \delta )}}}-differential privacy, \system derives a total privacy
cost of {{\color{\colorMATH}\ensuremath{(2\epsilon \sqrt {2k\log (1/\delta ^{\prime})},k\delta +\delta ^{\prime})}}}-differential privacy for this
implementation, which matches the total cost manually proven by
Bassily et al.~\cite{BST}. \system can also derive a total cost for
other privacy variants: the same program satisfies {{\color{\colorMATH}\ensuremath{k\rho }}}-zCDP, or
{{\color{\colorMATH}\ensuremath{(\alpha ,k\epsilon )}}}-RDP.

\subsection{Gradient Descent with Output Perturbation}
\label{sec:gd_output}

\begin{wrapfigure}{r}{2.6in}
\vspace{-1.4em}
\begin{minipage}{2.6in}
\begingroup \smallerplz
$$
\begingroup\renewcommand\colorMATH{\colorMATHP}\renewcommand\colorSYNTAX{\colorSYNTAXP}\begingroup\color{\colorMATH}
\begin{array}{l
} {{\color{\colorMATH}\ensuremath{\operatorname{gd-output-perturbation}}}}({\begingroup\renewcommand\colorMATH{\colorMATHM}\renewcommand\colorSYNTAX{\colorSYNTAXM}{{\color{\colorMATH}\ensuremath{xs}}}\endgroup },{\begingroup\renewcommand\colorMATH{\colorMATHM}\renewcommand\colorSYNTAX{\colorSYNTAXM}{{\color{\colorMATH}\ensuremath{ys}}}\endgroup },{\begingroup\renewcommand\colorMATH{\colorMATHM}\renewcommand\colorSYNTAX{\colorSYNTAXM}{{\color{\colorMATH}\ensuremath{k}}}\endgroup },{\begingroup\renewcommand\colorMATH{\colorMATHM}\renewcommand\colorSYNTAX{\colorSYNTAXM}{{\color{\colorMATH}\ensuremath{\eta }}}\endgroup },{\begingroup\renewcommand\colorMATH{\colorMATHM}\renewcommand\colorSYNTAX{\colorSYNTAXM}{{\color{\colorMATH}\ensuremath{\epsilon }}}\endgroup },{\begingroup\renewcommand\colorMATH{\colorMATHM}\renewcommand\colorSYNTAX{\colorSYNTAXM}{{\color{\colorMATH}\ensuremath{\delta }}}\endgroup }) \triangleq 
\cr  \hspace*{1.00em}{{\color{\colorSYNTAX}\texttt{let}}}\hspace*{0.33em}{\begingroup\renewcommand\colorMATH{\colorMATHM}\renewcommand\colorSYNTAX{\colorSYNTAXM}{{\color{\colorMATH}\ensuremath{m_{0}}}}\endgroup } = {\begingroup\renewcommand\colorMATH{\colorMATHS}\renewcommand\colorSYNTAX{\colorSYNTAXS}{{\color{\colorMATH}\ensuremath{{{\color{\colorMATH}\ensuremath{\operatorname{zeros}}}}\hspace*{0.33em}({{\color{\colorMATH}\ensuremath{\operatorname{cols}}}}\hspace*{0.33em}{\begingroup\renewcommand\colorMATH{\colorMATHM}\renewcommand\colorSYNTAX{\colorSYNTAXM}{{\color{\colorMATH}\ensuremath{X}}}\endgroup })}}}\endgroup }\hspace*{0.33em}{{\color{\colorSYNTAX}\texttt{in}}}
\cr
\end{array}\endgroup \endgroup 
$$ \endgroup
\end{minipage}
\vspace{-1.2em}
\end{wrapfigure}

An alternative to gradient perturbation is \emph{output perturbation}---adding noise to the final trained model, rather than during the training process. Wu et al.~\cite{PSGD} present a competitive algorithm based on this idea, which works by bounding the \emph{total sensitivity} (rather than privacy) of the iterative gradient descent process. Their algorithm leverages \emph{parallel composition} for sensitivity: it divides the dataset into small chunks called \emph{minibatches}, and each iteration of the algorithm processes one minibatch. A single pass over all minibatches (and thus, the whole dataset) is often called an \emph{epoch}. If the dataset has size {{\color{\colorMATH}\ensuremath{m}}} and each minibatch is of size {{\color{\colorMATH}\ensuremath{b}}}, then each epoch comprises {{\color{\colorMATH}\ensuremath{m/b}}} iterations of the training algorithm. This approach to minibatching is often used (without privacy) in deep learning. The sensitivity of a complete epoch in this technique is just {{\color{\colorMATH}\ensuremath{1/b}}}.

We encode parallel composition for sensitivity in \system using the {{\color{\colorMATH}\ensuremath{{\begingroup\renewcommand\colorMATH{\colorMATHS}\renewcommand\colorSYNTAX{\colorSYNTAXS}{{\color{\colorMATH}\ensuremath{{{\color{\colorSYNTAX}\texttt{mfold-row}}}}}}\endgroup }}}} function, defined in Section~\ref{sec:machine_learning}, whose type matches that of {{\color{\colorMATH}\ensuremath{{\begingroup\renewcommand\colorMATH{\colorMATHS}\renewcommand\colorSYNTAX{\colorSYNTAXS}{{\color{\colorMATH}\ensuremath{{{\color{\colorSYNTAX}\texttt{foldl}}}}}}\endgroup }}}} for lists in the \fuzz type system~\cite{reed2010distance}.
{{\color{\colorMATH}\ensuremath{{\begingroup\renewcommand\colorMATH{\colorMATHS}\renewcommand\colorSYNTAX{\colorSYNTAXS}{{\color{\colorMATH}\ensuremath{{{\color{\colorSYNTAX}\texttt{mfold-row}}}}}}\endgroup }}}} considers each row to be a ``minibatch'' of size 1, but is easily extended to consider multiple rows at a time (as in our encoding below). 
\system derives a sensitivity bound of {{\color{\colorMATH}\ensuremath{k / b}}} for the training process, and a total privacy cost of {{\color{\colorMATH}\ensuremath{(\epsilon , \delta )}}}-differential privacy, matching the manual analysis of Wu et al.~\cite{PSGD}.

\subsection{Noisy Frank-Wolfe}
\label{sec:noisy_fw}

\begin{wrapfigure}{r}{2.8in}
\vspace{-1.4em}
\begin{minipage}{2.8in}
\begingroup \smallerplz
$$
\begin{array}{l
} {\begingroup\renewcommand\colorMATH{\colorMATHP}\renewcommand\colorSYNTAX{\colorSYNTAXP}{{\color{\colorMATH}\ensuremath{{{\color{\colorMATH}\ensuremath{\operatorname{frank-wolfe}}}}\hspace*{0.33em}{\begingroup\renewcommand\colorMATH{\colorMATHM}\renewcommand\colorSYNTAX{\colorSYNTAXM}{{\color{\colorMATH}\ensuremath{X}}}\endgroup }\hspace*{0.33em}{\begingroup\renewcommand\colorMATH{\colorMATHM}\renewcommand\colorSYNTAX{\colorSYNTAXM}{{\color{\colorMATH}\ensuremath{y}}}\endgroup }\hspace*{0.33em}{\begingroup\renewcommand\colorMATH{\colorMATHM}\renewcommand\colorSYNTAX{\colorSYNTAXM}{{\color{\colorMATH}\ensuremath{k}}}\endgroup }\hspace*{0.33em}{\begingroup\renewcommand\colorMATH{\colorMATHM}\renewcommand\colorSYNTAX{\colorSYNTAXM}{{\color{\colorMATH}\ensuremath{\epsilon }}}\endgroup }\hspace*{0.33em}{\begingroup\renewcommand\colorMATH{\colorMATHM}\renewcommand\colorSYNTAX{\colorSYNTAXM}{{\color{\colorMATH}\ensuremath{\delta }}}\endgroup } \triangleq }}}\endgroup }
\cr  \hspace*{1.00em}  {\begingroup\renewcommand\colorMATH{\colorMATHP}\renewcommand\colorSYNTAX{\colorSYNTAXP}{{\color{\colorMATH}\ensuremath{{{\color{\colorSYNTAX}\texttt{\ensuremath{{{\color{\colorSYNTAX}\texttt{let}}}\hspace*{0.33em}{\begingroup\renewcommand\colorMATH{\colorMATHM}\renewcommand\colorSYNTAX{\colorSYNTAXM}{{\color{\colorMATH}\ensuremath{X_{1}}}}\endgroup } = {\begingroup\renewcommand\colorMATH{\colorMATHS}\renewcommand\colorSYNTAX{\colorSYNTAXS}{{\color{\colorMATH}\ensuremath{{{\color{\colorMATH}\ensuremath{\operatorname{clip-matrix}}}}_{L\infty }\hspace*{0.33em}{\begingroup\renewcommand\colorMATH{\colorMATHM}\renewcommand\colorSYNTAX{\colorSYNTAXM}{{\color{\colorMATH}\ensuremath{X}}}\endgroup }}}}\endgroup }\hspace*{0.33em}{{\color{\colorSYNTAX}\texttt{in}}}}}}}}}}\endgroup }
\cr
\end{array}
$$
\endgroup
\end{minipage}
\vspace{-1.2em}
\end{wrapfigure}

We next consider a variation on gradient perturbation called the private Frank-Wolfe algorithm~\cite{ttz16}. This algorithm has dimension-independent utility, making it useful for high-dimensional datasets.
In each iteration, the algorithm takes a step of fixed size in a \emph{single} dimension, using the exponential mechanism to choose the best direction based on the gradient. The sensitivity of each update is therefore dependent on the {{\color{\colorMATH}\ensuremath{L_{\infty }}}} norm of each sample, rather than the {{\color{\colorMATH}\ensuremath{L_{2}}}} norm.

Our implementation uses the exponential mechanism to select the direction in which the gradient has its maximum value, then updates {{\color{\colorMATH}\ensuremath{\theta }}} in only the selected dimension. To get the right sensitivity, we compute the gradient with {\begingroup\renewcommand\colorMATH{\colorMATHS}\renewcommand\colorSYNTAX{\colorSYNTAXS}{{\color{\colorMATH}\ensuremath{L{\mathrel{\nabla }}_{{\begingroup\renewcommand\colorMATH{\colorMATHM}\renewcommand\colorSYNTAX{\colorSYNTAXM}{{\color{\colorMATH}\ensuremath{{{\color{\colorSYNTAX}\texttt{\ensuremath{L\infty }}}}}}}\endgroup }}^{LR}}}}\endgroup }, which requires an {\begingroup\renewcommand\colorMATH{\colorMATHM}\renewcommand\colorSYNTAX{\colorSYNTAXM}{{\color{\colorMATH}\ensuremath{{{\color{\colorSYNTAX}\texttt{\ensuremath{L\infty }}}}}}}\endgroup } norm bound on its input and ensures bounded {{\color{\colorMATH}\ensuremath{L\infty }}} sensitivity.

We mix several variants of differential privacy in this implementation. Each use of the exponential mechanism provides {{\color{\colorMATH}\ensuremath{\epsilon }}}-differential privacy; each iteration of the loop satisfies {{\color{\colorMATH}\ensuremath{\frac{1}{2}\epsilon ^{2}}}}-zCDP, and the whole algorithm satisfies {{\color{\colorMATH}\ensuremath{(\frac{1}{2}\epsilon ^{2} + 2\sqrt {\frac{1}{2}\epsilon ^{2} \log (1/\delta )}, \delta )}}}-differential privacy. The use of zCDP for composition is an improvement over the manual analysis of Talwar et al.~\cite{ttz16}, which used advanced composition.

\subsection{Minibatching}
\label{sec:minibatching}

\begin{wrapfigure}{r}{3.1in}
\vspace{-1.0em}
\begin{minipage}{3.1in}
\begingroup \smallerplz
$$
\begingroup\renewcommand\colorMATH{\colorMATHP}\renewcommand\colorSYNTAX{\colorSYNTAXP}\begingroup\color{\colorMATH}
\begin{array}{l
} {{\color{\colorMATH}\ensuremath{\operatorname{minibatch-gradient-descent}}}}\hspace*{0.33em}{\begingroup\renewcommand\colorMATH{\colorMATHM}\renewcommand\colorSYNTAX{\colorSYNTAXM}{{\color{\colorMATH}\ensuremath{X}}}\endgroup }\hspace*{0.33em}{\begingroup\renewcommand\colorMATH{\colorMATHM}\renewcommand\colorSYNTAX{\colorSYNTAXM}{{\color{\colorMATH}\ensuremath{y}}}\endgroup }\hspace*{0.33em}{\begingroup\renewcommand\colorMATH{\colorMATHM}\renewcommand\colorSYNTAX{\colorSYNTAXM}{{\color{\colorMATH}\ensuremath{k}}}\endgroup }\hspace*{0.33em}{\begingroup\renewcommand\colorMATH{\colorMATHM}\renewcommand\colorSYNTAX{\colorSYNTAXM}{{\color{\colorMATH}\ensuremath{b}}}\endgroup }\hspace*{0.33em}{\begingroup\renewcommand\colorMATH{\colorMATHM}\renewcommand\colorSYNTAX{\colorSYNTAXM}{{\color{\colorMATH}\ensuremath{\eta }}}\endgroup }\hspace*{0.33em}{\begingroup\renewcommand\colorMATH{\colorMATHM}\renewcommand\colorSYNTAX{\colorSYNTAXM}{{\color{\colorMATH}\ensuremath{\epsilon }}}\endgroup }\hspace*{0.33em}{\begingroup\renewcommand\colorMATH{\colorMATHM}\renewcommand\colorSYNTAX{\colorSYNTAXM}{{\color{\colorMATH}\ensuremath{\delta }}}\endgroup } \triangleq 
\cr
\end{array}
  \endgroup \endgroup 
$$
\endgroup
\end{minipage}
\vspace{-1.0em}
\end{wrapfigure}

An alternative form of minibatching to the one discussed in Section~\ref{sec:gd_output} is to randomly sample a subset of of the data in each iteration. Bassily et al.~\cite{BST} present an algorithm for differentially private stochastic gradient descent based on this idea: their approach samples a single random example from the training to compute the gradient in each iteration, and leverages the idea of \emph{privacy amplification} to improve privacy cost. The privacy amplification lemma states that if mechanism {{\color{\colorMATH}\ensuremath{{\mathcal{M}}(D)}}} provides {{\color{\colorMATH}\ensuremath{(\epsilon , \delta )}}}-differential privacy for the datset {{\color{\colorMATH}\ensuremath{D}}} of size {{\color{\colorMATH}\ensuremath{n}}}, then running {{\color{\colorMATH}\ensuremath{{\mathcal{M}}}}} on uniformly random {{\color{\colorMATH}\ensuremath{\gamma n}}} entries of {{\color{\colorMATH}\ensuremath{D}}} (for {{\color{\colorMATH}\ensuremath{\gamma  \leq  1}}}) provides {{\color{\colorMATH}\ensuremath{(2\gamma \epsilon , \gamma \delta )}}}-differential privacy~\cite{BST, DBLP:journals/corr/abs-1808-00087} (this bound is loose, but used here for readability).

We encode the privacy amplification lemma in \system using the {{\color{\colorMATH}\ensuremath{{\begingroup\renewcommand\colorMATH{\colorMATHP}\renewcommand\colorSYNTAX{\colorSYNTAXP}{{\color{\colorMATH}\ensuremath{{{\color{\colorSYNTAX}\texttt{sample}}}}}}\endgroup }}}} construct defined in Section~\ref{sec:machine_learning}. Similar privacy amplification lemmas exist for RDP~\cite{DBLP:journals/corr/abs-1808-00087} and tCDP~\cite{bun2018composable}. Privacy amplification is not possible under zCDP.
We can use sampling with privacy amplification to implement minibatching SGD in \system.

Under {{\color{\colorMATH}\ensuremath{(\epsilon ,\delta )}}}-differential privacy with privacy amplification, \system derives a total privacy cost of {{\color{\colorMATH}\ensuremath{(4(b/m)\epsilon \sqrt {2k\log (1/\delta ^{\prime})},(b/m)k\delta +\delta ^{\prime})}}}-differential privacy for this algorithm, which is equivalent to the manual proof of Bassily et al.~\cite{BST}.


\subsection{Parallel-Composition Minibatching}
\label{sec:parallel_minibatching}

As a final form of minibatching, we consider extending the parallel composition approach used by Wu et al.~\cite{PSGD} for \emph{sensitivity} to parallel composition of \emph{privacy mechanisms} for minibatching in the gradient perturbation approach from Section~\ref{sec:noisy_gd}.
Since the minibatches are disjoint in this approach, we can leverage the {parallel composition} property for privacy mechanisms (McSherry~\cite{PINQ}, Theorem 4; Dwork \& Lei~\cite{dwork2009differential}, Corollary 20), which states that running an {{\color{\colorMATH}\ensuremath{(\epsilon ,\delta )}}}-differentially private mechanism {{\color{\colorMATH}\ensuremath{k}}} times on {{\color{\colorMATH}\ensuremath{k}}} disjoint subsets of a database yields {{\color{\colorMATH}\ensuremath{(\epsilon ,\delta )}}}-differential privacy (in contrast to sequential composition, which yields {{\color{\colorMATH}\ensuremath{(k\epsilon ,k\delta )}}}-differential privacy). The intuition matches that of the sensitivity case: the targeted individual falls into just one subset, and computations on the other subsets do not affect privacy.

We encode this concept in \system using the {{\color{\colorMATH}\ensuremath{{\begingroup\renewcommand\colorMATH{\colorMATHP}\renewcommand\colorSYNTAX{\colorSYNTAXP}{{\color{\colorMATH}\ensuremath{{{\color{\colorSYNTAX}\texttt{pfld-rows}}}}}}\endgroup }}}} construct defined in Section~\ref{sec:machine_learning}. The arguments to {{\color{\colorMATH}\ensuremath{{\begingroup\renewcommand\colorMATH{\colorMATHP}\renewcommand\colorSYNTAX{\colorSYNTAXP}{{\color{\colorMATH}\ensuremath{{{\color{\colorSYNTAX}\texttt{pfld-rows}}}}}}\endgroup }}}} include the dataset and a function representing an {{\color{\colorMATH}\ensuremath{(\epsilon ,\delta )}}}-differentially private mechanism, and {{\color{\colorMATH}\ensuremath{{\begingroup\renewcommand\colorMATH{\colorMATHP}\renewcommand\colorSYNTAX{\colorSYNTAXP}{{\color{\colorMATH}\ensuremath{{{\color{\colorSYNTAX}\texttt{pfld-rows}}}}}}\endgroup }}}} ensures {{\color{\colorMATH}\ensuremath{(\epsilon , \delta )}}}-differential privacy for the dataset. This version considers minibatches of size 1, and is easily extended to consider other sizes. We can use {{\color{\colorMATH}\ensuremath{{\begingroup\renewcommand\colorMATH{\colorMATHP}\renewcommand\colorSYNTAX{\colorSYNTAXP}{{\color{\colorMATH}\ensuremath{{{\color{\colorSYNTAX}\texttt{pfld-rows}}}}}}\endgroup }}}} to implement epoch-based minibatching with gradient perturbation, even for privacy variants like zCDP which do not admit sampling:

\vspace{-1.0em}


This algorithm is similar in concept to the output perturbation approach of Wu et al.~\cite{PSGD}, but leverages parallel composition of privacy mechanisms for gradient perturbation instead, and has not been previously published. The algorithm runs {{\color{\colorMATH}\ensuremath{k}}} epochs with a batch size of {{\color{\colorMATH}\ensuremath{b}}}, for a total of {{\color{\colorMATH}\ensuremath{kb}}} iterations. Duet derives a privacy cost of {{\color{\colorMATH}\ensuremath{k\rho }}}-zCDP for the algorithm.

\subsection{Gradient Clipping}
\label{sec:gradient_clipping}

The gradient functions we have used so far have bounded sensitivity (i.e. they are 1-Lipschitz). Many gradient functions used in practical machine learning (e.g. gradients of deep neural networks) have \emph{unbounded} sensitivity. 

A common approach for achieving differential privacy in this setting is to clip the \emph{output} of the ``black-box'' gradient function (called {{\color{\colorMATH}\ensuremath{{\begingroup\renewcommand\colorMATH{\colorMATHS}\renewcommand\colorSYNTAX{\colorSYNTAXS}{{\color{\colorMATH}\ensuremath{{{\color{\colorSYNTAX}\texttt{\ensuremath{U{\mathrel{\nabla }}}}}}}}}\endgroup }}}}), instead of clipping the input samples~\cite{SS15, DPDL}. This requires computing the gradient for each sample individually, clipping each gradient (using {{\color{\colorMATH}\ensuremath{{{\color{\colorSYNTAX}\texttt{clip}}}}}}), then calling {{\color{\colorMATH}\ensuremath{{{\color{\colorSYNTAX}\texttt{conv}}}}}} to transform the resulting matrix of {{\color{\colorMATH}\ensuremath{{\mathbb{D}}}}} elements to a matrix of  {{\color{\colorMATH}\ensuremath{{\mathbb{R}}}}} elements and taking its average. The {{\color{\colorMATH}\ensuremath{{{\color{\colorSYNTAX}\texttt{conv}}}}}} construct is typed by the {\sc Convert} rule, which encodes the fact that a {{\color{\colorMATH}\ensuremath{{\mathbb{D}}}}} matrix with bounded norm can be converted to a {{\color{\colorMATH}\ensuremath{{\mathbb{R}}}}} matrix with bounded \emph{sensitivity}.
This results in a sensitivity of {{\color{\colorMATH}\ensuremath{\frac{1}{n}}}} for the averaged clipped gradient, even though the sensitivity of {{\color{\colorMATH}\ensuremath{U{\mathrel{\nabla }}}}} is unbounded. 


\vspace{-1.0em}


Under {{\color{\colorMATH}\ensuremath{(\epsilon , \delta )}}}-differential privacy, \system derives a privacy cost of {{\color{\colorMATH}\ensuremath{(2\epsilon \sqrt {2k\log (1/\delta ^{\prime})},k\delta +\delta ^{\prime})}}}-differential privacy for this algorithm.
Using zero-concentrated differential privacy or R\'enyi differential privacy, \system matches the manually-derived privacy cost proven by Talwar et al.~\cite{DPDL}. This approach can be combined with the previously-discussed solutions for minibatching to produce practical training algorithms for deep learning.

\subsection{Adaptive Clipping}
\label{sec:adaptive_clipping}

Our {{\color{\colorMATH}\ensuremath{{{\color{\colorSYNTAX}\texttt{mclip[L2]}}}}}} construct clips samples to have an {{\color{\colorMATH}\ensuremath{{{\color{\colorMATH}\ensuremath{\operatorname{L2}}}}}}} norm of 1. However, the best clipping parameter for a dataset depends on the data, and data-independent clipping can eliminate useful information and reduce accuracy. An ideal solution would determine the scale of the given dataset first, and then re-scale the data before clipping so that the clipping process does not eliminate too much information from the data.

We can achieve this goal by considering several possible ways to scale the dataset and testing how many samples are modified during clipping for each possibility. We would like to ensure that only outliers (say, 10\% of the samples) are modified during clipping. We encode this test in a 1-sensitive \system function called {{\color{\colorMATH}\ensuremath{{{\color{\colorSYNTAX}\texttt{testScaleParam}}}}}}.

We would like to find a value for {{\color{\colorMATH}\ensuremath{b}}} (the scaling parameter) for which {{\color{\colorMATH}\ensuremath{{{\color{\colorSYNTAX}\texttt{testScaleParam}}}}}} returns at least {\begingroup\renewcommand\colorMATH{\colorMATHS}\renewcommand\colorSYNTAX{\colorSYNTAXS}{{\color{\colorMATH}\ensuremath{{\begingroup\renewcommand\colorMATH{\colorMATHM}\renewcommand\colorSYNTAX{\colorSYNTAXM}{{\color{\colorMATH}\ensuremath{0.9}}}\endgroup } \mathrel{\mathord{\cdotp }} {{\color{\colorMATH}\ensuremath{\operatorname{real}}}}\hspace*{0.33em}({{\color{\colorMATH}\ensuremath{\operatorname{rows}}}}\hspace*{0.33em}{\begingroup\renewcommand\colorMATH{\colorMATHM}\renewcommand\colorSYNTAX{\colorSYNTAXM}{{\color{\colorMATH}\ensuremath{xs}}}\endgroup })}}}\endgroup } (i.e. at least 90\% of the samples remain un-clipped). Since we can define such a numeric threshold as our goal, and {{\color{\colorMATH}\ensuremath{{{\color{\colorSYNTAX}\texttt{testScaleParam}}}}}} has sensitivity 1, we can apply the \emph{Sparse Vector Technique} in the form of the {{\color{\colorMATH}\ensuremath{{{\color{\colorMATH}\ensuremath{\operatorname{AboveThreshold}}}}}}} function (Dwork \& Roth~\cite{privacybook}, Algorithm 1). 
{{\color{\colorMATH}\ensuremath{{{\color{\colorMATH}\ensuremath{\operatorname{AboveThreshold}}}}}}} runs an \emph{arbitrary} stream of 1-sensitive queries under a \emph{fixed} privacy budget, returning the index of the first query whose (noisy) result is greater than a given threshold. This is perfect for our setting: it allows testing many values for {{\color{\colorMATH}\ensuremath{b}}} (hundreds or thousands, potentially), in increasing order, to find the smallest one which does not cause a large loss of information. We encode this concept in \system as {{\color{\colorMATH}\ensuremath{{{\color{\colorSYNTAX}\texttt{above-threshold}}}}}}, defined in Section~\ref{sec:machine_learning}, and we can use it to adaptively pick a scaling parameter which meets the criteria listed above.

\vspace{-1.0em}


This function builds the necessary stream of 1-sensitive queries using {{\color{\colorMATH}\ensuremath{{{\color{\colorSYNTAX}\texttt{testScaleParam}}}}}} for each of the possible values of {{\color{\colorMATH}\ensuremath{b}}}, then finds the first one which leaves at least 90\% of the samples untouched under clipping. For this program, \system derives a total privacy cost of {{\color{\colorMATH}\ensuremath{\epsilon }}}-differential privacy, no matter how many elements {{\color{\colorMATH}\ensuremath{bs}}} contains.

\subsection{Composing Privacy Variants to Build Complete Learning Systems}
\label{sec:example_sys}

Putting together the pieces we have described to build real machine learning systems that preserve differential privacy often requires mixing privacy variants in order to obtain optimal results. 
We can use \system's ability to mix variants of differential privacy to combine components in a way that optimizes the use of the privacy budget. 
We demonstrate this ability with an example that performs several data-dependent analyses as pre-processing steps before training a model.
Our example uses \system's ability to mix variants to compose z-score normalization (using both pure {{\color{\colorMATH}\ensuremath{\epsilon }}} and {{\color{\colorMATH}\ensuremath{(\epsilon , \delta )}}}-differential privacy), hyperparameter tuning (with {{\color{\colorMATH}\ensuremath{(\epsilon , \delta )}}}-differential privacy), and gradient descent (with zCDP), returning a total {{\color{\colorMATH}\ensuremath{(\epsilon ,\delta )}}} privacy cost.

\begin{wrapfigure}{r}{3.5in}
\vspace{-1.0em}
\begin{minipage}{3.7in}
\begingroup \smallerplz
$$
\begingroup\renewcommand\colorMATH{\colorMATHP}\renewcommand\colorSYNTAX{\colorSYNTAXP}
\begin{array}{l
} {\begingroup\renewcommand\colorMATH{\colorMATHP}\renewcommand\colorSYNTAX{\colorSYNTAXP}{{\color{\colorMATH}\ensuremath{{{\color{\colorMATH}\ensuremath{\operatorname{adaptiveClippingGradientDescent}}}}\hspace*{0.33em}{\begingroup\renewcommand\colorMATH{\colorMATHM}\renewcommand\colorSYNTAX{\colorSYNTAXM}{{\color{\colorMATH}\ensuremath{xs}}}\endgroup }\hspace*{0.33em}{\begingroup\renewcommand\colorMATH{\colorMATHM}\renewcommand\colorSYNTAX{\colorSYNTAXM}{{\color{\colorMATH}\ensuremath{ys}}}\endgroup }\hspace*{0.33em}{\begingroup\renewcommand\colorMATH{\colorMATHM}\renewcommand\colorSYNTAX{\colorSYNTAXM}{{\color{\colorMATH}\ensuremath{k}}}\endgroup }\hspace*{0.33em}{\begingroup\renewcommand\colorMATH{\colorMATHM}\renewcommand\colorSYNTAX{\colorSYNTAXM}{{\color{\colorMATH}\ensuremath{\epsilon }}}\endgroup }\hspace*{0.33em}{\begingroup\renewcommand\colorMATH{\colorMATHM}\renewcommand\colorSYNTAX{\colorSYNTAXM}{{\color{\colorMATH}\ensuremath{\delta }}}\endgroup }\hspace*{0.33em}{\begingroup\renewcommand\colorMATH{\colorMATHM}\renewcommand\colorSYNTAX{\colorSYNTAXM}{{\color{\colorMATH}\ensuremath{\eta s}}}\endgroup }\hspace*{0.33em}{\begingroup\renewcommand\colorMATH{\colorMATHM}\renewcommand\colorSYNTAX{\colorSYNTAXM}{{\color{\colorMATH}\ensuremath{bs}}}\endgroup } \triangleq }}}\endgroup }
\cr
\end{array}
\endgroup 
$$
\endgroup
\end{minipage}
\vspace{-1.0em}
\end{wrapfigure}

As this example demonstrates, the ability to combine variants is important for building systems composed of many different algorithms. In Section~\ref{sec:evaluation}, we leverage this facility of \system to evaluate the accuracy of gradient descent and Frank-Wolfe under different privacy variants, by converting the total privacy cost of each to {{\color{\colorMATH}\ensuremath{(\epsilon , \delta )}}} before comparing them.






\section{Implementation \& Evaluation}
\label{sec:evaluation}

This section describes our implementation of \system, and our empirical evaluation of \system's ability to produce accurate differentially private models.
Our results demonstrate that the state-of-the-art privacy bounds derivable by \system can result in huge gains in accuracy for a given level of privacy.

\subsection{Implementation \& Typechecking Performance}
\label{sec:implementation}

\begin{wrapfigure}{r}{2.8in}
  \vspace{-1.0em}
  \smallerplz
  \centering
    \begin{tabular}{l c c}

    \textbf{Technique} & \textbf{LOC} & \textbf{Time (ms)}\\
    \hline
    Noisy G.D. & 23 & 0.51ms \\
    G.D. + Output Pert. & 25 & 0.39ms\\
    Noisy Frank-Wolfe & 31 & 0.59ms\\
    Minibatching & 26 & 0.51ms\\
    Parallel minibatching & 42 & 0.65ms \\
    Gradient clipping  & 21 & 0.40ms\\
    Hyperparameter tuning  & 125 & 3.87ms\\
    Adaptive clipping & 68 & 1.01ms\\
    Z-Score normalization & 104 & 1.51ms\\
  \end{tabular}

  \caption{Summary of Typechecking Performance on Case Study Programs}
  \vspace{-0.5em}
  \label{tbl:timing_checker}
  \vspace{-1.0em}
\end{wrapfigure}

We have implemented a prototype of \system in Haskell that includes type
inference of privacy bounds, and an interpreter that runs on all examples
described in this paper. We do not implement Hindley-Milner-style
constraint-based type inference of quantified types; our type inference is
syntax-directed and limited to construction of privacy bounds as symbolic
formulas over input variables.
%
Our implementation of type inference roughly follows the bottom-up approach of
\dfuzz's implementation~\cite{dfuzz-impl}. Type checking requires solving
constraints over symbolic expressions containing log and square root
operations. Prior work (\dfuzz and \hoaresq) uses an SMT solver during
typechecking to check validity of these constraints, but SMT solvers typically
do not support operators like log and square root, and struggle in the presence
of non-linear formulas. Because of these limitations, we implement a custom solver for
inequalities over symbolic real expressions instead of relying on support from
off-the-shelf solvers. Our custom solver is based on a simple decidable (but
incomplete) theory which supports log and square
root operations, and a more general subset of non-linear (polynomial)
formulas than typical SMT theories.

The \system typechecker demonstrates very practical performance.
Figure~\ref{tbl:timing_checker} summarizes the number of lines of code and
typechecking time for each of our case study programs; even medium-size
programs with many functions typecheck in just a few milliseconds.
Our implementation is open source and freely available on GitHub
at: \ifarxiv %
\url{https://github.com/uvm-plaid/duet}.
\else %
\verb|<redacted for double blind reviewing>|.
\fi



\subsection{Evaluation of Private Gradient Descent and Private Frank-Wolfe}

We also study the accuracy of the models produced by the \system implementations of private gradient descent and private Frank-Wolfe in Section~\ref{sec:examples}. We evaluate both algorithms on 4 datasets. Details about the datasets can be found in Figure~\ref{tbl:datasets_alg}.


We ran both algorithms on each dataset with per-iteration {{\color{\colorMATH}\ensuremath{\epsilon _{i} \in  \{ 0.0001, 0.001, 0.01, 0.1\} }}} and then used \system to derive the corresponding total privacy cost. We fixed {{\color{\colorMATH}\ensuremath{\delta  = \frac{1}{n^{2}}}}}, where {{\color{\colorMATH}\ensuremath{n}}} is the size of the dataset. For private gradient descent, we set {{\color{\colorMATH}\ensuremath{\eta  = 1.0}}}, and for private Frank-Wolfe we set the size of each corner {{\color{\colorMATH}\ensuremath{c = 100}}}.

We randomly shuffled each dataset, then chose 80\% of the dataset as training data and reserved 20\% for testing. We ran each training algorithm 5 times on the training data, and take the average testing error over all 5, to account for the randomness in the training process.

\begin{wrapfigure}{r}{2in}
  \vspace{-1.0em}
  \smallerplz
  \centering
\begin{tabular}{r c c}

    \textbf{Dataset} & \textbf{Samples} & \textbf{Dim.}\\
    \hline
    Synthetic & 10,000 & 20\\
    Adult & 45,220 & 104\\
    KDDCup99 & 70,000 & 114\\
    Facebook & 40,949 & 54\\
  \end{tabular}

  \caption{Dataset Used in Accuracy Evaluation}
  \vspace{-0.5em}
  \label{tbl:datasets_alg}
  \vspace{-1.0em}
\end{wrapfigure}

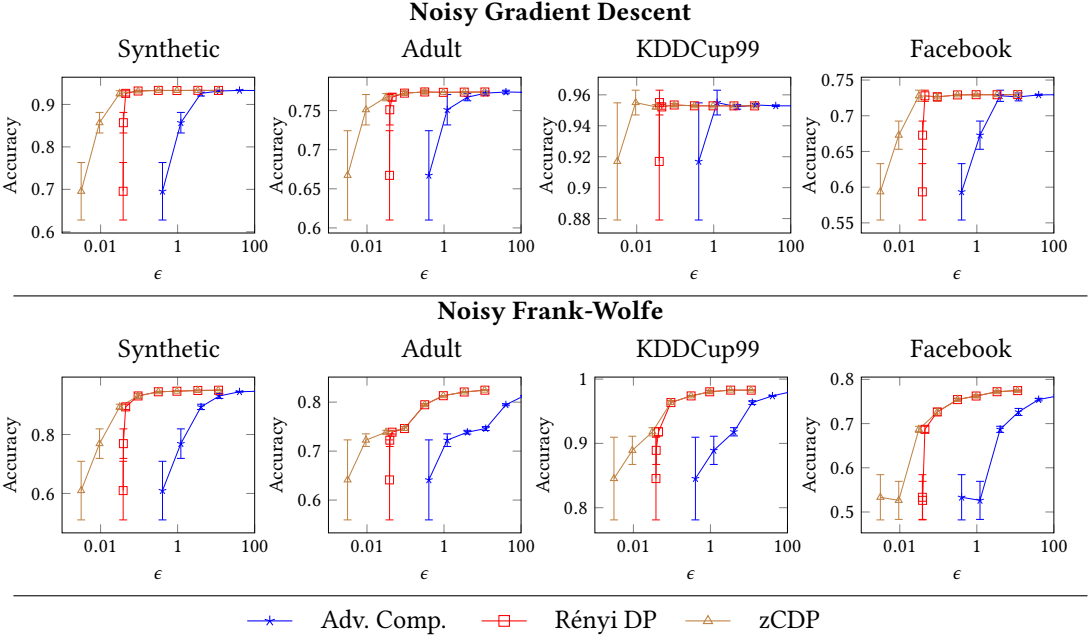
\begin{figure}
  \centering
\begin{tabular}{c c c c}
\hline
\multicolumn{4}{c}{\textbf{Noisy Gradient Descent}}\smallskip\\
    \hspace*{4mm}{Synthetic} & 
    \hspace*{1mm}{Adult} & 
    \hspace*{1mm}{KDDCup99} & 
    \hspace*{1mm}{Facebook} \\

    \hspace*{-4mm}\begin{tikzpicture}[scale=.75]
\begin{axis}[
  width=5cm,
  xmode=log,
  xmin=0.001,xmax=100,
  log ticks with fixed point,
  xlabel=$\epsilon$,
  ylabel=Accuracy,
  y label style={at={(axis description cs:0.1,.5)}}]
 \addplot[color=blue,mark=star,error bars/.cd,y dir=both,y explicit] coordinates {
(0.4046940586856042,0.6954400000000001) +- (6.77269104566272e-2,6.77269104566272e-2)
(1.2140821760568126,0.857) +- (2.4216027750231888e-2,2.4216027750231888e-2)
(4.046940586856042,0.92608) +- (6.1936741922707025e-3,6.1936741922707025e-3)
(12.140821760568127,0.93164) +- (1.5818975946627913e-3,1.5818975946627913e-3)
(40.46940586856042,0.93308) +- (4.833218389437701e-4,4.833218389437701e-4)
(121.40821760568126,0.93284) +- (2.939387691340034e-4,2.939387691340034e-4)
(404.6940586856042,0.9331600000000002) +- (7.999999999999118e-5,7.999999999999118e-5)
(1214.0821760568126,0.933) +- (2.1908902300208285e-4,2.1908902300208285e-4)

};
 \addplot[color=red,mark=square,error bars/.cd,y dir=both,y explicit] coordinates {
(3.856867299603084e-2,0.6954400000000001) +- (6.77269104566272e-2,6.77269104566272e-2)
(3.90799174333015e-2,0.857) +- (2.4216027750231888e-2,2.4216027750231888e-2)
(4.489532290725529e-2,0.92608) +- (6.1936741922707025e-3,6.1936741922707025e-3)
(9.415383184830123e-2,0.93164) +- (1.5818975946627913e-3,1.5818975946627913e-3)
(0.31534396049803126,0.93308) +- (4.833218389437701e-4,4.833218389437701e-4)
(0.9589803449377057,0.93284) +- (2.939387691340034e-4,2.939387691340034e-4)
(3.3488517620884335,0.9331600000000002) +- (7.999999999999118e-5,7.999999999999118e-5)
(11.506308380716687,0.933) +- (2.1908902300208285e-4,2.1908902300208285e-4)

};
 \addplot[color=brown,mark=triangle,error bars/.cd,y dir=both,y explicit] coordinates {
(3.132283446809938e-3,0.6954400000000001) +- (6.77269104566272e-2,6.77269104566272e-2)
(9.39812845152299e-3,0.857) +- (2.4216027750231888e-2,2.4216027750231888e-2)
(3.134200613449703e-2,0.92608) +- (6.1936741922707025e-3,6.1936741922707025e-3)
(9.415382951280875e-2,0.93164) +- (1.5818975946627913e-3,1.5818975946627913e-3)
(0.31533722798473524,0.93308) +- (4.833218389437701e-4,4.833218389437701e-4)
(0.9587927948859724,0.93284) +- (2.939387691340034e-4,2.939387691340034e-4)
(3.3450889438238516,0.9331600000000002) +- (7.999999999999118e-5,7.999999999999118e-5)
(11.31337792464821,0.933) +- (2.1908902300208285e-4,2.1908902300208285e-4)

};
\end{axis}
\end{tikzpicture} &
    \hspace*{-6mm}\begin{tikzpicture}[scale=.75]
\begin{axis}[
  width=5cm,
  xmode=log,
  xmin=0.001,xmax=100,
  log ticks with fixed point,
  xlabel=$\epsilon$,
  ylabel=Accuracy,
  y label style={at={(axis description cs:0.1,.5)}}]
 \addplot[color=blue,mark=star,error bars/.cd,y dir=both,y explicit] coordinates {
(0.4037782911072566,0.6671826625386996) +- (5.7100496003837084e-2,5.7100496003837084e-2)
(1.2113348733217695,0.7510835913312695) +- (1.943355872617613e-2,1.943355872617613e-2)
(4.037782911072566,0.7671384343210967) +- (4.5878115806775545e-3,4.5878115806775545e-3)
(12.113348733217697,0.772313135780628) +- (2.5198442340989406e-3,2.5198442340989406e-3)
(40.377829110725656,0.7738168951791243) +- (2.2526017829715894e-3,2.2526017829715894e-3)
(121.13348733217697,0.7733303847854931) +- (1.9779460216714506e-4,1.9779460216714506e-4)
(403.77829110725656,0.7735957540911101) +- (2.578926092368535e-4,2.578926092368535e-4)
(1211.3348733217697,0.7737726669615215) +- (1.3986190447451877e-4,1.3986190447451877e-4)

};
 \addplot[color=red,mark=square,error bars/.cd,y dir=both,y explicit] coordinates {
(3.856867299603084e-2,0.6671826625386996) +- (5.7100496003837084e-2,5.7100496003837084e-2)
(3.90799174333015e-2,0.7510835913312695) +- (1.943355872617613e-2,1.943355872617613e-2)
(4.489532290725529e-2,0.7671384343210967) +- (4.5878115806775545e-3,4.5878115806775545e-3)
(9.415383184830124e-2,0.772313135780628) +- (2.5198442340989406e-3,2.5198442340989406e-3)
(0.31534396049803126,0.7738168951791243) +- (2.2526017829715894e-3,2.2526017829715894e-3)
(0.9589803449377057,0.7733303847854931) +- (1.9779460216714506e-4,1.9779460216714506e-4)
(3.348851762088434,0.7735957540911101) +- (2.578926092368535e-4,2.578926092368535e-4)
(11.506308380716687,0.7737726669615215) +- (1.3986190447451877e-4,1.3986190447451877e-4)

};
 \addplot[color=brown,mark=triangle,error bars/.cd,y dir=both,y explicit] coordinates {
(3.1322834468099385e-3,0.6671826625386996) +- (5.7100496003837084e-2,5.7100496003837084e-2)
(9.39812845152299e-3,0.7510835913312695) +- (1.943355872617613e-2,1.943355872617613e-2)
(3.134200613449703e-2,0.7671384343210967) +- (4.5878115806775545e-3,4.5878115806775545e-3)
(9.415382951280875e-2,0.772313135780628) +- (2.5198442340989406e-3,2.5198442340989406e-3)
(0.3153372279847353,0.7738168951791243) +- (2.2526017829715894e-3,2.2526017829715894e-3)
(0.9587927948859724,0.7733303847854931) +- (1.9779460216714506e-4,1.9779460216714506e-4)
(3.345088943823852,0.7735957540911101) +- (2.578926092368535e-4,2.578926092368535e-4)
(11.31337792464821,0.7737726669615215) +- (1.3986190447451877e-4,1.3986190447451877e-4)

};
\end{axis}
\end{tikzpicture} &
    \hspace*{-6mm}\begin{tikzpicture}[scale=.75]
\begin{axis}[
  width=5cm,
  xmode=log,
  xmin=0.001,xmax=100,
  log ticks with fixed point,
  xlabel=$\epsilon$,
  ylabel=Accuracy,
  y label style={at={(axis description cs:0.1,.5)}}]
 \addplot[color=blue,mark=star,error bars/.cd,y dir=both,y explicit] coordinates {
(0.40774557436425757,0.9169428571428571) +- (3.787526251003524e-2,3.787526251003524e-2)
(1.2232367230927725,0.9550000000000001) +- (8.005100414926811e-3,8.005100414926811e-3)
(4.077455743642576,0.9522) +- (1.4752377878200545e-3,1.4752377878200545e-3)
(12.232367230927727,0.9534285714285714) +- (9.85610760609134e-4,9.85610760609134e-4)
(40.77455743642575,0.9529142857142856) +- (4.659858980085812e-4,4.659858980085812e-4)
(122.32367230927726,0.9528857142857141) +- (5.7142857142844504e-5,5.7142857142844504e-5)
(407.74557436425755,0.9528571428571428) +- (0.0,0.0)
(1223.2367230927725,0.9528571428571428) +- (0.0,0.0)

};
 \addplot[color=red,mark=square,error bars/.cd,y dir=both,y explicit] coordinates {
(3.856867299603084e-2,0.9169428571428571) +- (3.787526251003524e-2,3.787526251003524e-2)
(3.90799174333015e-2,0.9550000000000001) +- (8.005100414926811e-3,8.005100414926811e-3)
(4.489532290725529e-2,0.9522) +- (1.4752377878200545e-3,1.4752377878200545e-3)
(9.415383184830123e-2,0.9534285714285714) +- (9.85610760609134e-4,9.85610760609134e-4)
(0.31534396049803126,0.9529142857142856) +- (4.659858980085812e-4,4.659858980085812e-4)
(0.9589803449377057,0.9528857142857141) +- (5.7142857142844504e-5,5.7142857142844504e-5)
(3.348851762088433,0.9528571428571428) +- (0.0,0.0)
(11.506308380716687,0.9528571428571428) +- (0.0,0.0)

};
 \addplot[color=brown,mark=triangle,error bars/.cd,y dir=both,y explicit] coordinates {
(3.132283446809938e-3,0.9169428571428571) +- (3.787526251003524e-2,3.787526251003524e-2)
(9.398128451522988e-3,0.9550000000000001) +- (8.005100414926811e-3,8.005100414926811e-3)
(3.134200613449703e-2,0.9522) +- (1.4752377878200545e-3,1.4752377878200545e-3)
(9.415382951280875e-2,0.9534285714285714) +- (9.85610760609134e-4,9.85610760609134e-4)
(0.3153372279847353,0.9529142857142856) +- (4.659858980085812e-4,4.659858980085812e-4)
(0.9587927948859725,0.9528857142857141) +- (5.7142857142844504e-5,5.7142857142844504e-5)
(3.345088943823851,0.9528571428571428) +- (0.0,0.0)
(11.31337792464821,0.9528571428571428) +- (0.0,0.0)

};
\end{axis}
\end{tikzpicture}&
    \hspace*{-6mm}\begin{tikzpicture}[scale=.75]
\begin{axis}[
  width=5cm,
  xmode=log,
  xmin=0.001,xmax=100,
  log ticks with fixed point,
  xlabel=$\epsilon$,
  ylabel=Accuracy,
  y label style={at={(axis description cs:0.1,.5)}}]
 \addplot[color=blue,mark=star,error bars/.cd,y dir=both,y explicit] coordinates {
(0.4028720462655206,0.5934538348803126) +- (3.9474575145437715e-2,3.9474575145437715e-2)
(1.2086161387965615,0.6727894479726428) +- (1.9762783973859385e-2,1.9762783973859385e-2)
(4.028720462655206,0.7279433317049341) +- (7.829129703005599e-3,7.829129703005599e-3)
(12.086161387965618,0.7263800683927699) +- (5.564409765586338e-3,5.564409765586338e-3)
(40.28720462655206,0.7291646311675622) +- (7.785724206404134e-4,7.785724206404134e-4)
(120.86161387965616,0.7294577430385931) +- (7.662133434957633e-4,7.662133434957633e-4)
(402.87204626552057,0.7294088910600879) +- (3.9081582804106343e-4,3.9081582804106343e-4)
(1208.6161387965617,0.7296042989741084) +- (1.5448352028182307e-4,1.5448352028182307e-4)

};
 \addplot[color=red,mark=square,error bars/.cd,y dir=both,y explicit] coordinates {
(3.856867299603084e-2,0.5934538348803126) +- (3.9474575145437715e-2,3.9474575145437715e-2)
(3.90799174333015e-2,0.6727894479726428) +- (1.9762783973859385e-2,1.9762783973859385e-2)
(4.4895322907255286e-2,0.7279433317049341) +- (7.829129703005599e-3,7.829129703005599e-3)
(9.415383184830121e-2,0.7263800683927699) +- (5.564409765586338e-3,5.564409765586338e-3)
(0.3153439604980312,0.7291646311675622) +- (7.785724206404134e-4,7.785724206404134e-4)
(0.9589803449377055,0.7294577430385931) +- (7.662133434957633e-4,7.662133434957633e-4)
(3.348851762088433,0.7294088910600879) +- (3.9081582804106343e-4,3.9081582804106343e-4)
(11.506308380716685,0.7296042989741084) +- (1.5448352028182307e-4,1.5448352028182307e-4)

};
 \addplot[color=brown,mark=triangle,error bars/.cd,y dir=both,y explicit] coordinates {
(3.1322834468099376e-3,0.5934538348803126) +- (3.9474575145437715e-2,3.9474575145437715e-2)
(9.398128451522986e-3,0.6727894479726428) +- (1.9762783973859385e-2,1.9762783973859385e-2)
(3.134200613449702e-2,0.7279433317049341) +- (7.829129703005599e-3,7.829129703005599e-3)
(9.415382951280873e-2,0.7263800683927699) +- (5.564409765586338e-3,5.564409765586338e-3)
(0.31533722798473524,0.7291646311675622) +- (7.785724206404134e-4,7.785724206404134e-4)
(0.9587927948859722,0.7294577430385931) +- (7.662133434957633e-4,7.662133434957633e-4)
(3.345088943823851,0.7294088910600879) +- (3.9081582804106343e-4,3.9081582804106343e-4)
(11.31337792464821,0.7296042989741084) +- (1.5448352028182307e-4,1.5448352028182307e-4)

};
\end{axis}
\end{tikzpicture}\\

\hline
\multicolumn{4}{c}{\textbf{Noisy Frank-Wolfe}}\smallskip\\

    \hspace*{4mm}{Synthetic} & 
    \hspace*{1mm}{Adult} &
    \hspace*{1mm}{KDDCup99} & 
    \hspace*{1mm}{Facebook} \\

    \hspace*{-4mm}\begin{tikzpicture}[scale=.75]
\begin{axis}[
  width=5cm,
  xmode=log,
  xmin=0.001,xmax=100,
  log ticks with fixed point,
  xlabel=$\epsilon$,
  ylabel=Accuracy,
  y label style={at={(axis description cs:0.1,.5)}}]
 \addplot[color=blue,mark=star,error bars/.cd,y dir=both,y explicit] coordinates {
(0.4046940586856042,0.60968) +- (9.943026501020702e-2,9.943026501020702e-2)
(1.2140821760568126,0.76904) +- (5.0093257031261204e-2,5.0093257031261204e-2)
(4.046940586856042,0.89376) +- (9.018558643153578e-3,9.018558643153578e-3)
(12.140821760568127,0.9308400000000001) +- (8.490371016628194e-3,8.490371016628194e-3)
(40.46940586856042,0.9450000000000001) +- (1.3446189051177458e-3,1.3446189051177458e-3)
(121.40821760568126,0.9468) +- (2.529822128134776e-4,2.529822128134776e-4)
(404.6940586856042,0.9494400000000001) +- (1.1482160075525878e-3,1.1482160075525878e-3)
(1214.0821760568126,0.9501999999999999) +- (5.656854249492229e-4,5.656854249492229e-4)

};
 \addplot[color=red,mark=square,error bars/.cd,y dir=both,y explicit] coordinates {
(3.856867299603084e-2,0.60968) +- (9.943026501020702e-2,9.943026501020702e-2)
(3.90799174333015e-2,0.76904) +- (5.0093257031261204e-2,5.0093257031261204e-2)
(4.489532290725529e-2,0.89376) +- (9.018558643153578e-3,9.018558643153578e-3)
(9.415383184830123e-2,0.9308400000000001) +- (8.490371016628194e-3,8.490371016628194e-3)
(0.31534396049803126,0.9450000000000001) +- (1.3446189051177458e-3,1.3446189051177458e-3)
(0.9589803449377057,0.9468) +- (2.529822128134776e-4,2.529822128134776e-4)
(3.3488517620884335,0.9494400000000001) +- (1.1482160075525878e-3,1.1482160075525878e-3)
(11.506308380716687,0.9501999999999999) +- (5.656854249492229e-4,5.656854249492229e-4)

};
 \addplot[color=brown,mark=triangle,error bars/.cd,y dir=both,y explicit] coordinates {
(3.132283446809938e-3,0.60968) +- (9.943026501020702e-2,9.943026501020702e-2)
(9.39812845152299e-3,0.76904) +- (5.0093257031261204e-2,5.0093257031261204e-2)
(3.134200613449703e-2,0.89376) +- (9.018558643153578e-3,9.018558643153578e-3)
(9.415382951280875e-2,0.9308400000000001) +- (8.490371016628194e-3,8.490371016628194e-3)
(0.31533722798473524,0.9450000000000001) +- (1.3446189051177458e-3,1.3446189051177458e-3)
(0.9587927948859724,0.9468) +- (2.529822128134776e-4,2.529822128134776e-4)
(3.3450889438238516,0.9494400000000001) +- (1.1482160075525878e-3,1.1482160075525878e-3)
(11.31337792464821,0.9501999999999999) +- (5.656854249492229e-4,5.656854249492229e-4)

};
\end{axis}
\end{tikzpicture} &
    \hspace*{-6mm}\begin{tikzpicture}[scale=.75]
\begin{axis}[
  width=5cm,
  xmode=log,
  xmin=0.001,xmax=100,
  log ticks with fixed point,
  xlabel=$\epsilon$,
  ylabel=Accuracy,
  y label style={at={(axis description cs:0.1,.5)}}]
 \addplot[color=blue,mark=star,error bars/.cd,y dir=both,y explicit] coordinates {
(0.4037782911072566,0.6412206988058381) +- (8.169774294109675e-2,8.169774294109675e-2)
(1.2113348733217695,0.7223794781070324) +- (1.28609086267002e-2,1.28609086267002e-2)
(4.037782911072566,0.738655462184874) +- (3.8185117548252016e-3,3.8185117548252016e-3)
(12.113348733217697,0.7460858027421494) +- (4.023066356927683e-3,4.023066356927683e-3)
(40.377829110725656,0.7946483856700576) +- (1.9610614133034497e-3,1.9610614133034497e-3)
(121.13348733217697,0.8131357806280407) +- (4.195857134235212e-4,4.195857134235212e-4)
(403.77829110725656,0.8206988058381247) +- (8.999110990436067e-4,8.999110990436067e-4)
(1211.3348733217697,0.8244139761167626) +- (2.422479245931709e-4,2.422479245931709e-4)

};
 \addplot[color=red,mark=square,error bars/.cd,y dir=both,y explicit] coordinates {
(3.856867299603084e-2,0.6412206988058381) +- (8.169774294109675e-2,8.169774294109675e-2)
(3.90799174333015e-2,0.7223794781070324) +- (1.28609086267002e-2,1.28609086267002e-2)
(4.489532290725529e-2,0.738655462184874) +- (3.8185117548252016e-3,3.8185117548252016e-3)
(9.415383184830124e-2,0.7460858027421494) +- (4.023066356927683e-3,4.023066356927683e-3)
(0.31534396049803126,0.7946483856700576) +- (1.9610614133034497e-3,1.9610614133034497e-3)
(0.9589803449377057,0.8131357806280407) +- (4.195857134235212e-4,4.195857134235212e-4)
(3.348851762088434,0.8206988058381247) +- (8.999110990436067e-4,8.999110990436067e-4)
(11.506308380716687,0.8244139761167626) +- (2.422479245931709e-4,2.422479245931709e-4)

};
 \addplot[color=brown,mark=triangle,error bars/.cd,y dir=both,y explicit] coordinates {
(3.1322834468099385e-3,0.6412206988058381) +- (8.169774294109675e-2,8.169774294109675e-2)
(9.39812845152299e-3,0.7223794781070324) +- (1.28609086267002e-2,1.28609086267002e-2)
(3.134200613449703e-2,0.738655462184874) +- (3.8185117548252016e-3,3.8185117548252016e-3)
(9.415382951280875e-2,0.7460858027421494) +- (4.023066356927683e-3,4.023066356927683e-3)
(0.3153372279847353,0.7946483856700576) +- (1.9610614133034497e-3,1.9610614133034497e-3)
(0.9587927948859724,0.8131357806280407) +- (4.195857134235212e-4,4.195857134235212e-4)
(3.345088943823852,0.8206988058381247) +- (8.999110990436067e-4,8.999110990436067e-4)
(11.31337792464821,0.8244139761167626) +- (2.422479245931709e-4,2.422479245931709e-4)

};
\end{axis}
\end{tikzpicture} &
    \hspace*{-6mm}\begin{tikzpicture}[scale=.75]
\begin{axis}[
  width=5cm,
  xmode=log,
  xmin=0.001,xmax=100,
  log ticks with fixed point,
  xlabel=$\epsilon$,
  ylabel=Accuracy,
  y label style={at={(axis description cs:0.1,.5)}}]
 \addplot[color=blue,mark=star,error bars/.cd,y dir=both,y explicit] coordinates {
(0.40774557436425757,0.8453142857142856) +- (6.427589766318445e-2,6.427589766318445e-2)
(1.2232367230927725,0.8892) +- (2.19179546193042e-2,2.19179546193042e-2)
(4.077455743642576,0.9180571428571428) +- (6.412742416947881e-3,6.412742416947881e-3)
(12.232367230927727,0.9635142857142857) +- (2.913165050635615e-3,2.913165050635615e-3)
(40.77455743642575,0.9738571428571428) +- (7.559289460184257e-4,7.559289460184257e-4)
(122.32367230927726,0.9802857142857142) +- (2.209441706192172e-3,2.209441706192172e-3)
(407.74557436425755,0.9828571428571429) +- (9.035079029050514e-5,9.035079029050514e-5)
(1223.2367230927725,0.9828000000000001) +- (6.998542122236104e-5,6.998542122236104e-5)

};
 \addplot[color=red,mark=square,error bars/.cd,y dir=both,y explicit] coordinates {
(3.856867299603084e-2,0.8453142857142856) +- (6.427589766318445e-2,6.427589766318445e-2)
(3.90799174333015e-2,0.8892) +- (2.19179546193042e-2,2.19179546193042e-2)
(4.489532290725529e-2,0.9180571428571428) +- (6.412742416947881e-3,6.412742416947881e-3)
(9.415383184830123e-2,0.9635142857142857) +- (2.913165050635615e-3,2.913165050635615e-3)
(0.31534396049803126,0.9738571428571428) +- (7.559289460184257e-4,7.559289460184257e-4)
(0.9589803449377057,0.9802857142857142) +- (2.209441706192172e-3,2.209441706192172e-3)
(3.348851762088433,0.9828571428571429) +- (9.035079029050514e-5,9.035079029050514e-5)
(11.506308380716687,0.9828000000000001) +- (6.998542122236104e-5,6.998542122236104e-5)

};
 \addplot[color=brown,mark=triangle,error bars/.cd,y dir=both,y explicit] coordinates {
(3.132283446809938e-3,0.8453142857142856) +- (6.427589766318445e-2,6.427589766318445e-2)
(9.398128451522988e-3,0.8892) +- (2.19179546193042e-2,2.19179546193042e-2)
(3.134200613449703e-2,0.9180571428571428) +- (6.412742416947881e-3,6.412742416947881e-3)
(9.415382951280875e-2,0.9635142857142857) +- (2.913165050635615e-3,2.913165050635615e-3)
(0.3153372279847353,0.9738571428571428) +- (7.559289460184257e-4,7.559289460184257e-4)
(0.9587927948859725,0.9802857142857142) +- (2.209441706192172e-3,2.209441706192172e-3)
(3.345088943823851,0.9828571428571429) +- (9.035079029050514e-5,9.035079029050514e-5)
(11.31337792464821,0.9828000000000001) +- (6.998542122236104e-5,6.998542122236104e-5)

};
\end{axis}
\end{tikzpicture} &
    \hspace*{-6mm}\begin{tikzpicture}[scale=.75]
\begin{axis}[
  width=5cm,
  xmode=log,
  xmin=0.001,xmax=100,
  log ticks with fixed point,
  xlabel=$\epsilon$,
  ylabel=Accuracy,
  y label style={at={(axis description cs:0.1,.5)}}]
 \addplot[color=blue,mark=star,error bars/.cd,y dir=both,y explicit] coordinates {
(0.4028720462655206,0.5332681973619932) +- (5.1153970172115926e-2,5.1153970172115926e-2)
(1.2086161387965615,0.5262823644357597) +- (4.306423132905927e-2,4.306423132905927e-2)
(4.028720462655206,0.6870053737176356) +- (6.925274116920129e-3,6.925274116920129e-3)
(12.086161387965618,0.7263312164142649) +- (7.974402805840925e-3,7.974402805840925e-3)
(40.28720462655206,0.7543234000977039) +- (2.932746657509248e-3,2.932746657509248e-3)
(120.86161387965616,0.7623839765510503) +- (2.4494286539759207e-3,2.4494286539759207e-3)
(402.87204626552057,0.7721055202735709) +- (2.031913467814902e-3,2.031913467814902e-3)
(1208.6161387965617,0.7746458231558379) +- (1.2124742204192489e-3,1.2124742204192489e-3)

};
 \addplot[color=red,mark=square,error bars/.cd,y dir=both,y explicit] coordinates {
(3.856867299603084e-2,0.5332681973619932) +- (5.1153970172115926e-2,5.1153970172115926e-2)
(3.90799174333015e-2,0.5262823644357597) +- (4.306423132905927e-2,4.306423132905927e-2)
(4.4895322907255286e-2,0.6870053737176356) +- (6.925274116920129e-3,6.925274116920129e-3)
(9.415383184830121e-2,0.7263312164142649) +- (7.974402805840925e-3,7.974402805840925e-3)
(0.3153439604980312,0.7543234000977039) +- (2.932746657509248e-3,2.932746657509248e-3)
(0.9589803449377055,0.7623839765510503) +- (2.4494286539759207e-3,2.4494286539759207e-3)
(3.348851762088433,0.7721055202735709) +- (2.031913467814902e-3,2.031913467814902e-3)
(11.506308380716685,0.7746458231558379) +- (1.2124742204192489e-3,1.2124742204192489e-3)

};
 \addplot[color=brown,mark=triangle,error bars/.cd,y dir=both,y explicit] coordinates {
(3.1322834468099376e-3,0.5332681973619932) +- (5.1153970172115926e-2,5.1153970172115926e-2)
(9.398128451522986e-3,0.5262823644357597) +- (4.306423132905927e-2,4.306423132905927e-2)
(3.134200613449702e-2,0.6870053737176356) +- (6.925274116920129e-3,6.925274116920129e-3)
(9.415382951280873e-2,0.7263312164142649) +- (7.974402805840925e-3,7.974402805840925e-3)
(0.31533722798473524,0.7543234000977039) +- (2.932746657509248e-3,2.932746657509248e-3)
(0.9587927948859722,0.7623839765510503) +- (2.4494286539759207e-3,2.4494286539759207e-3)
(3.345088943823851,0.7721055202735709) +- (2.031913467814902e-3,2.031913467814902e-3)
(11.31337792464821,0.7746458231558379) +- (1.2124742204192489e-3,1.2124742204192489e-3)

};
\end{axis}
\end{tikzpicture}\\
\hline
\end{tabular}%

\rmspc \rmspc
  \caption{Accuracy Results for Noisy Gradient Descent (Top) and Noisy Frank-Wolfe (Bottom).}
  \vspace{-0.5em}
  \label{fig:accuracy_results_fw}
\end{figure}

We present the results in Figure~\ref{fig:accuracy_results_fw}.
Both algorithms are capable of generating accurate models at reasonable values of {{\color{\colorMATH}\ensuremath{\epsilon }}}. Note that all three models in the results provide \emph{exactly the same privacy guarantee} for a given value of {{\color{\colorMATH}\ensuremath{\epsilon }}}, yet their accuracies vary significantly. The results demonstrate the huge advantages afforded by recently developed variants of differential privacy---recent variants with tighter composition bounds yield \emph{significantly} better accuracy for a given level of privacy.

\section{Conclusion}

We have presented \system, a language and type system for expressing and statically verifying privacy-preserving programs. Unlike previous work, \system is agnostic to the underlying privacy definition, and requires only that it support sequential composition and post-processing. We have extended \system to support several recent variants of differential privacy, and our case studies demonstrate that \system derives state-of-the-art privacy bounds for a number of useful machine learning algorithms. We have implemented a prototype of \system, and our experimental results demonstrate the benefits of flexibility in privacy definition.

\ifarxiv
\section*{Acknowledgments}
The authors would like to thank Arthur Azevedo de Amorim, Justin Hsu, and Om
Thakkar for their helpful comments. This work was supported by the Center for
Long-Term Cybersecurity, Alibaba AIR, DARPA \& SPAWAR via N66001-15-C-4066, 
IARPA via 2019-1902070008, and NSF award 1901278. The U.S. Government is
authorized to reproduce and distribute reprints for Governmental purposes not
withstanding any copyright annotation therein. The views, opinions, and/or
findings expressed are those of the authors and should not be interpreted as
representing the official views or policies of any US Government agency.
\else
\fi

\bibliography{case-studies,dp,dpml}



\appendix

\section{Additional Case Studies}

\subsection{Hyperparameter Tuning}
\label{sec:hyperparameters}

Each of the gradient descent algorithms we have seen so far require a setting for the {{\color{\colorMATH}\ensuremath{\eta }}} hyperparameter, and many other hyperparameters (e.g. {{\color{\colorMATH}\ensuremath{\lambda }}} for regularization) are often used in practice. When privacy is not a concern, these hyperparameters are typically tuned by training many models on different hyperparameter combinations and picking the model with the best accuracy on the testing data (a \emph{grid search}). However, this approach cannot satisfy differential privacy, since selecting the best model based on the (sensitive) test data might reveal something about that data.

To tune hyperparameters with differential privacy, we must be careful to ensure that each trained model contributes to the overall privacy cost, and that the comparison of model accuracies is done in a differentially private way.
We can encode a simple algorithm for this purpose, originally due to Chaudhuri and Vinterbo~\cite{chaudhuri2013stability}, using the exponential mechanism:

\includegraphics[width=.5\textwidth]{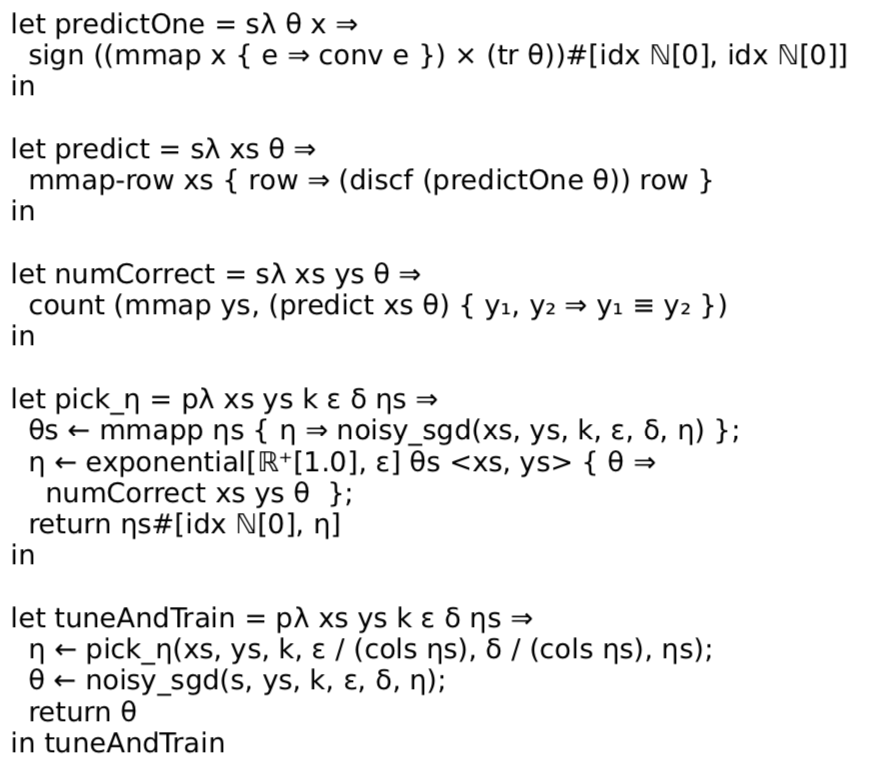}

This algorithm trains one model per value of {{\color{\colorMATH}\ensuremath{\eta }}} in the set {{\color{\colorMATH}\ensuremath{\eta s}}} and uses the exponential mechanism to select the value of {{\color{\colorMATH}\ensuremath{\eta }}} which maximizes the model's accuracy. Note that each prediction is \emph{infinitely} sensitive in the sample {{\color{\colorMATH}\ensuremath{x}}}, but the combination of {{\color{\colorMATH}\ensuremath{{\begingroup\renewcommand\colorMATH{\colorMATHS}\renewcommand\colorSYNTAX{\colorSYNTAXS}{{\color{\colorMATH}\ensuremath{{{\color{\colorSYNTAX}\texttt{mmap-row}}}}}}\endgroup }}}} and {{\color{\colorMATH}\ensuremath{{\begingroup\renewcommand\colorMATH{\colorMATHS}\renewcommand\colorSYNTAX{\colorSYNTAXS}{{\color{\colorMATH}\ensuremath{{{\color{\colorSYNTAX}\texttt{discf}}}}}}\endgroup }}}} ensures the resulting matrix of {{\color{\colorMATH}\ensuremath{{\mathbb{D}}}}} values is 1-sensitive in {{\color{\colorMATH}\ensuremath{xs}}}.

This algorithm re-trains the final model using the selected value of {{\color{\colorMATH}\ensuremath{\eta }}} so that a larger proportion of the privacy budget can be used for this model---each model trained during hyperparameter tuning only receives a budget of {{\color{\colorMATH}\ensuremath{(\epsilon /n, \delta /n)}}}, where {{\color{\colorMATH}\ensuremath{n}}} is the number of elements in {{\color{\colorMATH}\ensuremath{\eta s}}}; the final model is trained using a budget of {{\color{\colorMATH}\ensuremath{(\epsilon , \delta )}}}. \system derives a total privacy cost of {{\color{\colorMATH}\ensuremath{(2\epsilon \sqrt {2k\log (n/\delta )}+\epsilon /n+2\epsilon \sqrt {2k\log (1/\delta )}, 2k\delta +2\delta )}}}-differential privacy.


\subsection{Adaptive Z-Score Normalization}
\label{sec:normalization}

The adaptive clipping described in Section~\ref{sec:adaptive_clipping} results in a global scaling parameter for all features of the input dataset. More sophisticated preprocessing approaches---including those commonly used in non-private machine learning~\cite{sklearn_normalization}---modify each feature individually to ensure \emph{standardization}: zero mean, and unit variance. Standardization can be even more important in differentially-private machine learning, since a feature with very small variance may be especially susceptible to noise added during training.

Standardization for each feature is often achieved by \emph{z-score normalization} (in scikit-learn, provided by \textsf{preprocessing.StandardScaler}): subtract the feature's mean, and divide by its standard deviation. We can approximate this kind of normalization under differential privacy by computing a differentially private mean and scaling parameter for each feature individually, then normalizing based on these. Computing differentially private means requires first adaptively picking a clipping parameter for each column, to give bounded sensitivity for the mean itself.

\includegraphics[width=.6\textwidth]{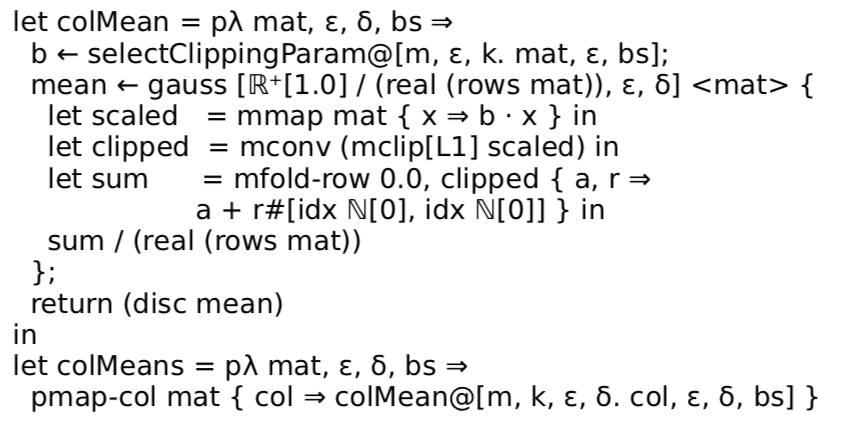}

Note that {{\color{\colorMATH}\ensuremath{{{\color{\colorSYNTAX}\texttt{pmap-col}}}}}} does \emph{not} provide parallel composition, because examining each of {{\color{\colorMATH}\ensuremath{n}}} columns individually allows examining some aspect of a single individual's data {{\color{\colorMATH}\ensuremath{n}}} times.
The next task is to determine the scaling parameter for each feature. This can be accomplished using {{\color{\colorMATH}\ensuremath{{{\color{\colorSYNTAX}\texttt{pickScaleParam}}}}}} from Section~\ref{sec:adaptive_clipping}, but we must first subtract the mean from each element of the column.

\includegraphics[width=.8\textwidth]{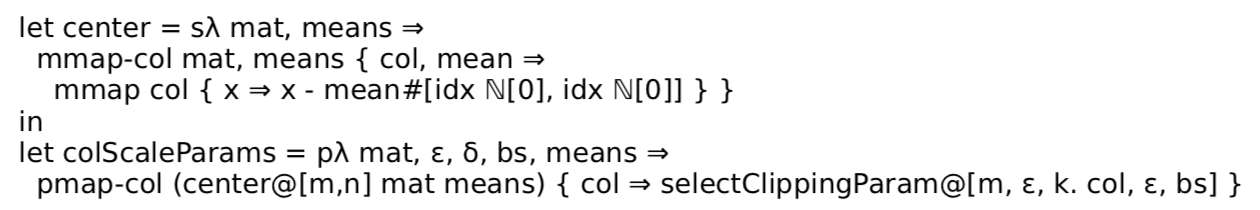}

Finally, we can define a {{\color{\colorMATH}\ensuremath{{{\color{\colorMATH}\ensuremath{\operatorname{normalize}}}}}}} function which normalizes the dataset, given the mean and scale of each column, and use this function to pre-process the data before training.

\includegraphics[width=.5\textwidth]{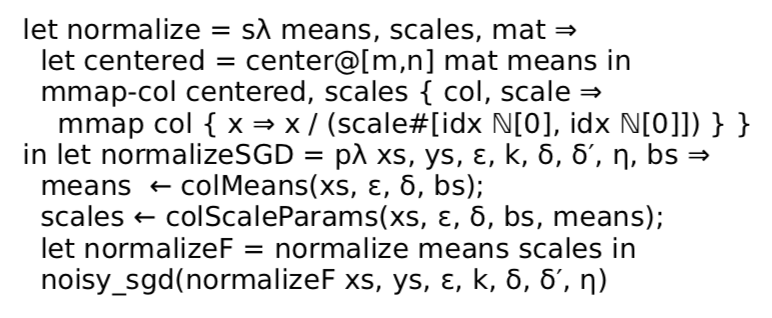}

Note that we partially apply {{\color{\colorMATH}\ensuremath{{{\color{\colorMATH}\ensuremath{\operatorname{normalize}}}}}}} to obtain {{\color{\colorMATH}\ensuremath{{{\color{\colorMATH}\ensuremath{\operatorname{normalizeF}}}}}}} because {{\color{\colorMATH}\ensuremath{{{\color{\colorMATH}\ensuremath{\operatorname{normalize}}}}}}} is not 1-sensitive in its first two arguments. Applying {{\color{\colorMATH}\ensuremath{{{\color{\colorMATH}\ensuremath{\operatorname{normalize}}}}}}} in the argument to {{\color{\colorMATH}\ensuremath{{{\color{\colorMATH}\ensuremath{\operatorname{noisySGD}}}}}}} would therefore not satisfy the requirement of {{\color{\colorMATH}\ensuremath{{\begingroup\renewcommand\colorMATH{\colorMATHP}\renewcommand\colorSYNTAX{\colorSYNTAXP}{{\color{\colorMATH}\ensuremath{p\lambda }}}\endgroup }}}} application that all arguments be 1-sensitive. Fortunately, we are not concerned about the privacy of the first two arguments to {{\color{\colorMATH}\ensuremath{{{\color{\colorMATH}\ensuremath{\operatorname{normalize}}}}}}} (they are already differentially private), so we partially apply the function and {{\color{\colorMATH}\ensuremath{{\begingroup\renewcommand\colorMATH{\colorMATHP}\renewcommand\colorSYNTAX{\colorSYNTAXP}{{\color{\colorMATH}\ensuremath{{{\color{\colorSYNTAX}\texttt{return}}}}}}\endgroup }}}} the result. \system derives a total privacy cost of {{\color{\colorMATH}\ensuremath{(3n\epsilon +2\epsilon \sqrt {2k\log (1/\delta ^{\prime})},k\delta +n\delta +\delta ^{\prime})}}}-differential privacy for {{\color{\colorMATH}\ensuremath{xs}}}, and {{\color{\colorMATH}\ensuremath{(2\epsilon \sqrt {2k\log (1/\delta ^{\prime})},k\delta +\delta ^{\prime})}}}-differential privacy for {{\color{\colorMATH}\ensuremath{ys}}} (since the labels do not participate in normalization).

\section{Full Type System and Formalism}
\label{sec:appendix_types}

Figure~\ref{fig:apx:types-kinds} shows the full type system. It includes a
type-level language of non-negative real-valued symbolic expressions {{\color{\colorMATH}\ensuremath{\eta }}}, in
particular including type-level variables {{\color{\colorMATH}\ensuremath{\beta }}}. Type-level variables {{\color{\colorMATH}\ensuremath{\beta }}}
(ranging over non-negative reals, not types) are quantified in privacy function
types. The function and application expression forms for privacy lambda
explicitly introduce and instantiate these quantified type-level
variables---they are not inferred through unification in our implementation.

Figure~\ref{fig:apx:type-metafunctions} shows type-level metafunctions used in
the typing rules. Each of the operations are over types, and return a triple
which encodes the resulting sensitivity ``cost'' of the operation in the
sensitivity language. {E.g.}, {{\color{\colorMATH}\ensuremath{{\begingroup\renewcommand\colorMATH{\colorMATHS}\renewcommand\colorSYNTAX{\colorSYNTAXS}{{\color{\colorMATH}\ensuremath{s_{1}}}}\endgroup },{\begingroup\renewcommand\colorMATH{\colorMATHS}\renewcommand\colorSYNTAX{\colorSYNTAXS}{{\color{\colorMATH}\ensuremath{s_{2}}}}\endgroup },\tau  = \tau _{1} + \tau _{2}}}} means that when you
add an expression at type {{\color{\colorMATH}\ensuremath{\tau _{1}}}} to an expression at type {{\color{\colorMATH}\ensuremath{\tau _{2}}}}, the resulting
type is {{\color{\colorMATH}\ensuremath{\tau _{3}}}}, the sensitivity of the left argument should be scaled by {\begingroup\renewcommand\colorMATH{\colorMATHS}\renewcommand\colorSYNTAX{\colorSYNTAXS}{{\color{\colorMATH}\ensuremath{s_{1}}}}\endgroup },
and the sensitivity of the second argument should be scaled by {\begingroup\renewcommand\colorMATH{\colorMATHS}\renewcommand\colorSYNTAX{\colorSYNTAXS}{{\color{\colorMATH}\ensuremath{s_{2}}}}\endgroup }. The mod
rules is special in that the sensitivities returned are interpreted as max
bounds on the arguments, not scaling factors. The main purpose of these
metafunctions are to support a form of ad-hoc polymorphism in the type system.
{E.g.}, adding two naturals returns a natural, adding two reals returns a
real, and adding two statically known numbers returns another statically known
number---known to be the sum of the arguments.

Rules for multiplication and mod are particularly important as a non-infinity
sensitivity can be given the left operand if the right operand is known
statically. Multiplication of a non-statically known value to the inverse of a
statically known value is commonly used in our examples to scale a result,
resulting in a lower scaled sensitivity cost.

Figure~\ref{fig:apx:syntax} shows the full language definition of sensitivity
and privacy languages. From this core language we derive helpers and typecheck
all case studies mentioned in the main body of the paper. Included in the
sensitivity language are standard linear logic connectives, as described by
Fuzz~\cite{reed2010distance}. We also include basic language features like
conditionals, and loops---both with static and non-static loop bounds.
Note that privacy lambdas include an explicit list of quantified type-level
variables {{\color{\colorMATH}\ensuremath{\beta }}} which range over non-negative real-valued expressions.
{\begingroup\renewcommand\colorMATH{\colorMATHS}\renewcommand\colorSYNTAX{\colorSYNTAXS}{{\color{\colorMATH}\ensuremath{{{\color{\colorSYNTAX}\texttt{conv}}}}}}\endgroup } is a conversion operation that matrices containing data clipped to
some norm {{\color{\colorMATH}\ensuremath{\ell }}} to matrices of real numbers, but with sensitivity {{\color{\colorMATH}\ensuremath{\ell }}}.

Figure~\ref{fig:apx:sens-priv-meta} shows arithmetic metafunctions for
sensitivities and privacy annotations. Note that privacy annotations support
{\begingroup\renewcommand\colorMATH{\colorMATHP}\renewcommand\colorSYNTAX{\colorSYNTAXP}{{\color{\colorMATH}\ensuremath{+}}}\endgroup } but not {\begingroup\renewcommand\colorMATH{\colorMATHP}\renewcommand\colorSYNTAX{\colorSYNTAXP}{{\color{\colorMATH}\ensuremath{\mathrel{\mathord{\cdotp }}}}}\endgroup }. The ``ceiling'' operator is defined in this figure in four
forms: sensitivity-to-sensitivity, privacy-to-sensitivity, privacy-to-privacy
and sensitivity-to-privacy. We write {{\color{\colorMATH}\ensuremath{\approx }}} instead of {{\color{\colorMATH}\ensuremath{=}}} because our
implementation uses an incomplete (but well-defined) decision procedure for
deciding when two real-valued expressions are equal.

Figure~\ref{fig:apx:kinding} shows the kinding system which establishes
well-formedness of type variables, symbolic non-negative real-valued
expressions, and types. The context {{\color{\colorMATH}\ensuremath{\Delta }}} is used to track kinds of type-level
variables,  which are either {{\color{\colorSYNTAX}\texttt{\ensuremath{{\mathbb{N}}}}}} or {{\color{\colorSYNTAX}\texttt{\ensuremath{{\mathbb{R}}^{+}}}}}. We implement a solver for
determining when two symbolic real expressions are equal or one is
less-than-or-equal to another, and we use the type information in the solver,
in addition to merely establishing type well-formedness.

Figure~\ref{fig:apx:sens-typing-lits-arith} shows the typing rules for the
fragment of the sensitivity language that concerns literals and operations over
basic primitive types. These typing rules use type-level metafunctions defined
in Figure~\ref{fig:apx:type-metafunctions}.

Figure~\ref{fig:apx:sens-typing-matrices} shows the typing rules for matrix
operations, including clipping, gradients, and aggregate operations like maps
and folds. We include both unary and binary variants of two different types of
maps---element-wise and row-wise.

Figure~\ref{fig:apx:sens-typing-connectives} shows the typing rules for
control flow---{e.g.}, loops and ifs---and compound types in the sensitivity
language. Compound types include sums, multiplicative products, additive
products, sensitivity functions and privacy functions.

Figure~\ref{fig:apx:priv-typing} shows the typing rules for the entire privacy
language. The running theme in each of these rules is to clip contexts up to
infinity in the conclusion of rules where no privacy guarantee can be made, and
to assume a clipped context up to some fixed value in the assumption of rules
where the privacy guarantee requires a bound on sensitivity or privacy on an
argument, written in curly brackets. Note that many rules take an
explicit list of variables to be considered for the purpose of the assumed
bound, written in ascii angle brackets.

Figures~\ref{fig:apx:metrics-1} and~\ref{fig:apx:metrics-2} define the metric
spaces used to interpret types. We call the model for types a ``domain'', because
we extend the usual notion of metric space with an additional operation which
captures how to interpret the norm of values in the underlying set. The
distance metric and the norm metric may not be related, and their
interpretations do not inter-depend on each other in the semantics.

Figure~\ref{fig:apx:typing-semantics} shows the semantics of types and typing
judgments in terms of domains, metrics and norms defined in
Figures~\ref{fig:apx:metrics-1} and~\ref{fig:apx:metrics-2}. Our soundness
result is that for well-typed terms there exists an interpretation for terms
which inhabits the interpretation of a typing context.

Each of the figures mentioned above is displayed next. After these figures, we
present key theorems from the literature which are used in our proof of type
soundness, after which we present key cases of the proof. After presenting key
proof cases, we re-presented rules which change when adapting the type system
to R\'enyi differential privacy, zero-concentrated differential privacy, and
truncated-concentrated differential privacy.

\FloatBarrier

\begin{figure*}
\begingroup\renewcommand\colorMATH{\colorMATHM}\renewcommand\colorSYNTAX{\colorSYNTAXM}
\vspace*{-0.25em}\begingroup\color{\colorMATH}\begin{gather*}

\vspace*{-1em}\end{gather*}\endgroup 
\endgroup 
\caption{Types and Kinds (shared)}
  \label{fig:apx:types-kinds}
\end{figure*}

\begin{figure*}
\begingroup\renewcommand\colorMATH{\colorMATHM}\renewcommand\colorSYNTAX{\colorSYNTAXM}
\begingroup\color{\colorMATH}\begin{mathpar}\begin{array}{rclclclcl
  } \underline{\hspace{0.66em}\vspace*{5ex}}&{}\sqcup {}&\underline{\hspace{0.66em}\vspace*{5ex}} &{}\in {}& \multicolumn{5}{l}{\tau  \times  \tau  \rightharpoonup  {\begingroup\renewcommand\colorMATH{\colorMATHS}\renewcommand\colorSYNTAX{\colorSYNTAXS}{{\color{\colorMATH}\ensuremath{{{\color{\colorMATH}\ensuremath{\operatorname{sens}}}}}}}\endgroup } \times  {\begingroup\renewcommand\colorMATH{\colorMATHS}\renewcommand\colorSYNTAX{\colorSYNTAXS}{{\color{\colorMATH}\ensuremath{{{\color{\colorMATH}\ensuremath{\operatorname{sens}}}}}}}\endgroup } \times  \tau }
  \cr  {{\color{\colorSYNTAX}\texttt{\ensuremath{{\mathbb{N}}[{{\color{\colorMATH}\ensuremath{\eta _{1}}}}]}}}} &{}\sqcup {}&{{\color{\colorSYNTAX}\texttt{\ensuremath{{\mathbb{N}}[{{\color{\colorMATH}\ensuremath{\eta _{2}}}}]}}}}  &{}\triangleq {}& \langle 0&{},{}&0&{},{}&{{\color{\colorSYNTAX}\texttt{\ensuremath{{\mathbb{N}}[{{\color{\colorMATH}\ensuremath{\eta _{1}}}}\sqcup {{\color{\colorMATH}\ensuremath{\eta _{2}}}}]}}}} \rangle 
  \cr  {{\color{\colorSYNTAX}\texttt{\ensuremath{{\mathbb{R}}^{+}[{{\color{\colorMATH}\ensuremath{\eta _{1}}}}]}}}}&{}\sqcup {}&{{\color{\colorSYNTAX}\texttt{\ensuremath{{\mathbb{R}}^{+}[{{\color{\colorMATH}\ensuremath{\eta _{2}}}}]}}}} &{}\triangleq {}& \langle 0&{},{}&0&{},{}&{{\color{\colorSYNTAX}\texttt{\ensuremath{{\mathbb{R}}^{+}[{{\color{\colorMATH}\ensuremath{\eta _{1}}}}\sqcup {{\color{\colorMATH}\ensuremath{\eta _{2}}}}]}}}}\rangle 
  \cr  {{\color{\colorSYNTAX}\texttt{\ensuremath{{\mathbb{N}}}}}}       &{}\sqcup {}&{{\color{\colorSYNTAX}\texttt{\ensuremath{{\mathbb{N}}}}}}        &{}\triangleq {}& \langle 1&{},{}&1&{},{}&{{\color{\colorSYNTAX}\texttt{\ensuremath{{\mathbb{N}}}}}}            \rangle 
  \cr
\end{array}
\and\begin{array}{rclclclcl
  } \underline{\hspace{0.66em}\vspace*{5ex}}&{}\sqcap {}&\underline{\hspace{0.66em}\vspace*{5ex}} &{}\in {}& \multicolumn{5}{l}{\tau  \times  \tau  \rightharpoonup  {\begingroup\renewcommand\colorMATH{\colorMATHS}\renewcommand\colorSYNTAX{\colorSYNTAXS}{{\color{\colorMATH}\ensuremath{{{\color{\colorMATH}\ensuremath{\operatorname{sens}}}}}}}\endgroup } \times  {\begingroup\renewcommand\colorMATH{\colorMATHS}\renewcommand\colorSYNTAX{\colorSYNTAXS}{{\color{\colorMATH}\ensuremath{{{\color{\colorMATH}\ensuremath{\operatorname{sens}}}}}}}\endgroup } \times  \tau }
  \cr  {{\color{\colorSYNTAX}\texttt{\ensuremath{{\mathbb{N}}[{{\color{\colorMATH}\ensuremath{\eta _{1}}}}]}}}} &{}\sqcap {}&{{\color{\colorSYNTAX}\texttt{\ensuremath{{\mathbb{N}}[{{\color{\colorMATH}\ensuremath{\eta _{2}}}}]}}}}  &{}\triangleq {}& \langle 0&{},{}&0&{},{}&{{\color{\colorSYNTAX}\texttt{\ensuremath{{\mathbb{N}}[{{\color{\colorMATH}\ensuremath{\eta _{1}}}}\sqcap {{\color{\colorMATH}\ensuremath{\eta _{2}}}}]}}}} \rangle 
  \cr  {{\color{\colorSYNTAX}\texttt{\ensuremath{{\mathbb{R}}^{+}[{{\color{\colorMATH}\ensuremath{\eta _{1}}}}]}}}}&{}\sqcap {}&{{\color{\colorSYNTAX}\texttt{\ensuremath{{\mathbb{R}}^{+}[{{\color{\colorMATH}\ensuremath{\eta _{2}}}}]}}}} &{}\triangleq {}& \langle 0&{},{}&0&{},{}&{{\color{\colorSYNTAX}\texttt{\ensuremath{{\mathbb{R}}^{+}[{{\color{\colorMATH}\ensuremath{\eta _{1}}}}\sqcap {{\color{\colorMATH}\ensuremath{\eta _{2}}}}]}}}}\rangle 
  \cr  {{\color{\colorSYNTAX}\texttt{\ensuremath{{\mathbb{N}}}}}}       &{}\sqcap {}&{{\color{\colorSYNTAX}\texttt{\ensuremath{{\mathbb{N}}}}}}        &{}\triangleq {}& \langle 1&{},{}&1&{},{}&{{\color{\colorSYNTAX}\texttt{\ensuremath{{\mathbb{N}}}}}}            \rangle 
  \cr
\end{array}
\and\begin{array}{rclclclcl
  } \underline{\hspace{0.66em}\vspace*{5ex}}&{}+{}&\underline{\hspace{0.66em}\vspace*{5ex}} &{}\in {}& \multicolumn{5}{l}{\tau  \times  \tau  \rightharpoonup  {\begingroup\renewcommand\colorMATH{\colorMATHS}\renewcommand\colorSYNTAX{\colorSYNTAXS}{{\color{\colorMATH}\ensuremath{{{\color{\colorMATH}\ensuremath{\operatorname{sens}}}}}}}\endgroup } \times  {\begingroup\renewcommand\colorMATH{\colorMATHS}\renewcommand\colorSYNTAX{\colorSYNTAXS}{{\color{\colorMATH}\ensuremath{{{\color{\colorMATH}\ensuremath{\operatorname{sens}}}}}}}\endgroup } \times  \tau }
  \cr  {{\color{\colorSYNTAX}\texttt{\ensuremath{{\mathbb{N}}[{{\color{\colorMATH}\ensuremath{\eta _{1}}}}]}}}} &{}+{}&{{\color{\colorSYNTAX}\texttt{\ensuremath{{\mathbb{N}}[{{\color{\colorMATH}\ensuremath{\eta _{2}}}}]}}}}  &{}\triangleq {}& \langle 0&{},{}&0&{},{}&{{\color{\colorSYNTAX}\texttt{\ensuremath{{\mathbb{N}}[{{\color{\colorMATH}\ensuremath{\eta _{1}}}}+{{\color{\colorMATH}\ensuremath{\eta _{2}}}}]}}}} \rangle 
  \cr  {{\color{\colorSYNTAX}\texttt{\ensuremath{{\mathbb{R}}^{+}[{{\color{\colorMATH}\ensuremath{\eta _{1}}}}]}}}}&{}+{}&{{\color{\colorSYNTAX}\texttt{\ensuremath{{\mathbb{R}}^{+}[{{\color{\colorMATH}\ensuremath{\eta _{2}}}}]}}}} &{}\triangleq {}& \langle 0&{},{}&0&{},{}&{{\color{\colorSYNTAX}\texttt{\ensuremath{{\mathbb{R}}^{+}[{{\color{\colorMATH}\ensuremath{\eta _{1}}}}+{{\color{\colorMATH}\ensuremath{\eta _{2}}}}]}}}}\rangle 
  \cr  {{\color{\colorSYNTAX}\texttt{\ensuremath{{\mathbb{N}}}}}}       &{}+{}&{{\color{\colorSYNTAX}\texttt{\ensuremath{{\mathbb{N}}}}}}        &{}\triangleq {}& \langle 1&{},{}&1&{},{}&{{\color{\colorSYNTAX}\texttt{\ensuremath{{\mathbb{N}}}}}}            \rangle 
  \cr  {{\color{\colorSYNTAX}\texttt{\ensuremath{{\mathbb{R}}}}}}       &{}+{}&{{\color{\colorSYNTAX}\texttt{\ensuremath{{\mathbb{R}}}}}}        &{}\triangleq {}& \langle 1&{},{}&1&{},{}&{{\color{\colorSYNTAX}\texttt{\ensuremath{{\mathbb{R}}}}}}            \rangle 
  \end{array}
\and\begin{array}{rclclclcl
  } \underline{\hspace{0.66em}\vspace*{5ex}}&{}\mathrel{\mathord{\cdotp }}{}&\underline{\hspace{0.66em}\vspace*{5ex}} &{}\in {}& \multicolumn{5}{l}{\tau  \times  \tau  \rightharpoonup  {\begingroup\renewcommand\colorMATH{\colorMATHS}\renewcommand\colorSYNTAX{\colorSYNTAXS}{{\color{\colorMATH}\ensuremath{{{\color{\colorMATH}\ensuremath{\operatorname{sens}}}}}}}\endgroup } \times  {\begingroup\renewcommand\colorMATH{\colorMATHS}\renewcommand\colorSYNTAX{\colorSYNTAXS}{{\color{\colorMATH}\ensuremath{{{\color{\colorMATH}\ensuremath{\operatorname{sens}}}}}}}\endgroup } \times  \tau }
  \cr  {{\color{\colorSYNTAX}\texttt{\ensuremath{{\mathbb{N}}[{{\color{\colorMATH}\ensuremath{\eta _{1}}}}]}}}}  &{}\mathrel{\mathord{\cdotp }}{}&{{\color{\colorSYNTAX}\texttt{\ensuremath{{\mathbb{N}}[{{\color{\colorMATH}\ensuremath{\eta _{2}}}}]}}}}  &{}\triangleq {}& \langle 0        &{},{}&0        &{},{}&{{\color{\colorSYNTAX}\texttt{\ensuremath{{\mathbb{N}}[{{\color{\colorMATH}\ensuremath{\eta _{1}}}}\mathrel{\mathord{\cdotp }}{{\color{\colorMATH}\ensuremath{\eta _{2}}}}]}}}} \rangle 
  \cr  {{\color{\colorSYNTAX}\texttt{\ensuremath{{\mathbb{R}}^{+}[{{\color{\colorMATH}\ensuremath{\eta _{1}}}}]}}}} &{}\mathrel{\mathord{\cdotp }}{}&{{\color{\colorSYNTAX}\texttt{\ensuremath{{\mathbb{R}}^{+}[{{\color{\colorMATH}\ensuremath{\eta _{2}}}}]}}}} &{}\triangleq {}& \langle 0        &{},{}&0        &{},{}&{{\color{\colorSYNTAX}\texttt{\ensuremath{{\mathbb{R}}^{+}[{{\color{\colorMATH}\ensuremath{\eta _{1}}}}\mathrel{\mathord{\cdotp }}{{\color{\colorMATH}\ensuremath{\eta _{2}}}}]}}}}\rangle 
  \cr  {{\color{\colorSYNTAX}\texttt{\ensuremath{{\mathbb{N}}[{{\color{\colorMATH}\ensuremath{\eta _{1}}}}]}}}}  &{}\mathrel{\mathord{\cdotp }}{}&{{\color{\colorSYNTAX}\texttt{\ensuremath{{\mathbb{N}}}}}}        &{}\triangleq {}& \langle 0        &{},{}&\eta _{1}       &{},{}&{{\color{\colorSYNTAX}\texttt{\ensuremath{{\mathbb{N}}}}}}            \rangle 
  \cr
\end{array}
\and\begin{array}{rclcl
  } 1/\underline{\hspace{0.66em}\vspace*{5ex}}           &{}\in {}& \multicolumn{3}{l}{\tau  \rightharpoonup  {\begingroup\renewcommand\colorMATH{\colorMATHS}\renewcommand\colorSYNTAX{\colorSYNTAXS}{{\color{\colorMATH}\ensuremath{{{\color{\colorMATH}\ensuremath{\operatorname{sens}}}}}}}\endgroup } \times  \tau }
  \cr  1/{{\color{\colorSYNTAX}\texttt{\ensuremath{{\mathbb{R}}^{+}[{{\color{\colorMATH}\ensuremath{\eta }}}]}}}} &{}\triangleq {}& \langle 0        &{},{}&{{\color{\colorSYNTAX}\texttt{\ensuremath{{\mathbb{R}}^{+}[1/{{\color{\colorMATH}\ensuremath{\eta }}}]}}}}\rangle 
  \cr  1/{{\color{\colorSYNTAX}\texttt{\ensuremath{{\mathbb{R}}}}}}       &{}\triangleq {}& \langle {\begingroup\renewcommand\colorMATH{\colorMATHS}\renewcommand\colorSYNTAX{\colorSYNTAXS}{{\color{\colorMATH}\ensuremath{{{\color{\colorSYNTAX}\texttt{\ensuremath{\infty }}}}}}}\endgroup }&{},{}&{{\color{\colorSYNTAX}\texttt{\ensuremath{{\mathbb{R}}}}}}        \rangle 
  \end{array}
\and\begin{array}{rclcl
  } \sqrt {\underline{\hspace{0.66em}\vspace*{5ex}}}           &{}\in {}& \multicolumn{3}{l}{\tau  \rightharpoonup  {\begingroup\renewcommand\colorMATH{\colorMATHS}\renewcommand\colorSYNTAX{\colorSYNTAXS}{{\color{\colorMATH}\ensuremath{{{\color{\colorMATH}\ensuremath{\operatorname{sens}}}}}}}\endgroup } \times  \tau }
  \cr  \sqrt {{{\color{\colorSYNTAX}\texttt{\ensuremath{{\mathbb{R}}^{+}[{{\color{\colorMATH}\ensuremath{\eta }}}]}}}}} &{}\triangleq {}& \langle 0        &{},{}&{{\color{\colorSYNTAX}\texttt{\ensuremath{{\mathbb{R}}^{+}[\sqrt {{{\color{\colorMATH}\ensuremath{\eta }}}}]}}}}\rangle 
  \cr  \sqrt {{{\color{\colorSYNTAX}\texttt{\ensuremath{{\mathbb{R}}}}}}}       &{}\triangleq {}& \langle {\begingroup\renewcommand\colorMATH{\colorMATHS}\renewcommand\colorSYNTAX{\colorSYNTAXS}{{\color{\colorMATH}\ensuremath{{{\color{\colorSYNTAX}\texttt{\ensuremath{\infty }}}}}}}\endgroup }&{},{}&{{\color{\colorSYNTAX}\texttt{\ensuremath{{\mathbb{R}}}}}}         \rangle 
  \end{array}
\and\begin{array}{rclcl
  } \ln \underline{\hspace{0.66em}\vspace*{5ex}}           &{}\in {}& \multicolumn{3}{l}{\tau  \rightharpoonup  {\begingroup\renewcommand\colorMATH{\colorMATHS}\renewcommand\colorSYNTAX{\colorSYNTAXS}{{\color{\colorMATH}\ensuremath{{{\color{\colorMATH}\ensuremath{\operatorname{sens}}}}}}}\endgroup } \times  \tau }
  \cr  \ln {{\color{\colorSYNTAX}\texttt{\ensuremath{{\mathbb{R}}^{+}[{{\color{\colorMATH}\ensuremath{\eta }}}]}}}} &{}\triangleq {}& \langle 0        &{},{}&{{\color{\colorSYNTAX}\texttt{\ensuremath{{\mathbb{R}}^{+}[\ln {{\color{\colorMATH}\ensuremath{\eta }}}]}}}}\rangle 
  \cr  \ln {{\color{\colorSYNTAX}\texttt{\ensuremath{{\mathbb{R}}}}}}       &{}\triangleq {}& \langle {\begingroup\renewcommand\colorMATH{\colorMATHS}\renewcommand\colorSYNTAX{\colorSYNTAXS}{{\color{\colorMATH}\ensuremath{{{\color{\colorSYNTAX}\texttt{\ensuremath{\infty }}}}}}}\endgroup }&{},{}&{{\color{\colorSYNTAX}\texttt{\ensuremath{{\mathbb{R}}}}}}        \rangle 
  \end{array}
\and\begin{array}{rclclclcl
  } \underline{\hspace{0.66em}\vspace*{5ex}}&{}\mathrel{{{\color{\colorMATH}\ensuremath{\operatorname{mod}}}}}{}&\underline{\hspace{0.66em}\vspace*{5ex}} &{}\in {}& \multicolumn{5}{l}{\tau  \times  \tau  \rightharpoonup  {\begingroup\renewcommand\colorMATH{\colorMATHS}\renewcommand\colorSYNTAX{\colorSYNTAXS}{{\color{\colorMATH}\ensuremath{{{\color{\colorMATH}\ensuremath{\operatorname{sens}}}}}}}\endgroup } \times  {\begingroup\renewcommand\colorMATH{\colorMATHS}\renewcommand\colorSYNTAX{\colorSYNTAXS}{{\color{\colorMATH}\ensuremath{{{\color{\colorMATH}\ensuremath{\operatorname{sens}}}}}}}\endgroup } \times  \tau }
  \cr  {{\color{\colorSYNTAX}\texttt{\ensuremath{{\mathbb{N}}}}}}           &{}\mathrel{{{\color{\colorMATH}\ensuremath{\operatorname{mod}}}}}{}&{{\color{\colorSYNTAX}\texttt{\ensuremath{{\mathbb{N}}[{{\color{\colorMATH}\ensuremath{\eta _{2}}}}]}}}}      &{}\triangleq {}& \langle \eta _{2}       &{},{}&0        &{},{}&{{\color{\colorSYNTAX}\texttt{\ensuremath{{\mathbb{N}}}}}}\rangle 
  \cr  {{\color{\colorSYNTAX}\texttt{\ensuremath{{{\color{\colorSYNTAX}\texttt{idx}}}[{{\color{\colorMATH}\ensuremath{\eta _{1}}}}]}}}} &{}\mathrel{{{\color{\colorMATH}\ensuremath{\operatorname{mod}}}}}{}&{{\color{\colorSYNTAX}\texttt{\ensuremath{{{\color{\colorSYNTAX}\texttt{idx}}}[{{\color{\colorMATH}\ensuremath{\eta _{2}}}}]}}}}  &{}\triangleq {}& \langle \eta _{2}       &{},{}&\eta _{1}       &{},{}&{{\color{\colorSYNTAX}\texttt{\ensuremath{{{\color{\colorSYNTAX}\texttt{idx}}}[{{\color{\colorMATH}\ensuremath{\eta _{1}}}}\sqcap {{\color{\colorMATH}\ensuremath{\eta _{2}}}}]}}}}\rangle 
  \cr
\end{array}
\and\begin{array}{rclclclcl
  } \underline{\hspace{0.66em}\vspace*{5ex}}&{}-{}&\underline{\hspace{0.66em}\vspace*{5ex}} &{}\in {}& \multicolumn{5}{l}{\tau  \times  \tau  \rightharpoonup  {\begingroup\renewcommand\colorMATH{\colorMATHS}\renewcommand\colorSYNTAX{\colorSYNTAXS}{{\color{\colorMATH}\ensuremath{{{\color{\colorMATH}\ensuremath{\operatorname{sens}}}}}}}\endgroup } \times  {\begingroup\renewcommand\colorMATH{\colorMATHS}\renewcommand\colorSYNTAX{\colorSYNTAXS}{{\color{\colorMATH}\ensuremath{{{\color{\colorMATH}\ensuremath{\operatorname{sens}}}}}}}\endgroup } \times  \tau }
  \cr  {{\color{\colorSYNTAX}\texttt{\ensuremath{{\mathbb{N}}}}}} &{}-{}& {{\color{\colorSYNTAX}\texttt{\ensuremath{{\mathbb{N}}}}}} &{}\triangleq {}& \langle 1&{},{}&1&{},{}&{{\color{\colorSYNTAX}\texttt{\ensuremath{{\mathbb{N}}}}}}\rangle 
  \cr  {{\color{\colorSYNTAX}\texttt{\ensuremath{{\mathbb{R}}}}}} &{}-{}& {{\color{\colorSYNTAX}\texttt{\ensuremath{{\mathbb{R}}}}}} &{}\triangleq {}& \langle 1&{},{}&1&{},{}&{{\color{\colorSYNTAX}\texttt{\ensuremath{{\mathbb{R}}}}}}\rangle 
  \end{array}
\and\begin{array}{rclclclcl
  } \underline{\hspace{0.66em}\vspace*{5ex}}&{}\mathrel{\overset{?}{\smash{=}}} {}&\underline{\hspace{0.66em}\vspace*{5ex}} &{}\in {}& \multicolumn{5}{l}{\tau  \times  \tau  \rightharpoonup  {\begingroup\renewcommand\colorMATH{\colorMATHS}\renewcommand\colorSYNTAX{\colorSYNTAXS}{{\color{\colorMATH}\ensuremath{{{\color{\colorMATH}\ensuremath{\operatorname{sens}}}}}}}\endgroup } \times  {\begingroup\renewcommand\colorMATH{\colorMATHS}\renewcommand\colorSYNTAX{\colorSYNTAXS}{{\color{\colorMATH}\ensuremath{{{\color{\colorMATH}\ensuremath{\operatorname{sens}}}}}}}\endgroup } \times  \tau }
  \cr  {{\color{\colorSYNTAX}\texttt{\ensuremath{{\mathbb{N}}}}}} &{}\mathrel{\overset{?}{\smash{=}}} {}& {{\color{\colorSYNTAX}\texttt{\ensuremath{{\mathbb{N}}}}}} &{}\triangleq {}& \langle 1&{},{}&1&{},{}&{{\color{\colorSYNTAX}\texttt{\ensuremath{{\mathbb{N}}}}}}\rangle 
  \cr  {{\color{\colorSYNTAX}\texttt{\ensuremath{{\mathbb{R}}}}}} &{}\mathrel{\overset{?}{\smash{=}}} {}& {{\color{\colorSYNTAX}\texttt{\ensuremath{{\mathbb{R}}}}}} &{}\triangleq {}& \langle 1&{},{}&1&{},{}&{{\color{\colorSYNTAX}\texttt{\ensuremath{{\mathbb{N}}}}}}\rangle 
  \end{array}
\end{mathpar}\endgroup 
\endgroup 
\caption{Type Metafunctions}
\label{fig:apx:type-metafunctions}
\end{figure*}

\begin{figure*}
\vspace*{-0.25em}\begingroup\color{\colorMATH}\begin{gather*}%
\vspace*{-1em}\end{gather*}\endgroup 
\caption{Sensitivity and Privacy Languages}
  \label{fig:apx:syntax}
\end{figure*}

\begin{figure*}
\begingroup\color{\colorMATH}\begin{mathpar}\begingroup\renewcommand\colorMATH{\colorMATHS}\renewcommand\colorSYNTAX{\colorSYNTAXS}\begingroup\color{\colorMATH}
  %
\right.
  \cr  (\Gamma _{1}\mathrel{\mathord{\cdotp }}\Gamma _{2})({\begingroup\renewcommand\colorMATH{\colorMATHM}\renewcommand\colorSYNTAX{\colorSYNTAXM}{{\color{\colorMATH}\ensuremath{x}}}\endgroup }) &{}\triangleq {}& {\begingroup\renewcommand\colorMATH{\colorMATHM}\renewcommand\colorSYNTAX{\colorSYNTAXM}{{\color{\colorMATH}\ensuremath{{{\color{\colorSYNTAX}\texttt{\ensuremath{{\begingroup\renewcommand\colorMATH{\colorMATHS}\renewcommand\colorSYNTAX{\colorSYNTAXS}{{\color{\colorMATH}\ensuremath{\Gamma _{1}({\begingroup\renewcommand\colorMATH{\colorMATHM}\renewcommand\colorSYNTAX{\colorSYNTAXM}{{\color{\colorMATH}\ensuremath{x}}}\endgroup })}}}\endgroup }\mathrel{\mathord{\cdotp }}{\begingroup\renewcommand\colorMATH{\colorMATHS}\renewcommand\colorSYNTAX{\colorSYNTAXS}{{\color{\colorMATH}\ensuremath{\Gamma _{2}({\begingroup\renewcommand\colorMATH{\colorMATHM}\renewcommand\colorSYNTAX{\colorSYNTAXM}{{\color{\colorMATH}\ensuremath{x}}}\endgroup })}}}\endgroup }}}}}}}}\endgroup }
  \end{array}
  \endgroup \endgroup 
\and\begingroup\renewcommand\colorMATH{\colorMATHP}\renewcommand\colorSYNTAX{\colorSYNTAXP}\begingroup\color{\colorMATH}
\right.
  \cr  {}\rceil {{\color{\colorSYNTAX}\texttt{\ensuremath{\infty }}}}\lceil {}^{s} &{}\triangleq {}& s
  \cr  {}\rceil \Gamma \lceil {}^{s}({\begingroup\renewcommand\colorMATH{\colorMATHM}\renewcommand\colorSYNTAX{\colorSYNTAXM}{{\color{\colorMATH}\ensuremath{x}}}\endgroup }) &{}\triangleq {}& {}\rceil \Gamma ({\begingroup\renewcommand\colorMATH{\colorMATHM}\renewcommand\colorSYNTAX{\colorSYNTAXM}{{\color{\colorMATH}\ensuremath{x}}}\endgroup })\lceil {}^{s}
  \end{array}
  \endgroup \endgroup 
\and\begingroup\renewcommand\colorMATH{\colorMATHS}\renewcommand\colorSYNTAX{\colorSYNTAXS}\begingroup\color{\colorMATH}
\right.
  \cr  {}\rceil {\begingroup\renewcommand\colorMATH{\colorMATHP}\renewcommand\colorSYNTAX{\colorSYNTAXP}{{\color{\colorMATH}\ensuremath{{{\color{\colorSYNTAX}\texttt{\ensuremath{\infty }}}}}}}\endgroup }\lceil {}^{s} &{}\triangleq {}& s
  \cr  {}\rceil {\begingroup\renewcommand\colorMATH{\colorMATHP}\renewcommand\colorSYNTAX{\colorSYNTAXP}{{\color{\colorMATH}\ensuremath{\Gamma }}}\endgroup }\lceil {}^{s}({\begingroup\renewcommand\colorMATH{\colorMATHM}\renewcommand\colorSYNTAX{\colorSYNTAXM}{{\color{\colorMATH}\ensuremath{x}}}\endgroup }) &{}\triangleq {}& {}\rceil {\begingroup\renewcommand\colorMATH{\colorMATHP}\renewcommand\colorSYNTAX{\colorSYNTAXP}{{\color{\colorMATH}\ensuremath{\Gamma ({\begingroup\renewcommand\colorMATH{\colorMATHM}\renewcommand\colorSYNTAX{\colorSYNTAXM}{{\color{\colorMATH}\ensuremath{x}}}\endgroup })}}}\endgroup }\lceil {}^{s}
  \end{array}
  \endgroup \endgroup 
\\\begingroup\renewcommand\colorMATH{\colorMATHP}\renewcommand\colorSYNTAX{\colorSYNTAXP}\begingroup\color{\colorMATH}
\right.
  \cr  {}\rceil {{\color{\colorSYNTAX}\texttt{\ensuremath{\infty }}}}\lceil {}^{p} &{}\triangleq {}& p
  \cr  {}\rceil \Gamma \lceil {}^{p}({\begingroup\renewcommand\colorMATH{\colorMATHM}\renewcommand\colorSYNTAX{\colorSYNTAXM}{{\color{\colorMATH}\ensuremath{x}}}\endgroup }) &{}\triangleq {}& {}\rceil \Gamma ({\begingroup\renewcommand\colorMATH{\colorMATHM}\renewcommand\colorSYNTAX{\colorSYNTAXM}{{\color{\colorMATH}\ensuremath{x}}}\endgroup })\lceil {}^{p}
  \end{array}
  \endgroup \endgroup 
\and\begingroup\renewcommand\colorMATH{\colorMATHP}\renewcommand\colorSYNTAX{\colorSYNTAXP}\begingroup\color{\colorMATH}
\right.
  \cr  {}\rceil {\begingroup\renewcommand\colorMATH{\colorMATHS}\renewcommand\colorSYNTAX{\colorSYNTAXS}{{\color{\colorMATH}\ensuremath{{{\color{\colorSYNTAX}\texttt{\ensuremath{\infty }}}}}}}\endgroup }\lceil {}^{p} &{}\triangleq {}& p
  \cr  {}\rceil {\begingroup\renewcommand\colorMATH{\colorMATHS}\renewcommand\colorSYNTAX{\colorSYNTAXS}{{\color{\colorMATH}\ensuremath{\Gamma }}}\endgroup }\lceil {}^{p}({\begingroup\renewcommand\colorMATH{\colorMATHM}\renewcommand\colorSYNTAX{\colorSYNTAXM}{{\color{\colorMATH}\ensuremath{x}}}\endgroup }) &{}\triangleq {}& {}\rceil {\begingroup\renewcommand\colorMATH{\colorMATHS}\renewcommand\colorSYNTAX{\colorSYNTAXS}{{\color{\colorMATH}\ensuremath{\Gamma ({\begingroup\renewcommand\colorMATH{\colorMATHM}\renewcommand\colorSYNTAX{\colorSYNTAXM}{{\color{\colorMATH}\ensuremath{x}}}\endgroup })}}}\endgroup }\lceil {}^{p}
  \end{array}
  \endgroup \endgroup 
\end{mathpar}\endgroup 
\caption{Sensitivity and Privacy Metafunctions}
\label{fig:apx:sens-priv-meta}
\end{figure*}

\begin{figure*}
\vspace*{-0.25em}\begingroup\color{\colorMATH}\begin{gather*}\begin{tabularx}{\linewidth}{>{\centering\arraybackslash\(}X<{\)}}\hfill\hspace{0pt}\begingroup\color{\colorTEXT}\boxed{\begingroup\color{\colorMATH} \Delta  \vdash  \eta  \mathrel{:} \kappa  \endgroup}\endgroup \end{tabularx}\vspace*{-1em}\end{gather*}\endgroup 
\begingroup\color{\colorMATH}\begin{mathpar}\inferrule*[lab={{\color{\colorTEXT}\textsc{\scriptsize RVar}}}
  ]{ \beta {\mathrel{:}}\kappa  \in  \Delta 
     }{
     \Delta  \vdash  \beta  \mathrel{:} \kappa 
  }
\and\inferrule*[lab={{\color{\colorTEXT}\textsc{\scriptsize Nat}}}
  ]{ }{
     \Delta  \vdash  n \mathrel{:} {{\color{\colorSYNTAX}\texttt{\ensuremath{{\mathbb{N}}}}}}
  }
\and\inferrule*[lab={{\color{\colorTEXT}\textsc{\scriptsize Real}}}
  ]{ }{
     \Delta  \vdash  r^{+} \mathrel{:} {{\color{\colorSYNTAX}\texttt{\ensuremath{{\mathbb{R}}^{+}}}}}
  }
\and\inferrule*[lab={{\color{\colorTEXT}\textsc{\scriptsize Max}}}
  ]{ \Delta  \vdash  \eta _{1} \mathrel{:} \kappa 
  \\ \Delta  \vdash  \eta _{2} \mathrel{:} \kappa 
     }{
     \Delta  \vdash  {{\color{\colorSYNTAX}\texttt{\ensuremath{{{\color{\colorMATH}\ensuremath{\eta _{1}}}} \sqcup  {{\color{\colorMATH}\ensuremath{\eta _{2}}}}}}}} \mathrel{:} \kappa 
  }
\and\inferrule*[lab={{\color{\colorTEXT}\textsc{\scriptsize Min}}}
  ]{ \Delta  \vdash  \eta _{1} \mathrel{:} \kappa 
  \\ \Delta  \vdash  \eta _{2} \mathrel{:} \kappa 
     }{
     \Delta  \vdash  {{\color{\colorSYNTAX}\texttt{\ensuremath{{{\color{\colorMATH}\ensuremath{\eta _{1}}}} \sqcap  {{\color{\colorMATH}\ensuremath{\eta _{2}}}}}}}} \mathrel{:} \kappa 
  }
\and\inferrule*[lab={{\color{\colorTEXT}\textsc{\scriptsize Plus}}}
  ]{ \Delta  \vdash  \eta _{1} \mathrel{:} \kappa 
  \\ \Delta  \vdash  \eta _{2} \mathrel{:} \kappa 
     }{
     \Delta  \vdash  {{\color{\colorSYNTAX}\texttt{\ensuremath{{{\color{\colorMATH}\ensuremath{\eta _{1}}}} + {{\color{\colorMATH}\ensuremath{\eta _{2}}}}}}}} \mathrel{:} \kappa 
  }
\and\inferrule*[lab={{\color{\colorTEXT}\textsc{\scriptsize Times}}}
  ]{ \Delta  \vdash  \eta _{1} \mathrel{:} \kappa 
  \\ \Delta  \vdash  \eta _{2} \mathrel{:} \kappa 
     }{
     \Delta  \vdash  {{\color{\colorSYNTAX}\texttt{\ensuremath{{{\color{\colorMATH}\ensuremath{\eta _{1}}}} \mathrel{\mathord{\cdotp }} {{\color{\colorMATH}\ensuremath{\eta _{2}}}}}}}} \mathrel{:} \kappa 
  }
\and\inferrule*[lab={{\color{\colorTEXT}\textsc{\scriptsize Div}}}
  ]{ \Delta  \vdash  \eta _{1} \mathrel{:} {{\color{\colorSYNTAX}\texttt{\ensuremath{{\mathbb{R}}^{+}}}}}
  \\ \Delta  \vdash  \eta _{2} \mathrel{:} {{\color{\colorSYNTAX}\texttt{\ensuremath{{\mathbb{R}}^{+}}}}}
     }{
     \Delta  \vdash  {{\color{\colorSYNTAX}\texttt{\ensuremath{{{\color{\colorMATH}\ensuremath{\eta _{1}}}} / {{\color{\colorMATH}\ensuremath{\eta _{2}}}}}}}} \mathrel{:} {{\color{\colorSYNTAX}\texttt{\ensuremath{{\mathbb{R}}^{+}}}}}
  }
\and\inferrule*[lab={{\color{\colorTEXT}\textsc{\scriptsize Root}}}
  ]{ \Delta  \vdash  \eta  \mathrel{:} {{\color{\colorSYNTAX}\texttt{\ensuremath{{\mathbb{R}}^{+}}}}}
     }{
     \Delta  \vdash  {{\color{\colorSYNTAX}\texttt{\ensuremath{\sqrt {{\color{\colorMATH}\ensuremath{\eta }}}}}}} \mathrel{:} {{\color{\colorSYNTAX}\texttt{\ensuremath{{\mathbb{R}}^{+}}}}}
  }
\and\inferrule*[lab={{\color{\colorTEXT}\textsc{\scriptsize Log}}}
  ]{ \Delta  \vdash  \eta  \mathrel{:} {{\color{\colorSYNTAX}\texttt{\ensuremath{{\mathbb{R}}^{+}}}}}
     }{
     \Delta  \vdash  {{\color{\colorSYNTAX}\texttt{\ensuremath{\ln {{\color{\colorMATH}\ensuremath{\eta }}}}}}} \mathrel{:} {{\color{\colorSYNTAX}\texttt{\ensuremath{{\mathbb{R}}^{+}}}}}
  }
\and\inferrule*[lab={{\color{\colorTEXT}\textsc{\scriptsize Weaken}}}
  ]{ \Delta  \vdash  \eta  \mathrel{:} {{\color{\colorSYNTAX}\texttt{\ensuremath{{\mathbb{N}}}}}}
     }{
     \Delta  \vdash  \eta  \mathrel{:} {{\color{\colorSYNTAX}\texttt{\ensuremath{{\mathbb{R}}^{+}}}}}
  }
\end{mathpar}\endgroup 

\vspace*{-0.25em}\begingroup\color{\colorMATH}\begin{gather*}\begin{tabularx}{\linewidth}{>{\centering\arraybackslash\(}X<{\)}}\hfill\hspace{0pt}\begingroup\color{\colorTEXT}\boxed{\begingroup\color{\colorMATH} \Delta  \vdash  {\begingroup\renewcommand\colorMATH{\colorMATHS}\renewcommand\colorSYNTAX{\colorSYNTAXS}{{\color{\colorMATH}\ensuremath{s}}}\endgroup } \endgroup}\endgroup \end{tabularx}\vspace*{-1em}\end{gather*}\endgroup 
\begingroup\color{\colorMATH}\begin{mathpar}\inferrule*[lab={{\color{\colorTEXT}\textsc{\scriptsize Sens}}}
  ]{ \Delta  \vdash  \eta  \mathrel{:} {{\color{\colorSYNTAX}\texttt{\ensuremath{{\mathbb{R}}^{+}}}}}
     }{
     \Delta  \vdash  \eta 
  }
\and\inferrule*[lab={{\color{\colorTEXT}\textsc{\scriptsize Inf}}}
  ]{ }{
     \Delta  \vdash  {\begingroup\renewcommand\colorMATH{\colorMATHS}\renewcommand\colorSYNTAX{\colorSYNTAXS}{{\color{\colorMATH}\ensuremath{{{\color{\colorSYNTAX}\texttt{\ensuremath{\infty }}}}}}}\endgroup }
  }
\end{mathpar}\endgroup 
\vspace*{-0.25em}\begingroup\color{\colorMATH}\begin{gather*}\begin{tabularx}{\linewidth}{>{\centering\arraybackslash\(}X<{\)}}\hfill\hspace{0pt}\begingroup\color{\colorTEXT}\boxed{\begingroup\color{\colorMATH} \Delta  \vdash  {\begingroup\renewcommand\colorMATH{\colorMATHP}\renewcommand\colorSYNTAX{\colorSYNTAXP}{{\color{\colorMATH}\ensuremath{p}}}\endgroup } \endgroup}\endgroup \end{tabularx}\vspace*{-1em}\end{gather*}\endgroup 
\begingroup\color{\colorMATH}\begin{mathpar}\inferrule*[lab={{\color{\colorTEXT}\textsc{\scriptsize Sens}}}
  ]{ \Delta  \vdash  \eta _{\epsilon } \mathrel{:} {{\color{\colorSYNTAX}\texttt{\ensuremath{{\mathbb{R}}^{+}}}}}
  \\ \Delta  \vdash  \eta _{\delta } \mathrel{:} {{\color{\colorSYNTAX}\texttt{\ensuremath{{\mathbb{R}}^{+}}}}}
     }{
     \Delta  \vdash  {\begingroup\renewcommand\colorMATH{\colorMATHP}\renewcommand\colorSYNTAX{\colorSYNTAXP}{{\color{\colorMATH}\ensuremath{{{\color{\colorSYNTAX}\texttt{\ensuremath{{\begingroup\renewcommand\colorMATH{\colorMATHM}\renewcommand\colorSYNTAX{\colorSYNTAXM}{{\color{\colorMATH}\ensuremath{\eta _{\epsilon }}}}\endgroup },{\begingroup\renewcommand\colorMATH{\colorMATHM}\renewcommand\colorSYNTAX{\colorSYNTAXM}{{\color{\colorMATH}\ensuremath{\eta _{\delta }}}}\endgroup }}}}}}}}\endgroup }
  }
\and\inferrule*[lab={{\color{\colorTEXT}\textsc{\scriptsize Inf}}}
  ]{ }{
     \Delta  \vdash  {\begingroup\renewcommand\colorMATH{\colorMATHP}\renewcommand\colorSYNTAX{\colorSYNTAXP}{{\color{\colorMATH}\ensuremath{{{\color{\colorSYNTAX}\texttt{\ensuremath{\infty }}}}}}}\endgroup }
  }
\end{mathpar}\endgroup 

\vspace*{-0.25em}\begingroup\color{\colorMATH}\begin{gather*}\begin{tabularx}{\linewidth}{>{\centering\arraybackslash\(}X<{\)}}\hfill\hspace{0pt}\begingroup\color{\colorTEXT}\boxed{\begingroup\color{\colorMATH} \Delta  \vdash  \tau  \endgroup}\endgroup \end{tabularx}\vspace*{-1em}\end{gather*}\endgroup 
\begingroup\color{\colorMATH}\begin{mathpar}\inferrule*[lab={{\color{\colorTEXT}\textsc{\scriptsize Static Nat}}}
  ]{ \Delta  \vdash  \eta  \mathrel{:} {{\color{\colorSYNTAX}\texttt{\ensuremath{{\mathbb{N}}}}}}
     }{
     \Delta  \vdash  {{\color{\colorSYNTAX}\texttt{\ensuremath{{\mathbb{N}}[{{\color{\colorMATH}\ensuremath{\eta }}}]}}}}
  }
\and\inferrule*[lab={{\color{\colorTEXT}\textsc{\scriptsize Static Real}}}
  ]{ \Delta  \vdash  \eta  \mathrel{:} {{\color{\colorSYNTAX}\texttt{\ensuremath{{\mathbb{R}}^{+}}}}}
     }{
     \Delta  \vdash  {{\color{\colorSYNTAX}\texttt{\ensuremath{{\mathbb{R}}^{+}[{{\color{\colorMATH}\ensuremath{\eta }}}]}}}}
  }
\and\inferrule*[lab={{\color{\colorTEXT}\textsc{\scriptsize Nat}}}
  ]{ }{
     \Delta  \vdash  {{\color{\colorSYNTAX}\texttt{\ensuremath{{\mathbb{N}}}}}}
  }
\and\inferrule*[lab={{\color{\colorTEXT}\textsc{\scriptsize Real}}}
  ]{ }{
     \Delta  \vdash  {{\color{\colorSYNTAX}\texttt{\ensuremath{{\mathbb{R}}}}}}
  }
\and\inferrule*[lab={{\color{\colorTEXT}\textsc{\scriptsize Data}}}
  ]{ }{
     \Delta  \vdash  {{\color{\colorSYNTAX}\texttt{data}}}
  }
\and\inferrule*[lab={{\color{\colorTEXT}\textsc{\scriptsize Index}}}
  ]{ \Delta  \vdash  \eta  \mathrel{:} {{\color{\colorSYNTAX}\texttt{\ensuremath{{\mathbb{N}}}}}}
     }{
     \Delta  \vdash  {{\color{\colorSYNTAX}\texttt{\ensuremath{{{\color{\colorSYNTAX}\texttt{idx}}}[{\begingroup\renewcommand\colorMATH{\colorMATHM}\renewcommand\colorSYNTAX{\colorSYNTAXM}{{\color{\colorMATH}\ensuremath{\eta }}}\endgroup }]}}}}
  }
\and\inferrule*[lab={{\color{\colorTEXT}\textsc{\scriptsize Matrix}}}
  ]{ \Delta  \vdash  \eta _{m} \mathrel{:} {{\color{\colorSYNTAX}\texttt{\ensuremath{{\mathbb{N}}}}}}
  \\ \Delta  \vdash  \eta _{n} \mathrel{:} {{\color{\colorSYNTAX}\texttt{\ensuremath{{\mathbb{N}}}}}}
  \\ \Delta  \vdash  \tau 
     }{
     \Delta  \vdash  {{\color{\colorSYNTAX}\texttt{\ensuremath{{{\color{\colorSYNTAX}\texttt{matrix}}}_{{{\color{\colorMATH}\ensuremath{\ell }}}}^{{{\color{\colorMATH}\ensuremath{c}}}}[{{\color{\colorMATH}\ensuremath{\eta _{m}}}},{{\color{\colorMATH}\ensuremath{\eta _{n}}}}]\hspace*{0.33em}{{\color{\colorMATH}\ensuremath{\tau }}}}}}}
  }
\and\inferrule*[lab={{\color{\colorSYNTAX}\texttt{\ensuremath{+}}}}{{\color{\colorTEXT}\textsc{\scriptsize -F}}}
  ]{ \Delta  \vdash  \tau _{1}
  \\ \Delta  \vdash  \tau _{2}
     }{
     \Delta  \vdash  {{\color{\colorSYNTAX}\texttt{\ensuremath{{{\color{\colorMATH}\ensuremath{\tau _{1}}}} + {{\color{\colorMATH}\ensuremath{\tau _{2}}}}}}}}
  }
\and\inferrule*[lab={{\color{\colorSYNTAX}\texttt{\ensuremath{\times }}}}{{\color{\colorTEXT}\textsc{\scriptsize -F}}}
  ]{ \Delta  \vdash  \tau _{1}
  \\ \Delta  \vdash  \tau _{2}
     }{
     \Delta  \vdash  {{\color{\colorSYNTAX}\texttt{\ensuremath{{{\color{\colorMATH}\ensuremath{\tau _{1}}}} \times  {{\color{\colorMATH}\ensuremath{\tau _{2}}}}}}}}
  }
\and\inferrule*[lab={{\color{\colorSYNTAX}\texttt{\ensuremath{\mathrel{\&}}}}}{{\color{\colorTEXT}\textsc{\scriptsize -F}}}
  ]{ \Delta  \vdash  \tau _{1}
  \\ \Delta  \vdash  \tau _{2}
     }{
     \Delta  \vdash  {{\color{\colorSYNTAX}\texttt{\ensuremath{{{\color{\colorMATH}\ensuremath{\tau _{1}}}} \mathrel{\&} {{\color{\colorMATH}\ensuremath{\tau _{2}}}}}}}}
  }
\and\inferrule*[lab={{\color{\colorMATH}\ensuremath{\multimap }}}{{\color{\colorTEXT}\textsc{\scriptsize -F}}}
  ]{ \Delta  \vdash  \tau _{1}
  \\ \Delta  \vdash  \tau _{2}
  \\ \Delta  \vdash  {\begingroup\renewcommand\colorMATH{\colorMATHS}\renewcommand\colorSYNTAX{\colorSYNTAXS}{{\color{\colorMATH}\ensuremath{s}}}\endgroup }
     }{
     \Delta  \vdash  {{\color{\colorSYNTAX}\texttt{\ensuremath{{{\color{\colorMATH}\ensuremath{\tau _{1}}}} \multimap _{{\begingroup\renewcommand\colorMATH{\colorMATHS}\renewcommand\colorSYNTAX{\colorSYNTAXS}{{\color{\colorMATH}\ensuremath{s}}}\endgroup }} {{\color{\colorMATH}\ensuremath{\tau _{2}}}}}}}}
  }
\and\inferrule*[lab={{\color{\colorMATH}\ensuremath{\multimap ^{*}}}}{{\color{\colorTEXT}\textsc{\scriptsize -F}}}
  ]{ \Delta ^{\prime} = \Delta \uplus \{ \beta _{1}{\mathrel{:}}\kappa _{1},{.}\hspace{-1pt}{.}\hspace{-1pt}{.},\beta _{n}{\mathrel{:}}\kappa _{n}\} 
  \\ \Delta ^{\prime} \vdash  \tau _{i}\hspace*{0.33em}(\forall i)
  \\ \Delta ^{\prime} \vdash  {\begingroup\renewcommand\colorMATH{\colorMATHP}\renewcommand\colorSYNTAX{\colorSYNTAXP}{{\color{\colorMATH}\ensuremath{p_{i}}}}\endgroup }\hspace*{0.33em}(\forall i)
  \\ \Delta ^{\prime} \vdash  \tau 
     }{
     \Delta  \vdash  {{\color{\colorSYNTAX}\texttt{\ensuremath{\forall \hspace*{0.33em}[{{\color{\colorMATH}\ensuremath{\beta _{1}}}}{\mathrel{:}}{{\color{\colorMATH}\ensuremath{\kappa _{1}}}},{.}\hspace{-1pt}{.}\hspace{-1pt}{.},{{\color{\colorMATH}\ensuremath{\beta _{n}}}}{\mathrel{:}}{{\color{\colorMATH}\ensuremath{\kappa _{n}}}}]\hspace*{0.33em}({{\color{\colorMATH}\ensuremath{\tau _{1}}}}@{\begingroup\renewcommand\colorMATH{\colorMATHP}\renewcommand\colorSYNTAX{\colorSYNTAXP}{{\color{\colorMATH}\ensuremath{p_{1}}}}\endgroup },{.}\hspace{-1pt}{.}\hspace{-1pt}{.},{{\color{\colorMATH}\ensuremath{\tau _{n}}}}@{\begingroup\renewcommand\colorMATH{\colorMATHP}\renewcommand\colorSYNTAX{\colorSYNTAXP}{{\color{\colorMATH}\ensuremath{p_{n}}}}\endgroup }) \multimap ^{*} {{\color{\colorMATH}\ensuremath{\tau }}}}}}}
  }
\end{mathpar}\endgroup 
\caption{Kinding and Type Formation}
\label{fig:apx:kinding}
\end{figure*}

\begin{figure*}
\begingroup\renewcommand\colorMATH{\colorMATHS}\renewcommand\colorSYNTAX{\colorSYNTAXS}
\vspace*{-0.25em}\begingroup\color{\colorMATH}\begin{gather*}\begin{tabularx}{\linewidth}{>{\centering\arraybackslash\(}X<{\)}}\hfill\hspace{0pt}\begingroup\color{\colorTEXT}\boxed{\begingroup\color{\colorMATH} {\begingroup\renewcommand\colorMATH{\colorMATHM}\renewcommand\colorSYNTAX{\colorSYNTAXM}{{\color{\colorMATH}\ensuremath{\Delta }}}\endgroup },\Gamma  \vdash  e \mathrel{:} {\begingroup\renewcommand\colorMATH{\colorMATHM}\renewcommand\colorSYNTAX{\colorSYNTAXM}{{\color{\colorMATH}\ensuremath{\tau }}}\endgroup } \endgroup}\endgroup \end{tabularx}\vspace*{-1em}\end{gather*}\endgroup 
\begingroup\color{\colorMATH}\begin{mathpar}\inferrule*[lab={{\color{\colorTEXT}\textsc{\scriptsize Singleton Nat}}}
  ]{ }{
     {\begingroup\renewcommand\colorMATH{\colorMATHM}\renewcommand\colorSYNTAX{\colorSYNTAXM}{{\color{\colorMATH}\ensuremath{\Delta }}}\endgroup },\Gamma  \vdash  {{\color{\colorSYNTAX}\texttt{\ensuremath{{\mathbb{N}}[{\begingroup\renewcommand\colorMATH{\colorMATHM}\renewcommand\colorSYNTAX{\colorSYNTAXM}{{\color{\colorMATH}\ensuremath{n}}}\endgroup }]}}}} \mathrel{:} {\begingroup\renewcommand\colorMATH{\colorMATHM}\renewcommand\colorSYNTAX{\colorSYNTAXM}{{\color{\colorMATH}\ensuremath{{{\color{\colorSYNTAX}\texttt{\ensuremath{{\mathbb{N}}[{{\color{\colorMATH}\ensuremath{n}}}]}}}}}}}\endgroup }
  }
\and\inferrule*[lab={{\color{\colorTEXT}\textsc{\scriptsize Singleton Real}}}
  ]{ }{
     {\begingroup\renewcommand\colorMATH{\colorMATHM}\renewcommand\colorSYNTAX{\colorSYNTAXM}{{\color{\colorMATH}\ensuremath{\Delta }}}\endgroup },\Gamma  \vdash  {{\color{\colorSYNTAX}\texttt{\ensuremath{{\mathbb{R}}^{+}[{\begingroup\renewcommand\colorMATH{\colorMATHM}\renewcommand\colorSYNTAX{\colorSYNTAXM}{{\color{\colorMATH}\ensuremath{r^{+}}}}\endgroup }]}}}} \mathrel{:} {\begingroup\renewcommand\colorMATH{\colorMATHM}\renewcommand\colorSYNTAX{\colorSYNTAXM}{{\color{\colorMATH}\ensuremath{{{\color{\colorSYNTAX}\texttt{\ensuremath{{\mathbb{R}}^{+}[{{\color{\colorMATH}\ensuremath{r^{+}}}}]}}}}}}}\endgroup }
  }
\and\inferrule*[lab={{\color{\colorTEXT}\textsc{\scriptsize Dynamic Nat}}}
  ]{ {\begingroup\renewcommand\colorMATH{\colorMATHM}\renewcommand\colorSYNTAX{\colorSYNTAXM}{{\color{\colorMATH}\ensuremath{\Delta }}}\endgroup },\Gamma  \vdash  e \mathrel{:} {\begingroup\renewcommand\colorMATH{\colorMATHM}\renewcommand\colorSYNTAX{\colorSYNTAXM}{{\color{\colorMATH}\ensuremath{{{\color{\colorSYNTAX}\texttt{\ensuremath{{\mathbb{N}}[{{\color{\colorMATH}\ensuremath{\eta }}}]}}}}}}}\endgroup }
     }{
     {\begingroup\renewcommand\colorMATH{\colorMATHM}\renewcommand\colorSYNTAX{\colorSYNTAXM}{{\color{\colorMATH}\ensuremath{\Delta }}}\endgroup },{\begingroup\renewcommand\colorMATH{\colorMATHM}\renewcommand\colorSYNTAX{\colorSYNTAXM}{{\color{\colorMATH}\ensuremath{0}}}\endgroup }\Gamma  \vdash  {{\color{\colorSYNTAX}\texttt{dyn}}}\hspace*{0.33em}e \mathrel{:} {\begingroup\renewcommand\colorMATH{\colorMATHM}\renewcommand\colorSYNTAX{\colorSYNTAXM}{{\color{\colorMATH}\ensuremath{{{\color{\colorSYNTAX}\texttt{\ensuremath{{\mathbb{N}}}}}}}}}\endgroup }
  }
\and\inferrule*[lab={{\color{\colorTEXT}\textsc{\scriptsize Dynamic Real}}}
  ]{ {\begingroup\renewcommand\colorMATH{\colorMATHM}\renewcommand\colorSYNTAX{\colorSYNTAXM}{{\color{\colorMATH}\ensuremath{\Delta }}}\endgroup },\Gamma  \vdash  e \mathrel{:} {\begingroup\renewcommand\colorMATH{\colorMATHM}\renewcommand\colorSYNTAX{\colorSYNTAXM}{{\color{\colorMATH}\ensuremath{{{\color{\colorSYNTAX}\texttt{\ensuremath{{\mathbb{R}}^{+}[{{\color{\colorMATH}\ensuremath{\eta }}}]}}}}}}}\endgroup }
     }{
     {\begingroup\renewcommand\colorMATH{\colorMATHM}\renewcommand\colorSYNTAX{\colorSYNTAXM}{{\color{\colorMATH}\ensuremath{\Delta }}}\endgroup },{\begingroup\renewcommand\colorMATH{\colorMATHM}\renewcommand\colorSYNTAX{\colorSYNTAXM}{{\color{\colorMATH}\ensuremath{0}}}\endgroup }\Gamma  \vdash  {{\color{\colorSYNTAX}\texttt{dyn}}}\hspace*{0.33em}e \mathrel{:} {\begingroup\renewcommand\colorMATH{\colorMATHM}\renewcommand\colorSYNTAX{\colorSYNTAXM}{{\color{\colorMATH}\ensuremath{{{\color{\colorSYNTAX}\texttt{\ensuremath{{\mathbb{R}}}}}}}}}\endgroup }
  }
\and\inferrule*[lab={{\color{\colorTEXT}\textsc{\scriptsize Dynamic Index}}}
  ]{ {\begingroup\renewcommand\colorMATH{\colorMATHM}\renewcommand\colorSYNTAX{\colorSYNTAXM}{{\color{\colorMATH}\ensuremath{\Delta }}}\endgroup },\Gamma  \vdash  e \mathrel{:} {\begingroup\renewcommand\colorMATH{\colorMATHM}\renewcommand\colorSYNTAX{\colorSYNTAXM}{{\color{\colorMATH}\ensuremath{{{\color{\colorSYNTAX}\texttt{\ensuremath{{{\color{\colorSYNTAX}\texttt{idx}}}[{{\color{\colorMATH}\ensuremath{\eta }}}]}}}}}}}\endgroup }
     }{
     {\begingroup\renewcommand\colorMATH{\colorMATHM}\renewcommand\colorSYNTAX{\colorSYNTAXM}{{\color{\colorMATH}\ensuremath{\Delta }}}\endgroup },{\begingroup\renewcommand\colorMATH{\colorMATHM}\renewcommand\colorSYNTAX{\colorSYNTAXM}{{\color{\colorMATH}\ensuremath{0}}}\endgroup }\Gamma  \vdash  {{\color{\colorSYNTAX}\texttt{dyn}}}\hspace*{0.33em}e \mathrel{:} {\begingroup\renewcommand\colorMATH{\colorMATHM}\renewcommand\colorSYNTAX{\colorSYNTAXM}{{\color{\colorMATH}\ensuremath{{{\color{\colorSYNTAX}\texttt{\ensuremath{{\mathbb{N}}}}}}}}}\endgroup }
  }
\and\inferrule*[lab={{\color{\colorTEXT}\textsc{\scriptsize Nat}}}
  ]{ }{
     {\begingroup\renewcommand\colorMATH{\colorMATHM}\renewcommand\colorSYNTAX{\colorSYNTAXM}{{\color{\colorMATH}\ensuremath{\Delta }}}\endgroup },\Gamma  \vdash  {\begingroup\renewcommand\colorMATH{\colorMATHM}\renewcommand\colorSYNTAX{\colorSYNTAXM}{{\color{\colorMATH}\ensuremath{n}}}\endgroup } \mathrel{:} {\begingroup\renewcommand\colorMATH{\colorMATHM}\renewcommand\colorSYNTAX{\colorSYNTAXM}{{\color{\colorMATH}\ensuremath{{{\color{\colorSYNTAX}\texttt{\ensuremath{{\mathbb{N}}}}}}}}}\endgroup }
  }
\and\inferrule*[lab={{\color{\colorTEXT}\textsc{\scriptsize Real}}}
  ]{ }{
     {\begingroup\renewcommand\colorMATH{\colorMATHM}\renewcommand\colorSYNTAX{\colorSYNTAXM}{{\color{\colorMATH}\ensuremath{\Delta }}}\endgroup },\Gamma  \vdash  {\begingroup\renewcommand\colorMATH{\colorMATHM}\renewcommand\colorSYNTAX{\colorSYNTAXM}{{\color{\colorMATH}\ensuremath{r}}}\endgroup } \mathrel{:} {\begingroup\renewcommand\colorMATH{\colorMATHM}\renewcommand\colorSYNTAX{\colorSYNTAXM}{{\color{\colorMATH}\ensuremath{{{\color{\colorSYNTAX}\texttt{\ensuremath{{\mathbb{R}}}}}}}}}\endgroup }
  }
\and\inferrule*[lab={{\color{\colorTEXT}\textsc{\scriptsize Singleton Real Nat}}}
  ]{ {\begingroup\renewcommand\colorMATH{\colorMATHM}\renewcommand\colorSYNTAX{\colorSYNTAXM}{{\color{\colorMATH}\ensuremath{\Delta }}}\endgroup },\Gamma  \vdash  e \mathrel{:} {\begingroup\renewcommand\colorMATH{\colorMATHM}\renewcommand\colorSYNTAX{\colorSYNTAXM}{{\color{\colorMATH}\ensuremath{{{\color{\colorSYNTAX}\texttt{\ensuremath{{\mathbb{N}}[{{\color{\colorMATH}\ensuremath{\eta }}}]}}}}}}}\endgroup }
    }{
     {\begingroup\renewcommand\colorMATH{\colorMATHM}\renewcommand\colorSYNTAX{\colorSYNTAXM}{{\color{\colorMATH}\ensuremath{\Delta }}}\endgroup },{\begingroup\renewcommand\colorMATH{\colorMATHM}\renewcommand\colorSYNTAX{\colorSYNTAXM}{{\color{\colorMATH}\ensuremath{0}}}\endgroup }\Gamma  \vdash  {{\color{\colorSYNTAX}\texttt{real}}}\hspace*{0.33em}e \mathrel{:} {\begingroup\renewcommand\colorMATH{\colorMATHM}\renewcommand\colorSYNTAX{\colorSYNTAXM}{{\color{\colorMATH}\ensuremath{{{\color{\colorSYNTAX}\texttt{\ensuremath{{\mathbb{R}}^{+}[{{\color{\colorMATH}\ensuremath{\eta }}}]}}}}}}}\endgroup }
  }
\and\inferrule*[lab={{\color{\colorTEXT}\textsc{\scriptsize Real Nat}}}
  ]{ {\begingroup\renewcommand\colorMATH{\colorMATHM}\renewcommand\colorSYNTAX{\colorSYNTAXM}{{\color{\colorMATH}\ensuremath{\Delta }}}\endgroup },\Gamma  \vdash  e \mathrel{:} {\begingroup\renewcommand\colorMATH{\colorMATHM}\renewcommand\colorSYNTAX{\colorSYNTAXM}{{\color{\colorMATH}\ensuremath{{{\color{\colorSYNTAX}\texttt{\ensuremath{{\mathbb{N}}}}}}}}}\endgroup }
    }{
     {\begingroup\renewcommand\colorMATH{\colorMATHM}\renewcommand\colorSYNTAX{\colorSYNTAXM}{{\color{\colorMATH}\ensuremath{\Delta }}}\endgroup },\Gamma  \vdash  {{\color{\colorSYNTAX}\texttt{real}}}\hspace*{0.33em}e \mathrel{:} {\begingroup\renewcommand\colorMATH{\colorMATHM}\renewcommand\colorSYNTAX{\colorSYNTAXM}{{\color{\colorMATH}\ensuremath{{{\color{\colorSYNTAX}\texttt{\ensuremath{{\mathbb{R}}}}}}}}}\endgroup }
  }
\and\inferrule*[lab={{\color{\colorTEXT}\textsc{\scriptsize Max}}}
  ]{ {\begingroup\renewcommand\colorMATH{\colorMATHM}\renewcommand\colorSYNTAX{\colorSYNTAXM}{{\color{\colorMATH}\ensuremath{\Delta }}}\endgroup },\Gamma _{1} \vdash  e_{1} \mathrel{:} {\begingroup\renewcommand\colorMATH{\colorMATHM}\renewcommand\colorSYNTAX{\colorSYNTAXM}{{\color{\colorMATH}\ensuremath{\tau _{1}}}}\endgroup }
  \\ {\begingroup\renewcommand\colorMATH{\colorMATHM}\renewcommand\colorSYNTAX{\colorSYNTAXM}{{\color{\colorMATH}\ensuremath{\Delta }}}\endgroup },\Gamma _{2} \vdash  e_{2} \mathrel{:} {\begingroup\renewcommand\colorMATH{\colorMATHM}\renewcommand\colorSYNTAX{\colorSYNTAXM}{{\color{\colorMATH}\ensuremath{\tau _{2}}}}\endgroup }
  \\ {\begingroup\renewcommand\colorMATH{\colorMATHM}\renewcommand\colorSYNTAX{\colorSYNTAXM}{{\color{\colorMATH}\ensuremath{\langle {\begingroup\renewcommand\colorMATH{\colorMATHS}\renewcommand\colorSYNTAX{\colorSYNTAXS}{{\color{\colorMATH}\ensuremath{s_{1}}}}\endgroup },{\begingroup\renewcommand\colorMATH{\colorMATHS}\renewcommand\colorSYNTAX{\colorSYNTAXS}{{\color{\colorMATH}\ensuremath{s_{2}}}}\endgroup },\tau \rangle  = \tau _{1}\sqcup \tau _{2}}}}\endgroup }
     }{
     {\begingroup\renewcommand\colorMATH{\colorMATHM}\renewcommand\colorSYNTAX{\colorSYNTAXM}{{\color{\colorMATH}\ensuremath{\Delta }}}\endgroup },{\begingroup\renewcommand\colorMATH{\colorMATHS}\renewcommand\colorSYNTAX{\colorSYNTAXS}{{\color{\colorMATH}\ensuremath{s_{1}}}}\endgroup }\Gamma _{1} + {\begingroup\renewcommand\colorMATH{\colorMATHS}\renewcommand\colorSYNTAX{\colorSYNTAXS}{{\color{\colorMATH}\ensuremath{s_{2}}}}\endgroup }\Gamma _{2} \vdash  {{\color{\colorSYNTAX}\texttt{\ensuremath{{{\color{\colorMATH}\ensuremath{e_{1}}}}\sqcup {{\color{\colorMATH}\ensuremath{e_{2}}}}}}}} \mathrel{:} {\begingroup\renewcommand\colorMATH{\colorMATHM}\renewcommand\colorSYNTAX{\colorSYNTAXM}{{\color{\colorMATH}\ensuremath{\tau }}}\endgroup }
  }
\and\inferrule*[lab={{\color{\colorTEXT}\textsc{\scriptsize Min}}}
  ]{ {\begingroup\renewcommand\colorMATH{\colorMATHM}\renewcommand\colorSYNTAX{\colorSYNTAXM}{{\color{\colorMATH}\ensuremath{\Delta }}}\endgroup },\Gamma _{1} \vdash  e_{1} \mathrel{:} {\begingroup\renewcommand\colorMATH{\colorMATHM}\renewcommand\colorSYNTAX{\colorSYNTAXM}{{\color{\colorMATH}\ensuremath{\tau _{1}}}}\endgroup }
  \\ {\begingroup\renewcommand\colorMATH{\colorMATHM}\renewcommand\colorSYNTAX{\colorSYNTAXM}{{\color{\colorMATH}\ensuremath{\Delta }}}\endgroup },\Gamma _{2} \vdash  e_{2} \mathrel{:} {\begingroup\renewcommand\colorMATH{\colorMATHM}\renewcommand\colorSYNTAX{\colorSYNTAXM}{{\color{\colorMATH}\ensuremath{\tau _{2}}}}\endgroup }
  \\ {\begingroup\renewcommand\colorMATH{\colorMATHM}\renewcommand\colorSYNTAX{\colorSYNTAXM}{{\color{\colorMATH}\ensuremath{\langle {\begingroup\renewcommand\colorMATH{\colorMATHS}\renewcommand\colorSYNTAX{\colorSYNTAXS}{{\color{\colorMATH}\ensuremath{s_{1}}}}\endgroup },{\begingroup\renewcommand\colorMATH{\colorMATHS}\renewcommand\colorSYNTAX{\colorSYNTAXS}{{\color{\colorMATH}\ensuremath{s_{2}}}}\endgroup },\tau \rangle  = \tau _{1}\sqcap \tau _{2}}}}\endgroup }
     }{
     {\begingroup\renewcommand\colorMATH{\colorMATHM}\renewcommand\colorSYNTAX{\colorSYNTAXM}{{\color{\colorMATH}\ensuremath{\Delta }}}\endgroup },{\begingroup\renewcommand\colorMATH{\colorMATHS}\renewcommand\colorSYNTAX{\colorSYNTAXS}{{\color{\colorMATH}\ensuremath{s_{1}}}}\endgroup }\Gamma _{1} + {\begingroup\renewcommand\colorMATH{\colorMATHS}\renewcommand\colorSYNTAX{\colorSYNTAXS}{{\color{\colorMATH}\ensuremath{s_{2}}}}\endgroup }\Gamma _{2} \vdash  {{\color{\colorSYNTAX}\texttt{\ensuremath{{{\color{\colorMATH}\ensuremath{e_{1}}}}\sqcap {{\color{\colorMATH}\ensuremath{e_{2}}}}}}}} \mathrel{:} {\begingroup\renewcommand\colorMATH{\colorMATHM}\renewcommand\colorSYNTAX{\colorSYNTAXM}{{\color{\colorMATH}\ensuremath{\tau }}}\endgroup }
  }
\and\inferrule*[lab={{\color{\colorTEXT}\textsc{\scriptsize Plus}}}
  ]{ {\begingroup\renewcommand\colorMATH{\colorMATHM}\renewcommand\colorSYNTAX{\colorSYNTAXM}{{\color{\colorMATH}\ensuremath{\Delta }}}\endgroup },\Gamma _{1} \vdash  e_{1} \mathrel{:} {\begingroup\renewcommand\colorMATH{\colorMATHM}\renewcommand\colorSYNTAX{\colorSYNTAXM}{{\color{\colorMATH}\ensuremath{\tau _{1}}}}\endgroup }
  \\ {\begingroup\renewcommand\colorMATH{\colorMATHM}\renewcommand\colorSYNTAX{\colorSYNTAXM}{{\color{\colorMATH}\ensuremath{\Delta }}}\endgroup },\Gamma _{2} \vdash  e_{2} \mathrel{:} {\begingroup\renewcommand\colorMATH{\colorMATHM}\renewcommand\colorSYNTAX{\colorSYNTAXM}{{\color{\colorMATH}\ensuremath{\tau _{2}}}}\endgroup }
  \\ {\begingroup\renewcommand\colorMATH{\colorMATHM}\renewcommand\colorSYNTAX{\colorSYNTAXM}{{\color{\colorMATH}\ensuremath{\langle {\begingroup\renewcommand\colorMATH{\colorMATHS}\renewcommand\colorSYNTAX{\colorSYNTAXS}{{\color{\colorMATH}\ensuremath{s_{1}}}}\endgroup },{\begingroup\renewcommand\colorMATH{\colorMATHS}\renewcommand\colorSYNTAX{\colorSYNTAXS}{{\color{\colorMATH}\ensuremath{s_{2}}}}\endgroup },\tau \rangle  = \tau _{1}+\tau _{2}}}}\endgroup }
     }{
     {\begingroup\renewcommand\colorMATH{\colorMATHM}\renewcommand\colorSYNTAX{\colorSYNTAXM}{{\color{\colorMATH}\ensuremath{\Delta }}}\endgroup },{\begingroup\renewcommand\colorMATH{\colorMATHS}\renewcommand\colorSYNTAX{\colorSYNTAXS}{{\color{\colorMATH}\ensuremath{s_{1}}}}\endgroup }\Gamma _{1} + {\begingroup\renewcommand\colorMATH{\colorMATHS}\renewcommand\colorSYNTAX{\colorSYNTAXS}{{\color{\colorMATH}\ensuremath{s_{2}}}}\endgroup }\Gamma _{2} \vdash  {{\color{\colorSYNTAX}\texttt{\ensuremath{{{\color{\colorMATH}\ensuremath{e_{1}}}}+{{\color{\colorMATH}\ensuremath{e_{2}}}}}}}} \mathrel{:} {\begingroup\renewcommand\colorMATH{\colorMATHM}\renewcommand\colorSYNTAX{\colorSYNTAXM}{{\color{\colorMATH}\ensuremath{\tau }}}\endgroup }
  }
\and\inferrule*[lab={{\color{\colorTEXT}\textsc{\scriptsize Times}}}
  ]{ {\begingroup\renewcommand\colorMATH{\colorMATHM}\renewcommand\colorSYNTAX{\colorSYNTAXM}{{\color{\colorMATH}\ensuremath{\Delta }}}\endgroup },\Gamma _{1} \vdash  e_{1} \mathrel{:} {\begingroup\renewcommand\colorMATH{\colorMATHM}\renewcommand\colorSYNTAX{\colorSYNTAXM}{{\color{\colorMATH}\ensuremath{\tau _{1}}}}\endgroup }
  \\ {\begingroup\renewcommand\colorMATH{\colorMATHM}\renewcommand\colorSYNTAX{\colorSYNTAXM}{{\color{\colorMATH}\ensuremath{\Delta }}}\endgroup },\Gamma _{2} \vdash  e_{2} \mathrel{:} {\begingroup\renewcommand\colorMATH{\colorMATHM}\renewcommand\colorSYNTAX{\colorSYNTAXM}{{\color{\colorMATH}\ensuremath{\tau _{2}}}}\endgroup }
  \\ {\begingroup\renewcommand\colorMATH{\colorMATHM}\renewcommand\colorSYNTAX{\colorSYNTAXM}{{\color{\colorMATH}\ensuremath{\langle {\begingroup\renewcommand\colorMATH{\colorMATHS}\renewcommand\colorSYNTAX{\colorSYNTAXS}{{\color{\colorMATH}\ensuremath{s_{1}}}}\endgroup },{\begingroup\renewcommand\colorMATH{\colorMATHS}\renewcommand\colorSYNTAX{\colorSYNTAXS}{{\color{\colorMATH}\ensuremath{s_{2}}}}\endgroup },\tau \rangle  = \tau _{1}\mathrel{\mathord{\cdotp }}\tau _{2}}}}\endgroup }
     }{
     {\begingroup\renewcommand\colorMATH{\colorMATHM}\renewcommand\colorSYNTAX{\colorSYNTAXM}{{\color{\colorMATH}\ensuremath{\Delta }}}\endgroup },{\begingroup\renewcommand\colorMATH{\colorMATHS}\renewcommand\colorSYNTAX{\colorSYNTAXS}{{\color{\colorMATH}\ensuremath{s_{1}}}}\endgroup }\Gamma _{1} + {\begingroup\renewcommand\colorMATH{\colorMATHS}\renewcommand\colorSYNTAX{\colorSYNTAXS}{{\color{\colorMATH}\ensuremath{s_{2}}}}\endgroup }\Gamma _{2} \vdash  {{\color{\colorSYNTAX}\texttt{\ensuremath{{{\color{\colorMATH}\ensuremath{e_{1}}}}\mathrel{\mathord{\cdotp }}{{\color{\colorMATH}\ensuremath{e_{2}}}}}}}} \mathrel{:} {\begingroup\renewcommand\colorMATH{\colorMATHM}\renewcommand\colorSYNTAX{\colorSYNTAXM}{{\color{\colorMATH}\ensuremath{\tau }}}\endgroup }
  }
\and\inferrule*[lab={{\color{\colorTEXT}\textsc{\scriptsize Inverse}}}
  ]{ {\begingroup\renewcommand\colorMATH{\colorMATHM}\renewcommand\colorSYNTAX{\colorSYNTAXM}{{\color{\colorMATH}\ensuremath{\Delta }}}\endgroup },\Gamma  \vdash  e \mathrel{:} {\begingroup\renewcommand\colorMATH{\colorMATHM}\renewcommand\colorSYNTAX{\colorSYNTAXM}{{\color{\colorMATH}\ensuremath{\tau }}}\endgroup }
  \\ {\begingroup\renewcommand\colorMATH{\colorMATHM}\renewcommand\colorSYNTAX{\colorSYNTAXM}{{\color{\colorMATH}\ensuremath{\langle {\begingroup\renewcommand\colorMATH{\colorMATHS}\renewcommand\colorSYNTAX{\colorSYNTAXS}{{\color{\colorMATH}\ensuremath{s}}}\endgroup },\tau ^{\prime}\rangle  = 1/\tau }}}\endgroup }
     }{
     {\begingroup\renewcommand\colorMATH{\colorMATHM}\renewcommand\colorSYNTAX{\colorSYNTAXM}{{\color{\colorMATH}\ensuremath{\Delta }}}\endgroup },s\Gamma  \vdash  {{\color{\colorSYNTAX}\texttt{\ensuremath{1/{{\color{\colorMATH}\ensuremath{e}}}}}}} \mathrel{:} {\begingroup\renewcommand\colorMATH{\colorMATHM}\renewcommand\colorSYNTAX{\colorSYNTAXM}{{\color{\colorMATH}\ensuremath{\tau ^{\prime}}}}\endgroup }
  }
\and\inferrule*[lab={{\color{\colorTEXT}\textsc{\scriptsize Root}}}
  ]{ {\begingroup\renewcommand\colorMATH{\colorMATHM}\renewcommand\colorSYNTAX{\colorSYNTAXM}{{\color{\colorMATH}\ensuremath{\Delta }}}\endgroup },\Gamma  \vdash  e \mathrel{:} {\begingroup\renewcommand\colorMATH{\colorMATHM}\renewcommand\colorSYNTAX{\colorSYNTAXM}{{\color{\colorMATH}\ensuremath{\tau }}}\endgroup }
  \\ {\begingroup\renewcommand\colorMATH{\colorMATHM}\renewcommand\colorSYNTAX{\colorSYNTAXM}{{\color{\colorMATH}\ensuremath{\langle {\begingroup\renewcommand\colorMATH{\colorMATHS}\renewcommand\colorSYNTAX{\colorSYNTAXS}{{\color{\colorMATH}\ensuremath{s}}}\endgroup },\tau ^{\prime}\rangle  = \sqrt \tau }}}\endgroup }
     }{
     {\begingroup\renewcommand\colorMATH{\colorMATHM}\renewcommand\colorSYNTAX{\colorSYNTAXM}{{\color{\colorMATH}\ensuremath{\Delta }}}\endgroup },s\Gamma  \vdash  {{\color{\colorSYNTAX}\texttt{\ensuremath{\sqrt {{{\color{\colorMATH}\ensuremath{e}}}}}}}} \mathrel{:} {\begingroup\renewcommand\colorMATH{\colorMATHM}\renewcommand\colorSYNTAX{\colorSYNTAXM}{{\color{\colorMATH}\ensuremath{\tau ^{\prime}}}}\endgroup }
  }
\and\inferrule*[lab={{\color{\colorTEXT}\textsc{\scriptsize Log}}}
  ]{ {\begingroup\renewcommand\colorMATH{\colorMATHM}\renewcommand\colorSYNTAX{\colorSYNTAXM}{{\color{\colorMATH}\ensuremath{\Delta }}}\endgroup },\Gamma  \vdash  e \mathrel{:} {\begingroup\renewcommand\colorMATH{\colorMATHM}\renewcommand\colorSYNTAX{\colorSYNTAXM}{{\color{\colorMATH}\ensuremath{\tau }}}\endgroup }
  \\ {\begingroup\renewcommand\colorMATH{\colorMATHM}\renewcommand\colorSYNTAX{\colorSYNTAXM}{{\color{\colorMATH}\ensuremath{\langle {\begingroup\renewcommand\colorMATH{\colorMATHS}\renewcommand\colorSYNTAX{\colorSYNTAXS}{{\color{\colorMATH}\ensuremath{s}}}\endgroup },\tau ^{\prime}\rangle  = \ln \tau }}}\endgroup }
     }{
     {\begingroup\renewcommand\colorMATH{\colorMATHM}\renewcommand\colorSYNTAX{\colorSYNTAXM}{{\color{\colorMATH}\ensuremath{\Delta }}}\endgroup },s\Gamma  \vdash  {{\color{\colorSYNTAX}\texttt{\ensuremath{\ln {{\color{\colorMATH}\ensuremath{e}}}}}}} \mathrel{:} {\begingroup\renewcommand\colorMATH{\colorMATHM}\renewcommand\colorSYNTAX{\colorSYNTAXM}{{\color{\colorMATH}\ensuremath{\tau ^{\prime}}}}\endgroup }
  }
\and\inferrule*[lab={{\color{\colorTEXT}\textsc{\scriptsize Mod}}}
  ]{ {\begingroup\renewcommand\colorMATH{\colorMATHM}\renewcommand\colorSYNTAX{\colorSYNTAXM}{{\color{\colorMATH}\ensuremath{\Delta }}}\endgroup },\Gamma _{1} \vdash  e_{1} \mathrel{:} {\begingroup\renewcommand\colorMATH{\colorMATHM}\renewcommand\colorSYNTAX{\colorSYNTAXM}{{\color{\colorMATH}\ensuremath{\tau _{1}}}}\endgroup }
  \\ {\begingroup\renewcommand\colorMATH{\colorMATHM}\renewcommand\colorSYNTAX{\colorSYNTAXM}{{\color{\colorMATH}\ensuremath{\Delta }}}\endgroup },\Gamma _{2} \vdash  e_{2} \mathrel{:} {\begingroup\renewcommand\colorMATH{\colorMATHM}\renewcommand\colorSYNTAX{\colorSYNTAXM}{{\color{\colorMATH}\ensuremath{\tau _{2}}}}\endgroup }
  \\ {\begingroup\renewcommand\colorMATH{\colorMATHM}\renewcommand\colorSYNTAX{\colorSYNTAXM}{{\color{\colorMATH}\ensuremath{\langle {\begingroup\renewcommand\colorMATH{\colorMATHS}\renewcommand\colorSYNTAX{\colorSYNTAXS}{{\color{\colorMATH}\ensuremath{s_{1}}}}\endgroup },{\begingroup\renewcommand\colorMATH{\colorMATHS}\renewcommand\colorSYNTAX{\colorSYNTAXS}{{\color{\colorMATH}\ensuremath{s_{2}}}}\endgroup },\tau \rangle  = \tau _{1}\mathrel{{{\color{\colorMATH}\ensuremath{\operatorname{mod}}}}}\tau _{2}}}}\endgroup }
     }{
     {\begingroup\renewcommand\colorMATH{\colorMATHM}\renewcommand\colorSYNTAX{\colorSYNTAXM}{{\color{\colorMATH}\ensuremath{\Delta }}}\endgroup },{}\rceil \Gamma _{1}\lceil {}^{{\begingroup\renewcommand\colorMATH{\colorMATHS}\renewcommand\colorSYNTAX{\colorSYNTAXS}{{\color{\colorMATH}\ensuremath{s_{1}}}}\endgroup }} + {\begingroup\renewcommand\colorMATH{\colorMATHS}\renewcommand\colorSYNTAX{\colorSYNTAXS}{{\color{\colorMATH}\ensuremath{s_{2}}}}\endgroup }\Gamma _{2} \vdash  {{\color{\colorSYNTAX}\texttt{\ensuremath{{{\color{\colorMATH}\ensuremath{e_{1}}}}\mathrel{{{\color{\colorSYNTAX}\texttt{mod}}}}{{\color{\colorMATH}\ensuremath{e_{2}}}}}}}} \mathrel{:} {\begingroup\renewcommand\colorMATH{\colorMATHM}\renewcommand\colorSYNTAX{\colorSYNTAXM}{{\color{\colorMATH}\ensuremath{\tau }}}\endgroup }
  }
\and\inferrule*[lab={{\color{\colorTEXT}\textsc{\scriptsize Minus}}}
  ]{ {\begingroup\renewcommand\colorMATH{\colorMATHM}\renewcommand\colorSYNTAX{\colorSYNTAXM}{{\color{\colorMATH}\ensuremath{\Delta }}}\endgroup },\Gamma _{1} \vdash  e_{1} \mathrel{:} {\begingroup\renewcommand\colorMATH{\colorMATHM}\renewcommand\colorSYNTAX{\colorSYNTAXM}{{\color{\colorMATH}\ensuremath{\tau _{1}}}}\endgroup }
  \\ {\begingroup\renewcommand\colorMATH{\colorMATHM}\renewcommand\colorSYNTAX{\colorSYNTAXM}{{\color{\colorMATH}\ensuremath{\Delta }}}\endgroup },\Gamma _{2} \vdash  e_{2} \mathrel{:} {\begingroup\renewcommand\colorMATH{\colorMATHM}\renewcommand\colorSYNTAX{\colorSYNTAXM}{{\color{\colorMATH}\ensuremath{\tau _{2}}}}\endgroup }
  \\ {\begingroup\renewcommand\colorMATH{\colorMATHM}\renewcommand\colorSYNTAX{\colorSYNTAXM}{{\color{\colorMATH}\ensuremath{\langle {\begingroup\renewcommand\colorMATH{\colorMATHS}\renewcommand\colorSYNTAX{\colorSYNTAXS}{{\color{\colorMATH}\ensuremath{s_{1}}}}\endgroup },{\begingroup\renewcommand\colorMATH{\colorMATHS}\renewcommand\colorSYNTAX{\colorSYNTAXS}{{\color{\colorMATH}\ensuremath{s_{2}}}}\endgroup },\tau \rangle  = \tau _{1}-\tau _{2}}}}\endgroup }
     }{
     {\begingroup\renewcommand\colorMATH{\colorMATHM}\renewcommand\colorSYNTAX{\colorSYNTAXM}{{\color{\colorMATH}\ensuremath{\Delta }}}\endgroup },{\begingroup\renewcommand\colorMATH{\colorMATHS}\renewcommand\colorSYNTAX{\colorSYNTAXS}{{\color{\colorMATH}\ensuremath{s_{1}}}}\endgroup }\Gamma _{1} + {\begingroup\renewcommand\colorMATH{\colorMATHS}\renewcommand\colorSYNTAX{\colorSYNTAXS}{{\color{\colorMATH}\ensuremath{s_{2}}}}\endgroup }\Gamma _{2} \vdash  {{\color{\colorSYNTAX}\texttt{\ensuremath{{{\color{\colorMATH}\ensuremath{e_{1}}}}-{{\color{\colorMATH}\ensuremath{e_{2}}}}}}}} \mathrel{:} {\begingroup\renewcommand\colorMATH{\colorMATHM}\renewcommand\colorSYNTAX{\colorSYNTAXM}{{\color{\colorMATH}\ensuremath{\tau }}}\endgroup }
  }
\and\inferrule*[lab={{\color{\colorTEXT}\textsc{\scriptsize Eq}}}
  ]{ {\begingroup\renewcommand\colorMATH{\colorMATHM}\renewcommand\colorSYNTAX{\colorSYNTAXM}{{\color{\colorMATH}\ensuremath{\Delta }}}\endgroup },\Gamma _{1} \vdash  e_{1} \mathrel{:} {\begingroup\renewcommand\colorMATH{\colorMATHM}\renewcommand\colorSYNTAX{\colorSYNTAXM}{{\color{\colorMATH}\ensuremath{\tau _{1}}}}\endgroup }
  \\ {\begingroup\renewcommand\colorMATH{\colorMATHM}\renewcommand\colorSYNTAX{\colorSYNTAXM}{{\color{\colorMATH}\ensuremath{\Delta }}}\endgroup },\Gamma _{2} \vdash  e_{2} \mathrel{:} {\begingroup\renewcommand\colorMATH{\colorMATHM}\renewcommand\colorSYNTAX{\colorSYNTAXM}{{\color{\colorMATH}\ensuremath{\tau _{2}}}}\endgroup }
  \\ {\begingroup\renewcommand\colorMATH{\colorMATHM}\renewcommand\colorSYNTAX{\colorSYNTAXM}{{\color{\colorMATH}\ensuremath{\langle {\begingroup\renewcommand\colorMATH{\colorMATHS}\renewcommand\colorSYNTAX{\colorSYNTAXS}{{\color{\colorMATH}\ensuremath{s_{1}}}}\endgroup },{\begingroup\renewcommand\colorMATH{\colorMATHS}\renewcommand\colorSYNTAX{\colorSYNTAXS}{{\color{\colorMATH}\ensuremath{s_{2}}}}\endgroup },\tau \rangle  = \tau _{1}\mathrel{\overset{?}{\smash{=}}} \tau _{2}}}}\endgroup }
     }{
     {\begingroup\renewcommand\colorMATH{\colorMATHM}\renewcommand\colorSYNTAX{\colorSYNTAXM}{{\color{\colorMATH}\ensuremath{\Delta }}}\endgroup },{\begingroup\renewcommand\colorMATH{\colorMATHS}\renewcommand\colorSYNTAX{\colorSYNTAXS}{{\color{\colorMATH}\ensuremath{s_{1}}}}\endgroup }\Gamma _{1} + {\begingroup\renewcommand\colorMATH{\colorMATHS}\renewcommand\colorSYNTAX{\colorSYNTAXS}{{\color{\colorMATH}\ensuremath{s_{2}}}}\endgroup }\Gamma _{2} \vdash  {{\color{\colorSYNTAX}\texttt{\ensuremath{{{\color{\colorMATH}\ensuremath{e_{1}}}}\mathrel{\overset{?}{\smash{=}}} {{\color{\colorMATH}\ensuremath{e_{2}}}}}}}} \mathrel{:} {\begingroup\renewcommand\colorMATH{\colorMATHM}\renewcommand\colorSYNTAX{\colorSYNTAXM}{{\color{\colorMATH}\ensuremath{\tau }}}\endgroup }
  }
\end{mathpar}\endgroup 
\endgroup 
\caption{Sensitivity Typing---Literals and Arithmetic}
\label{fig:apx:sens-typing-lits-arith}
\end{figure*}

\begin{figure*}
\begingroup\renewcommand\colorMATH{\colorMATHS}\renewcommand\colorSYNTAX{\colorSYNTAXS}
\vspace*{-0.25em}\begingroup\color{\colorMATH}\begin{gather*}\begin{tabularx}{\linewidth}{>{\centering\arraybackslash\(}X<{\)}}\hfill\hspace{0pt}\begingroup\color{\colorTEXT}\boxed{\begingroup\color{\colorMATH} {\begingroup\renewcommand\colorMATH{\colorMATHM}\renewcommand\colorSYNTAX{\colorSYNTAXM}{{\color{\colorMATH}\ensuremath{\Delta }}}\endgroup },\Gamma  \vdash  e \mathrel{:} {\begingroup\renewcommand\colorMATH{\colorMATHM}\renewcommand\colorSYNTAX{\colorSYNTAXM}{{\color{\colorMATH}\ensuremath{\tau }}}\endgroup } \endgroup}\endgroup \end{tabularx}\vspace*{-1em}\end{gather*}\endgroup 
\begingroup\color{\colorMATH}\begin{mathpar}\inferrule*[lab={{\color{\colorTEXT}\textsc{\scriptsize Matrix Creation}}}
  ]{ {\begingroup\renewcommand\colorMATH{\colorMATHM}\renewcommand\colorSYNTAX{\colorSYNTAXM}{{\color{\colorMATH}\ensuremath{\Delta }}}\endgroup },\Gamma _{1} \vdash  e_{1} \mathrel{:} {\begingroup\renewcommand\colorMATH{\colorMATHM}\renewcommand\colorSYNTAX{\colorSYNTAXM}{{\color{\colorMATH}\ensuremath{{{\color{\colorSYNTAX}\texttt{\ensuremath{{\mathbb{N}}[{{\color{\colorMATH}\ensuremath{\eta _{m}}}}]}}}}}}}\endgroup }
  \\ {\begingroup\renewcommand\colorMATH{\colorMATHM}\renewcommand\colorSYNTAX{\colorSYNTAXM}{{\color{\colorMATH}\ensuremath{\Delta }}}\endgroup },\Gamma _{2} \vdash  e_{2} \mathrel{:} {\begingroup\renewcommand\colorMATH{\colorMATHM}\renewcommand\colorSYNTAX{\colorSYNTAXM}{{\color{\colorMATH}\ensuremath{{{\color{\colorSYNTAX}\texttt{\ensuremath{{\mathbb{N}}[{{\color{\colorMATH}\ensuremath{\eta _{n}}}}]}}}}}}}\endgroup }
  \\ {\begingroup\renewcommand\colorMATH{\colorMATHM}\renewcommand\colorSYNTAX{\colorSYNTAXM}{{\color{\colorMATH}\ensuremath{\Delta }}}\endgroup },\Gamma _{3}\uplus \{ {\begingroup\renewcommand\colorMATH{\colorMATHM}\renewcommand\colorSYNTAX{\colorSYNTAXM}{{\color{\colorMATH}\ensuremath{x_{1}}}}\endgroup }{\mathrel{:}}_{{{\color{\colorSYNTAX}\texttt{\ensuremath{\infty }}}}}{\begingroup\renewcommand\colorMATH{\colorMATHM}\renewcommand\colorSYNTAX{\colorSYNTAXM}{{\color{\colorMATH}\ensuremath{{{\color{\colorSYNTAX}\texttt{\ensuremath{{{\color{\colorSYNTAX}\texttt{idx}}}[{{\color{\colorMATH}\ensuremath{\eta _{m}}}}]}}}}}}}\endgroup },{\begingroup\renewcommand\colorMATH{\colorMATHM}\renewcommand\colorSYNTAX{\colorSYNTAXM}{{\color{\colorMATH}\ensuremath{x_{2}}}}\endgroup }{\mathrel{:}}_{{{\color{\colorSYNTAX}\texttt{\ensuremath{\infty }}}}}{\begingroup\renewcommand\colorMATH{\colorMATHM}\renewcommand\colorSYNTAX{\colorSYNTAXM}{{\color{\colorMATH}\ensuremath{{{\color{\colorSYNTAX}\texttt{\ensuremath{{{\color{\colorSYNTAX}\texttt{idx}}}[{{\color{\colorMATH}\ensuremath{\eta _{n}}}}]}}}}}}}\endgroup }\}  \vdash  e_{3} \mathrel{:} {\begingroup\renewcommand\colorMATH{\colorMATHM}\renewcommand\colorSYNTAX{\colorSYNTAXM}{{\color{\colorMATH}\ensuremath{\tau }}}\endgroup }
     }{
     {\begingroup\renewcommand\colorMATH{\colorMATHM}\renewcommand\colorSYNTAX{\colorSYNTAXM}{{\color{\colorMATH}\ensuremath{\Delta }}}\endgroup },{\begingroup\renewcommand\colorMATH{\colorMATHM}\renewcommand\colorSYNTAX{\colorSYNTAXM}{{\color{\colorMATH}\ensuremath{0}}}\endgroup }(\Gamma _{1} + \Gamma _{2}) + {\begingroup\renewcommand\colorMATH{\colorMATHM}\renewcommand\colorSYNTAX{\colorSYNTAXM}{{\color{\colorMATH}\ensuremath{\eta _{m}\eta _{n}}}}\endgroup }\Gamma _{3} \mathrel{:} {{\color{\colorSYNTAX}\texttt{\ensuremath{{{\color{\colorSYNTAX}\texttt{mcreate}}}_{{\begingroup\renewcommand\colorMATH{\colorMATHM}\renewcommand\colorSYNTAX{\colorSYNTAXM}{{\color{\colorMATH}\ensuremath{\ell }}}\endgroup }}[{{\color{\colorMATH}\ensuremath{e_{1}}}},{{\color{\colorMATH}\ensuremath{e_{2}}}}]\{ {\begingroup\renewcommand\colorMATH{\colorMATHM}\renewcommand\colorSYNTAX{\colorSYNTAXM}{{\color{\colorMATH}\ensuremath{x_{1}}}}\endgroup },{\begingroup\renewcommand\colorMATH{\colorMATHM}\renewcommand\colorSYNTAX{\colorSYNTAXM}{{\color{\colorMATH}\ensuremath{x_{2}}}}\endgroup } \Rightarrow  {{\color{\colorMATH}\ensuremath{e_{3}}}}\} }}}} \mathrel{:} {\begingroup\renewcommand\colorMATH{\colorMATHM}\renewcommand\colorSYNTAX{\colorSYNTAXM}{{\color{\colorMATH}\ensuremath{{{\color{\colorSYNTAX}\texttt{\ensuremath{{{\color{\colorSYNTAX}\texttt{matrix}}}_{{{\color{\colorMATH}\ensuremath{\ell }}}}^{U}[{{\color{\colorMATH}\ensuremath{\eta _{m}}}},{{\color{\colorMATH}\ensuremath{\eta _{n}}}}]\hspace*{0.33em}{{\color{\colorMATH}\ensuremath{\tau }}}}}}}}}}\endgroup } 
  }
\and\inferrule*[lab={{\color{\colorTEXT}\textsc{\scriptsize Matrix Index}}}
  ]{ {\begingroup\renewcommand\colorMATH{\colorMATHM}\renewcommand\colorSYNTAX{\colorSYNTAXM}{{\color{\colorMATH}\ensuremath{\Delta }}}\endgroup },\Gamma _{1} \vdash  e_{1} \mathrel{:} {\begingroup\renewcommand\colorMATH{\colorMATHM}\renewcommand\colorSYNTAX{\colorSYNTAXM}{{\color{\colorMATH}\ensuremath{{{\color{\colorSYNTAX}\texttt{\ensuremath{{{\color{\colorSYNTAX}\texttt{matrix}}}_{{{\color{\colorMATH}\ensuremath{\ell }}}}^{{{\color{\colorMATH}\ensuremath{c}}}}[{{\color{\colorMATH}\ensuremath{\eta _{m}}}},{{\color{\colorMATH}\ensuremath{\eta _{n}}}}]\hspace*{0.33em}{{\color{\colorMATH}\ensuremath{\tau }}}}}}}}}}\endgroup }
  \\ {\begingroup\renewcommand\colorMATH{\colorMATHM}\renewcommand\colorSYNTAX{\colorSYNTAXM}{{\color{\colorMATH}\ensuremath{\Delta }}}\endgroup },\Gamma _{2} \vdash  e_{2} \mathrel{:} {\begingroup\renewcommand\colorMATH{\colorMATHM}\renewcommand\colorSYNTAX{\colorSYNTAXM}{{\color{\colorMATH}\ensuremath{{{\color{\colorSYNTAX}\texttt{\ensuremath{{{\color{\colorSYNTAX}\texttt{idx}}}[{{\color{\colorMATH}\ensuremath{\eta _{m}}}}]}}}}}}}\endgroup }
     }{
     {\begingroup\renewcommand\colorMATH{\colorMATHM}\renewcommand\colorSYNTAX{\colorSYNTAXM}{{\color{\colorMATH}\ensuremath{\Delta }}}\endgroup },\Gamma _{1} + {\begingroup\renewcommand\colorMATH{\colorMATHM}\renewcommand\colorSYNTAX{\colorSYNTAXM}{{\color{\colorMATH}\ensuremath{0}}}\endgroup }\Gamma _{2} \vdash  {{\color{\colorSYNTAX}\texttt{\ensuremath{{{\color{\colorMATH}\ensuremath{e_{1}}}}{\#}{{\color{\colorMATH}\ensuremath{e_{2}}}}}}}} \mathrel{:} {\begingroup\renewcommand\colorMATH{\colorMATHM}\renewcommand\colorSYNTAX{\colorSYNTAXM}{{\color{\colorMATH}\ensuremath{{{\color{\colorSYNTAX}\texttt{\ensuremath{{{\color{\colorSYNTAX}\texttt{matrix}}}_{{{\color{\colorMATH}\ensuremath{\ell }}}}^{{{\color{\colorMATH}\ensuremath{c}}}}[{{\color{\colorMATH}\ensuremath{1}}},{{\color{\colorMATH}\ensuremath{\eta _{n}}}}]\hspace*{0.33em}{{\color{\colorMATH}\ensuremath{\tau }}}}}}}}}}\endgroup }
  }
\and\inferrule*[lab={{\color{\colorTEXT}\textsc{\scriptsize Matrix Rows}}}
  ]{ {\begingroup\renewcommand\colorMATH{\colorMATHM}\renewcommand\colorSYNTAX{\colorSYNTAXM}{{\color{\colorMATH}\ensuremath{\Delta }}}\endgroup },\Gamma  \vdash  e \mathrel{:} {\begingroup\renewcommand\colorMATH{\colorMATHM}\renewcommand\colorSYNTAX{\colorSYNTAXM}{{\color{\colorMATH}\ensuremath{{{\color{\colorSYNTAX}\texttt{\ensuremath{{{\color{\colorSYNTAX}\texttt{matrix}}}_{{{\color{\colorMATH}\ensuremath{\bigstar }}}}^{{{\color{\colorMATH}\ensuremath{\bigstar }}}}[{{\color{\colorMATH}\ensuremath{\eta _{m}}}},{{\color{\colorMATH}\ensuremath{\eta _{n}}}}]\hspace*{0.33em}{{\color{\colorMATH}\ensuremath{\tau }}}}}}}}}}\endgroup }
     }{
     {\begingroup\renewcommand\colorMATH{\colorMATHM}\renewcommand\colorSYNTAX{\colorSYNTAXM}{{\color{\colorMATH}\ensuremath{\Delta }}}\endgroup },{\begingroup\renewcommand\colorMATH{\colorMATHM}\renewcommand\colorSYNTAX{\colorSYNTAXM}{{\color{\colorMATH}\ensuremath{0}}}\endgroup }\Gamma  \vdash  {{\color{\colorSYNTAX}\texttt{rows}}}\hspace*{0.33em}e \mathrel{:} {\begingroup\renewcommand\colorMATH{\colorMATHM}\renewcommand\colorSYNTAX{\colorSYNTAXM}{{\color{\colorMATH}\ensuremath{{{\color{\colorSYNTAX}\texttt{\ensuremath{{\mathbb{N}}[{{\color{\colorMATH}\ensuremath{\eta _{m}}}}]}}}}}}}\endgroup }
  }
\and\inferrule*[lab={{\color{\colorTEXT}\textsc{\scriptsize Matrix Cols}}}
  ]{ {\begingroup\renewcommand\colorMATH{\colorMATHM}\renewcommand\colorSYNTAX{\colorSYNTAXM}{{\color{\colorMATH}\ensuremath{\Delta }}}\endgroup },\Gamma  \vdash  e \mathrel{:} {\begingroup\renewcommand\colorMATH{\colorMATHM}\renewcommand\colorSYNTAX{\colorSYNTAXM}{{\color{\colorMATH}\ensuremath{{{\color{\colorSYNTAX}\texttt{\ensuremath{{{\color{\colorSYNTAX}\texttt{matrix}}}_{{{\color{\colorMATH}\ensuremath{\bigstar }}}}^{{{\color{\colorMATH}\ensuremath{\bigstar }}}}[{{\color{\colorMATH}\ensuremath{\eta _{m}}}},{{\color{\colorMATH}\ensuremath{\eta _{n}}}}]\hspace*{0.33em}{{\color{\colorMATH}\ensuremath{\tau }}}}}}}}}}\endgroup }
     }{
     {\begingroup\renewcommand\colorMATH{\colorMATHM}\renewcommand\colorSYNTAX{\colorSYNTAXM}{{\color{\colorMATH}\ensuremath{\Delta }}}\endgroup },{\begingroup\renewcommand\colorMATH{\colorMATHM}\renewcommand\colorSYNTAX{\colorSYNTAXM}{{\color{\colorMATH}\ensuremath{0}}}\endgroup }\Gamma  \vdash  {{\color{\colorSYNTAX}\texttt{cols}}}\hspace*{0.33em}e \mathrel{:} {\begingroup\renewcommand\colorMATH{\colorMATHM}\renewcommand\colorSYNTAX{\colorSYNTAXM}{{\color{\colorMATH}\ensuremath{{{\color{\colorSYNTAX}\texttt{\ensuremath{{\mathbb{N}}[{{\color{\colorMATH}\ensuremath{\eta _{n}}}}]}}}}}}}\endgroup }
  }
\and\inferrule*[lab={{\color{\colorTEXT}\textsc{\scriptsize Clip}}}
  ]{{{\color{\colorTEXT}\textit{\tiny (this could work for R, but we never need it at that type for examples)}}}
  \\\\ {\begingroup\renewcommand\colorMATH{\colorMATHM}\renewcommand\colorSYNTAX{\colorSYNTAXM}{{\color{\colorMATH}\ensuremath{\Delta }}}\endgroup },\Gamma  \vdash  e \mathrel{:} {\begingroup\renewcommand\colorMATH{\colorMATHM}\renewcommand\colorSYNTAX{\colorSYNTAXM}{{\color{\colorMATH}\ensuremath{{{\color{\colorSYNTAX}\texttt{\ensuremath{{{\color{\colorSYNTAX}\texttt{matrix}}}_{{{\color{\colorMATH}\ensuremath{\ell }}}}^{{{\color{\colorMATH}\ensuremath{\bigstar }}}}[{{\color{\colorMATH}\ensuremath{1}}},{{\color{\colorMATH}\ensuremath{\eta _{n}}}}]\hspace*{0.33em}{{\color{\colorSYNTAX}\texttt{data}}}}}}}}}}\endgroup }
     }{
     {\begingroup\renewcommand\colorMATH{\colorMATHM}\renewcommand\colorSYNTAX{\colorSYNTAXM}{{\color{\colorMATH}\ensuremath{\Delta }}}\endgroup },\Gamma  \vdash  {{\color{\colorSYNTAX}\texttt{clip}}}^{{\begingroup\renewcommand\colorMATH{\colorMATHM}\renewcommand\colorSYNTAX{\colorSYNTAXM}{{\color{\colorMATH}\ensuremath{c}}}\endgroup }}\hspace*{0.33em}e \mathrel{:} {\begingroup\renewcommand\colorMATH{\colorMATHM}\renewcommand\colorSYNTAX{\colorSYNTAXM}{{\color{\colorMATH}\ensuremath{{{\color{\colorSYNTAX}\texttt{\ensuremath{{{\color{\colorSYNTAX}\texttt{matrix}}}_{{{\color{\colorMATH}\ensuremath{\ell }}}}^{{{\color{\colorMATH}\ensuremath{c}}}}[{{\color{\colorMATH}\ensuremath{1}}},{{\color{\colorMATH}\ensuremath{\eta _{n}}}}]\hspace*{0.33em}{{\color{\colorSYNTAX}\texttt{data}}}}}}}}}}\endgroup }
  }
\and\inferrule*[lab={{\color{\colorTEXT}\textsc{\scriptsize Convert}}}
  ]{ {\begingroup\renewcommand\colorMATH{\colorMATHM}\renewcommand\colorSYNTAX{\colorSYNTAXM}{{\color{\colorMATH}\ensuremath{\Delta }}}\endgroup },\Gamma  \vdash  e \mathrel{:} {\begingroup\renewcommand\colorMATH{\colorMATHM}\renewcommand\colorSYNTAX{\colorSYNTAXM}{{\color{\colorMATH}\ensuremath{{{\color{\colorSYNTAX}\texttt{\ensuremath{{{\color{\colorSYNTAX}\texttt{matrix}}}_{{{\color{\colorMATH}\ensuremath{\bigstar }}}}^{{{\color{\colorMATH}\ensuremath{\ell }}}}[{{\color{\colorMATH}\ensuremath{1}}},{{\color{\colorMATH}\ensuremath{\eta _{n}}}}]\hspace*{0.33em}{{\color{\colorSYNTAX}\texttt{data}}}}}}}}}}\endgroup }
     }{
     {\begingroup\renewcommand\colorMATH{\colorMATHM}\renewcommand\colorSYNTAX{\colorSYNTAXM}{{\color{\colorMATH}\ensuremath{\Delta }}}\endgroup },\Gamma  \vdash  {{\color{\colorSYNTAX}\texttt{conv}}}\hspace*{0.33em}e \mathrel{:} {\begingroup\renewcommand\colorMATH{\colorMATHM}\renewcommand\colorSYNTAX{\colorSYNTAXM}{{\color{\colorMATH}\ensuremath{{{\color{\colorSYNTAX}\texttt{\ensuremath{{{\color{\colorSYNTAX}\texttt{matrix}}}_{{{\color{\colorMATH}\ensuremath{\ell }}}}^{U}[{{\color{\colorMATH}\ensuremath{1}}},{{\color{\colorMATH}\ensuremath{\eta _{n}}}}]\hspace*{0.33em}{\mathbb{R}}}}}}}}}\endgroup }
  }
\and\inferrule*[lab={{\color{\colorTEXT}\textsc{\scriptsize Lipschitz Gradient}}}
  ]{ {\begingroup\renewcommand\colorMATH{\colorMATHM}\renewcommand\colorSYNTAX{\colorSYNTAXM}{{\color{\colorMATH}\ensuremath{\Delta }}}\endgroup },\Gamma _{1} \vdash  e_{\theta } \mathrel{:} {\begingroup\renewcommand\colorMATH{\colorMATHM}\renewcommand\colorSYNTAX{\colorSYNTAXM}{{\color{\colorMATH}\ensuremath{{{\color{\colorSYNTAX}\texttt{\ensuremath{{{\color{\colorSYNTAX}\texttt{matrix}}}_{{{\color{\colorMATH}\ensuremath{\bigstar }}}}^{{{\color{\colorMATH}\ensuremath{\bigstar }}}}[{{\color{\colorMATH}\ensuremath{1}}},{{\color{\colorMATH}\ensuremath{\eta _{n}}}}]\hspace*{0.33em}{\mathbb{R}}}}}}}}}\endgroup }
  \\ {\begingroup\renewcommand\colorMATH{\colorMATHM}\renewcommand\colorSYNTAX{\colorSYNTAXM}{{\color{\colorMATH}\ensuremath{\Delta }}}\endgroup },\Gamma _{2} \vdash  e_{xs} \mathrel{:} {\begingroup\renewcommand\colorMATH{\colorMATHM}\renewcommand\colorSYNTAX{\colorSYNTAXM}{{\color{\colorMATH}\ensuremath{{{\color{\colorSYNTAX}\texttt{\ensuremath{{{\color{\colorSYNTAX}\texttt{matrix}}}_{{{\color{\colorMATH}\ensuremath{\bigstar }}}}^{{{\color{\colorMATH}\ensuremath{\ell }}}}[{{\color{\colorMATH}\ensuremath{1}}},{{\color{\colorMATH}\ensuremath{\eta _{n}}}}]\hspace*{0.33em}{{\color{\colorSYNTAX}\texttt{data}}}}}}}}}}\endgroup }
  \\ {\begingroup\renewcommand\colorMATH{\colorMATHM}\renewcommand\colorSYNTAX{\colorSYNTAXM}{{\color{\colorMATH}\ensuremath{\Delta }}}\endgroup },\Gamma _{3} \vdash  e_{y} \mathrel{:} {\begingroup\renewcommand\colorMATH{\colorMATHM}\renewcommand\colorSYNTAX{\colorSYNTAXM}{{\color{\colorMATH}\ensuremath{{{\color{\colorSYNTAX}\texttt{data}}}}}}\endgroup }
     }{
     {\begingroup\renewcommand\colorMATH{\colorMATHM}\renewcommand\colorSYNTAX{\colorSYNTAXM}{{\color{\colorMATH}\ensuremath{\Delta }}}\endgroup },\infty \Gamma _{1} + \Gamma _{2} + \Gamma _{3} \vdash  {{\color{\colorSYNTAX}\texttt{\ensuremath{L{\mathrel{\nabla }}_{{\begingroup\renewcommand\colorMATH{\colorMATHM}\renewcommand\colorSYNTAX{\colorSYNTAXM}{{\color{\colorMATH}\ensuremath{\ell }}}\endgroup }}^{{{\color{\colorMATH}\ensuremath{g}}}}[{{\color{\colorMATH}\ensuremath{e_{\theta }}}};{{\color{\colorMATH}\ensuremath{e_{xs}}}},{{\color{\colorMATH}\ensuremath{e_{ys}}}}]}}}} \mathrel{:} {\begingroup\renewcommand\colorMATH{\colorMATHM}\renewcommand\colorSYNTAX{\colorSYNTAXM}{{\color{\colorMATH}\ensuremath{{{\color{\colorSYNTAX}\texttt{\ensuremath{{{\color{\colorSYNTAX}\texttt{matrix}}}_{{{\color{\colorMATH}\ensuremath{\ell }}}}^{U}[{{\color{\colorMATH}\ensuremath{1}}},{{\color{\colorMATH}\ensuremath{\eta _{n}}}}]\hspace*{0.33em}{\mathbb{R}}}}}}}}}\endgroup }
  }
\and\inferrule*[lab={{\color{\colorTEXT}\textsc{\scriptsize Unbounded Gradient}}}
  ]{ {\begingroup\renewcommand\colorMATH{\colorMATHM}\renewcommand\colorSYNTAX{\colorSYNTAXM}{{\color{\colorMATH}\ensuremath{\Delta }}}\endgroup },\Gamma _{1} \vdash  e_{\theta }  \mathrel{:} {\begingroup\renewcommand\colorMATH{\colorMATHM}\renewcommand\colorSYNTAX{\colorSYNTAXM}{{\color{\colorMATH}\ensuremath{{{\color{\colorSYNTAX}\texttt{\ensuremath{{{\color{\colorSYNTAX}\texttt{matrix}}}_{{{\color{\colorMATH}\ensuremath{\bigstar }}}}^{{{\color{\colorMATH}\ensuremath{\bigstar }}}}[{{\color{\colorMATH}\ensuremath{1}}},{{\color{\colorMATH}\ensuremath{\eta _{n}}}}]\hspace*{0.33em}{\mathbb{R}}}}}}}}}\endgroup }
  \\ {\begingroup\renewcommand\colorMATH{\colorMATHM}\renewcommand\colorSYNTAX{\colorSYNTAXM}{{\color{\colorMATH}\ensuremath{\Delta }}}\endgroup },\Gamma _{2} \vdash  e_{xs} \mathrel{:} {\begingroup\renewcommand\colorMATH{\colorMATHM}\renewcommand\colorSYNTAX{\colorSYNTAXM}{{\color{\colorMATH}\ensuremath{{{\color{\colorSYNTAX}\texttt{\ensuremath{{{\color{\colorSYNTAX}\texttt{matrix}}}_{L\infty }^{{{\color{\colorMATH}\ensuremath{\bigstar }}}}[{{\color{\colorMATH}\ensuremath{1}}},{{\color{\colorMATH}\ensuremath{\eta _{n}}}}]\hspace*{0.33em}{{\color{\colorSYNTAX}\texttt{data}}}}}}}}}}\endgroup }
  \\ {\begingroup\renewcommand\colorMATH{\colorMATHM}\renewcommand\colorSYNTAX{\colorSYNTAXM}{{\color{\colorMATH}\ensuremath{\Delta }}}\endgroup },\Gamma _{3} \vdash  e_{y} \mathrel{:} {\begingroup\renewcommand\colorMATH{\colorMATHM}\renewcommand\colorSYNTAX{\colorSYNTAXM}{{\color{\colorMATH}\ensuremath{{{\color{\colorSYNTAX}\texttt{data}}}}}}\endgroup }
     }{
     {\begingroup\renewcommand\colorMATH{\colorMATHM}\renewcommand\colorSYNTAX{\colorSYNTAXM}{{\color{\colorMATH}\ensuremath{\Delta }}}\endgroup },\infty \Gamma _{1} + \Gamma _{2} + \Gamma _{3} \vdash  {{\color{\colorSYNTAX}\texttt{\ensuremath{U{\mathrel{\nabla }}[{{\color{\colorMATH}\ensuremath{e_{\theta }}}};{{\color{\colorMATH}\ensuremath{e_{xs}}}},{{\color{\colorMATH}\ensuremath{e_{y}}}}]}}}} \mathrel{:} {\begingroup\renewcommand\colorMATH{\colorMATHM}\renewcommand\colorSYNTAX{\colorSYNTAXM}{{\color{\colorMATH}\ensuremath{{{\color{\colorSYNTAX}\texttt{\ensuremath{{{\color{\colorSYNTAX}\texttt{matrix}}}_{L\infty }^{U}[{{\color{\colorMATH}\ensuremath{1}}},{{\color{\colorMATH}\ensuremath{\eta _{n}}}}]\hspace*{0.33em}{{\color{\colorSYNTAX}\texttt{data}}}}}}}}}}\endgroup }
  }
\and\inferrule*[lab={{\color{\colorTEXT}\textsc{\scriptsize Matrix Map}}}
  ]{ {\begingroup\renewcommand\colorMATH{\colorMATHM}\renewcommand\colorSYNTAX{\colorSYNTAXM}{{\color{\colorMATH}\ensuremath{\Delta }}}\endgroup },\Gamma _{1} \vdash  e_{1} \mathrel{:} {\begingroup\renewcommand\colorMATH{\colorMATHM}\renewcommand\colorSYNTAX{\colorSYNTAXM}{{\color{\colorMATH}\ensuremath{{{\color{\colorSYNTAX}\texttt{\ensuremath{{{\color{\colorSYNTAX}\texttt{matrix}}}_{{{\color{\colorMATH}\ensuremath{\ell }}}}^{{{\color{\colorMATH}\ensuremath{\bigstar }}}}[{{\color{\colorMATH}\ensuremath{\eta _{m}}}},{{\color{\colorMATH}\ensuremath{\eta _{n}}}}]\hspace*{0.33em}{{\color{\colorMATH}\ensuremath{\tau _{1}}}}}}}}}}}\endgroup }
  \\ {\begingroup\renewcommand\colorMATH{\colorMATHM}\renewcommand\colorSYNTAX{\colorSYNTAXM}{{\color{\colorMATH}\ensuremath{\Delta }}}\endgroup },\Gamma _{2}\uplus \{ {\begingroup\renewcommand\colorMATH{\colorMATHM}\renewcommand\colorSYNTAX{\colorSYNTAXM}{{\color{\colorMATH}\ensuremath{x}}}\endgroup }{\mathrel{:}}_{s}{\begingroup\renewcommand\colorMATH{\colorMATHM}\renewcommand\colorSYNTAX{\colorSYNTAXM}{{\color{\colorMATH}\ensuremath{\tau _{1}}}}\endgroup }\}  \vdash  e_{2} \mathrel{:} {\begingroup\renewcommand\colorMATH{\colorMATHM}\renewcommand\colorSYNTAX{\colorSYNTAXM}{{\color{\colorMATH}\ensuremath{\tau _{2}}}}\endgroup }
     }{
     {\begingroup\renewcommand\colorMATH{\colorMATHM}\renewcommand\colorSYNTAX{\colorSYNTAXM}{{\color{\colorMATH}\ensuremath{\Delta }}}\endgroup },s\Gamma _{1} + {\begingroup\renewcommand\colorMATH{\colorMATHM}\renewcommand\colorSYNTAX{\colorSYNTAXM}{{\color{\colorMATH}\ensuremath{\eta _{m}\eta _{n}}}}\endgroup }\Gamma _{2} \vdash  {{\color{\colorSYNTAX}\texttt{\ensuremath{{{\color{\colorSYNTAX}\texttt{mmap}}}\hspace*{0.33em}{{\color{\colorMATH}\ensuremath{e_{1}}}}\hspace*{0.33em}\{ {\begingroup\renewcommand\colorMATH{\colorMATHM}\renewcommand\colorSYNTAX{\colorSYNTAXM}{{\color{\colorMATH}\ensuremath{x}}}\endgroup }\Rightarrow {{\color{\colorMATH}\ensuremath{e_{2}}}}\} }}}} \mathrel{:} {\begingroup\renewcommand\colorMATH{\colorMATHM}\renewcommand\colorSYNTAX{\colorSYNTAXM}{{\color{\colorMATH}\ensuremath{{{\color{\colorSYNTAX}\texttt{\ensuremath{{{\color{\colorSYNTAX}\texttt{matrix}}}_{{{\color{\colorMATH}\ensuremath{\ell }}}}^{U}[{{\color{\colorMATH}\ensuremath{\eta _{m}}}},{{\color{\colorMATH}\ensuremath{\eta _{n}}}}]\hspace*{0.33em}{{\color{\colorMATH}\ensuremath{\tau _{2}}}}}}}}}}}\endgroup }
  }
\and\inferrule*[lab={{\color{\colorTEXT}\textsc{\scriptsize Matrix Binary Map}}}
  ]{ {\begingroup\renewcommand\colorMATH{\colorMATHM}\renewcommand\colorSYNTAX{\colorSYNTAXM}{{\color{\colorMATH}\ensuremath{\Delta }}}\endgroup },\Gamma _{1} \vdash  e_{1} \mathrel{:} {\begingroup\renewcommand\colorMATH{\colorMATHM}\renewcommand\colorSYNTAX{\colorSYNTAXM}{{\color{\colorMATH}\ensuremath{{{\color{\colorSYNTAX}\texttt{\ensuremath{{{\color{\colorSYNTAX}\texttt{matrix}}}_{{{\color{\colorMATH}\ensuremath{\ell }}}}^{{{\color{\colorMATH}\ensuremath{\bigstar }}}}[{{\color{\colorMATH}\ensuremath{\eta _{m}}}},{{\color{\colorMATH}\ensuremath{\eta _{n}}}}]\hspace*{0.33em}{{\color{\colorMATH}\ensuremath{\tau _{1}}}}}}}}}}}\endgroup }
  \\ {\begingroup\renewcommand\colorMATH{\colorMATHM}\renewcommand\colorSYNTAX{\colorSYNTAXM}{{\color{\colorMATH}\ensuremath{\Delta }}}\endgroup },\Gamma _{2} \vdash  e_{2} \mathrel{:} {\begingroup\renewcommand\colorMATH{\colorMATHM}\renewcommand\colorSYNTAX{\colorSYNTAXM}{{\color{\colorMATH}\ensuremath{{{\color{\colorSYNTAX}\texttt{\ensuremath{{{\color{\colorSYNTAX}\texttt{matrix}}}_{{{\color{\colorMATH}\ensuremath{\ell }}}}^{{{\color{\colorMATH}\ensuremath{\bigstar }}}}[{{\color{\colorMATH}\ensuremath{\eta _{m}}}},{{\color{\colorMATH}\ensuremath{\eta _{n}}}}]\hspace*{0.33em}{{\color{\colorMATH}\ensuremath{\tau _{2}}}}}}}}}}}\endgroup }
  \\ {\begingroup\renewcommand\colorMATH{\colorMATHM}\renewcommand\colorSYNTAX{\colorSYNTAXM}{{\color{\colorMATH}\ensuremath{\Delta }}}\endgroup },\Gamma _{3}\uplus \{ {\begingroup\renewcommand\colorMATH{\colorMATHM}\renewcommand\colorSYNTAX{\colorSYNTAXM}{{\color{\colorMATH}\ensuremath{x_{1}}}}\endgroup }{\mathrel{:}}_{s_{1}}{\begingroup\renewcommand\colorMATH{\colorMATHM}\renewcommand\colorSYNTAX{\colorSYNTAXM}{{\color{\colorMATH}\ensuremath{\tau _{1}}}}\endgroup },{\begingroup\renewcommand\colorMATH{\colorMATHM}\renewcommand\colorSYNTAX{\colorSYNTAXM}{{\color{\colorMATH}\ensuremath{x_{2}}}}\endgroup }{\mathrel{:}}_{s_{2}}{\begingroup\renewcommand\colorMATH{\colorMATHM}\renewcommand\colorSYNTAX{\colorSYNTAXM}{{\color{\colorMATH}\ensuremath{\tau _{2}}}}\endgroup }\}  \vdash  e_{3} \mathrel{:} {\begingroup\renewcommand\colorMATH{\colorMATHM}\renewcommand\colorSYNTAX{\colorSYNTAXM}{{\color{\colorMATH}\ensuremath{\tau _{3}}}}\endgroup }
     }{
     {\begingroup\renewcommand\colorMATH{\colorMATHM}\renewcommand\colorSYNTAX{\colorSYNTAXM}{{\color{\colorMATH}\ensuremath{\Delta }}}\endgroup },s_{1}\Gamma _{1} + s_{2}\Gamma _{2} + {\begingroup\renewcommand\colorMATH{\colorMATHM}\renewcommand\colorSYNTAX{\colorSYNTAXM}{{\color{\colorMATH}\ensuremath{\eta _{m}\eta _{n}}}}\endgroup }\Gamma _{3} \vdash  {{\color{\colorSYNTAX}\texttt{\ensuremath{{{\color{\colorSYNTAX}\texttt{mmap}}}\hspace*{0.33em}{{\color{\colorMATH}\ensuremath{e_{1}}}},{{\color{\colorMATH}\ensuremath{e_{2}}}}\hspace*{0.33em}\{ {\begingroup\renewcommand\colorMATH{\colorMATHM}\renewcommand\colorSYNTAX{\colorSYNTAXM}{{\color{\colorMATH}\ensuremath{x_{1}}}}\endgroup },{\begingroup\renewcommand\colorMATH{\colorMATHM}\renewcommand\colorSYNTAX{\colorSYNTAXM}{{\color{\colorMATH}\ensuremath{x_{2}}}}\endgroup }\Rightarrow {{\color{\colorMATH}\ensuremath{e_{3}}}}\} }}}} \mathrel{:} {\begingroup\renewcommand\colorMATH{\colorMATHM}\renewcommand\colorSYNTAX{\colorSYNTAXM}{{\color{\colorMATH}\ensuremath{{{\color{\colorSYNTAX}\texttt{\ensuremath{{{\color{\colorSYNTAX}\texttt{matrix}}}_{{{\color{\colorMATH}\ensuremath{\ell }}}}^{U}[{{\color{\colorMATH}\ensuremath{\eta _{m}}}},{{\color{\colorMATH}\ensuremath{\eta _{n}}}}]\hspace*{0.33em}{{\color{\colorMATH}\ensuremath{\tau _{3}}}}}}}}}}}\endgroup }
  }
\and\inferrule*[lab={{\color{\colorTEXT}\textsc{\scriptsize Matrix Map Rows}}}
  ]{ {\begingroup\renewcommand\colorMATH{\colorMATHM}\renewcommand\colorSYNTAX{\colorSYNTAXM}{{\color{\colorMATH}\ensuremath{\Delta }}}\endgroup },\Gamma _{1} \vdash  e_{1} \mathrel{:} {\begingroup\renewcommand\colorMATH{\colorMATHM}\renewcommand\colorSYNTAX{\colorSYNTAXM}{{\color{\colorMATH}\ensuremath{{{\color{\colorSYNTAX}\texttt{\ensuremath{{{\color{\colorSYNTAX}\texttt{matrix}}}^{{{\color{\colorMATH}\ensuremath{c_{1}}}}}_{{{\color{\colorMATH}\ensuremath{\ell _{1}}}}}[{{\color{\colorMATH}\ensuremath{\eta _{m}}}},{{\color{\colorMATH}\ensuremath{\eta _{n_{1}}}}}]\hspace*{0.33em}{{\color{\colorMATH}\ensuremath{\tau _{1}}}}}}}}}}}\endgroup }
  \\ {\begingroup\renewcommand\colorMATH{\colorMATHM}\renewcommand\colorSYNTAX{\colorSYNTAXM}{{\color{\colorMATH}\ensuremath{\Delta }}}\endgroup },\Gamma _{2}\uplus \{ {\begingroup\renewcommand\colorMATH{\colorMATHM}\renewcommand\colorSYNTAX{\colorSYNTAXM}{{\color{\colorMATH}\ensuremath{x}}}\endgroup }{\mathrel{:}}_{s}{\begingroup\renewcommand\colorMATH{\colorMATHM}\renewcommand\colorSYNTAX{\colorSYNTAXM}{{\color{\colorMATH}\ensuremath{{{\color{\colorSYNTAX}\texttt{\ensuremath{{{\color{\colorSYNTAX}\texttt{matrix}}}^{{{\color{\colorMATH}\ensuremath{c_{1}}}}}_{{{\color{\colorMATH}\ensuremath{\ell _{1}}}}}[{{\color{\colorMATH}\ensuremath{1}}},{{\color{\colorMATH}\ensuremath{\eta _{n_{1}}}}}]\hspace*{0.33em}{{\color{\colorMATH}\ensuremath{\tau _{1}}}}}}}}}}}\endgroup }\}  \vdash  e_{2} \mathrel{:} {\begingroup\renewcommand\colorMATH{\colorMATHM}\renewcommand\colorSYNTAX{\colorSYNTAXM}{{\color{\colorMATH}\ensuremath{{{\color{\colorSYNTAX}\texttt{\ensuremath{{{\color{\colorSYNTAX}\texttt{matrix}}}^{{{\color{\colorMATH}\ensuremath{c_{2}}}}}_{{{\color{\colorMATH}\ensuremath{\ell _{2}}}}}[{{\color{\colorMATH}\ensuremath{1}}},{{\color{\colorMATH}\ensuremath{\eta _{n_{2}}}}}]\hspace*{0.33em}{{\color{\colorMATH}\ensuremath{\tau _{2}}}}}}}}}}}\endgroup }
     }{
     {\begingroup\renewcommand\colorMATH{\colorMATHM}\renewcommand\colorSYNTAX{\colorSYNTAXM}{{\color{\colorMATH}\ensuremath{\Delta }}}\endgroup },s\Gamma _{1} + {\begingroup\renewcommand\colorMATH{\colorMATHM}\renewcommand\colorSYNTAX{\colorSYNTAXM}{{\color{\colorMATH}\ensuremath{\eta _{m}}}}\endgroup }\Gamma _{2} \vdash  {{\color{\colorSYNTAX}\texttt{\ensuremath{{{\color{\colorSYNTAX}\texttt{mmap-row}}}\hspace*{0.33em}{{\color{\colorMATH}\ensuremath{e_{1}}}}\hspace*{0.33em}\{ {\begingroup\renewcommand\colorMATH{\colorMATHM}\renewcommand\colorSYNTAX{\colorSYNTAXM}{{\color{\colorMATH}\ensuremath{x}}}\endgroup }\Rightarrow {{\color{\colorMATH}\ensuremath{e_{2}}}}\} }}}} \mathrel{:} {\begingroup\renewcommand\colorMATH{\colorMATHM}\renewcommand\colorSYNTAX{\colorSYNTAXM}{{\color{\colorMATH}\ensuremath{{{\color{\colorSYNTAX}\texttt{\ensuremath{{{\color{\colorSYNTAX}\texttt{matrix}}}^{{{\color{\colorMATH}\ensuremath{c_{2}}}}}_{{{\color{\colorMATH}\ensuremath{\ell _{2}}}}}[{{\color{\colorMATH}\ensuremath{\eta _{m}}}},{{\color{\colorMATH}\ensuremath{\eta _{n_{2}}}}}]\hspace*{0.33em}{{\color{\colorMATH}\ensuremath{\tau _{2}}}}}}}}}}}\endgroup }
  }
\and\inferrule*[lab={{\color{\colorTEXT}\textsc{\scriptsize Matrix Binary Map Rows}}}
  ]{ {\begingroup\renewcommand\colorMATH{\colorMATHM}\renewcommand\colorSYNTAX{\colorSYNTAXM}{{\color{\colorMATH}\ensuremath{\Delta }}}\endgroup },\Gamma _{1} \vdash  e_{1} \mathrel{:} {\begingroup\renewcommand\colorMATH{\colorMATHM}\renewcommand\colorSYNTAX{\colorSYNTAXM}{{\color{\colorMATH}\ensuremath{{{\color{\colorSYNTAX}\texttt{\ensuremath{{{\color{\colorSYNTAX}\texttt{matrix}}}^{{{\color{\colorMATH}\ensuremath{c_{1}}}}}_{{{\color{\colorMATH}\ensuremath{\ell _{1}}}}}[{{\color{\colorMATH}\ensuremath{\eta _{m}}}},{{\color{\colorMATH}\ensuremath{\eta _{n_{1}}}}}]\hspace*{0.33em}{{\color{\colorMATH}\ensuremath{\tau _{1}}}}}}}}}}}\endgroup }
  \\ {\begingroup\renewcommand\colorMATH{\colorMATHM}\renewcommand\colorSYNTAX{\colorSYNTAXM}{{\color{\colorMATH}\ensuremath{\Delta }}}\endgroup },\Gamma _{2} \vdash  e_{2} \mathrel{:} {\begingroup\renewcommand\colorMATH{\colorMATHM}\renewcommand\colorSYNTAX{\colorSYNTAXM}{{\color{\colorMATH}\ensuremath{{{\color{\colorSYNTAX}\texttt{\ensuremath{{{\color{\colorSYNTAX}\texttt{matrix}}}^{{{\color{\colorMATH}\ensuremath{c_{2}}}}}_{{{\color{\colorMATH}\ensuremath{\ell _{2}}}}}[{{\color{\colorMATH}\ensuremath{\eta _{m}}}},{{\color{\colorMATH}\ensuremath{\eta _{n_{2}}}}}]\hspace*{0.33em}{{\color{\colorMATH}\ensuremath{\tau _{2}}}}}}}}}}}\endgroup }
  \\ {\begingroup\renewcommand\colorMATH{\colorMATHM}\renewcommand\colorSYNTAX{\colorSYNTAXM}{{\color{\colorMATH}\ensuremath{\Delta }}}\endgroup },\Gamma _{3}\uplus \{ {\begingroup\renewcommand\colorMATH{\colorMATHM}\renewcommand\colorSYNTAX{\colorSYNTAXM}{{\color{\colorMATH}\ensuremath{x_{1}}}}\endgroup }{\mathrel{:}}_{s_{1}}{\begingroup\renewcommand\colorMATH{\colorMATHM}\renewcommand\colorSYNTAX{\colorSYNTAXM}{{\color{\colorMATH}\ensuremath{{{\color{\colorSYNTAX}\texttt{\ensuremath{{{\color{\colorSYNTAX}\texttt{matrix}}}^{{{\color{\colorMATH}\ensuremath{c_{1}}}}}_{{{\color{\colorMATH}\ensuremath{\ell _{1}}}}}[{{\color{\colorMATH}\ensuremath{1}}},{{\color{\colorMATH}\ensuremath{\eta _{n_{1}}}}}]\hspace*{0.33em}{{\color{\colorMATH}\ensuremath{\tau _{1}}}}}}}}}}}\endgroup },{\begingroup\renewcommand\colorMATH{\colorMATHM}\renewcommand\colorSYNTAX{\colorSYNTAXM}{{\color{\colorMATH}\ensuremath{x_{2}}}}\endgroup }{\mathrel{:}}_{s_{2}}{\begingroup\renewcommand\colorMATH{\colorMATHM}\renewcommand\colorSYNTAX{\colorSYNTAXM}{{\color{\colorMATH}\ensuremath{{{\color{\colorSYNTAX}\texttt{\ensuremath{{{\color{\colorSYNTAX}\texttt{matrix}}}^{{{\color{\colorMATH}\ensuremath{c_{2}}}}}_{{{\color{\colorMATH}\ensuremath{\ell _{2}}}}}[{{\color{\colorMATH}\ensuremath{1}}},{{\color{\colorMATH}\ensuremath{\eta _{n_{2}}}}}]\hspace*{0.33em}{{\color{\colorMATH}\ensuremath{\tau _{2}}}}}}}}}}}\endgroup }\}  
     \vdash  e_{3} \mathrel{:} {\begingroup\renewcommand\colorMATH{\colorMATHM}\renewcommand\colorSYNTAX{\colorSYNTAXM}{{\color{\colorMATH}\ensuremath{{{\color{\colorSYNTAX}\texttt{\ensuremath{{{\color{\colorSYNTAX}\texttt{matrix}}}^{{{\color{\colorMATH}\ensuremath{c_{3}}}}}_{{{\color{\colorMATH}\ensuremath{\ell _{3}}}}}[{{\color{\colorMATH}\ensuremath{1}}},{{\color{\colorMATH}\ensuremath{\eta _{n_{3}}}}}]\hspace*{0.33em}{{\color{\colorMATH}\ensuremath{\tau _{3}}}}}}}}}}}\endgroup }
     }{
     {\begingroup\renewcommand\colorMATH{\colorMATHM}\renewcommand\colorSYNTAX{\colorSYNTAXM}{{\color{\colorMATH}\ensuremath{\Delta }}}\endgroup },s_{1}\Gamma _{1} + s_{2}\Gamma _{2} + {\begingroup\renewcommand\colorMATH{\colorMATHM}\renewcommand\colorSYNTAX{\colorSYNTAXM}{{\color{\colorMATH}\ensuremath{\eta _{m}}}}\endgroup }\Gamma _{3} \vdash  {{\color{\colorSYNTAX}\texttt{\ensuremath{{{\color{\colorSYNTAX}\texttt{mmap-row}}}\hspace*{0.33em}{{\color{\colorMATH}\ensuremath{e_{1}}}},{{\color{\colorMATH}\ensuremath{e_{2}}}}\hspace*{0.33em}\{ {\begingroup\renewcommand\colorMATH{\colorMATHM}\renewcommand\colorSYNTAX{\colorSYNTAXM}{{\color{\colorMATH}\ensuremath{x_{1}}}}\endgroup },{\begingroup\renewcommand\colorMATH{\colorMATHM}\renewcommand\colorSYNTAX{\colorSYNTAXM}{{\color{\colorMATH}\ensuremath{x_{2}}}}\endgroup }\Rightarrow {{\color{\colorMATH}\ensuremath{e_{3}}}}\} }}}} \mathrel{:} {\begingroup\renewcommand\colorMATH{\colorMATHM}\renewcommand\colorSYNTAX{\colorSYNTAXM}{{\color{\colorMATH}\ensuremath{{{\color{\colorSYNTAX}\texttt{\ensuremath{{{\color{\colorSYNTAX}\texttt{matrix}}}^{{{\color{\colorMATH}\ensuremath{c_{3}}}}}_{{{\color{\colorMATH}\ensuremath{\ell _{3}}}}}[{{\color{\colorMATH}\ensuremath{\eta _{m}}}},{{\color{\colorMATH}\ensuremath{\eta _{n_{3}}}}}]\hspace*{0.33em}{{\color{\colorMATH}\ensuremath{\tau _{3}}}}}}}}}}}\endgroup }
  }
\and\inferrule*[lab={{\color{\colorTEXT}\textsc{\scriptsize Matrix Fold Rows}}}
  ]{ {\begingroup\renewcommand\colorMATH{\colorMATHM}\renewcommand\colorSYNTAX{\colorSYNTAXM}{{\color{\colorMATH}\ensuremath{\Delta }}}\endgroup },\Gamma _{1} \vdash  e_{1} \mathrel{:} {\begingroup\renewcommand\colorMATH{\colorMATHM}\renewcommand\colorSYNTAX{\colorSYNTAXM}{{\color{\colorMATH}\ensuremath{{{\color{\colorSYNTAX}\texttt{\ensuremath{{{\color{\colorSYNTAX}\texttt{matrix}}}_{{{\color{\colorMATH}\ensuremath{\ell }}}}^{{{\color{\colorMATH}\ensuremath{c}}}}[{{\color{\colorMATH}\ensuremath{\eta _{m}}}},{{\color{\colorMATH}\ensuremath{\eta _{n}}}}]\hspace*{0.33em}\tau _{1}}}}}}}}\endgroup }
  \\ {\begingroup\renewcommand\colorMATH{\colorMATHM}\renewcommand\colorSYNTAX{\colorSYNTAXM}{{\color{\colorMATH}\ensuremath{\Delta }}}\endgroup },\Gamma _{2} \vdash  e_{2} \mathrel{:} {\begingroup\renewcommand\colorMATH{\colorMATHM}\renewcommand\colorSYNTAX{\colorSYNTAXM}{{\color{\colorMATH}\ensuremath{\tau _{2}}}}\endgroup }
  \\ {\begingroup\renewcommand\colorMATH{\colorMATHM}\renewcommand\colorSYNTAX{\colorSYNTAXM}{{\color{\colorMATH}\ensuremath{\Delta }}}\endgroup },\Gamma _{3}\uplus \{ {\begingroup\renewcommand\colorMATH{\colorMATHM}\renewcommand\colorSYNTAX{\colorSYNTAXM}{{\color{\colorMATH}\ensuremath{x}}}\endgroup }{\mathrel{:}}_{s_{1}}{\begingroup\renewcommand\colorMATH{\colorMATHM}\renewcommand\colorSYNTAX{\colorSYNTAXM}{{\color{\colorMATH}\ensuremath{\tau _{1}}}}\endgroup },{\begingroup\renewcommand\colorMATH{\colorMATHM}\renewcommand\colorSYNTAX{\colorSYNTAXM}{{\color{\colorMATH}\ensuremath{y}}}\endgroup }{\mathrel{:}}_{s_{2}}{\begingroup\renewcommand\colorMATH{\colorMATHM}\renewcommand\colorSYNTAX{\colorSYNTAXM}{{\color{\colorMATH}\ensuremath{\tau _{2}}}}\endgroup }\}  \vdash  e_{3} \mathrel{:} {\begingroup\renewcommand\colorMATH{\colorMATHM}\renewcommand\colorSYNTAX{\colorSYNTAXM}{{\color{\colorMATH}\ensuremath{\tau _{2}}}}\endgroup }
     }{
     {\begingroup\renewcommand\colorMATH{\colorMATHM}\renewcommand\colorSYNTAX{\colorSYNTAXM}{{\color{\colorMATH}\ensuremath{\Delta }}}\endgroup },s_{1}\Gamma _{1} + s_{2}^{{\begingroup\renewcommand\colorMATH{\colorMATHM}\renewcommand\colorSYNTAX{\colorSYNTAXM}{{\color{\colorMATH}\ensuremath{\eta _{m}}}}\endgroup }}\Gamma _{2} + {\begingroup\renewcommand\colorMATH{\colorMATHM}\renewcommand\colorSYNTAX{\colorSYNTAXM}{{\color{\colorMATH}\ensuremath{\eta _{m}}}}\endgroup }\Gamma _{3} \vdash  {{\color{\colorSYNTAX}\texttt{\ensuremath{{{\color{\colorSYNTAX}\texttt{mfold-row}}}\hspace*{0.33em}{{\color{\colorMATH}\ensuremath{e_{1}}}}\hspace*{0.33em}{{\color{\colorSYNTAX}\texttt{on}}}\hspace*{0.33em}{{\color{\colorMATH}\ensuremath{e_{2}}}}\hspace*{0.33em}\{ {\begingroup\renewcommand\colorMATH{\colorMATHM}\renewcommand\colorSYNTAX{\colorSYNTAXM}{{\color{\colorMATH}\ensuremath{x}}}\endgroup },{\begingroup\renewcommand\colorMATH{\colorMATHM}\renewcommand\colorSYNTAX{\colorSYNTAXM}{{\color{\colorMATH}\ensuremath{y}}}\endgroup } \Rightarrow  {{\color{\colorMATH}\ensuremath{e_{3}}}}\} }}}} \mathrel{:} {\begingroup\renewcommand\colorMATH{\colorMATHM}\renewcommand\colorSYNTAX{\colorSYNTAXM}{{\color{\colorMATH}\ensuremath{{{\color{\colorSYNTAX}\texttt{\ensuremath{{{\color{\colorSYNTAX}\texttt{matrix}}}_{{{\color{\colorMATH}\ensuremath{\ell }}}}^{{{\color{\colorMATH}\ensuremath{c}}}}[{{\color{\colorMATH}\ensuremath{1}}},{{\color{\colorMATH}\ensuremath{\eta _{n}}}}]\hspace*{0.33em}{{\color{\colorMATH}\ensuremath{\tau _{2}}}}}}}}}}}\endgroup }
  }
\end{mathpar}\endgroup 
\endgroup 
\caption{Sensitivity Typing---Matrices}
\label{fig:apx:sens-typing-matrices}
\end{figure*}

\begin{figure*}
\begingroup\renewcommand\colorMATH{\colorMATHS}\renewcommand\colorSYNTAX{\colorSYNTAXS}
\vspace*{-0.25em}\begingroup\color{\colorMATH}\begin{gather*}\begin{tabularx}{\linewidth}{>{\centering\arraybackslash\(}X<{\)}}\hfill\hspace{0pt}\begingroup\color{\colorTEXT}\boxed{\begingroup\color{\colorMATH} {\begingroup\renewcommand\colorMATH{\colorMATHM}\renewcommand\colorSYNTAX{\colorSYNTAXM}{{\color{\colorMATH}\ensuremath{\Delta }}}\endgroup },\Gamma  \vdash  e \mathrel{:} {\begingroup\renewcommand\colorMATH{\colorMATHM}\renewcommand\colorSYNTAX{\colorSYNTAXM}{{\color{\colorMATH}\ensuremath{\tau }}}\endgroup } \endgroup}\endgroup \end{tabularx}\vspace*{-1em}\end{gather*}\endgroup 
\begingroup\color{\colorMATH}\begin{mathpar}\inferrule*[lab={{\color{\colorTEXT}\textsc{\scriptsize If}}}
  ]{ {\begingroup\renewcommand\colorMATH{\colorMATHM}\renewcommand\colorSYNTAX{\colorSYNTAXM}{{\color{\colorMATH}\ensuremath{\Delta }}}\endgroup },\Gamma _{1} \vdash  e_{1} \mathrel{:} {\begingroup\renewcommand\colorMATH{\colorMATHM}\renewcommand\colorSYNTAX{\colorSYNTAXM}{{\color{\colorMATH}\ensuremath{{{\color{\colorSYNTAX}\texttt{\ensuremath{{\mathbb{N}}}}}}}}}\endgroup }
  \\ {\begingroup\renewcommand\colorMATH{\colorMATHM}\renewcommand\colorSYNTAX{\colorSYNTAXM}{{\color{\colorMATH}\ensuremath{\Delta }}}\endgroup },\Gamma _{2} \vdash  e_{2} \mathrel{:} {\begingroup\renewcommand\colorMATH{\colorMATHM}\renewcommand\colorSYNTAX{\colorSYNTAXM}{{\color{\colorMATH}\ensuremath{\tau }}}\endgroup }
  \\ {\begingroup\renewcommand\colorMATH{\colorMATHM}\renewcommand\colorSYNTAX{\colorSYNTAXM}{{\color{\colorMATH}\ensuremath{\Delta }}}\endgroup },\Gamma _{3} \vdash  e_{3} \mathrel{:} {\begingroup\renewcommand\colorMATH{\colorMATHM}\renewcommand\colorSYNTAX{\colorSYNTAXM}{{\color{\colorMATH}\ensuremath{\tau }}}\endgroup }
     }{
     {\begingroup\renewcommand\colorMATH{\colorMATHM}\renewcommand\colorSYNTAX{\colorSYNTAXM}{{\color{\colorMATH}\ensuremath{\Delta }}}\endgroup },{{\color{\colorSYNTAX}\texttt{\ensuremath{\infty }}}}\Gamma _{1} + \Gamma _{2} + \Gamma _{3} \vdash  {{\color{\colorSYNTAX}\texttt{\ensuremath{{{\color{\colorSYNTAX}\texttt{if}}}\hspace*{0.33em}{{\color{\colorMATH}\ensuremath{e_{1}}}}\hspace*{0.33em}\{ {{\color{\colorMATH}\ensuremath{e_{2}}}}\} \hspace*{0.33em}\{ {{\color{\colorMATH}\ensuremath{e_{3}}}}\} }}}} \mathrel{:} {\begingroup\renewcommand\colorMATH{\colorMATHM}\renewcommand\colorSYNTAX{\colorSYNTAXM}{{\color{\colorMATH}\ensuremath{\tau }}}\endgroup }
  }
\and\inferrule*[lab={{\color{\colorTEXT}\textsc{\scriptsize Static Loop}}}
  ]{ {\begingroup\renewcommand\colorMATH{\colorMATHM}\renewcommand\colorSYNTAX{\colorSYNTAXM}{{\color{\colorMATH}\ensuremath{\Delta }}}\endgroup },\Gamma _{1} \vdash  e_{1} \mathrel{:} {\begingroup\renewcommand\colorMATH{\colorMATHM}\renewcommand\colorSYNTAX{\colorSYNTAXM}{{\color{\colorMATH}\ensuremath{{{\color{\colorSYNTAX}\texttt{\ensuremath{{\mathbb{N}}[{{\color{\colorMATH}\ensuremath{\eta }}}]}}}}}}}\endgroup }
  \\ {\begingroup\renewcommand\colorMATH{\colorMATHM}\renewcommand\colorSYNTAX{\colorSYNTAXM}{{\color{\colorMATH}\ensuremath{\Delta }}}\endgroup },\Gamma _{2} \vdash  e_{2} \mathrel{:} {\begingroup\renewcommand\colorMATH{\colorMATHM}\renewcommand\colorSYNTAX{\colorSYNTAXM}{{\color{\colorMATH}\ensuremath{\tau }}}\endgroup }
  \\ {\begingroup\renewcommand\colorMATH{\colorMATHM}\renewcommand\colorSYNTAX{\colorSYNTAXM}{{\color{\colorMATH}\ensuremath{\Delta }}}\endgroup },\Gamma _{3}\uplus \{ {\begingroup\renewcommand\colorMATH{\colorMATHM}\renewcommand\colorSYNTAX{\colorSYNTAXM}{{\color{\colorMATH}\ensuremath{x_{1}}}}\endgroup }{\mathrel{:}}_{{{\color{\colorSYNTAX}\texttt{\ensuremath{\infty }}}}}{\begingroup\renewcommand\colorMATH{\colorMATHM}\renewcommand\colorSYNTAX{\colorSYNTAXM}{{\color{\colorMATH}\ensuremath{{{\color{\colorSYNTAX}\texttt{\ensuremath{{{\color{\colorSYNTAX}\texttt{idx}}}[{{\color{\colorMATH}\ensuremath{\eta }}}]}}}}}}}\endgroup },{\begingroup\renewcommand\colorMATH{\colorMATHM}\renewcommand\colorSYNTAX{\colorSYNTAXM}{{\color{\colorMATH}\ensuremath{x_{2}}}}\endgroup }{\mathrel{:}}_{s}{\begingroup\renewcommand\colorMATH{\colorMATHM}\renewcommand\colorSYNTAX{\colorSYNTAXM}{{\color{\colorMATH}\ensuremath{\tau }}}\endgroup }\}  \vdash  e_{3} \mathrel{:} {\begingroup\renewcommand\colorMATH{\colorMATHM}\renewcommand\colorSYNTAX{\colorSYNTAXM}{{\color{\colorMATH}\ensuremath{\tau }}}\endgroup }
     }{
     {\begingroup\renewcommand\colorMATH{\colorMATHM}\renewcommand\colorSYNTAX{\colorSYNTAXM}{{\color{\colorMATH}\ensuremath{\Delta }}}\endgroup },{\begingroup\renewcommand\colorMATH{\colorMATHM}\renewcommand\colorSYNTAX{\colorSYNTAXM}{{\color{\colorMATH}\ensuremath{0}}}\endgroup }\Gamma _{1} + s^{{\begingroup\renewcommand\colorMATH{\colorMATHM}\renewcommand\colorSYNTAX{\colorSYNTAXM}{{\color{\colorMATH}\ensuremath{\eta }}}\endgroup }}\Gamma _{2} + {\begingroup\renewcommand\colorMATH{\colorMATHM}\renewcommand\colorSYNTAX{\colorSYNTAXM}{{\color{\colorMATH}\ensuremath{\eta }}}\endgroup }\Gamma _{3} \vdash  {{\color{\colorSYNTAX}\texttt{\ensuremath{{{\color{\colorSYNTAX}\texttt{sloop}}}\hspace*{0.33em}{{\color{\colorMATH}\ensuremath{e_{1}}}}\hspace*{0.33em}{{\color{\colorSYNTAX}\texttt{on}}}\hspace*{0.33em}{{\color{\colorMATH}\ensuremath{e_{2}}}}\hspace*{0.33em}\{ {\begingroup\renewcommand\colorMATH{\colorMATHM}\renewcommand\colorSYNTAX{\colorSYNTAXM}{{\color{\colorMATH}\ensuremath{x_{1}}}}\endgroup },{\begingroup\renewcommand\colorMATH{\colorMATHM}\renewcommand\colorSYNTAX{\colorSYNTAXM}{{\color{\colorMATH}\ensuremath{x_{2}}}}\endgroup } \Rightarrow  {{\color{\colorMATH}\ensuremath{e_{3}}}}\} }}}} \mathrel{:} {\begingroup\renewcommand\colorMATH{\colorMATHM}\renewcommand\colorSYNTAX{\colorSYNTAXM}{{\color{\colorMATH}\ensuremath{\tau }}}\endgroup }
  }
\and\inferrule*[lab={{\color{\colorTEXT}\textsc{\scriptsize Loop}}}
  ]{ {\begingroup\renewcommand\colorMATH{\colorMATHM}\renewcommand\colorSYNTAX{\colorSYNTAXM}{{\color{\colorMATH}\ensuremath{\Delta }}}\endgroup },\Gamma _{1} \vdash  e_{1} \mathrel{:} {\begingroup\renewcommand\colorMATH{\colorMATHM}\renewcommand\colorSYNTAX{\colorSYNTAXM}{{\color{\colorMATH}\ensuremath{{{\color{\colorSYNTAX}\texttt{\ensuremath{{\mathbb{N}}}}}}}}}\endgroup }
  \\ {\begingroup\renewcommand\colorMATH{\colorMATHM}\renewcommand\colorSYNTAX{\colorSYNTAXM}{{\color{\colorMATH}\ensuremath{\Delta }}}\endgroup },\Gamma _{2} \vdash  e_{2} \mathrel{:} {\begingroup\renewcommand\colorMATH{\colorMATHM}\renewcommand\colorSYNTAX{\colorSYNTAXM}{{\color{\colorMATH}\ensuremath{\tau }}}\endgroup }
  \\ {\begingroup\renewcommand\colorMATH{\colorMATHM}\renewcommand\colorSYNTAX{\colorSYNTAXM}{{\color{\colorMATH}\ensuremath{\Delta }}}\endgroup },\Gamma _{3}\uplus \{ {\begingroup\renewcommand\colorMATH{\colorMATHM}\renewcommand\colorSYNTAX{\colorSYNTAXM}{{\color{\colorMATH}\ensuremath{x_{1}}}}\endgroup }{\mathrel{:}}_{{{\color{\colorSYNTAX}\texttt{\ensuremath{\infty }}}}}{\begingroup\renewcommand\colorMATH{\colorMATHM}\renewcommand\colorSYNTAX{\colorSYNTAXM}{{\color{\colorMATH}\ensuremath{{{\color{\colorSYNTAX}\texttt{\ensuremath{{\mathbb{N}}}}}}}}}\endgroup },{\begingroup\renewcommand\colorMATH{\colorMATHM}\renewcommand\colorSYNTAX{\colorSYNTAXM}{{\color{\colorMATH}\ensuremath{x_{2}}}}\endgroup }{\mathrel{:}}_{{\begingroup\renewcommand\colorMATH{\colorMATHM}\renewcommand\colorSYNTAX{\colorSYNTAXM}{{\color{\colorMATH}\ensuremath{1}}}\endgroup }}{\begingroup\renewcommand\colorMATH{\colorMATHM}\renewcommand\colorSYNTAX{\colorSYNTAXM}{{\color{\colorMATH}\ensuremath{\tau }}}\endgroup }\}  \vdash  e_{3} \mathrel{:} {\begingroup\renewcommand\colorMATH{\colorMATHM}\renewcommand\colorSYNTAX{\colorSYNTAXM}{{\color{\colorMATH}\ensuremath{\tau }}}\endgroup }
     }{
     {\begingroup\renewcommand\colorMATH{\colorMATHM}\renewcommand\colorSYNTAX{\colorSYNTAXM}{{\color{\colorMATH}\ensuremath{\Delta }}}\endgroup },\Gamma _{1} + \Gamma _{2} + {{\color{\colorSYNTAX}\texttt{\ensuremath{\infty }}}}\Gamma _{3} \vdash  {{\color{\colorSYNTAX}\texttt{\ensuremath{{{\color{\colorSYNTAX}\texttt{loop}}}\hspace*{0.33em}{{\color{\colorMATH}\ensuremath{e_{1}}}}\hspace*{0.33em}{{\color{\colorSYNTAX}\texttt{on}}}\hspace*{0.33em}{{\color{\colorMATH}\ensuremath{e_{2}}}}\hspace*{0.33em}\{ {\begingroup\renewcommand\colorMATH{\colorMATHM}\renewcommand\colorSYNTAX{\colorSYNTAXM}{{\color{\colorMATH}\ensuremath{x_{1}}}}\endgroup },{\begingroup\renewcommand\colorMATH{\colorMATHM}\renewcommand\colorSYNTAX{\colorSYNTAXM}{{\color{\colorMATH}\ensuremath{x_{2}}}}\endgroup } \Rightarrow  {{\color{\colorMATH}\ensuremath{e_{3}}}}\} }}}} \mathrel{:} {\begingroup\renewcommand\colorMATH{\colorMATHM}\renewcommand\colorSYNTAX{\colorSYNTAXM}{{\color{\colorMATH}\ensuremath{\tau }}}\endgroup }
  }
\and\inferrule*[lab={{\color{\colorTEXT}\textsc{\scriptsize Var}}}
  ]{ }{
     {\begingroup\renewcommand\colorMATH{\colorMATHM}\renewcommand\colorSYNTAX{\colorSYNTAXM}{{\color{\colorMATH}\ensuremath{\Delta }}}\endgroup },\Gamma +\{ {\begingroup\renewcommand\colorMATH{\colorMATHM}\renewcommand\colorSYNTAX{\colorSYNTAXM}{{\color{\colorMATH}\ensuremath{x}}}\endgroup }{\mathrel{:}}_{{\begingroup\renewcommand\colorMATH{\colorMATHM}\renewcommand\colorSYNTAX{\colorSYNTAXM}{{\color{\colorMATH}\ensuremath{1}}}\endgroup }}{\begingroup\renewcommand\colorMATH{\colorMATHM}\renewcommand\colorSYNTAX{\colorSYNTAXM}{{\color{\colorMATH}\ensuremath{\tau }}}\endgroup }\}  \vdash  {\begingroup\renewcommand\colorMATH{\colorMATHM}\renewcommand\colorSYNTAX{\colorSYNTAXM}{{\color{\colorMATH}\ensuremath{x}}}\endgroup } \mathrel{:} {\begingroup\renewcommand\colorMATH{\colorMATHM}\renewcommand\colorSYNTAX{\colorSYNTAXM}{{\color{\colorMATH}\ensuremath{\tau }}}\endgroup }
  }
\and\inferrule*[lab={{\color{\colorTEXT}\textsc{\scriptsize Let}}}
  ]{ {\begingroup\renewcommand\colorMATH{\colorMATHM}\renewcommand\colorSYNTAX{\colorSYNTAXM}{{\color{\colorMATH}\ensuremath{\Delta }}}\endgroup },\Gamma _{1} \vdash  e_{1} \mathrel{:} {\begingroup\renewcommand\colorMATH{\colorMATHM}\renewcommand\colorSYNTAX{\colorSYNTAXM}{{\color{\colorMATH}\ensuremath{\tau _{1}}}}\endgroup }
  \\ {\begingroup\renewcommand\colorMATH{\colorMATHM}\renewcommand\colorSYNTAX{\colorSYNTAXM}{{\color{\colorMATH}\ensuremath{\Delta }}}\endgroup },\Gamma _{2}\uplus \{ {\begingroup\renewcommand\colorMATH{\colorMATHM}\renewcommand\colorSYNTAX{\colorSYNTAXM}{{\color{\colorMATH}\ensuremath{x}}}\endgroup }{\mathrel{:}}_{s}{\begingroup\renewcommand\colorMATH{\colorMATHM}\renewcommand\colorSYNTAX{\colorSYNTAXM}{{\color{\colorMATH}\ensuremath{\tau _{1}}}}\endgroup }\}  \vdash  e_{2} \mathrel{:} {\begingroup\renewcommand\colorMATH{\colorMATHM}\renewcommand\colorSYNTAX{\colorSYNTAXM}{{\color{\colorMATH}\ensuremath{\tau _{2}}}}\endgroup }
     }{
     {\begingroup\renewcommand\colorMATH{\colorMATHM}\renewcommand\colorSYNTAX{\colorSYNTAXM}{{\color{\colorMATH}\ensuremath{\Delta }}}\endgroup },s\Gamma _{1} + \Gamma _{2} \vdash  {{\color{\colorSYNTAX}\texttt{\ensuremath{{{\color{\colorSYNTAX}\texttt{let}}}\hspace*{0.33em}{\begingroup\renewcommand\colorMATH{\colorMATHM}\renewcommand\colorSYNTAX{\colorSYNTAXM}{{\color{\colorMATH}\ensuremath{x}}}\endgroup }={{\color{\colorMATH}\ensuremath{e_{1}}}}\hspace*{0.33em}{{\color{\colorSYNTAX}\texttt{in}}}\hspace*{0.33em}{{\color{\colorMATH}\ensuremath{e_{2}}}}}}}} \mathrel{:} {\begingroup\renewcommand\colorMATH{\colorMATHM}\renewcommand\colorSYNTAX{\colorSYNTAXM}{{\color{\colorMATH}\ensuremath{\tau _{2}}}}\endgroup }
  }
\and\inferrule*[flushleft,lab={{\color{\colorSYNTAX}\texttt{\ensuremath{\multimap }}}}{{\color{\colorTEXT}\textsc{\scriptsize -I}}}
  ]{ {\begingroup\renewcommand\colorMATH{\colorMATHM}\renewcommand\colorSYNTAX{\colorSYNTAXM}{{\color{\colorMATH}\ensuremath{\Delta  \vdash  \tau _{1}}}}\endgroup }
  \\ {\begingroup\renewcommand\colorMATH{\colorMATHM}\renewcommand\colorSYNTAX{\colorSYNTAXM}{{\color{\colorMATH}\ensuremath{\Delta }}}\endgroup },\Gamma \uplus \{ {\begingroup\renewcommand\colorMATH{\colorMATHM}\renewcommand\colorSYNTAX{\colorSYNTAXM}{{\color{\colorMATH}\ensuremath{x}}}\endgroup }{\mathrel{:}}_{s}{\begingroup\renewcommand\colorMATH{\colorMATHM}\renewcommand\colorSYNTAX{\colorSYNTAXM}{{\color{\colorMATH}\ensuremath{\tau _{1}}}}\endgroup }\}  \vdash  e \mathrel{:} {\begingroup\renewcommand\colorMATH{\colorMATHM}\renewcommand\colorSYNTAX{\colorSYNTAXM}{{\color{\colorMATH}\ensuremath{\tau _{2}}}}\endgroup }
     }{
     {\begingroup\renewcommand\colorMATH{\colorMATHM}\renewcommand\colorSYNTAX{\colorSYNTAXM}{{\color{\colorMATH}\ensuremath{\Delta }}}\endgroup },\Gamma  \vdash  ({{\color{\colorSYNTAX}\texttt{\ensuremath{\lambda \hspace*{0.33em}{\begingroup\renewcommand\colorMATH{\colorMATHM}\renewcommand\colorSYNTAX{\colorSYNTAXM}{{\color{\colorMATH}\ensuremath{x}}}\endgroup }{\mathrel{:}}{\begingroup\renewcommand\colorMATH{\colorMATHM}\renewcommand\colorSYNTAX{\colorSYNTAXM}{{\color{\colorMATH}\ensuremath{\tau _{1}}}}\endgroup }\Rightarrow {{\color{\colorMATH}\ensuremath{e}}}}}}}) \mathrel{:} {\begingroup\renewcommand\colorMATH{\colorMATHM}\renewcommand\colorSYNTAX{\colorSYNTAXM}{{\color{\colorMATH}\ensuremath{{{\color{\colorSYNTAX}\texttt{\ensuremath{{{\color{\colorMATH}\ensuremath{\tau _{1}}}} \multimap _{{\begingroup\renewcommand\colorMATH{\colorMATHS}\renewcommand\colorSYNTAX{\colorSYNTAXS}{{\color{\colorMATH}\ensuremath{s}}}\endgroup }} {{\color{\colorMATH}\ensuremath{\tau _{2}}}}}}}}}}}\endgroup }
  }
\and\inferrule*[flushleft,lab={{\color{\colorSYNTAX}\texttt{\ensuremath{\multimap }}}}{{\color{\colorTEXT}\textsc{\scriptsize -E}}}
  ]{ {\begingroup\renewcommand\colorMATH{\colorMATHM}\renewcommand\colorSYNTAX{\colorSYNTAXM}{{\color{\colorMATH}\ensuremath{\Delta }}}\endgroup },\Gamma _{1} \vdash  e_{1} \mathrel{:} {\begingroup\renewcommand\colorMATH{\colorMATHM}\renewcommand\colorSYNTAX{\colorSYNTAXM}{{\color{\colorMATH}\ensuremath{{{\color{\colorSYNTAX}\texttt{\ensuremath{{{\color{\colorMATH}\ensuremath{\tau _{1}}}} \multimap _{{\begingroup\renewcommand\colorMATH{\colorMATHS}\renewcommand\colorSYNTAX{\colorSYNTAXS}{{\color{\colorMATH}\ensuremath{s}}}\endgroup }} {{\color{\colorMATH}\ensuremath{\tau _{2}}}}}}}}}}}\endgroup }
  \\ {\begingroup\renewcommand\colorMATH{\colorMATHM}\renewcommand\colorSYNTAX{\colorSYNTAXM}{{\color{\colorMATH}\ensuremath{\Delta }}}\endgroup },\Gamma _{2} \vdash  e_{2} \mathrel{:} {\begingroup\renewcommand\colorMATH{\colorMATHM}\renewcommand\colorSYNTAX{\colorSYNTAXM}{{\color{\colorMATH}\ensuremath{\tau _{1}}}}\endgroup }
     }{
     {\begingroup\renewcommand\colorMATH{\colorMATHM}\renewcommand\colorSYNTAX{\colorSYNTAXM}{{\color{\colorMATH}\ensuremath{\Delta }}}\endgroup },\Gamma _{1} + s\Gamma _{2} \vdash  (e_{1}\hspace*{0.33em}e_{2}) \mathrel{:} {\begingroup\renewcommand\colorMATH{\colorMATHM}\renewcommand\colorSYNTAX{\colorSYNTAXM}{{\color{\colorMATH}\ensuremath{\tau _{2}}}}\endgroup }
  }
\and\inferrule*[flushleft,lab={{\color{\colorSYNTAX}\texttt{\ensuremath{\multimap ^{*}}}}}{{\color{\colorTEXT}\textsc{\scriptsize -I}}}
  ]{ {\begingroup\renewcommand\colorMATH{\colorMATHM}\renewcommand\colorSYNTAX{\colorSYNTAXM}{{\color{\colorMATH}\ensuremath{\Delta ^{\prime} = \Delta \uplus \{ \beta _{1}{\mathrel{:}}\kappa _{1},{.}\hspace{-1pt}{.}\hspace{-1pt}{.},\beta _{n}{\mathrel{:}}\kappa _{n}\} }}}\endgroup }
  \\ {\begingroup\renewcommand\colorMATH{\colorMATHM}\renewcommand\colorSYNTAX{\colorSYNTAXM}{{\color{\colorMATH}\ensuremath{\Delta ^{\prime} \vdash  \tau _{i}\hspace*{0.33em}(\forall i)}}}\endgroup }
  \\ {\begingroup\renewcommand\colorMATH{\colorMATHP}\renewcommand\colorSYNTAX{\colorSYNTAXP}{{\color{\colorMATH}\ensuremath{{\begingroup\renewcommand\colorMATH{\colorMATHM}\renewcommand\colorSYNTAX{\colorSYNTAXM}{{\color{\colorMATH}\ensuremath{\Delta ^{\prime}}}}\endgroup },\Gamma \uplus \{ {\begingroup\renewcommand\colorMATH{\colorMATHM}\renewcommand\colorSYNTAX{\colorSYNTAXM}{{\color{\colorMATH}\ensuremath{x_{1}}}}\endgroup },{\mathrel{:}}_{p_{1}}{\begingroup\renewcommand\colorMATH{\colorMATHM}\renewcommand\colorSYNTAX{\colorSYNTAXM}{{\color{\colorMATH}\ensuremath{\tau _{1}}}}\endgroup },{.}\hspace{-1pt}{.}\hspace{-1pt}{.},{\begingroup\renewcommand\colorMATH{\colorMATHM}\renewcommand\colorSYNTAX{\colorSYNTAXM}{{\color{\colorMATH}\ensuremath{x_{n}}}}\endgroup }{\mathrel{:}}_{p_{n}}{\begingroup\renewcommand\colorMATH{\colorMATHM}\renewcommand\colorSYNTAX{\colorSYNTAXM}{{\color{\colorMATH}\ensuremath{\tau _{n}}}}\endgroup }\}  \vdash  e \mathrel{:} {\begingroup\renewcommand\colorMATH{\colorMATHM}\renewcommand\colorSYNTAX{\colorSYNTAXM}{{\color{\colorMATH}\ensuremath{\tau }}}\endgroup }}}}\endgroup }
     }{
     {\begingroup\renewcommand\colorMATH{\colorMATHM}\renewcommand\colorSYNTAX{\colorSYNTAXM}{{\color{\colorMATH}\ensuremath{\Delta }}}\endgroup },{}\rceil {\begingroup\renewcommand\colorMATH{\colorMATHP}\renewcommand\colorSYNTAX{\colorSYNTAXP}{{\color{\colorMATH}\ensuremath{\Gamma }}}\endgroup }\lceil {}^{{{\color{\colorSYNTAX}\texttt{\ensuremath{\infty }}}}} \vdash  
     ({{\color{\colorSYNTAX}\texttt{\ensuremath{p\lambda \hspace*{0.33em}[{\begingroup\renewcommand\colorMATH{\colorMATHM}\renewcommand\colorSYNTAX{\colorSYNTAXM}{{\color{\colorMATH}\ensuremath{\beta _{1}}}}\endgroup }{\mathrel{:}}{\begingroup\renewcommand\colorMATH{\colorMATHM}\renewcommand\colorSYNTAX{\colorSYNTAXM}{{\color{\colorMATH}\ensuremath{\kappa _{1}}}}\endgroup },{.}\hspace{-1pt}{.}\hspace{-1pt}{.},{\begingroup\renewcommand\colorMATH{\colorMATHM}\renewcommand\colorSYNTAX{\colorSYNTAXM}{{\color{\colorMATH}\ensuremath{\beta _{n}}}}\endgroup }{\mathrel{:}}{\begingroup\renewcommand\colorMATH{\colorMATHM}\renewcommand\colorSYNTAX{\colorSYNTAXM}{{\color{\colorMATH}\ensuremath{\kappa _{n}}}}\endgroup }]\hspace*{0.33em}({\begingroup\renewcommand\colorMATH{\colorMATHM}\renewcommand\colorSYNTAX{\colorSYNTAXM}{{\color{\colorMATH}\ensuremath{x_{1}}}}\endgroup }{\mathrel{:}}{\begingroup\renewcommand\colorMATH{\colorMATHM}\renewcommand\colorSYNTAX{\colorSYNTAXM}{{\color{\colorMATH}\ensuremath{\tau _{1}}}}\endgroup },{.}\hspace{-1pt}{.}\hspace{-1pt}{.},{\begingroup\renewcommand\colorMATH{\colorMATHM}\renewcommand\colorSYNTAX{\colorSYNTAXM}{{\color{\colorMATH}\ensuremath{x_{n}}}}\endgroup }{\mathrel{:}}{\begingroup\renewcommand\colorMATH{\colorMATHM}\renewcommand\colorSYNTAX{\colorSYNTAXM}{{\color{\colorMATH}\ensuremath{\tau _{n}}}}\endgroup }) \Rightarrow  {\begingroup\renewcommand\colorMATH{\colorMATHP}\renewcommand\colorSYNTAX{\colorSYNTAXP}{{\color{\colorMATH}\ensuremath{e}}}\endgroup }}}}})
     \mathrel{:} {\begingroup\renewcommand\colorMATH{\colorMATHM}\renewcommand\colorSYNTAX{\colorSYNTAXM}{{\color{\colorMATH}\ensuremath{{{\color{\colorSYNTAX}\texttt{\ensuremath{\forall \hspace*{0.33em}[{{\color{\colorMATH}\ensuremath{\beta _{1}}}}{\mathrel{:}}{{\color{\colorMATH}\ensuremath{\kappa _{1}}}},{.}\hspace{-1pt}{.}\hspace{-1pt}{.},{{\color{\colorMATH}\ensuremath{\beta _{n}}}}{\mathrel{:}}{{\color{\colorMATH}\ensuremath{\kappa _{n}}}}]\hspace*{0.33em}({{\color{\colorMATH}\ensuremath{\tau _{1}}}}@{\begingroup\renewcommand\colorMATH{\colorMATHP}\renewcommand\colorSYNTAX{\colorSYNTAXP}{{\color{\colorMATH}\ensuremath{p_{1}}}}\endgroup },{.}\hspace{-1pt}{.}\hspace{-1pt}{.},{{\color{\colorMATH}\ensuremath{\tau _{n}}}}@{\begingroup\renewcommand\colorMATH{\colorMATHP}\renewcommand\colorSYNTAX{\colorSYNTAXP}{{\color{\colorMATH}\ensuremath{p_{n}}}}\endgroup }) \multimap ^{*} {{\color{\colorMATH}\ensuremath{\tau }}}}}}}}}}\endgroup }
  }
\and\inferrule*[lab={{\color{\colorSYNTAX}\texttt{\ensuremath{\uplus }}}}{{\color{\colorTEXT}\textsc{\scriptsize -I-1}}}
  ]{ {\begingroup\renewcommand\colorMATH{\colorMATHM}\renewcommand\colorSYNTAX{\colorSYNTAXM}{{\color{\colorMATH}\ensuremath{\Delta  \vdash  \tau _{2}}}}\endgroup }
  \\ {\begingroup\renewcommand\colorMATH{\colorMATHM}\renewcommand\colorSYNTAX{\colorSYNTAXM}{{\color{\colorMATH}\ensuremath{\Delta }}}\endgroup },\Gamma  \vdash  e \mathrel{:} {\begingroup\renewcommand\colorMATH{\colorMATHM}\renewcommand\colorSYNTAX{\colorSYNTAXM}{{\color{\colorMATH}\ensuremath{\tau _{1}}}}\endgroup }
     }{
     {\begingroup\renewcommand\colorMATH{\colorMATHM}\renewcommand\colorSYNTAX{\colorSYNTAXM}{{\color{\colorMATH}\ensuremath{\Delta }}}\endgroup },\Gamma  \vdash  {{\color{\colorSYNTAX}\texttt{\ensuremath{{{\color{\colorSYNTAX}\texttt{inl}}}[{\begingroup\renewcommand\colorMATH{\colorMATHM}\renewcommand\colorSYNTAX{\colorSYNTAXM}{{\color{\colorMATH}\ensuremath{\tau _{2}}}}\endgroup }]\hspace*{0.33em}{{\color{\colorMATH}\ensuremath{e}}}}}}} \mathrel{:} {\begingroup\renewcommand\colorMATH{\colorMATHM}\renewcommand\colorSYNTAX{\colorSYNTAXM}{{\color{\colorMATH}\ensuremath{{{\color{\colorSYNTAX}\texttt{\ensuremath{{{\color{\colorMATH}\ensuremath{\tau _{1}}}} \uplus  {{\color{\colorMATH}\ensuremath{\tau _{2}}}}}}}}}}}\endgroup }
  }
\and\inferrule*[lab={{\color{\colorSYNTAX}\texttt{\ensuremath{\uplus }}}}{{\color{\colorTEXT}\textsc{\scriptsize -I-2}}}
  ]{ {\begingroup\renewcommand\colorMATH{\colorMATHM}\renewcommand\colorSYNTAX{\colorSYNTAXM}{{\color{\colorMATH}\ensuremath{\Delta  \vdash  \tau _{1}}}}\endgroup }
  \\ {\begingroup\renewcommand\colorMATH{\colorMATHM}\renewcommand\colorSYNTAX{\colorSYNTAXM}{{\color{\colorMATH}\ensuremath{\Delta }}}\endgroup },\Gamma  \vdash  e \mathrel{:} {\begingroup\renewcommand\colorMATH{\colorMATHM}\renewcommand\colorSYNTAX{\colorSYNTAXM}{{\color{\colorMATH}\ensuremath{\tau _{2}}}}\endgroup }
     }{
     {\begingroup\renewcommand\colorMATH{\colorMATHM}\renewcommand\colorSYNTAX{\colorSYNTAXM}{{\color{\colorMATH}\ensuremath{\Delta }}}\endgroup },\Gamma  \vdash  {{\color{\colorSYNTAX}\texttt{\ensuremath{{{\color{\colorSYNTAX}\texttt{inr}}}[{\begingroup\renewcommand\colorMATH{\colorMATHM}\renewcommand\colorSYNTAX{\colorSYNTAXM}{{\color{\colorMATH}\ensuremath{\tau _{1}}}}\endgroup }]\hspace*{0.33em}{{\color{\colorMATH}\ensuremath{e}}}}}}} \mathrel{:} {\begingroup\renewcommand\colorMATH{\colorMATHM}\renewcommand\colorSYNTAX{\colorSYNTAXM}{{\color{\colorMATH}\ensuremath{{{\color{\colorSYNTAX}\texttt{\ensuremath{{{\color{\colorMATH}\ensuremath{\tau _{1}}}} \uplus  {{\color{\colorMATH}\ensuremath{\tau _{2}}}}}}}}}}}\endgroup }
  }
\and\inferrule*[lab={{\color{\colorSYNTAX}\texttt{\ensuremath{\uplus }}}}{{\color{\colorTEXT}\textsc{\scriptsize -E}}}
  ]{ {\begingroup\renewcommand\colorMATH{\colorMATHM}\renewcommand\colorSYNTAX{\colorSYNTAXM}{{\color{\colorMATH}\ensuremath{\Delta }}}\endgroup },\Gamma _{1} \vdash  e_{1} \mathrel{:} {\begingroup\renewcommand\colorMATH{\colorMATHM}\renewcommand\colorSYNTAX{\colorSYNTAXM}{{\color{\colorMATH}\ensuremath{{{\color{\colorSYNTAX}\texttt{\ensuremath{{{\color{\colorMATH}\ensuremath{\tau _{1}}}} \uplus  {{\color{\colorMATH}\ensuremath{\tau _{2}}}}}}}}}}}\endgroup }
  \\ {\begingroup\renewcommand\colorMATH{\colorMATHM}\renewcommand\colorSYNTAX{\colorSYNTAXM}{{\color{\colorMATH}\ensuremath{\Delta }}}\endgroup },\Gamma _{2}\uplus \{ {\begingroup\renewcommand\colorMATH{\colorMATHM}\renewcommand\colorSYNTAX{\colorSYNTAXM}{{\color{\colorMATH}\ensuremath{x_{1}}}}\endgroup }{\mathrel{:}}_{s}{\begingroup\renewcommand\colorMATH{\colorMATHM}\renewcommand\colorSYNTAX{\colorSYNTAXM}{{\color{\colorMATH}\ensuremath{\tau _{1}}}}\endgroup }\}  \vdash  e_{2} \mathrel{:} {\begingroup\renewcommand\colorMATH{\colorMATHM}\renewcommand\colorSYNTAX{\colorSYNTAXM}{{\color{\colorMATH}\ensuremath{\tau _{3}}}}\endgroup }
  \\ {\begingroup\renewcommand\colorMATH{\colorMATHM}\renewcommand\colorSYNTAX{\colorSYNTAXM}{{\color{\colorMATH}\ensuremath{\Delta }}}\endgroup },\Gamma _{2}\uplus \{ {\begingroup\renewcommand\colorMATH{\colorMATHM}\renewcommand\colorSYNTAX{\colorSYNTAXM}{{\color{\colorMATH}\ensuremath{x_{2}}}}\endgroup }{\mathrel{:}}_{s}{\begingroup\renewcommand\colorMATH{\colorMATHM}\renewcommand\colorSYNTAX{\colorSYNTAXM}{{\color{\colorMATH}\ensuremath{\tau _{2}}}}\endgroup }\}  \vdash  e_{3} \mathrel{:} {\begingroup\renewcommand\colorMATH{\colorMATHM}\renewcommand\colorSYNTAX{\colorSYNTAXM}{{\color{\colorMATH}\ensuremath{\tau _{3}}}}\endgroup }
     }{
     {\begingroup\renewcommand\colorMATH{\colorMATHM}\renewcommand\colorSYNTAX{\colorSYNTAXM}{{\color{\colorMATH}\ensuremath{\Delta }}}\endgroup },s\Gamma _{1} + \Gamma _{2} \vdash  {{\color{\colorSYNTAX}\texttt{\ensuremath{{{\color{\colorSYNTAX}\texttt{case}}}\hspace*{0.33em}{{\color{\colorMATH}\ensuremath{e_{1}}}}\hspace*{0.33em}\{ {\begingroup\renewcommand\colorMATH{\colorMATHM}\renewcommand\colorSYNTAX{\colorSYNTAXM}{{\color{\colorMATH}\ensuremath{x_{1}}}}\endgroup }\Rightarrow {{\color{\colorMATH}\ensuremath{e_{2}}}}\} \hspace*{0.33em}\{ {\begingroup\renewcommand\colorMATH{\colorMATHM}\renewcommand\colorSYNTAX{\colorSYNTAXM}{{\color{\colorMATH}\ensuremath{x_{2}}}}\endgroup }\Rightarrow {{\color{\colorMATH}\ensuremath{e_{3}}}}\} }}}} \mathrel{:} {\begingroup\renewcommand\colorMATH{\colorMATHM}\renewcommand\colorSYNTAX{\colorSYNTAXM}{{\color{\colorMATH}\ensuremath{\tau _{3}}}}\endgroup }
  }
\and\inferrule*[lab={{\color{\colorSYNTAX}\texttt{\ensuremath{\mathrel{\&}}}}}{{\color{\colorTEXT}\textsc{\scriptsize -I}}}
  ]{ {\begingroup\renewcommand\colorMATH{\colorMATHM}\renewcommand\colorSYNTAX{\colorSYNTAXM}{{\color{\colorMATH}\ensuremath{\Delta }}}\endgroup },\Gamma  \vdash  e_{1} \mathrel{:} {\begingroup\renewcommand\colorMATH{\colorMATHM}\renewcommand\colorSYNTAX{\colorSYNTAXM}{{\color{\colorMATH}\ensuremath{\tau _{1}}}}\endgroup }
  \\ {\begingroup\renewcommand\colorMATH{\colorMATHM}\renewcommand\colorSYNTAX{\colorSYNTAXM}{{\color{\colorMATH}\ensuremath{\Delta }}}\endgroup },\Gamma  \vdash  e_{2} \mathrel{:} {\begingroup\renewcommand\colorMATH{\colorMATHM}\renewcommand\colorSYNTAX{\colorSYNTAXM}{{\color{\colorMATH}\ensuremath{\tau _{2}}}}\endgroup }
     }{
     {\begingroup\renewcommand\colorMATH{\colorMATHM}\renewcommand\colorSYNTAX{\colorSYNTAXM}{{\color{\colorMATH}\ensuremath{\Delta }}}\endgroup },\Gamma  \vdash  {{\color{\colorSYNTAX}\texttt{\ensuremath{\langle {{\color{\colorMATH}\ensuremath{e_{1}}}},\!\!,{{\color{\colorMATH}\ensuremath{e_{2}}}}\rangle }}}} \mathrel{:} {\begingroup\renewcommand\colorMATH{\colorMATHM}\renewcommand\colorSYNTAX{\colorSYNTAXM}{{\color{\colorMATH}\ensuremath{{{\color{\colorSYNTAX}\texttt{\ensuremath{{{\color{\colorMATH}\ensuremath{\tau _{1}}}} \mathrel{\&} {{\color{\colorMATH}\ensuremath{\tau _{2}}}}}}}}}}}\endgroup }
  }
\and\inferrule*[lab={{\color{\colorSYNTAX}\texttt{\ensuremath{\mathrel{\&}}}}}{{\color{\colorTEXT}\textsc{\scriptsize -E-1}}}
  ]{ {\begingroup\renewcommand\colorMATH{\colorMATHM}\renewcommand\colorSYNTAX{\colorSYNTAXM}{{\color{\colorMATH}\ensuremath{\Delta }}}\endgroup },\Gamma  \vdash  e \mathrel{:} {\begingroup\renewcommand\colorMATH{\colorMATHM}\renewcommand\colorSYNTAX{\colorSYNTAXM}{{\color{\colorMATH}\ensuremath{{{\color{\colorSYNTAX}\texttt{\ensuremath{{{\color{\colorMATH}\ensuremath{\tau _{1}}}} \mathrel{\&} {{\color{\colorMATH}\ensuremath{\tau _{2}}}}}}}}}}}\endgroup }
     }{
     {\begingroup\renewcommand\colorMATH{\colorMATHM}\renewcommand\colorSYNTAX{\colorSYNTAXM}{{\color{\colorMATH}\ensuremath{\Delta }}}\endgroup },\Gamma  \vdash  {{\color{\colorSYNTAX}\texttt{prl}}}\hspace*{0.33em}e \mathrel{:} {\begingroup\renewcommand\colorMATH{\colorMATHM}\renewcommand\colorSYNTAX{\colorSYNTAXM}{{\color{\colorMATH}\ensuremath{\tau _{1}}}}\endgroup }
  }
\and\inferrule*[lab={{\color{\colorSYNTAX}\texttt{\ensuremath{\mathrel{\&}}}}}{{\color{\colorTEXT}\textsc{\scriptsize -E-2}}}
  ]{ {\begingroup\renewcommand\colorMATH{\colorMATHM}\renewcommand\colorSYNTAX{\colorSYNTAXM}{{\color{\colorMATH}\ensuremath{\Delta }}}\endgroup },\Gamma  \vdash  e \mathrel{:} {\begingroup\renewcommand\colorMATH{\colorMATHM}\renewcommand\colorSYNTAX{\colorSYNTAXM}{{\color{\colorMATH}\ensuremath{{{\color{\colorSYNTAX}\texttt{\ensuremath{{{\color{\colorMATH}\ensuremath{\tau _{1}}}} \mathrel{\&} {{\color{\colorMATH}\ensuremath{\tau _{2}}}}}}}}}}}\endgroup }
     }{
     {\begingroup\renewcommand\colorMATH{\colorMATHM}\renewcommand\colorSYNTAX{\colorSYNTAXM}{{\color{\colorMATH}\ensuremath{\Delta }}}\endgroup },\Gamma  \vdash  {{\color{\colorSYNTAX}\texttt{prr}}}\hspace*{0.33em}e \mathrel{:} {\begingroup\renewcommand\colorMATH{\colorMATHM}\renewcommand\colorSYNTAX{\colorSYNTAXM}{{\color{\colorMATH}\ensuremath{\tau _{2}}}}\endgroup }
  }
\and\inferrule*[lab={{\color{\colorSYNTAX}\texttt{\ensuremath{\times }}}}{{\color{\colorTEXT}\textsc{\scriptsize -I}}}
  ]{ {\begingroup\renewcommand\colorMATH{\colorMATHM}\renewcommand\colorSYNTAX{\colorSYNTAXM}{{\color{\colorMATH}\ensuremath{\Delta }}}\endgroup },\Gamma _{1} \vdash  e_{1} \mathrel{:} {\begingroup\renewcommand\colorMATH{\colorMATHM}\renewcommand\colorSYNTAX{\colorSYNTAXM}{{\color{\colorMATH}\ensuremath{\tau _{1}}}}\endgroup }
  \\ {\begingroup\renewcommand\colorMATH{\colorMATHM}\renewcommand\colorSYNTAX{\colorSYNTAXM}{{\color{\colorMATH}\ensuremath{\Delta }}}\endgroup },\Gamma _{2} \vdash  e_{2} \mathrel{:} {\begingroup\renewcommand\colorMATH{\colorMATHM}\renewcommand\colorSYNTAX{\colorSYNTAXM}{{\color{\colorMATH}\ensuremath{\tau _{2}}}}\endgroup }
     }{
     {\begingroup\renewcommand\colorMATH{\colorMATHM}\renewcommand\colorSYNTAX{\colorSYNTAXM}{{\color{\colorMATH}\ensuremath{\Delta }}}\endgroup },\Gamma _{1}+\Gamma _{2} \vdash  {{\color{\colorSYNTAX}\texttt{\ensuremath{\langle {{\color{\colorMATH}\ensuremath{e_{1}}}},{{\color{\colorMATH}\ensuremath{e_{2}}}}\rangle }}}} \mathrel{:} {\begingroup\renewcommand\colorMATH{\colorMATHM}\renewcommand\colorSYNTAX{\colorSYNTAXM}{{\color{\colorMATH}\ensuremath{{{\color{\colorSYNTAX}\texttt{\ensuremath{{{\color{\colorMATH}\ensuremath{\tau _{1}}}} \times  {{\color{\colorMATH}\ensuremath{\tau _{2}}}}}}}}}}}\endgroup }
  }
\and\inferrule*[lab={{\color{\colorSYNTAX}\texttt{\ensuremath{\times }}}}{{\color{\colorTEXT}\textsc{\scriptsize -E}}}
  ]{ {\begingroup\renewcommand\colorMATH{\colorMATHM}\renewcommand\colorSYNTAX{\colorSYNTAXM}{{\color{\colorMATH}\ensuremath{\Delta }}}\endgroup },\Gamma _{1} \vdash  e_{1} \mathrel{:} {\begingroup\renewcommand\colorMATH{\colorMATHM}\renewcommand\colorSYNTAX{\colorSYNTAXM}{{\color{\colorMATH}\ensuremath{{{\color{\colorSYNTAX}\texttt{\ensuremath{{{\color{\colorMATH}\ensuremath{\tau _{1}}}} \times  {{\color{\colorMATH}\ensuremath{\tau _{2}}}}}}}}}}}\endgroup }
  \\ {\begingroup\renewcommand\colorMATH{\colorMATHM}\renewcommand\colorSYNTAX{\colorSYNTAXM}{{\color{\colorMATH}\ensuremath{\Delta }}}\endgroup },\Gamma _{2}\uplus \{ {\begingroup\renewcommand\colorMATH{\colorMATHM}\renewcommand\colorSYNTAX{\colorSYNTAXM}{{\color{\colorMATH}\ensuremath{x_{1}}}}\endgroup }{\mathrel{:}}_{s}{\begingroup\renewcommand\colorMATH{\colorMATHM}\renewcommand\colorSYNTAX{\colorSYNTAXM}{{\color{\colorMATH}\ensuremath{\tau _{1}}}}\endgroup },{\begingroup\renewcommand\colorMATH{\colorMATHM}\renewcommand\colorSYNTAX{\colorSYNTAXM}{{\color{\colorMATH}\ensuremath{x_{2}}}}\endgroup }{\mathrel{:}}_{s}{\begingroup\renewcommand\colorMATH{\colorMATHM}\renewcommand\colorSYNTAX{\colorSYNTAXM}{{\color{\colorMATH}\ensuremath{\tau _{2}}}}\endgroup }\}  \vdash  e_{2} \mathrel{:} {\begingroup\renewcommand\colorMATH{\colorMATHM}\renewcommand\colorSYNTAX{\colorSYNTAXM}{{\color{\colorMATH}\ensuremath{\tau _{3}}}}\endgroup }
     }{
     {\begingroup\renewcommand\colorMATH{\colorMATHM}\renewcommand\colorSYNTAX{\colorSYNTAXM}{{\color{\colorMATH}\ensuremath{\Delta }}}\endgroup },s\Gamma _{1} + \Gamma _{2} \vdash  {{\color{\colorSYNTAX}\texttt{\ensuremath{{{\color{\colorSYNTAX}\texttt{let}}}\hspace*{0.33em}{\begingroup\renewcommand\colorMATH{\colorMATHM}\renewcommand\colorSYNTAX{\colorSYNTAXM}{{\color{\colorMATH}\ensuremath{x_{1}}}}\endgroup },{\begingroup\renewcommand\colorMATH{\colorMATHM}\renewcommand\colorSYNTAX{\colorSYNTAXM}{{\color{\colorMATH}\ensuremath{x_{2}}}}\endgroup }={{\color{\colorMATH}\ensuremath{e_{1}}}}\hspace*{0.33em}{{\color{\colorSYNTAX}\texttt{in}}}\hspace*{0.33em}{{\color{\colorMATH}\ensuremath{e_{2}}}}}}}} \mathrel{:} {\begingroup\renewcommand\colorMATH{\colorMATHM}\renewcommand\colorSYNTAX{\colorSYNTAXM}{{\color{\colorMATH}\ensuremath{\tau _{3}}}}\endgroup }
  }
\end{mathpar}\endgroup 
\endgroup 
\caption{Sensitivity Typing---Iteration and Connectives}
\label{fig:apx:sens-typing-connectives}
\end{figure*}

\begin{figure*}
\begingroup\renewcommand\colorMATH{\colorMATHP}\renewcommand\colorSYNTAX{\colorSYNTAXP}
\vspace*{-0.25em}\begingroup\color{\colorMATH}\begin{gather*}\begin{tabularx}{\linewidth}{>{\centering\arraybackslash\(}X<{\)}}\hfill\hspace{0pt}\begingroup\color{\colorTEXT}\boxed{\begingroup\color{\colorMATH} {\begingroup\renewcommand\colorMATH{\colorMATHM}\renewcommand\colorSYNTAX{\colorSYNTAXM}{{\color{\colorMATH}\ensuremath{\Delta }}}\endgroup },\Gamma  \vdash  e \mathrel{:} \tau  \endgroup}\endgroup \end{tabularx}\vspace*{-1em}\end{gather*}\endgroup 
\begingroup\color{\colorMATH}\begin{mathpar}\inferrule*[lab={{\color{\colorTEXT}\textsc{\scriptsize Return}}}
  ]{ {\begingroup\renewcommand\colorMATH{\colorMATHS}\renewcommand\colorSYNTAX{\colorSYNTAXS}{{\color{\colorMATH}\ensuremath{ {\begingroup\renewcommand\colorMATH{\colorMATHM}\renewcommand\colorSYNTAX{\colorSYNTAXM}{{\color{\colorMATH}\ensuremath{\Delta }}}\endgroup },\Gamma  \vdash  e \mathrel{:} {\begingroup\renewcommand\colorMATH{\colorMATHM}\renewcommand\colorSYNTAX{\colorSYNTAXM}{{\color{\colorMATH}\ensuremath{\tau }}}\endgroup } }}}\endgroup }
     }{
     {\begingroup\renewcommand\colorMATH{\colorMATHM}\renewcommand\colorSYNTAX{\colorSYNTAXM}{{\color{\colorMATH}\ensuremath{\Delta }}}\endgroup },{}\rceil {\begingroup\renewcommand\colorMATH{\colorMATHS}\renewcommand\colorSYNTAX{\colorSYNTAXS}{{\color{\colorMATH}\ensuremath{\Gamma }}}\endgroup }\lceil {}^{{{\color{\colorSYNTAX}\texttt{\ensuremath{\infty }}}}} \vdash  {{\color{\colorSYNTAX}\texttt{return}}}\hspace*{0.33em}{\begingroup\renewcommand\colorMATH{\colorMATHS}\renewcommand\colorSYNTAX{\colorSYNTAXS}{{\color{\colorMATH}\ensuremath{e}}}\endgroup } \mathrel{:} {\begingroup\renewcommand\colorMATH{\colorMATHM}\renewcommand\colorSYNTAX{\colorSYNTAXM}{{\color{\colorMATH}\ensuremath{\tau }}}\endgroup }
  }
\and\inferrule*[lab={{\color{\colorTEXT}\textsc{\scriptsize Bind}}}
  ]{ {\begingroup\renewcommand\colorMATH{\colorMATHM}\renewcommand\colorSYNTAX{\colorSYNTAXM}{{\color{\colorMATH}\ensuremath{\Delta }}}\endgroup },\Gamma _{1} \vdash  e_{1} \mathrel{:} {\begingroup\renewcommand\colorMATH{\colorMATHM}\renewcommand\colorSYNTAX{\colorSYNTAXM}{{\color{\colorMATH}\ensuremath{\tau _{1}}}}\endgroup }
  \\ {\begingroup\renewcommand\colorMATH{\colorMATHM}\renewcommand\colorSYNTAX{\colorSYNTAXM}{{\color{\colorMATH}\ensuremath{\Delta }}}\endgroup },\Gamma _{2}\uplus \{ {\begingroup\renewcommand\colorMATH{\colorMATHM}\renewcommand\colorSYNTAX{\colorSYNTAXM}{{\color{\colorMATH}\ensuremath{x}}}\endgroup }{\mathrel{:}}_{\infty }{\begingroup\renewcommand\colorMATH{\colorMATHM}\renewcommand\colorSYNTAX{\colorSYNTAXM}{{\color{\colorMATH}\ensuremath{\tau _{1}}}}\endgroup }\}  \vdash  e_{2} \mathrel{:} {\begingroup\renewcommand\colorMATH{\colorMATHM}\renewcommand\colorSYNTAX{\colorSYNTAXM}{{\color{\colorMATH}\ensuremath{\tau _{2}}}}\endgroup }
     }{
     {\begingroup\renewcommand\colorMATH{\colorMATHM}\renewcommand\colorSYNTAX{\colorSYNTAXM}{{\color{\colorMATH}\ensuremath{\Delta }}}\endgroup },\Gamma _{1} + \Gamma _{2} \vdash  {{\color{\colorSYNTAX}\texttt{\ensuremath{{\begingroup\renewcommand\colorMATH{\colorMATHM}\renewcommand\colorSYNTAX{\colorSYNTAXM}{{\color{\colorMATH}\ensuremath{x}}}\endgroup }\leftarrow {{\color{\colorMATH}\ensuremath{e_{1}}}}\mathrel{;}{{\color{\colorMATH}\ensuremath{e_{2}}}}}}}} \mathrel{:} {\begingroup\renewcommand\colorMATH{\colorMATHM}\renewcommand\colorSYNTAX{\colorSYNTAXM}{{\color{\colorMATH}\ensuremath{\tau _{2}}}}\endgroup }
  }
\and\inferrule*[flushleft,lab={{\color{\colorSYNTAX}\texttt{\ensuremath{\multimap ^{*}}}}}{{\color{\colorTEXT}\textsc{\scriptsize -E}}}
  ]{ {\begingroup\renewcommand\colorMATH{\colorMATHM}\renewcommand\colorSYNTAX{\colorSYNTAXM}{{\color{\colorMATH}\ensuremath{\Delta  \vdash  \eta _{i} \mathrel{:} \kappa _{i}\hspace*{0.33em}(\forall i)}}}\endgroup }
  \\ {\begingroup\renewcommand\colorMATH{\colorMATHS}\renewcommand\colorSYNTAX{\colorSYNTAXS}{{\color{\colorMATH}\ensuremath{{\begingroup\renewcommand\colorMATH{\colorMATHM}\renewcommand\colorSYNTAX{\colorSYNTAXM}{{\color{\colorMATH}\ensuremath{\Delta }}}\endgroup },\Gamma  \vdash  e \mathrel{:} {\begingroup\renewcommand\colorMATH{\colorMATHM}\renewcommand\colorSYNTAX{\colorSYNTAXM}{{\color{\colorMATH}\ensuremath{{{\color{\colorSYNTAX}\texttt{\ensuremath{\forall \hspace*{0.33em}[{{\color{\colorMATH}\ensuremath{\beta _{1}}}}{\mathrel{:}}{{\color{\colorMATH}\ensuremath{\kappa _{1}}}},{.}\hspace{-1pt}{.}\hspace{-1pt}{.},{{\color{\colorMATH}\ensuremath{\beta _{n}}}}{\mathrel{:}}{{\color{\colorMATH}\ensuremath{\kappa _{n}}}}]\hspace*{0.33em}({{\color{\colorMATH}\ensuremath{\tau _{1}}}}@{\begingroup\renewcommand\colorMATH{\colorMATHP}\renewcommand\colorSYNTAX{\colorSYNTAXP}{{\color{\colorMATH}\ensuremath{p_{1}}}}\endgroup },{.}\hspace{-1pt}{.}\hspace{-1pt}{.},{{\color{\colorMATH}\ensuremath{\tau _{n}}}}@{\begingroup\renewcommand\colorMATH{\colorMATHP}\renewcommand\colorSYNTAX{\colorSYNTAXP}{{\color{\colorMATH}\ensuremath{p_{n}}}}\endgroup }) \multimap ^{*} {{\color{\colorMATH}\ensuremath{\tau }}}}}}}}}}\endgroup }}}}\endgroup }
     }{
     {\begingroup\renewcommand\colorMATH{\colorMATHM}\renewcommand\colorSYNTAX{\colorSYNTAXM}{{\color{\colorMATH}\ensuremath{\Delta }}}\endgroup },{}\rceil {\begingroup\renewcommand\colorMATH{\colorMATHS}\renewcommand\colorSYNTAX{\colorSYNTAXS}{{\color{\colorMATH}\ensuremath{\Gamma }}}\endgroup }\lceil {}^{{{\color{\colorSYNTAX}\texttt{\ensuremath{\infty }}}}} + \{ {\begingroup\renewcommand\colorMATH{\colorMATHM}\renewcommand\colorSYNTAX{\colorSYNTAXM}{{\color{\colorMATH}\ensuremath{x_{1}}}}\endgroup }{\mathrel{:}}_{p_{1}}{\begingroup\renewcommand\colorMATH{\colorMATHM}\renewcommand\colorSYNTAX{\colorSYNTAXM}{{\color{\colorMATH}\ensuremath{\tau _{1}}}}\endgroup },{.}\hspace{-1pt}{.}\hspace{-1pt}{.},{\begingroup\renewcommand\colorMATH{\colorMATHM}\renewcommand\colorSYNTAX{\colorSYNTAXM}{{\color{\colorMATH}\ensuremath{x_{n}}}}\endgroup }{\mathrel{:}}_{p_{n}}{\begingroup\renewcommand\colorMATH{\colorMATHM}\renewcommand\colorSYNTAX{\colorSYNTAXM}{{\color{\colorMATH}\ensuremath{\tau _{n}}}}\endgroup }\}  \vdash  {{\color{\colorSYNTAX}\texttt{\ensuremath{{\begingroup\renewcommand\colorMATH{\colorMATHS}\renewcommand\colorSYNTAX{\colorSYNTAXS}{{\color{\colorMATH}\ensuremath{e}}}\endgroup }[{\begingroup\renewcommand\colorMATH{\colorMATHM}\renewcommand\colorSYNTAX{\colorSYNTAXM}{{\color{\colorMATH}\ensuremath{\eta _{1}}}}\endgroup },{.}\hspace{-1pt}{.}\hspace{-1pt}{.},{\begingroup\renewcommand\colorMATH{\colorMATHM}\renewcommand\colorSYNTAX{\colorSYNTAXM}{{\color{\colorMATH}\ensuremath{\eta _{n}}}}\endgroup }]({\begingroup\renewcommand\colorMATH{\colorMATHM}\renewcommand\colorSYNTAX{\colorSYNTAXM}{{\color{\colorMATH}\ensuremath{x_{1}}}}\endgroup },{.}\hspace{-1pt}{.}\hspace{-1pt}{.},{\begingroup\renewcommand\colorMATH{\colorMATHM}\renewcommand\colorSYNTAX{\colorSYNTAXM}{{\color{\colorMATH}\ensuremath{x_{n}}}}\endgroup })}}}} \mathrel{:} {\begingroup\renewcommand\colorMATH{\colorMATHM}\renewcommand\colorSYNTAX{\colorSYNTAXM}{{\color{\colorMATH}\ensuremath{\tau [\eta _{i}/\beta _{i}]}}}\endgroup }
  }
\and\inferrule*[flushleft,lab={{\color{\colorTEXT}\textsc{\scriptsize Static Loop}}} {{\color{\colorTEXT}\textit{(Advanced Composition)}}}
  ]{ {\begingroup\renewcommand\colorMATH{\colorMATHS}\renewcommand\colorSYNTAX{\colorSYNTAXS}{{\color{\colorMATH}\ensuremath{{\begingroup\renewcommand\colorMATH{\colorMATHM}\renewcommand\colorSYNTAX{\colorSYNTAXM}{{\color{\colorMATH}\ensuremath{\Delta }}}\endgroup },\Gamma _{1} \vdash  e_{1} \mathrel{:} {\begingroup\renewcommand\colorMATH{\colorMATHM}\renewcommand\colorSYNTAX{\colorSYNTAXM}{{\color{\colorMATH}\ensuremath{{{\color{\colorSYNTAX}\texttt{\ensuremath{{\mathbb{R}}^{+}[{{\color{\colorMATH}\ensuremath{\eta _{\delta ^{\prime}}}}}]}}}}}}}\endgroup }}}}\endgroup }
  \\ {\begingroup\renewcommand\colorMATH{\colorMATHS}\renewcommand\colorSYNTAX{\colorSYNTAXS}{{\color{\colorMATH}\ensuremath{{\begingroup\renewcommand\colorMATH{\colorMATHM}\renewcommand\colorSYNTAX{\colorSYNTAXM}{{\color{\colorMATH}\ensuremath{\Delta }}}\endgroup },\Gamma _{2} \vdash  e_{2} \mathrel{:} {\begingroup\renewcommand\colorMATH{\colorMATHM}\renewcommand\colorSYNTAX{\colorSYNTAXM}{{\color{\colorMATH}\ensuremath{{{\color{\colorSYNTAX}\texttt{\ensuremath{{\mathbb{R}}^{+}[{{\color{\colorMATH}\ensuremath{\eta _{n}}}}]}}}}}}}\endgroup }}}}\endgroup }
  \\ {\begingroup\renewcommand\colorMATH{\colorMATHS}\renewcommand\colorSYNTAX{\colorSYNTAXS}{{\color{\colorMATH}\ensuremath{{\begingroup\renewcommand\colorMATH{\colorMATHM}\renewcommand\colorSYNTAX{\colorSYNTAXM}{{\color{\colorMATH}\ensuremath{\Delta }}}\endgroup },\Gamma _{3} \vdash  e_{3} \mathrel{:} {\begingroup\renewcommand\colorMATH{\colorMATHM}\renewcommand\colorSYNTAX{\colorSYNTAXM}{{\color{\colorMATH}\ensuremath{\tau }}}\endgroup }}}}\endgroup }
  \\ {\begingroup\renewcommand\colorMATH{\colorMATHM}\renewcommand\colorSYNTAX{\colorSYNTAXM}{{\color{\colorMATH}\ensuremath{\Delta }}}\endgroup },\Gamma _{4} + \mathrlap{\hspace{-0.5pt}{}\rceil }\lfloor \Gamma _{5}\mathrlap{\hspace{-0.5pt}\rfloor }\lceil {}_{\{ {\begingroup\renewcommand\colorMATH{\colorMATHM}\renewcommand\colorSYNTAX{\colorSYNTAXM}{{\color{\colorMATH}\ensuremath{x_{1}^{\prime}}}}\endgroup },{.}\hspace{-1pt}{.}\hspace{-1pt}{.},{\begingroup\renewcommand\colorMATH{\colorMATHM}\renewcommand\colorSYNTAX{\colorSYNTAXM}{{\color{\colorMATH}\ensuremath{x_{n}^{\prime}}}}\endgroup }\} }^{{\begingroup\renewcommand\colorMATH{\colorMATHM}\renewcommand\colorSYNTAX{\colorSYNTAXM}{{\color{\colorMATH}\ensuremath{\eta _{\epsilon }}}}\endgroup },{\begingroup\renewcommand\colorMATH{\colorMATHM}\renewcommand\colorSYNTAX{\colorSYNTAXM}{{\color{\colorMATH}\ensuremath{\eta _{\delta }}}}\endgroup }} \uplus  \{ {\begingroup\renewcommand\colorMATH{\colorMATHM}\renewcommand\colorSYNTAX{\colorSYNTAXM}{{\color{\colorMATH}\ensuremath{x_{1}}}}\endgroup } {\mathrel{:}}_{\infty } {\begingroup\renewcommand\colorMATH{\colorMATHM}\renewcommand\colorSYNTAX{\colorSYNTAXM}{{\color{\colorMATH}\ensuremath{{{\color{\colorSYNTAX}\texttt{\ensuremath{{\mathbb{N}}}}}}}}}\endgroup },{\begingroup\renewcommand\colorMATH{\colorMATHM}\renewcommand\colorSYNTAX{\colorSYNTAXM}{{\color{\colorMATH}\ensuremath{x_{2}}}}\endgroup } {\mathrel{:}}_{\infty } {\begingroup\renewcommand\colorMATH{\colorMATHM}\renewcommand\colorSYNTAX{\colorSYNTAXM}{{\color{\colorMATH}\ensuremath{\tau }}}\endgroup }\}  \vdash  e_{4} \mathrel{:} {\begingroup\renewcommand\colorMATH{\colorMATHM}\renewcommand\colorSYNTAX{\colorSYNTAXM}{{\color{\colorMATH}\ensuremath{\tau }}}\endgroup }
     }{
     {\begingroup\renewcommand\colorMATH{\colorMATHM}\renewcommand\colorSYNTAX{\colorSYNTAXM}{{\color{\colorMATH}\ensuremath{\Delta }}}\endgroup },{}\rceil {\begingroup\renewcommand\colorMATH{\colorMATHS}\renewcommand\colorSYNTAX{\colorSYNTAXS}{{\color{\colorMATH}\ensuremath{\Gamma _{1} + \Gamma _{2}}}}\endgroup }\lceil {}^{{\begingroup\renewcommand\colorMATH{\colorMATHM}\renewcommand\colorSYNTAX{\colorSYNTAXM}{{\color{\colorMATH}\ensuremath{0}}}\endgroup },{\begingroup\renewcommand\colorMATH{\colorMATHM}\renewcommand\colorSYNTAX{\colorSYNTAXM}{{\color{\colorMATH}\ensuremath{0}}}\endgroup }} + {}\rceil {\begingroup\renewcommand\colorMATH{\colorMATHS}\renewcommand\colorSYNTAX{\colorSYNTAXS}{{\color{\colorMATH}\ensuremath{\Gamma _{3}}}}\endgroup }\lceil {}^{{{\color{\colorSYNTAX}\texttt{\ensuremath{\infty }}}}} + {}\rceil \Gamma _{4}\lceil {}^{{{\color{\colorSYNTAX}\texttt{\ensuremath{\infty }}}}} 
     + \mathrlap{\hspace{-0.5pt}{}\rceil }\lfloor \Gamma _{5}\mathrlap{\hspace{-0.5pt}\rfloor }\lceil {}_{\{ {\begingroup\renewcommand\colorMATH{\colorMATHM}\renewcommand\colorSYNTAX{\colorSYNTAXM}{{\color{\colorMATH}\ensuremath{x_{1}^{\prime}}}}\endgroup },{.}\hspace{-1pt}{.}\hspace{-1pt}{.},{\begingroup\renewcommand\colorMATH{\colorMATHM}\renewcommand\colorSYNTAX{\colorSYNTAXM}{{\color{\colorMATH}\ensuremath{x_{n}^{\prime}}}}\endgroup }\} }^{{\begingroup\renewcommand\colorMATH{\colorMATHM}\renewcommand\colorSYNTAX{\colorSYNTAXM}{{\color{\colorMATH}\ensuremath{{{\color{\colorSYNTAX}\texttt{\ensuremath{{{\color{\colorMATH}\ensuremath{2}}}{\mathrel{\mathord{\cdotp }}}{{\color{\colorMATH}\ensuremath{\eta _{\epsilon }}}}\sqrt {{{\color{\colorMATH}\ensuremath{2}}}{\mathrel{\mathord{\cdotp }}}{{\color{\colorMATH}\ensuremath{\eta _{n}}}}\ln (1/{{\color{\colorMATH}\ensuremath{\eta _{\delta ^{\prime}}}}})}}}}}}}}\endgroup },{\begingroup\renewcommand\colorMATH{\colorMATHM}\renewcommand\colorSYNTAX{\colorSYNTAXM}{{\color{\colorMATH}\ensuremath{{{\color{\colorSYNTAX}\texttt{\ensuremath{{{\color{\colorMATH}\ensuremath{\eta _{\delta ^{\prime}}}}}+{{\color{\colorMATH}\ensuremath{\eta _{n}\eta _{\delta }}}}}}}}}}}\endgroup }} 
     \vdash  {{\color{\colorSYNTAX}\texttt{loop}}}[{\begingroup\renewcommand\colorMATH{\colorMATHS}\renewcommand\colorSYNTAX{\colorSYNTAXS}{{\color{\colorMATH}\ensuremath{e_{1}}}}\endgroup }]\hspace*{0.33em}{\begingroup\renewcommand\colorMATH{\colorMATHS}\renewcommand\colorSYNTAX{\colorSYNTAXS}{{\color{\colorMATH}\ensuremath{e_{2}}}}\endgroup }\hspace*{0.33em}{{\color{\colorSYNTAX}\texttt{on}}}\hspace*{0.33em}{\begingroup\renewcommand\colorMATH{\colorMATHS}\renewcommand\colorSYNTAX{\colorSYNTAXS}{{\color{\colorMATH}\ensuremath{e_{3}}}}\endgroup }\hspace*{0.33em}{<}{\begingroup\renewcommand\colorMATH{\colorMATHM}\renewcommand\colorSYNTAX{\colorSYNTAXM}{{\color{\colorMATH}\ensuremath{x_{1}^{\prime}}}}\endgroup },{.}\hspace{-1pt}{.}\hspace{-1pt}{.},{\begingroup\renewcommand\colorMATH{\colorMATHM}\renewcommand\colorSYNTAX{\colorSYNTAXM}{{\color{\colorMATH}\ensuremath{x_{n}^{\prime}}}}\endgroup }{>}\hspace*{0.33em}\{ {\begingroup\renewcommand\colorMATH{\colorMATHM}\renewcommand\colorSYNTAX{\colorSYNTAXM}{{\color{\colorMATH}\ensuremath{x_{1}}}}\endgroup },{\begingroup\renewcommand\colorMATH{\colorMATHM}\renewcommand\colorSYNTAX{\colorSYNTAXM}{{\color{\colorMATH}\ensuremath{x_{2}}}}\endgroup } \Rightarrow  e_{4}\}  \mathrel{:} {\begingroup\renewcommand\colorMATH{\colorMATHM}\renewcommand\colorSYNTAX{\colorSYNTAXM}{{\color{\colorMATH}\ensuremath{\tau }}}\endgroup }
  }
\and\inferrule*[lab={{\color{\colorTEXT}\textsc{\scriptsize Gauss}}}
  ]{ {\begingroup\renewcommand\colorMATH{\colorMATHS}\renewcommand\colorSYNTAX{\colorSYNTAXS}{{\color{\colorMATH}\ensuremath{\Gamma _{1} \vdash  e_{1} \mathrel{:} {\begingroup\renewcommand\colorMATH{\colorMATHM}\renewcommand\colorSYNTAX{\colorSYNTAXM}{{\color{\colorMATH}\ensuremath{{{\color{\colorSYNTAX}\texttt{\ensuremath{{\mathbb{R}}^{+}[{{\color{\colorMATH}\ensuremath{\eta _{s}}}}]}}}}}}}\endgroup }}}}\endgroup }
  \\ {\begingroup\renewcommand\colorMATH{\colorMATHS}\renewcommand\colorSYNTAX{\colorSYNTAXS}{{\color{\colorMATH}\ensuremath{\Gamma _{2} \vdash  e_{2} \mathrel{:} {\begingroup\renewcommand\colorMATH{\colorMATHM}\renewcommand\colorSYNTAX{\colorSYNTAXM}{{\color{\colorMATH}\ensuremath{{{\color{\colorSYNTAX}\texttt{\ensuremath{{\mathbb{R}}^{+}[{{\color{\colorMATH}\ensuremath{\eta _{\epsilon }}}}]}}}}}}}\endgroup }}}}\endgroup }
  \\ {\begingroup\renewcommand\colorMATH{\colorMATHS}\renewcommand\colorSYNTAX{\colorSYNTAXS}{{\color{\colorMATH}\ensuremath{\Gamma _{3} \vdash  e_{3} \mathrel{:} {\begingroup\renewcommand\colorMATH{\colorMATHM}\renewcommand\colorSYNTAX{\colorSYNTAXM}{{\color{\colorMATH}\ensuremath{{{\color{\colorSYNTAX}\texttt{\ensuremath{{\mathbb{R}}^{+}[{{\color{\colorMATH}\ensuremath{\eta _{\delta }}}}]}}}}}}}\endgroup }}}}\endgroup }
  \\ {\begingroup\renewcommand\colorMATH{\colorMATHS}\renewcommand\colorSYNTAX{\colorSYNTAXS}{{\color{\colorMATH}\ensuremath{\Gamma _{4} + \mathrlap{\hspace{-0.5pt}{}\rceil }\lfloor \Gamma _{5}\mathrlap{\hspace{-0.5pt}\rfloor }\lceil {}_{\{ {\begingroup\renewcommand\colorMATH{\colorMATHM}\renewcommand\colorSYNTAX{\colorSYNTAXM}{{\color{\colorMATH}\ensuremath{x_{1}}}}\endgroup },{.}\hspace{-1pt}{.}\hspace{-1pt}{.},{\begingroup\renewcommand\colorMATH{\colorMATHM}\renewcommand\colorSYNTAX{\colorSYNTAXM}{{\color{\colorMATH}\ensuremath{x_{n}}}}\endgroup }\} }^{{\begingroup\renewcommand\colorMATH{\colorMATHM}\renewcommand\colorSYNTAX{\colorSYNTAXM}{{\color{\colorMATH}\ensuremath{\eta _{s}}}}\endgroup }} \vdash  e_{4} \mathrel{:} {\begingroup\renewcommand\colorMATH{\colorMATHM}\renewcommand\colorSYNTAX{\colorSYNTAXM}{{\color{\colorMATH}\ensuremath{{{\color{\colorSYNTAX}\texttt{\ensuremath{{\mathbb{R}}}}}}}}}\endgroup }}}}\endgroup }
     }{
     {}\rceil {\begingroup\renewcommand\colorMATH{\colorMATHS}\renewcommand\colorSYNTAX{\colorSYNTAXS}{{\color{\colorMATH}\ensuremath{\Gamma _{1} + \Gamma _{2} + \Gamma _{3}}}}\endgroup }\lceil {}^{{\begingroup\renewcommand\colorMATH{\colorMATHM}\renewcommand\colorSYNTAX{\colorSYNTAXM}{{\color{\colorMATH}\ensuremath{0}}}\endgroup },{\begingroup\renewcommand\colorMATH{\colorMATHM}\renewcommand\colorSYNTAX{\colorSYNTAXM}{{\color{\colorMATH}\ensuremath{0}}}\endgroup }} + {}\rceil {\begingroup\renewcommand\colorMATH{\colorMATHS}\renewcommand\colorSYNTAX{\colorSYNTAXS}{{\color{\colorMATH}\ensuremath{\Gamma _{4}}}}\endgroup }\lceil {}^{{{\color{\colorSYNTAX}\texttt{\ensuremath{\infty }}}}} + \mathrlap{\hspace{-0.5pt}{}\rceil }\lfloor {\begingroup\renewcommand\colorMATH{\colorMATHS}\renewcommand\colorSYNTAX{\colorSYNTAXS}{{\color{\colorMATH}\ensuremath{\Gamma _{5}}}}\endgroup }\mathrlap{\hspace{-0.5pt}\rfloor }\lceil {}_{\{ {\begingroup\renewcommand\colorMATH{\colorMATHM}\renewcommand\colorSYNTAX{\colorSYNTAXM}{{\color{\colorMATH}\ensuremath{x_{1}}}}\endgroup },{.}\hspace{-1pt}{.}\hspace{-1pt}{.},{\begingroup\renewcommand\colorMATH{\colorMATHM}\renewcommand\colorSYNTAX{\colorSYNTAXM}{{\color{\colorMATH}\ensuremath{x_{n}}}}\endgroup }\} }^{{\begingroup\renewcommand\colorMATH{\colorMATHM}\renewcommand\colorSYNTAX{\colorSYNTAXM}{{\color{\colorMATH}\ensuremath{\eta _{\epsilon }}}}\endgroup },{\begingroup\renewcommand\colorMATH{\colorMATHM}\renewcommand\colorSYNTAX{\colorSYNTAXM}{{\color{\colorMATH}\ensuremath{\eta _{\delta }}}}\endgroup }} 
     \vdash  {{\color{\colorSYNTAX}\texttt{\ensuremath{{{\color{\colorSYNTAX}\texttt{gauss}}}[{\begingroup\renewcommand\colorMATH{\colorMATHS}\renewcommand\colorSYNTAX{\colorSYNTAXS}{{\color{\colorMATH}\ensuremath{e_{1}}}}\endgroup },{\begingroup\renewcommand\colorMATH{\colorMATHS}\renewcommand\colorSYNTAX{\colorSYNTAXS}{{\color{\colorMATH}\ensuremath{e_{2}}}}\endgroup },{\begingroup\renewcommand\colorMATH{\colorMATHS}\renewcommand\colorSYNTAX{\colorSYNTAXS}{{\color{\colorMATH}\ensuremath{e_{3}}}}\endgroup }]\hspace*{0.33em}{<}{\begingroup\renewcommand\colorMATH{\colorMATHM}\renewcommand\colorSYNTAX{\colorSYNTAXM}{{\color{\colorMATH}\ensuremath{x_{1}}}}\endgroup },{.}\hspace{-1pt}{.}\hspace{-1pt}{.},{\begingroup\renewcommand\colorMATH{\colorMATHM}\renewcommand\colorSYNTAX{\colorSYNTAXM}{{\color{\colorMATH}\ensuremath{x_{n}}}}\endgroup }{>}\hspace*{0.33em}\{ {\begingroup\renewcommand\colorMATH{\colorMATHS}\renewcommand\colorSYNTAX{\colorSYNTAXS}{{\color{\colorMATH}\ensuremath{e_{4}}}}\endgroup }\} }}}} \mathrel{:} {\begingroup\renewcommand\colorMATH{\colorMATHM}\renewcommand\colorSYNTAX{\colorSYNTAXM}{{\color{\colorMATH}\ensuremath{{{\color{\colorSYNTAX}\texttt{\ensuremath{{\mathbb{R}}}}}}}}}\endgroup }
  }
\and\inferrule*[lab={{\color{\colorTEXT}\textsc{\scriptsize MGauss}}}
  ]{ {\begingroup\renewcommand\colorMATH{\colorMATHS}\renewcommand\colorSYNTAX{\colorSYNTAXS}{{\color{\colorMATH}\ensuremath{\Gamma _{1} \vdash  e_{1} \mathrel{:} {\begingroup\renewcommand\colorMATH{\colorMATHM}\renewcommand\colorSYNTAX{\colorSYNTAXM}{{\color{\colorMATH}\ensuremath{{{\color{\colorSYNTAX}\texttt{\ensuremath{{\mathbb{R}}^{+}[{{\color{\colorMATH}\ensuremath{\eta _{s}}}}]}}}}}}}\endgroup }}}}\endgroup }
  \\ {\begingroup\renewcommand\colorMATH{\colorMATHS}\renewcommand\colorSYNTAX{\colorSYNTAXS}{{\color{\colorMATH}\ensuremath{\Gamma _{2} \vdash  e_{2} \mathrel{:} {\begingroup\renewcommand\colorMATH{\colorMATHM}\renewcommand\colorSYNTAX{\colorSYNTAXM}{{\color{\colorMATH}\ensuremath{{{\color{\colorSYNTAX}\texttt{\ensuremath{{\mathbb{R}}^{+}[{{\color{\colorMATH}\ensuremath{\eta _{\epsilon }}}}]}}}}}}}\endgroup }}}}\endgroup }
  \\ {\begingroup\renewcommand\colorMATH{\colorMATHS}\renewcommand\colorSYNTAX{\colorSYNTAXS}{{\color{\colorMATH}\ensuremath{\Gamma _{3} \vdash  e_{3} \mathrel{:} {\begingroup\renewcommand\colorMATH{\colorMATHM}\renewcommand\colorSYNTAX{\colorSYNTAXM}{{\color{\colorMATH}\ensuremath{{{\color{\colorSYNTAX}\texttt{\ensuremath{{\mathbb{R}}^{+}[{{\color{\colorMATH}\ensuremath{\eta _{\delta }}}}]}}}}}}}\endgroup }}}}\endgroup }
  \\ {\begingroup\renewcommand\colorMATH{\colorMATHS}\renewcommand\colorSYNTAX{\colorSYNTAXS}{{\color{\colorMATH}\ensuremath{\Gamma _{4} + \mathrlap{\hspace{-0.5pt}{}\rceil }\lfloor \Gamma _{5}\mathrlap{\hspace{-0.5pt}\rfloor }\lceil {}_{\{ {\begingroup\renewcommand\colorMATH{\colorMATHM}\renewcommand\colorSYNTAX{\colorSYNTAXM}{{\color{\colorMATH}\ensuremath{x_{1}}}}\endgroup },{.}\hspace{-1pt}{.}\hspace{-1pt}{.},{\begingroup\renewcommand\colorMATH{\colorMATHM}\renewcommand\colorSYNTAX{\colorSYNTAXM}{{\color{\colorMATH}\ensuremath{x_{n}}}}\endgroup }\} }^{{\begingroup\renewcommand\colorMATH{\colorMATHM}\renewcommand\colorSYNTAX{\colorSYNTAXM}{{\color{\colorMATH}\ensuremath{\eta _{s}}}}\endgroup }} \vdash  e_{4} \mathrel{:} {\begingroup\renewcommand\colorMATH{\colorMATHM}\renewcommand\colorSYNTAX{\colorSYNTAXM}{{\color{\colorMATH}\ensuremath{{{\color{\colorSYNTAX}\texttt{\ensuremath{{{\color{\colorSYNTAX}\texttt{matrix}}}_{L2}^{{{\color{\colorMATH}\ensuremath{\bigstar }}}}[{{\color{\colorMATH}\ensuremath{\eta _{m}}}},{{\color{\colorMATH}\ensuremath{\eta _{n}}}}]\hspace*{0.33em}{\mathbb{R}}}}}}}}}\endgroup }}}}\endgroup }
     }{
     {}\rceil {\begingroup\renewcommand\colorMATH{\colorMATHS}\renewcommand\colorSYNTAX{\colorSYNTAXS}{{\color{\colorMATH}\ensuremath{\Gamma _{1} + \Gamma _{2} + \Gamma _{3}}}}\endgroup }\lceil {}^{{\begingroup\renewcommand\colorMATH{\colorMATHM}\renewcommand\colorSYNTAX{\colorSYNTAXM}{{\color{\colorMATH}\ensuremath{0}}}\endgroup },{\begingroup\renewcommand\colorMATH{\colorMATHM}\renewcommand\colorSYNTAX{\colorSYNTAXM}{{\color{\colorMATH}\ensuremath{0}}}\endgroup }} + {}\rceil {\begingroup\renewcommand\colorMATH{\colorMATHS}\renewcommand\colorSYNTAX{\colorSYNTAXS}{{\color{\colorMATH}\ensuremath{\Gamma _{4}}}}\endgroup }\lceil {}^{{{\color{\colorSYNTAX}\texttt{\ensuremath{\infty }}}}} + \mathrlap{\hspace{-0.5pt}{}\rceil }\lfloor {\begingroup\renewcommand\colorMATH{\colorMATHS}\renewcommand\colorSYNTAX{\colorSYNTAXS}{{\color{\colorMATH}\ensuremath{\Gamma _{5}}}}\endgroup }\mathrlap{\hspace{-0.5pt}\rfloor }\lceil {}_{\{ {\begingroup\renewcommand\colorMATH{\colorMATHM}\renewcommand\colorSYNTAX{\colorSYNTAXM}{{\color{\colorMATH}\ensuremath{x_{1}}}}\endgroup },{.}\hspace{-1pt}{.}\hspace{-1pt}{.},{\begingroup\renewcommand\colorMATH{\colorMATHM}\renewcommand\colorSYNTAX{\colorSYNTAXM}{{\color{\colorMATH}\ensuremath{x_{n}}}}\endgroup }\} }^{{\begingroup\renewcommand\colorMATH{\colorMATHM}\renewcommand\colorSYNTAX{\colorSYNTAXM}{{\color{\colorMATH}\ensuremath{\eta _{\epsilon }}}}\endgroup },{\begingroup\renewcommand\colorMATH{\colorMATHM}\renewcommand\colorSYNTAX{\colorSYNTAXM}{{\color{\colorMATH}\ensuremath{\eta _{\delta }}}}\endgroup }} 
     \vdash  {{\color{\colorSYNTAX}\texttt{\ensuremath{{{\color{\colorSYNTAX}\texttt{mgauss}}}[{\begingroup\renewcommand\colorMATH{\colorMATHS}\renewcommand\colorSYNTAX{\colorSYNTAXS}{{\color{\colorMATH}\ensuremath{e_{1}}}}\endgroup },{\begingroup\renewcommand\colorMATH{\colorMATHS}\renewcommand\colorSYNTAX{\colorSYNTAXS}{{\color{\colorMATH}\ensuremath{e_{2}}}}\endgroup },{\begingroup\renewcommand\colorMATH{\colorMATHS}\renewcommand\colorSYNTAX{\colorSYNTAXS}{{\color{\colorMATH}\ensuremath{e_{3}}}}\endgroup }]\hspace*{0.33em}{<}{\begingroup\renewcommand\colorMATH{\colorMATHM}\renewcommand\colorSYNTAX{\colorSYNTAXM}{{\color{\colorMATH}\ensuremath{x_{1}}}}\endgroup },{.}\hspace{-1pt}{.}\hspace{-1pt}{.},{\begingroup\renewcommand\colorMATH{\colorMATHM}\renewcommand\colorSYNTAX{\colorSYNTAXM}{{\color{\colorMATH}\ensuremath{x_{n}}}}\endgroup }{>}\hspace*{0.33em}\{ {\begingroup\renewcommand\colorMATH{\colorMATHS}\renewcommand\colorSYNTAX{\colorSYNTAXS}{{\color{\colorMATH}\ensuremath{e_{4}}}}\endgroup }\} }}}} \mathrel{:} {\begingroup\renewcommand\colorMATH{\colorMATHM}\renewcommand\colorSYNTAX{\colorSYNTAXM}{{\color{\colorMATH}\ensuremath{{{\color{\colorSYNTAX}\texttt{\ensuremath{{{\color{\colorSYNTAX}\texttt{matrix}}}_{L\infty }^{U}[{{\color{\colorMATH}\ensuremath{\eta _{m}}}},{{\color{\colorMATH}\ensuremath{\eta _{n}}}}]\hspace*{0.33em}{\mathbb{R}}}}}}}}}\endgroup }
  }
\and\inferrule*[lab={{\color{\colorTEXT}\textsc{\scriptsize Laplace}}}
  ]{ {\begingroup\renewcommand\colorMATH{\colorMATHS}\renewcommand\colorSYNTAX{\colorSYNTAXS}{{\color{\colorMATH}\ensuremath{\Gamma _{1} \vdash  e_{1} \mathrel{:} {\begingroup\renewcommand\colorMATH{\colorMATHM}\renewcommand\colorSYNTAX{\colorSYNTAXM}{{\color{\colorMATH}\ensuremath{{{\color{\colorSYNTAX}\texttt{\ensuremath{{\mathbb{R}}^{+}[{{\color{\colorMATH}\ensuremath{\eta _{s}}}}]}}}}}}}\endgroup }}}}\endgroup }
  \\ {\begingroup\renewcommand\colorMATH{\colorMATHS}\renewcommand\colorSYNTAX{\colorSYNTAXS}{{\color{\colorMATH}\ensuremath{\Gamma _{2} \vdash  e_{2} \mathrel{:} {\begingroup\renewcommand\colorMATH{\colorMATHM}\renewcommand\colorSYNTAX{\colorSYNTAXM}{{\color{\colorMATH}\ensuremath{{{\color{\colorSYNTAX}\texttt{\ensuremath{{\mathbb{R}}^{+}[{{\color{\colorMATH}\ensuremath{\eta _{\epsilon }}}}]}}}}}}}\endgroup }}}}\endgroup }
  \\ {\begingroup\renewcommand\colorMATH{\colorMATHS}\renewcommand\colorSYNTAX{\colorSYNTAXS}{{\color{\colorMATH}\ensuremath{\Gamma _{3} + \mathrlap{\hspace{-0.5pt}{}\rceil }\lfloor \Gamma _{4}\mathrlap{\hspace{-0.5pt}\rfloor }\lceil {}_{\{ {\begingroup\renewcommand\colorMATH{\colorMATHM}\renewcommand\colorSYNTAX{\colorSYNTAXM}{{\color{\colorMATH}\ensuremath{x_{1}}}}\endgroup },{.}\hspace{-1pt}{.}\hspace{-1pt}{.},{\begingroup\renewcommand\colorMATH{\colorMATHM}\renewcommand\colorSYNTAX{\colorSYNTAXM}{{\color{\colorMATH}\ensuremath{x_{n}}}}\endgroup }\} }^{{\begingroup\renewcommand\colorMATH{\colorMATHM}\renewcommand\colorSYNTAX{\colorSYNTAXM}{{\color{\colorMATH}\ensuremath{\eta _{s}}}}\endgroup }} \vdash  e_{3} \mathrel{:} {\begingroup\renewcommand\colorMATH{\colorMATHM}\renewcommand\colorSYNTAX{\colorSYNTAXM}{{\color{\colorMATH}\ensuremath{{{\color{\colorSYNTAX}\texttt{\ensuremath{{\mathbb{R}}}}}}}}}\endgroup } }}}\endgroup }
     }{
     {}\rceil {\begingroup\renewcommand\colorMATH{\colorMATHS}\renewcommand\colorSYNTAX{\colorSYNTAXS}{{\color{\colorMATH}\ensuremath{\Gamma _{1} + \Gamma _{2}}}}\endgroup }\lceil {}^{{\begingroup\renewcommand\colorMATH{\colorMATHM}\renewcommand\colorSYNTAX{\colorSYNTAXM}{{\color{\colorMATH}\ensuremath{0}}}\endgroup },{\begingroup\renewcommand\colorMATH{\colorMATHM}\renewcommand\colorSYNTAX{\colorSYNTAXM}{{\color{\colorMATH}\ensuremath{0}}}\endgroup }} + {}\rceil {\begingroup\renewcommand\colorMATH{\colorMATHS}\renewcommand\colorSYNTAX{\colorSYNTAXS}{{\color{\colorMATH}\ensuremath{\Gamma _{3}}}}\endgroup }\lceil {}^{{{\color{\colorSYNTAX}\texttt{\ensuremath{\infty }}}}} + \mathrlap{\hspace{-0.5pt}{}\rceil }\lfloor {\begingroup\renewcommand\colorMATH{\colorMATHS}\renewcommand\colorSYNTAX{\colorSYNTAXS}{{\color{\colorMATH}\ensuremath{\Gamma _{4}}}}\endgroup }\mathrlap{\hspace{-0.5pt}\rfloor }\lceil {}_{\{ {\begingroup\renewcommand\colorMATH{\colorMATHM}\renewcommand\colorSYNTAX{\colorSYNTAXM}{{\color{\colorMATH}\ensuremath{x_{1}}}}\endgroup },{.}\hspace{-1pt}{.}\hspace{-1pt}{.},{\begingroup\renewcommand\colorMATH{\colorMATHM}\renewcommand\colorSYNTAX{\colorSYNTAXM}{{\color{\colorMATH}\ensuremath{x_{n}}}}\endgroup }\} }^{{\begingroup\renewcommand\colorMATH{\colorMATHM}\renewcommand\colorSYNTAX{\colorSYNTAXM}{{\color{\colorMATH}\ensuremath{\eta _{\epsilon }}}}\endgroup },{\begingroup\renewcommand\colorMATH{\colorMATHM}\renewcommand\colorSYNTAX{\colorSYNTAXM}{{\color{\colorMATH}\ensuremath{0}}}\endgroup }} 
     \vdash  {{\color{\colorSYNTAX}\texttt{\ensuremath{{{\color{\colorSYNTAX}\texttt{laplace}}}[{\begingroup\renewcommand\colorMATH{\colorMATHS}\renewcommand\colorSYNTAX{\colorSYNTAXS}{{\color{\colorMATH}\ensuremath{e_{1}}}}\endgroup },{\begingroup\renewcommand\colorMATH{\colorMATHS}\renewcommand\colorSYNTAX{\colorSYNTAXS}{{\color{\colorMATH}\ensuremath{e_{2}}}}\endgroup }]\hspace*{0.33em}{<}{\begingroup\renewcommand\colorMATH{\colorMATHM}\renewcommand\colorSYNTAX{\colorSYNTAXM}{{\color{\colorMATH}\ensuremath{x_{1}}}}\endgroup },{.}\hspace{-1pt}{.}\hspace{-1pt}{.},{\begingroup\renewcommand\colorMATH{\colorMATHM}\renewcommand\colorSYNTAX{\colorSYNTAXM}{{\color{\colorMATH}\ensuremath{x_{n}}}}\endgroup }{>}\hspace*{0.33em}\{ {\begingroup\renewcommand\colorMATH{\colorMATHS}\renewcommand\colorSYNTAX{\colorSYNTAXS}{{\color{\colorMATH}\ensuremath{e_{3}}}}\endgroup }\} }}}} \mathrel{:} {\begingroup\renewcommand\colorMATH{\colorMATHM}\renewcommand\colorSYNTAX{\colorSYNTAXM}{{\color{\colorMATH}\ensuremath{{{\color{\colorSYNTAX}\texttt{\ensuremath{{\mathbb{R}}}}}}}}}\endgroup }
  }
\and\inferrule*[lab={{\color{\colorTEXT}\textsc{\scriptsize Exponential}}}
  ]{ {\begingroup\renewcommand\colorMATH{\colorMATHS}\renewcommand\colorSYNTAX{\colorSYNTAXS}{{\color{\colorMATH}\ensuremath{\Gamma _{1} \vdash  e_{1} \mathrel{:} {\begingroup\renewcommand\colorMATH{\colorMATHM}\renewcommand\colorSYNTAX{\colorSYNTAXM}{{\color{\colorMATH}\ensuremath{{{\color{\colorSYNTAX}\texttt{\ensuremath{{\mathbb{R}}^{+}[{{\color{\colorMATH}\ensuremath{\eta _{s}}}}]}}}}}}}\endgroup }}}}\endgroup }
  \\ {\begingroup\renewcommand\colorMATH{\colorMATHS}\renewcommand\colorSYNTAX{\colorSYNTAXS}{{\color{\colorMATH}\ensuremath{\Gamma _{2} \vdash  e_{2} \mathrel{:} {\begingroup\renewcommand\colorMATH{\colorMATHM}\renewcommand\colorSYNTAX{\colorSYNTAXM}{{\color{\colorMATH}\ensuremath{{{\color{\colorSYNTAX}\texttt{\ensuremath{{\mathbb{R}}^{+}[{{\color{\colorMATH}\ensuremath{\eta _{\epsilon }}}}]}}}}}}}\endgroup }}}}\endgroup }
  \\ {\begingroup\renewcommand\colorMATH{\colorMATHS}\renewcommand\colorSYNTAX{\colorSYNTAXS}{{\color{\colorMATH}\ensuremath{\Gamma _{3} \vdash  e_{3} \mathrel{:} {\begingroup\renewcommand\colorMATH{\colorMATHM}\renewcommand\colorSYNTAX{\colorSYNTAXM}{{\color{\colorMATH}\ensuremath{{{\color{\colorSYNTAX}\texttt{\ensuremath{{{\color{\colorSYNTAX}\texttt{matrix}}}_{{{\color{\colorMATH}\ensuremath{\bigstar }}}}^{{{\color{\colorMATH}\ensuremath{\bigstar }}}}[{{\color{\colorMATH}\ensuremath{1}}},{{\color{\colorMATH}\ensuremath{\eta _{m}}}}]({{\color{\colorMATH}\ensuremath{\tau }}})}}}}}}}\endgroup }}}}\endgroup }
  \\ {\begingroup\renewcommand\colorMATH{\colorMATHS}\renewcommand\colorSYNTAX{\colorSYNTAXS}{{\color{\colorMATH}\ensuremath{\Gamma _{4} + \mathrlap{\hspace{-0.5pt}{}\rceil }\lfloor \Gamma _{5}\mathrlap{\hspace{-0.5pt}\rfloor }\lceil {}_{\{ {\begingroup\renewcommand\colorMATH{\colorMATHM}\renewcommand\colorSYNTAX{\colorSYNTAXM}{{\color{\colorMATH}\ensuremath{x_{1}}}}\endgroup },{.}\hspace{-1pt}{.}\hspace{-1pt}{.},{\begingroup\renewcommand\colorMATH{\colorMATHM}\renewcommand\colorSYNTAX{\colorSYNTAXM}{{\color{\colorMATH}\ensuremath{x_{n}}}}\endgroup }\} }^{{\begingroup\renewcommand\colorMATH{\colorMATHM}\renewcommand\colorSYNTAX{\colorSYNTAXM}{{\color{\colorMATH}\ensuremath{\eta _{s}}}}\endgroup }}\uplus \{ {\begingroup\renewcommand\colorMATH{\colorMATHM}\renewcommand\colorSYNTAX{\colorSYNTAXM}{{\color{\colorMATH}\ensuremath{x}}}\endgroup }{\mathrel{:}}_{{{\color{\colorSYNTAX}\texttt{\ensuremath{\infty }}}}}{\begingroup\renewcommand\colorMATH{\colorMATHM}\renewcommand\colorSYNTAX{\colorSYNTAXM}{{\color{\colorMATH}\ensuremath{\tau }}}\endgroup }\}  \vdash  e_{4} \mathrel{:} {\begingroup\renewcommand\colorMATH{\colorMATHM}\renewcommand\colorSYNTAX{\colorSYNTAXM}{{\color{\colorMATH}\ensuremath{{{\color{\colorSYNTAX}\texttt{\ensuremath{{\mathbb{R}}}}}}}}}\endgroup } }}}\endgroup }
     }{
     {}\rceil {\begingroup\renewcommand\colorMATH{\colorMATHS}\renewcommand\colorSYNTAX{\colorSYNTAXS}{{\color{\colorMATH}\ensuremath{\Gamma _{1} + \Gamma _{2}}}}\endgroup }\lceil {}^{{\begingroup\renewcommand\colorMATH{\colorMATHM}\renewcommand\colorSYNTAX{\colorSYNTAXM}{{\color{\colorMATH}\ensuremath{0}}}\endgroup },{\begingroup\renewcommand\colorMATH{\colorMATHM}\renewcommand\colorSYNTAX{\colorSYNTAXM}{{\color{\colorMATH}\ensuremath{0}}}\endgroup }} + {}\rceil {\begingroup\renewcommand\colorMATH{\colorMATHS}\renewcommand\colorSYNTAX{\colorSYNTAXS}{{\color{\colorMATH}\ensuremath{\Gamma _{3} + \Gamma _{4}}}}\endgroup }\lceil {}^{{{\color{\colorSYNTAX}\texttt{\ensuremath{\infty }}}}} + \mathrlap{\hspace{-0.5pt}{}\rceil }\lfloor {\begingroup\renewcommand\colorMATH{\colorMATHS}\renewcommand\colorSYNTAX{\colorSYNTAXS}{{\color{\colorMATH}\ensuremath{\Gamma _{5}}}}\endgroup }\mathrlap{\hspace{-0.5pt}\rfloor }\lceil {}_{\{ {\begingroup\renewcommand\colorMATH{\colorMATHM}\renewcommand\colorSYNTAX{\colorSYNTAXM}{{\color{\colorMATH}\ensuremath{x_{1}}}}\endgroup },{.}\hspace{-1pt}{.}\hspace{-1pt}{.},{\begingroup\renewcommand\colorMATH{\colorMATHM}\renewcommand\colorSYNTAX{\colorSYNTAXM}{{\color{\colorMATH}\ensuremath{x_{n}}}}\endgroup }\} }^{{\begingroup\renewcommand\colorMATH{\colorMATHM}\renewcommand\colorSYNTAX{\colorSYNTAXM}{{\color{\colorMATH}\ensuremath{\eta _{\epsilon }}}}\endgroup },{\begingroup\renewcommand\colorMATH{\colorMATHM}\renewcommand\colorSYNTAX{\colorSYNTAXM}{{\color{\colorMATH}\ensuremath{0}}}\endgroup }} 
     \vdash  {{\color{\colorSYNTAX}\texttt{\ensuremath{{{\color{\colorSYNTAX}\texttt{exponential}}}[{\begingroup\renewcommand\colorMATH{\colorMATHS}\renewcommand\colorSYNTAX{\colorSYNTAXS}{{\color{\colorMATH}\ensuremath{e_{1}}}}\endgroup },{\begingroup\renewcommand\colorMATH{\colorMATHS}\renewcommand\colorSYNTAX{\colorSYNTAXS}{{\color{\colorMATH}\ensuremath{e_{2}}}}\endgroup }]\hspace*{0.33em}{<}{\begingroup\renewcommand\colorMATH{\colorMATHM}\renewcommand\colorSYNTAX{\colorSYNTAXM}{{\color{\colorMATH}\ensuremath{x_{1}}}}\endgroup },{.}\hspace{-1pt}{.}\hspace{-1pt}{.},{\begingroup\renewcommand\colorMATH{\colorMATHM}\renewcommand\colorSYNTAX{\colorSYNTAXM}{{\color{\colorMATH}\ensuremath{x_{n}}}}\endgroup }{>}\hspace*{0.33em}{\begingroup\renewcommand\colorMATH{\colorMATHS}\renewcommand\colorSYNTAX{\colorSYNTAXS}{{\color{\colorMATH}\ensuremath{e_{3}}}}\endgroup }\hspace*{0.33em}\{ {\begingroup\renewcommand\colorMATH{\colorMATHM}\renewcommand\colorSYNTAX{\colorSYNTAXM}{{\color{\colorMATH}\ensuremath{x}}}\endgroup } \Rightarrow  {\begingroup\renewcommand\colorMATH{\colorMATHS}\renewcommand\colorSYNTAX{\colorSYNTAXS}{{\color{\colorMATH}\ensuremath{e_{4}}}}\endgroup }\} }}}} \mathrel{:} {\begingroup\renewcommand\colorMATH{\colorMATHM}\renewcommand\colorSYNTAX{\colorSYNTAXM}{{\color{\colorMATH}\ensuremath{\tau }}}\endgroup }
  }
\and\inferrule*[lab={{\color{\colorTEXT}\textsc{\scriptsize Rand-Resp}}}
  ]{ {\begingroup\renewcommand\colorMATH{\colorMATHS}\renewcommand\colorSYNTAX{\colorSYNTAXS}{{\color{\colorMATH}\ensuremath{\Gamma _{1} \vdash  e_{1} \mathrel{:} {\begingroup\renewcommand\colorMATH{\colorMATHM}\renewcommand\colorSYNTAX{\colorSYNTAXM}{{\color{\colorMATH}\ensuremath{{{\color{\colorSYNTAX}\texttt{\ensuremath{{\mathbb{N}}[{{\color{\colorMATH}\ensuremath{\eta _{n}}}}]}}}}}}}\endgroup }}}}\endgroup }
  \\ {\begingroup\renewcommand\colorMATH{\colorMATHS}\renewcommand\colorSYNTAX{\colorSYNTAXS}{{\color{\colorMATH}\ensuremath{\Gamma _{2} \vdash  e_{2} \mathrel{:} {\begingroup\renewcommand\colorMATH{\colorMATHM}\renewcommand\colorSYNTAX{\colorSYNTAXM}{{\color{\colorMATH}\ensuremath{{{\color{\colorSYNTAX}\texttt{\ensuremath{{\mathbb{R}}^{+}[{{\color{\colorMATH}\ensuremath{\eta _{\epsilon }}}}]}}}}}}}\endgroup }}}}\endgroup }
  \\ {\begingroup\renewcommand\colorMATH{\colorMATHS}\renewcommand\colorSYNTAX{\colorSYNTAXS}{{\color{\colorMATH}\ensuremath{\Gamma _{3} + \mathrlap{\hspace{-0.5pt}{}\rceil }\lfloor \Gamma _{4}\mathrlap{\hspace{-0.5pt}\rfloor }\lceil {}^{{\begingroup\renewcommand\colorMATH{\colorMATHM}\renewcommand\colorSYNTAX{\colorSYNTAXM}{{\color{\colorMATH}\ensuremath{\eta _{n}}}}\endgroup }}_{\{ {\begingroup\renewcommand\colorMATH{\colorMATHM}\renewcommand\colorSYNTAX{\colorSYNTAXM}{{\color{\colorMATH}\ensuremath{x_{1}}}}\endgroup },{.}\hspace{-1pt}{.}\hspace{-1pt}{.},{\begingroup\renewcommand\colorMATH{\colorMATHM}\renewcommand\colorSYNTAX{\colorSYNTAXM}{{\color{\colorMATH}\ensuremath{x_{n}}}}\endgroup }\} } \vdash  e_{3} \mathrel{:} {\begingroup\renewcommand\colorMATH{\colorMATHM}\renewcommand\colorSYNTAX{\colorSYNTAXM}{{\color{\colorMATH}\ensuremath{{{\color{\colorSYNTAX}\texttt{\ensuremath{{\mathbb{N}}}}}}}}}\endgroup }}}}\endgroup }
     }{
     {}\rceil {\begingroup\renewcommand\colorMATH{\colorMATHS}\renewcommand\colorSYNTAX{\colorSYNTAXS}{{\color{\colorMATH}\ensuremath{\Gamma _{1} + \Gamma _{2}}}}\endgroup }\lceil {}^{{\begingroup\renewcommand\colorMATH{\colorMATHM}\renewcommand\colorSYNTAX{\colorSYNTAXM}{{\color{\colorMATH}\ensuremath{0}}}\endgroup },{\begingroup\renewcommand\colorMATH{\colorMATHM}\renewcommand\colorSYNTAX{\colorSYNTAXM}{{\color{\colorMATH}\ensuremath{0}}}\endgroup }} + {}\rceil {\begingroup\renewcommand\colorMATH{\colorMATHS}\renewcommand\colorSYNTAX{\colorSYNTAXS}{{\color{\colorMATH}\ensuremath{\Gamma _{3}}}}\endgroup }\lceil {}^{{{\color{\colorSYNTAX}\texttt{\ensuremath{\infty }}}}} + \mathrlap{\hspace{-0.5pt}{}\rceil }\lfloor {\begingroup\renewcommand\colorMATH{\colorMATHS}\renewcommand\colorSYNTAX{\colorSYNTAXS}{{\color{\colorMATH}\ensuremath{\Gamma _{4}}}}\endgroup }\mathrlap{\hspace{-0.5pt}\rfloor }\lceil {}_{\{ {\begingroup\renewcommand\colorMATH{\colorMATHM}\renewcommand\colorSYNTAX{\colorSYNTAXM}{{\color{\colorMATH}\ensuremath{x_{1}}}}\endgroup },{.}\hspace{-1pt}{.}\hspace{-1pt}{.},{\begingroup\renewcommand\colorMATH{\colorMATHM}\renewcommand\colorSYNTAX{\colorSYNTAXM}{{\color{\colorMATH}\ensuremath{x_{n}}}}\endgroup }\} }^{{{\color{\colorSYNTAX}\texttt{\ensuremath{{\begingroup\renewcommand\colorMATH{\colorMATHM}\renewcommand\colorSYNTAX{\colorSYNTAXM}{{\color{\colorMATH}\ensuremath{\eta _{\epsilon }}}}\endgroup },{\begingroup\renewcommand\colorMATH{\colorMATHM}\renewcommand\colorSYNTAX{\colorSYNTAXM}{{\color{\colorMATH}\ensuremath{0}}}\endgroup }}}}}}
     \vdash  {{\color{\colorSYNTAX}\texttt{\ensuremath{{{\color{\colorSYNTAX}\texttt{rand-resp}}}[{\begingroup\renewcommand\colorMATH{\colorMATHS}\renewcommand\colorSYNTAX{\colorSYNTAXS}{{\color{\colorMATH}\ensuremath{e_{1}}}}\endgroup },{\begingroup\renewcommand\colorMATH{\colorMATHS}\renewcommand\colorSYNTAX{\colorSYNTAXS}{{\color{\colorMATH}\ensuremath{e_{2}}}}\endgroup }]\hspace*{0.33em}{<}{\begingroup\renewcommand\colorMATH{\colorMATHM}\renewcommand\colorSYNTAX{\colorSYNTAXM}{{\color{\colorMATH}\ensuremath{x_{1}}}}\endgroup },{.}\hspace{-1pt}{.}\hspace{-1pt}{.},{\begingroup\renewcommand\colorMATH{\colorMATHM}\renewcommand\colorSYNTAX{\colorSYNTAXM}{{\color{\colorMATH}\ensuremath{x_{n}}}}\endgroup }{>}\hspace*{0.33em}\{ {\begingroup\renewcommand\colorMATH{\colorMATHS}\renewcommand\colorSYNTAX{\colorSYNTAXS}{{\color{\colorMATH}\ensuremath{e_{3}}}}\endgroup }\} }}}} \mathrel{:} {\begingroup\renewcommand\colorMATH{\colorMATHM}\renewcommand\colorSYNTAX{\colorSYNTAXM}{{\color{\colorMATH}\ensuremath{{{\color{\colorSYNTAX}\texttt{\ensuremath{{\mathbb{N}}}}}}}}}\endgroup }
  }
\and\inferrule*[flushleft,lab={{\color{\colorTEXT}\textsc{\scriptsize Sample}}}
  ]{ {\begingroup\renewcommand\colorMATH{\colorMATHS}\renewcommand\colorSYNTAX{\colorSYNTAXS}{{\color{\colorMATH}\ensuremath{{\begingroup\renewcommand\colorMATH{\colorMATHM}\renewcommand\colorSYNTAX{\colorSYNTAXM}{{\color{\colorMATH}\ensuremath{\Delta }}}\endgroup },\Gamma _{1} \vdash  e_{1} \mathrel{:} {\begingroup\renewcommand\colorMATH{\colorMATHM}\renewcommand\colorSYNTAX{\colorSYNTAXM}{{\color{\colorMATH}\ensuremath{{{\color{\colorSYNTAX}\texttt{\ensuremath{{\mathbb{N}}[{{\color{\colorMATH}\ensuremath{\eta _{m_{2}}}}}]}}}}}}}\endgroup }}}}\endgroup } 
  \\ {\begingroup\renewcommand\colorMATH{\colorMATHM}\renewcommand\colorSYNTAX{\colorSYNTAXM}{{\color{\colorMATH}\ensuremath{\eta _{m_{2}} \leq  \eta _{m_{1}}}}}\endgroup }
  \\ {\begingroup\renewcommand\colorMATH{\colorMATHM}\renewcommand\colorSYNTAX{\colorSYNTAXM}{{\color{\colorMATH}\ensuremath{\eta _{\epsilon ^{\prime}} = {{\color{\colorSYNTAX}\texttt{\ensuremath{{{\color{\colorMATH}\ensuremath{2}}}{\mathrel{\mathord{\cdotp }}}{{\color{\colorMATH}\ensuremath{\eta _{m_{2}}}}}{\mathrel{\mathord{\cdotp }}}1/{{\color{\colorMATH}\ensuremath{\eta _{m_{1}}}}}}}}}}}}\endgroup }
  \\ {\begingroup\renewcommand\colorMATH{\colorMATHM}\renewcommand\colorSYNTAX{\colorSYNTAXM}{{\color{\colorMATH}\ensuremath{\eta _{\delta ^{\prime}} = {{\color{\colorSYNTAX}\texttt{\ensuremath{{{\color{\colorMATH}\ensuremath{\eta _{m_{2}}}}}{\mathrel{\mathord{\cdotp }}}1/{{\color{\colorMATH}\ensuremath{\eta _{m_{1}}}}}}}}}}}}\endgroup }
  \\\\ \Gamma _{2} \supseteq  \{ {\begingroup\renewcommand\colorMATH{\colorMATHM}\renewcommand\colorSYNTAX{\colorSYNTAXM}{{\color{\colorMATH}\ensuremath{x_{1}}}}\endgroup } \mathrel{:}_{({\begingroup\renewcommand\colorMATH{\colorMATHM}\renewcommand\colorSYNTAX{\colorSYNTAXM}{{\color{\colorMATH}\ensuremath{\eta _{\epsilon ^{\prime}}\eta _{\epsilon _{1}}}}}\endgroup },{\begingroup\renewcommand\colorMATH{\colorMATHM}\renewcommand\colorSYNTAX{\colorSYNTAXM}{{\color{\colorMATH}\ensuremath{\eta _{\delta ^{\prime}}\eta _{\delta _{1}}}}}\endgroup })} {\begingroup\renewcommand\colorMATH{\colorMATHM}\renewcommand\colorSYNTAX{\colorSYNTAXM}{{\color{\colorMATH}\ensuremath{{{\color{\colorSYNTAX}\texttt{\ensuremath{{\mathbb{M}}^{{{\color{\colorMATH}\ensuremath{c_{1}}}}}_{{{\color{\colorMATH}\ensuremath{\ell _{1}}}}}[{{\color{\colorMATH}\ensuremath{m_{1}}}},{{\color{\colorMATH}\ensuremath{n_{1}}}}]\hspace*{0.33em}{{\color{\colorMATH}\ensuremath{\tau _{1}}}}}}}}}}}\endgroup },{\begingroup\renewcommand\colorMATH{\colorMATHM}\renewcommand\colorSYNTAX{\colorSYNTAXM}{{\color{\colorMATH}\ensuremath{x_{2}}}}\endgroup } \mathrel{:}_{({\begingroup\renewcommand\colorMATH{\colorMATHM}\renewcommand\colorSYNTAX{\colorSYNTAXM}{{\color{\colorMATH}\ensuremath{\eta _{\epsilon ^{\prime}}\eta _{\epsilon _{2}}}}}\endgroup },{\begingroup\renewcommand\colorMATH{\colorMATHM}\renewcommand\colorSYNTAX{\colorSYNTAXM}{{\color{\colorMATH}\ensuremath{\eta _{\delta ^{\prime}}\eta _{\delta _{2}}}}}\endgroup })} {\begingroup\renewcommand\colorMATH{\colorMATHM}\renewcommand\colorSYNTAX{\colorSYNTAXM}{{\color{\colorMATH}\ensuremath{{{\color{\colorSYNTAX}\texttt{\ensuremath{{\mathbb{M}}^{{{\color{\colorMATH}\ensuremath{c_{2}}}}}_{{{\color{\colorMATH}\ensuremath{\ell _{2}}}}}[{{\color{\colorMATH}\ensuremath{m_{1}}}},{{\color{\colorMATH}\ensuremath{n_{2}}}}]\hspace*{0.33em}{{\color{\colorMATH}\ensuremath{\tau _{2}}}}}}}}}}}\endgroup }\} 
  \\ {\begingroup\renewcommand\colorMATH{\colorMATHM}\renewcommand\colorSYNTAX{\colorSYNTAXM}{{\color{\colorMATH}\ensuremath{\Delta }}}\endgroup },\Gamma _{3}\uplus \{ {\begingroup\renewcommand\colorMATH{\colorMATHM}\renewcommand\colorSYNTAX{\colorSYNTAXM}{{\color{\colorMATH}\ensuremath{y_{1}}}}\endgroup }{\mathrel{:}}_{({\begingroup\renewcommand\colorMATH{\colorMATHM}\renewcommand\colorSYNTAX{\colorSYNTAXM}{{\color{\colorMATH}\ensuremath{\eta _{\epsilon _{1}}}}}\endgroup },{\begingroup\renewcommand\colorMATH{\colorMATHM}\renewcommand\colorSYNTAX{\colorSYNTAXM}{{\color{\colorMATH}\ensuremath{\eta _{\delta _{1}}}}}\endgroup })}{\begingroup\renewcommand\colorMATH{\colorMATHM}\renewcommand\colorSYNTAX{\colorSYNTAXM}{{\color{\colorMATH}\ensuremath{{{\color{\colorSYNTAX}\texttt{\ensuremath{{\mathbb{M}}^{{{\color{\colorMATH}\ensuremath{c_{1}}}}}_{{{\color{\colorMATH}\ensuremath{\ell _{1}}}}}[{{\color{\colorMATH}\ensuremath{m_{2}}}},{{\color{\colorMATH}\ensuremath{n_{1}}}}]\hspace*{0.33em}{{\color{\colorMATH}\ensuremath{\tau _{1}}}}}}}}}}}\endgroup },{\begingroup\renewcommand\colorMATH{\colorMATHM}\renewcommand\colorSYNTAX{\colorSYNTAXM}{{\color{\colorMATH}\ensuremath{y_{2}}}}\endgroup }{\mathrel{:}}_{({\begingroup\renewcommand\colorMATH{\colorMATHM}\renewcommand\colorSYNTAX{\colorSYNTAXM}{{\color{\colorMATH}\ensuremath{\eta _{\epsilon _{2}}}}}\endgroup },{\begingroup\renewcommand\colorMATH{\colorMATHM}\renewcommand\colorSYNTAX{\colorSYNTAXM}{{\color{\colorMATH}\ensuremath{\eta _{\delta _{2}}}}}\endgroup })}{\begingroup\renewcommand\colorMATH{\colorMATHM}\renewcommand\colorSYNTAX{\colorSYNTAXM}{{\color{\colorMATH}\ensuremath{{{\color{\colorSYNTAX}\texttt{\ensuremath{{\mathbb{M}}^{{{\color{\colorMATH}\ensuremath{c_{2}}}}}_{{{\color{\colorMATH}\ensuremath{\ell _{2}}}}}[{{\color{\colorMATH}\ensuremath{m_{2}}}},{{\color{\colorMATH}\ensuremath{n_{2}}}}]\hspace*{0.33em}{{\color{\colorMATH}\ensuremath{\tau _{2}}}}}}}}}}}\endgroup }\}  \vdash  e_{2} \mathrel{:} {\begingroup\renewcommand\colorMATH{\colorMATHM}\renewcommand\colorSYNTAX{\colorSYNTAXM}{{\color{\colorMATH}\ensuremath{\tau _{3}}}}\endgroup }
     }{
     {}\rceil {\begingroup\renewcommand\colorMATH{\colorMATHS}\renewcommand\colorSYNTAX{\colorSYNTAXS}{{\color{\colorMATH}\ensuremath{\Gamma _{1}}}}\endgroup }\lceil {}^{{\begingroup\renewcommand\colorMATH{\colorMATHM}\renewcommand\colorSYNTAX{\colorSYNTAXM}{{\color{\colorMATH}\ensuremath{0}}}\endgroup },{\begingroup\renewcommand\colorMATH{\colorMATHM}\renewcommand\colorSYNTAX{\colorSYNTAXM}{{\color{\colorMATH}\ensuremath{0}}}\endgroup }} + \Gamma _{2} + \Gamma _{3} \vdash  {{\color{\colorSYNTAX}\texttt{\ensuremath{{{\color{\colorSYNTAX}\texttt{sample}}}\hspace*{0.33em}{\begingroup\renewcommand\colorMATH{\colorMATHS}\renewcommand\colorSYNTAX{\colorSYNTAXS}{{\color{\colorMATH}\ensuremath{e_{1}}}}\endgroup }\hspace*{0.33em}{{\color{\colorSYNTAX}\texttt{on}}}\hspace*{0.33em}{\begingroup\renewcommand\colorMATH{\colorMATHM}\renewcommand\colorSYNTAX{\colorSYNTAXM}{{\color{\colorMATH}\ensuremath{x_{1}}}}\endgroup },{\begingroup\renewcommand\colorMATH{\colorMATHM}\renewcommand\colorSYNTAX{\colorSYNTAXM}{{\color{\colorMATH}\ensuremath{x_{2}}}}\endgroup }\hspace*{0.33em}\{ {\begingroup\renewcommand\colorMATH{\colorMATHM}\renewcommand\colorSYNTAX{\colorSYNTAXM}{{\color{\colorMATH}\ensuremath{x_{3}}}}\endgroup },{\begingroup\renewcommand\colorMATH{\colorMATHM}\renewcommand\colorSYNTAX{\colorSYNTAXM}{{\color{\colorMATH}\ensuremath{x_{4}}}}\endgroup } \Rightarrow  {{\color{\colorMATH}\ensuremath{e_{2}}}}\} }}}} \mathrel{:} {\begingroup\renewcommand\colorMATH{\colorMATHM}\renewcommand\colorSYNTAX{\colorSYNTAXM}{{\color{\colorMATH}\ensuremath{\tau _{3}}}}\endgroup }
  }
\and\inferrule*[flushleft,lab={{\color{\colorTEXT}\textsc{\scriptsize Rand-Nat}}}
  ]{ {\begingroup\renewcommand\colorMATH{\colorMATHS}\renewcommand\colorSYNTAX{\colorSYNTAXS}{{\color{\colorMATH}\ensuremath{\Gamma _{1} \vdash  e_{1} \mathrel{:} {\begingroup\renewcommand\colorMATH{\colorMATHM}\renewcommand\colorSYNTAX{\colorSYNTAXM}{{\color{\colorMATH}\ensuremath{{{\color{\colorSYNTAX}\texttt{\ensuremath{{\mathbb{N}}}}}}}}}\endgroup }}}}\endgroup }
  \\ {\begingroup\renewcommand\colorMATH{\colorMATHS}\renewcommand\colorSYNTAX{\colorSYNTAXS}{{\color{\colorMATH}\ensuremath{\Gamma _{2} \vdash  e_{2} \mathrel{:} {\begingroup\renewcommand\colorMATH{\colorMATHM}\renewcommand\colorSYNTAX{\colorSYNTAXM}{{\color{\colorMATH}\ensuremath{{{\color{\colorSYNTAX}\texttt{\ensuremath{{\mathbb{N}}}}}}}}}\endgroup }}}}\endgroup }
     }{
     {}\rceil {\begingroup\renewcommand\colorMATH{\colorMATHS}\renewcommand\colorSYNTAX{\colorSYNTAXS}{{\color{\colorMATH}\ensuremath{\Gamma _{1} + \Gamma _{2}}}}\endgroup }\lceil {}^{{{\color{\colorSYNTAX}\texttt{\ensuremath{\infty }}}}} \vdash  {{\color{\colorSYNTAX}\texttt{\ensuremath{{{\color{\colorSYNTAX}\texttt{rand-nat}}}({\begingroup\renewcommand\colorMATH{\colorMATHS}\renewcommand\colorSYNTAX{\colorSYNTAXS}{{\color{\colorMATH}\ensuremath{e_{1}}}}\endgroup },{\begingroup\renewcommand\colorMATH{\colorMATHS}\renewcommand\colorSYNTAX{\colorSYNTAXS}{{\color{\colorMATH}\ensuremath{e_{2}}}}\endgroup })}}}} \mathrel{:} {\begingroup\renewcommand\colorMATH{\colorMATHM}\renewcommand\colorSYNTAX{\colorSYNTAXM}{{\color{\colorMATH}\ensuremath{{{\color{\colorSYNTAX}\texttt{\ensuremath{{\mathbb{N}}}}}}}}}\endgroup }
  }
\end{mathpar}\endgroup 
\endgroup 
\caption{Privacy Type Systems}
\label{fig:apx:priv-typing}
\end{figure*}

\begin{figure*}
\vspace*{-0.25em}\begingroup\color{\colorMATH}\begin{gather*}

\vspace*{-1em}\end{gather*}\endgroup                 
\caption{Metrics (1)}
\label{fig:apx:metrics-1}
\end{figure*}

\begin{figure*}
\vspace*{-0.25em}\begingroup\color{\colorMATH}\begin{gather*}%
\vspace*{-1em}\end{gather*}\endgroup                 
\caption{Metrics (2)}
\label{fig:apx:metrics-2}
\end{figure*}

\begin{figure*}
\vspace*{-0.25em}\begingroup\color{\colorMATH}\begin{gather*}\begin{tabularx}{\linewidth}{>{\centering\arraybackslash\(}X<{\)}}\hfill\hspace{0pt}\begingroup\color{\colorTEXT}\boxed{\begingroup\color{\colorMATH} \llbracket \underline{\hspace{0.66em}\vspace*{5ex}}\rrbracket ^{\underline{\hspace{0.66em}\vspace*{5ex}}} \in  {{\color{\colorMATH}\ensuremath{\operatorname{rexp}}}} \times  {{\color{\colorMATH}\ensuremath{\operatorname{kenv}}}} \rightarrow  {\mathbb{R}}_{\infty }^{+}\endgroup}\endgroup \end{tabularx}\vspace*{-1em}\end{gather*}\endgroup 
\vspace*{-0.25em}\begingroup\color{\colorMATH}\begin{gather*}\begin{tabularx}{\linewidth}{>{\centering\arraybackslash\(}X<{\)}}\hfill\hspace{0pt}
  \begin{array}{rcl
  } \llbracket \beta \rrbracket ^{\gamma } &{}\triangleq {}& \gamma (\beta )
  \cr  \llbracket n\rrbracket ^{\gamma } &{}\triangleq {}& n
  \cr  \llbracket r^{+}\rrbracket ^{\gamma } &{}\triangleq {}& r^{+}
  \cr  \llbracket \eta _{1}\sqcup \eta _{2}\rrbracket ^{\gamma } &{}\triangleq {}& \llbracket \eta _{1}\rrbracket ^{\gamma }\sqcup \llbracket \eta _{2}\rrbracket ^{\gamma }
  \cr  \llbracket \eta _{1}\sqcap \eta _{2}\rrbracket ^{\gamma } &{}\triangleq {}& \llbracket \eta _{1}\rrbracket ^{\gamma }\sqcap \llbracket \eta _{2}\rrbracket ^{\gamma }
  \cr  \llbracket \eta _{1}+\eta _{2}\rrbracket ^{\gamma } &{}\triangleq {}& \llbracket \eta _{1}\rrbracket ^{\gamma }+\llbracket \eta _{2}\rrbracket ^{\gamma }
  \cr  \llbracket \eta _{1}\mathrel{\mathord{\cdotp }}\eta _{2}\rrbracket ^{\gamma } &{}\triangleq {}& \llbracket \eta _{1}\rrbracket ^{\gamma }\mathrel{\mathord{\cdotp }}\llbracket \eta _{2}\rrbracket ^{\gamma }
  \cr  \llbracket 1/\eta \rrbracket ^{\gamma }   &{}\triangleq {}& 1/\llbracket \eta \rrbracket ^{\gamma }
  \cr  \llbracket \sqrt \eta \rrbracket ^{\gamma }    &{}\triangleq {}& \sqrt {\llbracket \eta \rrbracket ^{\gamma }}
  \cr  \llbracket \ln \eta \rrbracket ^{\gamma }   &{}\triangleq {}& \ln \llbracket \eta \rrbracket ^{\gamma }
  \end{array}
  \hfill\hspace{0pt}\end{tabularx}
\vspace*{-1em}\end{gather*}\endgroup 
\vspace*{-0.25em}\begingroup\color{\colorMATH}\begin{gather*}\begin{tabularx}{\linewidth}{>{\centering\arraybackslash\(}X<{\)}}\hfill\hspace{0pt}\begingroup\color{\colorTEXT}\boxed{\begingroup\color{\colorMATH} \llbracket {\begingroup\renewcommand\colorMATH{\colorMATHS}\renewcommand\colorSYNTAX{\colorSYNTAXS}{{\color{\colorMATH}\ensuremath{\underline{\hspace{0.66em}\vspace*{5ex}}}}}\endgroup }\rrbracket ^{\underline{\hspace{0.66em}\vspace*{5ex}}} \in  {\begingroup\renewcommand\colorMATH{\colorMATHS}\renewcommand\colorSYNTAX{\colorSYNTAXS}{{\color{\colorMATH}\ensuremath{{{\color{\colorMATH}\ensuremath{\operatorname{sens}}}}}}}\endgroup } \times  {{\color{\colorMATH}\ensuremath{\operatorname{kenv}}}} \rightarrow  {\mathbb{R}}_{\infty }^{+}\endgroup}\endgroup \end{tabularx}\vspace*{-1em}\end{gather*}\endgroup 
\vspace*{-0.25em}\begingroup\color{\colorMATH}\begin{gather*}\begin{tabularx}{\linewidth}{>{\centering\arraybackslash\(}X<{\)}}\hfill\hspace{0pt}
  \begin{array}{rcl
  } \llbracket \eta \rrbracket ^{\gamma }         &{}\triangleq {}& \llbracket \eta \rrbracket ^{\gamma }
  \cr  \llbracket {\begingroup\renewcommand\colorMATH{\colorMATHS}\renewcommand\colorSYNTAX{\colorSYNTAXS}{{\color{\colorMATH}\ensuremath{{{\color{\colorSYNTAX}\texttt{\ensuremath{\infty }}}}}}}\endgroup }\rrbracket ^{\gamma } &{}\triangleq {}& \infty 
  \end{array}
  \hfill\hspace{0pt}\end{tabularx}
\vspace*{-1em}\end{gather*}\endgroup 
\vspace*{-0.25em}\begingroup\color{\colorMATH}\begin{gather*}\begin{tabularx}{\linewidth}{>{\centering\arraybackslash\(}X<{\)}}\hfill\hspace{0pt}\begingroup\color{\colorTEXT}\boxed{\begingroup\color{\colorMATH} \llbracket {\begingroup\renewcommand\colorMATH{\colorMATHP}\renewcommand\colorSYNTAX{\colorSYNTAXP}{{\color{\colorMATH}\ensuremath{\underline{\hspace{0.66em}\vspace*{5ex}}}}}\endgroup }\rrbracket ^{\underline{\hspace{0.66em}\vspace*{5ex}}} \in  {\begingroup\renewcommand\colorMATH{\colorMATHP}\renewcommand\colorSYNTAX{\colorSYNTAXP}{{\color{\colorMATH}\ensuremath{{{\color{\colorMATH}\ensuremath{\operatorname{priv}}}}}}}\endgroup } \times  {{\color{\colorMATH}\ensuremath{\operatorname{kenv}}}} \rightarrow  {\mathbb{R}}_{\infty }^{+} \times  {\mathbb{R}}_{\infty }^{+}\endgroup}\endgroup \end{tabularx}\vspace*{-1em}\end{gather*}\endgroup 
\vspace*{-0.25em}\begingroup\color{\colorMATH}\begin{gather*}\begin{tabularx}{\linewidth}{>{\centering\arraybackslash\(}X<{\)}}\hfill\hspace{0pt}
  \begin{array}{rcl
  } \llbracket {\begingroup\renewcommand\colorMATH{\colorMATHP}\renewcommand\colorSYNTAX{\colorSYNTAXP}{{\color{\colorMATH}\ensuremath{{{\color{\colorSYNTAX}\texttt{\ensuremath{{\begingroup\renewcommand\colorMATH{\colorMATHM}\renewcommand\colorSYNTAX{\colorSYNTAXM}{{\color{\colorMATH}\ensuremath{\eta _{\epsilon }}}}\endgroup },{\begingroup\renewcommand\colorMATH{\colorMATHM}\renewcommand\colorSYNTAX{\colorSYNTAXM}{{\color{\colorMATH}\ensuremath{\eta _{\delta }}}}\endgroup }}}}}}}}\endgroup }\rrbracket ^{\gamma } &{}\triangleq {}& \llbracket \eta _{\epsilon }\rrbracket ^{\gamma },\llbracket \eta _{\delta }\rrbracket ^{\gamma }
  \cr  \llbracket {\begingroup\renewcommand\colorMATH{\colorMATHP}\renewcommand\colorSYNTAX{\colorSYNTAXP}{{\color{\colorMATH}\ensuremath{{{\color{\colorSYNTAX}\texttt{\ensuremath{\infty }}}}}}}\endgroup }\rrbracket ^{\gamma }                 &{}\triangleq {}& \infty ,0
  \end{array}
  \hfill\hspace{0pt}\end{tabularx}
\vspace*{-1em}\end{gather*}\endgroup 
\vspace*{-0.25em}\begingroup\color{\colorMATH}\begin{gather*}\begin{tabularx}{\linewidth}{>{\centering\arraybackslash\(}X<{\)}}\hfill\hspace{0pt}\begingroup\color{\colorTEXT}\boxed{\begingroup\color{\colorMATH} \llbracket \underline{\hspace{0.66em}\vspace*{5ex}}\rrbracket ^{\underline{\hspace{0.66em}\vspace*{5ex}}} \in  {{\color{\colorMATH}\ensuremath{\operatorname{type}}}} \times  {{\color{\colorMATH}\ensuremath{\operatorname{kenv}}}} \rightarrow  {{\color{\colorMATH}\ensuremath{\operatorname{dom}}}}\endgroup}\endgroup \end{tabularx}\vspace*{-1em}\end{gather*}\endgroup 
\begingroup\color{\colorMATH}\begin{mathpar}\begin{array}{rcl
  } \llbracket {{\color{\colorSYNTAX}\texttt{\ensuremath{{\mathbb{N}}[{{\color{\colorMATH}\ensuremath{\eta }}}]}}}}\rrbracket ^{\gamma }       &{}\triangleq {}& \hat {\mathbb{N}}[\llbracket \eta \rrbracket ^{\gamma }]
  \cr  \llbracket {{\color{\colorSYNTAX}\texttt{\ensuremath{{\mathbb{R}}^{+}[{{\color{\colorMATH}\ensuremath{\eta }}}]}}}}\rrbracket ^{\gamma }      &{}\triangleq {}& \hat {\mathbb{R}}^{+}[\llbracket \eta \rrbracket ^{\gamma }]
  \cr  \llbracket {{\color{\colorSYNTAX}\texttt{\ensuremath{{\mathbb{N}}}}}}\rrbracket ^{\gamma }            &{}\triangleq {}& \hat {\mathbb{N}}
  \cr  \llbracket {{\color{\colorSYNTAX}\texttt{\ensuremath{{\mathbb{R}}}}}}\rrbracket ^{\gamma }            &{}\triangleq {}& \hat {\mathbb{R}}
  \cr  \llbracket {{\color{\colorSYNTAX}\texttt{data}}}\rrbracket ^{\gamma }           &{}\triangleq {}& {{\color{\colorMATH}\ensuremath{\operatorname{data}}}}
  \cr  \llbracket {{\color{\colorSYNTAX}\texttt{\ensuremath{{{\color{\colorSYNTAX}\texttt{idx}}}[{{\color{\colorMATH}\ensuremath{\eta }}}]}}}}\rrbracket ^{\gamma }   &{}\triangleq {}& {{\color{\colorMATH}\ensuremath{\operatorname{idx}}}}[\llbracket \eta \rrbracket ^{\gamma }]
  \end{array}
\and\begin{array}{rcl
  } \llbracket {{\color{\colorSYNTAX}\texttt{\ensuremath{{{\color{\colorSYNTAX}\texttt{matrix}}}_{{{\color{\colorMATH}\ensuremath{\ell }}}}^{{{\color{\colorMATH}\ensuremath{c}}}}[{{\color{\colorMATH}\ensuremath{\eta _{m}}}},{{\color{\colorMATH}\ensuremath{\eta _{m}}}}]\hspace*{0.33em}{{\color{\colorMATH}\ensuremath{\tau }}}}}}}\rrbracket ^{\gamma }        &{}\triangleq {}& {{\color{\colorMATH}\ensuremath{\operatorname{m}}}}\widehat {{{\color{\colorMATH}\ensuremath{\operatorname{atri}}}}}{{\color{\colorMATH}\ensuremath{\operatorname{x}}}}_{\ell }^{c}[\llbracket \eta _{m}\rrbracket ^{\gamma },\llbracket \eta _{n}\rrbracket ^{\gamma }](\llbracket \tau \rrbracket ^{\gamma })
  \cr  \llbracket {{\color{\colorSYNTAX}\texttt{\ensuremath{{{\color{\colorMATH}\ensuremath{\tau _{1}}}}\uplus {{\color{\colorMATH}\ensuremath{\tau _{2}}}}}}}}\rrbracket ^{\gamma }                                &{}\triangleq {}& \llbracket \tau _{1}\rrbracket ^{\gamma } \hat \uplus   \llbracket \tau _{2}\rrbracket ^{\gamma }
  \cr  \llbracket {{\color{\colorSYNTAX}\texttt{\ensuremath{{{\color{\colorMATH}\ensuremath{\tau _{1}}}}\times {{\color{\colorMATH}\ensuremath{\tau _{2}}}}}}}}\rrbracket ^{\gamma }                                &{}\triangleq {}& \llbracket \tau _{1}\rrbracket ^{\gamma } \hat \times   \llbracket \tau _{2}\rrbracket ^{\gamma }
  \cr  \llbracket {{\color{\colorSYNTAX}\texttt{\ensuremath{{{\color{\colorMATH}\ensuremath{\tau _{1}}}}\mathrel{\&}{{\color{\colorMATH}\ensuremath{\tau _{2}}}}}}}}\rrbracket ^{\gamma }                                &{}\triangleq {}& \llbracket \tau _{1}\rrbracket ^{\gamma } \mathrel{\hat \&}  \llbracket \tau _{2}\rrbracket ^{\gamma }
  \cr
\end{array}
\end{mathpar}\endgroup 
\begingroup\renewcommand\colorMATH{\colorMATHS}\renewcommand\colorSYNTAX{\colorSYNTAXS}
\vspace*{-0.25em}\begingroup\color{\colorMATH}\begin{gather*}\begin{tabularx}{\linewidth}{>{\centering\arraybackslash\(}X<{\)}}\hfill\hspace{0pt}\begingroup\color{\colorTEXT}\boxed{\begingroup\color{\colorMATH} \llbracket {\begingroup\renewcommand\colorMATH{\colorMATHM}\renewcommand\colorSYNTAX{\colorSYNTAXM}{{\color{\colorMATH}\ensuremath{\underline{\hspace{0.66em}\vspace*{5ex}}}}}\endgroup },\underline{\hspace{0.66em}\vspace*{5ex}}\vdash {\begingroup\renewcommand\colorMATH{\colorMATHM}\renewcommand\colorSYNTAX{\colorSYNTAXM}{{\color{\colorMATH}\ensuremath{\underline{\hspace{0.66em}\vspace*{5ex}}}}}\endgroup }\rrbracket  \in  {\begingroup\renewcommand\colorMATH{\colorMATHM}\renewcommand\colorSYNTAX{\colorSYNTAXM}{{\color{\colorMATH}\ensuremath{{{\color{\colorMATH}\ensuremath{\operatorname{kcxt}}}}}}}\endgroup } \times  {{\color{\colorMATH}\ensuremath{\operatorname{tcxt}}}} \times  {\begingroup\renewcommand\colorMATH{\colorMATHM}\renewcommand\colorSYNTAX{\colorSYNTAXM}{{\color{\colorMATH}\ensuremath{{{\color{\colorMATH}\ensuremath{\operatorname{type}}}}}}}\endgroup } \rightarrow  {\begingroup\renewcommand\colorMATH{\colorMATHM}\renewcommand\colorSYNTAX{\colorSYNTAXM}{{\color{\colorMATH}\ensuremath{{{\color{\colorMATH}\ensuremath{\operatorname{dom}}}}}}}\endgroup } \endgroup}\endgroup \end{tabularx}
\vspace*{-1em}\end{gather*}\endgroup 
\endgroup 
\begingroup\renewcommand\colorMATH{\colorMATHP}\renewcommand\colorSYNTAX{\colorSYNTAXP}
\vspace*{-0.25em}\begingroup\color{\colorMATH}\begin{gather*}\begin{tabularx}{\linewidth}{>{\centering\arraybackslash\(}X<{\)}}\hfill\hspace{0pt}\begingroup\color{\colorTEXT}\boxed{\begingroup\color{\colorMATH} \llbracket {\begingroup\renewcommand\colorMATH{\colorMATHM}\renewcommand\colorSYNTAX{\colorSYNTAXM}{{\color{\colorMATH}\ensuremath{\underline{\hspace{0.66em}\vspace*{5ex}}}}}\endgroup },\underline{\hspace{0.66em}\vspace*{5ex}}\vdash {\begingroup\renewcommand\colorMATH{\colorMATHM}\renewcommand\colorSYNTAX{\colorSYNTAXM}{{\color{\colorMATH}\ensuremath{\underline{\hspace{0.66em}\vspace*{5ex}}}}}\endgroup }\rrbracket  \in  {\begingroup\renewcommand\colorMATH{\colorMATHM}\renewcommand\colorSYNTAX{\colorSYNTAXM}{{\color{\colorMATH}\ensuremath{{{\color{\colorMATH}\ensuremath{\operatorname{kcxt}}}}}}}\endgroup } \times  {{\color{\colorMATH}\ensuremath{\operatorname{tcxt}}}} \times  {\begingroup\renewcommand\colorMATH{\colorMATHM}\renewcommand\colorSYNTAX{\colorSYNTAXM}{{\color{\colorMATH}\ensuremath{{{\color{\colorMATH}\ensuremath{\operatorname{type}}}}}}}\endgroup } \rightarrow  {\begingroup\renewcommand\colorMATH{\colorMATHM}\renewcommand\colorSYNTAX{\colorSYNTAXM}{{\color{\colorMATH}\ensuremath{{{\color{\colorMATH}\ensuremath{\operatorname{dom}}}}}}}\endgroup } \endgroup}\endgroup \end{tabularx}%
\vspace*{-1em}\end{gather*}\endgroup 
\endgroup 
\caption{Semantics of Typing}
\label{fig:apx:typing-semantics}
\end{figure*}

\FloatBarrier
\clearpage

\section{Theorems \& Lemmas}

\begin{theorem}[Gaussian mechanism]
  \label{thm:gauss}
  
  If {{\color{\colorMATH}\ensuremath{ | f(\gamma ) - f(\gamma [x_{i}\mapsto d]) | \leq  r }}}, then for {{\color{\colorMATH}\ensuremath{ f^{\prime} = \lambda  \gamma  .\hspace*{0.33em} f(\gamma ) + {\mathcal{N}}(0, \sigma ^{2}) }}} and {{\color{\colorMATH}\ensuremath{\sigma ^{2} = \frac{2 \log (1.25/\delta )r^{2}}{\epsilon ^{2}}}}}, and all {{\color{\colorMATH}\ensuremath{\epsilon _{i}, \delta _{i} > 0}}}:

  \vspace*{-0.25em}\begingroup\color{\colorMATH}\begin{gather*} {{\color{\colorMATH}\ensuremath{\operatorname{Pr}}}}[f^{\prime}(\gamma ) = d] \leq  e^{\epsilon _{i}} {{\color{\colorMATH}\ensuremath{\operatorname{Pr}}}}[f^{\prime}(\gamma [x_{i}\mapsto d^{\prime}]) = d] + \delta _{i}
  \vspace*{-1em}\end{gather*}\endgroup 
\end{theorem}

\begin{proof}
  By the assumption, {{\color{\colorMATH}\ensuremath{ \| f(\gamma ) - f(\gamma [x_{i}\mapsto d])\| _{2} \leq  r }}}. The conclusion follows by Dwork and Roth~\cite{privacybook}, Theorem A.1.
\end{proof}

\begin{theorem}[Advanced composition]
  \label{thm:adv_comp}
  
  If:

  \begin{itemize}[leftmargin=*,label=\textbf{{{\color{\colorMATH}\ensuremath{\mathrel{\mathord{\cdotp }}}}}
  }]\item  {{\color{\colorMATH}\ensuremath{{{\color{\colorMATH}\ensuremath{\operatorname{Pr}}}}[f(\gamma ) = d] \leq  e^{\epsilon _{i}} {{\color{\colorMATH}\ensuremath{\operatorname{Pr}}}}[f(\gamma [x_{i}\mapsto d^{\prime}]) = d] + \delta _{i}}}}
  \item  {{\color{\colorMATH}\ensuremath{\delta ^{\prime} > 0}}}
  \item  {{\color{\colorMATH}\ensuremath{\hat \epsilon  = e^{2\epsilon _{i}\sqrt {2n\log (1/\delta ^{\prime})}}}}}
  \item  {{\color{\colorMATH}\ensuremath{\hat \delta  = \delta ^{\prime}+n\delta _{i}}}}
  \end{itemize}

  \noindent Then:

  \vspace*{-0.25em}\begingroup\color{\colorMATH}\begin{gather*} {{\color{\colorMATH}\ensuremath{\operatorname{Pr}}}}[{{\color{\colorMATH}\ensuremath{\operatorname{iter}}}}(k, f(\gamma ), \iota ) = d^{\prime \prime}] \leq 
  \cr  \hspace*{1.00em} \hspace*{1.00em} e^{\hat \epsilon } {{\color{\colorMATH}\ensuremath{\operatorname{Pr}}}}[{{\color{\colorMATH}\ensuremath{\operatorname{iter}}}}(k, f(\gamma [x_{i}\mapsto d^{\prime}]), \iota ) = d^{\prime \prime}] + \hat \delta 
  \vspace*{-1em}\end{gather*}\endgroup 

  \noindent Where {{\color{\colorMATH}\ensuremath{{{\color{\colorMATH}\ensuremath{\operatorname{iter}}}}}}} is defined as:
  
  \vspace*{-0.25em}\begingroup\color{\colorMATH}\begin{gather*} \begin{array}{rcl
  } {{\color{\colorMATH}\ensuremath{\operatorname{iter}}}}(0, f, d) &{}\triangleq {}& d
  \cr  {{\color{\colorMATH}\ensuremath{\operatorname{iter}}}}(n, f, d) &{}\triangleq {}& {{\color{\colorMATH}\ensuremath{\operatorname{iter}}}}(n-1, f, f(d))
  \end{array}
  \vspace*{-1em}\end{gather*}\endgroup 
\end{theorem}

\begin{proof}
   By its definition, {{\color{\colorMATH}\ensuremath{ {{\color{\colorMATH}\ensuremath{\operatorname{iter}}}}(n, f, d) }}} is an instance of {{\color{\colorMATH}\ensuremath{n}}}-fold adaptive composition of the {{\color{\colorMATH}\ensuremath{(\epsilon _{i}, \delta _{i})}}}-differentially private mechanism {{\color{\colorMATH}\ensuremath{f}}}. The conclusion follows from Dwork and Roth~\cite{privacybook}, Theorem 3.20.
\end{proof}

\begin{theorem}[Sequential Composition]
  \label{thm:let}
  
  \noindent  If:
     \vspace*{-0.25em}\begingroup\color{\colorMATH}\begin{gather*} |\gamma [x_{i}] - d| \leq  1 \implies   
     \cr  {{\color{\colorMATH}\ensuremath{\operatorname{Pr}}}}[f_{1}(\gamma )=d^{\prime}] \leq  e^{\epsilon _{i}}{{\color{\colorMATH}\ensuremath{\operatorname{Pr}}}}[f_{1}(\gamma [x_{i}\mapsto d])=d^{\prime}] + \delta _{i}
     \vspace*{-1em}\end{gather*}\endgroup 
  \\\noindent  and:
     \vspace*{-0.25em}\begingroup\color{\colorMATH}\begin{gather*} |\gamma [x_{i}] - d| \leq  1 \implies   
     \cr  {{\color{\colorMATH}\ensuremath{\operatorname{Pr}}}}[f_{2}(\gamma [x\mapsto d^{\prime}])=d^{\prime \prime}] \leq  
     \cr  e^{\epsilon _{i}^{\prime}}{{\color{\colorMATH}\ensuremath{\operatorname{Pr}}}}[f_{2}(\gamma [x\mapsto d^{\prime},x_{i}\mapsto d])=d^{\prime \prime}] + \delta _{i}^{\prime}
     \vspace*{-1em}\end{gather*}\endgroup 
  \\\noindent  then:
     \vspace*{-0.25em}\begingroup\color{\colorMATH}\begin{gather*} |\gamma [x_{i}] - d| \leq  1 \implies   
     \cr  {{\color{\colorMATH}\ensuremath{\operatorname{Pr}}}}[f_{1}(\gamma )=d^{\prime},f_{2}(\gamma [x\mapsto d^{\prime}])=d^{\prime \prime}] \leq  
     \cr  e^{\epsilon _{i}+\epsilon _{i}^{\prime}}{{\color{\colorMATH}\ensuremath{\operatorname{Pr}}}}[f_{1}(\gamma [x_{i}\mapsto d])=d^{\prime},f_{2}(\gamma [x\mapsto d^{\prime},x_{i}\mapsto d])=d^{\prime \prime}] + \delta _{i} + \delta _{i}^{\prime}
     \vspace*{-1em}\end{gather*}\endgroup 
  
\end{theorem}

\begin{proof}
  The conclusion follows from Dwork and Roth~\cite{privacybook}, Theorem~B.1.





\end{proof}

\section{Proofs}

We define the semantics of each language simultaneously with its proof of
correctness. That is, we give a denotation for well typed terms that inhabits
the interpretation of the typing judgment. We simplify the proof slightly to
not consider symbolic real expressions in singleton and matrix types---the
denotation for those expressions is straightforward as elements of real numbers
extended with infinity, so we just consider this set in the proof below.
(In this section, we use the same variable names at different colors to
indicate distinct metavariables, {e.g.}, {\begingroup\renewcommand\colorMATH{\colorMATHS}\renewcommand\colorSYNTAX{\colorSYNTAXS}{{\color{\colorMATH}\ensuremath{\Gamma }}}\endgroup } and {\begingroup\renewcommand\colorMATH{\colorMATHP}\renewcommand\colorSYNTAX{\colorSYNTAXP}{{\color{\colorMATH}\ensuremath{\Gamma }}}\endgroup } are not the same.)
\begin{theorem}[Soundness]
  There exists an interpretation of well typed terms {\begingroup\renewcommand\colorMATH{\colorMATHS}\renewcommand\colorSYNTAX{\colorSYNTAXS}{{\color{\colorMATH}\ensuremath{\Gamma  \vdash  e \mathrel{:} {\begingroup\renewcommand\colorMATH{\colorMATHM}\renewcommand\colorSYNTAX{\colorSYNTAXM}{{\color{\colorMATH}\ensuremath{\tau }}}\endgroup }}}}\endgroup } and {\begingroup\renewcommand\colorMATH{\colorMATHP}\renewcommand\colorSYNTAX{\colorSYNTAXP}{{\color{\colorMATH}\ensuremath{\Gamma 
  \vdash  e \mathrel{:} {\begingroup\renewcommand\colorMATH{\colorMATHM}\renewcommand\colorSYNTAX{\colorSYNTAXM}{{\color{\colorMATH}\ensuremath{\tau }}}\endgroup }}}}\endgroup }, notated {\begingroup\renewcommand\colorMATH{\colorMATHS}\renewcommand\colorSYNTAX{\colorSYNTAXS}{{\color{\colorMATH}\ensuremath{\llbracket e\rrbracket }}}\endgroup } and {\begingroup\renewcommand\colorMATH{\colorMATHP}\renewcommand\colorSYNTAX{\colorSYNTAXP}{{\color{\colorMATH}\ensuremath{\llbracket e\rrbracket }}}\endgroup }, such that {{\color{\colorMATH}\ensuremath{{\begingroup\renewcommand\colorMATH{\colorMATHP}\renewcommand\colorSYNTAX{\colorSYNTAXP}{{\color{\colorMATH}\ensuremath{\llbracket e\rrbracket }}}\endgroup } \in  {\begingroup\renewcommand\colorMATH{\colorMATHP}\renewcommand\colorSYNTAX{\colorSYNTAXP}{{\color{\colorMATH}\ensuremath{\llbracket \Gamma  \vdash 
  {\begingroup\renewcommand\colorMATH{\colorMATHM}\renewcommand\colorSYNTAX{\colorSYNTAXM}{{\color{\colorMATH}\ensuremath{\tau }}}\endgroup }\rrbracket }}}\endgroup }}}} and {{\color{\colorMATH}\ensuremath{{\begingroup\renewcommand\colorMATH{\colorMATHS}\renewcommand\colorSYNTAX{\colorSYNTAXS}{{\color{\colorMATH}\ensuremath{\llbracket e\rrbracket }}}\endgroup } \in  {\begingroup\renewcommand\colorMATH{\colorMATHS}\renewcommand\colorSYNTAX{\colorSYNTAXS}{{\color{\colorMATH}\ensuremath{\llbracket \Gamma  \vdash  {\begingroup\renewcommand\colorMATH{\colorMATHM}\renewcommand\colorSYNTAX{\colorSYNTAXM}{{\color{\colorMATH}\ensuremath{\tau }}}\endgroup }\rrbracket }}}\endgroup }}}}.
\end{theorem}
\begin{proof}

We construct the interpretaion by induction on typing derivations {\begingroup\renewcommand\colorMATH{\colorMATHS}\renewcommand\colorSYNTAX{\colorSYNTAXS}{{\color{\colorMATH}\ensuremath{\Gamma  \vdash  e \mathrel{:}
{\begingroup\renewcommand\colorMATH{\colorMATHM}\renewcommand\colorSYNTAX{\colorSYNTAXM}{{\color{\colorMATH}\ensuremath{\tau }}}\endgroup }}}}\endgroup } and {\begingroup\renewcommand\colorMATH{\colorMATHP}\renewcommand\colorSYNTAX{\colorSYNTAXP}{{\color{\colorMATH}\ensuremath{\Gamma  \vdash  e \mathrel{:} {\begingroup\renewcommand\colorMATH{\colorMATHM}\renewcommand\colorSYNTAX{\colorSYNTAXM}{{\color{\colorMATH}\ensuremath{\tau }}}\endgroup }}}}\endgroup } which are defined mutually inductively.

We show key cases for the proof in each language. For the sensitivity language
{\begingroup\renewcommand\colorMATH{\colorMATHS}\renewcommand\colorSYNTAX{\colorSYNTAXS}{{\color{\colorMATH}\ensuremath{\Gamma  \vdash  e \mathrel{:} {\begingroup\renewcommand\colorMATH{\colorMATHM}\renewcommand\colorSYNTAX{\colorSYNTAXM}{{\color{\colorMATH}\ensuremath{\tau }}}\endgroup }}}}\endgroup }, there is only one interesting case that is not already
  considered in prior work (Fuzz and DFuzz): the introduction rule for the
  privacy lambda.

\vspace*{-0.25em}\begingroup\color{\colorMATH}\begin{gather*}\begin{tabularx}{\linewidth}{>{\centering\arraybackslash\(}X<{\)}}\hfill\hspace{0pt}\begingroup\color{\colorTEXT}\boxed{\begingroup\color{\colorMATH} {\begingroup\renewcommand\colorMATH{\colorMATHS}\renewcommand\colorSYNTAX{\colorSYNTAXS}{{\color{\colorMATH}\ensuremath{\llbracket \underline{\hspace{0.66em}\vspace*{5ex}}\rrbracket  \in  \{ e \in  {{\color{\colorMATH}\ensuremath{\operatorname{exp}}}} \mathrel{|} \Gamma  \vdash  e \mathrel{:} {\begingroup\renewcommand\colorMATH{\colorMATHM}\renewcommand\colorSYNTAX{\colorSYNTAXM}{{\color{\colorMATH}\ensuremath{\tau }}}\endgroup }\}  \rightarrow  \llbracket \Gamma  \vdash  {\begingroup\renewcommand\colorMATH{\colorMATHM}\renewcommand\colorSYNTAX{\colorSYNTAXM}{{\color{\colorMATH}\ensuremath{\tau }}}\endgroup }\rrbracket }}}\endgroup } \endgroup}\endgroup \end{tabularx}\vspace*{-1em}\end{gather*}\endgroup 
\begin{itemize}[leftmargin=*,label=\textbf{-
}]\item  \noindent  Case {\begingroup\renewcommand\colorMATH{\colorMATHS}\renewcommand\colorSYNTAX{\colorSYNTAXS}{{\color{\colorMATH}\ensuremath{\overbracketarg \Gamma {{}\rceil {\begingroup\renewcommand\colorMATH{\colorMATHP}\renewcommand\colorSYNTAX{\colorSYNTAXP}{{\color{\colorMATH}\ensuremath{\Gamma }}}\endgroup }\lceil {}^{{{\color{\colorSYNTAX}\texttt{\ensuremath{\infty }}}}}} \vdash  ({{\color{\colorSYNTAX}\texttt{\ensuremath{p\lambda \hspace*{0.33em}({.}\hspace{-1pt}{.}\hspace{-1pt}{.}{,}{\begingroup\renewcommand\colorMATH{\colorMATHM}\renewcommand\colorSYNTAX{\colorSYNTAXM}{{\color{\colorMATH}\ensuremath{x_{n}}}}\endgroup }{\mathrel{:}}{\begingroup\renewcommand\colorMATH{\colorMATHM}\renewcommand\colorSYNTAX{\colorSYNTAXM}{{\color{\colorMATH}\ensuremath{\tau _{n}}}}\endgroup }) \Rightarrow  {\begingroup\renewcommand\colorMATH{\colorMATHP}\renewcommand\colorSYNTAX{\colorSYNTAXP}{{\color{\colorMATH}\ensuremath{e}}}\endgroup }}}}}) \mathrel{:} {\begingroup\renewcommand\colorMATH{\colorMATHM}\renewcommand\colorSYNTAX{\colorSYNTAXM}{{\color{\colorMATH}\ensuremath{{{\color{\colorSYNTAX}\texttt{\ensuremath{({.}\hspace{-1pt}{.}\hspace{-1pt}{.}{,}{{\color{\colorMATH}\ensuremath{\tau _{n}}}}@{\begingroup\renewcommand\colorMATH{\colorMATHP}\renewcommand\colorSYNTAX{\colorSYNTAXP}{{\color{\colorMATH}\ensuremath{p_{n}}}}\endgroup }) \multimap ^{*} {{\color{\colorMATH}\ensuremath{\tau }}}}}}}}}}\endgroup } }}}\endgroup }:
   \\\noindent  By inversion:
      \vspace*{-0.25em}\begingroup\color{\colorMATH}\begin{gather*} {\begingroup\renewcommand\colorMATH{\colorMATHP}\renewcommand\colorSYNTAX{\colorSYNTAXP}{{\color{\colorMATH}\ensuremath{\Gamma \uplus \{ {\begingroup\renewcommand\colorMATH{\colorMATHM}\renewcommand\colorSYNTAX{\colorSYNTAXM}{{\color{\colorMATH}\ensuremath{x_{1}}}}\endgroup },{\mathrel{:}}_{p_{1}}{\begingroup\renewcommand\colorMATH{\colorMATHM}\renewcommand\colorSYNTAX{\colorSYNTAXM}{{\color{\colorMATH}\ensuremath{\tau _{1}}}}\endgroup },{.}\hspace{-1pt}{.}\hspace{-1pt}{.},{\begingroup\renewcommand\colorMATH{\colorMATHM}\renewcommand\colorSYNTAX{\colorSYNTAXM}{{\color{\colorMATH}\ensuremath{x_{n}}}}\endgroup }{\mathrel{:}}_{p_{n}}{\begingroup\renewcommand\colorMATH{\colorMATHM}\renewcommand\colorSYNTAX{\colorSYNTAXM}{{\color{\colorMATH}\ensuremath{\tau _{n}}}}\endgroup }\}  \vdash  e \mathrel{:} {\begingroup\renewcommand\colorMATH{\colorMATHM}\renewcommand\colorSYNTAX{\colorSYNTAXM}{{\color{\colorMATH}\ensuremath{\tau }}}\endgroup }}}}\endgroup } 
      \vspace*{-1em}\end{gather*}\endgroup 
   \\\noindent  By induction hypothesis (IH):
      \vspace*{-0.25em}\begingroup\color{\colorMATH}\begin{gather*} {\begingroup\renewcommand\colorMATH{\colorMATHP}\renewcommand\colorSYNTAX{\colorSYNTAXP}{{\color{\colorMATH}\ensuremath{\llbracket e\rrbracket }}}\endgroup } = f \in  {\begingroup\renewcommand\colorMATH{\colorMATHP}\renewcommand\colorSYNTAX{\colorSYNTAXP}{{\color{\colorMATH}\ensuremath{\llbracket \Gamma \uplus \{ {\begingroup\renewcommand\colorMATH{\colorMATHM}\renewcommand\colorSYNTAX{\colorSYNTAXM}{{\color{\colorMATH}\ensuremath{x_{1}}}}\endgroup },{\mathrel{:}}_{p_{1}}{\begingroup\renewcommand\colorMATH{\colorMATHM}\renewcommand\colorSYNTAX{\colorSYNTAXM}{{\color{\colorMATH}\ensuremath{\tau _{1}}}}\endgroup },{.}\hspace{-1pt}{.}\hspace{-1pt}{.},{\begingroup\renewcommand\colorMATH{\colorMATHM}\renewcommand\colorSYNTAX{\colorSYNTAXM}{{\color{\colorMATH}\ensuremath{x_{n}}}}\endgroup }{\mathrel{:}}_{p_{n}}{\begingroup\renewcommand\colorMATH{\colorMATHM}\renewcommand\colorSYNTAX{\colorSYNTAXM}{{\color{\colorMATH}\ensuremath{\tau _{n}}}}\endgroup }\}  \vdash  {\begingroup\renewcommand\colorMATH{\colorMATHM}\renewcommand\colorSYNTAX{\colorSYNTAXM}{{\color{\colorMATH}\ensuremath{\tau }}}\endgroup }\rrbracket }}}\endgroup } 
      \vspace*{-1em}\end{gather*}\endgroup 
   \\\noindent  Define:
      \vspace*{-0.25em}\begingroup\color{\colorMATH}\begin{gather*} \begin{array}{l
         } {\begingroup\renewcommand\colorMATH{\colorMATHS}\renewcommand\colorSYNTAX{\colorSYNTAXS}{{\color{\colorMATH}\ensuremath{\llbracket {{\color{\colorSYNTAX}\texttt{\ensuremath{p\lambda \hspace*{0.33em}({.}\hspace{-1pt}{.}\hspace{-1pt}{.}{,}{\begingroup\renewcommand\colorMATH{\colorMATHM}\renewcommand\colorSYNTAX{\colorSYNTAXM}{{\color{\colorMATH}\ensuremath{x_{n}}}}\endgroup }{\mathrel{:}}{\begingroup\renewcommand\colorMATH{\colorMATHM}\renewcommand\colorSYNTAX{\colorSYNTAXM}{{\color{\colorMATH}\ensuremath{\tau _{n}}}}\endgroup }) \Rightarrow  {\begingroup\renewcommand\colorMATH{\colorMATHP}\renewcommand\colorSYNTAX{\colorSYNTAXP}{{\color{\colorMATH}\ensuremath{e}}}\endgroup }}}}}\rrbracket }}}\endgroup } \triangleq  f^{\prime} \hspace*{1.00em}{{\color{\colorTEXT}\textit{where}}}
         \cr  \hspace*{1.00em}f^{\prime}(\gamma ) = \lambda \gamma ^{\prime}.\hspace*{0.33em}f^{\prime}(\gamma \uplus \gamma ^{\prime})
         \end{array}
      \vspace*{-1em}\end{gather*}\endgroup 
   \\\noindent  To show {{\color{\colorMATH}\ensuremath{f^{\prime} \in  {\begingroup\renewcommand\colorMATH{\colorMATHS}\renewcommand\colorSYNTAX{\colorSYNTAXS}{{\color{\colorMATH}\ensuremath{\llbracket \Gamma  \vdash  {\begingroup\renewcommand\colorMATH{\colorMATHM}\renewcommand\colorSYNTAX{\colorSYNTAXM}{{\color{\colorMATH}\ensuremath{{{\color{\colorSYNTAX}\texttt{\ensuremath{({.}\hspace{-1pt}{.}\hspace{-1pt}{.}{,}{{\color{\colorMATH}\ensuremath{\tau _{n}}}}@{\begingroup\renewcommand\colorMATH{\colorMATHP}\renewcommand\colorSYNTAX{\colorSYNTAXP}{{\color{\colorMATH}\ensuremath{p_{n}}}}\endgroup }) \multimap ^{*} {{\color{\colorMATH}\ensuremath{\tau }}}}}}}}}}\endgroup }\rrbracket }}}\endgroup }}}}, {i.e.}:
      \vspace*{-0.25em}\begingroup\color{\colorMATH}\begin{gather*} (1)\hspace*{1.00em}

             \end{array}
      \vspace*{-1em}\end{gather*}\endgroup 
   \\\noindent  \begin{itemize}[leftmargin=*,label=\textbf{*
      }]\item  \noindent  (1) Subcase: {{\color{\colorMATH}\ensuremath{{\begingroup\renewcommand\colorMATH{\colorMATHS}\renewcommand\colorSYNTAX{\colorSYNTAXS}{{\color{\colorMATH}\ensuremath{\dot r_{i}^{\prime}}}}\endgroup } = {{\color{\colorSYNTAX}\texttt{\ensuremath{\infty }}}}}}}
         \\\noindent  The property holds trivially when {{\color{\colorMATH}\ensuremath{{\begingroup\renewcommand\colorMATH{\colorMATHS}\renewcommand\colorSYNTAX{\colorSYNTAXS}{{\color{\colorMATH}\ensuremath{\dot r_{i}^{\prime}}}}\endgroup } = {{\color{\colorSYNTAX}\texttt{\ensuremath{\infty }}}}}}}.
         
      \item  \noindent  (1) Subcase: {{\color{\colorMATH}\ensuremath{{\begingroup\renewcommand\colorMATH{\colorMATHS}\renewcommand\colorSYNTAX{\colorSYNTAXS}{{\color{\colorMATH}\ensuremath{\dot r_{i}^{\prime}}}}\endgroup } = 0}}}
         \\\noindent  It must be that {{\color{\colorMATH}\ensuremath{\{ x_{i}^{\prime}{\mathrel{:}}_{0,0}\tau _{i}^{\prime}\} \in  {\begingroup\renewcommand\colorMATH{\colorMATHP}\renewcommand\colorSYNTAX{\colorSYNTAXP}{{\color{\colorMATH}\ensuremath{\Gamma }}}\endgroup }}}}.
         \\\noindent  By IH, {{\color{\colorMATH}\ensuremath{f}}} is constant {w.r.t.} {{\color{\colorMATH}\ensuremath{x_{i}^{\prime}}}} {i.e.} {{\color{\colorMATH}\ensuremath{{{\color{\colorMATH}\ensuremath{\operatorname{Pr}}}}[f^{\prime}(\gamma )(\gamma ^{\prime})=d] = {{\color{\colorMATH}\ensuremath{\operatorname{Pr}}}}[f^{\prime}(\gamma )(\gamma ^{\prime}[x_{i}^{\prime}{\mapsto }d])=d]}}}
         \\\noindent  Therefore {{\color{\colorMATH}\ensuremath{|f(\gamma ) -f(\gamma [x_{i}^{\prime}{\mapsto }d])|_{{\begingroup\renewcommand\colorMATH{\colorMATHM}\renewcommand\colorSYNTAX{\colorSYNTAXM}{{\color{\colorMATH}\ensuremath{\llbracket \tau _{i}^{\prime}\rrbracket }}}\endgroup }} = 0}}}.
         
      \item  \noindent  (2) follows immediately from IH and the definition of {\begingroup\renewcommand\colorMATH{\colorMATHM}\renewcommand\colorSYNTAX{\colorSYNTAXM}{{\color{\colorMATH}\ensuremath{\llbracket {{\color{\colorSYNTAX}\texttt{\ensuremath{({.}\hspace{-1pt}{.}\hspace{-1pt}{.}{,}{{\color{\colorMATH}\ensuremath{\tau _{n}}}}@{\begingroup\renewcommand\colorMATH{\colorMATHP}\renewcommand\colorSYNTAX{\colorSYNTAXP}{{\color{\colorMATH}\ensuremath{p_{n}}}}\endgroup }) {\multimap ^{*}} {{\color{\colorMATH}\ensuremath{\tau }}}}}}}\rrbracket }}}\endgroup }
         
      \end{itemize}
   
\end{itemize}

For the privacy language {\begingroup\renewcommand\colorMATH{\colorMATHP}\renewcommand\colorSYNTAX{\colorSYNTAXP}{{\color{\colorMATH}\ensuremath{\Gamma  \vdash  {\begingroup\renewcommand\colorMATH{\colorMATHM}\renewcommand\colorSYNTAX{\colorSYNTAXM}{{\color{\colorMATH}\ensuremath{\tau }}}\endgroup }}}}\endgroup } we repeat the {{\color{\colorTEXT}\textsc{\scriptsize Return}}} and {{\color{\colorTEXT}\textsc{\scriptsize Bind}}} rules
(from Fuzz and DFuzz) in addition to our novel rules for privacy lambda
application, mgauss, and loop. The rest of the rules in the privacy language
are straightforward adaptations of these proofs.

\vspace*{-0.25em}\begingroup\color{\colorMATH}\begin{gather*}\begin{tabularx}{\linewidth}{>{\centering\arraybackslash\(}X<{\)}}\hfill\hspace{0pt}\begingroup\color{\colorTEXT}\boxed{\begingroup\color{\colorMATH} {\begingroup\renewcommand\colorMATH{\colorMATHP}\renewcommand\colorSYNTAX{\colorSYNTAXP}{{\color{\colorMATH}\ensuremath{\llbracket \underline{\hspace{0.66em}\vspace*{5ex}}\rrbracket  \in  \{ e \in  {{\color{\colorMATH}\ensuremath{\operatorname{exp}}}} \mathrel{|} \Gamma  \vdash  e \mathrel{:} {\begingroup\renewcommand\colorMATH{\colorMATHM}\renewcommand\colorSYNTAX{\colorSYNTAXM}{{\color{\colorMATH}\ensuremath{\tau }}}\endgroup }\}  \rightarrow  \llbracket \Gamma  \vdash  {\begingroup\renewcommand\colorMATH{\colorMATHM}\renewcommand\colorSYNTAX{\colorSYNTAXM}{{\color{\colorMATH}\ensuremath{\tau }}}\endgroup }\rrbracket }}}\endgroup } \endgroup}\endgroup \end{tabularx}\vspace*{-1em}\end{gather*}\endgroup 
\begin{itemize}[leftmargin=*,label=\textbf{-
}]\item  \noindent  Case: {\begingroup\renewcommand\colorMATH{\colorMATHP}\renewcommand\colorSYNTAX{\colorSYNTAXP}{{\color{\colorMATH}\ensuremath{\overbracketarg \Gamma {{}\rceil {\begingroup\renewcommand\colorMATH{\colorMATHS}\renewcommand\colorSYNTAX{\colorSYNTAXS}{{\color{\colorMATH}\ensuremath{\Gamma }}}\endgroup }\lceil {}^{{{\color{\colorSYNTAX}\texttt{\ensuremath{\infty }}}}}} \vdash  {{\color{\colorSYNTAX}\texttt{return}}}\hspace*{0.33em}{\begingroup\renewcommand\colorMATH{\colorMATHS}\renewcommand\colorSYNTAX{\colorSYNTAXS}{{\color{\colorMATH}\ensuremath{e}}}\endgroup } \mathrel{:} {\begingroup\renewcommand\colorMATH{\colorMATHM}\renewcommand\colorSYNTAX{\colorSYNTAXM}{{\color{\colorMATH}\ensuremath{\tau }}}\endgroup }}}}\endgroup }
   \\\noindent  By inversion:
      \vspace*{-0.25em}\begingroup\color{\colorMATH}\begin{gather*} {\begingroup\renewcommand\colorMATH{\colorMATHS}\renewcommand\colorSYNTAX{\colorSYNTAXS}{{\color{\colorMATH}\ensuremath{ \Gamma  \vdash  e \mathrel{:} {\begingroup\renewcommand\colorMATH{\colorMATHM}\renewcommand\colorSYNTAX{\colorSYNTAXM}{{\color{\colorMATH}\ensuremath{\tau }}}\endgroup }}}}\endgroup } \vspace*{-1em}\end{gather*}\endgroup 
   \\\noindent  By Induction Hypothesis (IH): 
      \vspace*{-0.25em}\begingroup\color{\colorMATH}\begin{gather*} {{\color{\colorMATH}\ensuremath{{\begingroup\renewcommand\colorMATH{\colorMATHS}\renewcommand\colorSYNTAX{\colorSYNTAXS}{{\color{\colorMATH}\ensuremath{\llbracket e\rrbracket }}}\endgroup } = {\begingroup\renewcommand\colorMATH{\colorMATHM}\renewcommand\colorSYNTAX{\colorSYNTAXM}{{\color{\colorMATH}\ensuremath{f}}}\endgroup } \in  {\begingroup\renewcommand\colorMATH{\colorMATHS}\renewcommand\colorSYNTAX{\colorSYNTAXS}{{\color{\colorMATH}\ensuremath{ \llbracket  \Gamma  \vdash  {\begingroup\renewcommand\colorMATH{\colorMATHM}\renewcommand\colorSYNTAX{\colorSYNTAXM}{{\color{\colorMATH}\ensuremath{\tau }}}\endgroup } \rrbracket }}}\endgroup }}}} \vspace*{-1em}\end{gather*}\endgroup 
   \\\noindent  Define:
      \vspace*{-0.25em}\begingroup\color{\colorMATH}\begin{gather*} 

      \vspace*{-1em}\end{gather*}\endgroup 
   \\\noindent  \begin{itemize}[leftmargin=*,label=\textbf{*
      }]\item  \noindent  Subcase: {{\color{\colorMATH}\ensuremath{\epsilon _{i},\delta _{i} = \infty ,0}}}
         \\\noindent  The property holds trivially when {{\color{\colorMATH}\ensuremath{\epsilon _{i},\delta _{i}}}} = {{\color{\colorMATH}\ensuremath{\infty ,0}}}
         
      \item  \noindent  Subcase: {{\color{\colorMATH}\ensuremath{\epsilon _{i},\delta _{i} = 0,0}}}
         \\\noindent  It must be that {{\color{\colorMATH}\ensuremath{\{ x_{i}^{\prime}{\mathrel{:}}_{0}\tau _{i}^{\prime}\} \in  {\begingroup\renewcommand\colorMATH{\colorMATHS}\renewcommand\colorSYNTAX{\colorSYNTAXS}{{\color{\colorMATH}\ensuremath{\Gamma }}}\endgroup }}}}.
         \\\noindent  By IH, {{\color{\colorMATH}\ensuremath{f}}} is constant {w.r.t.} {{\color{\colorMATH}\ensuremath{x_{i}}}} {i.e.} 
         \\\noindent  {{\color{\colorMATH}\ensuremath{|f(\gamma ) -f(\gamma [x_{i}{\mapsto }d])|_{{\begingroup\renewcommand\colorMATH{\colorMATHM}\renewcommand\colorSYNTAX{\colorSYNTAXM}{{\color{\colorMATH}\ensuremath{\llbracket \tau _{i}\rrbracket }}}\endgroup }} = 0}}}
         \\\noindent  Therefore {{\color{\colorMATH}\ensuremath{{{\color{\colorMATH}\ensuremath{\operatorname{Pr}}}}[f^{\prime}(\gamma )=d] = {{\color{\colorMATH}\ensuremath{\operatorname{Pr}}}}[f^{\prime}(\gamma [x_{i}{\mapsto }d])=d]}}}
         
      \end{itemize}

\item  \noindent  Case: {\begingroup\renewcommand\colorMATH{\colorMATHP}\renewcommand\colorSYNTAX{\colorSYNTAXP}{{\color{\colorMATH}\ensuremath{\overbracketarg \Gamma {{}\rceil {\begingroup\renewcommand\colorMATH{\colorMATHS}\renewcommand\colorSYNTAX{\colorSYNTAXS}{{\color{\colorMATH}\ensuremath{\Gamma _{1}}}}\endgroup }\lceil {}^{{{\color{\colorSYNTAX}\texttt{\ensuremath{\infty }}}}} + \{ {\begingroup\renewcommand\colorMATH{\colorMATHM}\renewcommand\colorSYNTAX{\colorSYNTAXM}{{\color{\colorMATH}\ensuremath{x_{1}}}}\endgroup }{\mathrel{:}}_{p_{1}}{\begingroup\renewcommand\colorMATH{\colorMATHM}\renewcommand\colorSYNTAX{\colorSYNTAXM}{{\color{\colorMATH}\ensuremath{\tau _{1}}}}\endgroup },{.}\hspace{-1pt}{.}\hspace{-1pt}{.},{\begingroup\renewcommand\colorMATH{\colorMATHM}\renewcommand\colorSYNTAX{\colorSYNTAXM}{{\color{\colorMATH}\ensuremath{x_{n}}}}\endgroup }{\mathrel{:}}_{p_{n}}{\begingroup\renewcommand\colorMATH{\colorMATHM}\renewcommand\colorSYNTAX{\colorSYNTAXM}{{\color{\colorMATH}\ensuremath{\tau _{n}}}}\endgroup }\} } \vdash  {{\color{\colorSYNTAX}\texttt{\ensuremath{{\begingroup\renewcommand\colorMATH{\colorMATHS}\renewcommand\colorSYNTAX{\colorSYNTAXS}{{\color{\colorMATH}\ensuremath{e}}}\endgroup }({\begingroup\renewcommand\colorMATH{\colorMATHM}\renewcommand\colorSYNTAX{\colorSYNTAXM}{{\color{\colorMATH}\ensuremath{x_{1}}}}\endgroup }{,}...{,}{\begingroup\renewcommand\colorMATH{\colorMATHM}\renewcommand\colorSYNTAX{\colorSYNTAXM}{{\color{\colorMATH}\ensuremath{x_{n}}}}\endgroup })}}}} \mathrel{:} {\begingroup\renewcommand\colorMATH{\colorMATHM}\renewcommand\colorSYNTAX{\colorSYNTAXM}{{\color{\colorMATH}\ensuremath{\tau }}}\endgroup }}}}\endgroup }
   \\\noindent  By inversion:
      \vspace*{-0.25em}\begingroup\color{\colorMATH}\begin{gather*} {\begingroup\renewcommand\colorMATH{\colorMATHS}\renewcommand\colorSYNTAX{\colorSYNTAXS}{{\color{\colorMATH}\ensuremath{\Gamma  \vdash  e \mathrel{:} {\begingroup\renewcommand\colorMATH{\colorMATHM}\renewcommand\colorSYNTAX{\colorSYNTAXM}{{\color{\colorMATH}\ensuremath{{{\color{\colorSYNTAX}\texttt{\ensuremath{({{\color{\colorMATH}\ensuremath{\tau _{1}}}}@{\begingroup\renewcommand\colorMATH{\colorMATHP}\renewcommand\colorSYNTAX{\colorSYNTAXP}{{\color{\colorMATH}\ensuremath{p_{1}}}}\endgroup },{.}\hspace{-1pt}{.}\hspace{-1pt}{.},{{\color{\colorMATH}\ensuremath{\tau _{n}}}}@{\begingroup\renewcommand\colorMATH{\colorMATHP}\renewcommand\colorSYNTAX{\colorSYNTAXP}{{\color{\colorMATH}\ensuremath{p_{n}}}}\endgroup }) \multimap ^{*} {{\color{\colorMATH}\ensuremath{\tau }}}}}}}}}}\endgroup }}}}\endgroup } \vspace*{-1em}\end{gather*}\endgroup 
   \\\noindent  By Induction Hypothesis (IH): 
      \vspace*{-0.25em}\begingroup\color{\colorMATH}\begin{gather*} {\begingroup\renewcommand\colorMATH{\colorMATHS}\renewcommand\colorSYNTAX{\colorSYNTAXS}{{\color{\colorMATH}\ensuremath{\llbracket e\rrbracket }}}\endgroup } = f \in  {\begingroup\renewcommand\colorMATH{\colorMATHS}\renewcommand\colorSYNTAX{\colorSYNTAXS}{{\color{\colorMATH}\ensuremath{\llbracket \Gamma  \vdash  {\begingroup\renewcommand\colorMATH{\colorMATHM}\renewcommand\colorSYNTAX{\colorSYNTAXM}{{\color{\colorMATH}\ensuremath{{{\color{\colorSYNTAX}\texttt{\ensuremath{({{\color{\colorMATH}\ensuremath{\tau _{1}}}}@{\begingroup\renewcommand\colorMATH{\colorMATHP}\renewcommand\colorSYNTAX{\colorSYNTAXP}{{\color{\colorMATH}\ensuremath{p_{1}}}}\endgroup },{.}\hspace{-1pt}{.}\hspace{-1pt}{.},{{\color{\colorMATH}\ensuremath{\tau _{n}}}}@{\begingroup\renewcommand\colorMATH{\colorMATHP}\renewcommand\colorSYNTAX{\colorSYNTAXP}{{\color{\colorMATH}\ensuremath{p_{n}}}}\endgroup }) \multimap ^{*} {{\color{\colorMATH}\ensuremath{\tau }}}}}}}}}}\endgroup }\rrbracket }}}\endgroup } \vspace*{-1em}\end{gather*}\endgroup 
   \\\noindent  Define:
      \vspace*{-0.25em}\begingroup\color{\colorMATH}\begin{gather*} 
\begin{array}{l
         } {\begingroup\renewcommand\colorMATH{\colorMATHP}\renewcommand\colorSYNTAX{\colorSYNTAXP}{{\color{\colorMATH}\ensuremath{\llbracket {{\color{\colorSYNTAX}\texttt{\ensuremath{{\begingroup\renewcommand\colorMATH{\colorMATHS}\renewcommand\colorSYNTAX{\colorSYNTAXS}{{\color{\colorMATH}\ensuremath{e}}}\endgroup }({\begingroup\renewcommand\colorMATH{\colorMATHM}\renewcommand\colorSYNTAX{\colorSYNTAXM}{{\color{\colorMATH}\ensuremath{x_{1}}}}\endgroup }{,}...{,}{\begingroup\renewcommand\colorMATH{\colorMATHM}\renewcommand\colorSYNTAX{\colorSYNTAXM}{{\color{\colorMATH}\ensuremath{x_{n}}}}\endgroup })}}}}\rrbracket }}}\endgroup } \triangleq  f^{\prime}\hspace*{1.00em}{{\color{\colorTEXT}\textit{where}}}
         \cr  \hspace*{1.00em} {{\color{\colorMATH}\ensuremath{\operatorname{Pr}}}}[f^{\prime}(\gamma )=d] \triangleq  {{\color{\colorMATH}\ensuremath{\operatorname{Pr}}}}[f(\gamma )(\gamma [x_{1}], {.}\hspace{-1pt}{.}\hspace{-1pt}{.}, \gamma [x_{n}]) = d]
         \end{array}%
      \vspace*{-1em}\end{gather*}\endgroup 
   \\\noindent  By definition of {\begingroup\renewcommand\colorMATH{\colorMATHP}\renewcommand\colorSYNTAX{\colorSYNTAXP}{{\color{\colorMATH}\ensuremath{+}}}\endgroup } for privacy contexts:
   \\\noindent  {{\color{\colorMATH}\ensuremath{\epsilon _{i},\delta _{i} = \epsilon _{i}^{\prime} + \epsilon _{i}^{\prime \prime},\delta _{i}^{\prime} + \delta _{i}^{\prime \prime}}}} for 
   \\\noindent  {{\color{\colorMATH}\ensuremath{\{ x_{i}\mathrel{:}_{\epsilon _{i}^{\prime},\delta _{i}^{\prime}}\tau _{i}\}  \in  {\begingroup\renewcommand\colorMATH{\colorMATHP}\renewcommand\colorSYNTAX{\colorSYNTAXP}{{\color{\colorMATH}\ensuremath{{}\rceil {\begingroup\renewcommand\colorMATH{\colorMATHS}\renewcommand\colorSYNTAX{\colorSYNTAXS}{{\color{\colorMATH}\ensuremath{\Gamma _{1}}}}\endgroup }\lceil {}^{{{\color{\colorSYNTAX}\texttt{\ensuremath{\infty }}}}}}}}\endgroup }}}} and {{\color{\colorMATH}\ensuremath{\epsilon _{i}^{\prime \prime},\delta _{i}^{\prime \prime} = {\begingroup\renewcommand\colorMATH{\colorMATHP}\renewcommand\colorSYNTAX{\colorSYNTAXP}{{\color{\colorMATH}\ensuremath{\llbracket p_{i}\rrbracket }}}\endgroup }}}}.
   \\\noindent  \begin{itemize}[leftmargin=*,label=\textbf{*
      }]\item  \noindent  Subcase {{\color{\colorMATH}\ensuremath{\epsilon _{i}^{\prime},\delta _{i}^{\prime} = \infty ,0}}}:
         \\\noindent  {{\color{\colorMATH}\ensuremath{\epsilon _{i},\delta _{i} = \infty ,0}}}; the property holds trivially
         
       \item  \noindent  Subcase {{\color{\colorMATH}\ensuremath{\epsilon _{i}^{\prime},\delta _{i}^{\prime} = 0,0}}}:
         \\\noindent  {{\color{\colorMATH}\ensuremath{\epsilon _{i},\delta _{i}}}} = {{\color{\colorMATH}\ensuremath{\epsilon _{i}^{\prime \prime},\delta _{i}^{\prime \prime}}}}; the property follows by IH and definition of
            {{\color{\colorSYNTAX}\texttt{\ensuremath{\multimap ^{*}}}}} instiated to index {{\color{\colorMATH}\ensuremath{i}}}.
         
      \end{itemize}

\item  \noindent  Case {\begingroup\renewcommand\colorMATH{\colorMATHP}\renewcommand\colorSYNTAX{\colorSYNTAXP}{{\color{\colorMATH}\ensuremath{\overbracketarg \Gamma {\Gamma _{1} + \Gamma _{2}} \vdash  {{\color{\colorSYNTAX}\texttt{\ensuremath{{\begingroup\renewcommand\colorMATH{\colorMATHM}\renewcommand\colorSYNTAX{\colorSYNTAXM}{{\color{\colorMATH}\ensuremath{x}}}\endgroup } \leftarrow  {{\color{\colorMATH}\ensuremath{e_{1}}}}\mathrel{;}{{\color{\colorMATH}\ensuremath{e_{2}}}}}}}} \mathrel{:} {\begingroup\renewcommand\colorMATH{\colorMATHM}\renewcommand\colorSYNTAX{\colorSYNTAXM}{{\color{\colorMATH}\ensuremath{\tau _{2}}}}\endgroup }}}}\endgroup }
   \\\noindent  By inversion:
      \vspace*{-0.25em}\begingroup\color{\colorMATH}\begin{gather*} 

      \vspace*{-1em}\end{gather*}\endgroup 
   \\\noindent  By Induction Hypothesis (IH):
      \vspace*{-0.25em}\begingroup\color{\colorMATH}\begin{gather*} %
      \vspace*{-1em}\end{gather*}\endgroup 
   \\\noindent  By definition of {\begingroup\renewcommand\colorMATH{\colorMATHP}\renewcommand\colorSYNTAX{\colorSYNTAXP}{{\color{\colorMATH}\ensuremath{+}}}\endgroup } for privacy contexts:
   \\\noindent  {{\color{\colorMATH}\ensuremath{\epsilon _{i},\delta _{i} =  \epsilon _{i}^{\prime} + \epsilon _{i}^{\prime \prime},\delta _{i}^{\prime} + \delta _{i}^{\prime \prime}}}} for
   \\\noindent  {{\color{\colorMATH}\ensuremath{{\begingroup\renewcommand\colorMATH{\colorMATHP}\renewcommand\colorSYNTAX{\colorSYNTAXP}{{\color{\colorMATH}\ensuremath{\{ {\begingroup\renewcommand\colorMATH{\colorMATHM}\renewcommand\colorSYNTAX{\colorSYNTAXM}{{\color{\colorMATH}\ensuremath{x_{i}}}}\endgroup }{\mathrel{:}}_{{\begingroup\renewcommand\colorMATH{\colorMATHM}\renewcommand\colorSYNTAX{\colorSYNTAXM}{{\color{\colorMATH}\ensuremath{\epsilon _{i}^{\prime}}}}\endgroup },{\begingroup\renewcommand\colorMATH{\colorMATHM}\renewcommand\colorSYNTAX{\colorSYNTAXM}{{\color{\colorMATH}\ensuremath{\delta _{i}^{\prime}}}}\endgroup }} {\begingroup\renewcommand\colorMATH{\colorMATHM}\renewcommand\colorSYNTAX{\colorSYNTAXM}{{\color{\colorMATH}\ensuremath{\tau }}}\endgroup }\} }}}\endgroup } \in  {\begingroup\renewcommand\colorMATH{\colorMATHP}\renewcommand\colorSYNTAX{\colorSYNTAXP}{{\color{\colorMATH}\ensuremath{\Gamma _{1}}}}\endgroup }}}} and {{\color{\colorMATH}\ensuremath{{\begingroup\renewcommand\colorMATH{\colorMATHP}\renewcommand\colorSYNTAX{\colorSYNTAXP}{{\color{\colorMATH}\ensuremath{\{ {\begingroup\renewcommand\colorMATH{\colorMATHM}\renewcommand\colorSYNTAX{\colorSYNTAXM}{{\color{\colorMATH}\ensuremath{x_{i}}}}\endgroup }{\mathrel{:}}_{{\begingroup\renewcommand\colorMATH{\colorMATHM}\renewcommand\colorSYNTAX{\colorSYNTAXM}{{\color{\colorMATH}\ensuremath{\epsilon _{i}^{\prime \prime}}}}\endgroup },{\begingroup\renewcommand\colorMATH{\colorMATHM}\renewcommand\colorSYNTAX{\colorSYNTAXM}{{\color{\colorMATH}\ensuremath{\delta _{i}^{\prime \prime}}}}\endgroup }} {\begingroup\renewcommand\colorMATH{\colorMATHM}\renewcommand\colorSYNTAX{\colorSYNTAXM}{{\color{\colorMATH}\ensuremath{\tau }}}\endgroup }\} }}}\endgroup } \in  {\begingroup\renewcommand\colorMATH{\colorMATHP}\renewcommand\colorSYNTAX{\colorSYNTAXP}{{\color{\colorMATH}\ensuremath{\Gamma _{2}}}}\endgroup }}}}. 
   \\\noindent  Property holds via IH.1, IH.2 and theorem~\ref{thm:let} instantiated with
      {{\color{\colorMATH}\ensuremath{(\epsilon _{i}^{\prime},\delta _{i}^{\prime})}}} and {{\color{\colorMATH}\ensuremath{(\epsilon _{i}^{\prime \prime},\delta _{i}^{\prime \prime})}}}.
   
\item  \noindent  Case: 
   \\\noindent  {{\color{\colorMATH}\ensuremath{\hspace*{1.00em}

      \vspace*{-1em}\end{gather*}\endgroup 
   \\\noindent  By Induction Hypothesis (IH):
      \vspace*{-0.25em}\begingroup\color{\colorMATH}\begin{gather*} %
      \vspace*{-1em}\end{gather*}\endgroup 
   \\\noindent  By IH, we have that {{\color{\colorMATH}\ensuremath{\sigma ^{2} = \frac{2 \log (1.25/\delta ) \dot r^{2}}{\epsilon ^{2}}}}}
   \\\noindent  By definition of {\begingroup\renewcommand\colorMATH{\colorMATHP}\renewcommand\colorSYNTAX{\colorSYNTAXP}{{\color{\colorMATH}\ensuremath{+}}}\endgroup } for privacy contexts:
   \\\noindent  {{\color{\colorMATH}\ensuremath{\epsilon _{i},\delta _{i} =  \epsilon _{i}^{\prime} + \epsilon _{i}^{\prime \prime} + \epsilon _{i}^{\prime \prime \prime} + \epsilon _{i}^{\prime 4} + \epsilon _{i}^{\prime 5},\delta _{i}^{\prime} + \delta _{i}^{\prime \prime} + \delta _{i}^{\prime \prime \prime} + \delta _{i}^{\prime 4} + \delta _{i}^{\prime 5}}}} for
   \\\noindent  {{\color{\colorMATH}\ensuremath{{\begingroup\renewcommand\colorMATH{\colorMATHP}\renewcommand\colorSYNTAX{\colorSYNTAXP}{{\color{\colorMATH}\ensuremath{\{ {\begingroup\renewcommand\colorMATH{\colorMATHM}\renewcommand\colorSYNTAX{\colorSYNTAXM}{{\color{\colorMATH}\ensuremath{x_{i}}}}\endgroup }{\mathrel{:}}_{{\begingroup\renewcommand\colorMATH{\colorMATHM}\renewcommand\colorSYNTAX{\colorSYNTAXM}{{\color{\colorMATH}\ensuremath{\epsilon _{i}^{\prime}}}}\endgroup },{\begingroup\renewcommand\colorMATH{\colorMATHM}\renewcommand\colorSYNTAX{\colorSYNTAXM}{{\color{\colorMATH}\ensuremath{\delta _{i}^{\prime}}}}\endgroup }} {\begingroup\renewcommand\colorMATH{\colorMATHM}\renewcommand\colorSYNTAX{\colorSYNTAXM}{{\color{\colorMATH}\ensuremath{\tau }}}\endgroup }\} }}}\endgroup } \in  {\begingroup\renewcommand\colorMATH{\colorMATHP}\renewcommand\colorSYNTAX{\colorSYNTAXP}{{\color{\colorMATH}\ensuremath{{}\rceil {\begingroup\renewcommand\colorMATH{\colorMATHS}\renewcommand\colorSYNTAX{\colorSYNTAXS}{{\color{\colorMATH}\ensuremath{\Gamma _{1}}}}\endgroup }\lceil {}^{{\begingroup\renewcommand\colorMATH{\colorMATHM}\renewcommand\colorSYNTAX{\colorSYNTAXM}{{\color{\colorMATH}\ensuremath{0,0}}}\endgroup }}}}}\endgroup }}}} 
   \\\noindent  {{\color{\colorMATH}\ensuremath{{\begingroup\renewcommand\colorMATH{\colorMATHP}\renewcommand\colorSYNTAX{\colorSYNTAXP}{{\color{\colorMATH}\ensuremath{\{ {\begingroup\renewcommand\colorMATH{\colorMATHM}\renewcommand\colorSYNTAX{\colorSYNTAXM}{{\color{\colorMATH}\ensuremath{x_{i}}}}\endgroup }{\mathrel{:}}_{{\begingroup\renewcommand\colorMATH{\colorMATHM}\renewcommand\colorSYNTAX{\colorSYNTAXM}{{\color{\colorMATH}\ensuremath{\epsilon _{i}^{\prime \prime}}}}\endgroup },{\begingroup\renewcommand\colorMATH{\colorMATHM}\renewcommand\colorSYNTAX{\colorSYNTAXM}{{\color{\colorMATH}\ensuremath{\delta _{i}^{\prime \prime}}}}\endgroup }} {\begingroup\renewcommand\colorMATH{\colorMATHM}\renewcommand\colorSYNTAX{\colorSYNTAXM}{{\color{\colorMATH}\ensuremath{\tau }}}\endgroup }\} }}}\endgroup } \in  {\begingroup\renewcommand\colorMATH{\colorMATHP}\renewcommand\colorSYNTAX{\colorSYNTAXP}{{\color{\colorMATH}\ensuremath{{}\rceil {\begingroup\renewcommand\colorMATH{\colorMATHS}\renewcommand\colorSYNTAX{\colorSYNTAXS}{{\color{\colorMATH}\ensuremath{\Gamma _{2}}}}\endgroup }\lceil {}^{{\begingroup\renewcommand\colorMATH{\colorMATHM}\renewcommand\colorSYNTAX{\colorSYNTAXM}{{\color{\colorMATH}\ensuremath{0,0}}}\endgroup }}}}}\endgroup }}}} 
   \\\noindent  {{\color{\colorMATH}\ensuremath{{\begingroup\renewcommand\colorMATH{\colorMATHP}\renewcommand\colorSYNTAX{\colorSYNTAXP}{{\color{\colorMATH}\ensuremath{\{ {\begingroup\renewcommand\colorMATH{\colorMATHM}\renewcommand\colorSYNTAX{\colorSYNTAXM}{{\color{\colorMATH}\ensuremath{x_{i}}}}\endgroup }{\mathrel{:}}_{{\begingroup\renewcommand\colorMATH{\colorMATHM}\renewcommand\colorSYNTAX{\colorSYNTAXM}{{\color{\colorMATH}\ensuremath{\epsilon _{i}^{\prime \prime \prime}}}}\endgroup },{\begingroup\renewcommand\colorMATH{\colorMATHM}\renewcommand\colorSYNTAX{\colorSYNTAXM}{{\color{\colorMATH}\ensuremath{\delta _{i}^{\prime \prime \prime}}}}\endgroup }} {\begingroup\renewcommand\colorMATH{\colorMATHM}\renewcommand\colorSYNTAX{\colorSYNTAXM}{{\color{\colorMATH}\ensuremath{\tau }}}\endgroup }\} }}}\endgroup } \in  {\begingroup\renewcommand\colorMATH{\colorMATHP}\renewcommand\colorSYNTAX{\colorSYNTAXP}{{\color{\colorMATH}\ensuremath{{}\rceil {\begingroup\renewcommand\colorMATH{\colorMATHS}\renewcommand\colorSYNTAX{\colorSYNTAXS}{{\color{\colorMATH}\ensuremath{\Gamma _{3}}}}\endgroup }\lceil {}^{{\begingroup\renewcommand\colorMATH{\colorMATHM}\renewcommand\colorSYNTAX{\colorSYNTAXM}{{\color{\colorMATH}\ensuremath{0,0}}}\endgroup }}}}}\endgroup }}}} 
   \\\noindent  {{\color{\colorMATH}\ensuremath{{\begingroup\renewcommand\colorMATH{\colorMATHP}\renewcommand\colorSYNTAX{\colorSYNTAXP}{{\color{\colorMATH}\ensuremath{\{ {\begingroup\renewcommand\colorMATH{\colorMATHM}\renewcommand\colorSYNTAX{\colorSYNTAXM}{{\color{\colorMATH}\ensuremath{x_{i}}}}\endgroup }{\mathrel{:}}_{{\begingroup\renewcommand\colorMATH{\colorMATHM}\renewcommand\colorSYNTAX{\colorSYNTAXM}{{\color{\colorMATH}\ensuremath{\epsilon _{i}^{\prime 4}}}}\endgroup },{\begingroup\renewcommand\colorMATH{\colorMATHM}\renewcommand\colorSYNTAX{\colorSYNTAXM}{{\color{\colorMATH}\ensuremath{\delta _{i}^{\prime 4}}}}\endgroup }} {\begingroup\renewcommand\colorMATH{\colorMATHM}\renewcommand\colorSYNTAX{\colorSYNTAXM}{{\color{\colorMATH}\ensuremath{\tau }}}\endgroup }\} }}}\endgroup } \in  {\begingroup\renewcommand\colorMATH{\colorMATHP}\renewcommand\colorSYNTAX{\colorSYNTAXP}{{\color{\colorMATH}\ensuremath{{}\rceil {\begingroup\renewcommand\colorMATH{\colorMATHS}\renewcommand\colorSYNTAX{\colorSYNTAXS}{{\color{\colorMATH}\ensuremath{\Gamma _{4}}}}\endgroup }\lceil {}^{{{\color{\colorSYNTAX}\texttt{\ensuremath{\infty }}}}}}}}\endgroup }}}} 
   \\\noindent  {{\color{\colorMATH}\ensuremath{{\begingroup\renewcommand\colorMATH{\colorMATHP}\renewcommand\colorSYNTAX{\colorSYNTAXP}{{\color{\colorMATH}\ensuremath{\{ {\begingroup\renewcommand\colorMATH{\colorMATHM}\renewcommand\colorSYNTAX{\colorSYNTAXM}{{\color{\colorMATH}\ensuremath{x_{i}}}}\endgroup }{\mathrel{:}}_{{\begingroup\renewcommand\colorMATH{\colorMATHM}\renewcommand\colorSYNTAX{\colorSYNTAXM}{{\color{\colorMATH}\ensuremath{\epsilon _{i}^{\prime 5}}}}\endgroup },{\begingroup\renewcommand\colorMATH{\colorMATHM}\renewcommand\colorSYNTAX{\colorSYNTAXM}{{\color{\colorMATH}\ensuremath{\delta _{i}^{\prime 5}}}}\endgroup }} {\begingroup\renewcommand\colorMATH{\colorMATHM}\renewcommand\colorSYNTAX{\colorSYNTAXM}{{\color{\colorMATH}\ensuremath{\tau }}}\endgroup }\} }}}\endgroup } \in  {\begingroup\renewcommand\colorMATH{\colorMATHP}\renewcommand\colorSYNTAX{\colorSYNTAXP}{{\color{\colorMATH}\ensuremath{\mathrlap{\hspace{-0.5pt}{}\rceil }\lfloor {\begingroup\renewcommand\colorMATH{\colorMATHS}\renewcommand\colorSYNTAX{\colorSYNTAXS}{{\color{\colorMATH}\ensuremath{\Gamma _{5}}}}\endgroup }\mathrlap{\hspace{-0.5pt}\rfloor }\lceil {}_{\{ {\begingroup\renewcommand\colorMATH{\colorMATHM}\renewcommand\colorSYNTAX{\colorSYNTAXM}{{\color{\colorMATH}\ensuremath{x_{1}}}}\endgroup },{.}\hspace{-1pt}{.}\hspace{-1pt}{.},{\begingroup\renewcommand\colorMATH{\colorMATHM}\renewcommand\colorSYNTAX{\colorSYNTAXM}{{\color{\colorMATH}\ensuremath{x_{n}}}}\endgroup }\} }^{{\begingroup\renewcommand\colorMATH{\colorMATHM}\renewcommand\colorSYNTAX{\colorSYNTAXM}{{\color{\colorMATH}\ensuremath{\epsilon }}}\endgroup },{\begingroup\renewcommand\colorMATH{\colorMATHM}\renewcommand\colorSYNTAX{\colorSYNTAXM}{{\color{\colorMATH}\ensuremath{\delta }}}\endgroup }}}}}\endgroup }}}} 
   \\\noindent  Each of {{\color{\colorMATH}\ensuremath{\epsilon _{i}^{\prime}}}}, {{\color{\colorMATH}\ensuremath{\epsilon _{i}^{\prime \prime}}}}, {{\color{\colorMATH}\ensuremath{\epsilon _{i}^{\prime \prime \prime}}}}, {{\color{\colorMATH}\ensuremath{\delta _{i}^{\prime}}}}, {{\color{\colorMATH}\ensuremath{\delta _{i}^{\prime \prime}}}}, {{\color{\colorMATH}\ensuremath{\delta _{i}^{\prime \prime \prime}}}} must be {{\color{\colorMATH}\ensuremath{0}}}.
   \\\noindent  \begin{itemize}[leftmargin=*,label=\textbf{*
      }]\item  \noindent  Subcase {{\color{\colorMATH}\ensuremath{\epsilon _{i}^{\prime 4},\delta _{i}^{\prime 4} = \infty ,0}}}:
        \\\noindent  {{\color{\colorMATH}\ensuremath{ \epsilon _{i},\delta _{i} = \infty ,0 }}}; the property holds trivially
        
      \item  \noindent  Subcase {{\color{\colorMATH}\ensuremath{\epsilon ^{\prime 4},\delta ^{\prime 4} = 0,0}}}:
         \\\noindent  {{\color{\colorMATH}\ensuremath{\epsilon _{i},\delta _{i}}}} = {{\color{\colorMATH}\ensuremath{\epsilon _{i}^{\prime 5},\delta _{i}^{\prime 5}}}} = {{\color{\colorMATH}\ensuremath{\epsilon ,\delta }}}
         \\\noindent  By IH, {{\color{\colorMATH}\ensuremath{ |f_{4}(\gamma ) - f_{4}(\gamma [x_{i}\mapsto d])|_{\llbracket {\mathbb{R}}\rrbracket } \leq  r }}}
         \\\noindent  The property follows from Theorem~\ref{thm:gauss}.
         
      \end{itemize}

\item  \noindent  Case: 
   \\\noindent  {{\color{\colorMATH}\ensuremath{\hspace*{1.00em}

      \vspace*{-1em}\end{gather*}\endgroup 
   \\\noindent  By Induction Hypothesis (IH):
      \vspace*{-0.25em}\begingroup\color{\colorMATH}\begin{gather*} %
      \vspace*{-1em}\end{gather*}\endgroup 
   \\\noindent  By definition of {\begingroup\renewcommand\colorMATH{\colorMATHP}\renewcommand\colorSYNTAX{\colorSYNTAXP}{{\color{\colorMATH}\ensuremath{+}}}\endgroup } for privacy contexts:
   \\\noindent  {{\color{\colorMATH}\ensuremath{\epsilon _{i},\delta _{i} =  \epsilon _{i}^{\prime} + \epsilon _{i}^{\prime \prime} + \epsilon _{i}^{\prime \prime \prime} + \epsilon _{i}^{\prime 4} + \epsilon _{i}^{\prime 5},\delta _{i}^{\prime} + \delta _{i}^{\prime \prime} + \delta _{i}^{\prime \prime \prime} + \delta _{i}^{\prime 4} + \delta _{i}^{\prime 5}}}} for
   \\\noindent  {{\color{\colorMATH}\ensuremath{{\begingroup\renewcommand\colorMATH{\colorMATHP}\renewcommand\colorSYNTAX{\colorSYNTAXP}{{\color{\colorMATH}\ensuremath{\{ {\begingroup\renewcommand\colorMATH{\colorMATHM}\renewcommand\colorSYNTAX{\colorSYNTAXM}{{\color{\colorMATH}\ensuremath{x_{i}}}}\endgroup }{\mathrel{:}}_{{\begingroup\renewcommand\colorMATH{\colorMATHM}\renewcommand\colorSYNTAX{\colorSYNTAXM}{{\color{\colorMATH}\ensuremath{\epsilon _{i}^{\prime}}}}\endgroup },{\begingroup\renewcommand\colorMATH{\colorMATHM}\renewcommand\colorSYNTAX{\colorSYNTAXM}{{\color{\colorMATH}\ensuremath{\delta _{i}^{\prime}}}}\endgroup }} {\begingroup\renewcommand\colorMATH{\colorMATHM}\renewcommand\colorSYNTAX{\colorSYNTAXM}{{\color{\colorMATH}\ensuremath{\tau }}}\endgroup }\} }}}\endgroup } \in  {\begingroup\renewcommand\colorMATH{\colorMATHP}\renewcommand\colorSYNTAX{\colorSYNTAXP}{{\color{\colorMATH}\ensuremath{{}\rceil {\begingroup\renewcommand\colorMATH{\colorMATHS}\renewcommand\colorSYNTAX{\colorSYNTAXS}{{\color{\colorMATH}\ensuremath{\Gamma _{1}}}}\endgroup }\lceil {}^{{\begingroup\renewcommand\colorMATH{\colorMATHM}\renewcommand\colorSYNTAX{\colorSYNTAXM}{{\color{\colorMATH}\ensuremath{0,0}}}\endgroup }}}}}\endgroup }}}} 
   \\\noindent  {{\color{\colorMATH}\ensuremath{{\begingroup\renewcommand\colorMATH{\colorMATHP}\renewcommand\colorSYNTAX{\colorSYNTAXP}{{\color{\colorMATH}\ensuremath{\{ {\begingroup\renewcommand\colorMATH{\colorMATHM}\renewcommand\colorSYNTAX{\colorSYNTAXM}{{\color{\colorMATH}\ensuremath{x_{i}}}}\endgroup }{\mathrel{:}}_{{\begingroup\renewcommand\colorMATH{\colorMATHM}\renewcommand\colorSYNTAX{\colorSYNTAXM}{{\color{\colorMATH}\ensuremath{\epsilon _{i}^{\prime \prime}}}}\endgroup },{\begingroup\renewcommand\colorMATH{\colorMATHM}\renewcommand\colorSYNTAX{\colorSYNTAXM}{{\color{\colorMATH}\ensuremath{\delta _{i}^{\prime \prime}}}}\endgroup }} {\begingroup\renewcommand\colorMATH{\colorMATHM}\renewcommand\colorSYNTAX{\colorSYNTAXM}{{\color{\colorMATH}\ensuremath{\tau }}}\endgroup }\} }}}\endgroup } \in  {\begingroup\renewcommand\colorMATH{\colorMATHP}\renewcommand\colorSYNTAX{\colorSYNTAXP}{{\color{\colorMATH}\ensuremath{{}\rceil {\begingroup\renewcommand\colorMATH{\colorMATHS}\renewcommand\colorSYNTAX{\colorSYNTAXS}{{\color{\colorMATH}\ensuremath{\Gamma _{2}}}}\endgroup }\lceil {}^{{\begingroup\renewcommand\colorMATH{\colorMATHM}\renewcommand\colorSYNTAX{\colorSYNTAXM}{{\color{\colorMATH}\ensuremath{0,0}}}\endgroup }}}}}\endgroup }}}} 
   \\\noindent  {{\color{\colorMATH}\ensuremath{{\begingroup\renewcommand\colorMATH{\colorMATHP}\renewcommand\colorSYNTAX{\colorSYNTAXP}{{\color{\colorMATH}\ensuremath{\{ {\begingroup\renewcommand\colorMATH{\colorMATHM}\renewcommand\colorSYNTAX{\colorSYNTAXM}{{\color{\colorMATH}\ensuremath{x_{i}}}}\endgroup }{\mathrel{:}}_{{\begingroup\renewcommand\colorMATH{\colorMATHM}\renewcommand\colorSYNTAX{\colorSYNTAXM}{{\color{\colorMATH}\ensuremath{\epsilon _{i}^{\prime \prime \prime}}}}\endgroup },{\begingroup\renewcommand\colorMATH{\colorMATHM}\renewcommand\colorSYNTAX{\colorSYNTAXM}{{\color{\colorMATH}\ensuremath{\delta _{i}^{\prime \prime \prime}}}}\endgroup }} {\begingroup\renewcommand\colorMATH{\colorMATHM}\renewcommand\colorSYNTAX{\colorSYNTAXM}{{\color{\colorMATH}\ensuremath{\tau }}}\endgroup }\} }}}\endgroup } \in  {\begingroup\renewcommand\colorMATH{\colorMATHP}\renewcommand\colorSYNTAX{\colorSYNTAXP}{{\color{\colorMATH}\ensuremath{{}\rceil {\begingroup\renewcommand\colorMATH{\colorMATHS}\renewcommand\colorSYNTAX{\colorSYNTAXS}{{\color{\colorMATH}\ensuremath{\Gamma _{3}}}}\endgroup }\lceil {}^{{{\color{\colorSYNTAX}\texttt{\ensuremath{\infty }}}}}}}}\endgroup }}}} 
   \\\noindent  {{\color{\colorMATH}\ensuremath{{\begingroup\renewcommand\colorMATH{\colorMATHP}\renewcommand\colorSYNTAX{\colorSYNTAXP}{{\color{\colorMATH}\ensuremath{\{ {\begingroup\renewcommand\colorMATH{\colorMATHM}\renewcommand\colorSYNTAX{\colorSYNTAXM}{{\color{\colorMATH}\ensuremath{x_{i}}}}\endgroup }{\mathrel{:}}_{{\begingroup\renewcommand\colorMATH{\colorMATHM}\renewcommand\colorSYNTAX{\colorSYNTAXM}{{\color{\colorMATH}\ensuremath{\epsilon _{i}^{\prime 4}}}}\endgroup },{\begingroup\renewcommand\colorMATH{\colorMATHM}\renewcommand\colorSYNTAX{\colorSYNTAXM}{{\color{\colorMATH}\ensuremath{\delta _{i}^{\prime 4}}}}\endgroup }} {\begingroup\renewcommand\colorMATH{\colorMATHM}\renewcommand\colorSYNTAX{\colorSYNTAXM}{{\color{\colorMATH}\ensuremath{\tau }}}\endgroup }\} }}}\endgroup } \in  {\begingroup\renewcommand\colorMATH{\colorMATHP}\renewcommand\colorSYNTAX{\colorSYNTAXP}{{\color{\colorMATH}\ensuremath{{}\rceil \Gamma _{4}\lceil {}^{{{\color{\colorSYNTAX}\texttt{\ensuremath{\infty }}}}}}}}\endgroup }}}} 
   \\\noindent  {{\color{\colorMATH}\ensuremath{{\begingroup\renewcommand\colorMATH{\colorMATHP}\renewcommand\colorSYNTAX{\colorSYNTAXP}{{\color{\colorMATH}\ensuremath{\{ {\begingroup\renewcommand\colorMATH{\colorMATHM}\renewcommand\colorSYNTAX{\colorSYNTAXM}{{\color{\colorMATH}\ensuremath{x_{i}}}}\endgroup }{\mathrel{:}}_{{\begingroup\renewcommand\colorMATH{\colorMATHM}\renewcommand\colorSYNTAX{\colorSYNTAXM}{{\color{\colorMATH}\ensuremath{\epsilon _{i}^{\prime 5}}}}\endgroup },{\begingroup\renewcommand\colorMATH{\colorMATHM}\renewcommand\colorSYNTAX{\colorSYNTAXM}{{\color{\colorMATH}\ensuremath{\delta _{i}^{\prime 5}}}}\endgroup }} {\begingroup\renewcommand\colorMATH{\colorMATHM}\renewcommand\colorSYNTAX{\colorSYNTAXM}{{\color{\colorMATH}\ensuremath{\tau }}}\endgroup }\} }}}\endgroup } \in  {\begingroup\renewcommand\colorMATH{\colorMATHP}\renewcommand\colorSYNTAX{\colorSYNTAXP}{{\color{\colorMATH}\ensuremath{\mathrlap{\hspace{-0.5pt}{}\rceil }\lfloor \Gamma _{5}\mathrlap{\hspace{-0.5pt}\rfloor }\lceil {}_{\{ {\begingroup\renewcommand\colorMATH{\colorMATHM}\renewcommand\colorSYNTAX{\colorSYNTAXM}{{\color{\colorMATH}\ensuremath{x_{1}}}}\endgroup },{.}\hspace{-1pt}{.}\hspace{-1pt}{.},{\begingroup\renewcommand\colorMATH{\colorMATHM}\renewcommand\colorSYNTAX{\colorSYNTAXM}{{\color{\colorMATH}\ensuremath{x_{n}}}}\endgroup }\} }^{{\begingroup\renewcommand\colorMATH{\colorMATHM}\renewcommand\colorSYNTAX{\colorSYNTAXM}{{\color{\colorMATH}\ensuremath{\epsilon }}}\endgroup },{\begingroup\renewcommand\colorMATH{\colorMATHM}\renewcommand\colorSYNTAX{\colorSYNTAXM}{{\color{\colorMATH}\ensuremath{\delta }}}\endgroup }}}}}\endgroup }}}} 
   \\\noindent  Each of {{\color{\colorMATH}\ensuremath{\epsilon _{i}^{\prime}}}}, {{\color{\colorMATH}\ensuremath{\epsilon _{i}^{\prime \prime}}}}, {{\color{\colorMATH}\ensuremath{\delta _{i}^{\prime}}}}, {{\color{\colorMATH}\ensuremath{\delta _{i}^{\prime \prime}}}} must be {{\color{\colorMATH}\ensuremath{0}}}.
   \\\noindent  \begin{itemize}[leftmargin=*,label=\textbf{*
      }]\item  \noindent  Subcase {{\color{\colorMATH}\ensuremath{\epsilon _{i}^{\prime \prime \prime},\delta _{i}^{\prime \prime \prime} = \infty ,0}}} or {{\color{\colorMATH}\ensuremath{\epsilon _{i}^{\prime 4},\delta _{i}^{\prime 4} = \infty ,0}}}:
        \\\noindent  {{\color{\colorMATH}\ensuremath{ \epsilon _{i},\delta _{i} = \infty ,0 }}}; the property holds trivially
        
      \item  \noindent  Subcase {{\color{\colorMATH}\ensuremath{\epsilon _{i}^{\prime \prime \prime},\delta _{i}^{\prime \prime \prime} = 0,0}}} and {{\color{\colorMATH}\ensuremath{\epsilon ^{\prime 4},\delta ^{\prime 4} = 0,0}}}:
         \\\noindent  {{\color{\colorMATH}\ensuremath{\epsilon _{i},\delta _{i} = \epsilon _{i}^{\prime 5},\delta _{i}^{\prime 5} = 2\epsilon \sqrt {2n\ln (1/\delta ^{\prime})},\delta ^{\prime}+n\delta }}}
         \\\noindent  By IH: {{\color{\colorMATH}\ensuremath{{{\color{\colorMATH}\ensuremath{\operatorname{Pr}}}}[f^{\prime}(\gamma ) = d] \leq  e^{\epsilon }{{\color{\colorMATH}\ensuremath{\operatorname{Pr}}}}[f^{\prime}(\gamma [x_{i}\mapsto d]) = d] + \delta }}}
         \\\noindent  The property holds by Theorem~\ref{thm:adv_comp}: 
         \\\noindent  {{\color{\colorMATH}\ensuremath{{{\color{\colorMATH}\ensuremath{\operatorname{Pr}}}}[f^{\prime}(\gamma ) = d] \leq  e^{2\epsilon \sqrt {2n\log (1/\delta ^{\prime})}}{{\color{\colorMATH}\ensuremath{\operatorname{Pr}}}}[f^{\prime}(\gamma [x_{i}{\mapsto }d]) = d] {+} \delta }}}
         
      \end{itemize}
   
\end{itemize}
\end{proof}

\section{R\'enyi Differential Privacy}

\begin{theorem}[Gaussian mechanism (R\'enyi Differential Privacy)]
  \label{thm:renyi_gauss}
  
  If {{\color{\colorMATH}\ensuremath{ | f(\gamma ) - f(\gamma [x_{i}\mapsto d]) | \leq  r }}}, then for {{\color{\colorMATH}\ensuremath{ f^{\prime} = \lambda  \gamma  .\hspace*{0.33em} f(\gamma ) + {\mathcal{N}}(0, \sigma ^{2}) }}} and {{\color{\colorMATH}\ensuremath{ \sigma ^{2} = \alpha  r^{2} / (2\epsilon ) }}}, and all {{\color{\colorMATH}\ensuremath{\alpha , \epsilon  > 0}}}:

  \vspace*{-0.25em}\begingroup\color{\colorMATH}\begin{gather*} D_{\alpha }(f^{\prime}(\gamma ) \|  f^{\prime}(\gamma [x_{i}\mapsto d^{\prime}])) \leq  \epsilon 
  \vspace*{-1em}\end{gather*}\endgroup 
\end{theorem}

\begin{proof}
  By Mironov~\cite{mironov2017renyi}, Proposition 7, we have that:

  \vspace*{-0.25em}\begingroup\color{\colorMATH}\begin{gather*} D_{\alpha }(f^{\prime}(\gamma ) \|  f^{\prime}(\gamma [x_{i}\mapsto d^{\prime}])) \leq  \alpha  r^{2} / (2\sigma ^{2}) = \epsilon 
  \vspace*{-1em}\end{gather*}\endgroup 
\end{proof}

\begin{theorem}[Adaptive sequential composition (R\'enyi differential privacy)]
  \label{thm:renyi_let}
  \noindent  If:
     \vspace*{-0.25em}\begingroup\color{\colorMATH}\begin{gather*} |\gamma [x_{i}] - d| \leq  1 \implies   
     \cr  D_{\alpha }(f_{1}(\gamma ) \|  f_{1}(\gamma [x_{i}\mapsto d])) \leq  \epsilon _{i}
     \vspace*{-1em}\end{gather*}\endgroup 
  \\\noindent  and:
     \vspace*{-0.25em}\begingroup\color{\colorMATH}\begin{gather*} |\gamma [x_{i}] - d| \leq  1 \implies   
     \cr  D_{\alpha }(f_{2}(\gamma [x\mapsto d^{\prime}]) \|  f_{2}(\gamma [x\mapsto d^{\prime},x_{i}\mapsto d])) \leq  \epsilon _{i}^{\prime}
     \vspace*{-1em}\end{gather*}\endgroup 
  \\\noindent  and:
     \vspace*{-0.25em}\begingroup\color{\colorMATH}\begin{gather*} {{\color{\colorMATH}\ensuremath{\operatorname{Pr}}}}[f_{3}(\gamma ) = d^{\prime \prime}] = {{\color{\colorMATH}\ensuremath{\operatorname{Pr}}}}[f_{1}(\gamma ) = d^{\prime}, f_{2}(\gamma [x\mapsto d^{\prime}]) = d^{\prime \prime}] \vspace*{-1em}\end{gather*}\endgroup 
  \\\noindent  then:
     \vspace*{-0.25em}\begingroup\color{\colorMATH}\begin{gather*} |\gamma [x_{i}] - d| \leq  1 \implies   
     \cr  D_{\alpha }(f_{3}(\gamma ) \|  f_{3}(\gamma [x_{i}\mapsto d])) \leq  \epsilon _{i} + \epsilon _{i}^{\prime}
     \vspace*{-1em}\end{gather*}\endgroup 
  
\end{theorem}

\begin{proof}
  The result follows directly from Mironov~\cite{mironov2017renyi}, Proposition 1, setting {{\color{\colorMATH}\ensuremath{f = f_{1}}}} and {{\color{\colorMATH}\ensuremath{g = f_{2}}}}.
\end{proof}

\begin{figure*}
\begingroup\renewcommand\colorMATH{\colorMATHP}\renewcommand\colorSYNTAX{\colorSYNTAXP}
\vspace*{-0.25em}\begingroup\color{\colorMATH}\begin{gather*}\begin{tabularx}{\linewidth}{>{\centering\arraybackslash\(}X<{\)}}\hfill\hspace{0pt}\begingroup\color{\colorTEXT}\boxed{\begingroup\color{\colorMATH} {\begingroup\renewcommand\colorMATH{\colorMATHM}\renewcommand\colorSYNTAX{\colorSYNTAXM}{{\color{\colorMATH}\ensuremath{\Delta }}}\endgroup },\Gamma  \vdash  e \mathrel{:} \tau  \endgroup}\endgroup \end{tabularx}\vspace*{-1em}\end{gather*}\endgroup 
\begingroup\color{\colorMATH}\begin{mathpar}\inferrule*[flushleft,lab={{\color{\colorTEXT}\textsc{\scriptsize Static Loop}}}
  ]{ {\begingroup\renewcommand\colorMATH{\colorMATHS}\renewcommand\colorSYNTAX{\colorSYNTAXS}{{\color{\colorMATH}\ensuremath{{\begingroup\renewcommand\colorMATH{\colorMATHM}\renewcommand\colorSYNTAX{\colorSYNTAXM}{{\color{\colorMATH}\ensuremath{\Delta }}}\endgroup },\Gamma _{1} \vdash  e_{1} \mathrel{:} {\begingroup\renewcommand\colorMATH{\colorMATHM}\renewcommand\colorSYNTAX{\colorSYNTAXM}{{\color{\colorMATH}\ensuremath{{{\color{\colorSYNTAX}\texttt{\ensuremath{{\mathbb{R}}^{+}[{{\color{\colorMATH}\ensuremath{\eta _{n}}}}]}}}}}}}\endgroup }}}}\endgroup }
  \\ {\begingroup\renewcommand\colorMATH{\colorMATHS}\renewcommand\colorSYNTAX{\colorSYNTAXS}{{\color{\colorMATH}\ensuremath{{\begingroup\renewcommand\colorMATH{\colorMATHM}\renewcommand\colorSYNTAX{\colorSYNTAXM}{{\color{\colorMATH}\ensuremath{\Delta }}}\endgroup },\Gamma _{2} \vdash  e_{2} \mathrel{:} {\begingroup\renewcommand\colorMATH{\colorMATHM}\renewcommand\colorSYNTAX{\colorSYNTAXM}{{\color{\colorMATH}\ensuremath{\tau }}}\endgroup }}}}\endgroup }
  \\ {\begingroup\renewcommand\colorMATH{\colorMATHM}\renewcommand\colorSYNTAX{\colorSYNTAXM}{{\color{\colorMATH}\ensuremath{\Delta }}}\endgroup },\Gamma _{3} + \mathrlap{\hspace{-0.5pt}{}\rceil }\lfloor \Gamma _{4}\mathrlap{\hspace{-0.5pt}\rfloor }\lceil {}_{\{ {\begingroup\renewcommand\colorMATH{\colorMATHM}\renewcommand\colorSYNTAX{\colorSYNTAXM}{{\color{\colorMATH}\ensuremath{x_{1}^{\prime}}}}\endgroup },{.}\hspace{-1pt}{.}\hspace{-1pt}{.},{\begingroup\renewcommand\colorMATH{\colorMATHM}\renewcommand\colorSYNTAX{\colorSYNTAXM}{{\color{\colorMATH}\ensuremath{x_{n}^{\prime}}}}\endgroup }\} }^{{\begingroup\renewcommand\colorMATH{\colorMATHM}\renewcommand\colorSYNTAX{\colorSYNTAXM}{{\color{\colorMATH}\ensuremath{\eta _{\alpha }}}}\endgroup },{\begingroup\renewcommand\colorMATH{\colorMATHM}\renewcommand\colorSYNTAX{\colorSYNTAXM}{{\color{\colorMATH}\ensuremath{\eta _{\epsilon }}}}\endgroup }} \uplus  \{ {\begingroup\renewcommand\colorMATH{\colorMATHM}\renewcommand\colorSYNTAX{\colorSYNTAXM}{{\color{\colorMATH}\ensuremath{x_{1}}}}\endgroup } {\mathrel{:}}_{\infty } {\begingroup\renewcommand\colorMATH{\colorMATHM}\renewcommand\colorSYNTAX{\colorSYNTAXM}{{\color{\colorMATH}\ensuremath{{{\color{\colorSYNTAX}\texttt{\ensuremath{{\mathbb{N}}}}}}}}}\endgroup },{\begingroup\renewcommand\colorMATH{\colorMATHM}\renewcommand\colorSYNTAX{\colorSYNTAXM}{{\color{\colorMATH}\ensuremath{x_{2}}}}\endgroup } {\mathrel{:}}_{\infty } {\begingroup\renewcommand\colorMATH{\colorMATHM}\renewcommand\colorSYNTAX{\colorSYNTAXM}{{\color{\colorMATH}\ensuremath{\tau }}}\endgroup }\}  \vdash  e_{3} \mathrel{:} {\begingroup\renewcommand\colorMATH{\colorMATHM}\renewcommand\colorSYNTAX{\colorSYNTAXM}{{\color{\colorMATH}\ensuremath{\tau }}}\endgroup }
     }{
     {\begingroup\renewcommand\colorMATH{\colorMATHM}\renewcommand\colorSYNTAX{\colorSYNTAXM}{{\color{\colorMATH}\ensuremath{\Delta }}}\endgroup },{}\rceil {\begingroup\renewcommand\colorMATH{\colorMATHS}\renewcommand\colorSYNTAX{\colorSYNTAXS}{{\color{\colorMATH}\ensuremath{\Gamma _{1} + \Gamma _{1}}}}\endgroup }\lceil {}^{{\begingroup\renewcommand\colorMATH{\colorMATHM}\renewcommand\colorSYNTAX{\colorSYNTAXM}{{\color{\colorMATH}\ensuremath{0}}}\endgroup },{\begingroup\renewcommand\colorMATH{\colorMATHM}\renewcommand\colorSYNTAX{\colorSYNTAXM}{{\color{\colorMATH}\ensuremath{0}}}\endgroup }} + {}\rceil {\begingroup\renewcommand\colorMATH{\colorMATHS}\renewcommand\colorSYNTAX{\colorSYNTAXS}{{\color{\colorMATH}\ensuremath{\Gamma _{2}}}}\endgroup }\lceil {}^{{{\color{\colorSYNTAX}\texttt{\ensuremath{\infty }}}}} + {}\rceil \Gamma _{3}\lceil {}^{{{\color{\colorSYNTAX}\texttt{\ensuremath{\infty }}}}} 
     + \mathrlap{\hspace{-0.5pt}{}\rceil }\lfloor \Gamma _{4}\mathrlap{\hspace{-0.5pt}\rfloor }\lceil {}_{\{ {\begingroup\renewcommand\colorMATH{\colorMATHM}\renewcommand\colorSYNTAX{\colorSYNTAXM}{{\color{\colorMATH}\ensuremath{x_{1}^{\prime}}}}\endgroup },{.}\hspace{-1pt}{.}\hspace{-1pt}{.},{\begingroup\renewcommand\colorMATH{\colorMATHM}\renewcommand\colorSYNTAX{\colorSYNTAXM}{{\color{\colorMATH}\ensuremath{x_{n}^{\prime}}}}\endgroup }\} }^{{\begingroup\renewcommand\colorMATH{\colorMATHM}\renewcommand\colorSYNTAX{\colorSYNTAXM}{{\color{\colorMATH}\ensuremath{\eta _{\alpha }}}}\endgroup },{\begingroup\renewcommand\colorMATH{\colorMATHM}\renewcommand\colorSYNTAX{\colorSYNTAXM}{{\color{\colorMATH}\ensuremath{\eta _{n}{\mathrel{\mathord{\cdotp }}}\eta _{\epsilon }}}}\endgroup }}
     \vdash  {{\color{\colorSYNTAX}\texttt{loop}}}\hspace*{0.33em}{\begingroup\renewcommand\colorMATH{\colorMATHS}\renewcommand\colorSYNTAX{\colorSYNTAXS}{{\color{\colorMATH}\ensuremath{e_{1}}}}\endgroup }\hspace*{0.33em}{{\color{\colorSYNTAX}\texttt{on}}}\hspace*{0.33em}{\begingroup\renewcommand\colorMATH{\colorMATHS}\renewcommand\colorSYNTAX{\colorSYNTAXS}{{\color{\colorMATH}\ensuremath{e_{2}}}}\endgroup }\hspace*{0.33em}{<}{\begingroup\renewcommand\colorMATH{\colorMATHM}\renewcommand\colorSYNTAX{\colorSYNTAXM}{{\color{\colorMATH}\ensuremath{x_{1}^{\prime}}}}\endgroup },{.}\hspace{-1pt}{.}\hspace{-1pt}{.},{\begingroup\renewcommand\colorMATH{\colorMATHM}\renewcommand\colorSYNTAX{\colorSYNTAXM}{{\color{\colorMATH}\ensuremath{x_{n}^{\prime}}}}\endgroup }{>}\hspace*{0.33em}\{ {\begingroup\renewcommand\colorMATH{\colorMATHM}\renewcommand\colorSYNTAX{\colorSYNTAXM}{{\color{\colorMATH}\ensuremath{x_{1}}}}\endgroup },{\begingroup\renewcommand\colorMATH{\colorMATHM}\renewcommand\colorSYNTAX{\colorSYNTAXM}{{\color{\colorMATH}\ensuremath{x_{2}}}}\endgroup } \Rightarrow  e_{3}\}  \mathrel{:} {\begingroup\renewcommand\colorMATH{\colorMATHM}\renewcommand\colorSYNTAX{\colorSYNTAXM}{{\color{\colorMATH}\ensuremath{\tau }}}\endgroup }
  }
\and\inferrule*[lab={{\color{\colorTEXT}\textsc{\scriptsize Gauss}}}
  ]{ {\begingroup\renewcommand\colorMATH{\colorMATHS}\renewcommand\colorSYNTAX{\colorSYNTAXS}{{\color{\colorMATH}\ensuremath{\Gamma _{1} \vdash  e_{1} \mathrel{:} {\begingroup\renewcommand\colorMATH{\colorMATHM}\renewcommand\colorSYNTAX{\colorSYNTAXM}{{\color{\colorMATH}\ensuremath{{{\color{\colorSYNTAX}\texttt{\ensuremath{{\mathbb{R}}^{+}[{{\color{\colorMATH}\ensuremath{\eta _{s}}}}]}}}}}}}\endgroup }}}}\endgroup }
  \\ {\begingroup\renewcommand\colorMATH{\colorMATHS}\renewcommand\colorSYNTAX{\colorSYNTAXS}{{\color{\colorMATH}\ensuremath{\Gamma _{2} \vdash  e_{2} \mathrel{:} {\begingroup\renewcommand\colorMATH{\colorMATHM}\renewcommand\colorSYNTAX{\colorSYNTAXM}{{\color{\colorMATH}\ensuremath{{{\color{\colorSYNTAX}\texttt{\ensuremath{{\mathbb{R}}^{+}[{{\color{\colorMATH}\ensuremath{\eta _{\alpha }}}}]}}}}}}}\endgroup }}}}\endgroup }
  \\ {\begingroup\renewcommand\colorMATH{\colorMATHS}\renewcommand\colorSYNTAX{\colorSYNTAXS}{{\color{\colorMATH}\ensuremath{\Gamma _{3} \vdash  e_{3} \mathrel{:} {\begingroup\renewcommand\colorMATH{\colorMATHM}\renewcommand\colorSYNTAX{\colorSYNTAXM}{{\color{\colorMATH}\ensuremath{{{\color{\colorSYNTAX}\texttt{\ensuremath{{\mathbb{R}}^{+}[{{\color{\colorMATH}\ensuremath{\eta _{\epsilon }}}}]}}}}}}}\endgroup }}}}\endgroup }
  \\ {\begingroup\renewcommand\colorMATH{\colorMATHS}\renewcommand\colorSYNTAX{\colorSYNTAXS}{{\color{\colorMATH}\ensuremath{\Gamma _{4} + \mathrlap{\hspace{-0.5pt}{}\rceil }\lfloor \Gamma _{5}\mathrlap{\hspace{-0.5pt}\rfloor }\lceil {}_{\{ {\begingroup\renewcommand\colorMATH{\colorMATHM}\renewcommand\colorSYNTAX{\colorSYNTAXM}{{\color{\colorMATH}\ensuremath{x_{1}}}}\endgroup },{.}\hspace{-1pt}{.}\hspace{-1pt}{.},{\begingroup\renewcommand\colorMATH{\colorMATHM}\renewcommand\colorSYNTAX{\colorSYNTAXM}{{\color{\colorMATH}\ensuremath{x_{n}}}}\endgroup }\} }^{{\begingroup\renewcommand\colorMATH{\colorMATHM}\renewcommand\colorSYNTAX{\colorSYNTAXM}{{\color{\colorMATH}\ensuremath{\eta _{s}}}}\endgroup }} \vdash  e_{4} \mathrel{:} {\begingroup\renewcommand\colorMATH{\colorMATHM}\renewcommand\colorSYNTAX{\colorSYNTAXM}{{\color{\colorMATH}\ensuremath{{{\color{\colorSYNTAX}\texttt{\ensuremath{{\mathbb{R}}}}}}}}}\endgroup }}}}\endgroup }
     }{
     {}\rceil {\begingroup\renewcommand\colorMATH{\colorMATHS}\renewcommand\colorSYNTAX{\colorSYNTAXS}{{\color{\colorMATH}\ensuremath{\Gamma _{1} + \Gamma _{2} + \Gamma _{3}}}}\endgroup }\lceil {}^{{\begingroup\renewcommand\colorMATH{\colorMATHM}\renewcommand\colorSYNTAX{\colorSYNTAXM}{{\color{\colorMATH}\ensuremath{0}}}\endgroup },{\begingroup\renewcommand\colorMATH{\colorMATHM}\renewcommand\colorSYNTAX{\colorSYNTAXM}{{\color{\colorMATH}\ensuremath{0}}}\endgroup }} + {}\rceil {\begingroup\renewcommand\colorMATH{\colorMATHS}\renewcommand\colorSYNTAX{\colorSYNTAXS}{{\color{\colorMATH}\ensuremath{\Gamma _{4}}}}\endgroup }\lceil {}^{{{\color{\colorSYNTAX}\texttt{\ensuremath{\infty }}}}} + \mathrlap{\hspace{-0.5pt}{}\rceil }\lfloor {\begingroup\renewcommand\colorMATH{\colorMATHS}\renewcommand\colorSYNTAX{\colorSYNTAXS}{{\color{\colorMATH}\ensuremath{\Gamma _{5}}}}\endgroup }\mathrlap{\hspace{-0.5pt}\rfloor }\lceil {}_{\{ {\begingroup\renewcommand\colorMATH{\colorMATHM}\renewcommand\colorSYNTAX{\colorSYNTAXM}{{\color{\colorMATH}\ensuremath{x_{1}}}}\endgroup },{.}\hspace{-1pt}{.}\hspace{-1pt}{.},{\begingroup\renewcommand\colorMATH{\colorMATHM}\renewcommand\colorSYNTAX{\colorSYNTAXM}{{\color{\colorMATH}\ensuremath{x_{n}}}}\endgroup }\} }^{{\begingroup\renewcommand\colorMATH{\colorMATHM}\renewcommand\colorSYNTAX{\colorSYNTAXM}{{\color{\colorMATH}\ensuremath{\eta _{\alpha }}}}\endgroup },{\begingroup\renewcommand\colorMATH{\colorMATHM}\renewcommand\colorSYNTAX{\colorSYNTAXM}{{\color{\colorMATH}\ensuremath{\eta _{\epsilon }}}}\endgroup }} 
     \vdash  {{\color{\colorSYNTAX}\texttt{\ensuremath{{{\color{\colorSYNTAX}\texttt{gauss}}}[{\begingroup\renewcommand\colorMATH{\colorMATHS}\renewcommand\colorSYNTAX{\colorSYNTAXS}{{\color{\colorMATH}\ensuremath{e_{1}}}}\endgroup },{\begingroup\renewcommand\colorMATH{\colorMATHS}\renewcommand\colorSYNTAX{\colorSYNTAXS}{{\color{\colorMATH}\ensuremath{e_{2}}}}\endgroup },{\begingroup\renewcommand\colorMATH{\colorMATHS}\renewcommand\colorSYNTAX{\colorSYNTAXS}{{\color{\colorMATH}\ensuremath{e_{3}}}}\endgroup }]\hspace*{0.33em}{<}{\begingroup\renewcommand\colorMATH{\colorMATHM}\renewcommand\colorSYNTAX{\colorSYNTAXM}{{\color{\colorMATH}\ensuremath{x_{1}}}}\endgroup },{.}\hspace{-1pt}{.}\hspace{-1pt}{.},{\begingroup\renewcommand\colorMATH{\colorMATHM}\renewcommand\colorSYNTAX{\colorSYNTAXM}{{\color{\colorMATH}\ensuremath{x_{n}}}}\endgroup }{>}\hspace*{0.33em}\{ {\begingroup\renewcommand\colorMATH{\colorMATHS}\renewcommand\colorSYNTAX{\colorSYNTAXS}{{\color{\colorMATH}\ensuremath{e_{4}}}}\endgroup }\} }}}} \mathrel{:} {\begingroup\renewcommand\colorMATH{\colorMATHM}\renewcommand\colorSYNTAX{\colorSYNTAXM}{{\color{\colorMATH}\ensuremath{{{\color{\colorSYNTAX}\texttt{\ensuremath{{\mathbb{R}}}}}}}}}\endgroup }
  }
\and\inferrule*[lab={{\color{\colorTEXT}\textsc{\scriptsize MGauss}}}
  ]{ {\begingroup\renewcommand\colorMATH{\colorMATHS}\renewcommand\colorSYNTAX{\colorSYNTAXS}{{\color{\colorMATH}\ensuremath{\Gamma _{1} \vdash  e_{1} \mathrel{:} {\begingroup\renewcommand\colorMATH{\colorMATHM}\renewcommand\colorSYNTAX{\colorSYNTAXM}{{\color{\colorMATH}\ensuremath{{{\color{\colorSYNTAX}\texttt{\ensuremath{{\mathbb{R}}^{+}[{{\color{\colorMATH}\ensuremath{\eta _{s}}}}]}}}}}}}\endgroup }}}}\endgroup }
  \\ {\begingroup\renewcommand\colorMATH{\colorMATHS}\renewcommand\colorSYNTAX{\colorSYNTAXS}{{\color{\colorMATH}\ensuremath{\Gamma _{2} \vdash  e_{2} \mathrel{:} {\begingroup\renewcommand\colorMATH{\colorMATHM}\renewcommand\colorSYNTAX{\colorSYNTAXM}{{\color{\colorMATH}\ensuremath{{{\color{\colorSYNTAX}\texttt{\ensuremath{{\mathbb{R}}^{+}[{{\color{\colorMATH}\ensuremath{\eta _{\alpha }}}}]}}}}}}}\endgroup }}}}\endgroup }
  \\ {\begingroup\renewcommand\colorMATH{\colorMATHS}\renewcommand\colorSYNTAX{\colorSYNTAXS}{{\color{\colorMATH}\ensuremath{\Gamma _{3} \vdash  e_{3} \mathrel{:} {\begingroup\renewcommand\colorMATH{\colorMATHM}\renewcommand\colorSYNTAX{\colorSYNTAXM}{{\color{\colorMATH}\ensuremath{{{\color{\colorSYNTAX}\texttt{\ensuremath{{\mathbb{R}}^{+}[{{\color{\colorMATH}\ensuremath{\eta _{\epsilon }}}}]}}}}}}}\endgroup }}}}\endgroup }
  \\ {\begingroup\renewcommand\colorMATH{\colorMATHS}\renewcommand\colorSYNTAX{\colorSYNTAXS}{{\color{\colorMATH}\ensuremath{\Gamma _{4} + \mathrlap{\hspace{-0.5pt}{}\rceil }\lfloor \Gamma _{5}\mathrlap{\hspace{-0.5pt}\rfloor }\lceil {}_{\{ {\begingroup\renewcommand\colorMATH{\colorMATHM}\renewcommand\colorSYNTAX{\colorSYNTAXM}{{\color{\colorMATH}\ensuremath{x_{1}}}}\endgroup },{.}\hspace{-1pt}{.}\hspace{-1pt}{.},{\begingroup\renewcommand\colorMATH{\colorMATHM}\renewcommand\colorSYNTAX{\colorSYNTAXM}{{\color{\colorMATH}\ensuremath{x_{n}}}}\endgroup }\} }^{{\begingroup\renewcommand\colorMATH{\colorMATHM}\renewcommand\colorSYNTAX{\colorSYNTAXM}{{\color{\colorMATH}\ensuremath{\eta _{s}}}}\endgroup }} \vdash  e_{4} \mathrel{:} {\begingroup\renewcommand\colorMATH{\colorMATHM}\renewcommand\colorSYNTAX{\colorSYNTAXM}{{\color{\colorMATH}\ensuremath{{{\color{\colorSYNTAX}\texttt{\ensuremath{{{\color{\colorSYNTAX}\texttt{matrix}}}_{L2}^{{{\color{\colorMATH}\ensuremath{\bigstar }}}}[{{\color{\colorMATH}\ensuremath{\eta _{m}}}},{{\color{\colorMATH}\ensuremath{\eta _{n}}}}]\hspace*{0.33em}{\mathbb{R}}}}}}}}}\endgroup }}}}\endgroup }
     }{
     {}\rceil {\begingroup\renewcommand\colorMATH{\colorMATHS}\renewcommand\colorSYNTAX{\colorSYNTAXS}{{\color{\colorMATH}\ensuremath{\Gamma _{1} + \Gamma _{2} + \Gamma _{3}}}}\endgroup }\lceil {}^{{\begingroup\renewcommand\colorMATH{\colorMATHM}\renewcommand\colorSYNTAX{\colorSYNTAXM}{{\color{\colorMATH}\ensuremath{0}}}\endgroup },{\begingroup\renewcommand\colorMATH{\colorMATHM}\renewcommand\colorSYNTAX{\colorSYNTAXM}{{\color{\colorMATH}\ensuremath{0}}}\endgroup }} + {}\rceil {\begingroup\renewcommand\colorMATH{\colorMATHS}\renewcommand\colorSYNTAX{\colorSYNTAXS}{{\color{\colorMATH}\ensuremath{\Gamma _{4}}}}\endgroup }\lceil {}^{{{\color{\colorSYNTAX}\texttt{\ensuremath{\infty }}}}} + \mathrlap{\hspace{-0.5pt}{}\rceil }\lfloor {\begingroup\renewcommand\colorMATH{\colorMATHS}\renewcommand\colorSYNTAX{\colorSYNTAXS}{{\color{\colorMATH}\ensuremath{\Gamma _{5}}}}\endgroup }\mathrlap{\hspace{-0.5pt}\rfloor }\lceil {}_{\{ {\begingroup\renewcommand\colorMATH{\colorMATHM}\renewcommand\colorSYNTAX{\colorSYNTAXM}{{\color{\colorMATH}\ensuremath{x_{1}}}}\endgroup },{.}\hspace{-1pt}{.}\hspace{-1pt}{.},{\begingroup\renewcommand\colorMATH{\colorMATHM}\renewcommand\colorSYNTAX{\colorSYNTAXM}{{\color{\colorMATH}\ensuremath{x_{n}}}}\endgroup }\} }^{{\begingroup\renewcommand\colorMATH{\colorMATHM}\renewcommand\colorSYNTAX{\colorSYNTAXM}{{\color{\colorMATH}\ensuremath{\eta _{\alpha }}}}\endgroup },{\begingroup\renewcommand\colorMATH{\colorMATHM}\renewcommand\colorSYNTAX{\colorSYNTAXM}{{\color{\colorMATH}\ensuremath{\eta _{\epsilon }}}}\endgroup }} 
     \vdash  {{\color{\colorSYNTAX}\texttt{\ensuremath{{{\color{\colorSYNTAX}\texttt{mgauss}}}[{\begingroup\renewcommand\colorMATH{\colorMATHS}\renewcommand\colorSYNTAX{\colorSYNTAXS}{{\color{\colorMATH}\ensuremath{e_{1}}}}\endgroup },{\begingroup\renewcommand\colorMATH{\colorMATHS}\renewcommand\colorSYNTAX{\colorSYNTAXS}{{\color{\colorMATH}\ensuremath{e_{2}}}}\endgroup },{\begingroup\renewcommand\colorMATH{\colorMATHS}\renewcommand\colorSYNTAX{\colorSYNTAXS}{{\color{\colorMATH}\ensuremath{e_{3}}}}\endgroup }]\hspace*{0.33em}{<}{\begingroup\renewcommand\colorMATH{\colorMATHM}\renewcommand\colorSYNTAX{\colorSYNTAXM}{{\color{\colorMATH}\ensuremath{x_{1}}}}\endgroup },{.}\hspace{-1pt}{.}\hspace{-1pt}{.},{\begingroup\renewcommand\colorMATH{\colorMATHM}\renewcommand\colorSYNTAX{\colorSYNTAXM}{{\color{\colorMATH}\ensuremath{x_{n}}}}\endgroup }{>}\hspace*{0.33em}\{ {\begingroup\renewcommand\colorMATH{\colorMATHS}\renewcommand\colorSYNTAX{\colorSYNTAXS}{{\color{\colorMATH}\ensuremath{e_{4}}}}\endgroup }\} }}}} \mathrel{:} {\begingroup\renewcommand\colorMATH{\colorMATHM}\renewcommand\colorSYNTAX{\colorSYNTAXM}{{\color{\colorMATH}\ensuremath{{{\color{\colorSYNTAX}\texttt{\ensuremath{{{\color{\colorSYNTAX}\texttt{matrix}}}_{L\infty }^{U}[{{\color{\colorMATH}\ensuremath{\eta _{m}}}},{{\color{\colorMATH}\ensuremath{\eta _{n}}}}]\hspace*{0.33em}{\mathbb{R}}}}}}}}}\endgroup }
  }
\end{mathpar}\endgroup 
\endgroup 
\caption{Privacy Type System Modifications, R\'enyi Differential Privacy}
\end{figure*}

\begin{theorem}[Soundness]
  There exists an interpretation of well typed terms {\begingroup\renewcommand\colorMATH{\colorMATHS}\renewcommand\colorSYNTAX{\colorSYNTAXS}{{\color{\colorMATH}\ensuremath{\Gamma  \vdash  e \mathrel{:} {\begingroup\renewcommand\colorMATH{\colorMATHM}\renewcommand\colorSYNTAX{\colorSYNTAXM}{{\color{\colorMATH}\ensuremath{\tau }}}\endgroup }}}}\endgroup } and {\begingroup\renewcommand\colorMATH{\colorMATHP}\renewcommand\colorSYNTAX{\colorSYNTAXP}{{\color{\colorMATH}\ensuremath{\Gamma 
  \vdash  e \mathrel{:} {\begingroup\renewcommand\colorMATH{\colorMATHM}\renewcommand\colorSYNTAX{\colorSYNTAXM}{{\color{\colorMATH}\ensuremath{\tau }}}\endgroup }}}}\endgroup }, notated {\begingroup\renewcommand\colorMATH{\colorMATHS}\renewcommand\colorSYNTAX{\colorSYNTAXS}{{\color{\colorMATH}\ensuremath{\llbracket e\rrbracket }}}\endgroup } and {\begingroup\renewcommand\colorMATH{\colorMATHP}\renewcommand\colorSYNTAX{\colorSYNTAXP}{{\color{\colorMATH}\ensuremath{\llbracket e\rrbracket }}}\endgroup }, such that {{\color{\colorMATH}\ensuremath{{\begingroup\renewcommand\colorMATH{\colorMATHP}\renewcommand\colorSYNTAX{\colorSYNTAXP}{{\color{\colorMATH}\ensuremath{\llbracket e\rrbracket }}}\endgroup } \in  {\begingroup\renewcommand\colorMATH{\colorMATHP}\renewcommand\colorSYNTAX{\colorSYNTAXP}{{\color{\colorMATH}\ensuremath{\llbracket \Gamma  \vdash 
  {\begingroup\renewcommand\colorMATH{\colorMATHM}\renewcommand\colorSYNTAX{\colorSYNTAXM}{{\color{\colorMATH}\ensuremath{\tau }}}\endgroup }\rrbracket }}}\endgroup }}}} and {{\color{\colorMATH}\ensuremath{{\begingroup\renewcommand\colorMATH{\colorMATHS}\renewcommand\colorSYNTAX{\colorSYNTAXS}{{\color{\colorMATH}\ensuremath{\llbracket e\rrbracket }}}\endgroup } \in  {\begingroup\renewcommand\colorMATH{\colorMATHS}\renewcommand\colorSYNTAX{\colorSYNTAXS}{{\color{\colorMATH}\ensuremath{\llbracket \Gamma  \vdash  {\begingroup\renewcommand\colorMATH{\colorMATHM}\renewcommand\colorSYNTAX{\colorSYNTAXM}{{\color{\colorMATH}\ensuremath{\tau }}}\endgroup }\rrbracket }}}\endgroup }}}}.
\end{theorem}
\begin{proof}
  We include the cases that are different from the proof for {{\color{\colorMATH}\ensuremath{(\epsilon , \delta )}}}-differential privacy: {\sc Bind} and {\sc Gauss}.

\vspace*{-0.25em}\begingroup\color{\colorMATH}\begin{gather*}\begin{tabularx}{\linewidth}{>{\centering\arraybackslash\(}X<{\)}}\hfill\hspace{0pt}\begingroup\color{\colorTEXT}\boxed{\begingroup\color{\colorMATH} {\begingroup\renewcommand\colorMATH{\colorMATHP}\renewcommand\colorSYNTAX{\colorSYNTAXP}{{\color{\colorMATH}\ensuremath{\llbracket \underline{\hspace{0.66em}\vspace*{5ex}}\rrbracket  \in  \{ e \in  {{\color{\colorMATH}\ensuremath{\operatorname{exp}}}} \mathrel{|} \Gamma  \vdash  e \mathrel{:} {\begingroup\renewcommand\colorMATH{\colorMATHM}\renewcommand\colorSYNTAX{\colorSYNTAXM}{{\color{\colorMATH}\ensuremath{\tau }}}\endgroup }\}  \rightarrow  \llbracket \Gamma  \vdash  {\begingroup\renewcommand\colorMATH{\colorMATHM}\renewcommand\colorSYNTAX{\colorSYNTAXM}{{\color{\colorMATH}\ensuremath{\tau }}}\endgroup }\rrbracket }}}\endgroup } \endgroup}\endgroup \end{tabularx}\vspace*{-1em}\end{gather*}\endgroup 
\begin{itemize}[leftmargin=*,label=\textbf{-
}]\item  \noindent  Case {\begingroup\renewcommand\colorMATH{\colorMATHP}\renewcommand\colorSYNTAX{\colorSYNTAXP}{{\color{\colorMATH}\ensuremath{\overbracketarg \Gamma {\Gamma _{1} + \Gamma _{2}} \vdash  {{\color{\colorSYNTAX}\texttt{\ensuremath{{\begingroup\renewcommand\colorMATH{\colorMATHM}\renewcommand\colorSYNTAX{\colorSYNTAXM}{{\color{\colorMATH}\ensuremath{x}}}\endgroup } \leftarrow  {{\color{\colorMATH}\ensuremath{e_{1}}}}\mathrel{;}{{\color{\colorMATH}\ensuremath{e_{2}}}}}}}} \mathrel{:} {\begingroup\renewcommand\colorMATH{\colorMATHM}\renewcommand\colorSYNTAX{\colorSYNTAXM}{{\color{\colorMATH}\ensuremath{\tau _{2}}}}\endgroup }}}}\endgroup }
   \\\noindent  By inversion:
      \vspace*{-0.25em}\begingroup\color{\colorMATH}\begin{gather*} 

      \vspace*{-1em}\end{gather*}\endgroup 
   \\\noindent  By Induction Hypothesis (IH):
      \vspace*{-0.25em}\begingroup\color{\colorMATH}\begin{gather*} %
      \vspace*{-1em}\end{gather*}\endgroup 
   \\\noindent  By definition of {\begingroup\renewcommand\colorMATH{\colorMATHP}\renewcommand\colorSYNTAX{\colorSYNTAXP}{{\color{\colorMATH}\ensuremath{+}}}\endgroup } for privacy contexts:
   \\\noindent  {{\color{\colorMATH}\ensuremath{\epsilon _{i} =  \epsilon _{i}^{\prime} + \epsilon _{i}^{\prime \prime}}}} for
   \\\noindent  {{\color{\colorMATH}\ensuremath{{\begingroup\renewcommand\colorMATH{\colorMATHP}\renewcommand\colorSYNTAX{\colorSYNTAXP}{{\color{\colorMATH}\ensuremath{\{ {\begingroup\renewcommand\colorMATH{\colorMATHM}\renewcommand\colorSYNTAX{\colorSYNTAXM}{{\color{\colorMATH}\ensuremath{x_{i}}}}\endgroup }{\mathrel{:}}_{{\begingroup\renewcommand\colorMATH{\colorMATHM}\renewcommand\colorSYNTAX{\colorSYNTAXM}{{\color{\colorMATH}\ensuremath{\alpha _{i}}}}\endgroup },{\begingroup\renewcommand\colorMATH{\colorMATHM}\renewcommand\colorSYNTAX{\colorSYNTAXM}{{\color{\colorMATH}\ensuremath{\epsilon _{i}^{\prime}}}}\endgroup }} {\begingroup\renewcommand\colorMATH{\colorMATHM}\renewcommand\colorSYNTAX{\colorSYNTAXM}{{\color{\colorMATH}\ensuremath{\tau }}}\endgroup }\} }}}\endgroup } \in  {\begingroup\renewcommand\colorMATH{\colorMATHP}\renewcommand\colorSYNTAX{\colorSYNTAXP}{{\color{\colorMATH}\ensuremath{\Gamma _{1}}}}\endgroup }}}} and {{\color{\colorMATH}\ensuremath{{\begingroup\renewcommand\colorMATH{\colorMATHP}\renewcommand\colorSYNTAX{\colorSYNTAXP}{{\color{\colorMATH}\ensuremath{\{ {\begingroup\renewcommand\colorMATH{\colorMATHM}\renewcommand\colorSYNTAX{\colorSYNTAXM}{{\color{\colorMATH}\ensuremath{x_{i}}}}\endgroup }{\mathrel{:}}_{{\begingroup\renewcommand\colorMATH{\colorMATHM}\renewcommand\colorSYNTAX{\colorSYNTAXM}{{\color{\colorMATH}\ensuremath{\alpha _{i}}}}\endgroup },{\begingroup\renewcommand\colorMATH{\colorMATHM}\renewcommand\colorSYNTAX{\colorSYNTAXM}{{\color{\colorMATH}\ensuremath{\epsilon _{i}^{\prime \prime}}}}\endgroup }} {\begingroup\renewcommand\colorMATH{\colorMATHM}\renewcommand\colorSYNTAX{\colorSYNTAXM}{{\color{\colorMATH}\ensuremath{\tau }}}\endgroup }\} }}}\endgroup } \in  {\begingroup\renewcommand\colorMATH{\colorMATHP}\renewcommand\colorSYNTAX{\colorSYNTAXP}{{\color{\colorMATH}\ensuremath{\Gamma _{2}}}}\endgroup }}}}. 
   \\\noindent  Property holds via IH.1, IH.2 and theorem~\ref{thm:renyi_let} instantiated with
      {{\color{\colorMATH}\ensuremath{(\alpha _{i},\epsilon _{i}^{\prime})}}} and {{\color{\colorMATH}\ensuremath{(\alpha _{i},\epsilon _{i}^{\prime \prime})}}}.
   
\item  \noindent  Case: 
   \\\noindent  {{\color{\colorMATH}\ensuremath{\hspace*{1.00em}

      \vspace*{-1em}\end{gather*}\endgroup 
   \\\noindent  By Induction Hypothesis (IH):
      \vspace*{-0.25em}\begingroup\color{\colorMATH}\begin{gather*} %
      \vspace*{-1em}\end{gather*}\endgroup 
   \\\noindent  By IH, we have that {{\color{\colorMATH}\ensuremath{\sigma ^{2} = \frac{r^{2} \alpha }{2\epsilon }}}}
   \\\noindent  By definition of {\begingroup\renewcommand\colorMATH{\colorMATHP}\renewcommand\colorSYNTAX{\colorSYNTAXP}{{\color{\colorMATH}\ensuremath{+}}}\endgroup } for privacy contexts:
   \\\noindent  {{\color{\colorMATH}\ensuremath{\epsilon _{i} = \epsilon _{i}^{\prime} + \epsilon _{i}^{\prime \prime} + \epsilon _{i}^{\prime \prime \prime} + \epsilon _{i}^{\prime 4} + \epsilon _{i}^{\prime 5}}}} for
   \\\noindent  {{\color{\colorMATH}\ensuremath{{\begingroup\renewcommand\colorMATH{\colorMATHP}\renewcommand\colorSYNTAX{\colorSYNTAXP}{{\color{\colorMATH}\ensuremath{\{ {\begingroup\renewcommand\colorMATH{\colorMATHM}\renewcommand\colorSYNTAX{\colorSYNTAXM}{{\color{\colorMATH}\ensuremath{x_{i}}}}\endgroup }{\mathrel{:}}_{{\begingroup\renewcommand\colorMATH{\colorMATHM}\renewcommand\colorSYNTAX{\colorSYNTAXM}{{\color{\colorMATH}\ensuremath{\alpha _{i}}}}\endgroup },{\begingroup\renewcommand\colorMATH{\colorMATHM}\renewcommand\colorSYNTAX{\colorSYNTAXM}{{\color{\colorMATH}\ensuremath{\epsilon _{i}^{\prime}}}}\endgroup }} {\begingroup\renewcommand\colorMATH{\colorMATHM}\renewcommand\colorSYNTAX{\colorSYNTAXM}{{\color{\colorMATH}\ensuremath{\tau }}}\endgroup }\} }}}\endgroup } \in  {\begingroup\renewcommand\colorMATH{\colorMATHP}\renewcommand\colorSYNTAX{\colorSYNTAXP}{{\color{\colorMATH}\ensuremath{{}\rceil {\begingroup\renewcommand\colorMATH{\colorMATHS}\renewcommand\colorSYNTAX{\colorSYNTAXS}{{\color{\colorMATH}\ensuremath{\Gamma _{1}}}}\endgroup }\lceil {}^{{\begingroup\renewcommand\colorMATH{\colorMATHM}\renewcommand\colorSYNTAX{\colorSYNTAXM}{{\color{\colorMATH}\ensuremath{0,0}}}\endgroup }}}}}\endgroup }}}} 
   \\\noindent  {{\color{\colorMATH}\ensuremath{{\begingroup\renewcommand\colorMATH{\colorMATHP}\renewcommand\colorSYNTAX{\colorSYNTAXP}{{\color{\colorMATH}\ensuremath{\{ {\begingroup\renewcommand\colorMATH{\colorMATHM}\renewcommand\colorSYNTAX{\colorSYNTAXM}{{\color{\colorMATH}\ensuremath{x_{i}}}}\endgroup }{\mathrel{:}}_{{\begingroup\renewcommand\colorMATH{\colorMATHM}\renewcommand\colorSYNTAX{\colorSYNTAXM}{{\color{\colorMATH}\ensuremath{\alpha _{i}}}}\endgroup },{\begingroup\renewcommand\colorMATH{\colorMATHM}\renewcommand\colorSYNTAX{\colorSYNTAXM}{{\color{\colorMATH}\ensuremath{\epsilon _{i}^{\prime \prime}}}}\endgroup }} {\begingroup\renewcommand\colorMATH{\colorMATHM}\renewcommand\colorSYNTAX{\colorSYNTAXM}{{\color{\colorMATH}\ensuremath{\tau }}}\endgroup }\} }}}\endgroup } \in  {\begingroup\renewcommand\colorMATH{\colorMATHP}\renewcommand\colorSYNTAX{\colorSYNTAXP}{{\color{\colorMATH}\ensuremath{{}\rceil {\begingroup\renewcommand\colorMATH{\colorMATHS}\renewcommand\colorSYNTAX{\colorSYNTAXS}{{\color{\colorMATH}\ensuremath{\Gamma _{2}}}}\endgroup }\lceil {}^{{\begingroup\renewcommand\colorMATH{\colorMATHM}\renewcommand\colorSYNTAX{\colorSYNTAXM}{{\color{\colorMATH}\ensuremath{0,0}}}\endgroup }}}}}\endgroup }}}} 
   \\\noindent  {{\color{\colorMATH}\ensuremath{{\begingroup\renewcommand\colorMATH{\colorMATHP}\renewcommand\colorSYNTAX{\colorSYNTAXP}{{\color{\colorMATH}\ensuremath{\{ {\begingroup\renewcommand\colorMATH{\colorMATHM}\renewcommand\colorSYNTAX{\colorSYNTAXM}{{\color{\colorMATH}\ensuremath{x_{i}}}}\endgroup }{\mathrel{:}}_{{\begingroup\renewcommand\colorMATH{\colorMATHM}\renewcommand\colorSYNTAX{\colorSYNTAXM}{{\color{\colorMATH}\ensuremath{\alpha _{i}}}}\endgroup },{\begingroup\renewcommand\colorMATH{\colorMATHM}\renewcommand\colorSYNTAX{\colorSYNTAXM}{{\color{\colorMATH}\ensuremath{\epsilon _{i}^{\prime \prime \prime}}}}\endgroup }} {\begingroup\renewcommand\colorMATH{\colorMATHM}\renewcommand\colorSYNTAX{\colorSYNTAXM}{{\color{\colorMATH}\ensuremath{\tau }}}\endgroup }\} }}}\endgroup } \in  {\begingroup\renewcommand\colorMATH{\colorMATHP}\renewcommand\colorSYNTAX{\colorSYNTAXP}{{\color{\colorMATH}\ensuremath{{}\rceil {\begingroup\renewcommand\colorMATH{\colorMATHS}\renewcommand\colorSYNTAX{\colorSYNTAXS}{{\color{\colorMATH}\ensuremath{\Gamma _{3}}}}\endgroup }\lceil {}^{{\begingroup\renewcommand\colorMATH{\colorMATHM}\renewcommand\colorSYNTAX{\colorSYNTAXM}{{\color{\colorMATH}\ensuremath{0,0}}}\endgroup }}}}}\endgroup }}}} 
   \\\noindent  {{\color{\colorMATH}\ensuremath{{\begingroup\renewcommand\colorMATH{\colorMATHP}\renewcommand\colorSYNTAX{\colorSYNTAXP}{{\color{\colorMATH}\ensuremath{\{ {\begingroup\renewcommand\colorMATH{\colorMATHM}\renewcommand\colorSYNTAX{\colorSYNTAXM}{{\color{\colorMATH}\ensuremath{x_{i}}}}\endgroup }{\mathrel{:}}_{{\begingroup\renewcommand\colorMATH{\colorMATHM}\renewcommand\colorSYNTAX{\colorSYNTAXM}{{\color{\colorMATH}\ensuremath{\alpha _{i}}}}\endgroup },{\begingroup\renewcommand\colorMATH{\colorMATHM}\renewcommand\colorSYNTAX{\colorSYNTAXM}{{\color{\colorMATH}\ensuremath{\epsilon _{i}^{\prime 4}}}}\endgroup }} {\begingroup\renewcommand\colorMATH{\colorMATHM}\renewcommand\colorSYNTAX{\colorSYNTAXM}{{\color{\colorMATH}\ensuremath{\tau }}}\endgroup }\} }}}\endgroup } \in  {\begingroup\renewcommand\colorMATH{\colorMATHP}\renewcommand\colorSYNTAX{\colorSYNTAXP}{{\color{\colorMATH}\ensuremath{{}\rceil {\begingroup\renewcommand\colorMATH{\colorMATHS}\renewcommand\colorSYNTAX{\colorSYNTAXS}{{\color{\colorMATH}\ensuremath{\Gamma _{4}}}}\endgroup }\lceil {}^{{{\color{\colorSYNTAX}\texttt{\ensuremath{\infty }}}}}}}}\endgroup }}}} 
   \\\noindent  {{\color{\colorMATH}\ensuremath{{\begingroup\renewcommand\colorMATH{\colorMATHP}\renewcommand\colorSYNTAX{\colorSYNTAXP}{{\color{\colorMATH}\ensuremath{\{ {\begingroup\renewcommand\colorMATH{\colorMATHM}\renewcommand\colorSYNTAX{\colorSYNTAXM}{{\color{\colorMATH}\ensuremath{x_{i}}}}\endgroup }{\mathrel{:}}_{{\begingroup\renewcommand\colorMATH{\colorMATHM}\renewcommand\colorSYNTAX{\colorSYNTAXM}{{\color{\colorMATH}\ensuremath{\alpha _{i}}}}\endgroup },{\begingroup\renewcommand\colorMATH{\colorMATHM}\renewcommand\colorSYNTAX{\colorSYNTAXM}{{\color{\colorMATH}\ensuremath{\epsilon _{i}^{\prime 5}}}}\endgroup }} {\begingroup\renewcommand\colorMATH{\colorMATHM}\renewcommand\colorSYNTAX{\colorSYNTAXM}{{\color{\colorMATH}\ensuremath{\tau }}}\endgroup }\} }}}\endgroup } \in  {\begingroup\renewcommand\colorMATH{\colorMATHP}\renewcommand\colorSYNTAX{\colorSYNTAXP}{{\color{\colorMATH}\ensuremath{\mathrlap{\hspace{-0.5pt}{}\rceil }\lfloor {\begingroup\renewcommand\colorMATH{\colorMATHS}\renewcommand\colorSYNTAX{\colorSYNTAXS}{{\color{\colorMATH}\ensuremath{\Gamma _{5}}}}\endgroup }\mathrlap{\hspace{-0.5pt}\rfloor }\lceil {}_{\{ {\begingroup\renewcommand\colorMATH{\colorMATHM}\renewcommand\colorSYNTAX{\colorSYNTAXM}{{\color{\colorMATH}\ensuremath{x_{1}}}}\endgroup },{.}\hspace{-1pt}{.}\hspace{-1pt}{.},{\begingroup\renewcommand\colorMATH{\colorMATHM}\renewcommand\colorSYNTAX{\colorSYNTAXM}{{\color{\colorMATH}\ensuremath{x_{n}}}}\endgroup }\} }^{{\begingroup\renewcommand\colorMATH{\colorMATHM}\renewcommand\colorSYNTAX{\colorSYNTAXM}{{\color{\colorMATH}\ensuremath{\alpha }}}\endgroup },{\begingroup\renewcommand\colorMATH{\colorMATHM}\renewcommand\colorSYNTAX{\colorSYNTAXM}{{\color{\colorMATH}\ensuremath{\epsilon }}}\endgroup }}}}}\endgroup }}}} 
   \\\noindent  Each of {{\color{\colorMATH}\ensuremath{\epsilon _{i}^{\prime}}}}, {{\color{\colorMATH}\ensuremath{\epsilon _{i}^{\prime \prime}}}}, {{\color{\colorMATH}\ensuremath{\epsilon _{i}^{\prime \prime \prime}}}} must be {{\color{\colorMATH}\ensuremath{0}}}.
   \\\noindent  \begin{itemize}[leftmargin=*,label=\textbf{*
      }]\item  \noindent  Subcase {{\color{\colorMATH}\ensuremath{\epsilon _{i}^{\prime 4} = \infty }}}:
        \\\noindent  {{\color{\colorMATH}\ensuremath{ \epsilon _{i} = \infty  }}}; the property holds trivially
        
      \item  \noindent  Subcase {{\color{\colorMATH}\ensuremath{\epsilon ^{\prime 4} = 0}}}:
         \\\noindent  {{\color{\colorMATH}\ensuremath{\alpha _{i},\epsilon _{i}}}} = {{\color{\colorMATH}\ensuremath{\alpha _{i},\epsilon _{i}^{\prime 5}}}} = {{\color{\colorMATH}\ensuremath{\alpha ,\epsilon }}}
         \\\noindent  By IH, {{\color{\colorMATH}\ensuremath{ |f_{4}(\gamma ) - f_{4}(\gamma [x_{i}\mapsto d])|_{\llbracket {\mathbb{R}}\rrbracket } \leq  r }}}
         \\\noindent  The property follows from Theorem~\ref{thm:renyi_gauss}.
         
      \end{itemize}
   
\end{itemize}

\end{proof}

\section{Zero-Concentrated Differential Privacy}

\begin{theorem}[Gaussian mechanism (Zero-Concentrated Differential Privacy)]
  \label{thm:zcdp_gauss}
  
  If {{\color{\colorMATH}\ensuremath{ | f(\gamma ) - f(\gamma [x_{i}\mapsto d]) | \leq  r }}}, then for {{\color{\colorMATH}\ensuremath{ f^{\prime} = \lambda  \gamma  .\hspace*{0.33em} f(\gamma ) + {\mathcal{N}}(0, \sigma ^{2}) }}} and {{\color{\colorMATH}\ensuremath{ \sigma ^{2} = r^{2} / (2\rho ) }}}, and all {{\color{\colorMATH}\ensuremath{\rho  > 0}}} and {{\color{\colorMATH}\ensuremath{\alpha  \geq  1}}}:

  \vspace*{-0.25em}\begingroup\color{\colorMATH}\begin{gather*} D_{\alpha }(f^{\prime}(\gamma ) \|  f^{\prime}(\gamma [x_{i}\mapsto d^{\prime}])) \leq  \alpha  \rho 
  \vspace*{-1em}\end{gather*}\endgroup 
\end{theorem}

\begin{proof}
  By Bun and Steinke~\cite{bun2016concentrated}, Lemma 2.4, we have that:

  \vspace*{-0.25em}\begingroup\color{\colorMATH}\begin{gather*} D_{\alpha }(f^{\prime}(\gamma ) \|  f^{\prime}(\gamma [x_{i}\mapsto d^{\prime}])) \leq  \alpha  r^{2} / (2\sigma ^{2}) = \alpha \rho 
  \vspace*{-1em}\end{gather*}\endgroup 
\end{proof}

\begin{theorem}[Adaptive sequential composition (Zero-Concentrated differential privacy)]
  \label{thm:zcdp_let}
  \noindent  For all {{\color{\colorMATH}\ensuremath{\alpha  \geq  1}}}, if:
     \vspace*{-0.25em}\begingroup\color{\colorMATH}\begin{gather*} |\gamma [x_{i}] - d| \leq  1 \implies   
     \cr  D_{\alpha }(f_{1}(\gamma ) \|  f_{1}(\gamma [x_{i}\mapsto d])) \leq  \alpha  \rho _{i}
     \vspace*{-1em}\end{gather*}\endgroup 
  \\\noindent  and:
     \vspace*{-0.25em}\begingroup\color{\colorMATH}\begin{gather*} |\gamma [x_{i}] - d| \leq  1 \implies   
     \cr  D_{\alpha }(f_{2}(\gamma [x\mapsto d^{\prime}]) \|  f_{2}(\gamma [x\mapsto d^{\prime},x_{i}\mapsto d])) \leq  \alpha  \rho _{i}^{\prime}
     \vspace*{-1em}\end{gather*}\endgroup 
  \\\noindent  and:
     \vspace*{-0.25em}\begingroup\color{\colorMATH}\begin{gather*} {{\color{\colorMATH}\ensuremath{\operatorname{Pr}}}}[f_{3}(\gamma ) = d^{\prime \prime}] = {{\color{\colorMATH}\ensuremath{\operatorname{Pr}}}}[f_{1}(\gamma ) = d^{\prime}, f_{2}(\gamma [x\mapsto d^{\prime}]) = d^{\prime \prime}] \vspace*{-1em}\end{gather*}\endgroup 
  \\\noindent  then:
     \vspace*{-0.25em}\begingroup\color{\colorMATH}\begin{gather*} |\gamma [x_{i}] - d| \leq  1 \implies   
     \cr  D_{\alpha }(f_{3}(\gamma ) \|  f_{3}(\gamma [x_{i}\mapsto d])) \leq  \alpha (\rho _{i} + \rho _{i}^{\prime})
     \vspace*{-1em}\end{gather*}\endgroup 
  
\end{theorem}

\begin{proof}
  The result follows directly from Bun and Steinke~\cite{bun2016concentrated}, Lemma 2.3, setting {{\color{\colorMATH}\ensuremath{M = f_{1}}}} and {{\color{\colorMATH}\ensuremath{M' = f_{2}}}}.
\end{proof}

\begin{figure*}
\begingroup\renewcommand\colorMATH{\colorMATHP}\renewcommand\colorSYNTAX{\colorSYNTAXP}
\vspace*{-0.25em}\begingroup\color{\colorMATH}\begin{gather*}\begin{tabularx}{\linewidth}{>{\centering\arraybackslash\(}X<{\)}}\hfill\hspace{0pt}\begingroup\color{\colorTEXT}\boxed{\begingroup\color{\colorMATH} {\begingroup\renewcommand\colorMATH{\colorMATHM}\renewcommand\colorSYNTAX{\colorSYNTAXM}{{\color{\colorMATH}\ensuremath{\Delta }}}\endgroup },\Gamma  \vdash  e \mathrel{:} \tau  \endgroup}\endgroup \end{tabularx}\vspace*{-1em}\end{gather*}\endgroup 
\begingroup\color{\colorMATH}\begin{mathpar}\inferrule*[flushleft,lab={{\color{\colorTEXT}\textsc{\scriptsize Static Loop}}}
  ]{ {\begingroup\renewcommand\colorMATH{\colorMATHS}\renewcommand\colorSYNTAX{\colorSYNTAXS}{{\color{\colorMATH}\ensuremath{{\begingroup\renewcommand\colorMATH{\colorMATHM}\renewcommand\colorSYNTAX{\colorSYNTAXM}{{\color{\colorMATH}\ensuremath{\Delta }}}\endgroup },\Gamma _{1} \vdash  e_{1} \mathrel{:} {\begingroup\renewcommand\colorMATH{\colorMATHM}\renewcommand\colorSYNTAX{\colorSYNTAXM}{{\color{\colorMATH}\ensuremath{{{\color{\colorSYNTAX}\texttt{\ensuremath{{\mathbb{R}}^{+}[{{\color{\colorMATH}\ensuremath{\eta _{n}}}}]}}}}}}}\endgroup }}}}\endgroup }
  \\ {\begingroup\renewcommand\colorMATH{\colorMATHS}\renewcommand\colorSYNTAX{\colorSYNTAXS}{{\color{\colorMATH}\ensuremath{{\begingroup\renewcommand\colorMATH{\colorMATHM}\renewcommand\colorSYNTAX{\colorSYNTAXM}{{\color{\colorMATH}\ensuremath{\Delta }}}\endgroup },\Gamma _{2} \vdash  e_{2} \mathrel{:} {\begingroup\renewcommand\colorMATH{\colorMATHM}\renewcommand\colorSYNTAX{\colorSYNTAXM}{{\color{\colorMATH}\ensuremath{\tau }}}\endgroup }}}}\endgroup }
  \\ {\begingroup\renewcommand\colorMATH{\colorMATHM}\renewcommand\colorSYNTAX{\colorSYNTAXM}{{\color{\colorMATH}\ensuremath{\Delta }}}\endgroup },\Gamma _{3} + \mathrlap{\hspace{-0.5pt}{}\rceil }\lfloor \Gamma _{4}\mathrlap{\hspace{-0.5pt}\rfloor }\lceil {}_{\{ {\begingroup\renewcommand\colorMATH{\colorMATHM}\renewcommand\colorSYNTAX{\colorSYNTAXM}{{\color{\colorMATH}\ensuremath{x_{1}^{\prime}}}}\endgroup },{.}\hspace{-1pt}{.}\hspace{-1pt}{.},{\begingroup\renewcommand\colorMATH{\colorMATHM}\renewcommand\colorSYNTAX{\colorSYNTAXM}{{\color{\colorMATH}\ensuremath{x_{n}^{\prime}}}}\endgroup }\} }^{{\begingroup\renewcommand\colorMATH{\colorMATHM}\renewcommand\colorSYNTAX{\colorSYNTAXM}{{\color{\colorMATH}\ensuremath{\eta _{\rho }}}}\endgroup }} \uplus  \{ {\begingroup\renewcommand\colorMATH{\colorMATHM}\renewcommand\colorSYNTAX{\colorSYNTAXM}{{\color{\colorMATH}\ensuremath{x_{1}}}}\endgroup } {\mathrel{:}}_{\infty } {\begingroup\renewcommand\colorMATH{\colorMATHM}\renewcommand\colorSYNTAX{\colorSYNTAXM}{{\color{\colorMATH}\ensuremath{{{\color{\colorSYNTAX}\texttt{\ensuremath{{\mathbb{N}}}}}}}}}\endgroup },{\begingroup\renewcommand\colorMATH{\colorMATHM}\renewcommand\colorSYNTAX{\colorSYNTAXM}{{\color{\colorMATH}\ensuremath{x_{2}}}}\endgroup } {\mathrel{:}}_{\infty } {\begingroup\renewcommand\colorMATH{\colorMATHM}\renewcommand\colorSYNTAX{\colorSYNTAXM}{{\color{\colorMATH}\ensuremath{\tau }}}\endgroup }\}  \vdash  e_{3} \mathrel{:} {\begingroup\renewcommand\colorMATH{\colorMATHM}\renewcommand\colorSYNTAX{\colorSYNTAXM}{{\color{\colorMATH}\ensuremath{\tau }}}\endgroup }
     }{
     {\begingroup\renewcommand\colorMATH{\colorMATHM}\renewcommand\colorSYNTAX{\colorSYNTAXM}{{\color{\colorMATH}\ensuremath{\Delta }}}\endgroup },{}\rceil {\begingroup\renewcommand\colorMATH{\colorMATHS}\renewcommand\colorSYNTAX{\colorSYNTAXS}{{\color{\colorMATH}\ensuremath{\Gamma _{1} + \Gamma _{1}}}}\endgroup }\lceil {}^{{\begingroup\renewcommand\colorMATH{\colorMATHM}\renewcommand\colorSYNTAX{\colorSYNTAXM}{{\color{\colorMATH}\ensuremath{0}}}\endgroup },{\begingroup\renewcommand\colorMATH{\colorMATHM}\renewcommand\colorSYNTAX{\colorSYNTAXM}{{\color{\colorMATH}\ensuremath{0}}}\endgroup }} + {}\rceil {\begingroup\renewcommand\colorMATH{\colorMATHS}\renewcommand\colorSYNTAX{\colorSYNTAXS}{{\color{\colorMATH}\ensuremath{\Gamma _{2}}}}\endgroup }\lceil {}^{{{\color{\colorSYNTAX}\texttt{\ensuremath{\infty }}}}} + {}\rceil \Gamma _{3}\lceil {}^{{{\color{\colorSYNTAX}\texttt{\ensuremath{\infty }}}}} 
     + \mathrlap{\hspace{-0.5pt}{}\rceil }\lfloor \Gamma _{4}\mathrlap{\hspace{-0.5pt}\rfloor }\lceil {}_{\{ {\begingroup\renewcommand\colorMATH{\colorMATHM}\renewcommand\colorSYNTAX{\colorSYNTAXM}{{\color{\colorMATH}\ensuremath{x_{1}^{\prime}}}}\endgroup },{.}\hspace{-1pt}{.}\hspace{-1pt}{.},{\begingroup\renewcommand\colorMATH{\colorMATHM}\renewcommand\colorSYNTAX{\colorSYNTAXM}{{\color{\colorMATH}\ensuremath{x_{n}^{\prime}}}}\endgroup }\} }^{{\begingroup\renewcommand\colorMATH{\colorMATHM}\renewcommand\colorSYNTAX{\colorSYNTAXM}{{\color{\colorMATH}\ensuremath{\eta _{n}{\mathrel{\mathord{\cdotp }}}\eta _{\rho }}}}\endgroup }}
     \vdash  {{\color{\colorSYNTAX}\texttt{loop}}}\hspace*{0.33em}{\begingroup\renewcommand\colorMATH{\colorMATHS}\renewcommand\colorSYNTAX{\colorSYNTAXS}{{\color{\colorMATH}\ensuremath{e_{1}}}}\endgroup }\hspace*{0.33em}{{\color{\colorSYNTAX}\texttt{on}}}\hspace*{0.33em}{\begingroup\renewcommand\colorMATH{\colorMATHS}\renewcommand\colorSYNTAX{\colorSYNTAXS}{{\color{\colorMATH}\ensuremath{e_{2}}}}\endgroup }\hspace*{0.33em}{<}{\begingroup\renewcommand\colorMATH{\colorMATHM}\renewcommand\colorSYNTAX{\colorSYNTAXM}{{\color{\colorMATH}\ensuremath{x_{1}^{\prime}}}}\endgroup },{.}\hspace{-1pt}{.}\hspace{-1pt}{.},{\begingroup\renewcommand\colorMATH{\colorMATHM}\renewcommand\colorSYNTAX{\colorSYNTAXM}{{\color{\colorMATH}\ensuremath{x_{n}^{\prime}}}}\endgroup }{>}\hspace*{0.33em}\{ {\begingroup\renewcommand\colorMATH{\colorMATHM}\renewcommand\colorSYNTAX{\colorSYNTAXM}{{\color{\colorMATH}\ensuremath{x_{1}}}}\endgroup },{\begingroup\renewcommand\colorMATH{\colorMATHM}\renewcommand\colorSYNTAX{\colorSYNTAXM}{{\color{\colorMATH}\ensuremath{x_{2}}}}\endgroup } \Rightarrow  e_{3}\}  \mathrel{:} {\begingroup\renewcommand\colorMATH{\colorMATHM}\renewcommand\colorSYNTAX{\colorSYNTAXM}{{\color{\colorMATH}\ensuremath{\tau }}}\endgroup }
  }
\and\inferrule*[lab={{\color{\colorTEXT}\textsc{\scriptsize Gauss}}}
  ]{ {\begingroup\renewcommand\colorMATH{\colorMATHS}\renewcommand\colorSYNTAX{\colorSYNTAXS}{{\color{\colorMATH}\ensuremath{\Gamma _{1} \vdash  e_{1} \mathrel{:} {\begingroup\renewcommand\colorMATH{\colorMATHM}\renewcommand\colorSYNTAX{\colorSYNTAXM}{{\color{\colorMATH}\ensuremath{{{\color{\colorSYNTAX}\texttt{\ensuremath{{\mathbb{R}}^{+}[{{\color{\colorMATH}\ensuremath{\eta _{s}}}}]}}}}}}}\endgroup }}}}\endgroup }
  \\ {\begingroup\renewcommand\colorMATH{\colorMATHS}\renewcommand\colorSYNTAX{\colorSYNTAXS}{{\color{\colorMATH}\ensuremath{\Gamma _{2} \vdash  e_{2} \mathrel{:} {\begingroup\renewcommand\colorMATH{\colorMATHM}\renewcommand\colorSYNTAX{\colorSYNTAXM}{{\color{\colorMATH}\ensuremath{{{\color{\colorSYNTAX}\texttt{\ensuremath{{\mathbb{R}}^{+}[{{\color{\colorMATH}\ensuremath{\eta _{\rho }}}}]}}}}}}}\endgroup }}}}\endgroup }
  \\ {\begingroup\renewcommand\colorMATH{\colorMATHS}\renewcommand\colorSYNTAX{\colorSYNTAXS}{{\color{\colorMATH}\ensuremath{\Gamma _{3} + \mathrlap{\hspace{-0.5pt}{}\rceil }\lfloor \Gamma _{4}\mathrlap{\hspace{-0.5pt}\rfloor }\lceil {}_{\{ {\begingroup\renewcommand\colorMATH{\colorMATHM}\renewcommand\colorSYNTAX{\colorSYNTAXM}{{\color{\colorMATH}\ensuremath{x_{1}}}}\endgroup },{.}\hspace{-1pt}{.}\hspace{-1pt}{.},{\begingroup\renewcommand\colorMATH{\colorMATHM}\renewcommand\colorSYNTAX{\colorSYNTAXM}{{\color{\colorMATH}\ensuremath{x_{n}}}}\endgroup }\} }^{{\begingroup\renewcommand\colorMATH{\colorMATHM}\renewcommand\colorSYNTAX{\colorSYNTAXM}{{\color{\colorMATH}\ensuremath{\eta _{s}}}}\endgroup }} \vdash  e_{3} \mathrel{:} {\begingroup\renewcommand\colorMATH{\colorMATHM}\renewcommand\colorSYNTAX{\colorSYNTAXM}{{\color{\colorMATH}\ensuremath{{{\color{\colorSYNTAX}\texttt{\ensuremath{{\mathbb{R}}}}}}}}}\endgroup }}}}\endgroup }
     }{
     {}\rceil {\begingroup\renewcommand\colorMATH{\colorMATHS}\renewcommand\colorSYNTAX{\colorSYNTAXS}{{\color{\colorMATH}\ensuremath{\Gamma _{1} + \Gamma _{2}}}}\endgroup }\lceil {}^{{\begingroup\renewcommand\colorMATH{\colorMATHM}\renewcommand\colorSYNTAX{\colorSYNTAXM}{{\color{\colorMATH}\ensuremath{0}}}\endgroup }} + {}\rceil {\begingroup\renewcommand\colorMATH{\colorMATHS}\renewcommand\colorSYNTAX{\colorSYNTAXS}{{\color{\colorMATH}\ensuremath{\Gamma _{3}}}}\endgroup }\lceil {}^{{{\color{\colorSYNTAX}\texttt{\ensuremath{\infty }}}}} + \mathrlap{\hspace{-0.5pt}{}\rceil }\lfloor {\begingroup\renewcommand\colorMATH{\colorMATHS}\renewcommand\colorSYNTAX{\colorSYNTAXS}{{\color{\colorMATH}\ensuremath{\Gamma _{4}}}}\endgroup }\mathrlap{\hspace{-0.5pt}\rfloor }\lceil {}_{\{ {\begingroup\renewcommand\colorMATH{\colorMATHM}\renewcommand\colorSYNTAX{\colorSYNTAXM}{{\color{\colorMATH}\ensuremath{x_{1}}}}\endgroup },{.}\hspace{-1pt}{.}\hspace{-1pt}{.},{\begingroup\renewcommand\colorMATH{\colorMATHM}\renewcommand\colorSYNTAX{\colorSYNTAXM}{{\color{\colorMATH}\ensuremath{x_{n}}}}\endgroup }\} }^{{\begingroup\renewcommand\colorMATH{\colorMATHM}\renewcommand\colorSYNTAX{\colorSYNTAXM}{{\color{\colorMATH}\ensuremath{\eta _{\rho }}}}\endgroup }} 
     \vdash  {{\color{\colorSYNTAX}\texttt{\ensuremath{{{\color{\colorSYNTAX}\texttt{gauss}}}[{\begingroup\renewcommand\colorMATH{\colorMATHS}\renewcommand\colorSYNTAX{\colorSYNTAXS}{{\color{\colorMATH}\ensuremath{e_{1}}}}\endgroup },{\begingroup\renewcommand\colorMATH{\colorMATHS}\renewcommand\colorSYNTAX{\colorSYNTAXS}{{\color{\colorMATH}\ensuremath{e_{2}}}}\endgroup }]\hspace*{0.33em}{<}{\begingroup\renewcommand\colorMATH{\colorMATHM}\renewcommand\colorSYNTAX{\colorSYNTAXM}{{\color{\colorMATH}\ensuremath{x_{1}}}}\endgroup },{.}\hspace{-1pt}{.}\hspace{-1pt}{.},{\begingroup\renewcommand\colorMATH{\colorMATHM}\renewcommand\colorSYNTAX{\colorSYNTAXM}{{\color{\colorMATH}\ensuremath{x_{n}}}}\endgroup }{>}\hspace*{0.33em}\{ {\begingroup\renewcommand\colorMATH{\colorMATHS}\renewcommand\colorSYNTAX{\colorSYNTAXS}{{\color{\colorMATH}\ensuremath{e_{3}}}}\endgroup }\} }}}} \mathrel{:} {\begingroup\renewcommand\colorMATH{\colorMATHM}\renewcommand\colorSYNTAX{\colorSYNTAXM}{{\color{\colorMATH}\ensuremath{{{\color{\colorSYNTAX}\texttt{\ensuremath{{\mathbb{R}}}}}}}}}\endgroup }
  }
\and\inferrule*[lab={{\color{\colorTEXT}\textsc{\scriptsize MGauss}}}
  ]{ {\begingroup\renewcommand\colorMATH{\colorMATHS}\renewcommand\colorSYNTAX{\colorSYNTAXS}{{\color{\colorMATH}\ensuremath{\Gamma _{1} \vdash  e_{1} \mathrel{:} {\begingroup\renewcommand\colorMATH{\colorMATHM}\renewcommand\colorSYNTAX{\colorSYNTAXM}{{\color{\colorMATH}\ensuremath{{{\color{\colorSYNTAX}\texttt{\ensuremath{{\mathbb{R}}^{+}[{{\color{\colorMATH}\ensuremath{\eta _{s}}}}]}}}}}}}\endgroup }}}}\endgroup }
  \\ {\begingroup\renewcommand\colorMATH{\colorMATHS}\renewcommand\colorSYNTAX{\colorSYNTAXS}{{\color{\colorMATH}\ensuremath{\Gamma _{2} \vdash  e_{2} \mathrel{:} {\begingroup\renewcommand\colorMATH{\colorMATHM}\renewcommand\colorSYNTAX{\colorSYNTAXM}{{\color{\colorMATH}\ensuremath{{{\color{\colorSYNTAX}\texttt{\ensuremath{{\mathbb{R}}^{+}[{{\color{\colorMATH}\ensuremath{\eta _{\rho }}}}]}}}}}}}\endgroup }}}}\endgroup }
  \\ {\begingroup\renewcommand\colorMATH{\colorMATHS}\renewcommand\colorSYNTAX{\colorSYNTAXS}{{\color{\colorMATH}\ensuremath{\Gamma _{3} + \mathrlap{\hspace{-0.5pt}{}\rceil }\lfloor \Gamma _{4}\mathrlap{\hspace{-0.5pt}\rfloor }\lceil {}_{\{ {\begingroup\renewcommand\colorMATH{\colorMATHM}\renewcommand\colorSYNTAX{\colorSYNTAXM}{{\color{\colorMATH}\ensuremath{x_{1}}}}\endgroup },{.}\hspace{-1pt}{.}\hspace{-1pt}{.},{\begingroup\renewcommand\colorMATH{\colorMATHM}\renewcommand\colorSYNTAX{\colorSYNTAXM}{{\color{\colorMATH}\ensuremath{x_{n}}}}\endgroup }\} }^{{\begingroup\renewcommand\colorMATH{\colorMATHM}\renewcommand\colorSYNTAX{\colorSYNTAXM}{{\color{\colorMATH}\ensuremath{\eta _{s}}}}\endgroup }} \vdash  e_{3} \mathrel{:} {\begingroup\renewcommand\colorMATH{\colorMATHM}\renewcommand\colorSYNTAX{\colorSYNTAXM}{{\color{\colorMATH}\ensuremath{{{\color{\colorSYNTAX}\texttt{\ensuremath{{{\color{\colorSYNTAX}\texttt{matrix}}}_{L2}^{{{\color{\colorMATH}\ensuremath{\bigstar }}}}[{{\color{\colorMATH}\ensuremath{\eta _{m}}}},{{\color{\colorMATH}\ensuremath{\eta _{n}}}}]\hspace*{0.33em}{\mathbb{R}}}}}}}}}\endgroup }}}}\endgroup }
     }{
     {}\rceil {\begingroup\renewcommand\colorMATH{\colorMATHS}\renewcommand\colorSYNTAX{\colorSYNTAXS}{{\color{\colorMATH}\ensuremath{\Gamma _{1} + \Gamma _{2}}}}\endgroup }\lceil {}^{{\begingroup\renewcommand\colorMATH{\colorMATHM}\renewcommand\colorSYNTAX{\colorSYNTAXM}{{\color{\colorMATH}\ensuremath{0}}}\endgroup }} + {}\rceil {\begingroup\renewcommand\colorMATH{\colorMATHS}\renewcommand\colorSYNTAX{\colorSYNTAXS}{{\color{\colorMATH}\ensuremath{\Gamma _{3}}}}\endgroup }\lceil {}^{{{\color{\colorSYNTAX}\texttt{\ensuremath{\infty }}}}} + \mathrlap{\hspace{-0.5pt}{}\rceil }\lfloor {\begingroup\renewcommand\colorMATH{\colorMATHS}\renewcommand\colorSYNTAX{\colorSYNTAXS}{{\color{\colorMATH}\ensuremath{\Gamma _{4}}}}\endgroup }\mathrlap{\hspace{-0.5pt}\rfloor }\lceil {}_{\{ {\begingroup\renewcommand\colorMATH{\colorMATHM}\renewcommand\colorSYNTAX{\colorSYNTAXM}{{\color{\colorMATH}\ensuremath{x_{1}}}}\endgroup },{.}\hspace{-1pt}{.}\hspace{-1pt}{.},{\begingroup\renewcommand\colorMATH{\colorMATHM}\renewcommand\colorSYNTAX{\colorSYNTAXM}{{\color{\colorMATH}\ensuremath{x_{n}}}}\endgroup }\} }^{{\begingroup\renewcommand\colorMATH{\colorMATHM}\renewcommand\colorSYNTAX{\colorSYNTAXM}{{\color{\colorMATH}\ensuremath{\eta _{\rho }}}}\endgroup }} 
     \vdash  {{\color{\colorSYNTAX}\texttt{\ensuremath{{{\color{\colorSYNTAX}\texttt{mgauss}}}[{\begingroup\renewcommand\colorMATH{\colorMATHS}\renewcommand\colorSYNTAX{\colorSYNTAXS}{{\color{\colorMATH}\ensuremath{e_{1}}}}\endgroup },{\begingroup\renewcommand\colorMATH{\colorMATHS}\renewcommand\colorSYNTAX{\colorSYNTAXS}{{\color{\colorMATH}\ensuremath{e_{2}}}}\endgroup }]\hspace*{0.33em}{<}{\begingroup\renewcommand\colorMATH{\colorMATHM}\renewcommand\colorSYNTAX{\colorSYNTAXM}{{\color{\colorMATH}\ensuremath{x_{1}}}}\endgroup },{.}\hspace{-1pt}{.}\hspace{-1pt}{.},{\begingroup\renewcommand\colorMATH{\colorMATHM}\renewcommand\colorSYNTAX{\colorSYNTAXM}{{\color{\colorMATH}\ensuremath{x_{n}}}}\endgroup }{>}\hspace*{0.33em}\{ {\begingroup\renewcommand\colorMATH{\colorMATHS}\renewcommand\colorSYNTAX{\colorSYNTAXS}{{\color{\colorMATH}\ensuremath{e_{3}}}}\endgroup }\} }}}} \mathrel{:} {\begingroup\renewcommand\colorMATH{\colorMATHM}\renewcommand\colorSYNTAX{\colorSYNTAXM}{{\color{\colorMATH}\ensuremath{{{\color{\colorSYNTAX}\texttt{\ensuremath{{{\color{\colorSYNTAX}\texttt{matrix}}}_{L\infty }^{U}[{{\color{\colorMATH}\ensuremath{\eta _{m}}}},{{\color{\colorMATH}\ensuremath{\eta _{n}}}}]\hspace*{0.33em}{\mathbb{R}}}}}}}}}\endgroup }
  }
\end{mathpar}\endgroup 
\endgroup 
\caption{Privacy Type System Modifications, Zero-Concentrated Differential Privacy}
\end{figure*}

\begin{theorem}[Soundness]
  There exists an interpretation of well typed terms {\begingroup\renewcommand\colorMATH{\colorMATHS}\renewcommand\colorSYNTAX{\colorSYNTAXS}{{\color{\colorMATH}\ensuremath{\Gamma  \vdash  e \mathrel{:} {\begingroup\renewcommand\colorMATH{\colorMATHM}\renewcommand\colorSYNTAX{\colorSYNTAXM}{{\color{\colorMATH}\ensuremath{\tau }}}\endgroup }}}}\endgroup } and {\begingroup\renewcommand\colorMATH{\colorMATHP}\renewcommand\colorSYNTAX{\colorSYNTAXP}{{\color{\colorMATH}\ensuremath{\Gamma 
  \vdash  e \mathrel{:} {\begingroup\renewcommand\colorMATH{\colorMATHM}\renewcommand\colorSYNTAX{\colorSYNTAXM}{{\color{\colorMATH}\ensuremath{\tau }}}\endgroup }}}}\endgroup }, notated {\begingroup\renewcommand\colorMATH{\colorMATHS}\renewcommand\colorSYNTAX{\colorSYNTAXS}{{\color{\colorMATH}\ensuremath{\llbracket e\rrbracket }}}\endgroup } and {\begingroup\renewcommand\colorMATH{\colorMATHP}\renewcommand\colorSYNTAX{\colorSYNTAXP}{{\color{\colorMATH}\ensuremath{\llbracket e\rrbracket }}}\endgroup }, such that {{\color{\colorMATH}\ensuremath{{\begingroup\renewcommand\colorMATH{\colorMATHP}\renewcommand\colorSYNTAX{\colorSYNTAXP}{{\color{\colorMATH}\ensuremath{\llbracket e\rrbracket }}}\endgroup } \in  {\begingroup\renewcommand\colorMATH{\colorMATHP}\renewcommand\colorSYNTAX{\colorSYNTAXP}{{\color{\colorMATH}\ensuremath{\llbracket \Gamma  \vdash 
  {\begingroup\renewcommand\colorMATH{\colorMATHM}\renewcommand\colorSYNTAX{\colorSYNTAXM}{{\color{\colorMATH}\ensuremath{\tau }}}\endgroup }\rrbracket }}}\endgroup }}}} and {{\color{\colorMATH}\ensuremath{{\begingroup\renewcommand\colorMATH{\colorMATHS}\renewcommand\colorSYNTAX{\colorSYNTAXS}{{\color{\colorMATH}\ensuremath{\llbracket e\rrbracket }}}\endgroup } \in  {\begingroup\renewcommand\colorMATH{\colorMATHS}\renewcommand\colorSYNTAX{\colorSYNTAXS}{{\color{\colorMATH}\ensuremath{\llbracket \Gamma  \vdash  {\begingroup\renewcommand\colorMATH{\colorMATHM}\renewcommand\colorSYNTAX{\colorSYNTAXM}{{\color{\colorMATH}\ensuremath{\tau }}}\endgroup }\rrbracket }}}\endgroup }}}}.
\end{theorem}
\begin{proof}
  We include the cases that are different from the proof for {{\color{\colorMATH}\ensuremath{(\epsilon , \delta )}}}-differential privacy: {\sc Bind} and {\sc Gauss}.

\vspace*{-0.25em}\begingroup\color{\colorMATH}\begin{gather*}\begin{tabularx}{\linewidth}{>{\centering\arraybackslash\(}X<{\)}}\hfill\hspace{0pt}\begingroup\color{\colorTEXT}\boxed{\begingroup\color{\colorMATH} {\begingroup\renewcommand\colorMATH{\colorMATHP}\renewcommand\colorSYNTAX{\colorSYNTAXP}{{\color{\colorMATH}\ensuremath{\llbracket \underline{\hspace{0.66em}\vspace*{5ex}}\rrbracket  \in  \{ e \in  {{\color{\colorMATH}\ensuremath{\operatorname{exp}}}} \mathrel{|} \Gamma  \vdash  e \mathrel{:} {\begingroup\renewcommand\colorMATH{\colorMATHM}\renewcommand\colorSYNTAX{\colorSYNTAXM}{{\color{\colorMATH}\ensuremath{\tau }}}\endgroup }\}  \rightarrow  \llbracket \Gamma  \vdash  {\begingroup\renewcommand\colorMATH{\colorMATHM}\renewcommand\colorSYNTAX{\colorSYNTAXM}{{\color{\colorMATH}\ensuremath{\tau }}}\endgroup }\rrbracket }}}\endgroup } \endgroup}\endgroup \end{tabularx}\vspace*{-1em}\end{gather*}\endgroup 
\begin{itemize}[leftmargin=*,label=\textbf{-
}]\item  \noindent  Case {\begingroup\renewcommand\colorMATH{\colorMATHP}\renewcommand\colorSYNTAX{\colorSYNTAXP}{{\color{\colorMATH}\ensuremath{\overbracketarg \Gamma {\Gamma _{1} + \Gamma _{2}} \vdash  {{\color{\colorSYNTAX}\texttt{\ensuremath{{\begingroup\renewcommand\colorMATH{\colorMATHM}\renewcommand\colorSYNTAX{\colorSYNTAXM}{{\color{\colorMATH}\ensuremath{x}}}\endgroup } \leftarrow  {{\color{\colorMATH}\ensuremath{e_{1}}}}\mathrel{;}{{\color{\colorMATH}\ensuremath{e_{2}}}}}}}} \mathrel{:} {\begingroup\renewcommand\colorMATH{\colorMATHM}\renewcommand\colorSYNTAX{\colorSYNTAXM}{{\color{\colorMATH}\ensuremath{\tau _{2}}}}\endgroup }}}}\endgroup }
   \\\noindent  By inversion:
      \vspace*{-0.25em}\begingroup\color{\colorMATH}\begin{gather*} 

      \vspace*{-1em}\end{gather*}\endgroup 
   \\\noindent  By Induction Hypothesis (IH):
      \vspace*{-0.25em}\begingroup\color{\colorMATH}\begin{gather*} %
      \vspace*{-1em}\end{gather*}\endgroup 
   \\\noindent  By definition of {\begingroup\renewcommand\colorMATH{\colorMATHP}\renewcommand\colorSYNTAX{\colorSYNTAXP}{{\color{\colorMATH}\ensuremath{+}}}\endgroup } for privacy contexts:
   \\\noindent  {{\color{\colorMATH}\ensuremath{\rho _{i} =  \rho _{i}^{\prime} + \rho _{i}^{\prime \prime}}}} for
   \\\noindent  {{\color{\colorMATH}\ensuremath{{\begingroup\renewcommand\colorMATH{\colorMATHP}\renewcommand\colorSYNTAX{\colorSYNTAXP}{{\color{\colorMATH}\ensuremath{\{ {\begingroup\renewcommand\colorMATH{\colorMATHM}\renewcommand\colorSYNTAX{\colorSYNTAXM}{{\color{\colorMATH}\ensuremath{x_{i}}}}\endgroup }{\mathrel{:}}_{{\begingroup\renewcommand\colorMATH{\colorMATHM}\renewcommand\colorSYNTAX{\colorSYNTAXM}{{\color{\colorMATH}\ensuremath{\rho _{i}^{\prime}}}}\endgroup }} {\begingroup\renewcommand\colorMATH{\colorMATHM}\renewcommand\colorSYNTAX{\colorSYNTAXM}{{\color{\colorMATH}\ensuremath{\tau }}}\endgroup }\} }}}\endgroup } \in  {\begingroup\renewcommand\colorMATH{\colorMATHP}\renewcommand\colorSYNTAX{\colorSYNTAXP}{{\color{\colorMATH}\ensuremath{\Gamma _{1}}}}\endgroup }}}} and {{\color{\colorMATH}\ensuremath{{\begingroup\renewcommand\colorMATH{\colorMATHP}\renewcommand\colorSYNTAX{\colorSYNTAXP}{{\color{\colorMATH}\ensuremath{\{ {\begingroup\renewcommand\colorMATH{\colorMATHM}\renewcommand\colorSYNTAX{\colorSYNTAXM}{{\color{\colorMATH}\ensuremath{x_{i}}}}\endgroup }{\mathrel{:}}_{{\begingroup\renewcommand\colorMATH{\colorMATHM}\renewcommand\colorSYNTAX{\colorSYNTAXM}{{\color{\colorMATH}\ensuremath{\rho _{i}^{\prime \prime}}}}\endgroup }} {\begingroup\renewcommand\colorMATH{\colorMATHM}\renewcommand\colorSYNTAX{\colorSYNTAXM}{{\color{\colorMATH}\ensuremath{\tau }}}\endgroup }\} }}}\endgroup } \in  {\begingroup\renewcommand\colorMATH{\colorMATHP}\renewcommand\colorSYNTAX{\colorSYNTAXP}{{\color{\colorMATH}\ensuremath{\Gamma _{2}}}}\endgroup }}}}. 
   \\\noindent  Property holds via IH.1, IH.2 and theorem~\ref{thm:zcdp_let} instantiated with
      {{\color{\colorMATH}\ensuremath{\rho _{i}^{\prime}}}} and {{\color{\colorMATH}\ensuremath{\rho _{i}^{\prime \prime}}}}.
   
\item  \noindent  Case: 
   \\\noindent  {{\color{\colorMATH}\ensuremath{\hspace*{1.00em}

      \vspace*{-1em}\end{gather*}\endgroup 
   \\\noindent  By Induction Hypothesis (IH):
      \vspace*{-0.25em}\begingroup\color{\colorMATH}\begin{gather*} %
      \vspace*{-1em}\end{gather*}\endgroup 
   \\\noindent  By IH, we have that {{\color{\colorMATH}\ensuremath{\sigma ^{2} = \frac{r^{2}}{2\rho }}}}
   \\\noindent  By definition of {\begingroup\renewcommand\colorMATH{\colorMATHP}\renewcommand\colorSYNTAX{\colorSYNTAXP}{{\color{\colorMATH}\ensuremath{+}}}\endgroup } for privacy contexts:
   \\\noindent  {{\color{\colorMATH}\ensuremath{\rho _{i} = \rho _{i}^{\prime} + \rho _{i}^{\prime \prime} + \rho _{i}^{\prime \prime \prime} + \rho _{i}^{\prime 4}}}} for
   \\\noindent  {{\color{\colorMATH}\ensuremath{{\begingroup\renewcommand\colorMATH{\colorMATHP}\renewcommand\colorSYNTAX{\colorSYNTAXP}{{\color{\colorMATH}\ensuremath{\{ {\begingroup\renewcommand\colorMATH{\colorMATHM}\renewcommand\colorSYNTAX{\colorSYNTAXM}{{\color{\colorMATH}\ensuremath{x_{i}}}}\endgroup }{\mathrel{:}}_{{\begingroup\renewcommand\colorMATH{\colorMATHM}\renewcommand\colorSYNTAX{\colorSYNTAXM}{{\color{\colorMATH}\ensuremath{\rho _{i}^{\prime}}}}\endgroup }} {\begingroup\renewcommand\colorMATH{\colorMATHM}\renewcommand\colorSYNTAX{\colorSYNTAXM}{{\color{\colorMATH}\ensuremath{\tau }}}\endgroup }\} }}}\endgroup } \in  {\begingroup\renewcommand\colorMATH{\colorMATHP}\renewcommand\colorSYNTAX{\colorSYNTAXP}{{\color{\colorMATH}\ensuremath{{}\rceil {\begingroup\renewcommand\colorMATH{\colorMATHS}\renewcommand\colorSYNTAX{\colorSYNTAXS}{{\color{\colorMATH}\ensuremath{\Gamma _{1}}}}\endgroup }\lceil {}^{{\begingroup\renewcommand\colorMATH{\colorMATHM}\renewcommand\colorSYNTAX{\colorSYNTAXM}{{\color{\colorMATH}\ensuremath{0,0}}}\endgroup }}}}}\endgroup }}}} 
   \\\noindent  {{\color{\colorMATH}\ensuremath{{\begingroup\renewcommand\colorMATH{\colorMATHP}\renewcommand\colorSYNTAX{\colorSYNTAXP}{{\color{\colorMATH}\ensuremath{\{ {\begingroup\renewcommand\colorMATH{\colorMATHM}\renewcommand\colorSYNTAX{\colorSYNTAXM}{{\color{\colorMATH}\ensuremath{x_{i}}}}\endgroup }{\mathrel{:}}_{{\begingroup\renewcommand\colorMATH{\colorMATHM}\renewcommand\colorSYNTAX{\colorSYNTAXM}{{\color{\colorMATH}\ensuremath{\rho _{i}^{\prime \prime}}}}\endgroup }} {\begingroup\renewcommand\colorMATH{\colorMATHM}\renewcommand\colorSYNTAX{\colorSYNTAXM}{{\color{\colorMATH}\ensuremath{\tau }}}\endgroup }\} }}}\endgroup } \in  {\begingroup\renewcommand\colorMATH{\colorMATHP}\renewcommand\colorSYNTAX{\colorSYNTAXP}{{\color{\colorMATH}\ensuremath{{}\rceil {\begingroup\renewcommand\colorMATH{\colorMATHS}\renewcommand\colorSYNTAX{\colorSYNTAXS}{{\color{\colorMATH}\ensuremath{\Gamma _{2}}}}\endgroup }\lceil {}^{{\begingroup\renewcommand\colorMATH{\colorMATHM}\renewcommand\colorSYNTAX{\colorSYNTAXM}{{\color{\colorMATH}\ensuremath{0,0}}}\endgroup }}}}}\endgroup }}}} 
   \\\noindent  {{\color{\colorMATH}\ensuremath{{\begingroup\renewcommand\colorMATH{\colorMATHP}\renewcommand\colorSYNTAX{\colorSYNTAXP}{{\color{\colorMATH}\ensuremath{\{ {\begingroup\renewcommand\colorMATH{\colorMATHM}\renewcommand\colorSYNTAX{\colorSYNTAXM}{{\color{\colorMATH}\ensuremath{x_{i}}}}\endgroup }{\mathrel{:}}_{{\begingroup\renewcommand\colorMATH{\colorMATHM}\renewcommand\colorSYNTAX{\colorSYNTAXM}{{\color{\colorMATH}\ensuremath{\rho _{i}^{\prime \prime \prime}}}}\endgroup }} {\begingroup\renewcommand\colorMATH{\colorMATHM}\renewcommand\colorSYNTAX{\colorSYNTAXM}{{\color{\colorMATH}\ensuremath{\tau }}}\endgroup }\} }}}\endgroup } \in  {\begingroup\renewcommand\colorMATH{\colorMATHP}\renewcommand\colorSYNTAX{\colorSYNTAXP}{{\color{\colorMATH}\ensuremath{{}\rceil {\begingroup\renewcommand\colorMATH{\colorMATHS}\renewcommand\colorSYNTAX{\colorSYNTAXS}{{\color{\colorMATH}\ensuremath{\Gamma _{3}}}}\endgroup }\lceil {}^{{{\color{\colorSYNTAX}\texttt{\ensuremath{\infty }}}}}}}}\endgroup }}}} 
   \\\noindent  {{\color{\colorMATH}\ensuremath{{\begingroup\renewcommand\colorMATH{\colorMATHP}\renewcommand\colorSYNTAX{\colorSYNTAXP}{{\color{\colorMATH}\ensuremath{\{ {\begingroup\renewcommand\colorMATH{\colorMATHM}\renewcommand\colorSYNTAX{\colorSYNTAXM}{{\color{\colorMATH}\ensuremath{x_{i}}}}\endgroup }{\mathrel{:}}_{{\begingroup\renewcommand\colorMATH{\colorMATHM}\renewcommand\colorSYNTAX{\colorSYNTAXM}{{\color{\colorMATH}\ensuremath{\rho _{i}^{\prime 4}}}}\endgroup }} {\begingroup\renewcommand\colorMATH{\colorMATHM}\renewcommand\colorSYNTAX{\colorSYNTAXM}{{\color{\colorMATH}\ensuremath{\tau }}}\endgroup }\} }}}\endgroup } \in  {\begingroup\renewcommand\colorMATH{\colorMATHP}\renewcommand\colorSYNTAX{\colorSYNTAXP}{{\color{\colorMATH}\ensuremath{\mathrlap{\hspace{-0.5pt}{}\rceil }\lfloor {\begingroup\renewcommand\colorMATH{\colorMATHS}\renewcommand\colorSYNTAX{\colorSYNTAXS}{{\color{\colorMATH}\ensuremath{\Gamma _{4}}}}\endgroup }\mathrlap{\hspace{-0.5pt}\rfloor }\lceil {}_{\{ {\begingroup\renewcommand\colorMATH{\colorMATHM}\renewcommand\colorSYNTAX{\colorSYNTAXM}{{\color{\colorMATH}\ensuremath{x_{1}}}}\endgroup },{.}\hspace{-1pt}{.}\hspace{-1pt}{.},{\begingroup\renewcommand\colorMATH{\colorMATHM}\renewcommand\colorSYNTAX{\colorSYNTAXM}{{\color{\colorMATH}\ensuremath{x_{n}}}}\endgroup }\} }^{{\begingroup\renewcommand\colorMATH{\colorMATHM}\renewcommand\colorSYNTAX{\colorSYNTAXM}{{\color{\colorMATH}\ensuremath{\rho }}}\endgroup }}}}}\endgroup }}}} 
   \\\noindent  Each of {{\color{\colorMATH}\ensuremath{\rho _{i}^{\prime}}}}, {{\color{\colorMATH}\ensuremath{\rho _{i}^{\prime \prime}}}} must be {{\color{\colorMATH}\ensuremath{0}}}.
   \\\noindent  \begin{itemize}[leftmargin=*,label=\textbf{*
      }]\item  \noindent  Subcase {{\color{\colorMATH}\ensuremath{\rho _{i}^{\prime \prime \prime} = \infty }}}:
        \\\noindent  {{\color{\colorMATH}\ensuremath{ \rho _{i} = \infty  }}}; the property holds trivially
        
      \item  \noindent  Subcase {{\color{\colorMATH}\ensuremath{\rho _{i}^{\prime \prime \prime} = 0}}}:
         \\\noindent  {{\color{\colorMATH}\ensuremath{\rho _{i}}}} = {{\color{\colorMATH}\ensuremath{\rho _{i}^{\prime 4}}}} = {{\color{\colorMATH}\ensuremath{\rho }}}
         \\\noindent  By IH, {{\color{\colorMATH}\ensuremath{ |f_{3}(\gamma ) - f_{3}(\gamma [x_{i}\mapsto d])|_{\llbracket {\mathbb{R}}\rrbracket } \leq  r }}}
         \\\noindent  The property follows from Theorem~\ref{thm:zcdp_gauss}.
         
      \end{itemize}
   
\end{itemize}

\end{proof}

\section{Truncated Concentrated Differential Privacy}

\begin{theorem}[Sinh-normal mechanism (Truncated Concentrated Differential Privacy)]
  \label{thm:tcdp_gauss}
  
  If {{\color{\colorMATH}\ensuremath{ | f(\gamma ) - f(\gamma [x_{i}\mapsto d]) | \leq  r }}}, {{\color{\colorMATH}\ensuremath{\rho  > 0}}}, and {{\color{\colorMATH}\ensuremath{\omega  > 1/\sqrt \rho }}}, then for
  \vspace*{-0.25em}\begingroup\color{\colorMATH}\begin{gather*}f^{\prime} = \lambda  \gamma  .\hspace*{0.33em} f(\gamma ) + \omega  r\hspace*{0.33em}{{\color{\colorMATH}\ensuremath{\operatorname{arsinh}}}}(\frac{1}{\omega  r} {\mathcal{N}}(0, \sigma ^{2})) \vspace*{-1em}\end{gather*}\endgroup 
  and {{\color{\colorMATH}\ensuremath{ \sigma ^{2} = r^{2} / (2\rho ) }}}, we have:

  \vspace*{-0.25em}\begingroup\color{\colorMATH}\begin{gather*} \forall  \alpha  \in  (1, \omega ) .\hspace*{0.33em} D_{\alpha }(f^{\prime}(\gamma ) \|  f^{\prime}(\gamma [x_{i}\mapsto d^{\prime}])) \leq  \alpha  \rho 
  \vspace*{-1em}\end{gather*}\endgroup 
\end{theorem}

\begin{proof}
  The result follows directly from Bun et al.\cite{bun2018composable}, Proposition 4.
\end{proof}

\begin{theorem}[Adaptive sequential composition (Truncated Concentrated Differential Privacy)]
  \label{thm:tcdp_let}
  \noindent  If:
     \vspace*{-0.25em}\begingroup\color{\colorMATH}\begin{gather*} |\gamma [x_{i}] - d| \leq  1 \implies   
     \cr  \forall  \alpha  \in  (1, \omega _{i}) .\hspace*{0.33em} D_{\alpha }(f_{1}(\gamma ) \|  f_{1}(\gamma [x_{i}\mapsto d])) \leq  \alpha  \rho _{i}
     \vspace*{-1em}\end{gather*}\endgroup 
  \\\noindent  and:
     \vspace*{-0.25em}\begingroup\color{\colorMATH}\begin{gather*} |\gamma [x_{i}] - d| \leq  1 \implies   
     \cr  \forall  \alpha  \in  (1, \omega _{i}^{\prime}) .\hspace*{0.33em} D_{\alpha }(f_{2}(\gamma [x\mapsto d^{\prime}]) \|  f_{2}(\gamma [x\mapsto d^{\prime},x_{i}\mapsto d])) \leq  \alpha  \rho _{i}^{\prime}
     \vspace*{-1em}\end{gather*}\endgroup 
  \\\noindent  and:
     \vspace*{-0.25em}\begingroup\color{\colorMATH}\begin{gather*} {{\color{\colorMATH}\ensuremath{\operatorname{Pr}}}}[f_{3}(\gamma ) = d^{\prime \prime}] = {{\color{\colorMATH}\ensuremath{\operatorname{Pr}}}}[f_{1}(\gamma ) = d^{\prime}, f_{2}(\gamma [x\mapsto d^{\prime}]) = d^{\prime \prime}] \vspace*{-1em}\end{gather*}\endgroup 
  \\\noindent  then:
     \vspace*{-0.25em}\begingroup\color{\colorMATH}\begin{gather*} |\gamma [x_{i}] - d| \leq  1 \implies   
     \cr  \forall  \alpha  \in  (1, \omega _{i} \sqcap  \omega _{i}^{\prime}) .\hspace*{0.33em} D_{\alpha }(f_{3}(\gamma ) \|  f_{3}(\gamma [x_{i}\mapsto d])) \leq  \alpha (\rho _{i} + \rho _{i}^{\prime})
     \vspace*{-1em}\end{gather*}\endgroup 
  
\end{theorem}

\begin{proof}
  The result follows directly from Bun et al.~\cite{bun2018composable}, Lemma 2, setting {{\color{\colorMATH}\ensuremath{M = f_{1}}}} and {{\color{\colorMATH}\ensuremath{M^{\prime} = f_{2}}}}.
\end{proof}

\begin{figure*}
\begingroup\renewcommand\colorMATH{\colorMATHP}\renewcommand\colorSYNTAX{\colorSYNTAXP}
\vspace*{-0.25em}\begingroup\color{\colorMATH}\begin{gather*}\begin{tabularx}{\linewidth}{>{\centering\arraybackslash\(}X<{\)}}\hfill\hspace{0pt}\begingroup\color{\colorTEXT}\boxed{\begingroup\color{\colorMATH} {\begingroup\renewcommand\colorMATH{\colorMATHM}\renewcommand\colorSYNTAX{\colorSYNTAXM}{{\color{\colorMATH}\ensuremath{\Delta }}}\endgroup },\Gamma  \vdash  e \mathrel{:} \tau  \endgroup}\endgroup \end{tabularx}\vspace*{-1em}\end{gather*}\endgroup 
\begingroup\color{\colorMATH}\begin{mathpar}\inferrule*[flushleft,lab={{\color{\colorTEXT}\textsc{\scriptsize Static Loop}}}
  ]{ {\begingroup\renewcommand\colorMATH{\colorMATHS}\renewcommand\colorSYNTAX{\colorSYNTAXS}{{\color{\colorMATH}\ensuremath{{\begingroup\renewcommand\colorMATH{\colorMATHM}\renewcommand\colorSYNTAX{\colorSYNTAXM}{{\color{\colorMATH}\ensuremath{\Delta }}}\endgroup },\Gamma _{1} \vdash  e_{1} \mathrel{:} {\begingroup\renewcommand\colorMATH{\colorMATHM}\renewcommand\colorSYNTAX{\colorSYNTAXM}{{\color{\colorMATH}\ensuremath{{{\color{\colorSYNTAX}\texttt{\ensuremath{{\mathbb{R}}^{+}[{{\color{\colorMATH}\ensuremath{\eta _{n}}}}]}}}}}}}\endgroup }}}}\endgroup }
  \\ {\begingroup\renewcommand\colorMATH{\colorMATHS}\renewcommand\colorSYNTAX{\colorSYNTAXS}{{\color{\colorMATH}\ensuremath{{\begingroup\renewcommand\colorMATH{\colorMATHM}\renewcommand\colorSYNTAX{\colorSYNTAXM}{{\color{\colorMATH}\ensuremath{\Delta }}}\endgroup },\Gamma _{2} \vdash  e_{2} \mathrel{:} {\begingroup\renewcommand\colorMATH{\colorMATHM}\renewcommand\colorSYNTAX{\colorSYNTAXM}{{\color{\colorMATH}\ensuremath{\tau }}}\endgroup }}}}\endgroup }
  \\ {\begingroup\renewcommand\colorMATH{\colorMATHM}\renewcommand\colorSYNTAX{\colorSYNTAXM}{{\color{\colorMATH}\ensuremath{\Delta }}}\endgroup },\Gamma _{3} + \mathrlap{\hspace{-0.5pt}{}\rceil }\lfloor \Gamma _{4}\mathrlap{\hspace{-0.5pt}\rfloor }\lceil {}_{\{ {\begingroup\renewcommand\colorMATH{\colorMATHM}\renewcommand\colorSYNTAX{\colorSYNTAXM}{{\color{\colorMATH}\ensuremath{x_{1}^{\prime}}}}\endgroup },{.}\hspace{-1pt}{.}\hspace{-1pt}{.},{\begingroup\renewcommand\colorMATH{\colorMATHM}\renewcommand\colorSYNTAX{\colorSYNTAXM}{{\color{\colorMATH}\ensuremath{x_{n}^{\prime}}}}\endgroup }\} }^{{\begingroup\renewcommand\colorMATH{\colorMATHM}\renewcommand\colorSYNTAX{\colorSYNTAXM}{{\color{\colorMATH}\ensuremath{\eta _{\rho }}}}\endgroup },{\begingroup\renewcommand\colorMATH{\colorMATHM}\renewcommand\colorSYNTAX{\colorSYNTAXM}{{\color{\colorMATH}\ensuremath{\eta _{\omega }}}}\endgroup }} \uplus  \{ {\begingroup\renewcommand\colorMATH{\colorMATHM}\renewcommand\colorSYNTAX{\colorSYNTAXM}{{\color{\colorMATH}\ensuremath{x_{1}}}}\endgroup } {\mathrel{:}}_{\infty } {\begingroup\renewcommand\colorMATH{\colorMATHM}\renewcommand\colorSYNTAX{\colorSYNTAXM}{{\color{\colorMATH}\ensuremath{{{\color{\colorSYNTAX}\texttt{\ensuremath{{\mathbb{N}}}}}}}}}\endgroup },{\begingroup\renewcommand\colorMATH{\colorMATHM}\renewcommand\colorSYNTAX{\colorSYNTAXM}{{\color{\colorMATH}\ensuremath{x_{2}}}}\endgroup } {\mathrel{:}}_{\infty } {\begingroup\renewcommand\colorMATH{\colorMATHM}\renewcommand\colorSYNTAX{\colorSYNTAXM}{{\color{\colorMATH}\ensuremath{\tau }}}\endgroup }\}  \vdash  e_{3} \mathrel{:} {\begingroup\renewcommand\colorMATH{\colorMATHM}\renewcommand\colorSYNTAX{\colorSYNTAXM}{{\color{\colorMATH}\ensuremath{\tau }}}\endgroup }
     }{
     {\begingroup\renewcommand\colorMATH{\colorMATHM}\renewcommand\colorSYNTAX{\colorSYNTAXM}{{\color{\colorMATH}\ensuremath{\Delta }}}\endgroup },{}\rceil {\begingroup\renewcommand\colorMATH{\colorMATHS}\renewcommand\colorSYNTAX{\colorSYNTAXS}{{\color{\colorMATH}\ensuremath{\Gamma _{1} + \Gamma _{1}}}}\endgroup }\lceil {}^{{\begingroup\renewcommand\colorMATH{\colorMATHM}\renewcommand\colorSYNTAX{\colorSYNTAXM}{{\color{\colorMATH}\ensuremath{0}}}\endgroup },{\begingroup\renewcommand\colorMATH{\colorMATHM}\renewcommand\colorSYNTAX{\colorSYNTAXM}{{\color{\colorMATH}\ensuremath{0}}}\endgroup }} + {}\rceil {\begingroup\renewcommand\colorMATH{\colorMATHS}\renewcommand\colorSYNTAX{\colorSYNTAXS}{{\color{\colorMATH}\ensuremath{\Gamma _{2}}}}\endgroup }\lceil {}^{{{\color{\colorSYNTAX}\texttt{\ensuremath{\infty }}}}} + {}\rceil \Gamma _{3}\lceil {}^{{{\color{\colorSYNTAX}\texttt{\ensuremath{\infty }}}}} 
     + \mathrlap{\hspace{-0.5pt}{}\rceil }\lfloor \Gamma _{4}\mathrlap{\hspace{-0.5pt}\rfloor }\lceil {}_{\{ {\begingroup\renewcommand\colorMATH{\colorMATHM}\renewcommand\colorSYNTAX{\colorSYNTAXM}{{\color{\colorMATH}\ensuremath{x_{1}^{\prime}}}}\endgroup },{.}\hspace{-1pt}{.}\hspace{-1pt}{.},{\begingroup\renewcommand\colorMATH{\colorMATHM}\renewcommand\colorSYNTAX{\colorSYNTAXM}{{\color{\colorMATH}\ensuremath{x_{n}^{\prime}}}}\endgroup }\} }^{{\begingroup\renewcommand\colorMATH{\colorMATHM}\renewcommand\colorSYNTAX{\colorSYNTAXM}{{\color{\colorMATH}\ensuremath{\eta _{n}{\mathrel{\mathord{\cdotp }}}\eta _{\rho }}}}\endgroup },{\begingroup\renewcommand\colorMATH{\colorMATHM}\renewcommand\colorSYNTAX{\colorSYNTAXM}{{\color{\colorMATH}\ensuremath{\eta _{\omega }}}}\endgroup }}
     \vdash  {{\color{\colorSYNTAX}\texttt{loop}}}\hspace*{0.33em}{\begingroup\renewcommand\colorMATH{\colorMATHS}\renewcommand\colorSYNTAX{\colorSYNTAXS}{{\color{\colorMATH}\ensuremath{e_{1}}}}\endgroup }\hspace*{0.33em}{{\color{\colorSYNTAX}\texttt{on}}}\hspace*{0.33em}{\begingroup\renewcommand\colorMATH{\colorMATHS}\renewcommand\colorSYNTAX{\colorSYNTAXS}{{\color{\colorMATH}\ensuremath{e_{2}}}}\endgroup }\hspace*{0.33em}{<}{\begingroup\renewcommand\colorMATH{\colorMATHM}\renewcommand\colorSYNTAX{\colorSYNTAXM}{{\color{\colorMATH}\ensuremath{x_{1}^{\prime}}}}\endgroup },{.}\hspace{-1pt}{.}\hspace{-1pt}{.},{\begingroup\renewcommand\colorMATH{\colorMATHM}\renewcommand\colorSYNTAX{\colorSYNTAXM}{{\color{\colorMATH}\ensuremath{x_{n}^{\prime}}}}\endgroup }{>}\hspace*{0.33em}\{ {\begingroup\renewcommand\colorMATH{\colorMATHM}\renewcommand\colorSYNTAX{\colorSYNTAXM}{{\color{\colorMATH}\ensuremath{x_{1}}}}\endgroup },{\begingroup\renewcommand\colorMATH{\colorMATHM}\renewcommand\colorSYNTAX{\colorSYNTAXM}{{\color{\colorMATH}\ensuremath{x_{2}}}}\endgroup } \Rightarrow  e_{3}\}  \mathrel{:} {\begingroup\renewcommand\colorMATH{\colorMATHM}\renewcommand\colorSYNTAX{\colorSYNTAXM}{{\color{\colorMATH}\ensuremath{\tau }}}\endgroup }
  }
\and\inferrule*[lab={{\color{\colorTEXT}\textsc{\scriptsize SinhNormal}}}
  ]{ {\begingroup\renewcommand\colorMATH{\colorMATHS}\renewcommand\colorSYNTAX{\colorSYNTAXS}{{\color{\colorMATH}\ensuremath{\Gamma _{1} \vdash  e_{1} \mathrel{:} {\begingroup\renewcommand\colorMATH{\colorMATHM}\renewcommand\colorSYNTAX{\colorSYNTAXM}{{\color{\colorMATH}\ensuremath{{{\color{\colorSYNTAX}\texttt{\ensuremath{{\mathbb{R}}^{+}[{{\color{\colorMATH}\ensuremath{\eta _{s}}}}]}}}}}}}\endgroup }}}}\endgroup }
  \\ {\begingroup\renewcommand\colorMATH{\colorMATHS}\renewcommand\colorSYNTAX{\colorSYNTAXS}{{\color{\colorMATH}\ensuremath{\Gamma _{2} \vdash  e_{2} \mathrel{:} {\begingroup\renewcommand\colorMATH{\colorMATHM}\renewcommand\colorSYNTAX{\colorSYNTAXM}{{\color{\colorMATH}\ensuremath{{{\color{\colorSYNTAX}\texttt{\ensuremath{{\mathbb{R}}^{+}[{{\color{\colorMATH}\ensuremath{\eta _{\rho }}}}]}}}}}}}\endgroup }}}}\endgroup }
  \\ {\begingroup\renewcommand\colorMATH{\colorMATHS}\renewcommand\colorSYNTAX{\colorSYNTAXS}{{\color{\colorMATH}\ensuremath{\Gamma _{3} \vdash  e_{3} \mathrel{:} {\begingroup\renewcommand\colorMATH{\colorMATHM}\renewcommand\colorSYNTAX{\colorSYNTAXM}{{\color{\colorMATH}\ensuremath{{{\color{\colorSYNTAX}\texttt{\ensuremath{{\mathbb{R}}^{+}[{{\color{\colorMATH}\ensuremath{\eta _{\omega }}}}]}}}}}}}\endgroup }}}}\endgroup }
  \\ {\begingroup\renewcommand\colorMATH{\colorMATHS}\renewcommand\colorSYNTAX{\colorSYNTAXS}{{\color{\colorMATH}\ensuremath{\Gamma _{4} + \mathrlap{\hspace{-0.5pt}{}\rceil }\lfloor \Gamma _{5}\mathrlap{\hspace{-0.5pt}\rfloor }\lceil {}_{\{ {\begingroup\renewcommand\colorMATH{\colorMATHM}\renewcommand\colorSYNTAX{\colorSYNTAXM}{{\color{\colorMATH}\ensuremath{x_{1}}}}\endgroup },{.}\hspace{-1pt}{.}\hspace{-1pt}{.},{\begingroup\renewcommand\colorMATH{\colorMATHM}\renewcommand\colorSYNTAX{\colorSYNTAXM}{{\color{\colorMATH}\ensuremath{x_{n}}}}\endgroup }\} }^{{\begingroup\renewcommand\colorMATH{\colorMATHM}\renewcommand\colorSYNTAX{\colorSYNTAXM}{{\color{\colorMATH}\ensuremath{\eta _{s}}}}\endgroup }} \vdash  e_{4} \mathrel{:} {\begingroup\renewcommand\colorMATH{\colorMATHM}\renewcommand\colorSYNTAX{\colorSYNTAXM}{{\color{\colorMATH}\ensuremath{{{\color{\colorSYNTAX}\texttt{\ensuremath{{\mathbb{R}}}}}}}}}\endgroup }}}}\endgroup }
     }{
     {}\rceil {\begingroup\renewcommand\colorMATH{\colorMATHS}\renewcommand\colorSYNTAX{\colorSYNTAXS}{{\color{\colorMATH}\ensuremath{\Gamma _{1} + \Gamma _{2} + \Gamma _{3}}}}\endgroup }\lceil {}^{{\begingroup\renewcommand\colorMATH{\colorMATHM}\renewcommand\colorSYNTAX{\colorSYNTAXM}{{\color{\colorMATH}\ensuremath{0}}}\endgroup },{\begingroup\renewcommand\colorMATH{\colorMATHM}\renewcommand\colorSYNTAX{\colorSYNTAXM}{{\color{\colorMATH}\ensuremath{0}}}\endgroup }} + {}\rceil {\begingroup\renewcommand\colorMATH{\colorMATHS}\renewcommand\colorSYNTAX{\colorSYNTAXS}{{\color{\colorMATH}\ensuremath{\Gamma _{4}}}}\endgroup }\lceil {}^{{{\color{\colorSYNTAX}\texttt{\ensuremath{\infty }}}}} + \mathrlap{\hspace{-0.5pt}{}\rceil }\lfloor {\begingroup\renewcommand\colorMATH{\colorMATHS}\renewcommand\colorSYNTAX{\colorSYNTAXS}{{\color{\colorMATH}\ensuremath{\Gamma _{5}}}}\endgroup }\mathrlap{\hspace{-0.5pt}\rfloor }\lceil {}_{\{ {\begingroup\renewcommand\colorMATH{\colorMATHM}\renewcommand\colorSYNTAX{\colorSYNTAXM}{{\color{\colorMATH}\ensuremath{x_{1}}}}\endgroup },{.}\hspace{-1pt}{.}\hspace{-1pt}{.},{\begingroup\renewcommand\colorMATH{\colorMATHM}\renewcommand\colorSYNTAX{\colorSYNTAXM}{{\color{\colorMATH}\ensuremath{x_{n}}}}\endgroup }\} }^{{\begingroup\renewcommand\colorMATH{\colorMATHM}\renewcommand\colorSYNTAX{\colorSYNTAXM}{{\color{\colorMATH}\ensuremath{\eta _{\rho }}}}\endgroup },{\begingroup\renewcommand\colorMATH{\colorMATHM}\renewcommand\colorSYNTAX{\colorSYNTAXM}{{\color{\colorMATH}\ensuremath{\eta _{\omega }}}}\endgroup }} 
     \vdash  {{\color{\colorSYNTAX}\texttt{\ensuremath{{{\color{\colorSYNTAX}\texttt{sinh-normal}}}[{\begingroup\renewcommand\colorMATH{\colorMATHS}\renewcommand\colorSYNTAX{\colorSYNTAXS}{{\color{\colorMATH}\ensuremath{e_{1}}}}\endgroup },{\begingroup\renewcommand\colorMATH{\colorMATHS}\renewcommand\colorSYNTAX{\colorSYNTAXS}{{\color{\colorMATH}\ensuremath{e_{2}}}}\endgroup },{\begingroup\renewcommand\colorMATH{\colorMATHS}\renewcommand\colorSYNTAX{\colorSYNTAXS}{{\color{\colorMATH}\ensuremath{e_{3}}}}\endgroup }]\hspace*{0.33em}{<}{\begingroup\renewcommand\colorMATH{\colorMATHM}\renewcommand\colorSYNTAX{\colorSYNTAXM}{{\color{\colorMATH}\ensuremath{x_{1}}}}\endgroup },{.}\hspace{-1pt}{.}\hspace{-1pt}{.},{\begingroup\renewcommand\colorMATH{\colorMATHM}\renewcommand\colorSYNTAX{\colorSYNTAXM}{{\color{\colorMATH}\ensuremath{x_{n}}}}\endgroup }{>}\hspace*{0.33em}\{ {\begingroup\renewcommand\colorMATH{\colorMATHS}\renewcommand\colorSYNTAX{\colorSYNTAXS}{{\color{\colorMATH}\ensuremath{e_{4}}}}\endgroup }\} }}}} \mathrel{:} {\begingroup\renewcommand\colorMATH{\colorMATHM}\renewcommand\colorSYNTAX{\colorSYNTAXM}{{\color{\colorMATH}\ensuremath{{{\color{\colorSYNTAX}\texttt{\ensuremath{{\mathbb{R}}}}}}}}}\endgroup }
  }
\and\inferrule*[lab={{\color{\colorTEXT}\textsc{\scriptsize MSinhNormal}}}
  ]{ {\begingroup\renewcommand\colorMATH{\colorMATHS}\renewcommand\colorSYNTAX{\colorSYNTAXS}{{\color{\colorMATH}\ensuremath{\Gamma _{1} \vdash  e_{1} \mathrel{:} {\begingroup\renewcommand\colorMATH{\colorMATHM}\renewcommand\colorSYNTAX{\colorSYNTAXM}{{\color{\colorMATH}\ensuremath{{{\color{\colorSYNTAX}\texttt{\ensuremath{{\mathbb{R}}^{+}[{{\color{\colorMATH}\ensuremath{\eta _{s}}}}]}}}}}}}\endgroup }}}}\endgroup }
  \\ {\begingroup\renewcommand\colorMATH{\colorMATHS}\renewcommand\colorSYNTAX{\colorSYNTAXS}{{\color{\colorMATH}\ensuremath{\Gamma _{2} \vdash  e_{2} \mathrel{:} {\begingroup\renewcommand\colorMATH{\colorMATHM}\renewcommand\colorSYNTAX{\colorSYNTAXM}{{\color{\colorMATH}\ensuremath{{{\color{\colorSYNTAX}\texttt{\ensuremath{{\mathbb{R}}^{+}[{{\color{\colorMATH}\ensuremath{\eta _{\rho }}}}]}}}}}}}\endgroup }}}}\endgroup }
  \\ {\begingroup\renewcommand\colorMATH{\colorMATHS}\renewcommand\colorSYNTAX{\colorSYNTAXS}{{\color{\colorMATH}\ensuremath{\Gamma _{3} \vdash  e_{3} \mathrel{:} {\begingroup\renewcommand\colorMATH{\colorMATHM}\renewcommand\colorSYNTAX{\colorSYNTAXM}{{\color{\colorMATH}\ensuremath{{{\color{\colorSYNTAX}\texttt{\ensuremath{{\mathbb{R}}^{+}[{{\color{\colorMATH}\ensuremath{\eta _{\omega }}}}]}}}}}}}\endgroup }}}}\endgroup }
  \\ {\begingroup\renewcommand\colorMATH{\colorMATHS}\renewcommand\colorSYNTAX{\colorSYNTAXS}{{\color{\colorMATH}\ensuremath{\Gamma _{4} + \mathrlap{\hspace{-0.5pt}{}\rceil }\lfloor \Gamma _{5}\mathrlap{\hspace{-0.5pt}\rfloor }\lceil {}_{\{ {\begingroup\renewcommand\colorMATH{\colorMATHM}\renewcommand\colorSYNTAX{\colorSYNTAXM}{{\color{\colorMATH}\ensuremath{x_{1}}}}\endgroup },{.}\hspace{-1pt}{.}\hspace{-1pt}{.},{\begingroup\renewcommand\colorMATH{\colorMATHM}\renewcommand\colorSYNTAX{\colorSYNTAXM}{{\color{\colorMATH}\ensuremath{x_{n}}}}\endgroup }\} }^{{\begingroup\renewcommand\colorMATH{\colorMATHM}\renewcommand\colorSYNTAX{\colorSYNTAXM}{{\color{\colorMATH}\ensuremath{\eta _{s}}}}\endgroup }} \vdash  e_{4} \mathrel{:} {\begingroup\renewcommand\colorMATH{\colorMATHM}\renewcommand\colorSYNTAX{\colorSYNTAXM}{{\color{\colorMATH}\ensuremath{{{\color{\colorSYNTAX}\texttt{\ensuremath{{{\color{\colorSYNTAX}\texttt{matrix}}}_{L2}^{{{\color{\colorMATH}\ensuremath{\bigstar }}}}[{{\color{\colorMATH}\ensuremath{\eta _{m}}}},{{\color{\colorMATH}\ensuremath{\eta _{n}}}}]\hspace*{0.33em}{\mathbb{R}}}}}}}}}\endgroup }}}}\endgroup }
     }{
     {}\rceil {\begingroup\renewcommand\colorMATH{\colorMATHS}\renewcommand\colorSYNTAX{\colorSYNTAXS}{{\color{\colorMATH}\ensuremath{\Gamma _{1} + \Gamma _{2} + \Gamma _{3}}}}\endgroup }\lceil {}^{{\begingroup\renewcommand\colorMATH{\colorMATHM}\renewcommand\colorSYNTAX{\colorSYNTAXM}{{\color{\colorMATH}\ensuremath{0}}}\endgroup },{\begingroup\renewcommand\colorMATH{\colorMATHM}\renewcommand\colorSYNTAX{\colorSYNTAXM}{{\color{\colorMATH}\ensuremath{0}}}\endgroup }} + {}\rceil {\begingroup\renewcommand\colorMATH{\colorMATHS}\renewcommand\colorSYNTAX{\colorSYNTAXS}{{\color{\colorMATH}\ensuremath{\Gamma _{4}}}}\endgroup }\lceil {}^{{{\color{\colorSYNTAX}\texttt{\ensuremath{\infty }}}}} + \mathrlap{\hspace{-0.5pt}{}\rceil }\lfloor {\begingroup\renewcommand\colorMATH{\colorMATHS}\renewcommand\colorSYNTAX{\colorSYNTAXS}{{\color{\colorMATH}\ensuremath{\Gamma _{5}}}}\endgroup }\mathrlap{\hspace{-0.5pt}\rfloor }\lceil {}_{\{ {\begingroup\renewcommand\colorMATH{\colorMATHM}\renewcommand\colorSYNTAX{\colorSYNTAXM}{{\color{\colorMATH}\ensuremath{x_{1}}}}\endgroup },{.}\hspace{-1pt}{.}\hspace{-1pt}{.},{\begingroup\renewcommand\colorMATH{\colorMATHM}\renewcommand\colorSYNTAX{\colorSYNTAXM}{{\color{\colorMATH}\ensuremath{x_{n}}}}\endgroup }\} }^{{\begingroup\renewcommand\colorMATH{\colorMATHM}\renewcommand\colorSYNTAX{\colorSYNTAXM}{{\color{\colorMATH}\ensuremath{\eta _{\rho }}}}\endgroup },{\begingroup\renewcommand\colorMATH{\colorMATHM}\renewcommand\colorSYNTAX{\colorSYNTAXM}{{\color{\colorMATH}\ensuremath{\eta _{\omega }}}}\endgroup }} 
     \vdash  {{\color{\colorSYNTAX}\texttt{\ensuremath{{{\color{\colorSYNTAX}\texttt{msinh-normal}}}[{\begingroup\renewcommand\colorMATH{\colorMATHS}\renewcommand\colorSYNTAX{\colorSYNTAXS}{{\color{\colorMATH}\ensuremath{e_{1}}}}\endgroup },{\begingroup\renewcommand\colorMATH{\colorMATHS}\renewcommand\colorSYNTAX{\colorSYNTAXS}{{\color{\colorMATH}\ensuremath{e_{2}}}}\endgroup },{\begingroup\renewcommand\colorMATH{\colorMATHS}\renewcommand\colorSYNTAX{\colorSYNTAXS}{{\color{\colorMATH}\ensuremath{e_{3}}}}\endgroup }]\hspace*{0.33em}{<}{\begingroup\renewcommand\colorMATH{\colorMATHM}\renewcommand\colorSYNTAX{\colorSYNTAXM}{{\color{\colorMATH}\ensuremath{x_{1}}}}\endgroup },{.}\hspace{-1pt}{.}\hspace{-1pt}{.},{\begingroup\renewcommand\colorMATH{\colorMATHM}\renewcommand\colorSYNTAX{\colorSYNTAXM}{{\color{\colorMATH}\ensuremath{x_{n}}}}\endgroup }{>}\hspace*{0.33em}\{ {\begingroup\renewcommand\colorMATH{\colorMATHS}\renewcommand\colorSYNTAX{\colorSYNTAXS}{{\color{\colorMATH}\ensuremath{e_{4}}}}\endgroup }\} }}}} \mathrel{:} {\begingroup\renewcommand\colorMATH{\colorMATHM}\renewcommand\colorSYNTAX{\colorSYNTAXM}{{\color{\colorMATH}\ensuremath{{{\color{\colorSYNTAX}\texttt{\ensuremath{{{\color{\colorSYNTAX}\texttt{matrix}}}_{L\infty }^{U}[{{\color{\colorMATH}\ensuremath{\eta _{m}}}},{{\color{\colorMATH}\ensuremath{\eta _{n}}}}]\hspace*{0.33em}{\mathbb{R}}}}}}}}}\endgroup }
  }
\end{mathpar}\endgroup 
\endgroup 
\caption{Privacy Type System Modifications, Truncated Concentrated Differential Privacy}
\end{figure*}

\begin{theorem}[Soundness]
  There exists an interpretation of well typed terms {\begingroup\renewcommand\colorMATH{\colorMATHS}\renewcommand\colorSYNTAX{\colorSYNTAXS}{{\color{\colorMATH}\ensuremath{\Gamma  \vdash  e \mathrel{:} {\begingroup\renewcommand\colorMATH{\colorMATHM}\renewcommand\colorSYNTAX{\colorSYNTAXM}{{\color{\colorMATH}\ensuremath{\tau }}}\endgroup }}}}\endgroup } and {\begingroup\renewcommand\colorMATH{\colorMATHP}\renewcommand\colorSYNTAX{\colorSYNTAXP}{{\color{\colorMATH}\ensuremath{\Gamma 
  \vdash  e \mathrel{:} {\begingroup\renewcommand\colorMATH{\colorMATHM}\renewcommand\colorSYNTAX{\colorSYNTAXM}{{\color{\colorMATH}\ensuremath{\tau }}}\endgroup }}}}\endgroup }, notated {\begingroup\renewcommand\colorMATH{\colorMATHS}\renewcommand\colorSYNTAX{\colorSYNTAXS}{{\color{\colorMATH}\ensuremath{\llbracket e\rrbracket }}}\endgroup } and {\begingroup\renewcommand\colorMATH{\colorMATHP}\renewcommand\colorSYNTAX{\colorSYNTAXP}{{\color{\colorMATH}\ensuremath{\llbracket e\rrbracket }}}\endgroup }, such that {{\color{\colorMATH}\ensuremath{{\begingroup\renewcommand\colorMATH{\colorMATHP}\renewcommand\colorSYNTAX{\colorSYNTAXP}{{\color{\colorMATH}\ensuremath{\llbracket e\rrbracket }}}\endgroup } \in  {\begingroup\renewcommand\colorMATH{\colorMATHP}\renewcommand\colorSYNTAX{\colorSYNTAXP}{{\color{\colorMATH}\ensuremath{\llbracket \Gamma  \vdash 
  {\begingroup\renewcommand\colorMATH{\colorMATHM}\renewcommand\colorSYNTAX{\colorSYNTAXM}{{\color{\colorMATH}\ensuremath{\tau }}}\endgroup }\rrbracket }}}\endgroup }}}} and {{\color{\colorMATH}\ensuremath{{\begingroup\renewcommand\colorMATH{\colorMATHS}\renewcommand\colorSYNTAX{\colorSYNTAXS}{{\color{\colorMATH}\ensuremath{\llbracket e\rrbracket }}}\endgroup } \in  {\begingroup\renewcommand\colorMATH{\colorMATHS}\renewcommand\colorSYNTAX{\colorSYNTAXS}{{\color{\colorMATH}\ensuremath{\llbracket \Gamma  \vdash  {\begingroup\renewcommand\colorMATH{\colorMATHM}\renewcommand\colorSYNTAX{\colorSYNTAXM}{{\color{\colorMATH}\ensuremath{\tau }}}\endgroup }\rrbracket }}}\endgroup }}}}.
\end{theorem}
\begin{proof}
  We include the cases that are different from the proof for {{\color{\colorMATH}\ensuremath{(\epsilon , \delta )}}}-differential privacy: {\sc Bind} and {\sc SinhNormal}.

\vspace*{-0.25em}\begingroup\color{\colorMATH}\begin{gather*}\begin{tabularx}{\linewidth}{>{\centering\arraybackslash\(}X<{\)}}\hfill\hspace{0pt}\begingroup\color{\colorTEXT}\boxed{\begingroup\color{\colorMATH} {\begingroup\renewcommand\colorMATH{\colorMATHP}\renewcommand\colorSYNTAX{\colorSYNTAXP}{{\color{\colorMATH}\ensuremath{\llbracket \underline{\hspace{0.66em}\vspace*{5ex}}\rrbracket  \in  \{ e \in  {{\color{\colorMATH}\ensuremath{\operatorname{exp}}}} \mathrel{|} \Gamma  \vdash  e \mathrel{:} {\begingroup\renewcommand\colorMATH{\colorMATHM}\renewcommand\colorSYNTAX{\colorSYNTAXM}{{\color{\colorMATH}\ensuremath{\tau }}}\endgroup }\}  \rightarrow  \llbracket \Gamma  \vdash  {\begingroup\renewcommand\colorMATH{\colorMATHM}\renewcommand\colorSYNTAX{\colorSYNTAXM}{{\color{\colorMATH}\ensuremath{\tau }}}\endgroup }\rrbracket }}}\endgroup } \endgroup}\endgroup \end{tabularx}\vspace*{-1em}\end{gather*}\endgroup 
\begin{itemize}[leftmargin=*,label=\textbf{-
}]\item  \noindent  Case {\begingroup\renewcommand\colorMATH{\colorMATHP}\renewcommand\colorSYNTAX{\colorSYNTAXP}{{\color{\colorMATH}\ensuremath{\overbracketarg \Gamma {\Gamma _{1} + \Gamma _{2}} \vdash  {{\color{\colorSYNTAX}\texttt{\ensuremath{{\begingroup\renewcommand\colorMATH{\colorMATHM}\renewcommand\colorSYNTAX{\colorSYNTAXM}{{\color{\colorMATH}\ensuremath{x}}}\endgroup } \leftarrow  {{\color{\colorMATH}\ensuremath{e_{1}}}}\mathrel{;}{{\color{\colorMATH}\ensuremath{e_{2}}}}}}}} \mathrel{:} {\begingroup\renewcommand\colorMATH{\colorMATHM}\renewcommand\colorSYNTAX{\colorSYNTAXM}{{\color{\colorMATH}\ensuremath{\tau _{2}}}}\endgroup }}}}\endgroup }
   \\\noindent  By inversion:
      \vspace*{-0.25em}\begingroup\color{\colorMATH}\begin{gather*} 

      \vspace*{-1em}\end{gather*}\endgroup 
   \\\noindent  By Induction Hypothesis (IH):
      \vspace*{-0.25em}\begingroup\color{\colorMATH}\begin{gather*} %
      \vspace*{-1em}\end{gather*}\endgroup 
   \\\noindent  By definition of {\begingroup\renewcommand\colorMATH{\colorMATHP}\renewcommand\colorSYNTAX{\colorSYNTAXP}{{\color{\colorMATH}\ensuremath{+}}}\endgroup } for privacy contexts:
   \\\noindent  {{\color{\colorMATH}\ensuremath{\rho _{i},\omega _{i} =  \rho _{i}^{\prime} + \rho _{i}^{\prime \prime}, \omega _{i}^{\prime} \sqcap  \omega _{i}^{\prime \prime}}}} for
   \\\noindent  {{\color{\colorMATH}\ensuremath{{\begingroup\renewcommand\colorMATH{\colorMATHP}\renewcommand\colorSYNTAX{\colorSYNTAXP}{{\color{\colorMATH}\ensuremath{\{ {\begingroup\renewcommand\colorMATH{\colorMATHM}\renewcommand\colorSYNTAX{\colorSYNTAXM}{{\color{\colorMATH}\ensuremath{x_{i}}}}\endgroup }{\mathrel{:}}_{{\begingroup\renewcommand\colorMATH{\colorMATHM}\renewcommand\colorSYNTAX{\colorSYNTAXM}{{\color{\colorMATH}\ensuremath{\rho _{i}}}}\endgroup },{\begingroup\renewcommand\colorMATH{\colorMATHM}\renewcommand\colorSYNTAX{\colorSYNTAXM}{{\color{\colorMATH}\ensuremath{\omega _{i}^{\prime}}}}\endgroup }} {\begingroup\renewcommand\colorMATH{\colorMATHM}\renewcommand\colorSYNTAX{\colorSYNTAXM}{{\color{\colorMATH}\ensuremath{\tau }}}\endgroup }\} }}}\endgroup } \in  {\begingroup\renewcommand\colorMATH{\colorMATHP}\renewcommand\colorSYNTAX{\colorSYNTAXP}{{\color{\colorMATH}\ensuremath{\Gamma _{1}}}}\endgroup }}}} and {{\color{\colorMATH}\ensuremath{{\begingroup\renewcommand\colorMATH{\colorMATHP}\renewcommand\colorSYNTAX{\colorSYNTAXP}{{\color{\colorMATH}\ensuremath{\{ {\begingroup\renewcommand\colorMATH{\colorMATHM}\renewcommand\colorSYNTAX{\colorSYNTAXM}{{\color{\colorMATH}\ensuremath{x_{i}}}}\endgroup }{\mathrel{:}}_{{\begingroup\renewcommand\colorMATH{\colorMATHM}\renewcommand\colorSYNTAX{\colorSYNTAXM}{{\color{\colorMATH}\ensuremath{\rho _{i}}}}\endgroup },{\begingroup\renewcommand\colorMATH{\colorMATHM}\renewcommand\colorSYNTAX{\colorSYNTAXM}{{\color{\colorMATH}\ensuremath{\omega _{i}^{\prime \prime}}}}\endgroup }} {\begingroup\renewcommand\colorMATH{\colorMATHM}\renewcommand\colorSYNTAX{\colorSYNTAXM}{{\color{\colorMATH}\ensuremath{\tau }}}\endgroup }\} }}}\endgroup } \in  {\begingroup\renewcommand\colorMATH{\colorMATHP}\renewcommand\colorSYNTAX{\colorSYNTAXP}{{\color{\colorMATH}\ensuremath{\Gamma _{2}}}}\endgroup }}}}. 
   \\\noindent  Property holds via IH.1, IH.2 and theorem~\ref{thm:tcdp_let} instantiated with
      {{\color{\colorMATH}\ensuremath{(\rho _{i},\omega _{i}^{\prime})}}} and {{\color{\colorMATH}\ensuremath{(\rho _{i},\omega _{i}^{\prime \prime})}}}.
   
\item  \noindent  Case: 
   \\\noindent  {{\color{\colorMATH}\ensuremath{\hspace*{1.00em}

      \vspace*{-1em}\end{gather*}\endgroup 
   \\\noindent  By Induction Hypothesis (IH):
      \vspace*{-0.25em}\begingroup\color{\colorMATH}\begin{gather*} %
      \vspace*{-1em}\end{gather*}\endgroup 
   \\\noindent  By IH, we have that {{\color{\colorMATH}\ensuremath{\sigma ^{2} = \frac{r^{2}}{2\rho }}}}
   \\\noindent  By definition of {\begingroup\renewcommand\colorMATH{\colorMATHP}\renewcommand\colorSYNTAX{\colorSYNTAXP}{{\color{\colorMATH}\ensuremath{+}}}\endgroup } for privacy contexts:
   \\\noindent  {{\color{\colorMATH}\ensuremath{\rho _{i},\omega _{i} =  \rho _{i}^{\prime} + \rho _{i}^{\prime \prime} + \rho _{i}^{\prime \prime \prime} + \rho _{i}^{\prime 4} + \rho _{i}^{\prime 5},\omega _{i}^{\prime} + \omega _{i}^{\prime \prime} + \omega _{i}^{\prime \prime \prime} + \omega _{i}^{\prime 4} + \omega _{i}^{\prime 5}}}} for
   \\\noindent  {{\color{\colorMATH}\ensuremath{{\begingroup\renewcommand\colorMATH{\colorMATHP}\renewcommand\colorSYNTAX{\colorSYNTAXP}{{\color{\colorMATH}\ensuremath{\{ {\begingroup\renewcommand\colorMATH{\colorMATHM}\renewcommand\colorSYNTAX{\colorSYNTAXM}{{\color{\colorMATH}\ensuremath{x_{i}}}}\endgroup }{\mathrel{:}}_{{\begingroup\renewcommand\colorMATH{\colorMATHM}\renewcommand\colorSYNTAX{\colorSYNTAXM}{{\color{\colorMATH}\ensuremath{\rho _{i}}}}\endgroup },{\begingroup\renewcommand\colorMATH{\colorMATHM}\renewcommand\colorSYNTAX{\colorSYNTAXM}{{\color{\colorMATH}\ensuremath{\omega _{i}^{\prime}}}}\endgroup }} {\begingroup\renewcommand\colorMATH{\colorMATHM}\renewcommand\colorSYNTAX{\colorSYNTAXM}{{\color{\colorMATH}\ensuremath{\tau }}}\endgroup }\} }}}\endgroup } \in  {\begingroup\renewcommand\colorMATH{\colorMATHP}\renewcommand\colorSYNTAX{\colorSYNTAXP}{{\color{\colorMATH}\ensuremath{{}\rceil {\begingroup\renewcommand\colorMATH{\colorMATHS}\renewcommand\colorSYNTAX{\colorSYNTAXS}{{\color{\colorMATH}\ensuremath{\Gamma _{1}}}}\endgroup }\lceil {}^{{\begingroup\renewcommand\colorMATH{\colorMATHM}\renewcommand\colorSYNTAX{\colorSYNTAXM}{{\color{\colorMATH}\ensuremath{0,0}}}\endgroup }}}}}\endgroup }}}} 
   \\\noindent  {{\color{\colorMATH}\ensuremath{{\begingroup\renewcommand\colorMATH{\colorMATHP}\renewcommand\colorSYNTAX{\colorSYNTAXP}{{\color{\colorMATH}\ensuremath{\{ {\begingroup\renewcommand\colorMATH{\colorMATHM}\renewcommand\colorSYNTAX{\colorSYNTAXM}{{\color{\colorMATH}\ensuremath{x_{i}}}}\endgroup }{\mathrel{:}}_{{\begingroup\renewcommand\colorMATH{\colorMATHM}\renewcommand\colorSYNTAX{\colorSYNTAXM}{{\color{\colorMATH}\ensuremath{\rho _{i}}}}\endgroup },{\begingroup\renewcommand\colorMATH{\colorMATHM}\renewcommand\colorSYNTAX{\colorSYNTAXM}{{\color{\colorMATH}\ensuremath{\omega _{i}^{\prime \prime}}}}\endgroup }} {\begingroup\renewcommand\colorMATH{\colorMATHM}\renewcommand\colorSYNTAX{\colorSYNTAXM}{{\color{\colorMATH}\ensuremath{\tau }}}\endgroup }\} }}}\endgroup } \in  {\begingroup\renewcommand\colorMATH{\colorMATHP}\renewcommand\colorSYNTAX{\colorSYNTAXP}{{\color{\colorMATH}\ensuremath{{}\rceil {\begingroup\renewcommand\colorMATH{\colorMATHS}\renewcommand\colorSYNTAX{\colorSYNTAXS}{{\color{\colorMATH}\ensuremath{\Gamma _{2}}}}\endgroup }\lceil {}^{{\begingroup\renewcommand\colorMATH{\colorMATHM}\renewcommand\colorSYNTAX{\colorSYNTAXM}{{\color{\colorMATH}\ensuremath{0,0}}}\endgroup }}}}}\endgroup }}}} 
   \\\noindent  {{\color{\colorMATH}\ensuremath{{\begingroup\renewcommand\colorMATH{\colorMATHP}\renewcommand\colorSYNTAX{\colorSYNTAXP}{{\color{\colorMATH}\ensuremath{\{ {\begingroup\renewcommand\colorMATH{\colorMATHM}\renewcommand\colorSYNTAX{\colorSYNTAXM}{{\color{\colorMATH}\ensuremath{x_{i}}}}\endgroup }{\mathrel{:}}_{{\begingroup\renewcommand\colorMATH{\colorMATHM}\renewcommand\colorSYNTAX{\colorSYNTAXM}{{\color{\colorMATH}\ensuremath{\rho _{i}}}}\endgroup },{\begingroup\renewcommand\colorMATH{\colorMATHM}\renewcommand\colorSYNTAX{\colorSYNTAXM}{{\color{\colorMATH}\ensuremath{\omega _{i}^{\prime \prime \prime}}}}\endgroup }} {\begingroup\renewcommand\colorMATH{\colorMATHM}\renewcommand\colorSYNTAX{\colorSYNTAXM}{{\color{\colorMATH}\ensuremath{\tau }}}\endgroup }\} }}}\endgroup } \in  {\begingroup\renewcommand\colorMATH{\colorMATHP}\renewcommand\colorSYNTAX{\colorSYNTAXP}{{\color{\colorMATH}\ensuremath{{}\rceil {\begingroup\renewcommand\colorMATH{\colorMATHS}\renewcommand\colorSYNTAX{\colorSYNTAXS}{{\color{\colorMATH}\ensuremath{\Gamma _{3}}}}\endgroup }\lceil {}^{{\begingroup\renewcommand\colorMATH{\colorMATHM}\renewcommand\colorSYNTAX{\colorSYNTAXM}{{\color{\colorMATH}\ensuremath{0,0}}}\endgroup }}}}}\endgroup }}}} 
   \\\noindent  {{\color{\colorMATH}\ensuremath{{\begingroup\renewcommand\colorMATH{\colorMATHP}\renewcommand\colorSYNTAX{\colorSYNTAXP}{{\color{\colorMATH}\ensuremath{\{ {\begingroup\renewcommand\colorMATH{\colorMATHM}\renewcommand\colorSYNTAX{\colorSYNTAXM}{{\color{\colorMATH}\ensuremath{x_{i}}}}\endgroup }{\mathrel{:}}_{{\begingroup\renewcommand\colorMATH{\colorMATHM}\renewcommand\colorSYNTAX{\colorSYNTAXM}{{\color{\colorMATH}\ensuremath{\rho _{i}}}}\endgroup },{\begingroup\renewcommand\colorMATH{\colorMATHM}\renewcommand\colorSYNTAX{\colorSYNTAXM}{{\color{\colorMATH}\ensuremath{\omega _{i}^{\prime 4}}}}\endgroup }} {\begingroup\renewcommand\colorMATH{\colorMATHM}\renewcommand\colorSYNTAX{\colorSYNTAXM}{{\color{\colorMATH}\ensuremath{\tau }}}\endgroup }\} }}}\endgroup } \in  {\begingroup\renewcommand\colorMATH{\colorMATHP}\renewcommand\colorSYNTAX{\colorSYNTAXP}{{\color{\colorMATH}\ensuremath{{}\rceil {\begingroup\renewcommand\colorMATH{\colorMATHS}\renewcommand\colorSYNTAX{\colorSYNTAXS}{{\color{\colorMATH}\ensuremath{\Gamma _{4}}}}\endgroup }\lceil {}^{{{\color{\colorSYNTAX}\texttt{\ensuremath{\infty }}}}}}}}\endgroup }}}} 
   \\\noindent  {{\color{\colorMATH}\ensuremath{{\begingroup\renewcommand\colorMATH{\colorMATHP}\renewcommand\colorSYNTAX{\colorSYNTAXP}{{\color{\colorMATH}\ensuremath{\{ {\begingroup\renewcommand\colorMATH{\colorMATHM}\renewcommand\colorSYNTAX{\colorSYNTAXM}{{\color{\colorMATH}\ensuremath{x_{i}}}}\endgroup }{\mathrel{:}}_{{\begingroup\renewcommand\colorMATH{\colorMATHM}\renewcommand\colorSYNTAX{\colorSYNTAXM}{{\color{\colorMATH}\ensuremath{\rho _{i}}}}\endgroup },{\begingroup\renewcommand\colorMATH{\colorMATHM}\renewcommand\colorSYNTAX{\colorSYNTAXM}{{\color{\colorMATH}\ensuremath{\omega _{i}^{\prime 5}}}}\endgroup }} {\begingroup\renewcommand\colorMATH{\colorMATHM}\renewcommand\colorSYNTAX{\colorSYNTAXM}{{\color{\colorMATH}\ensuremath{\tau }}}\endgroup }\} }}}\endgroup } \in  {\begingroup\renewcommand\colorMATH{\colorMATHP}\renewcommand\colorSYNTAX{\colorSYNTAXP}{{\color{\colorMATH}\ensuremath{\mathrlap{\hspace{-0.5pt}{}\rceil }\lfloor {\begingroup\renewcommand\colorMATH{\colorMATHS}\renewcommand\colorSYNTAX{\colorSYNTAXS}{{\color{\colorMATH}\ensuremath{\Gamma _{5}}}}\endgroup }\mathrlap{\hspace{-0.5pt}\rfloor }\lceil {}_{\{ {\begingroup\renewcommand\colorMATH{\colorMATHM}\renewcommand\colorSYNTAX{\colorSYNTAXM}{{\color{\colorMATH}\ensuremath{x_{1}}}}\endgroup },{.}\hspace{-1pt}{.}\hspace{-1pt}{.},{\begingroup\renewcommand\colorMATH{\colorMATHM}\renewcommand\colorSYNTAX{\colorSYNTAXM}{{\color{\colorMATH}\ensuremath{x_{n}}}}\endgroup }\} }^{{\begingroup\renewcommand\colorMATH{\colorMATHM}\renewcommand\colorSYNTAX{\colorSYNTAXM}{{\color{\colorMATH}\ensuremath{\rho }}}\endgroup },{\begingroup\renewcommand\colorMATH{\colorMATHM}\renewcommand\colorSYNTAX{\colorSYNTAXM}{{\color{\colorMATH}\ensuremath{\omega }}}\endgroup }}}}}\endgroup }}}} 
   \\\noindent  Each of {{\color{\colorMATH}\ensuremath{\rho _{i}^{\prime}}}}, {{\color{\colorMATH}\ensuremath{\rho _{i}^{\prime \prime}}}}, {{\color{\colorMATH}\ensuremath{\rho _{i}^{\prime \prime \prime}}}}, {{\color{\colorMATH}\ensuremath{\omega _{i}^{\prime}}}}, {{\color{\colorMATH}\ensuremath{\omega _{i}^{\prime \prime}}}}, {{\color{\colorMATH}\ensuremath{\omega _{i}^{\prime \prime \prime}}}} must be {{\color{\colorMATH}\ensuremath{0}}}.
   \\\noindent  \begin{itemize}[leftmargin=*,label=\textbf{*
      }]\item  \noindent  Subcase {{\color{\colorMATH}\ensuremath{\rho _{i}^{\prime 4} = \infty }}}:
        \\\noindent  {{\color{\colorMATH}\ensuremath{ \rho _{i} = \infty  }}}; the property holds trivially
        
      \item  \noindent  Subcase {{\color{\colorMATH}\ensuremath{\rho ^{\prime 4} = 0}}}:
         \\\noindent  {{\color{\colorMATH}\ensuremath{\rho _{i},\omega _{i}}}} = {{\color{\colorMATH}\ensuremath{\rho _{i},\omega _{i}^{\prime 5}}}} = {{\color{\colorMATH}\ensuremath{\rho ,\omega }}}
         \\\noindent  By IH, {{\color{\colorMATH}\ensuremath{ |f_{4}(\gamma ) - f_{4}(\gamma [x_{i}\mapsto d])|_{\llbracket {\mathbb{R}}\rrbracket } \leq  r }}}
         \\\noindent  The property follows from Theorem~\ref{thm:tcdp_gauss}.
         
      \end{itemize}
   
\end{itemize}

\end{proof}



\end{document}